\documentstyle[epsfig,graphicx,amssymb]{elsart}

\newcommand{\Pomeron}{I\!\!P}
\newcommand{\Reggeon}{I\!\!R}

\begin{document}
\begin{frontmatter}

\vspace*{-0.5cm}
\hspace*{10cm}
JLAB-THY-11-1379

\vspace*{0.5cm}

\title{Leading twist nuclear shadowing phenomena in hard processes with nuclei}

\author{L. Frankfurt}
\address{Nuclear Physics Department, School of Physics and Astronomy, Tel Aviv
University, 69978 Tel Aviv, Israel}
\ead{frankfur@tauphy.tau.ac.il}
\author{V. Guzey} 
\address{Theory Center, Thomas Jefferson National Accelerator Facility, 
Newport News, VA 23606, USA}
\ead{vguzey@jlab.org}
\author{M. Strikman}
\address{Department of Physics, the Pennsylvania State University, State
  College,\\ PA 16802, USA}
\ead{strikman@phys.psu.edu}

\bigskip
\bigskip
\bigskip
\bigskip

\begin{abstract} 
\noindent

We present and discuss the theory and phenomenology of the leading twist theory of 
nuclear shadowing which is based on the combination of
the generalization of the Gribov-Glauber theory, QCD factorization theorems, 
and the HERA QCD analysis of diffraction in lepton-proton deep inelastic scattering (DIS).
We apply this technique for the analysis of a wide range of
hard processes with nuclei---inclusive DIS on deuterons, medium-range and heavy nuclei,
coherent and incoherent diffractive DIS with nuclei, and hard diffraction in 
proton-nucleus scattering---and make predictions for the effect of nuclear shadowing in the
corresponding sea quark and gluon parton distributions. 
 We also analyze the role of the leading twist 
nuclear shadowing in generalized parton distributions in nuclei and in certain 
characteristics of final states in nuclear DIS. 
We discuss the limits of applicability
of the leading twist approximation for small $x$ scattering off nuclei and the
onset of the black disk regime and methods of detecting it.
It will be possible to check many of our predictions in the near future 
in the studies of the ultraperipheral collisions at the Large Hadron Collider (LHC). 
Further checks will be possible in $pA$ collisions at the LHC and forward hadron 
production at the Relativistic Heavy Ion Collider (RHIC). 
Detailed tests will be possible at an 
Electron-Ion Collider (EIC) in the USA and at the Large Hadron-Electron Collider 
(LHeC) at CERN.
\end{abstract}

\end{frontmatter}

\tableofcontents 
\section{Introduction}
\label{sec:hab_intro}

The most straightforward 
way to analyze the microscopic structure of atomic nuclei is to study
distributions of quarks and gluons in nuclei
using high-energy hard processes with nuclear targets. 
The term
''hard'' means that such processes contain a large momentum transfer 
(scale), which allows one to resolve the parton content of the target.
The presence of the large scale allows one to use collinear factorization 
theorems which introduce various universal distributions of partons in the hadronic target 
and determine their QCD evolution when the hard scale is varied
(DGLAP evolution).
In another limit, when the hard scale $Q^2$ is fixed  and the energy tends to infinity,
the DGLAP evolution breaks down and ultimately the black disk regime sets in.

In this review, we discuss nuclear parton distribution functions (nPDFs),
nuclear diffractive parton distribution functions (nDPDFs), 
and nuclear generalized parton distributions (nGPDs)
at small values of Bjorken $x$.

Nuclear PDFs is one of the key elements in the evaluation of numerous high 
energy hard phenomena involving nuclei.
Their understanding is also
important for the searches of the new phases of the QCD matter. 
Hence, the aim of this paper is to review legitimate calculations of the nuclear shadowing phenomena, analyze the range of the applicability of the theory, and 
provide numerical results.
These results are topical and increasingly important because the phenomenon of 
nuclear shadowing is involved in
the interpretation of Relativistic Heavy Ion Collider (RHIC) data,
the evaluation of hard phenomena in proton-nucleus
and nucleus-nucleus collisions at the Large Hadron Collider (LHC),
the estimation of the onset of
the black disk limit in ultra high-energy proton-nucleus and real photon-nucleus 
interactions at the LHC and also in DIS of leptons off nuclei at a future 
Electron-Ion Collider (EIC).

So far most of the information on nPDFs has been obtained from the measurements 
of the ratio of the nuclear to nucleon structure functions, $F_{2A}(x,Q^2)/[AF_{2N}(x,Q^2)]$,
in 
inclusive deep inelastic scattering (DIS) of leptons on nuclear targets~\cite{Aubert:1983xm,Bodek:1983qn,Bodek:1983ec,Arnold:1983mw,Dasu:1988ru,Bari:1985ga,Benvenuti:1987az,Ashman:1988bf,Arneodo:1989sy,Amaudruz:1991cc,Amaudruz:1991dj,Amaudruz:1992wn,Adams:1992vm,Adams:1992nf,Adams:1995is,Gomez:1993ri,Amaudruz:1995tq,Arneodo:1995cs,Arneodo:1995cq,Arneodo:1996rv,Arneodo:1996ru,Ackerstaff:1999ac}.
Complimentary information on nPDFs has also been obtained in $J/\psi$ production in 
DIS~\cite{Amaudruz:1991sr}
and in proton-nucleus high-mass dimuon production (nuclear Drell-Yan)~\cite{Alde:1990im,Vasilev:1999fa}.
Also, nPDFs can be constrained by dijet production in DIS and 
photoproduction and by $W$ and $Z$ production in proton-nucleus and 
nucleus-nucleus scattering. 
 
Nuclear DPDFs can be studied in diffractive DIS of leptons on nuclear targets~\cite{Frankfurt:2003gx,Kowalski:2008sa}
and, to some extent, in hard proton-nucleus diffraction.
 
Nuclear GPDs enter the theoretical description of hard exclusive 
processes with nuclei such as deeply virtual Compton scattering (DVCS)~\cite{Ellinghaus:2002zw,Freund:2003wm,Freund:2003ix,Goeke:2008rn,Goeke:2008jz},
coherent exclusive production of vector mesons in lepton-nucleus 
DIS, coherent photoproduction of
heavy vector mesons ($J/\psi$ and $\Upsilon$) in ultraperipheral nucleus-nucleus collisions~\cite{Frankfurt:2003qy,Frankfurt:2003wv,Baltz:2007kq}.
All three types of nuclear distributions---nPDFs, nDPDFs and nGPDs---can
be studied at a future Electron-Ion Collider~\cite{Deshpande:2005wd,eic}.

The experimental data on lepton-nucleus scattering obtained 
in a series of fixed target experiments have established 
that the cross section of scattering off nuclei is significantly smaller 
than the sum of free nucleon cross sections for $x\le 0.03$, 
see the review in~\cite{Arneodo:1992wf}. This phenomenon is called nuclear shadowing;
it is expected to be present for all parton flavors (sea and valence quarks and gluons),
i.e., the ratio of the nuclear PDF of flavor $j$,
$f_{j/A}(x,Q^2)$, to the corresponding 
PDF of a free nucleon, $f_{j/N}(x,Q^2)$, is suppressed for $x\le 0.03$,
 $f_{j/A}(x,Q^2)/[A f_{j/N}(x,Q^2)] <1$
($A$ is the number of nucleons).
As one increases $x$, the suppression is followed by an enhancement, 
$f_{j/A}(x,Q^2)/[A f_{j/N}(x,Q^2)] >1$ for $0.05 \leq x \leq 0.2$, which is called
antishadowing. 
The analyses of the data indicate that antishadowing takes place for valence quarks and
gluons and is absent for sea quarks.

Practically all presently available information on nPDFs in the shadowing region 
comes from DIS experiments.
The major obstacle that prevents the reliable determination of nPDFs
 at small $x$ from these experiments is the fact that these are 
fixed (stationary) target experiments. In the fixed target kinematics,
the values of $x$ and $Q^2$ are strongly correlated,
and, hence, one measures nPDFs essentially in a narrow band in the $x-Q^2$ plane 
rather than  on the entire plane.
Moreover, requiring that $Q^2$ is sufficiently large for the application of
perturbative QCD (factorization theorem), e.g.,~$Q^2 > 1$ GeV$^2$, 
the data cover the region $x > 5 \times 10^{-3}$ 
(for the NMC energies), 
where the effect of nuclear shadowing is rather small.

This makes the extraction of nuclear quark PDFs problematic for $x\le 0.01$ and impossible for $x\le 0.005$, where the maximal value of $Q^2$ is of the order of 1 GeV$^2$.
The situation is even worse for the gluon nPDF since 
it is extracted indirectly
using
the scaling violations of 
$F_{2A}(x,Q^2)$. 
As a result, the fits to the current data do not have predictive 
power for nPDFs for $ x < 0.01$.
Indeed, the  global fits to  the available data performed by various groups 
 by modeling nPDFs at some initial scale $Q_0^2$,
produce very  different 
results because nPDFs at small $x$ are not constrained by the available 
data~\cite{Eskola:1998iy,Eskola:1998df,Eskola:2002us,Paukkunen:2010qi,Hirai:2001np,Hirai:2004wq,Hirai:2007sx,deFlorian:2003qf,Li:2001xa,Eskola:2003cc,Eskola:2007my,Eskola:2008ca,Eskola:2009uj,Schienbein:2009kk}.
Moreover, the uncertainties of the resulting predictions for nPDFs are very large, especially in the 
gluon channel~\cite{Hirai:2004wq,Hirai:2007sx,Eskola:2009uj}.
It  would take a lepton-nucleus collider---a future EIC---to open up the kinematics and probe deep in the
shadowing region while keeping sufficiently large $Q^2$ to reliably determine
nPDFs at small $x$.

Significant nuclear shadowing  is also theoretically 
predicted for nuclear diffractive PDFs~\cite{Frankfurt:2003gx} and nuclear GPDs~\cite{Freund:2003wm,Freund:2003ix,Goeke:2008rn,Goeke:2008jz}. 
In addition, large shadowing is predicted for various cross sections in the formalisms which do not separate leading twist (LT) and non-LT contributions  
(see below).

Note that besides DIS off nuclear targets, the data on proton-nucleus high-mass dimuon production (nuclear Drell-Yan) is also used in the global QCD fits. 
These data mostly provide additional constraints on the sea quark nPDFs 
for $x \geq 0.03$ and $Q^2 \geq 16$ GeV$^2$.
In the gluon channel,
additional constraints on the nuclear gluon distribution can be inferred from
inclusive prompt photon~\cite{Arleo:2007js} and 
$J/\psi$~\cite{Arleo:2008zc} production in proton-nucleus scattering.
A recent analysis of nPDFs also included in the fits the RHIC data on inclusive  
high-$p_T$ hadron production in deuteron-gold  scattering at RHIC~\cite{Eskola:2008ca,Eskola:2009uj}. 
However, it is not clear how to separate various effects in nPDF in this case since the LT approximation is strongly violated in the considered RHIC kinematics
(see the discussion in Sec.~\ref{sec:bdr}).

The situation with the low-$x$ uncertainty in the extraction of nPDFs
will soon change with the start of
the studies of
ultra-peripheral nucleus-nucleus collisions
at the LHC, which will allow one to probe nPDFs
and nGPDs (especially in the gluon channel) down to $x=10^{-5}$ and for large 
virtualities. Also, single hadron production in proton-nucleus ($^{208}$Pb) 
scattering at the LHC will
provide stringent tests on nPDFs~\cite{QuirogaArias:2010wh} provided one would find
the kinematics where the factorization assumption underlying global nPDF fits holds and the nuclear shadowing effect is still significant.

Hence, at present, the only realistic way to determine nuclear PDFs at small $x$ 
is to  build the theory of nuclear shadowing based on the well understood properties 
of QCD; this is the aim of the present paper.  
We use the approach based on the leading twist approximation to 
the theory of nuclear shadowing
in which 
nuclear shadowing  is expressed 
in terms of elementary diffraction, i.e.,
nuclear shadowing in $eA$ scattering is expressed in terms of 
$ep$ diffraction. 
Our formalism---which we call the leading twist theory of nuclear shadowing---is 
based on the combination of the following ingredients:
\begin{itemize}
\item[(i)] The generalization 
of the technique developed by Gribov for the case of 
 nuclear shadowing in soft processes (pion-deuteron scattering)~\cite{Gribov:1968jf}
to hard processes, notably, 
to DIS with arbitrary nuclei~\cite{Frankfurt:1998ym},
\item[(ii)]  QCD factorization theorems (the leading twist approximation) for 
inclusive cross section of DIS~\cite{Brock:1993sz} and  diffraction in DIS~\cite{Collins:1997sr},
\item[(iii)]  
QCD analyses of the Hadron-Electron Ring Accelerator (HERA) 
data on  diffraction in 
DIS~\cite{Adloff:1997sc,Aktas:2006hy,Aktas:2006hx,Aktas:2007bv,Aktas:2007hn,Aktas:2006up,Breitweg:1997aa,Breitweg:1998gc,Chekanov:2003gt,Chekanov:2004hy,Chekanov:2007yw,Chekanov:2008cw,Chekanov:2008fh,Chekanov:2009qja}
which confirmed the validity of QCD factorization for diffraction and determined 
 diffractive PDFs of the nucleon.
\end{itemize}
Some of the ingredients listed above are formulated in the nucleus rest frame, 
others---in the nucleus fast frame.
It is the QCD factorization theorems which allow one to connect 
them in an unambiguous way.

Several theoretical phenomena complicate calculations of 
nuclear shadowing.\\ 
(i) The dominance of particle production in high energy processes invalidates the eikonal approximation, which is legitimate within the  framework of non-relativistic 
quantum mechanics.  A famous  example is the Glauber correction for hadron-deuteron 
cross section which disappears at high energies since the 
space-time evolution of high energy processes is different in a quantum field theory and quantum mechanics, see the discussion in Sec.~\ref{subsect:gribov_glauber}. 
V.~Gribov has demonstrated how to account for the rapid increase 
of the longitudinal distances with an increase of energy 
(Lorentz time dilation) and calculated the shadowing correction~\cite{Gribov:1968jf} 
that matches well the Glauber correction valid at lower energies.\\
(ii) The eikonal approximation strongly violates energy-momentum conservation at the energies where inelastic processes dominate, see
Sec.~\ref{subsec:cem}. 
An account of the diffractive processes within the method of color fluctuations allows one to overcome this problem by taking into account the splitting of the virtual photon energy between the interacting partons long before the collision. 
The resulting series in terms of the number of collisions with the target nucleons
has the same combinatoric structure as the Glauber series, 
where 
the factors $(\sigma_{\rm tot})^n$  are replaced by the interaction averaged over the color fluctuations, $\langle \sigma^n \rangle$.\\  
(iii) 
It is necessary to account properly for the $Q^2$ evolution of the nuclear PDFs
which mixes contributions of small and large $x$. This problem is naturally solved 
by using the QCD factorization theorems.

The leading twist theory of nuclear shadowing was proposed 
in~\cite{Frankfurt:1998ym} and 
later developed and
elaborated on in~\cite{Frankfurt:2000ty,Frankfurt:2002kd,Frankfurt:2003zd}.
The theory is consistent with the leading twist evolution equations to all orders in
the strong coupling constant $\alpha_s$ and 
predicts next-to-leading order (NLO) 
nPDFs and nDPDFs of different parton flavors
(quarks and gluons) as a function of Bjorken $x$ and the impact parameter $b$
at some initial scale $Q_0^2$. The $Q^2$ dependence of  nPDFs and nDPDFs is given by the 
Dokshitzer-Gribov-Lipatov-Altarelli-Parisi (DGLAP) evolution 
equations~\cite{dglap}.
The approach also allows us to predict nGPDs in a special limit.

We point out that
the crucial part of the leading twist theory of nuclear shadowing is
the use of the QCD factorization theorem for hard diffraction in 
DIS~\cite{Collins:1997sr}. 
Only this does allow us to make predictions for
nPDFs of different parton flavors, i.e., separately for quarks and gluons.
 This key feature distinguishes the 
leading twist theory of nuclear shadowing from all other theoretical 
approaches
to nuclear shadowing (see the discussion below).

Before the first HERA data on  diffraction in DIS on hydrogen was obtained and
analyzed in the middle of the 1990s, a large number of model calculations 
of nuclear shadowing in lepton-nucleus scattering
based on the Gribov work on nuclear shadowing~\cite{Gribov:1968jf} 
were performed. Initially, nuclear shadowing was estimated for the interaction
of real and quasi-real (small $Q^2$) photons with nuclei, where the 
generalized vector dominance
model gives a good description of diffraction, see e.g.,~\cite{Schildknecht:1973gi,Ditsas:1976yv,Bauer:1977iq}.
Later nuclear shadowing in deep inelastic lepton-nucleus 
scattering was evaluated using models of 
diffraction in the virtual photon-nucleon scattering, some of which in part
were based on the first HERA data on hard diffraction in DIS.
These models include 
the QCD-improved
aligned jet model~\cite{Frankfurt:1988nt,Frankfurt:1988zg},
the two-gluon exchange model which neglected the Sudakov suppression of the 
$q{\bar q}$ aligned jet contribution~\cite{Nikolaev:1990yw,Nikolaev:1990ja},
the two-component model based on the sum of the vector
meson (for low-mass diffraction) and continuum (for 
high-mass diffraction) contributions~\cite{Kwiecinski:1988ys,Badelek:1991qa,Badelek:1994qg,Piller:1995kh,Melnitchouk:1992eu,Melnitchouk:1993vc,Melnitchouk:1995am,Melnitchouk:2002ud,Kopeliovich:1995ju},
the generalized vector dominance model with off-diagonal transitions~\cite{Bilchak:1988zn,Shaw:1993gx},
the model based on the Pomeron and Reggeon contributions to the 
$\gamma^{\ast}p$ diffraction~\cite{Brodsky:1989qz}.

Among recent approaches to nuclear shadowing, one should mention the 
one~\cite{Capella:1997yv,Armesto:2003fi,Armesto:2010kr} based on the combination of
the Regge-motivated model for the diffractive structure 
function $F_2^{D(3)}$ measured by HERA in hard 
diffractive DIS with 
the Gribov connection between shadowing and diffraction.
 The approach gives
a good description of the data on nuclear
shadowing at intermediate values of $Q^2$.

Nuclear shadowing in quasi-real photon-nucleus scattering
is successfully described using 
the Gribov theory of nuclear shadowing combined with the phenomenological
fit to inclusive diffraction in photon-nucleon scattering~\cite{Piller:1997ny,Piller:1999wx,Adeluyi:2006xy}.  

In general, most of the models mentioned above use the connection between nuclear shadowing and diffraction,
but lack the consistency with the QCD evolution equations at large $Q^2$ and do not account for the difference between the gluon-induced and quark-induced diffraction.

A number of reviews on nuclear shadowing exists in the literature.
The 1988 review of Frankfurt and Strikman~\cite{Frankfurt:1988nt} discusses 
a wide range of hard 
processes with nuclei and corresponding phenomena, including nuclear shadowing. 
The review of Arneodo~\cite{Arneodo:1992wf} discusses the phenomenology of modifications
of nuclear structure functions known by 1994, with an emphasis on the EMC effect.
The review of Piller and Weise~\cite{Piller:1999wx} is a thorough account of
experimental and theoretical understanding of nuclear structure functions
achieved by 1999. The recent work of Armesto~\cite{Armesto:2006ph} reviews
several approaches to nuclear shadowing, including the
kinematics where the decomposition over twists is not valid.

The present review focuses on two limits which are comparatively well understood 
now. These are the limit of sufficiently large $Q^2$ and small but fixed $x$, 
where the leading twist (LT) 
approximation is justified, and the opposite regime---$x \to 0$ and $Q^2$ is large but fixed---where the interaction reaches 
the maximally possible strength allowed by the probability conservation
and the black disk regime (BDR) sets in.
In the LT approximation, the QCD factorization theorems allow us to separate 
the hard  interaction of partons with a given probe from the soft interactions 
with the target nucleus, which have the same structure as in soft dynamics.
Hence, the expected accuracy of the treatment of nuclear shadowing in DIS 
is as 
good as in the soft hadron-nucleus interactions, i.e., a few percent;
a comparable uncertainty comes from the uncertainty in the $ep$ diffractive data.
In the BDR limit, predictions for the total cross section are rather straightforward 
and do not depend on the details of the BDR dynamics. At the same time, 
the suppression of the leading particle production in the current fragmentation 
region provides an early signal for the onset of the BDR and is sensitive to the
details of the BDR dynamics, in particular, to the expectation of the emergence of
the three $x$-layers of different symmetries in the hadron wave function.

The gluon nuclear shadowing affects numerous observables in heavy ion collisions 
for all rapidities in the LHC kinematics and forward rapidities at RHIC. 
In the near future, the best tool to study the nuclear shadowing related phenomena will be ultraperipheral heavy ion collisions at the LHC, including forward jet production, hard diffraction and exclusive production of  $J/ \psi$ and $\Upsilon$. 
Feasibility of these studies,  in which $x$ down to $10^{-4 }$ could be covered, was demonstrated  recently in the review~\cite{Baltz:2007kq}, so there is no need to repeat it here.
Studies of $pA$ collisions at the LHC  will also be promising; 
a challenge in this case would be to disentangle the effects due to onset of the black disk regime  and the leading twist effects. In a longer run, an Electron-Ion 
Collider in USA, which can reach down $x \sim 10^{-3}$, would provide a perfect tool to study the kinematics of the onset of nuclear shadowing and to test a number of model-independent predictions made in this review. A plan of a Large Hadron-Electron Collider (LHeC) discussed now at CERN  
would be perfect for checking our predictions for smaller $x$,  where one explores both the kinematics, where leading twist effects dominate, and the kinematics, where
an  onset of the black disk regime is expected for rather large virtualities.

This review is structured as follows.
In Sec.~\ref{sec:gribov}, we present the Gribov picture of 
high-energy hadron-nucleus scattering and the results for nuclear
shadowing in hadron-deuteron scattering.
The generalization of the Gribov work to DIS off arbitrary nuclei and the derivation of
master equations for nuclear PDFs are presented in Sec.~\ref{sec:hab_pdfs}.
The section also contains a mini-review of hard inclusive diffraction in DIS at HERA.
Nuclear shadowing in DIS off the deuteron is presented in Sec.~\ref{sec:deuteron}.
Section~\ref{sec:phen} concerns with  numerous
applications of the theory of the leading 
twist nuclear shadowing. In this section, among several results,
we give our predictions for nPDFs and nuclear
structure functions, including their dependence on the impact parameter,
 discuss the accuracy of our predictions, compare our results to the 
eikonal approximation and to the selected available fixed-target data.
In Sec.~\ref{sec:final_states}, we present our predictions for nuclear shadowing in 
nuclear diffractive PDFs and structure functions, 
nuclear GPDs and exclusive diffraction at small $x$ as well as
  for nuclear effects in inclusive spectra in $eA$ collisions and in hadron production
at central rapidities. 
Leading twist nuclear shadowing and suppression of hard coherent 
diffraction in proton-nucleus scattering is discussed in Sec.~\ref{sec:pA}.
Finally, we present and discuss our results on the high-energy black disk regime in 
Sec.~\ref{sec:bdr}, which include the results for the total $\gamma^{\ast}A$ cross section and the post-selection effect of the suppression of the leading hadron production in $eA$ and $pA$ collisions.
We conclude our review with a short summary 
in Sec.~\ref{sec:conclusions}.

\section{Gribov results for nuclear shadowing}
\label{sec:gribov}

In this section, we describe the Gribov results for nuclear shadowing in 
hadron-nucleus scattering which are based on the space-time picture of the strong
interactions at high energies. Since this approach was formulated before the 
development of QCD as the theory of the strong interactions, it does not 
distinguish between soft and hard QCD processes.

\subsection{Gribov picture of space-time evolution of high-energy processes}
\label{subsect:gribov_picture}

The space-time picture of the strong interactions developed by Gribov is based on the
observation that the distances important in the strong interactions at high energies rapidly increase with energy~\cite{Feinberg56,Gribov:1965hf,Ioffe:1969kf}.

The following discussion is based on~\cite{Gribov:book,Gribov:1973jg}.
Let us consider, for example, pion-nucleus scattering in the laboratory reference 
frame. Since we would like to consider creation of particles, we should take into account 
that the pion has
a partonic structure (in the sense of the Feynman parton model~\cite{Feynman}).
When the momentum of the incoming pion is large, the pion fluctuates into its partons
and these fluctuations exist for a long time. The fluctuation time is called the 
coherence length, $l_c$. 

Let us consider for certainty the case when  the incoming pion with the large momentum $\vec{p}$
(it is convenient to take  $\vec{p}$ along the $z$-direction) and 
mass $\mu$ fluctuates into
two partons with momenta $\vec{p}_1$ and $\vec{p}-\vec{p}_1$ and mass $\mu_1$.
We also assume that the 
transverse component $\vec{p}_{1t}$ is energy-independent (as is the case of soft QCD) 
and introduce
$\alpha \equiv p_{1z}/|\vec{p}_1|$, 
$0 < \alpha  <1$.
The corresponding non-conservation of energy, $\Delta E$, is
\begin{eqnarray}
\Delta E&=&\sqrt{\mu_1^2+\vec{p}_1^{\,2}}+\sqrt{\mu_1^2+(\vec{p}-\vec{p}_1)^2}
-\sqrt{\mu^2+\vec{p}^{\,2}} \approx \frac{\mu_1^2+\vec{p}_{1t}^{\,2}}{2|p_{1z}|}+\frac{\mu_1^2+\vec{p}_{1t}^{\,2}}{2|p_z-p_{1z}|}- \frac{\mu^2}{2|p_z|}
\nonumber\\
&=&\frac{\mu_1^2 +\vec{p}_{1t}^{\,2}}{2|p_{z}|} \frac{1}{\alpha(1-\alpha)}-\frac{\mu^2}{2|p_z|}
\,.
\label{eq:lc}
\end{eqnarray}
Because of the uncertainty principle, 
the lifetime of the considered fluctuation is inversely 
proportional to $\Delta E$. Correspondingly, the length over which 
the considered partonic configuration remains coherent is  $l_c=1/\Delta E$.
It follows from Eq.~(\ref{eq:lc}) that 
\begin{equation}
l_c \equiv \frac{1}{\Delta E}=\frac{2 |\vec{p}|}{\frac{\mu_1^2 +\vec{p}_{1t}^2}{
\alpha(1-\alpha)}-\mu^2}
 \propto |\vec{p}| \,,
\label{eq:lc2}
\end{equation}
and, therefore, $l_c$ grows with energy and, 
at sufficiently large energies (large $|\vec{p}|$), 
$l_c$ exceeds by far the diameter of the target.
As a result, the pion interacts with all nucleons located at the same impact 
parameter in the same partonic configuration.

\begin{figure}[t]
\begin{center}
\epsfig{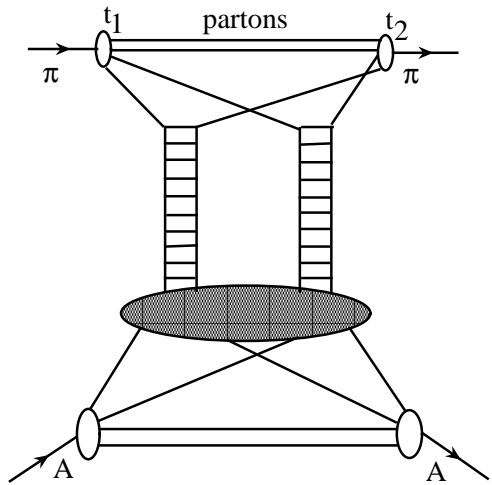}
\caption{The space-time picture of the interaction of a fast pion with two nucleons
of a nuclear target.}
\label{fig:pion-nucleus}
\end{center}
\end{figure}

The process of the parton branching described above continues 
and
leads to the following physical picture.
At high energies, the incoming pion with a large momentum can be 
described as a coherent ensemble of long-lived non-interacting partons and 
some small-momenta partons.
The
pion-nucleus interaction is described by the 
diagram presented in Fig.~\ref{fig:pion-nucleus}.
At a time $t_1$, the pion fluctuates into its partonic configurations.
The configurations strongly interact with the nuclear target denoted by $A$ 
and live for the time
$l_c=t_2-t_1$. At a time $t_2$, the final pion (or some other
final state) is formed.

In order to numerically estimate  the applicability of the space-time 
picture presented in Fig.~\ref{fig:pion-nucleus}, let us
 assume that the target nucleus is deuteron and the intermediate parton
configuration has the mass comparable to that of the $\rho$ meson~\cite{Gribov:1968jf}.
Then from the requirement that $l_c=R_d$, where $R_d \approx 4$ fm is the 
radius of the deuteron (the average distance between the proton and neutron), we obtain from Eq.~(\ref{eq:lc}):
\begin{equation}
\label{eq:cl3}
|\vec{p}| \geq\frac{R_d}{2} \left(m_{\rho}^2-\mu^2\right)=5.8 \quad {\rm GeV/c} \,.
\end{equation}
Therefore, in pion-deuteron scattering, the Gribov space-time picture starts to 
be applicable when the incoming momentum is
larger than  6 GeV/c.
 
\subsection{Nuclear shadowing in pion-deuteron scattering}
\label{subsect:pion_deuteron}

In this subsection, we present key steps in the derivation of the connection
between the nuclear shadowing correction to the total hadron-deuteron 
cross section and the hadron-nucleon diffractive cross section;
this connection
was first derived  
by Gribov in 1968~\cite{Gribov:1968jf}. 
Note that the Gribov derivation is justified by the small binding energy of the deuteron;
it has been realized long time ago that many characteristics of the deuteron 
can be understood within this approximation~\cite{Akhiezer}.
This example of soft QCD dynamics will be of use for us in the consideration of 
hard nuclear processes in Sec.~\ref{sec:hab_pdfs}.

\begin{figure}[h]
\begin{center}
\epsfig{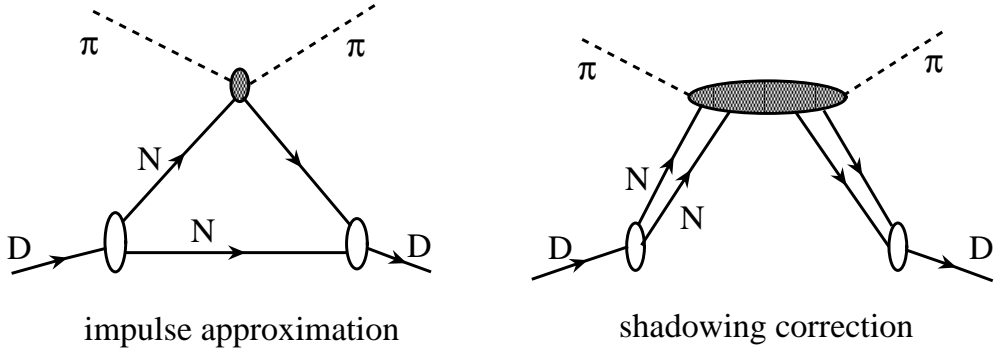}
\caption{Graphs for pion-deuteron scattering.}
\label{fig:piDFeynman}
\end{center}
\end{figure}

For certainty, let us consider 
scattering of a pion with high momentum $p$ on a deuteron at rest.
The corresponding scattering amplitude is given by the sum of the diagrams in 
Fig.~\ref{fig:piDFeynman}. The left graph corresponds to the interaction with one
nucleon of the target; this contribution is called the impulse approximation. 
The right graph corresponds 
to the simultaneous interaction with both nucleons of the target and leads to
a small negative contribution to the total pion-deuteron cross section, which is
called the nuclear shadowing correction. 

\vspace*{0.5cm}

Below we consider each graph in detail, assuming for simplicity 
that all involved particles and
the deuteron are spinless and the proton and the neutron are 
indistinguishable.

\begin{figure}[h]
\begin{center}
\epsfig{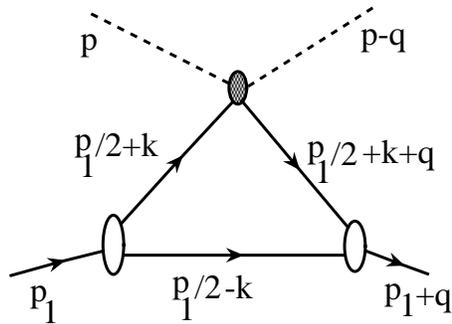}
\caption{The momentum flow in the left graph in Fig.~\ref{fig:piDFeynman}
and in Eq.~(\ref{eq:ia1}).}
\label{fig:piD1}
\end{center}
\end{figure}

The contribution of the impulse approximation to the pion-deuteron scattering
amplitude, $F_D^{\rm imp}(s,q)$, is
\begin{eqnarray}
F_D^{\rm imp}(s,q)&=&i \int \frac{d^4 k}{(2\pi)^4} \frac{1}{ 
[(\frac{p_1}{2}+k)^2-m^2+i \epsilon] [(\frac{p_1}{2}-k)^2-m^2+i \epsilon]
[(\frac{p_1}{2}+q+k)^2-m^2+i \epsilon]} \nonumber\\
&\times&
\Gamma\left(\left(\frac{p_1}{2}-k\right)^2,\left(\frac{p_1}{2}+k\right)^2\right) \Gamma\left(\left(\frac{p_1}{2}-k\right)^2,\left(\frac{p_1}{2}+q+k\right)^2\right) \nonumber\\
&\times&f_N \left(\left(p+\frac{p_1}{2}+k\right)^2,q^2,\left(\frac{p_1}{2}+k\right)^2,
\left(\frac{p_1}{2}+q+k\right)^2\right) \,,
\label{eq:ia1}
\end{eqnarray}
\noindent
where $\Gamma$ is the $D \to NN$ vertex;
$f_N$ is the pion-nucleon scattering amplitude; $m$ is the nucleon mass;
$q$ is the momentum transfer; $p_1$ is the momentum of the initial deuteron.
The momentum flow used in Eq.~(\ref{eq:ia1}) is depicted in Fig.~\ref{fig:piD1}.
\vspace*{0.5cm}

In the deuteron rest frame, the inverse nucleon propagators in Eq.~(\ref{eq:ia1}) are
\begin{samepage}
\begin{eqnarray}
\left(\frac{p_1}{2}+k\right)^2-m^2+i \epsilon&=&-\Delta^2+M k^0+(k^0)^2-\vec{k}^2+i\epsilon \,, \nonumber\\
\left(\frac{p_1}{2}-k\right)^2-m^2+i \epsilon&=&-\Delta^2-M k^0+(k^0)^2-\vec{k}^2+ i\epsilon\,, \nonumber\\
\left(\frac{p_1}{2}+q+k\right)^2-m^2+i \epsilon&=&-\Delta^2+M(q^0+k^0)-(\vec{q}+\vec{k})^2 
+i\epsilon \,,
\label{eq:ia2}
\end{eqnarray}
\end{samepage}
\noindent
where $\Delta^2=m^2-M^2/4$ and $M$ is the deuteron mass. 
The nucleon motion within the deuteron is effectively non-relativistic 
since $k^0\approx \Delta^2/(2m)$. 
Hence, the contribution of $k^0$ to Eq.~(\ref{eq:ia2})
 can be neglected. 
The small value of the deuteron binding energy, $\epsilon_D=2.2$ MeV, 
leads to the dominance of large internucleon distances in the impulse approximation 
[Eq.~(\ref{eq:ia1})] of the order of $1/ \sqrt{\epsilon_Dm}$, 
which are significantly larger than the radius of the strong interaction 
for  small momentum transfers, $\vec{q}^{\,2} \sim \Delta^2$. In this case,
all nucleons in the loop in Fig.~\ref{fig:piD1} are near the mass-shell,
\begin{eqnarray}
\left(\frac{p_1}{2}\pm k\right)^2=m^2 +{\cal O}(\Delta^2) \,, \nonumber\\
\left(\frac{p_1}{2}+q+k\right)^2=m^2 +{\cal O}(\Delta^2) \,.
\label{eq:ia3}
\end{eqnarray}
Therefore, the pion-nucleon scattering amplitude $f_N$ in Eq.~(\ref{eq:ia1}) 
can be replaced by the on-shell scattering amplitude evaluated at the invariant
mass squared $s_1$ and the momentum transfer $q$, where
\begin{equation}
s_1=\left(p+\frac{p_1}{2}+k\right)^2=\mu^2+p^0 M +m^2+{\cal O}(\Delta^2) \approx
p^0 M \approx \frac{s}{2} \,,
\label{eq:ia4}
\end{equation}
and $s=(k+p_1)^2$ is the pion-deuteron invariant mass squared. 
Since now $f_N$ does not depend
on the integration variables, it can be taken out the integral. The remaining
integral is proportional to the deuteron form factor of the baryon density current
(evaluated in the theory where the nucleon fields  are scalar), $\rho(q^2)$,
\begin{eqnarray}
\rho(q^2)&=&\frac{i}{2} \int \frac{d^4 k}{(2\pi)^4} \frac{1}{ 
[(\frac{p_1}{2}+k)^2-m^2+i \epsilon] [(\frac{p_1}{2}-k)^2-m^2+i \epsilon]
[(\frac{p_1}{2}+q+k)^2-m^2+i \epsilon]} \nonumber\\
&\times&
\Gamma\left(\left(\frac{p_1}{2}-k\right)^2,\left(\frac{p_1}{2}+k\right)^2\right) \Gamma\left(\left(\frac{p_1}{2}-k\right)^2,\left(\frac{p_1}{2}+q+k\right)^2\right)
\,.
\label{eq:ia5}
\end{eqnarray}
With this definition, $\rho(0)=1$.
Therefore, the contribution of the left graph in Fig.~\ref{fig:piDFeynman} is equal to
\begin{equation}
F_D^{\rm imp}(s,q)=2 f_N\left(\frac{s}{2},q^2\right) \,\rho(q^2) \,.
\label{eq:ia6}
\end{equation}
 Note that in the derivation, we neglected  for simplicity the difference between the pion-proton
and pion-neutron scattering amplitudes.

\begin{figure}[h]
\begin{center}
\epsfig{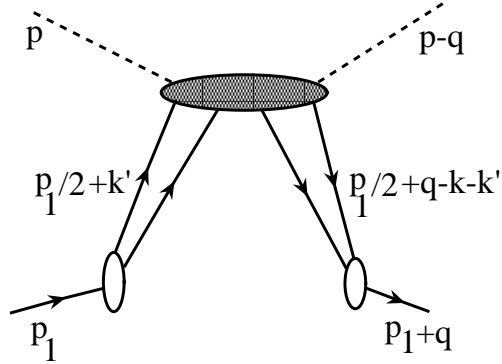}
\caption{The momentum flow in the right graph in Fig.~\ref{fig:piDFeynman}
and in Eq.~(\ref{eq:np1}).}
\label{fig:piD2}
\end{center}
\end{figure}

Turning to the evaluation of the right graph in Fig.~\ref{fig:piDFeynman}, we find that
its contribution reads:
\begin{eqnarray}
F_{D}^{\rm shad}(s,q)&=&- \int \frac{d^4 k}{(2\pi)^4} \frac{d^4 k^{\prime}}{(2\pi)^4}
\frac{1}{ 
[(\frac{p_1}{2}+k^{\prime})^2-m^2+i \epsilon] [(\frac{p_1}{2}-k^{\prime})^2-m^2+i \epsilon]
}\nonumber\\
&\times& \frac{1}{[(\frac{p_1}{2}+k+k^{\prime})^2-m^2+i \epsilon]
[(\frac{p_1}{2}+q-k-k^{\prime})^2-m^2+i \epsilon]} \nonumber\\
&\times& \Gamma\left(\left(\frac{p_1}{2}+k^{\prime} \right)^2,\left(\frac{p_1}{2}-k^{\prime}\right)^2\right) \Gamma\left(\left(\frac{p_1}{2}+k+k^{\prime}\right)^2,\left(\frac{p_1}{2}+q-k-k^{\prime}\right)^2\right) \nonumber\\
&\times& f \left(s_1, \vec{k}^2, (\vec{q}-\vec{k})^2, \vec{q}^{\,2}, s^{\prime} \right) \,,
\label{eq:np1}
\end{eqnarray}
where $s^{\prime}=(p-k)^2=\mu^2-2p^0 k^0+2 |\vec{p}| k_z-k^2$. 
The momentum flow used in Eq.~(\ref{eq:np1}) is depicted in Fig.~\ref{fig:piD2}.
\vspace*{0.5cm}

An analysis similar to the one presented above for  $F_{D}^{\rm imp}(s,q)$
leads to the conclusion that
the dominant contribution to
the loop integrals in Eq.~(\ref{eq:np1}) is given by the values
of
$k^0, \, k^{\prime\,0} \sim k^2/M$ and  $\vec{k}^2, \, \vec{k}^{\prime\,2} \gg\Delta^2$. 
The integration over $\vec{k}^2$ is cut off by the wave function of the deuteron.
As a result, the shadowing contribution is controlled by the 
small
internucleon distances 
$\ll r_D=1/\sqrt{\epsilon_Dm}$ [cf.~Eq.~(\ref{eq:g2})]. 
Consequently,  in the numerical calculations of nuclear shadowing, the nucleon
momenta up to  $\sim$ 400 MeV/c are important.
 In particular, the D-wave gives a large fraction of the shadowing correction in spite 
of the small probability of the D-wave in the deuteron, 
cf.~the discussion in Sec.~\ref{sec:deuteron}.

In the non-relativistic approximation, the vertex functions $\Gamma$ depend only on the absolute value 
of the relative three-momentum of the nucleons,
\begin{eqnarray}
\Gamma\left(\left(\frac{p_1}{2}+k^{\prime} \right)^2,\left(\frac{p_1}{2}-k^{\prime}\right)^2\right)&=&
\Gamma\left(\vec{k}^2 \right) \,, \nonumber\\
\Gamma\left(\left(\frac{p_1}{2}+k+k^{\prime}\right)^2,\left(\frac{p_1}{2}+q-k-k^{\prime}\right)^2\right)&=&\Gamma\left((\vec{k}+\vec{k^{\prime}}+\frac{\vec{q}}{2})^2 \right) \,.
\label{eq:np2}
\end{eqnarray}

Another quantity in Eq.~(\ref{eq:np1}),
the scattering  amplitude $f$, 
depends only on the five indicated variables,
which
 is a consequence of the assumption that $f$ depends only on the momentum transfer to
the nucleons. This approximation means that we neglected the effects of the motion of
the nucleons (Fermi motion), see e.g.,~\cite{Bertocchi:1972cj}.
In addition, in the non-relativistic approximation,  
the term proportional to $k^0$ in the expression for
$s^{\prime}$ can be neglected. 

Integration over $k^0$ and $k^{\prime \,0}$ in Eq.~(\ref{eq:np1}) gives
\begin{eqnarray}
F_{D}^{\rm shad}(s,q)&=&\frac{1}{(2 M)^2} \int \frac{d^3 \vec{k}}{(2\pi)^3} \frac{d^3 \vec{k^{\prime}}}{(2\pi)^3}
\frac{1}{ 
[\Delta^2+(\vec{k^{\prime}})^2] [\Delta^2+(\vec{k}+\vec{k^{\prime}}-\frac{\vec{q}}{2})^2]}
\nonumber\\
&\times&\Gamma\left(\vec{k}^2 \right)
\Gamma\left(\left(\vec{k}+\vec{k^{\prime}}+\frac{\vec{q}}{2}\right)^2 \right)
 f \left(s_1, \vec{k}^2, (\vec{q}-\vec{k})^2, \vec{q}^{\,2}, s^{\prime} \right) \,.
\label{eq:np3}
\end{eqnarray}
Equation~(\ref{eq:np3}) can be written in a compact form by introducing the deuteron form
factor~$\rho(q^2)$:
\begin{equation}
F_{D}^{\rm shad}(s,q)= \frac{2}{M} \int \frac{d^3 \vec{k}}{(2\pi)^3} \rho\left((2 \vec{k}+q)^2\right) f \left(s_1, \vec{k}^2, (\vec{q}-\vec{k})^2, \vec{q}^{\,2}, s^{\prime} \right) \,.
\label{eq:np4}
\end{equation}
Indeed, integrating over $k^0$ in Eq.~(\ref{eq:ia5}), one obtains
\begin{equation}
\rho(q^2)=\frac{1}{8 M} \int \frac{d^3 \vec{k^{\prime}}}{(2\pi)^3} \frac{1}{ 
[\Delta^2+(\vec{k^{\prime}})^2] [ \Delta^2+(\vec{k^{\prime}}+\frac{\vec{q}}{2})^2]}
\Gamma\left((\vec{k^{\prime}})^2\right) \Gamma\left((\vec{k^{\prime}}+\frac{\vec{q}}{2})^2
\right)
\,.
\label{eq:np5}
\end{equation}
A comparison of Eqs.~(\ref{eq:np5}) and (\ref{eq:np3}) leads to Eq.~(\ref{eq:np4}).

Let us now consider the $q=0$ forward scattering case. The shadowing correction becomes:
\begin{samepage}
\begin{eqnarray}
F_{D}^{\rm shad}(s)\equiv F_{D}^{\rm shad}(s,0)&=& \frac{2}{M} \int \frac{d^3 \vec{k}}{(2\pi)^3} \rho\left(4 \vec{k}^2\right) f \left(s_1,\vec{k}^2,s^{\prime} \right) \nonumber\\
&=& \frac{1}{M} \int \frac{d\vec{k}^2}{4\pi^2} \rho\left(4 \vec{k}^2\right)
\int^{2 |\vec{p}||\vec{k}|+\mu^2}_{-2 |\vec{p}||\vec{k}|+\mu^2}  \frac{d s^{\prime}}{2|\vec{p}|} f \left(s_1,\vec{k}^2,s^{\prime} \right) \nonumber\\
&=& \frac{2}{M} \int \frac{d\vec{k}^2}{4\pi^2} \rho\left(4 \vec{k}^2\right)
\int^{2 |\vec{p}||\vec{k}|+\mu^2}_{0}  \frac{d s^{\prime}}{2|\vec{p}|} f \left(s_1,\vec{k}^2,s^{\prime} \right)
\,.
\label{eq:np6}
\end{eqnarray}
\end{samepage}

\begin{figure}[t]
\begin{center}
\epsfig{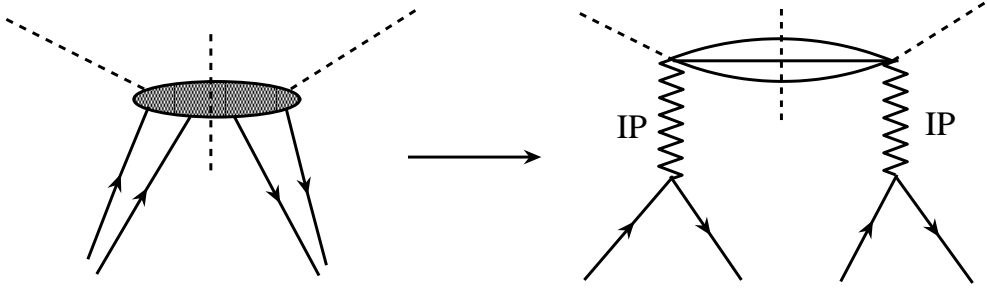}
\caption{Graphical representation of the imaginary part of the scattering amplitude $f$
in terms of Pomeron exchanges in the $t$-channel.}
\label{fig:Imaginary}
\end{center}
\end{figure}

The optical theorem relates the imaginary part of the scattering amplitude $f$ to the 
$\pi N \to X N$  cross section. 
Since at high energies inelastic processes are determined by the Pomeron exchange in 
the $t$-channel, $\Im m f$ is determined by the diagram presented in
Fig.~\ref{fig:Imaginary}. 
A direct evaluation gives
\begin{equation}
\Im m f(s_1,\vec{k}^2,s^{\prime})=- 4 p^0 m^2 (2 \pi)^3 \frac{d^3 \sigma^{\pi N}_{{\rm diff}}(\vec{k})}{d^3 \vec{k}} \,,
\label{eq:np7}
\end{equation}
where $\sigma^{\pi N}_{{\rm diff}}$ is the cross section of all diffractive processes
($\pi N \to X N$)
 with a small momentum transfer $\vec{k}$ to the 
nucleon. Note that $\Im m f <0$ since each 
of the Pomeron exchange amplitudes is purely imaginary.

Applying the optical theorem to the pion-deuteron scattering amplitude at $q=0$ and 
using Eqs.~(\ref{eq:ia6}) and (\ref{eq:np7}), 
we obtain the total pion-deuteron cross section,
\begin{equation}
\sigma_{{\rm tot}}^{\pi D}=2\sigma_{{\rm tot}}^{\pi N}-2\int d \vec{k}^2 
\rho\left(4 \vec{k}^2 \right) \frac{d\sigma^{\pi N}_{{\rm diff}}(\vec{k})}{d \vec{k}^2} \,.
\label{eq:f1}
\end{equation}
Equation~(\ref{eq:f1}) expresses the shadowing correction to the total hadron-deuteron
cross section in terms of the hadron-nucleon diffractive cross section.

As derived by Gribov, Eq.~(\ref{eq:f1}) assumes that the real 
part of the scattering amplitude $f$ is zero 
(this corresponds to the intercept of the Pomeron trajectory $\alpha_{\Pomeron}(0)=1$.)
 However, this assumption 
is not necessary and Eq.~(\ref{eq:f1}) can be straightforwardly generalized: 
\begin{equation}
\sigma_{{\rm tot}}^{\pi D}=2\sigma_{{\rm tot}}^{\pi N}-2\frac{1-\eta^2}{1+\eta^2}\int d \vec{k}^2 
\rho\left(4 \vec{k}^2 \right) \frac{d\sigma^{\pi N}_{{\rm diff}}(\vec{k})}{d \vec{k}^2} \,,
\label{eq:f1b}
\end{equation}
where $\eta$ is the ratio of the real to imaginary parts of the scattering
amplitude $f$.
The fast convergence of the integral over $d \vec{k}^2$ in Eq.~(\ref{eq:f1b})
allows us to neglect
a weak dependence of $\eta$ on  $k^2$.

\begin{figure}[h]
\begin{center}
\epsfig{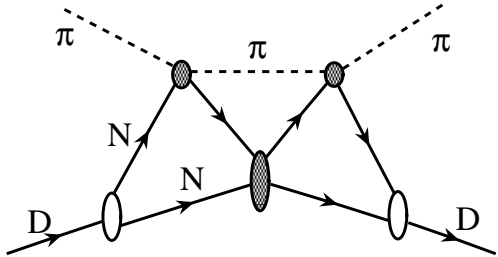}
\caption{An example of the contribution to the pion-deuteron cross section that vanishes at
large energies.}
\label{fig:Mandelstam}
\end{center}
\end{figure}

It should be noted that the graphs in Fig.~\ref{fig:piDFeynman}
give the complete answer for the pion-deuteron scattering amplitude at high
pion momenta. Other contributions, for instance, the diagram presented in Fig.~\ref{fig:Mandelstam},
vanish as $p \to \infty$~\cite{Mandelstam:1963cw}.
The physical reason for the negligibly small contribution of the diagram in
Fig.~\ref{fig:Mandelstam} is that during the short time required for the pion
to cover the distance between the two nucleons, the slow nucleons
in the deuteron cannot (do not have enough time to) interact.

It is possible to extend the Gribov analysis to include the relativistic motion of 
the nucleons using the light-cone formalism. One finds that the corrections due to the 
nucleon Fermi motion are very small due to the dominance of the $pn$ intermediate states
 in the deuteron wave function up to the internal momenta $\sim$ 500 MeV/c. 
Note here that a small value of the  admixture of non-nucleonic states in the nucleus wave function 
is confirmed by the smallness of the EMC effect due to hadronic effects up to $x\sim 0.55$, 
see the discussion in Sec.~\ref{subsec:EMC_effect}.

\subsection{Comparison of Gribov and Glauber results for nuclear shadowing}
\label{subsect:gribov_glauber}

\begin{figure}[h]
\begin{center}
\epsfig{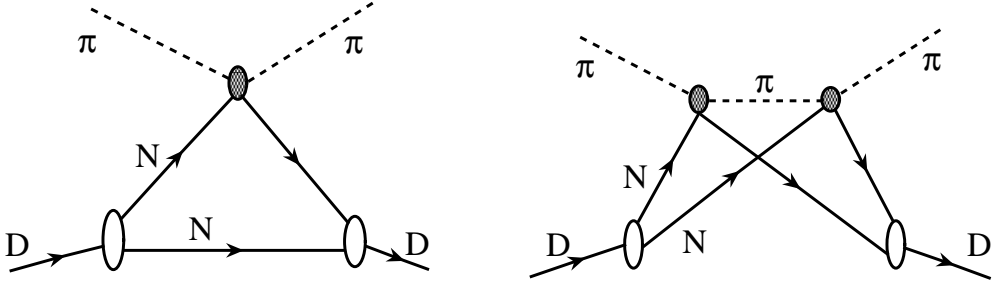}
\caption{Graphs for pion-deuteron scattering in the Glauber approach.}
\label{fig:Glauber}
\end{center}
\end{figure}

Originally the nuclear shadowing correction to the pion-deuteron cross section
was calculated by Glauber in 1955~\cite{Glauber:1955qq} 
for the energy range $E_{\pi}\sim $ 1 GeV, where the Lorentz dilation was not important. 
In the Glauber approach,
the pion-deuteron scattering amplitude receives contributions 
from the impulse approximation term and from the term corresponding to the subsequent 
interactions of the pion with the two nucleons
of the target; 
the both terms are presented in Fig.~\ref{fig:Glauber}.

The corresponding expression for the total pion-deuteron cross section reads~\cite{Glauber:1955qq}:
\begin{equation}
\sigma_{{\rm tot}}^{\pi D}=2\sigma_{{\rm tot}}^{\pi N}-\frac{\left(\sigma_{{\rm tot}}^{\pi N}\right)^2}{4 \pi} \left\langle \frac{1}{r^2}\right\rangle_D \,, 
\label{eq:g1}
\end{equation}
where $\langle 1/r^2 \rangle_D$ is the average inverse radius squared of the deuteron,
\begin{equation}
\left\langle \frac{1}{r^2}\right\rangle_D=\int d^3 \vec{r} \,|\psi_D(\vec{r})|^2 \frac{1}{\vec{r}^2} \,,
\label{eq:g2}
\end{equation}
with $\psi_D(\vec{r})$ the deuteron wave function.

The Gribov formula for the nuclear shadowing correction~(\ref{eq:f1}) is the generalization 
of that of Glauber~(\ref{eq:g1}) to high energies. Noticing that
in Eq.~(\ref{eq:f1}),
the $|\vec{k}|^2$ dependence of the deuteron form factor is much faster than that of
the diffractive cross section and assuming that only the elastic intermediate state
contributes, Eq.~(\ref{eq:f1}) can be written as
\begin{equation}
\sigma_{{\rm tot}}^{\pi D} \approx 2\sigma_{{\rm tot}}^{\pi N}-
\frac{d\sigma^{\pi N}_{{\rm el}}(\vec{k})}{d \vec{k}^2}\Bigg|_{|\vec{k}|^2=0}
\,2\int d \vec{k}^2 
\rho\left(4 \vec{k}^2 \right) \,.
\label{eq:f2}
\end{equation}
Using the $S$-matrix  unitarity condition,
\begin{equation}
\frac{d\sigma^{\pi N}_{{\rm el}}(\vec{k})}{d \vec{k}^2}\Bigg|_{|\vec{k}|^2=0}=
\frac{\left(\sigma_{{\rm tot}}^{\pi N}\right)^2}{16 \pi}  
 \,,
\label{eq:f3}
\end{equation}
and the expression for $\langle 1/r^2 \rangle_D$ in the momentum representation,
\begin{equation}
\int d \vec{k}^2  \rho\left(4 \vec{k}^2 \right)=
2 \left\langle \frac{1}{r^2} \right\rangle_D \,,
\label{eq:f4}
\end{equation}
one readily sees that the Gribov~(\ref{eq:f2}) and Glauber~(\ref{eq:g1}) formulas coincide, 
if  the intermediate state is purely elastic. 
However, when inelastic diffraction is important, the Gribov formula 
leads to larger shadowing.

Despite the similarity of the results obtained within the 
Gribov and Glauber approaches, the two approaches are based on very different pictures
of high-energy hadron-nucleus scattering. The Glauber approach neglects the
 Lorentz time dilation effects related to the hadron production.
Indeed,  the method is essentially quantum-mechanical and the
creation of particles in the intermediate states is not possible.  
 As a result, the incoming hadron is formed after each interaction 
and scatters {\it successively} on the target 
nucleons, see Fig.~\ref{fig:Glauber}.

More generally, in the $p \to \infty$ limit, the shadowing correction in the Glauber approach
(the right graph in Fig.~\ref{fig:Glauber}) vanishes.
This can be proven by exact calculations in any quantum field theory which 
accounts for particle production.
Using analytic properties of the scattering amplitude 
with respect to the mass squared of the produced state, one can demonstrate
the  exact cancellation of the diagrams with the eikonal 
topology~\cite{Mandelstam:1963cw,Gribov:1968fc} (the right graph in Fig.~\ref{fig:Glauber}
is an example of such diagrams).
The physical reason for this cancellation is that during the finite time 
it takes for the partonic fluctuation to traverse the nucleus, the fluctuation 
does not have enough time (which is of the order of $l_c\propto p$) 
to form back into the projectile.

In the Gribov approach, the 
projectile
 interacts with the target as a superposition of
 different configurations that
interact with different  
strengths, but which evolve very little during the passage through  
the nucleus. These configurations emerge behind the nucleus as a
  distorted---but still a coherent---superposition of configurations, 
which, when decomposed over the eigenstates of the strong  
Hamiltonian, contains both the original hadron (elastic scattering) 
as well as diffractively excited states (coherent diffraction).
 The Gribov approach is essentially field-theoretical and the creation of
particles in the intermediate state is properly taken into account, see 
Figs.~\ref{fig:piDFeynman}~and~\ref{fig:Imaginary}.
Hence, although the final answer
for nuclear shadowing in the Glauber and Gribov approaches
 is expressed through topologically different diagrams,
it has the structure of the sum of the eikonal term and the same-sign term
corresponding to the contribution of other diffractive states.

{\it Comment}. 
A simple picture of the scattering eigenstates by Feinberg and Pomeranchuk~\cite{Feinberg56} and
Good and Walker~\cite{Good:1960ba} 
provides  an $s$-channel model for 
the picture of high-energy scattering employed in the Gribov approach.
In particular,
 a projectile being in different eigenstates interacts with the two nucleons of the deuteron. 
The contribution of this interactions to the elastic scattering amplitude 
at $t=0$ is given by the overlapping integral between the final state and projectile wave functions.
When expressed through  the cross section of diffractive $hN$ scattering at $t=0$ with help 
of the Miettinen-Pumplin relation~\cite{Miettinen:1978jb}, 
one finds~\cite{Kopeliovich:1978qz} the same expression 
as found by Gribov, see Eq.~(\ref{eq:f1}). We will further discuss the Good--Walker picture later on.

It is worth noting that in the Gribov-Glauber approximation, the nucleus
is  treated as a dilute system. Namely, it is assumed that the characteristic impact
parameters for the projectile-nucleon interaction are much smaller than the typical
transverse  distance between the interacting nucleon and its neighbor. 
The corrections to this approximation
are difficult to estimate in a model-independent way, although they may become important at the LHC energies, where
the typical impact parameters 
in the $pp$ interaction are as large as 1.5 fm, which is close to the average
distance to the nearest neighbor. However, phenomenological analyses indicate that
the Gribov-Glauber approximation works well for fixed-target energies in 
nucleon-nucleus scattering at the beam energies $E_N\le$ 400 GeV, 
for a recent analysis, see Ref.~\cite{Alvioli:2009iw}.
Since in the energy range that we discuss in the present review 
the impact parameters in $\gamma^{\ast} p$
diffraction do not exceed those in $NN$ scattering at fixed-target  energies, we
will neglect  these effects in our analysis.

\subsection{The AGK cutting rules and nuclear shadowing}
\label{subsect:agk}

In the Gribov approach, the nuclear shadowing correction to the total 
pion-deuteron cross section is given by the diffractive cut 
of the  
graph, where the fast pion exchanges two Pomerons with the target,
see Fig.~\ref{fig:Imaginary}. The resulting shadowing correction is negative and 
given in terms of the pion-nucleon diffractive cross section.
These two features of the Gribov result can be understood using the 
Abramovsky-Gribov-Kancheli (AGK) cutting rules in the Reggeon field 
theory~\cite{Abramovsky:1973fm}.

Let us consider the part of the pion-deuteron scattering 
amplitude that gives rise to the shadowing correction by assuming that the high-energy pion interacts with the
target nucleons by the Pomeron exchanges. In the symbolic form 
(omitting the integration over the transverse momentum
of exchanged Pomerons in the loop which does not change the AGK rules), the amplitude reads:
\begin{equation}
F_{np} = -i N (iD_1) N (iD_2) \,,
\label{eq:agk1}
\end{equation}
where $D_{1,2}$ denote the complex Reggeon amplitudes; 
$N$ is the real-valued particle-Reggeon vertex function
which is an operator in the space of diffractively produced particles (see below).
The imaginary part of $F_{np}$ is then readily found:
\begin{equation}
\Im m F_{np} = -2 N^2 \left(\Im m D_1 \Im m D_2-\Re e D_1 \Re e D_2\right) \,,
\label{eq:agk2}
\end{equation}
where $N^2 =\sum_n \langle i|N|n \rangle \langle n|N|f \rangle d\tau_n$
(in this expression, $|n \rangle$ denotes the diffractively produced state; $d\tau_n$ is its
phase volume).
The additional factor of two 
originates from the fact that the deuteron consists of two nucleons.

\vspace*{0.5cm}
\begin{figure}[h]
\begin{center}
\epsfig{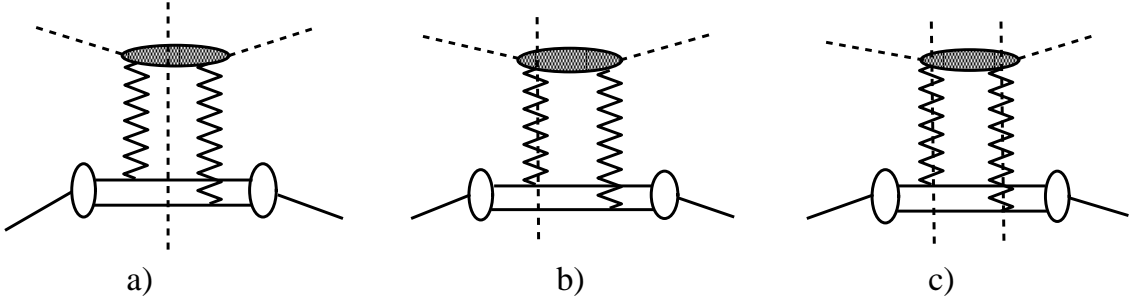}
\caption{The cuts of $F_{np}$ that contribute to $\Im m F_{np}$.}
\label{fig:AGK}
\end{center}
\end{figure}
Alternatively, the imaginary part of $F_{np}$ can be evaluated by summing
all possible cuts of the diagram corresponding to $F_{np}$, see Fig.~\ref{fig:AGK}.
Graph $a$ corresponds to the diffractive final state in the 
$\pi N \to X N$ reaction, when the pion  
diffractively dissociates into the hadronic states $X$. 
Hence, this cut is called diffractive. Graph $b$ corresponds to the
single multiplicity of the final state $Y$ in the $\pi D \to Y$ reaction; 
graph $c$ corresponds to the double multiplicity in the 
$\pi D \to Y$ reaction.

Denoting the results of the cutting of graphs $a$, $b$ and $c$ in Fig.~\ref{fig:AGK}
as $\Im m F_{np}^a$, $\Im m F_{np}^b$ and $\Im m F_{np}^c$, respectively,
a direct evaluation gives~\cite{Abramovsky:1973fm}:
\begin{eqnarray}
\Im m F_{np}^a &=& 2 N^2 \left(\Im m D_1 \Im m D_2+\Re e D_1 \Re e D_2\right)
=2 N^2 |D_1 D_2^{\ast}|
 \,, \nonumber \\
\Im m F_{np}^b & = & -8 N^2 \,\Im m D_1 \Im m D_2 \,, \nonumber \\
\Im m F_{np}^c & = & 4 N^2 \,\Im m D_1 \Im m D_2 \,. 
\label{eq:agk3_a}
\end{eqnarray}
The sum of these contributions leads to Eq.~(\ref{eq:agk2}). Indeed,
\begin{eqnarray}
\Im m F_{np} &\equiv & \Im m F_{np}^a+\Im m F_{np}^b+\Im m F_{np}^c =-2 N^2 \left(\Im m D_1 \Im m D_2-\Re e D_1 \Re e D_2\right) \nonumber\\
&=& -2 \frac{1-\eta^2}{1+\eta^2} N^2 |D_1 D_2^{\ast}| \,,
\label{eq:agk3}
\end{eqnarray}
where $\eta \equiv \Re e D_{1,2}/ \Im m D_{1,2}$.

Equation~(\ref{eq:agk3}) demonstrates that the shadowing correction, which is 
proportional to $\Im m F_{np}$, is negative and expressed in terms of
the pion-nucleon diffractive cross section (the latter is proportional to
$|D_1 D_2^{\ast}|$). The real part of the pion-nucleon 
scattering amplitude is accounted for by the factor $(1-\eta^2)/(1+\eta^2)$, see also 
Eq.~(\ref{eq:f1b}).

A different derivation of the AGK cutting rules 
for  hadron-nucleus scattering was proposed in~\cite{Bertocchi:1976bq}. 
Assuming validity of the eikonal (Glauber) approximation,
the following expression for the total inelastic hadron-nucleus cross section,
$\sigma_{\rm summed}^{hA, {\rm inel}}$,
was derived,
\begin{equation}
\sigma_{\rm summed}^{hA, {\rm inel}}=\sum_{n=1}^{A} \sigma_n
\label{eq:bertocchi_b}
\,,
\end{equation} 
where
$\sigma_n$ is the inelastic cross section that corresponds to inelastic
production on $n$ nucleons and no inelastic absorption on $A-n$ nucleons.
Further, one can show that 
in any multiple scattering
theory which is unitary,
\begin{equation}
\sum_{n=1}^A n \sigma_n=A \sigma_{{\rm inel}} \,,
\label{eq:bertocchi}
\end{equation}
where 
$\sigma_{{\rm inel}}$ is the 
hadron-nucleon inelastic cross section.
Equation~(\ref{eq:bertocchi}) gives an example of the so-called
AGK cancellation.

\section{Leading twist theory of nuclear shadowing for quark and gluon nuclear parton 
distribution functions}
\label{sec:hab_pdfs}

The derivation of the leading twist theory of nuclear
shadowing is based on combining the Gribov 
technique (discussed in the previous chapter), the factorization theorem for  diffraction in DIS~\cite{Collins:1997sr,Frankfurt:1997ij},
and the QCD analyses of the HERA data on  diffraction 
in lepton-nucleon DIS~\cite{Adloff:1997sc,Aktas:2006hy,Aktas:2006hx,Aktas:2007bv,Aktas:2007hn,Aktas:2006up,Breitweg:1997aa,Breitweg:1998gc,Chekanov:2003gt,Chekanov:2004hy,Chekanov:2007yw,Chekanov:2008cw,Chekanov:2008fh,Chekanov:2009qja}.   The name {\it leading twist nuclear shadowing} derives from the fact that the model-independent contribution to nuclear shadowing coming from the
interaction with two nucleons of the target is given in terms of the
diffractive structure functions which are leading twist quantities 
that were determined in a series of
the analyses of the HERA data on diffraction in DIS.  
In this section, we discuss each component of the leading twist theory
of nuclear shadowing separately and in detail and present the derivation of the 
master equation for nuclear shadowing in nuclear PDFs.

\subsection{Derivation of the master equation for nuclear parton distribution
functions}
\label{subsec:derivation}

In this subsection, we give
 the derivation of the equation that expresses nuclear shadowing in nuclear parton 
distribution functions (nPDFs) in the small $x$ shadowing region
in terms of the 
proton (nucleon) diffractive PDFs (DPDFs). The derived master equation
constitutes the key expression of the theory of leading twist 
nuclear shadowing.
The leading twist theory of nuclear shadowing was proposed 
and developed by Frankfurt and Strikman
in 1998~\cite{Frankfurt:1998ym} 
by  exploring the topology of the AGK cutting rules
and was further
elaborated later on in~\cite{Frankfurt:2000ty,Frankfurt:2002kd,Frankfurt:2003zd}.

\subsubsection{Contributions of the interaction with one and two nucleons of the target}

The starting point of the derivation is the generalization of the Gribov
theory of nuclear shadowing in hadron-deuteron scattering to the case
of inclusive deep inelastic scattering (DIS) of leptons off an arbitrary
nucleus with $A$ nucleons.
According to the space-time picture of the strong interactions 
presented in the Introduction, the virtual photon with the large momentum
$|\vec{q}|$ interacts with the target 
by fluctuating into strongly interacting states. The lifetime of such 
fluctuations, which is called the coherence length $l_c$, can be estimated 
in the limit of small $x$
using Eq.~(\ref{eq:lc}) by assuming that the invariant mass squared of the fluctuation,
$M_X^2$, approximately
equals the virtuality of the photon $Q^2$,
\begin{eqnarray}
\Delta E & = & \sqrt{\vec{q}^{\,2}+M_X^2}-\sqrt{\vec{q}^{\,2}-Q^2} \approx \frac{M_X^2+Q^2}{2 |\vec{q}|} \approx  \frac{Q^2}{|\vec{q}|}  \simeq 2 m_N x \,, 
\nonumber\\
l_c & \equiv & \frac {1}{\Delta E}=\frac{1}{2m_N x} \,,
\label{eq:m1}
\end{eqnarray}
where $m_N$ is the nucleon mass; $x$ is the usual Bjorken variable,
$x=Q^2/(2 m_N q_0)$ in the laboratory frame.
Note that in this derivation, we used the estimate of the aligned jet model 
that $\left<M_X^2\right>= Q^2$
and also that $|\vec{q}| \approx q_0$ at small $x$.
At very small $x$ corresponding to $l_c\gg 2 R_A$,
$\left<M_X^2\right> /Q^2$ gradually increases with a decrease of $x$, see 
also the
discussion in Sec.~\ref{sec:bdr}.

When $l_c$ is larger than the diameter of the nucleus, 
$2 R_A$, the virtual photon
coherently (``simultaneously'') interacts with all nucleons of the target
located at the same impact parameter.
For instance, for the nucleus of $^{40}$Ca, this happens for 
$x \leq 0.01$.  On the other hand, when $l_c$ decreases and becomes compatible to the average  distance between two nucleons in the nucleus, $r_{NN} \approx 1.7$ fm, all 
effects associated with large $l_c$ are expected to disappear.
Therefore, the nuclear effects of shadowing and antishadowing 
disappear for $x > 0.2$ (see also the discussion in Sec.~\ref{subsect:fast}
where this is discussed in the reference frame of the fast moving nucleus).

The wave function of the projectile virtual photon is characterized by the distribution
 over components (fluctuations) that widely differ in the strength of the interaction
 with the target: the fluctuations of a small transverse size correspond to
the small interaction strength and the large phase volume, while the fluctuations of a large 
transverse size correspond to the large interaction strength but the small phase volume.
A proper account of the  interplay between the phase volume of different configurations and their strength of interactions shows~\cite{Bj71} 
that these components  lead to the contributions characterized by the same power of $Q^2$: $\sigma_{\gamma^{\ast} T}\propto 1/Q^2$.~\footnote{This parton-model reasoning is modified
in QCD where the configurations with almost on-mass-shell quarks are suppressed at large $Q^2$ 
by the Sudakov form factor. An account of radiation ($Q^2$ evolution) leads to the appearance of
hard gluons (in addition to the near on-mass-shell quarks) in the wave function of the virtual photon. This property of QCD is important
for the theoretical analysis of hard diffractive processes considered in Sec.~\ref{sec:final_states}. 
} 
Hence, at moderately small $x$, nuclear shadowing is a predominantly non-perturbative QCD phenomenon complicated by the leading twist
$Q^2$ evolution.
 At extremely  small $x$, perturbative QCD (pQCD) interactions become strong which leads to a change 
of the dynamics of nuclear shadowing, see the discussion in Sec.~\ref{sec:bdr}.

At sufficiently high energies (small Bjorken $x$), when the virtual photon  interacts with many nucleons of the target, the lepton-nucleus scattering amplitude receives contributions from the graphs presented in Fig.~\ref{fig:Master1}. Considering the forward scattering and taking the imaginary part of the graphs 
in Fig.~\ref{fig:Master1} (presented by the vertical dashed lines), 
one obtains the graphical representation for the 
total virtual photon-nucleus cross section, $\sigma_{\gamma^{\ast}A}$. 
Note that there are other 
graphs, corresponding to the interaction with  four and more nucleons of the target, which are not shown in Fig.~\ref{fig:Master1};
the contribution of these graphs to $\sigma_{\gamma^{\ast}A}$ is insignificant. However, 
they appear to be
important in the case of the events with the multiplicity significantly larger than the average.
\vspace*{0.5cm}
\begin{figure}[h]
\begin{center}
\epsfig{file=Master1b_2011.epsi,scale=0.9}
\caption{
Graphs for to the total virtual photon-nucleus cross section, $\sigma_{\gamma^{\ast}A}$.
Graph $a$ gives the impulse approximation;  
graphs $b$ and $c$ give the shadowing correction arising from the interaction with
two and three nucleons of the target, respectively.
}
\label{fig:Master1}
\end{center}
\end{figure} 

Graph $a$ in Fig.~\ref{fig:Master1}, which is a generalization of
the left graph in Fig.~\ref{fig:piDFeynman} to the case of DIS,
corresponds to the interaction with one nucleon of the target
(the impulse approximation). The contribution of graph $a$ 
to $\sigma_{\gamma^{\ast}A}$, which we denote $\sigma_{\gamma^{\ast}A}^{(a)}$, is
\begin{equation}
\sigma_{\gamma^{\ast}A}^{(a)}=A \sigma_{\gamma^{\ast}N} \,,
\label{eq:m2_cs}
\end{equation}
where $\sigma_{\gamma^{\ast}N}$ is the total virtual photon-nucleon cross section.
The proton and neutron total cross sections (structure functions) 
are  very close at small $x$, and, therefore,
unless specified, we shall not distinguish between 
protons and neutrons.
Also, in Eq.~(\ref{eq:m2_cs}), we employed  the non-relativistic approximation for the nucleus wave function. A more accurate treatment would involve the light-cone many-nucleon  
approximation for the description of nuclei which leads to tiny  corrections to 
Eq.~(\ref{eq:m2_cs}) for small $x$ due to the Fermi motion effect, see
Sec.~\ref{subsect:fast}.
The good accuracy of this approximation has been tested by numerous studies of elastic and total
hadron-nucleus scattering cross sections at intermediate energies.

The total cross section in Eq.~(\ref{eq:m2_cs}) corresponds to the sum of the
cross sections with the transverse ($\sigma_{\gamma_T^{\ast}N}$) and 
longitudinal ($\sigma_{\gamma_L^{\ast}N}$) polarizations of the virtual photon. 
These cross sections can be expressed in terms of the isospin-averaged inclusive (unpolarized) structure function $F_{2N}(x,Q^2)$ and longitudinal
structure function $F_{L}(x,Q^2)$, see, e.g.~\cite{Piller:1999wx}:
\begin{eqnarray}
&&\sigma_{\gamma_T^{\ast}N}+\sigma_{\gamma_L^{\ast}N}=\sigma_{\gamma^{\ast}N}=
\frac{4 \pi^2 \alpha_{\rm em}}{Q^2 (1-x)} F_{2N}(x,Q^2) \,,
 \nonumber\\
&&\sigma_{\gamma_L^{\ast}N}=
\frac{4 \pi^2 \alpha_{\rm em}}{Q^2 (1-x)} F_{L}(x,Q^2) \,,
\label{eq:sigma_TL}
\end{eqnarray}
where $\alpha_{\rm em}$ is the fine-structure constant. The structure functions
$F_{2N}(x,Q^2)$ and $F_{L}(x,Q^2)$ parameterize the unpolarized lepton-hadron cross section, 
$d^2 \sigma/(dx dQ^2)$, using the standard expression~\cite{Brock:1993sz}:
\begin{equation}
\frac{d^2 \sigma}{dx dQ^2}=\frac{2 \pi \alpha^2_{\rm em}}{xQ^4}\left[\left(1+(1-y)^2\right) F_2(x,Q^2)-y^2F_L(x,Q^2) \right] \,,
\label{eq:definition_F2}
\end{equation}
where $y=(p \cdot q)/(p \cdot k)$, $p$ is the momentum of the hadron, $k$ is the momentum
of the initial lepton, and $q$ is the momentum of the exchanged virtual photon.
Note that the contribution of 
$F_L(x,Q^2)$ to the cross section
is generally significantly smaller than that of $F_2(x,Q^2)$.

Using Eqs.~(\ref{eq:m2_cs}) and (\ref{eq:sigma_TL}), we obtain the following
connection between the structure functions:
\begin{equation}
F_{2A}^{(a)}(x,Q^2)=A F_{2N}(x,Q^2) \,,
\label{eq:m2}
\end{equation}
where $F_{2A}^{(a)}(x,Q^2)$ is the contribution of graph $a$ in Fig.~\ref{fig:Master1}
(impulse approximation) to the nuclear structure function $F_{2A}(x,Q^2)$.
A similar relation is also valid for the longitudinal structure functions.

Graph $b$ in Fig.~\ref{fig:Master1} is the generalization
of the right graph in Fig.~\ref{fig:piDFeynman}
(see also Fig.~\ref{fig:Imaginary})   to the case of 
DIS off an arbitrary nucleus. The intermediate state $X$ denotes 
the diffractive final state of the $\gamma^{\ast}N \to X N$ reaction.
Therefore, 
the answer for  the contribution of graph $b$ to 
$\sigma_{\gamma^{\ast}A}$, which we shall call $\sigma_{\gamma^{\ast}A}^{(b)}$,
should have the
structure of the shadowing term in the Gribov formula~(\ref{eq:f1}). 
The generalization to the case of the photon interaction with more than two nucleons
will be considered 
in Sec.~\ref{subsubsec:cf}.

The calculation of graph $b$ in Fig.~\ref{fig:Master1} is significantly simplified if 
one observes that the nuclear part of this interaction (graph) 
weakly depends on energy and has the same structure 
as in the Glauber multiple scattering formalism~\cite{Glauber:1955qq}.  
Indeed, restricting the projectile-nucleon intermediate states by the
diffractive ones 
and taking into account that at high energies 
the interaction with the target nucleons depends only on the transverse part of 
the momentum transfer, 
the expression for graph $b$ in Fig.~\ref{fig:Master1} should have the same
form as that in the multiple scattering formalism, 
except for the additional
 effect of the longitudinal momentum transfer to the nucleons which cuts off 
the contribution of large-mass intermediate states 
[the last factor in Eq.~(\ref{eq:m3}) below].
The effects of Fermi motion and the
dependence of the amplitudes of diffractive processes on energy
can be easily taken into account within light-cone quantum mechanics of 
nuclei; these effects lead to small corrections~\cite{Frankfurt:1981mk}. 

However, 
in contrast to the Glauber multiple scattering formalism,
 QCD predicts the existence of the contribution of the diffractively produced inelastic states 
relevant for the  color coherent phenomena and triple Pomeron diagrams. 
Besides one needs to implement another QCD phenomenon, namely, energy-momentum conservation, 
which is impossible to enforce within the eikonal approximation,  see the discussion 
in Sec.~\ref{subsec:eikonal}.   
At sufficiently large energies, where the pQCD interaction becomes strong (possibly, at the LHC), 
the nuclear part of graph $b$ in Fig.~\ref{fig:Master1} will loose its
universality and one would need to explore the approximation of the black disk limit 
in order 
to 
do the calculations in a model-independent way.   

In summary, for the calculation of graph $b$ in  Fig.~\ref{fig:Master1}, 
at moderately small $x$,
one can use the Glauber multiple scattering formalism generalized to include inelastic diffractive intermediate states, i.e., coherent phenomena
and energy-momentum conservation.
 Expressing the scattering amplitude corresponding to graph $b$ 
in Fig.~\ref{fig:Master1} in the momentum representation and performing 
the  Fourier transform to the coordinate space, one 
obtains  an 
operator whose matrix element between the initial and final nuclear states
integrated over the positions of nucleons 
gives the contribution to $\sigma_{\gamma^{\ast}A}$ that
we seek~\cite{Bauer:1977iq,Bertocchi:1972cj,Alvero:1998bz} (an account of the energy-momentum
conservation will be discussed in Sec.~\ref{sec:phen}):
\begin{eqnarray}
&&\sigma_{\gamma^{\ast}A}^{(b)}
= \nonumber\\
&-&2 \Re e \int d^2 \vec{b} \sum_X A(A-1)\left\langle
\Theta(z_2-z_1) \Gamma_{\gamma^{\ast}X}(\vec{b}-\vec{r}_{1 \perp})
 \Gamma_{X\gamma^{\ast}}(\vec{b}-\vec{r}_{2 \perp}) e^{i (z_1-z_2) 
\Delta_{\gamma^{\ast}X}} \right\rangle  \,,
\label{eq:m3}
\end{eqnarray}
 where $\sum_X$ denotes the sum over all diffractive intermediate states (see Fig.~\ref{fig:Master1});  $A(A-1)$ is the number of the nucleon pairs;  $(\vec{r}_{i \perp},z_{i})$ are the transverse and longitudinal (with respect  to the direction of the momentum of 
$\gamma^{\ast}$, $\vec{q}$) coordinates of the involved nucleons;
$\Theta(z_2-z_1)$ is the step-function reflecting the underlying 
space-time evolution  of the process;
$\Gamma_{\gamma^{\ast}X}$ is the $\gamma^{\ast}N \to XN$ scattering amplitude 
in the space of the impact-parameter $\vec{b}$; 
the brackets denote the matrix element between
the nuclear ground-states; $\Delta_{\gamma^{\ast}X}$ is the longitudinal momentum transfer,
or, equivalently,
the inverse coherence length for the
$\gamma^{\ast} \to X$ fluctuation,
\begin{equation}
\Delta_{\gamma^{\ast}X}=\frac{M_X^2+Q^2}{2 |\vec{q}|} \,.
\label{eq:m4}
\end{equation}

For sufficiently heavy nuclei, the $t$ dependence of the $\gamma^{\ast}N \to XN$ 
scattering amplitude is much slower than that of the nuclear form factor and, hence,
can be safely neglected. Therefore,
 $\Gamma_{\gamma^{\ast}X}$ in Eq.~(\ref{eq:m3}) 
can be used in the following approximate form,
 see e.g.,~\cite{Bauer:1977iq,Yennie:1986gm}:
\begin{equation}
\Gamma_{\gamma^{\ast}X}(\vec{b}-\vec{r}_{1 \perp})=\frac{1-i\eta}{2} 
\sqrt{\frac{16 \pi \frac{d\sigma_{\gamma^{\ast}N \to XN}}{dt}(t_{\rm min})}{1+\eta^2}}\,\delta^2(\vec{b}-\vec{r}_{1 \perp}) \,,
\label{eq:m5}
\end{equation}
where $d \sigma_{\gamma^{\ast}N \to XN}/dt$ is the differential cross section
of the $\gamma^{\ast} +N\to X+N$ process;
$t_{\rm min}\approx -x ^2 m_N^2(1+M_X^2/Q^2)^2$ is the minimal momentum transfer  defined by kinematics;
$\eta$ is the ratio of the real to the imaginary parts of the 
$\gamma^{\ast}N \to X N$ scattering amplitude.
The normalization of $\Gamma_{\gamma^{\ast}X}$ in Eq.~(\ref{eq:m5}) is 
fixed 
by the S-matrix unitarity condition for the hadronic fluctuation $X$ 
of the virtual photon, 
\begin{equation}
\frac{d \sigma_{XN \to X N}}{dt}(t_{\rm min})=(1+\eta^2)\frac{\sigma_{XN \to NX}^2}{16 \pi} \,.
\label{eq:m8}
\end{equation}
Note that for DIS on deuterium and other light nuclei such as e.g., $^{3}$He and $^{4}$He,  one cannot neglect the $t$ dependence of the
elementary $\gamma^{\ast}N \to XN$ amplitude, see Sec.~\ref{sec:deuteron}.

Unless specified, we consider sufficiently large nuclei, whose 
ground-state wave function  squared can be approximated by the product of independent, one-particle nuclear densities $\rho_A$,
\begin{equation}
|\psi_A(\vec{r}_1,\vec{r}_2,\dots,\vec{r}_A)|^2=\prod_{i=1}^A \rho_A(\vec{r}_i) \,.
\label{eq:m6}
\end{equation}
The nuclear density
$\rho_A$ is normalized to unity,
$\int d^3 r \rho_A(\vec{r})=1$.
The approximation of independent nucleons is used only for simplification: 
nucleon-nucleon correlations can be straightforwardly introduced and this will 
not noticeably change our results. For instance, corrections due to short-range correlations
between nucleons is a few percent effect for the total hadron-nucleus cross sections~\cite{Alvioli:2008rw}.
 For the case of the deuteron target, we use directly the deuteron wave function, 
see details in Sec.~\ref{sec:deuteron}.

Substituting Eq.~(\ref{eq:m5}) in Eq.~(\ref{eq:m3}) and integrating over the nucleon coordinates
using Eq.~(\ref{eq:m6}), we obtain:
\begin{eqnarray}
\sigma_{\gamma^{\ast}A}^{(b)}&=&
-8 \pi A(A-1) \Re e \int d^2 \vec{b} \sum_X \frac{(1-i\eta)^2}{1+\eta^2} \nonumber\\
&\times&
\frac{d\sigma_{\gamma^{\ast}N \to XN}}{dt}(t_{\rm min})
\int^{\infty}_{-\infty}d z_1 \int^{\infty}_{z_1}d z_2 \,\rho_A(\vec{b},z_1) \rho_A(\vec{b},z_2) 
e^{i (z_1-z_2) 
\Delta_{\gamma^{\ast}X}}   \,.
\label{eq:m9}
\end{eqnarray}
The $d \sigma_{\gamma^{\ast}N \to X N}/dt$ 
differential cross section can be expressed in terms of the diffractive structure functions $F_2^{D(4)}$
and  $F_L^{D(4)}$ [compare to Eq.~(\ref{eq:sigma_TL})]
which
parameterize the cross section of 
inclusive diffraction $ep\to e + p +X$ [compare to Eq.~(\ref{eq:definition_F2})]:
\begin{equation}
\frac{d^4 \sigma_{ep}^D}{dx_{\Pomeron}\,dt\, dx\, dQ^2}=\frac{2 \pi \alpha_{\rm em}^2}{x Q^4} 
\left[\left(1+(1-y)^2\right) F_2^{D(4)}(x,Q^2,x_{\Pomeron},t)-y^2F_L^{D(4)}(x,Q^2,x_{\Pomeron},t)
\right] \,.
\label{eq:ft1_0}
\end{equation} 
The diffractive structure functions depend 
on the virtuality $Q^2$, Bjorken $x$, the invariant momentum transfer $t$, and the 
light-cone fraction $x_{\Pomeron}$,
\begin{equation}
x_{\Pomeron}=\frac{M_X^2+Q^2}{W^2+Q^2} \,,
\label{eq:xPom}
\end{equation}
where $W^2=(q+p)^2$. 
For a mini-review of hard diffraction in
lepton-nucleon DIS, we refer the reader  to Sec.~\ref{subsec:ft} and 
\ref{subsec:diffdata}.

Using the connection between the total and diffractive cross sections and 
the corresponding structure functions
[Eqs.~(\ref{eq:sigma_TL}), (\ref{eq:definition_F2}), and (\ref{eq:ft1_0})]
and replacing the sum over the diffractive states $X$ in Eq.~(\ref{eq:m9}) by the integration
over $x_{\Pomeron}$, we obtain our final expression for 
the contribution of graph $b$ in Fig.~\ref{fig:Master1} to the nuclear
structure function $F_{2A}(x,Q^2)$, which we denote
$F_{2A}^{(b)}(x,Q^2)$:
\begin{eqnarray}
F_{2A}^{(b)}(x,Q^2)&=&
-8 \pi A(A-1) \Re e \frac{(1-i\eta)^2}{1+\eta^2} \int^{0.1}_x d x_{\Pomeron}
F_2^{D(4)}(x,Q^2,x_{\Pomeron},t_{{\rm min}}) \nonumber\\
&\times& \int d^2 \vec{b}  
\int^{\infty}_{-\infty}d z_1 \int^{\infty}_{z_1}d z_2 \,\rho_A(\vec{b},z_1) \rho_A(\vec{b},z_2) 
e^{i (z_1-z_2) x_{\Pomeron} m_N}   \,.
\label{eq:m11}
\end{eqnarray}
The lower limit of integration over $x_{\Pomeron}$ corresponds to
$M_X=0$ [see Eq.~(\ref{eq:xPom})]; the upper limit of integration is defined by the typical cut on the diffractively produced 
masses $M_X^2 \leq 0.1 W^2$. The contribution of large diffractive masses, $M_X^2 \gtrsim 0.1 W^2$, is automatically suppressed by the 
$e^{i (z_1-z_2) x_{\Pomeron} m_N}$ factor in the above integrand.
A similar expression is also valid for the longitudinal structure functions:
\begin{eqnarray}
F_{L}^{A(b)}(x,Q^2)&=&
-8 \pi A(A-1) \Re e \frac{(1-i\eta)^2}{1+\eta^2} \int^{0.1}_x d x_{\Pomeron}
F_L^{D(4)}(x,Q^2,x_{\Pomeron},t_{{\rm min}}) \nonumber\\
&\times& \int d^2 \vec{b}  
\int^{\infty}_{-\infty}d z_1 \int^{\infty}_{z_1}d z_2 \,\rho_A(\vec{b},z_1) \rho_A(\vec{b},z_2) 
e^{i (z_1-z_2) x_{\Pomeron} m_N}   \,.
\label{eq:m11_FL}
\end{eqnarray}

It is important to point out that Eqs.~(\ref{eq:m11}) and (\ref{eq:m11_FL}) give the complete 
and model-independent answer for the shadowing correction 
for the interaction with
two nucleons of the target, which is the case in the low nuclear density limit 
and in the case of the deuteron target.
One should also note that Eqs.~(\ref{eq:m11}) and (\ref{eq:m11_FL})
do not require the 
decomposition over twists. The only requirement is that the nucleus is a system of color neutral objects---nucleons.
The data on the EMC ratio $F_{2A}(x,Q^2)/[A F_{2N}(x,Q^2)]$ for $x > 0.1$ 
indicate that the corrections to the multinucleon picture of 
the nucleus do not exceed few percent for $x \leq 0.5$,
see the discussion in Sec.~\ref{subsect:fast}.

\begin{figure}[h]
\begin{center}
\epsfig{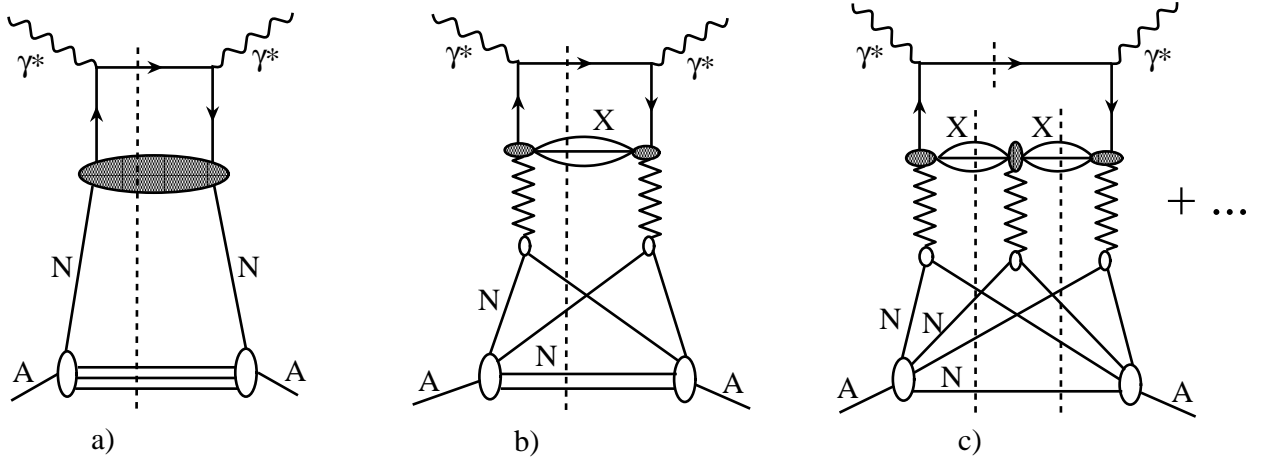}
\caption{Graphs corresponding to sea quark nuclear PDFs. 
Graphs $a$, $b$, and $c$ correspond to the 
interaction with one, two, and three  nucleons, respectively. 
Graph $a$ gives the impulse approximation;
graphs $b$ and $c$ contribute to the shadowing correction.
}
\label{fig:Master1_quarks}
\end{center}
\end{figure}

\begin{figure}[h]
\begin{center}
\epsfig{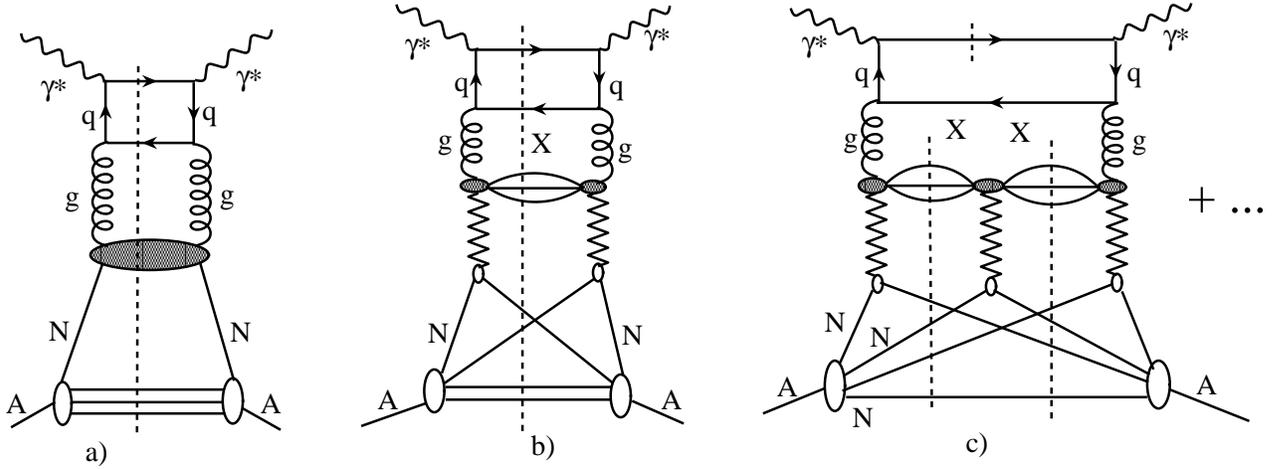}
\caption{Graphs corresponding to the gluon nuclear PDF.
For the legend, see Fig.~\ref{fig:Master1_quarks}.
}
\label{fig:Master1_gluons}
\end{center}
\end{figure}

The next crucial step in the derivation of our master equation for
nuclear PDFs is the use of the QCD factorization theorems for inclusive DIS  
and hard diffraction in  DIS.
According to the QCD factorization theorem for 
inclusive DIS (for a review, see, e.g., \cite{Brock:1993sz})
the inclusive structure 
function $F_{2}(x,Q^2)$ (of any target) is given by the convolution of hard scattering coefficients   $C_j$ with the parton distribution functions of the target $f_j$  ($j$ is the parton flavor):
\begin{equation}
F_2(x,Q^2)=x\sum_{j=q,\bar{q},g} \int_{x}^{1} \frac{d y}{y}C_j (\frac{x}{y},Q^2) f_j(y,Q^2)
 \,.
\label{eq:m2_b}
\end{equation}
Since the coefficient functions $C_j$ do not depend on the target, Eq.~(\ref{eq:m2}) leads to the relation between
nuclear PDFs of flavor $j$, which are evaluated in the impulse
approximation, $f_{j/A}^{(a)}$, and
the nucleon PDFs~$f_{j/N}$,
\begin{equation}
x f_{j/A}^{(a)}(x,Q^2)=A \,x f_{j/N}(x,Q^2) \,.
\label{eq:m2_c}
\end{equation}
In the graphical form, $f_{j/A}^{(a)}$ is given by graph $a$ in 
Figs.~\ref{fig:Master1_quarks} and \ref{fig:Master1_gluons}.

Note also that one can take into account the difference between the proton and neutron  
PDFs by replacing $Af_{j/N} \to Zf_{j/p}+(A-Z)f_{j/n}$, where 
$Z$ is the number of protons, and
the subscripts $p$ and $n$ refer to the free proton and neutron,
respectively.
\vspace*{0.5cm}

Similarly to the inclusive case, the factorization theorem for hard diffraction in DIS states that,  at given fixed $t$ and $x_{\Pomeron}$ and in the leading twist (LT) approximation, the diffractive structure function $F_2^{D(4)}$ can be written as the convolution of the same hard scattering coefficient functions $C_j$ with universal diffractive parton distributions $f_j^{D(4)}$:
\begin{equation}
F_2^{D(4)}(x,Q^2,x_{\Pomeron},t)=\beta\sum_{j=q,\bar{q},g}  \int_{\beta}^{1} \frac{d y}{y}C_j (\frac{\beta}{y},Q^2) f_j^{D(4)}(y,Q^2,x_{\Pomeron},t) \,,
\label{eq:ft3_a1}
\end{equation}
where $\beta=x/x_{\Pomeron}$.
The diffractive PDFs $f_j^{D(4)}$ are conditional probabilities 
to find a parton of flavor $j$ with a light-cone fraction $\beta$
in the proton that undergoes 
diffractive scattering characterized by the longitudinal momentum fraction $x_{\Pomeron}$ and the momentum transfer $t$, see Sec.~\ref{subsec:ft} and 
\ref{subsec:diffdata} for details.

Since the inclusive and diffractive structure functions in 
Eq.~(\ref{eq:m11}) are given by the convolution of the corresponding PDFs with
the same hard scattering
coefficients $C_j$, 
Eq.~(\ref{eq:m11}) can be turned into the relation between nuclear 
PDFs, $f_{j/A}^{(b)}$, and the diffractive PDFs of
the nucleon, $f_j^{D(4)}$:
\begin{eqnarray}
x f_{j/A}^{(b)}(x,Q^2)&=&
-8 \pi A(A-1) \Re e \frac{(1-i\eta)^2}{1+\eta^2} \int^{0.1}_x d x_{\Pomeron}
\beta f_j^{D(4)}(\beta,Q^2,x_{\Pomeron},t_{{\rm min}}) \nonumber\\
&\times& \int d^2 \vec{b}  
\int^{\infty}_{-\infty}d z_1 \int^{\infty}_{z_1}d z_2 \,\rho_A(\vec{b},z_1) \rho_A(\vec{b},z_2) 
e^{i (z_1-z_2) x_{\Pomeron} m_N}   \,.
\label{eq:m12}
\end{eqnarray}
In the graphical form, $f_{j/A}^{(b)}$ is given by 
graph $b$ in Figs.~\ref{fig:Master1_quarks} and \ref{fig:Master1_gluons}
(the superscript $(b)$ indicates that we took into account only the contribution of
graph $b$
in Figs.~\ref{fig:Master1}, \ref{fig:Master1_quarks}, and \ref{fig:Master1_gluons}).
Again, similarly to the case of Eq.~(\ref{eq:m11}), it is important to 
note that Eq.~(\ref{eq:m12}) gives the complete answer 
for the shadowing correction to nuclear PDFs when the interaction 
with only two nucleons of the target is important
(see the discussion above). Since according to the factorization theorem
diffractive PDFs have the same anomalous  dimensions  as the 
usual PDFs, the left-hand and right-hand sides of  Eq.~(\ref{eq:m12}) 
satisfy the same evolution equations at any order in the strong coupling constant $\alpha_s$. 
 We also simplified the expression in Eq.~(\ref{eq:m12}) by pulling 
$(1-i\eta)^2/(1+\eta^2)$ out of the integral over $x_{\Pomeron}$. 
This can be done because, experimentally,  
$\eta$ is practically $x_{\Pomeron}$-independent 
and the $(1-i\eta)^2/(1+\eta^2)$ factor
itself is rather close to unity. 
This follows from the Gribov-Migdal formula which was derived within 
the Pomeron exchange framework~\cite{Gribov:1968uy}:
\begin{equation}
\eta\equiv \frac{\Re e A_{\gamma^{\ast} N \to XN}}{\Im m A_{\gamma^{\ast} N \to XN}} 
=\frac{\pi}{2} {\partial \ln \Im m A_{\gamma^{\ast} N \to XN} \over \partial \ln (1/x)}
\approx \frac{\pi}{2} (\alpha_{\Pomeron}(0)-1)=0.174 \,,
\end{equation}
where in the last step we used the relation between the energy dependence of
the imaginary part of the diffractive amplitude $\Im m A_{\gamma^{\ast} N \to XN}$ and
the intercept of the Pomeron trajectory, $\alpha_{\Pomeron}(0)$, for which we used 
the phenomenological value $\alpha_{\Pomeron}(0)=1.111$, see the detailed discussion in
Sec.~\ref{subsec:diffdata} and Sec.~\ref{sec:phen}.

The derivation of the expressions for $f_{j/A}^{(a)}$ and 
$f_{j/A}^{(b)}$ is general and model-independent: we only made 
a simplifying approximation neglecting
nucleon correlations in the nuclear wave function
and a small correction associated with the $t$ dependence
of the elementary diffractive $\gamma^{\ast}N \to X N$ amplitude. 
(For the deuteron target, the latter two approximations are not used, see Sec.~\ref{sec:deuteron}.)  By the virtue of the factorization theorem
the derived results do not depend on the specific current 
used to probe nuclear PDFs and
are valid for any kind of current: transversely and longitudinally polarized virtual photons $\gamma^{\ast}_{T}$ and $\gamma^{\ast}_{L}$, respectively,
gauge boson $Z^{0}$, or any leading twist operator that couples directly to gluons.

\subsubsection{Contribution to nuclear shadowing of the interactions 
with $N \geq 3$ nucleons of the target: the cross section (color) fluctuation formalism}
\label{subsubsec:cf}

We derived the  model-independent expressions for the interaction 
of the virtual photon with  one and two nucleons of the nuclear 
target and expressed them in terms of measurable  quantities, see 
Eqs.~(\ref{eq:m2_c}) and (\ref{eq:m12}).  
To generalize the above results
to the interaction with three and more nucleons of the target, 
graph $c$ in Figs.~\ref{fig:Master1_quarks} and 
\ref{fig:Master1_gluons}, requires invoking additional ideas. 
(The $A$-dependence of the contributions of the interaction with  a given number 
of nucleons can still be calculated in a model-independent way).
Below we explain the  problem and how to resolve it.
We suggest 
an approximation  for treating the interactions with $N> 2$ nucleons that circumvents the problem,
 takes into account main features of the diffractive dynamics and  enables us to express 
the shadowing correction in terms of the parton distributions.

The problem of nuclear shadowing in DIS can be reformulated as 
the  problem of propagation of QCD color singlet states in a nuclear medium. 
While at each step of our calculations we are 
dealing with colorless objects, the strength of the interactions of 
the virtual photon with the nucleons 
and, in particular, the presence of point-like configurations, is 
an unambiguous consequence of the QCD color dynamics. 
The QCD factorization theorem for diffraction in DIS is also valid   for
the interactions with $N=3$ nucleons (graph $c$ of Figs.~\ref{fig:Master1_quarks} and \ref{fig:Master1_gluons}) and $N >3$ nucleons. Hence, one
can easily derive  the general expression for the nuclear shadowing 
correction for a heavy nucleus. However,
this expression is not calculable in terms of DIS diffraction off a nucleon since 
the configurations in the photon wave function, which lead to diffractive final states 
and those which do not, enter in a different proportion in 
the interactions with $N=2$ and $N \geq 3$ nucleons.  
Therefore, we have to 
elaborate further our approach.

In order to take into account the sum over diffractively produced states in DIS, 
we use  the formalism of cross section  
fluctuations~\cite{Feinberg56,Good:1960ba,Blaettel:1993ah,Frankfurt:2000tya}. 
In this formalism, the wave function of a fast projectile (virtual photon)  is represented 
as a superposition of 
the eigenstates of the scattering operator, $|\sigma \rangle$. 
Each eigenstate interacts with a target nucleon with a certain cross section 
$\sigma$. 
The usefulness of such a decomposition follows from the 
well understood property of QCD that the wave function 
of a virtual photon (hadron) is a superposition of  quark-gluon configurations of different transverse sizes whose interaction is proportional  the transverse area occupied by 
color, see, e.g., \cite{Frankfurt:1994hf}.
The existence and important role of small transverse size configurations 
in the wave functions of photons and pions has been confirmed by the observation 
of the color transparency phenomenon
in a variety of different processes; for a review, see~\cite{Frankfurt:1994hf}.
The probability of the incoming virtual photon to fluctuate 
into a given eigenstate is given by the distribution $P_j(\sigma)$. 
We explicitly show the dependence of $P_j(\sigma)$ on parton 
flavor $j$ as a reminder that DIS  probes a particular parton distribution of the target. 
 In soft hadron interactions,
the formalism of cross section fluctuations provides a good 
description of the total hadron-nucleus cross sections 
and the coherent inelastic diffraction in hadron-nucleus scattering, for a review and references, see Ref.~\cite{Frankfurt:2000tya}. The latter is far less trivial as the coherent 
inelastic diffraction would have been absent if the fluctuations were not present.

Note also that the validity of the formalism of cross section fluctuations for
the virtual photon
is supported by the  observation of the low value of the intercept of the Pomeron trajectory,  $\alpha_{\Pomeron}(0)=1.111 \pm 0.007$, see Sec.~\ref{subsec:diffdata}.  The closeness of $\alpha_{\Pomeron}(0)$ extracted from the HERA data on diffraction in DIS to  
$\alpha_{\Pomeron}(0)=1.0808$ extracted from the fits to soft hadron-hadron cross
sections~\cite{Donnachie:1992ny} and to $\alpha_{\Pomeron}(0) \approx 1.08$ extracted 
from the energy dependence of elastic $\rho_0$ photoproduction at 
HERA~\cite{Breitweg:1997ed}
indicates that 
our approximation should work in lepton-nucleus DIS approximately as well as in high-energy hadron-nucleus 
scattering~\cite{Frankfurt:1993qi,Guzey:2005tk}.

The entire series of the interactions  with the target nucleons shown 
in Figs.~\ref{fig:Master1_quarks} and \ref{fig:Master1_gluons}
can be summed as in  the standard Glauber formalism, with the substitution of  
$\sigma^k$ in the term corresponding to the interaction with $k$ nucleons by 
$\langle \sigma^k \rangle_j$,
\begin{equation}
\langle \sigma^k \rangle_j=\int_0^{\infty} d \sigma P_j(\sigma) \sigma^k \,,
\label{eq:Pj_moments}
\end{equation}
which accounts for the  color fluctuations of the strength of the interaction, 
 see, e.g., Ref.~\cite{Frankfurt:2000tya}.
Assuming that $A \gg 1$ such that the interactions can be exponentiated, one obtains~\cite{Guzey:2009jr}:
\begin{eqnarray}
xf_{j/A}(x,Q^2)&=&\frac{xf_{j/N}(x,Q^2)}{\langle \sigma \rangle_j}2\, \Re e \int d^2 b \left\langle 
\left(1-e^{-\frac{A}{2}(1-i\eta) \sigma T_A(b)}\right) \right\rangle_j
\nonumber\\
&=&A xf_{j/N}(x,Q^2)-xf_{j/N}(x,Q^2) \frac{A^2 \langle \sigma^2 \rangle_j}{4\langle \sigma \rangle_j}
\Re e (1-i\eta)^2 \int d^2 b \,T_A^2(b)
\nonumber\\
&-& xf_{j/N}(x,Q^2) 2\, \Re e \int d^2b \frac{\sum_{k=3}^{\infty}(-\frac{A}{2}(1-i\eta) T_A(b))^k \langle \sigma^k \rangle_j}{k! \,\langle \sigma  \rangle_j} \,,
\label{eq:fluct1}
\end{eqnarray}
where $T_A(b)=\int^{\infty}_{-\infty} dz \rho_A(b,z)$.
In the second and third lines of Eq.~(\ref{eq:fluct1}), we made an expansion in the number of the interactions with the target nucleons.
The interaction with $k$ nucleons of the nuclear target probes 
the $k$-th moment of the distribution $P_j(\sigma)$.  
Note that the above equation has no evident problems with energy-momentum conservation and 
causality that are characteristic for the eikonal 
approximation since the energy is split between the constituents of the projectile 
well before the interaction and 
different configurations are practically frozen during the propagation
of the wave packet through the nucleus.

Equation~(\ref{eq:fluct1}) is valid at high energies (small $x$), when the effect of 
the finite coherence length (the coherence length is proportional
to the lifetime of the fluctuations $|\sigma \rangle$) is unimportant. In this case, 
all factors associated with the space-time development of the scattering, such as, e.g.,
the $e^{i(z_1-z_2) m_N x_{\Pomeron}}$ factor, should be set to unity.
Note that our numerical analysis shows that the $e^{i(z_1-z_2) m_N x_{\Pomeron}}$ factor
can be safely set to unity for $x \leq 10^{-2}$, see Fig.~\ref{fig:LT2009_pb208_nolc}.

In Eq.~(\ref{eq:fluct1}),  the first term corresponds to the interaction with one nucleon  of the target, and, hence, is equal to 
$A xf_{j/N}^{(a)}(x,Q^2)$. The second term describes the interaction with two nucleons, and, hence, should be equal to 
$xf_{j/N}^{(b)}(x,Q^2)$ after $e^{i(z_1-z_2) m_N x_{\Pomeron}}$ 
is set to unity in  Eq.~(\ref{eq:m12}).
[Note that we take into account  the effect of the finite coherence length in our final expression below,  see Eq.~(\ref{eq:fluct2_c}).]
Indeed,  as follows from the formalism of cross section fluctuations,
the second moment $\langle \sigma^2 \rangle_j$ is proportional to the differential cross section of diffractive dissociation~\cite{Blaettel:1993ah}.  In the case of DIS and in our notation (normalization),
$\langle \sigma^2 \rangle_j$ is related
to the diffractive parton distribution
$f_j^{D(4)}$~\cite{Frankfurt:1998ym,Frankfurt:2003zd}:
\begin{equation}
\frac{\langle \sigma^2 \rangle_j}{\langle \sigma \rangle_j} \equiv \sigma_2^j(x,Q^2)
=\frac{16 \pi}{(1+\eta^2) x f_{j/N}(x,Q^2)}
\int^{0.1}_{x} d x_{\Pomeron} \beta f_j^{D(4)}(\beta,Q^2,x_{\Pomeron},t_{\rm min}) \,.
\label{eq:m17}
\end{equation}
Equation~(\ref{eq:m17}) is similar to 
the Miettinen-Pumplin relation~\cite{Miettinen:1978jb} generalized 
to include the real part of the diffractive amplitude.

Also, one notices that $\int^{\infty}_{-\infty} dz_1 \int^{\infty}_{z_1} dz_2 \rho_A(\vec{b},z_1)\rho_A(\vec{b},z_2)=(1/2) T_A^2(\vec{b})$. Therefore, 
in the discussed high-energy limit, we obtain from Eq.~(\ref{eq:m12}): 
\begin{equation}
x f_{j/A}^{(b)}(x,Q^2)=-
xf_{j/N}(x,Q^2) \frac{A^2}{4} \sigma_{2}^j(x,Q^2)
\Re e (1-i\eta)^2 \int d^2 b \,T_A^2(b) \,,
\label{eq:b-term}
\end{equation}
which coincides with the second term in Eq.~(\ref{eq:fluct1}).

The last term in Eq.~(\ref{eq:fluct1}) describes
the interaction with three and more nucleons of the target. It corresponds to graph $c$ and implied (not shown)
 higher rescattering terms in Figs.~\ref{fig:Master1_quarks} and \ref{fig:Master1_gluons}.
Denoting the contribution of the last term in Eq.~(\ref{eq:fluct1}) 
by $xf_{j/A}^{(c)}(x,Q^2)$, we obtain:
\begin{eqnarray}
&&xf_{j/A}^{(c)}(x,Q^2)= -xf_{j/N}(x,Q^2) \sigma_{2}^j(x,Q^2) 2 \Re e \int d^2b \frac{\sum_{k=3}^{\infty}(-\frac{A}{2}(1-i\eta) T_A(b))^k \langle \sigma^k \rangle_j}{k! \,\langle \sigma^2  \rangle_j} \nonumber\\
&=&-xf_{j/N}(x,Q^2) \sigma_{2}^j(x,Q^2)2\, \Re e \int d^2 b \nonumber\\
&\times&
\frac{\left\langle 
\left(e^{-\frac{A}{2}(1-i\eta) \sigma T_A(b)}-1+\frac{A}{2}(1-i\eta) \sigma T_A(b)
-\frac{1}{2}[\frac{A}{2}(1-i\eta) \sigma T_A(b)]^2\right) \right\rangle_j}{\langle \sigma^2 \rangle_j}
\,.
\label{eq:c-term}
\end{eqnarray}
Therefore, the full expression for the nuclear parton distribution, 
$xf_{j/A}=xf_{j/A}^{(a)}+xf_{j/A}^{(b)}+xf_{j/A}^{(c)}$, is
\begin{eqnarray}
&&xf_{j/A}(x,Q^2)=Axf_{j/N}(x,Q^2) \nonumber\\
&-&xf_{j/N}(x,Q^2)\sigma_{2}^j(x,Q^2)2\, \Re e \int d^2 b 
\frac{\left\langle 
\left(e^{-\frac{A}{2}(1-i\eta) \sigma T_A(b)}-1+\frac{A}{2}(1-i\eta) \sigma T_A(b)\right) \right\rangle_j}{\langle \sigma^2 \rangle_j}
\,.
\label{eq:fluct2}
\end{eqnarray}
To cast Eq.~(\ref{eq:fluct2}) into the more standard form, which reflects the space-time development of the scattering process~\cite{Frankfurt:2003zd},   we reintroduce the dependence on the longitudinal coordinates $z_1$ and $z_2$,  use the definition of $\sigma_{2}^j(x,Q^2)$ from Eq.~(\ref{eq:m17}),  and 
equivalently
rewrite Eq.~(\ref{eq:fluct2}) in the following form:
\begin{eqnarray}
xf_{j/A}(x,Q^2)&=&Axf_{j/N}(x,Q^2) \nonumber\\
&-&8 \pi A^2 \Re e \frac{(1-i\eta)^2}{1+\eta^2}
\int^{0.1}_{x} d x_{\Pomeron} \beta f_j^{D(4)}(\beta,Q^2,x_{\Pomeron},t_{\rm min})
\nonumber\\
&\times&
\int d^2 b  \int^{\infty}_{-\infty}d z_1 \int^{\infty}_{z_1}d z_2\, \rho_A(\vec{b},z_1) \rho_A(\vec{b},z_2) 
 \frac{\left\langle \sigma^2 e^{-\frac{A}{2} (1-i\eta) \sigma \int_{z_1}^{z_2} dz^{\prime} \rho_A(\vec{b},z^{\prime})} \right\rangle_j}{\langle \sigma^2 \rangle_j}
\,.
\label{eq:fluct2_b}
\end{eqnarray}
Equation~(\ref{eq:fluct2_b}) is derived in the limit of high energies when the 
coherence  length $l_c$ is infinite.   
To derive the formula applicable for moderately small  $x$ corresponding to 
$l_c \sim R_A$, we need  to take into account the effect of the 
finite $l_c$, which was  correctly included in Eq.~(\ref{eq:m12}) for 
$f_{j/A}^{(b)}$. Therefore, we restore the effect of the finite 
coherence length by reintroducing the 
$e^{i (z_1-z_2) x_{\Pomeron} m_N}$ factor in the integral over $x_{\Pomeron}$. In addition,
we replace $A^2$ by $A(A-1)$ to account for the 
well-understood
$1/A$ corrections  in the double scattering term 
[cf.~Eq.~(\ref{eq:m3})].
Our resulting expression for the nuclear parton distribution reads:
\begin{eqnarray}
&&xf_{j/A}(x,Q^2)=Axf_{j/N}(x,Q^2) \nonumber\\
&-&8 \pi A (A-1) \Re e \frac{(1-i\eta)^2}{1+\eta^2}
\int^{0.1}_{x} d x_{\Pomeron} \beta f_j^{D(4)}(\beta,Q^2,x_{\Pomeron},t_{\rm min})
\nonumber\\
&\times&
\int d^2 b  \int^{\infty}_{-\infty}d z_1 \int^{\infty}_{z_1}d z_2 \rho_A(\vec{b},z_1) \rho_A(\vec{b},z_2) e^{i (z_1-z_2) x_{\Pomeron} m_N}
 \frac{\left\langle \sigma^2 e^{-\frac{A}{2} (1-i\eta) \sigma \int_{z_1}^{z_2} dz^{\prime} \rho_A(\vec{b},z^{\prime})} \right\rangle_j}{\langle \sigma^2 \rangle_j}
\,.
\label{eq:fluct2_c_n}
\end{eqnarray}
It is important to note here that the key input in Eq.~(\ref{eq:fluct2_c_n}) is the diffractive
PDFs $f_j^{D(4)}(\beta,Q^2,x_{\Pomeron},t_{\rm min})$ evaluated at $t=t_{\min}$
because the slope of the $t$ dependence of the nuclear form factor 
is much larger than  the slope of the diffractive structure function.  
In the cases when the $t$ dependence of the diffractive structure functions 
(diffractive  PDFs) was measured at HERA, it was fitted to the exponential form 
(see Sec.~\ref{subsec:f2d4}),
\begin{equation}
f_j^{D(4)}(\beta,Q^2,x_{\Pomeron},t)=e^{B_{{\rm diff}}(t-t_{{\rm min}})} f_j^{D(4)}(\beta,Q^2,x_{\Pomeron},t_{{\rm min}}) \,,
\label{eq:m13_b}
\end{equation}
where $B_{{\rm diff}} \approx 6$ GeV$^{-2}$~\cite{Aktas:2006hx}.
Note that the recent ZEUS analysis reports a rather close value,
$B_{{\rm diff}} \approx 7 \pm 0.3$ GeV$^{-2}$~\cite{Chekanov:2008fh}.   Integrating Eq.~(\ref{eq:m13_b}) over $t$, one obtains 
the simple relation between
$f_j^{D(4)}(\beta,Q^2,x_{\Pomeron},t_{{\rm min}})$ and the diffractive PDFs $f_j^{D(3)}(\beta,Q^2,x_{\Pomeron})$ obtained from the fits to the $t$-integrated diffractive
structure function $F_2^{D(3)}(\beta,Q^2,x_{\Pomeron})$,
\begin{equation}
f_j^{D(4)}(\beta,Q^2,x_{\Pomeron},t_{{\rm min}})=B_{{\rm diff}} f_j^{D(3)}(\beta,Q^2,x_{\Pomeron}) \,,
\label{eq:m13_c}
\end{equation}
where 
\begin{equation}
f_j^{D(3)}(\beta,Q^2,x_{\Pomeron}) \equiv \int_{-1\, {\rm GeV}^2}^{t_{{\rm min}}} dt
f_j^{D(4)}(\beta,Q^2,x_{\Pomeron},t_{{\rm min}}) \,.
\label{eq:f3f4}
\end{equation}
The use of $f_j^{D(3)}$ enables us to express the shadowing correction in terms of the 
quantities known to date from the QCD analysis of  diffraction in $ep$ DIS. The final expression for $xf_{j/A}(x,Q^2)$ obtained with help of the color fluctuation formalism reads:
\begin{samepage}
\begin{eqnarray}
&&xf_{j/A}(x,Q^2)=Axf_{j/N}(x,Q^2) \nonumber\\
&-&8 \pi A (A-1) \Re e \frac{(1-i\eta)^2}{1+\eta^2}
B_{\rm diff}
\int^{0.1}_{x} d x_{\Pomeron} \beta f_j^{D(3)}(\beta,Q^2,x_{\Pomeron})
\nonumber\\
&\times&
\int d^2 b  \int^{\infty}_{-\infty}d z_1 \int^{\infty}_{z_1}d z_2 \,\rho_A(\vec{b},z_1) \rho_A(\vec{b},z_2) e^{i (z_1-z_2) x_{\Pomeron} m_N}
 \frac{\left\langle \sigma^2 e^{-\frac{A}{2} (1-i\eta) \sigma \int_{z_1}^{z_2} dz^{\prime} \rho_A(\vec{b},z^{\prime})} \right\rangle_j}{\langle \sigma^2 \rangle_j}
\,.
\label{eq:fluct2_c}
\end{eqnarray}
\end{samepage}

\subsubsection{The color fluctuation approximation within the color fluctuation formalism}

The  distribution $P_j(\sigma)$ that enters 
Eq.~(\ref{eq:fluct2_c}) is restricted by the general properties of QCD. 
For small $\sigma$, it is calculable in pQCD and leads to a singular behavior $P_j(\sigma)\propto 1/\sigma$~\cite{Frankfurt:1996ri,Frankfurt:1997zk}. 
Such a behavior follows from the factorization theorem, the value of the cross
section for the spatially small wave packet of quarks and gluons, and the form
of the light-cone wave function of the virtual photon in the case of large
transverse momenta of the constituents.
For large $\sigma$, 
$P_j(\sigma)$ has to decrease with an increase of $\sigma$ to ensure the convergence 
of the moments.  
In contrast to DIS, in the case of hadron projectiles, the small-$\sigma$ 
behavior of the corresponding distribution $P_{h}(\sigma)$
is not singular. For example, for pions (mesons), $P_{\pi}(\sigma)_{\left|\sigma\to 0\right.} \to {\rm const}$. Also, in the case of hadrons, several first moments of
$P_{h}(\sigma)$ can be extracted from the data and used to model $P_{\pi}(\sigma)$
and  $P_{N}(\sigma)$. These models could in turn be tested using coherent nuclear diffraction data (see the discussion and references in Sec.~\ref{sec:pA}).

In the case of hadrons, color fluctuations are known to lead only to small corrections to the total cross section of hadron-nucleus scattering~\cite{Frankfurt:2000ty,Alvioli:2009iw,Frankfurt:1994hf}. 
In the case of virtual photons, we are faced with
a new situation since in this case  $P_j(\sigma)$ is very broad and includes the states $|\sigma \rangle$ that correspond to both small and large cross sections 
$\sigma$~\cite{Frankfurt:1996ri,Frankfurt:1997zk}. 
The states (fluctuations) with small cross sections 
constitute the perturbative contribution to the
photon-nucleon cross section; the fluctuations with large cross sections  correspond to the hadronic component of the virtual photon.
In practice,  only hadronic-size configurations contribute to 
$\langle \sigma^2 \rangle$.  This expectation is 
based on the QCD-improved
 aligned jet model 
which takes into account the $Q^2$ evolution~\cite{Abramowicz:1995hb} and
agrees well with the final analyses of the HERA data on diffraction in DIS that find that the energy dependence of the diffractive amplitudes is practically the same as in the soft  QCD  
processes~\cite{Donnachie:1992ny} and is given by 
$\alpha_{\Pomeron} (t=0) =1.111 \pm 0.007$~\cite{Aktas:2006hy,Aktas:2006hx}. Hence,
the diffractive state $X$ in Figs.~\ref{fig:Master1}, \ref{fig:Master1_quarks} and \ref{fig:Master1_gluons} 
is dominated by the large-$\sigma$ hadron-like fluctuations.
At the same time, 
 weakly interacting configurations give an important contribution to 
$\langle \sigma \rangle$ down to very small values of $x$. 
One of the sources of such weakly interacting configurations is the QCD evolution 
which generates small-$x$ partons at $Q^2 \sim$ few GeV$^2$ from the configurations 
with $x \ge 0.1$ and $Q_0^2\sim 1$  GeV$^2$ which do not lead to diffractive states since in such processes production of nucleons with  $x_{\Pomeron} < 0.1 $ is kinematically forbidden.

The key feature of Eq.~(\ref{eq:fluct2_c})  is that it separates the contributions of the small and large cross sections
[it was the main purpose of rewriting Eq.~(\ref{eq:fluct1}) in the 
form of Eq.~(\ref{eq:fluct2}) 
which led to Eq.~(\ref{eq:fluct2_c})]. 
While the fluctuations with large cross sections contribute 
to all moments $\langle \sigma^k \rangle$,
the fluctuations with small cross sections contribute
significantly only to 
$\langle \sigma \rangle$ and $\langle \sigma^2 \rangle$, i.e.,
to the $A xf_{j/N}(x,Q^2)$ term and the double scattering term proportional to $f_j^{D(3)}$. 
Therefore, since the 
last term in Eq.~(\ref{eq:fluct2_c}) proportional to
\begin{equation}
\frac{\left\langle \sigma^2 e^{-\frac{A}{2} (1-i\eta) \sigma \int_{z_1}^{z_2} dz^{\prime} \rho_A(\vec{b},z^{\prime})} \right\rangle_j}{\langle \sigma^2 \rangle_j} \equiv \eta_j 
\label{eq:eta_j}
\end{equation}
probes the higher moments of $P_j(\sigma)$,
$\langle \sigma^k \rangle/\langle \sigma^2 \rangle$ with $k \geq 3$, 
it  can be evaluated with the distribution $P_j(\sigma)$ that 
neglects the small-$\sigma$ perturbative contribution and uses 
only the information on cross section fluctuations
from soft hadron-hadron scattering.

The factor $\eta_j$ 
in 
Eqs.~(\ref{eq:fluct2_c}) and (\ref{eq:eta_j})
can be identically expanded in terms of $\langle \sigma^{k} \rangle_j/\langle \sigma^{2} \rangle_j$ with
$k \geq 3$. We have just explained that the required component of the distribution $P_j(\sigma)$ is dominated
by soft hadron-like fluctuations.  
Similarly to the case of such fluctuations for the total 
hadron-nucleus cross sections mentioned above, 
the dispersion of $P_j(\sigma)$ does not lead to 
significant modifications of 
$\eta_j$
in Eq.~(\ref{eq:fluct2_c}). 
In particular, our numerical studies using 
$P_j(\sigma) \propto P_{\pi}(\sigma)$ have found that 
it is a good approximation to use 
$\langle \sigma^{k} \rangle_j/\langle \sigma^{2} \rangle_j=(\langle \sigma^{3} \rangle_j/\langle \sigma^{2} \rangle_j)^{k-2}$ for all $k\geq 3$, which we shall call the 
{\it color fluctuation approximation}.
 Therefore, the 
$\eta_j$
 term 
in Eqs.~(\ref{eq:fluct2_c}) and (\ref{eq:eta_j}) can be 
expressed in terms of a single cross section, $\sigma_{\rm soft}^j(x,Q^2)$,
\begin{equation}
\sigma_{\rm soft}^j(x,Q^2) \equiv \langle \sigma^{3} \rangle_j/\langle \sigma^{2} \rangle_j=
\left(\langle \sigma^{k} \rangle_j/\langle \sigma^{2} \rangle_j \right)^{1/(k-2)} \,, \ 
{\rm for } \ k\ge 3 \,.
\label{eq:s_soft}
\end{equation}
We can estimate $\sigma_{\rm soft}^j(x,Q^2)$ based on the analysis of the inelastic diffraction in the pion-nucleon scattering. In our numerical 
predictions for nuclear shadowing,
we will use two models for $\sigma_{\rm soft}^j(x,Q^2)$, which are based on the 
assumption that soft physics dominates the interaction of the configurations leading to diffraction, see 
the discussion in Sec.~\ref{subsubsec:color_fluct}. 
It is also worth emphasizing here that
for realistic nuclei, a typical number of interactions even at small impact parameters does not exceed three. As a result, the uncertainties in the predictions 
(which are quite small in the case of nuclear PDFs) are dominated by the uncertainties in
the value of  $\sigma_{\rm soft}^j$ rather than by the color fluctuation approximation 
for the $k\ge 4 $ moments~(\ref{eq:s_soft}).

Note that the factor
$1 - \sigma_{2}^j(x,Q^2)/ \sigma_{\rm soft}^j(x,Q^2)$ 
can be interpreted
as the fraction of the DIS cross section (PDF) originating from
 the point-like configurations---it is the parameter $\lambda$ of the 
QCD-improved aligned jet model~\cite{Frankfurt:1998ym,Frankfurt:1988nt}.  
As one  can see from our numerical studies described 
below, $\lambda$ decreases with decreasing $x$, which reflects 
the onset of the strong interaction regime for the increasing 
fraction of the configurations contributing to the 
PDFs. 

We shall postpone the detailed discussion of $\sigma_{\rm soft}^j$
until Sec.~\ref{subsubsec:color_fluct}.
At this point, to get the feeling about the meaning and magnitude of
$\sigma_{\rm soft}^j$, we note that if diffraction were described by the 
aligned jet model, we would expect the typical strength of the interaction of a  
large-size $q\bar q$ configuration with the nucleon to be compatible to that 
for pions ($\rho$ mesons, etc.), i.e.,
$\sigma_{\rm {aligned\, jet} - N} \approx 25$ mb at $x=0.01$ and 
$\sigma_{{\rm aligned\, jet} - N}  \approx  40$ mb at $x=10^{-5}$. 

Applying the color fluctuation approximation 
to Eq.~(\ref{eq:fluct2_c}), we obtain our final
 expression for the nuclear parton distribution modified by nuclear shadowing,
\begin{eqnarray}
&&xf_{j/A}(x,Q_0^2)=Axf_{j/N}(x,Q_0^2) \nonumber\\
&-&8 \pi A (A-1)\, \Re e \frac{(1-i \eta)^2}{1+\eta^2} B_{\rm diff}
\int^{0.1}_{x} d x_{\Pomeron} \beta f_j^{D(3)}(\beta,Q_0^2,x_{\Pomeron})
 \nonumber\\
&\times& \int d^2 b \int^{\infty}_{-\infty}d z_1 \int^{\infty}_{z_1}d z_2 \rho_A(\vec{b},z_1) \rho_A(\vec{b},z_2) e^{i (z_1-z_2) x_{\Pomeron} m_N}
 e^{-\frac{A}{2} (1-i\eta) \sigma_{\rm soft}^j(x,Q_0^2) \int_{z_1}^{z_2} dz^{\prime} \rho_A(\vec{b},z^{\prime})} \,,
\label{eq:m13master}
\end{eqnarray}
where $Af_{j/N} \equiv Zf_{j/p}+(A-Z)f_{j/n}$; 
$Q_0^2$ is a low scale at which the color fluctuation approximation is applicable (see below).
The nuclear PDFs $f_{j/A}$ given by Eq.~(\ref{eq:m13master}) are 
next-to-leading (NLO) PDFs since the nucleon diffractive PDFs 
$f_j^{D(3)}$ are obtained from the NLO QCD fit.

Our master equation~(\ref{eq:m13master}) determines the nuclear PDFs $f_{j/A}$
at a particular input scale $Q^2=Q_0^2$, which is explicitly present in $f_{j/N}$, $f_j^{D(3)}$
and $\sigma_{\rm soft}^j$. 
The color fluctuation approximation is more accurate if the fluctuations 
are more hadron-like, i.e., when the contribution of the point-like configurations (PLCs) is small. 
This demands that $Q_0^2$ is not too large. At the same time, 
we would like to stay within the perturbative regime, where higher twist contributions 
to the diffractive structure functions are still small and  where the 
fits to diffractive PDFs do not
have to be extrapolated
too strongly. 
(In the extraction of the diffractive PDFs from the HERA data on diffraction,
only the data with $Q^2 > 8.5$ GeV$^2$ were used~\cite{Aktas:2006hy}. 
However, it has been checked that the extrapolation down  to $Q^2=4$ GeV$^2$ works with a good accuracy.)
Accordingly, in our numerical analysis, we use $Q_0^2=4$ GeV$^2$. 
We will demonstrate that our results depend weakly on the choice of $Q_0^2$, even if we keep $\sigma_{\rm soft}^j$ fixed. This is because the approximations discussed above 
are needed only for the interactions with three and more nucleons of the target;
the double rescattering contribution is evaluated in a model-independent way.

It is important to emphasize that while Eq.~(\ref{eq:fluct2_c}) gives a general
expression for the effect of cross section (color) fluctuations on the multiple
interactions, Eq.~(\ref{eq:m13master}) presents a particular approximation---the
color fluctuation approximation. In this approximation, the interaction cross section
with $N \geq 3$ nucleons is 
$\sigma_{\rm soft}^j(x,Q^2)=\langle \sigma^3 \rangle_j/ \langle \sigma^2 \rangle_j$, 
see Eq.~(\ref{eq:s_soft}).  
Equation~(\ref{eq:m13master}) allows for a simple interpretation: 
the factor $B_{\rm diff} \int^{0.1}_{x} d x_{\Pomeron} \beta f_j^{D(3)}(\beta,Q^2,x_{\Pomeron})$ describes the probability for a photon to diffract into diffractive states in the interaction with a target nucleon at point $(z_1, \vec b)$ and to be absorbed in the interaction with 
another nucleon at  point $(z_2, \vec b)$, while the factor in the third line  
of Eq.~(\ref{eq:m13master})
describes the interaction of the diffractive states with other nucleons of the nucleus with the cross section $\sigma_{\rm soft}^j$ between points $z_1$ and $z_2$.

It is important to note that  $\sigma_{\rm soft}^j(x,Q^2)$ can be 
determined experimentally by measuring
nuclear shadowing with a light nucleus, for instance, with $^4$He.
Alternatively, $\sigma_{\rm soft}^j(x,Q^2)$ can be extracted directly from coherent diffraction in 
DIS on deuterium~\cite{Blaettel:1993ah}. After $\sigma_{\rm soft}^j(x,Q^2)$ will have been determined, 
the leading twist theory will
contain no model-dependent parameters and can be used to predict nuclear
shadowing for an arbitrary nucleus in a completely model-independent way. The discussed
measurements can be carried out at a future 
Electron-Ion Collider.

In the treatment of multiple rescatterings in the leading twist theory 
of nuclear shadowing in Ref.~\cite{Frankfurt:2003zd}, we used the 
so-called quasi-eikonal approximation, which neglects color fluctuations and, hence, uses  $\sigma_{\rm soft}^j(x,Q^2)=\sigma_{2}^j(x,Q^2)\equiv \langle \sigma^2 \rangle_j/ \langle \sigma \rangle_j$ 
in Eq.~(\ref{eq:m13master}). 
Such an approximation gives the results identical to 
Eq.~(\ref{eq:m13master})
 for the interaction with one and two nucleons of the nuclear target.
 However, it neglects the presence of point-like configurations in the virtual photon
wave function
 and, hence, overestimates shadowing at $x\sim 10^{-3}$, where the 
contribution of the interactions with 
$N > 2$ is already important, while the contribution of the point-like configurations is still significant. 
 We will use   a comparison between the
color fluctuation and quasi-eikonal approximations to illustrate the role of color fluctuations in  Sec.~\ref{subsec:cf_qe}.
(Note that the quasi-eikonal approximation is popular in the literature in spite of 
its deep shortcomings discussed above and also in Sec.~\ref{subsec:cem}.)

In the very small-$x$ limit, which for practical purposes means 
$x <10^{-2}$
(see Fig.~\ref{fig:LT2009_pb208_nolc}), 
the factor $e^{i (z_1-z_2) x_{\Pomeron} m_N}$ in Eq.~(\ref{eq:m13master}) can be 
safely neglected. This results in a significant simplification of the master
formula after the integration by parts two times (cf.~\cite{Bauer:1977iq}):
\begin{eqnarray}
x f_{j/A}(x,Q_0^2)&=& A\, x f_{j/N}(x,Q_0^2)
-8 \pi A(A-1) B_{{\rm diff}}\,\Re e \frac{(1-i\eta)^2}{1+\eta^2} \int^{0.1}_x d x_{\Pomeron}
\beta f_j^{D(3)}(\beta,Q_0^2,x_{\Pomeron}) \nonumber\\
&\times& \int d^2 \vec{b} \, \frac{e^{-L T_A(b)}-1+LT_A(b)}{L^2} \,,
\label{eq:m13master_approx}
\end{eqnarray}
where $L=A/2\,(1-i \eta) \sigma_{\rm soft}^j(x,Q_0^2)$; 
$T_A(b)=\int^{\infty}_{-\infty} dz\, \rho_A(z)$.

We will discuss the energy and $A$ dependence of nuclear shadowing in nuclear PDFs and
structure functions in Sec.~\ref{sec:phen}.
At this point, we note that in the limit of $x={\rm const}$ and $A \to \infty$,
Eq.~(\ref{eq:m13master_approx}) predicts that nuclear shadowing tends to a constant, i.e.,
\begin{equation}
\frac{xf_{j/A}(x,Q_0^2)}{Axf_{j/N}(x,Q_0^2)}=1-\frac{\sigma_2^j(x,Q_0^2)}{\sigma_{\rm soft}^j(x,Q_0^2)} 
\equiv \lambda \,,
\label{eq:m13master_approx_b}
\end{equation}
where $\lambda=0-0.13$ for gluons and $\lambda=0.25-0.50$ for quarks, and 
decreasing with a decrease of $x$ for fixed $Q^2=Q_0^2$, see our 
results in Sec.~\ref{sec:phen}.
This is a consequence of the fact that in our approach, we effectively allow for the 
presence of unshadowed PLCs in the virtual photon wave function
and the $A \to \infty$ limit chooses only those. 
Note that the PLCs  can be present 
even in this limit due to the QCD evolution from larger $x \ge 0.01$ at the input scale $Q_0^2$,  
see the discussion of the evolution trajectories in Sec.~\ref{subsec:qcd_curve}.
In any case, the $A\to \infty $ limit is far from the realistic one for realistic nuclei, 
where very few nucleons are involved in the shadowing and, consequently, the 
suppression due to nuclear shadowing does not exceed approximately a factor of three, 
even for small impact parameters,  see our
predictions in  Sec.~\ref{sec:phen}.

The color fluctuation pattern is  changing  with an increase of $Q^2$: 
the configuration that interacted with the strength $\sigma$
at the resolution $Q^2_0$ at $x_0$, at higher $Q^2$ will interact with the same strength 
at smaller $x$. As a result, the strength of the fluctuations at fixed $x$ grows with
an  increase of $Q^2$. In particular, it leads  to an increase of the contribution 
of 
the PLCs. 
These effects are automatically taken into account by the QCD evolution and they could be visualized by inspecting the trajectories of the QCD evolution which we present in Sec.~\ref{subsec:qcd_curve}.

\subsubsection{Eikonal approximation violates constraints due to   
energy-momentum conservation}
\label{subsec:cem}

In this section we explain caveats of the eikonal approximation 
to QCD amplitudes of high energy processes and how they are resolved within the color fluctuation approach.  As a consequence 
of the Lorentz dilation, transitions between different configurations in the wave function of the energetic projectile occur in the target rest frame at the distances comparable to the scale characterized by the coherence length,
\begin{equation}
l_c=\frac{2E_h}{M^2_n-M^2_h} \,,
\label{coherencelength}
\end{equation}
where $E_h$ is the energy of the projectile; $M_{n}$ is the mass of the intermediate state 
$|n\rangle$; $M_h$ is the projectile mass.
 At sufficiently large $E_h$, when coherence length exceeds internucleon distances
in nuclei, the approximation of consecutive multiple collisions (the Glauber approximation) becomes inapplicable, see the discussion in Sec.~\ref{sec:gribov}.
Instead, the projectile interacts at the same time with all 
nucleons located at the same impact parameter (the Gribov approximation).  
Thus, the quark-gluon configurations in the wave 
function of the projectile are frozen during collisions.

At high energies, diffractive processes are a shadow of inelastic ones.
Indeed, the Abramovsky-Gribov-Kancheli cutting rules~\cite{Abramovsky:1973fm}
allow one  to calculate the amplitudes of diffractive processes in terms of inelastic ones. 
This relationship allows one to visualize the exact 
constraints imposed by energy-momentum conservation 
and also suggests how to satisfy them.

To visualize the constraints due to energy-momentum conservation,
let us consider  the double and triple scattering of a colorless dipole
of the projectile 
off a nucleus target. Each constituent of the dipole carries either the fraction $z$ 
or $1-z$ of the projectile momentum. 
In the case of the double  inelastic collision, the dipole scatters off different nucleons.   
The total energy released 
into the final state is $z\, s+(1-z)\, s =s$, 
where $s$ is the total invariant energy of the initial state,
i.e., the energy-momentum constraint is fulfilled. 
In the case of the triple scattering, the  invariant energy of the final state is
$ z\, s +(1-z)\, s+z\, s=(1+z)\, s$ or 
 $(1-z)\, s +(1-z)\, s+z\, s=(2-z)\, s$, i.e.,
the eikonal approximation neglects the energy lost by the projectile during the collision.
Thus, the eikonal approximation has fundamental problems in the application to the high-energy processes when inelastic processes dominate.  
The constraint imposed by the energy-momentum conservation is stronger and more specific
than the cancellation of the eikonal term due to causality found 
in~\cite{Mandelstam:1963cw,Gribov:1968fc}.

The resolution of this puzzle is that the projectile dipole 
develops other components containing more partons 
which participate in the triple, quadruple, etc. collisions.   
However, this quantum field theory effect is beyond the framework of the eikonal approximation, but it is well consistent with the color fluctuation approach,
which also includes large-mass diffraction (contribution of the triple Pomeron limit).

\subsection{Space-time picture of leading twist shadowing in the nucleus 
infinite momentum
 frame
and the transverse structure of the nuclear wave function}
\label{subsect:fast}

As we already mentioned above, 
the starting point of the discussion of nuclear shadowing in DIS 
is the observation that, to a very good approximation, a nucleus 
can be described as a many-nucleon nonrelativistic system. 
This condition is easier to implement in the reference frame where the nucleus 
is at rest. Hence, many calculations, including those of Sec.~\ref{subsec:derivation}, 
 are performed in that frame. 
However, in the rest frame, one does not explicitly use  
quark and gluon degrees of freedom in the nucleus.  
Hence,  in order to explain what processes lead 
to leading twist shadowing, it is instructive to consider the frame 
where the target nucleus is fast (the infinite momentum frame).
\begin{figure}[h]
\begin{center}
\epsfig{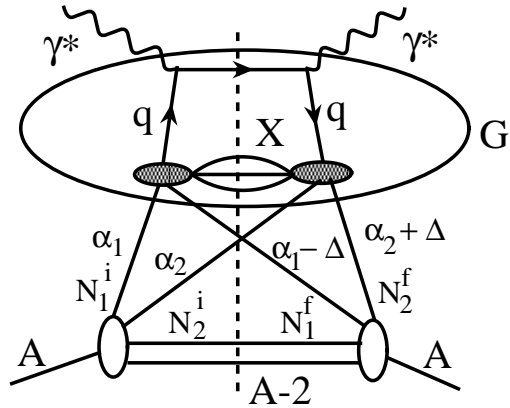}
\caption{The interchange (interference) diagram corresponding to the leading twist 
contribution to the diffractive final state.
}
\label{fig:Master1_interference}
\end{center}
\end{figure} 

For certainty,  we will consider the interaction with two nucleons of the target,
see Fig.~\ref{fig:Master1_interference}.
Since $\gamma^{\ast}$'s are attached to the same quark,
 it appears at the first glance that it is impossible to break the 
additivity of the 
interaction since, naively, the quark should belong to the
same nucleon in the $\left|in\right>$ and $  \left<out\right|$
states (similarly to graph $a$ of Fig.~\ref{fig:Master1_quarks}).
 Let us determine the necessary conditions to avoid this conclusion.  
In Fig.~\ref{fig:Master1_interference}, the nucleon that interacts in the  $\left|in\right>$ 
state is denoted as $N_1^i$ and the one that interacts in the final state as $N_2^f$. 
For each nucleon in Fig.~\ref{fig:Master1_interference}, we introduce
the light-cone fractions $\alpha \equiv Ap_N^+/p_A^+$, where
$p_N^+$ and $p_A^+$ are the plus-momenta of the nucleon and
the nucleus, respectively (the plus-momentum is defined as
$p^+=(p^0+p^3)/\sqrt{2}$). For the nucleon at rest,
 $\alpha=1$.
It follows from the DIS kinematics that
\begin{eqnarray}
&& \alpha_1^f \le \alpha_1^i -x \,, \nonumber\\
&& \alpha_2^i \le \alpha_2^f -x \,,
\label{eq:st1}
\end{eqnarray}
where $x$ is the Bjorken $x$.
 Since  $\alpha -1 \approx p_3/m_N$, where $p_3$ is the projection of the nucleon momentum on the reaction axis, and typical momenta of nucleons are less than $k_F \sim 250$~MeV/c, the interference (interchange) diagram in Fig.~\ref{fig:Master1_interference} is automatically suppressed for $x\ge 0.1$ since 
this requires scattering off nucleons with very large momenta.
Also, there is an additional suppression since only a small fraction of nucleons is produced with Feynman $x_F=\alpha/(1-x)$ close to unity. 
In fact, one expects that the $\gamma^{\ast}N \to N X$ inclusive cross section
behaves as~\cite{Frankfurt:1981mk,Frankfurt:1977vc}:
\begin{eqnarray}
&&{d\sigma^{\gamma^{\ast}N \to N X} (x_F)\over dx_F/x_F}\Big|_{x_F\to 1} \propto (1-x_F)^{n(x)} \,,
\nonumber\\  
&& n(x\ge 0.2) \sim 1\,, \quad n(0.02< x< 0.1) \sim 0\,, \quad  n(x<0.01) \sim -1 \,,
\label{eq:st2}
\end{eqnarray}
which further suppresses the contribution of the $x\ge 0.05 $ region and leads to
the dominance of the diffractive contribution for $x\le 0.01$.

The above analysis demonstrates that the interference effects are restricted to the region 
of small $x$ and any deviations from the additivity for $x\ge 0.2$ should 
be due to the presence of non-nucleonic degrees of freedom in the nucleus wave function. 
These effects have indeed been 
observed---the EMC effect for this kinematics---but they become significant
 only for rare configurations of quarks  in the nucleons  with $x\ge 0.5$.

Introducing the light-cone wave function of the nucleus, $\psi_A$,
we can write the contribution due to 
 the diffractive cut of the interference diagram
to the total cross section as
 (see Fig.~\ref{fig:Master1_interference}):
\begin{eqnarray} 
&& \Delta^{\rm diff} F_{2A}(x,Q^2)={1\over 16 \pi}
\int \prod_{i=1,2} {d \alpha_i\over \alpha_i}\, d^2 p_{\perp i} \, d \Delta \,d^2q_{\perp}\,  G(\alpha_1, \alpha_2, \Delta, p_{\perp 1}, p_{\perp 2} + q_{\perp}, x, Q^2)
\nonumber\\
&\times&
\psi_A(\alpha_i, p_{\perp i})\psi_A^{\ast}(\alpha_1-\Delta,\alpha_2 +\Delta, \alpha_3,\dots, p_{\perp 1} + q_t, p_{\perp 2} - q_{\perp}, p_{\perp 3},\dots) \,,
\label{eq:st3}
\end{eqnarray}
where $\Delta$ is related to the invariant mass of the system produced in the intermediate state,
$M_X^2=-Q^2+ \Delta W^2$; $q_{\perp}$ is the transverse momentum of the intermediate state; $G$ describes the upper part of the diagram 
associated with
the $\gamma^{\ast} NN \to \gamma^{\ast} NN$ interaction.
Note that  $\Delta=x_ {\Pomeron}$ [see Eq.~(\ref{eq:xPom})].

Since the nucleon momenta are small,  the integration is symmetric with respect to the
$p_3 \to -p_3$ transformation,
and $G$ only weakly depends on the incident energy, we can neglect the dependence of
the factor $G$ on  $\alpha_1$, $\alpha_2$,  $p_{\perp 1}$ and $p_{\perp 2}$. 
In this approximation (implicit in the Gribov derivation of shadowing for 
hadron-deuteron scattering, see Sec.~\ref{subsect:pion_deuteron}),
if we neglect the Fermi motion and take $\alpha_i \simeq 1$ and $\Delta \ll 1$, 
 $G \equiv G(x,Q^2,\Delta, q_{\perp})$ is the cross section of 
the nucleon production for  $x_F \sim 1$. Indeed,
the Fermi motion leads to the rescaling of the invariant energy $\hat{s}$ in
the diffractive amplitude by the factor $\alpha_1^n (\alpha_2+\Delta)^n$, 
where $n \approx 0.11$. Since $\Delta$ is small, the correction factor 
for $\alpha_1 \simeq  \alpha_2 \simeq 1$
is $(1+ n \Delta)$ in the integrand where $G \equiv G(x,Q^2,\Delta, q_{\perp})$.
The terms proportional to $(\alpha_i-1)$ cancel out due to the $\alpha \rightarrow 2 -\alpha$
symmetry for $\alpha \sim 1$.
As a result, the Fermi motion correction is proportional to $n\left<\vec{p}^2\right>/(3m_N^2)$
and is smaller than 1\%. Hence, overall the approximation of Eq.~(\ref{eq:st4}) is 
accurate
to better than 1\%.

As a result,  we can rewrite Eq.~(\ref{eq:st3}) as 
\begin{equation} 
\Delta^{\rm diff} F_{2A}(x,Q^2)=
{1\over 16 \pi}
\int d \Delta \,d^2q_{\perp}\, G(x,Q^2,  \Delta,  q_{\perp}) F_A( \Delta m_N,  q_{\perp}) \,,
\label{eq:st4}
\end{equation}
where $F_A$ is the two-nucleon form factor of the nucleus. 
In the case of the calculation of the nuclear shadowing phenomenon, 
it is sufficient to have  $F_A$ in the non-relativistic limit:
\begin{equation}
F_A(\Delta m_N, q_{\perp})= \int d^3k_i \,\psi_A(k_i) \psi_A^{\ast}(k_1+\vec{q}, k_2-\vec{q}, k_3,\dots) \,,
\label{eq:st5}
\end{equation}
where $\vec{q}=(\Delta m_N, q_{\perp})$.
The application of the AGK cutting rules allows one to connect
$\Delta^{\rm diff} F_{2A}(x,Q^2)$ with the shadowing contribution to 
 the nuclear structure function $F_{2A}(x,Q^2)$. The result  is
\begin{equation}
\Delta^{\rm diff} F_{2A}(x,Q^2)=-F_{2A}^{(b)}(x,Q^2) \,,
\label{eq:st6}
\end{equation}
where $F_{2A}^{(b)}(x,Q^2)$ is derived in the nucleus rest frame and is
given by Eq.~(\ref{eq:m11}).

The main contribution to nuclear shadowing originates from diffraction at  
small $t$, which is dominated by non-spin-flip hadron production.
The  spin-flip contribution is small for $t\sim 0$
(due to the helicity conservation at $t=0$),  becoming  
important for larger $t$ and leading  to the slope of the inelastic  
diffractive cross section,
which
 is smaller than that of the elastic cross section. 
Hence, the  slope of the non-spin-flip term, $B$, is likely to be somewhat larger  
than $B_{\rm diff} \sim 6$ GeV$^2$, probably closer to the slope of the elastic  
meson-nucleon cross section, $B \sim 10$ GeV$^2$.
 One can express the average distance in the transverse plane  between
 the centers of the two nucleons which contribute to nuclear shadowing as
\begin{equation}
\left<(r_{\perp 1} - r_{\perp 2})^2\right>  
= 4\,B \,.
\end{equation}
Therefore, the average transverse distance between the nucleon centers 
contributing to 
nuclear shadowing is  
of the order of $1 - 1.2$ fm, so that the nucleons overlap rather strongly in  
the transverse plane.

As  Bjorken $x$ is decreased, the strength of the interaction 
increases, and
an increasing number of nucleons  screen each other within the cylinder of the radius of $\sim \sqrt{2\,B} \approx  0.9$ fm (although this radius  should  slowly increase with decreasing $x$ for $x\le 10^{-3}$, the current data 
does not find a significant change of the slope in the HERA range of energies).
Therefore,
 a transverse slice of the wave function of a heavy nucleus  
for  $x \sim 5 \times 10^{-3}$ looks like as a system of colorless (white) clusters with some clusters ($\sim 30 \%$ -- cf.~a numerical study below) built of two rather than of one nucleon, with a gradual increase of the number of two-nucleon, three-nucleon, etc.~clusters with decreasing $x$.

\begin{figure}
\begin{center}
\epsfig{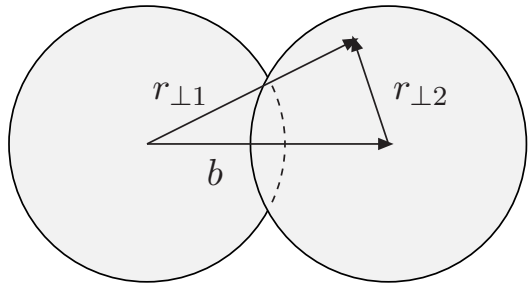}
\caption{Geometry of the parton overlap in the transverse plane.}
\label{fig:Strikman_sketch}
\end{center}
\end{figure}

The microscopic picture of nuclear
shadowing described above allows one
to address also the question of at what transverse distances from the centers of two nucleons, 
$\rho_1=r_{\perp 1}$ and $\rho_2=r_{\perp 2}$, for a given transverse internucleon distance, $b$,  shadowing occurs, see Fig.~\ref{fig:Strikman_sketch}.
First we observe that experimentally  the $t$ dependence of inclusive diffraction and deeply virtual Compton scattering (DVCS) for similar values of  $x$ and $Q^2$ are very close,  
$B_{\rm DVCS}=6.02 \pm 0.35 \pm 0.39$ GeV$^{-2}$ in the H1 2005 analysis~\cite{Aktas:2005ty}
and  $B_{\rm DVCS}=5.45 \pm 0.19 \pm 0.34$ GeV$^{-2}$ in the H1 2007 analysis~\cite{Aaron:2007cz},
so that $|B_{\rm DVCS} - B| <2$ GeV$^{-2}$.
This implies that the parton removed from the 
initial
nucleon and the parton in the final nucleon are located at very close impact parameters. 
As a result, the screening effect occurs very locally in the transverse plane,
 mostly in the region along the axis between the two nucleons. If we neglect the small difference between the slopes of DVCS and diffraction, we obtain:
\begin{equation}
f(b)= \int p(r_{\perp 1})\, p(r_{\perp 2})\,\delta (r_{\perp 1} -r_{\perp 2}-b)\, d^2r_{\perp 1} \,d^2 r_{\perp 2} \,,
\end{equation}
where $f(b)$ is the Fourier transform of the $t$ dependence of the diffractive cross section; $p(r_{\perp i})$ are transverse distributions of partons. 

In our derivations,
the global and local color neutrality are satisfied at every step.
This is very different from the approaches where the nucleus is initially built from 
free
quarks and 
the color neutrality is achieved by imposing
additional conditions at a later stage.

\subsection{On nuclear shadowing of valence quark parton distributions} 
\label{subsec:valence}

In this subsection we discuss nuclear shadowing effects in the parton distributions
in the channels with non-vacuum quantum numbers 
in the crossed channel, 
like the structure function $F_{3A}(x,Q^2)$ which could be measured for example in the 
$\nu(\bar \nu)A$ 
scattering.

\begin{figure}[ht]
\begin{center}
\epsfig{file=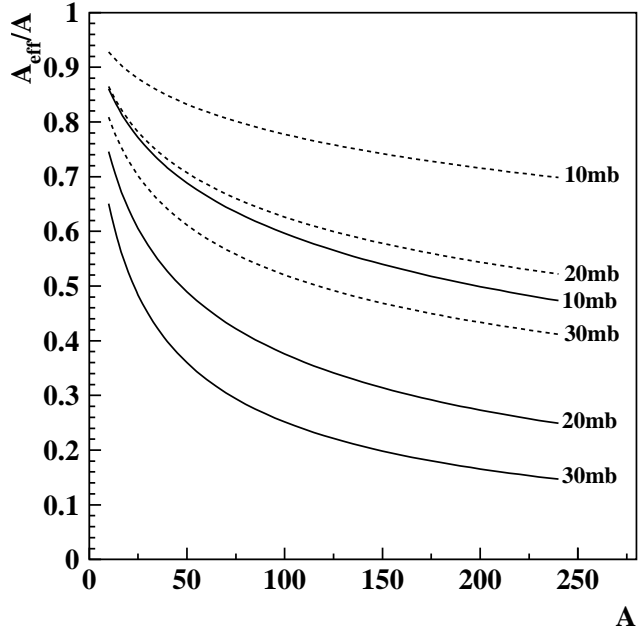,scale=0.52}
\vskip 0cm
\caption{Comparison of shadowing in the non-vacuum  channel (solid curves) and vacuum channel (dashed curves)
for different values of $\left<\sigma\right>$.}
\label{valsea}
\end{center}
\end{figure}

In the case of soft QCD dynamics, the shadowing of cross sections 
with non-vacuum quantum numbers in the crossed channel
is significantly stronger than in the vacuum channel. 
Indeed, let us consider the difference of the scattering cross sections of a particle $h$ ($h=p,K,...$) and
its antiparticle $\bar h$  off a nucleus,   
$\Delta \sigma_{hA} =  \sigma_{hA} -  \sigma_{\bar hA}$.
It is straightforward to  
derive within the eikonal approximation~\cite{Frankfurt:1988nt,Strikman:1985bu}
  (neglecting inelastic screening effects):
\begin{equation}
\Delta \sigma_{hA}= \Delta \sigma_{hN}\int d^2b AT_A(b) e^{-A/2 \langle \sigma \rangle T_A(b)} \,.
\label{eikval}
\end{equation}
Here we introduced the notation $\left<\sigma\right> = (\sigma_{hA} + \sigma_{\bar hA})/2$ and took the limit 
$\Delta \sigma_{hA}\ll \left<\sigma\right>$. It is easy to see from Eq.~(\ref{eikval}) that the shadowing for $\Delta \sigma_{hA}$ is  much larger than for the total cross sections of $hA$ scattering, see also Fig.~\ref{valsea}.   The physical  reason for the enhancement of shadowing in the non-vacuum channel is that scattering at small impact parameters does not contribute to the difference since the  interaction is essentially black 
(i.e., equal)
for small  $|\vec b|$. In terms of the Gribov Reggeon Calculus, the reason is that 
for 
the $\Reggeon - n\Pomeron$ exchange, the factor $1/n!$ is changed to $1/(n-1)!$ since only $n-1$ exchanges are identical.

A qualitative difference of 
the nuclear shadowing in the LT limit  
from that in  soft QCD
 is that in  
the former 
case the 
 masses 
 produced by a current,
 which typically contribute to shadowing, are at least of the order of $Q^2$
 and increasing with energy.   
 Moreover, the spectrum of masses produced in the valence and vacuum channels are vastly different.  For the non-vacuum channel, 
 $N(z)=(d\sigma (z)/d z)/\sigma \propto (1-z)^n$ with $n\ge 0$ in difference from the vacuum exchange where $n \sim -1$. 
 (Here $z$ is the fraction of nucleon momentum carried by the interacting parton.)  
Thus, shadowing given by the overlapping integral between the diffractive amplitudes in the non-vacuum and vacuum channels should be 
 suppressed. 
 
As we discussed in Sec.~\ref{subsect:fast}
only the final states, where a small momentum is transferred to the 
target
nucleon of the nucleus, can contribute to shadowing.
Only in this case an effective interference between 
scattering off  one nucleon in the $\left| in\right> $ state  
 and another nucleon in the $\left< out \right| $ state is possible.
In the infinite momentum frame this corresponds to the  requirement that 
$ 1- z\le 0.03$.  This is because in
the case of the removal of a valence quark, one expects that 
$N(z)=(d\sigma (z)/d z)\sigma \propto (1-z)^n$ with $n\ge 0$.

To estimate the suppression factor for nuclear shadowing of the valence quark distribution we may explore qualitative properties
of parton distributions.
Although there are no measurements of $N(z)$ for the $e+ p \to e+ p +X$
cross section where the electron interacts with a valence quark, we 
use as a guide  
the cross section of the process $e+p\to e +n +X$, which is dominated for small $x$ by 
the 
scattering off the sea quarks.
The HERA data 
\cite{Chekanov:2007tv}
indicate that  the neutron spectrum $dN/dz$ in the $ep$ DIS scattering is nearly constant for $z< 0.80$. For larger $z$, it drops and the $z$ dependence is  consistent with $\propto (1-z)$ (which corresponds to the QCD quark counting rules). If we assume a similar $dN/dz$  for the valence quark removal case,
we find that the ratio of the probability to produce a proton (neutron) in the scattering off  a valence quark of the proton (neutron) and the corresponding probability for the scattering off a sea quark, $r_{v/s}$,  is of the order $ \sim 10^{-2}$. This estimate does not include an additional small factor due to the contribution of the spin-flip processes to the nucleon spectrum since such a contribution  to the shadowing is strongly suppressed  by the nuclear overlapping integral. 
Since we are considering here the interference of the production in the valence and sea channels, 
the relevant  suppression factor for  the ratio of LT shadowing in the 
vacuum and non-vacuum channels is $\sqrt{r_{v/s}}$. 
 
As a result, we obtain an estimate for the suppression of the valence quark shadowing in  the double scattering approximation (in the approximation of the interaction with two nucleons of the nuclear
target):  
\begin{equation}
{1- V_{A}/AV_{N}\over 1- F_{2A}/AF_{2N}} \le 2 \sqrt{r_{v/s}}  \,,
\end{equation}
where $V_A$ and $V_N$ are the valence quark distributions in the nucleus and nucleon, respectively;
the factor of two comes from combinatorics [cf.~Eq.~(\ref{eikval})].
Thus we conclude that 
nuclear shadowing in the valence quark channel (for 
the double scattering contribution)
is strongly suppressed as compared to the sea channel 
in spite of the combinatorial factor of two. 

Therefore, we expect a rather small shadowing for $V_A$ for realistic nuclei ($\le 10\%$ for $A\sim 200$), though the $A$ dependence of $1- V_{A}/(AV_{N})$
should be stronger than 
that for the 
sea quark distribution 
due to combinatorial factors similar to the one we discussed in relation to Eq.~(\ref{eikval}).

At the same time, no such suppression is expected for the higher twist (HT) effects due to transitions to the low mass intermediate states which are not suppressed for $Q^2 \le 1 \div 2$ GeV$^2$;
these contributions would result in an enhanced 
 HT shadowing in the non-vacuum channel. Note also that the HT shadowing  may extend to higher $x$ than 
the leading twist shadowing. Indeed, the characteristic coherence length in the case
of the vector meson contribution, $l_c=2\nu/(Q^2+ m_V^2)$, 
 is somewhat larger than the average 
 $l_c=2\nu/(Q^2+M^2)$
  for $Q^2> m_V^2$, where $M^2\approx Q^2$ .
 
{\it Comment}.
There exists another example 
where nuclear shadowing effects in neutrino and electron  interactions should be different. 
Indeed, it follows from 
the Adler theorem~\cite{Adler:1964yx} that, in the limit of very small $Q^2$,
\begin{equation}
{\sigma^{\nu +A \to \mu + X}(E_{\nu},Q^2) \over \sigma^{\nu +N \to \mu + X}(E_{\nu},Q^2)}
=
{\sigma_{\rm tot}(\pi A)\over \sigma_{\rm tot}(\pi N)} \,.
\label{eq:adler}
\end{equation}
Since shadowing for $\sigma_{\rm tot}(\gamma A)$ 
 is known to be smaller than for $\sigma_{tot}(\pi A)$  by about 
$20\div 30\%$, we conclude that for sufficiently low $Q^2$
neutrino scattering, the higher twist shadowing in $\nu A$ scattering is
likely to  be 
significantly
larger  than in the photon case.

\subsection{The upper limit on nuclear shadowing in nuclear PDFs}
\label{susec:upper_limit}

We discussed in Sec.~\ref{subsect:fast} that the 
leading twist
shadowing  is dominated by the interaction of partons which are located at close impact parameters. 
 In this approximation,
the use of the infinite momentum nuclear frame allows one to derive a lower limit for the nuclear PDFs (the maximal value of nuclear shadowing)~\cite{Alvioli:2009ab}. Below we sketch the derivation and present the final results.

Our  starting point is the observation that 
nuclear shadowing in the scattering off the deuteron cannot reduce the cross section to the value smaller than the cross section of scattering off one nucleon. This is basically because one nucleon can screen another one, but not itself.

Similarly, it is natural to expect that the gluon density
(the quark case can be worked out similarly) 
at a given impact parameter $b$, $g_A(x, b, Q^2)$, cannot be less 
than the maximum of the gluon densities of the nucleon, $g_N(x,\rho, Q^2)$.
Therefore, 
\begin{equation}
 g_A(x,b, Q^2)  \ge \langle \max_{i=1, A} \{g_N^{(i)}(x,r_\perp - b, Q^2)\}\rangle \, ,
 \label{averageconf}
\end{equation}
where $\rho = r_\perp - b$ are the transverse coordinates of partons in the c.m.~of the corresponding nucleons, 
see Fig.~\ref{fig:Strikman_sketch}; 
the brackets denote taking the average over the nucleon configurations.
While at the moment the physical meaning of the impact parameter dependent PDFs
$g_A(x,b, Q^2)$ and $g_N(x,\rho, Q^2)$ is intuitively clear, they will be defined and discussed
in Sec.~\ref{subsec:impact}. Those PDFs are nothing else but the 
generalized parton distributions (GPDs) in the $\xi=0$ limit
in the mixed momentum-coordinate (impact parameter) representation.

In the $A\to \infty $ limit,  at a fixed $b$ there will be at least one nucleon with 
$\rho=r_\perp - b$ close to zero.
Hence, in this limit, Eq.~(\ref{averageconf}) gives:
\begin{equation}
g_A(x,b, Q^2)_{|{\rm min}} \ge g_N(x,\rho=0, Q^2) \,.
\label{eq:gA_min}
\end{equation}
It is convenient to analyze the gluon GPD of the nucleon in the factorized form,
$g_N(x,\rho, Q^2) = g_N(x, Q^2) F_g(x,\rho, Q^2)$,
where $g_N(x, Q^2)$ is the usual forward gluon distribution and
$F_g(x,\rho, Q^2)$ is the so-called two-gluon form factor defining the skewness of the gluon GPD.
The onset of the limiting behavior depends on the transverse shape of the gluon GPD. 
In Ref.~\cite{Frankfurt:2006jp}, two parameterizations of the two-gluon form factor $F_g(x,\rho,Q^2)$
were discussed and fitted to the $J/\psi$ photoproduction data~\cite{Chekanov:2002xi},
which were taken in the form of an exponential and a dipole, respectively. 
The corresponding transverse
spatial distributions of the gluon GPD are
\begin{eqnarray}
F^{(1)}_g(\vec{\rho}, Q^2)&=&\frac{1}{2\pi B_g(Q^2)}e^{-\rho^2/(2B_g(Q^2))}\,,\nonumber\\
\label{gluondens}
F^{(2)}_g(\vec{\rho},Q^2)&=&\frac{m^2_g}{2\pi}\frac{m_g(Q^2)\rho}{2}K_1(m_g(Q^2) \rho) \,,
\label{eq:two_gluon_ff}
\end{eqnarray}
where $B_g(Q^2)=3.24/m^2_g(Q^2)$;
$K_1$ is the modified Bessel function; $m^2_g(Q^2= 3\ {\rm GeV}^2)=0.6$ GeV$^{2}$ for $x\sim 10^{-4}$.
The average over the nucleon configurations in a nucleus was 
calculated using the Monte Carlo generator of \cite{Alvioli:2009ab}
and taking into account short-range  correlations of nucleons 
(this effect is numerically small for the quantities discussed here).

\begin{figure}[h]
\begin{center}
\epsfig{file=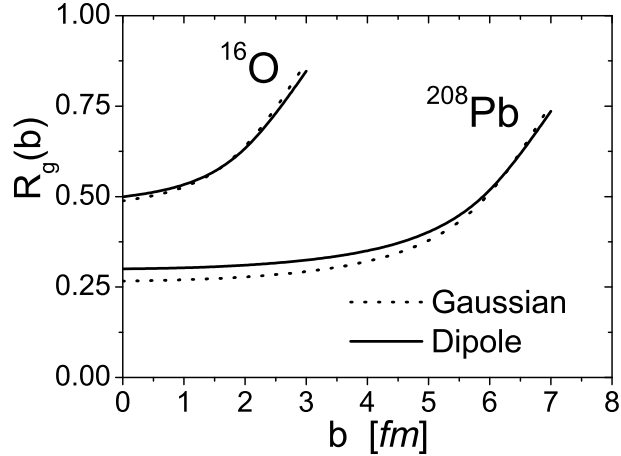,scale=0.55}
\caption{The maximal value of nuclear shadowing in the gluon channel,
 $R_g(b)$, see Eq.~(\ref{eq:R_g}), for $^{16}$O and $^{208}$Pb
    calculated using the Gaussian (dotted curves) and dipole (solid curves)
forms for the gluon density distribution 
and correlated configurations.
$R_g(b)$ is plotted
 as a function of the impact parameter $b$
    for $m^2_g=0.6$ GeV$^2$ corresponding to $x\sim 10^{-4}$ and $Q^2= 4$ GeV$^2$.}
\label{Fig6}
\end{center}
\end{figure}

\begin{figure}[!ht]
\begin{center}
\epsfig{file=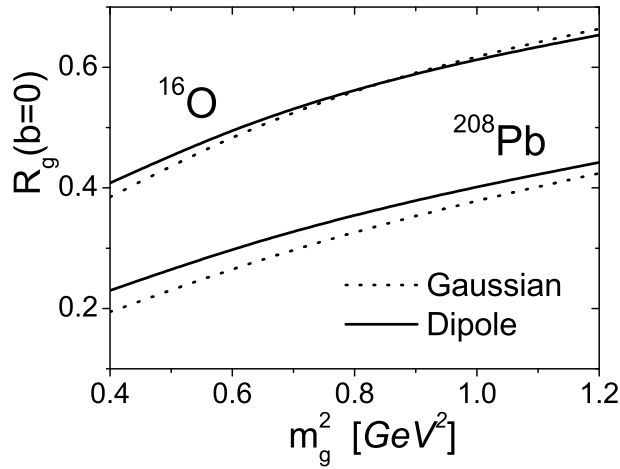,scale=0.55}
\caption{$R_g(b)$ as a function of $m^2_g$ at $b=0$. For the details, see the 
caption of Fig.~\ref{Fig6}.}
\label{Fig7}
\end{center}
\end{figure}

The described procedure allows one to determine the maximal value of the gluon shadowing at a given
$b$, $R_g(b)$,
which is given by
the ratio of the lower limit on $g_A(x,b)$ 
to its value in the impulse
approximation, $g_A(x,b)=g_N(x)T_A(b)$:
\begin{equation}
R_g(b)=\frac{g_A(x,b, Q^2)_{|{\rm min}}}{g_N(x)T_A(b)} \,.
\label{eq:R_g}
\end{equation}
The results for 
$R_g(b)$ for $^{16}$O and $^{208}$Pb and
the two models of the gluon GPD of Eq.~(\ref{eq:two_gluon_ff})
are presented in Figs.~\ref{Fig6} and \ref{Fig7}. 
The dotted curves correspond to the calculation with $F^{(1)}_g(\vec{\rho})$
(Gaussian form); the solid curves correspond to $F^{(2)}_g(\vec{\rho})$
(dipole form).
Figure~\ref{Fig6} presents $R_g(b)$
 as a function of the impact parameter $b$ and plotted
 for $m^2_g=0.6$ GeV$^2$.

Figure~\ref{Fig7} presents $R_g(b)$ as a function of 
$m^2_g$ at $b=0$. [Note that $m^2_g$
is a parameter in Eq.~(\ref{eq:R_g}); its variation matches the change of $x$.]

In the $A\rightarrow\infty$ limit, the results of the two parameterizations differ by
$F^{(1)}_g(0)/F^{(2)}_g(0)=2/3.24=0.62$. However, this limit is reached at extremely
large $A\gg 10^3$.
The reasons for this are
 a rather small radius of the gluon density
in the transverse plane, 
$r^{\rm tr}_g\le 0.5$ fm, and low nuclear density, which 
in combination lead
to a small probability for more than three nucleons to significantly screen each other
up to very large $A$. 

It is important to point out that the results of our calculations of nuclear shadowing
in the impact parameter dependent nuclear gluon PDF discussed in Sec.~\ref{subsec:impact}
are consistent with the upper limit on nuclear shadowing, $R_g(b)$,
discussed in this subsection and presented in Figs.~\ref{Fig6} and \ref{Fig7}.
Indeed, at $Q^2 \sim 4$ GeV$^2$ and 
for $x\sim 10^{-4}$, where $m_g^2\sim 0.6$ GeV$^2$, the limit for $^{208}$Pb from 
Fig.~\ref{Fig6} is $ R_g(x=10^{-4},b=0) \ge  0.3$. At the same time, our calculation 
for $^{208}$Pb
in the framework of the leading twist theory of nuclear shadowing predicts 
for $x=10^{-4}$ and $b=0$
that
$g_A(x,b,Q^2)/[AT_A(b) g_N(x,Q^2)]=0.33$ (FGS10\_H) and
$g_A(x,b,Q^2)/[AT_A(b) g_N(x,Q^2)]=0.51$ (FGS10\_L), see Fig.~\ref{fig:LT2009_ca40_impact}.

The discussed 
results
give another illustration of the observation that realistic nuclei can be treated 
as rather dilute systems in the processes involving nuclear shadowing with large fluctuations of the number of involved nucleons, even at small impact parameters.

\subsection{Diffraction in DIS and
the QCD factorization theorem}
\label{subsec:ft}

\subsubsection{Nucleon fragmentation in DIS}

In 
DIS a struck parton is removed from the nucleon and moves with a large momentum relative to the spectator system. The struck parton and spectator system
fragment into 
separate groups of 
hadrons. (Hadrons at the central  rapidities
may belong to either of the  groups.)  It is convenient to consider the process in 
the Breit frame where the nucleon momentum 
$P\to \infty $ 
and the photon momentum is aligned along the same axis: $\vec{q}=-2x\vec{P}$
and $q_{\mu}= 0$ for all other components.
In the parton model approximation, 
 the final quark flies with the momentum $-xP$ in the opposite direction 
with respect to the residual system
that carries the momentum $(1-x)P$. 
As a result, a hadron in the target fragmentation region can be produced with the maximal 
light-cone fraction $z$ relative to the incident nucleon: $z \le (1-x)$.
For large $x\ge 0.1$, the process corresponds to the removal of the valence quark from the nucleon and creation of a color flow between the current and target fragmentation regions. 
As a result, 
for such $x$, the distribution in the variable 
$x_F= z/(1-x)$ should go to zero at the kinematic limit $x_F\to 1$~\cite{Frankfurt:1981mk,Frankfurt:1977vc}.
(This kinematic limit follows from the requirement that the minus component of the four momentum of 
the system $X$ should be positive.
The actual dependence on $x_F$ follows from details of the QCD dynamics and is often
parameterized in terms of quark counting rules.)
With a decrease of $x$, the dynamics changes; 
hence, the shape of the distribution $z(x_F)$ should depend on $x$.

\subsubsection{Diffractive structure functions and diffractive PDFs }

\begin{figure}[h]
\begin{center}
\epsfig{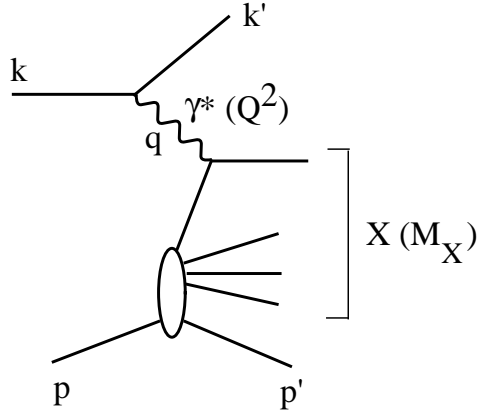}
\caption{Diffractive production of a hadron with momentum p$^{\prime}$ in the nucleon fragmentation region in  DIS.}
\label{fig:Diffraction}
\end{center}
\end{figure}

Most of the HERA experimental studies were performed at small $x$. In this case,
 one often uses the variable $x_{\Pomeron}=1-z$. 
The cross section for the process $ep\to e + p +X$ (or production of any other hadron), 
see Fig.~\ref{fig:Diffraction},
is usually parameterized in the following form:
\begin{equation}
\frac{d^4 \sigma_{ep}^D}{dx_{\Pomeron}\,dt\, dx\, dQ^2}=\frac{2 \pi \alpha^2}{x Q^4} 
\left[\left(1+(1-y)^2\right) F_2^{D(4)}(x,Q^2,x_{\Pomeron},t)-y^2F_L^{D(4)}(x,Q^2,x_{\Pomeron},t)
\right] \,,
\label{eq:ft1}
\end{equation}
where $Q^2$ is the virtuality of the exchanged photon; 
$x=Q^2/(2 p\cdot q)$ is the Bjorken
variable; 
$y=(p\cdot q)/(p \cdot k)$ is the fractional energy loss of the incoming
lepton.  
We follow here the notations commonly used  
for the description of phenomena in the small $x$ kinematics;
in order to emphasize the role of small $x_{\Pomeron}$ processes, 
one introduces the superscript ''$D$'' denoting
$F_2^{D(4)}$ and $F_L^{D(4)}$ as the diffractive structure functions
(the superscript ''(4)'' denotes that the structure 
functions depend on four variables). 
(Note that in the case of  generic 
$x$ and $z$, 
these
quantities are often referred to 
as fracture functions~\cite{Trentadue:1993ka}.)
The variables $x_{\Pomeron}$ and $t$ are expressed through the experimentally 
measured quantities:
\begin{eqnarray}
t&=&(p^{\prime}-p)^2 \,, \nonumber\\
x_{\Pomeron}&=&\frac{q \cdot (p-p^{\prime})}{q \cdot p} \approx \frac{M_X^2+Q^2}{W^2+Q^2} \,, \nonumber\\
\beta &=&\frac{Q^2}{2q \cdot (p-p^{\prime})}=\frac{x}{x_{\Pomeron}} \approx 
\frac{Q^2}{Q^2+M_X^2} \,,
\label{eq:ft2}
\end{eqnarray}
where $M_X$ is the invariant mass of the  produced final state;
$W^2$ is the invariant mass squared of the $\gamma^{\ast} p$ system (see Fig.\ref{fig:Diffraction}).
The variable $x_{\Pomeron}$ describes the fractional loss of the proton
longitudinal momentum; we also defined here $\beta$ which is the longitudinal momentum fraction 
with respect to $x_{\Pomeron}$
carried by the 
interacting parton (to the leading order in $\alpha_s$).
Note that the 
contribution of the 
term proportional to $F_L^{D(4)}$ 
in Eq.~(\ref{eq:ft1}) is
kinematically suppressed and usually neglected in the analysis of 
diffraction.

In pQCD a parton with a virtuality $Q_0^2$ is resolved at higher 
$Q^2$ leading to the scaling violations. If a parton at the resolution scale  $(x,Q^2)$ 
is removed,  
the final state in the fragmentation region will be changed as compared to the
removal of a parent parton at the scale $(x_0, Q^2_0)$.  
The difference  is due to the emission of partons in the evolution process and 
fragmentation  of the struck quark. However,  partons produced in the hard process of 
the evolution from scale $Q_0$ to scale $Q$  have the transverse momenta 
$\ge Q_0$ and, hence, their overlapping integral with a low $p_t$ 
and 
finite $z$ hadron 
is suppressed by a power of $Q_0^2$~\cite{Frankfurt:1997ij}. 
The quark-gluon system produced in the hard interaction is well  localized in 
the transverse directions and, hence, should  interact with the target 
in the same way as the parton at  $(x_0, Q_0^2)$. 
As a result, the $Q^2$ evolution of the fragmentation functions for fixed $t$ and $z$
 is given by the same DGLAP equations as those for the nucleon PDFs~\cite{Collins:1997sr,Frankfurt:1997ij}.  This result follows from the fact that QCD evolution occurs in both cases off a single parton. 
The kinematical window appropriate for the onset of the applicability of the QCD
factorization theorem depends on the
interplay between $z$ and $x$:  (i) the selection of smaller $x$ increases the contribution
of higher-twist effects, and (ii) the products of the
hard parton fragmentation tend to fill the rapidity gap between the photon and target
fragmentation regions,
especially in the case when this parton carries a small fraction $z$ of the photon
momentum. Thus,
larger $Q_0$ is necessary to suppress the both effects.

Similarly to the inclusive case, 
the factorization theorem for  diffraction (production of a hadron with  fixed $z$ and $t$) 
in DIS states that,
at given fixed $t$ and $x_{\Pomeron}$ and in the leading twist approximation, the diffractive structure function $F_2^{D(4)}$ 
is given by 
the convolution of the same hard scattering coefficient functions $C_j$ with universal diffractive parton distributions $f_j^{D(4)}$:
\begin{equation}
F_2^{D(4)}(x,Q^2,x_{\Pomeron},t)=\beta\sum_{j=q,\bar{q},g}  \int_{\beta}^{1} \frac{d y}{y}C_j (\frac{\beta}{y},Q^2) f_j^{D(4)}(y,Q^2,x_{\Pomeron},t) \,,
\label{eq:ft3_a}
\end{equation}
where $\beta=x/x_{\Pomeron}$.
The diffractive PDFs $f_j^{D(4)}$ are conditional probabilities to find a parton of
flavor $j$ with a light-cone fraction $\beta$ in the proton that undergoes 
diffractive scattering characterized by the longitudinal momentum fraction $x_{\Pomeron}$ and the momentum transfer $t$, see Sec.~\ref{subsec:ft} and 
\ref{subsec:diffdata} for details.

\subsubsection{Diffractive dynamics in DIS}

DIS at finite $x$ creates a color flow between the current and target fragmentation regions leading to 
a strong break-up of the nucleon since a typical nucleon carries a relatively small 
light-cone fraction of the initial nucleon momentum (remember that $z > 1-x $ 
is kinematically forbidden in this case).  Hence,  the HERA observation of the significant diffraction in DIS at small $x$  came as a surprise to the theorists not used to the small $x$ dynamics
since pQCD  and the confinement of color
do not allow rapidity gaps.

The key to resolving   this puzzle has been provided long time ago by the 
aligned jet model (AJM)~\cite{Bj71}. 
The model was proposed to address the  
Gribov paradox consisting in the observation that
if all configurations in the virtual 
photon wave  function interacted with large hadronic strengths with nuclei,
the  Bjorken scaling would be grossly violated at small $x$. 
Bjorken has demonstrated  that if one follows the spirit of the parton model and allows 
only the interactions of the partons with small $k_t$, the scaling is restored. 
 The dominant configurations in the photon wave function are the
 $q\bar q$ pairs with the invariant masses of the order of $Q^2$ and transverse momenta 
$k_{\rm soft}$. In the rest frame of the target, the partons carry the
momenta  $p_1 \sim q_0$ and $p_2= k_{\rm soft}^2/(2xm_N)$. 
In coordinate space, the process proceeds as follows: 
$\gamma^{\ast}$ transforms into a $q\bar q$ pair  with the momenta $\pm k_{\rm soft}$ 
at  a large distance $1/(2m_Nx)$ from the target.
After covering this distance to the target, the $q\bar q$ pair
 has the transverse separation which is of the order of $1/k_{\rm soft}$ and the 
system can interact with the typical hadronic strength.

 In QCD one needs to modify the AJM to account for two effects~\cite{Frankfurt:1988nt}. 
One is the Sudakov form factor:  $\gamma^{\ast}$ cannot transform into a $q\bar q$
pair with small $k_t$ without gluon radiation. 
This effect  is taken into account by the pQCD evolution (change of $x$ of the parton).
It does not change the transverse size of the system and,  as a result, 
the system interacts with the same strength at large $Q^2$.
The second modification  is the presence of large $k_t$ configurations 
that have small transverse sizes. Their interaction is suppressed by the factor $\alpha_s(k_t)^2/k_t^2$---the color transparency effect. 
 However, due to a large phase volume, these configurations give a contribution 
comparable to that of the AJM.  (The estimate of \cite{Frankfurt:1988nt,Frankfurt:1988zg} suggested that the AJM contributes about 70\% to $F_{2p}(x\sim 10^{-2}, Q_0^2\sim 2 \div 3\ {\rm GeV}^2)$.)

While diffraction for the AJM 
configurations
is expected to be comparable to that of hadrons, 
it is  strongly suppressed for small size configurations 
for moderate $x > 10^{-3}$ 
since the strength of the interaction enters quadratically in the diffractive cross section.

 The dominance of the AJM configurations leads to the expectation that 
the $W$ dependence of diffraction at fixed $Q^2$ and $M_X^2$ should be close 
to that  for soft processes~\cite{Abramowicz:1995hb}.  
 Another  important
contribution to diffraction
is 
due to large size color octet dipoles ($q\bar q g$ configurations in the virtual photon).  These predictions are  in a good agreement with the current HERA data, see below. 
 
It is also instructive to consider diffraction in the Breit frame. It is easy to see that 
the AJM contribution corresponds to the following process: a parton with the light-cone 
fraction $x$ absorbs $\gamma^{\ast}$ and  turns around so that it has the momentum $(xP,-xP)$.
 To produce a color neutral system with the typical mass squared $M_X^2\approx Q^2$, 
it has to pick up a parton with the momentum $(x^{\prime}P,x^{\prime}P)$ leading to $M_X^2=Q^2(x^{\prime}/x)$ and pull it out of the nucleon. This implies that although the diffraction involves the absorption of $\gamma^{\ast}$ by one parton,    
it requires the presence of a strong short-range  correlation in rapidity between the partons in the nucleon light-cone wave function~\cite{Abramowicz:1995hb}. 
A nearly hadron-level strength of the diffraction indicates that a strong color screening 
takes place in the  proton wave function  for small $x$ {\it locally} in $x$ 
(in  rapidity $\Delta Y=\ln x^{\prime}/x$).

\subsection{Summary of QCD analysis of the data on hard diffraction at HERA}
\label{subsec:diffdata}

\subsubsection{Diffractive structure function $F_2^{D(3)}$} 

The bulk of the data on diffraction in DIS at HERA comes from 
inclusive measurements performed by 
 H1 and ZEUS collaborations~\cite{Adloff:1997sc,Aktas:2006hy,Aktas:2006hx,Aktas:2007bv,Aktas:2007hn,Aktas:2006up,Breitweg:1997aa,Breitweg:1998gc,Chekanov:2003gt,Chekanov:2004hy,Chekanov:2007yw,Chekanov:2008cw,Chekanov:2008fh,Chekanov:2009qja}.
When the $t$ dependence of the diffractive cross section is not 
measured~\cite{Adloff:1997sc,Aktas:2006hy,Breitweg:1998gc,Chekanov:2008cw,Chekanov:2009qja}, the data 
are analyzed in 
terms of the diffractive structure function~$F_2^{D(3)}$:
\begin{equation}
F_2^{D(3)}(x,Q^2,x_{\Pomeron})=\int_{-1\, {\rm GeV}^2}^{t_{{\rm min}}} dt F_2^{D(4)}(x,Q^2,x_{\Pomeron},t) \,,
\label{eq:data1}
\end{equation}
where $F_2^{D(4)}$ is defined by Eq.~(\ref{eq:ft1});
$t_{{\rm min}} = -m_N^2 x_{\Pomeron}^2/(1-x_{\Pomeron}) \approx -m_N^2 x^2 (1+M_X^2/Q^2)^2$ 
with $m_N$ the nucleon mass.

The weak (logarithmic) $Q^2$ dependence of $F_2^{D(3)}$, which follows from
the QCD evolution equations for diffractive PDFs,  was observed 
experimentally, see, e.g., Fig.~\ref{fig:FitAdata_scaled} below.

As we discussed above the diffractive structure function $F_2^{D(3)}$ is given in terms of the diffractive PDFs $f_j^{D(3)}$:
\begin{equation}
F_2^{D(3)}(x,Q^2,x_{\Pomeron})=\beta\sum_{j=q,\bar{q},g}  \int_
{\beta}^{1} \frac{d y}{y}C_j (\frac{\beta}{y},Q^2) f_j^{D(3)}(y,Q^2,x_
{\Pomeron})  \,.
\label{eq:data2}
\end{equation}
Extensive studies of hard inclusive diffraction at HERA were performed both by H1 and 
ZEUS collaborations~\cite{Adloff:1997sc,Aktas:2006hy,Aktas:2006hx,Aktas:2007bv,Aktas:2007hn,Aktas:2006up,Breitweg:1997aa,Breitweg:1998gc,Chekanov:2003gt,Chekanov:2004hy,Chekanov:2007yw,Chekanov:2008cw,Chekanov:2008fh,Chekanov:2009qja}. 
Within the normalization uncertainties, 
the measurements of the two collaborations are in good agreement, 
see, e.g., the comparison in Ref.~\cite{Chekanov:2008fh}. 

It was suggested in \cite{Ingelman:1984ns} that diffraction in hard process can be treated as scattering off a $t$-channel exchange---Pomeron---which has the same properties for different $x_{\Pomeron}$. We have argued above that the dominant source of the diffraction in DIS 
is the AJM-like configurations in the virtual photon. 
In a wide energy range, these hadron-like configurations should interact through a coupling 
to a soft ladder.  
The properties of such a ladder (or a multiladder system), which are local in rapidity,
 should weakly depend on its length  in rapidity proportional to $ \ln (x_0/x_{\Pomeron})$, 
where $x_0\sim 0.01$.

 In line with the suggestion of \cite{Ingelman:1984ns}, the QCD 
analyses of the HERA diffractive data make an additional {\it soft / Regge factorization} assumption 
(which does not contradict the data)
that 
DPDFs $f_j^{D(3)}$  can be  presented as a sum of 
the leading Pomeron-exchange term and the subleading Reggeon-exchange term (the latter plays a role only at large $x_{\Pomeron}$). 
Each of the terms is given as the product of the corresponding flux factors
and the parton distribution functions,
\begin{equation}
f_j^{D(3)}(\beta,Q^2,x_{\Pomeron})=f_{\Pomeron/p}(x_{\Pomeron}) 
f_{j/\Pomeron}(\beta,Q^2)+n_{\Reggeon} f_{\Reggeon/p}(x_{\Pomeron}) 
f_{j/\Reggeon}(\beta,Q^2) \,,
 \label{eq:data3}
\end{equation}
where $f_{\Pomeron/p}(x_{\Pomeron})$ is the Pomeron flux factor;
$f_{\Reggeon/p}$ is the Reggeon flux factor; $f_{j/\Pomeron}(\beta,Q^2)$ 
can be interpreted as the PDF of flavor $j$ of the Pomeron;  
$f_{j/\Reggeon}(\beta,Q^2)$ are PDFs of the subleading Reggeon;
$n_{\Reggeon}$ is a small free parameter determined from the fit to the data.
The $Q^2$ dependence of $f_{j/\Pomeron}(\beta,Q^2)$ is given by the DGLAP evolution
equations.

Note that Eq.~(\ref{eq:data3}) does not follow from the  
QCD factorization theorem, but 
it is rather a hypothesis of the soft matching to the non-perturbative QCD, which is supported by
the data (see the discussion below).

\vspace*{0.5cm}
\begin{figure}[h]
\begin{center}
\epsfig{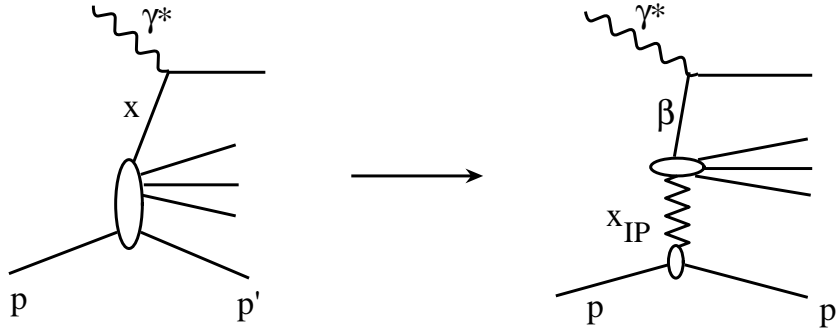}
\caption{A schematic representation of the  factorization of the diffractive PDFs into the product
 of the Pomeron or Reggeon flux factor and the corresponding PDFs, see Eq.~(\ref{eq:data3}).}
\label{fig:ReggeFactorization}
\end{center}
\end{figure}

The schematic view of the separation of $f_j^{D(3)}$ into the flux factors and the corresponding
PDFs used in Eq.~(\ref{eq:data3}) is presented in Fig.~\ref{fig:ReggeFactorization}.
The figure also illustrates the physical interpretation of the variable $\beta$: 
$\beta$ is the light-cone fraction of the Pomeron (or Reggeon) momentum
carried by the interacting parton.

It is important to emphasize that the words ''Pomeron'' and ''Reggeon'' are used 
in the analysis of the hard diffraction in DIS only as bookeeping terms
since those terms are reserved for soft hadron-hadron interactions.
The parameters (intercepts, slopes, etc.) of the Pomeron and Reggeon
exchanges as determined from the phenomenology of  soft hadron-hadron
interactions may differ from the parameters obtained from the fits to the
hard diffractive data at HERA.

In Eq.~(\ref{eq:data3}), the Pomeron and Reggeon flux factors have the following form:
\begin{eqnarray}
f_{\Pomeron/p}(x_{\Pomeron})&=&\int_{-1\, {\rm GeV}^2}^{t_{{\rm min}}} dt\,A_{\Pomeron} \frac{e^{B_{\Pomeron}t}}{x_{\Pomeron}^{2 \alpha_{\Pomeron}(t)-1}} \,,
\quad \alpha_{\Pomeron}(t)=\alpha_{\Pomeron}(0)+\alpha_{\Pomeron}^{\prime} t \,,
\nonumber\\
f_{\Reggeon/p}(x_{\Pomeron})&=&\int_{-1\, {\rm GeV}^2}^{t_{{\rm min}}} dt \,A_{\Reggeon} \frac{e^{B_{\Reggeon}t}}{x_{\Pomeron}^{2 \alpha_{\Reggeon}(t)-1}} 
\,,
\quad \alpha_{\Reggeon}(t)=\alpha_{\Reggeon}(0)+\alpha_{\Reggeon}^{\prime} t \,.
\label{eq:data4}
\end{eqnarray}
In the following, we concentrate on the result of the QCD analysis of the
hard inclusive diffraction at HERA by the 
H1 collaboration~\cite{Aktas:2006hy,Aktas:2006hx}
since we used the H1 Fit B as 
an input for our calculations of nuclear shadowing
(the QCD analysis of hard diffraction by the ZEUS collaboration will be discussed
in the end of this subsection).

The H1 QCD fit gives
$B_{\Pomeron}=5.5$ GeV$^{-2}$;
 $\alpha_{\Pomeron}^{\prime}=0.06$ GeV$^{-2}$;
$B_{\Reggeon}=1.6$ GeV$^{-2}$;
$\alpha_{\Reggeon}(0)=0.5$;
$\alpha_{\Reggeon}^{\prime}=0.3$ GeV$^{-2}$.
The coefficients $A_{\Pomeron}$ and $A_{\Reggeon}$ are 
found from the conditions
$x_{\Pomeron} f_{\Pomeron/p}(x_{\Pomeron})=1$ and 
$x_{\Pomeron} f_{\Reggeon/p}(x_{\Pomeron})=1$ at 
$x_{\Pomeron}=0.003$.
The intercept of the Pomeron trajectory, $\alpha_{\Pomeron}(0)$, is a free parameter of the fit to the data.

The fit to the HERA data on hard diffraction in DIS is carried out as follows.  One assumes a particular shape of $f_{j/\Pomeron}$ at a certain value of $Q^2=Q^2_0$  ($Q_0^2=1.75-2.5$ GeV$^2$),
\begin{equation}
\beta f_{j/\Pomeron}(\beta,Q^2_0)=A_j \beta^{B_j} (1-\beta)^{C_j} \,,
\label{eq:data5}
\end{equation}
where $A_j$, $B_j$ and $C_j$ are free parameters. Since the Pomeron exchange 
is a flavor-singlet, it is assumed that $f_{u/\Pomeron}=f_{{\bar u}/\Pomeron}=
f_{d/\Pomeron}=f_{{\bar d}/\Pomeron}=f_{s/\Pomeron}=f_{{\bar s}/\Pomeron}$.
The theoretical prediction for the diffractive structure function $F_2^{D(3)}$
at given $x$, $Q^2$ and $x_{\Pomeron}$ is obtained using Eqs.~(\ref{eq:data2}),
(\ref{eq:data3}), (\ref{eq:data4}) and  (\ref{eq:data5}).
The $\chi^2$ fit to the experimental values of $F_2^{D(3)}$ determines the free parameters
of the fit: $n_{\Reggeon}$, $\alpha_{\Pomeron}(0)$, $A_j$, $B_j$ and $C_j$.

The 2006 H1 data on diffraction in $ep \to eXY$ DIS ($Y$ denotes
products of the proton dissociation)~\cite{Aktas:2006hy,Aktas:2006hx} 
covers the following kinematics:
$3.5 \leq Q^2 < 1600$ GeV$^2$, $0.0003 < x_{\Pomeron} < 0.03$,
$0.0017 < \beta < 0.8$, $|t| <1$ GeV$^2$.
Since the diffractive events were reconstructed using the 
rapidity gap selection method, the proton was allowed to dissociate into
states with a low invariant mass,
 $M_Y < 1.6$ GeV. In order to avoid the kinematic regions which are most likely to be 
influenced by higher twist contributions, only the data with $Q^2 \geq 8.5$ GeV$^2$ and
$M_X^2 > 2$ GeV were included in the QCD analysis (fit).

\begin{figure}[h]
\begin{center}
\epsfig{file=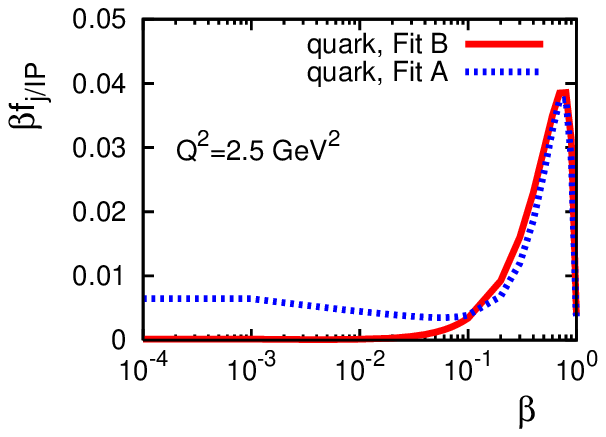,scale=1.2}
\epsfig{file=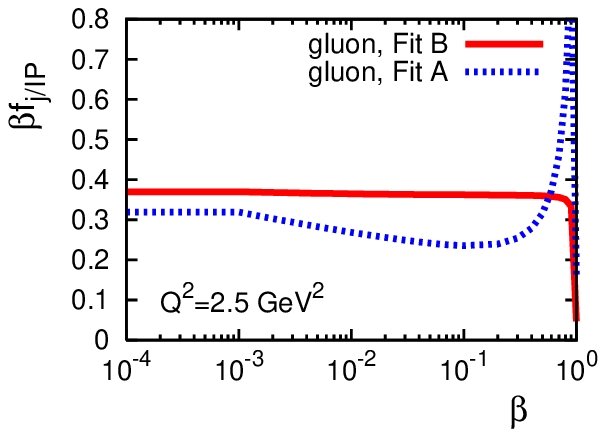,scale=1.2}
\caption{The diffractive quark and gluon PDFs $f_{j/\Pomeron}(\beta,Q^2)$  at $Q^2=2.5$ GeV$^2$ as functions of $\beta$.}
\label{fig:FitAB}
\end{center}
\end{figure}

The results of the H1 QCD fit in terms of 
the diffractive 
quark and gluon PDFs, $f_{u/\Pomeron}(\beta,Q^2)$ and 
$f_{g/\Pomeron}(\beta,Q^2)$, at $Q^2=2.5$ GeV$^2$ as functions of $\beta$ 
are presented in Fig.~\ref{fig:FitAB}. The solid curves correspond to fit~B; 
the dotted curves correspond to fit~A.
The difference between fits A and B is that while the parameters $A_j$,
$B_j$ and $C_j$ in Eq.~(\ref{eq:data5}) are free in fit~A, $C_g=0$ for
the gluon PDF in Fit~B.

The need to have two types of fits is explained by the fact that the gluon 
diffractive PDF is determined from the scaling violations of 
$F_2^{D(3)}$. However, at large $\beta$, the scaling violations of 
$F_2^{D(3)}$ are predominantly determined by the quark diffractive PDFs. 
Therefore, the gluon diffractive PDF at large $\beta$ 
is very weakly constrained by the data, which allows one (requires) to consider
two scenarios (fits A and B) of the gluon diffractive PDFs with 
a different
behavior in the large-$\beta$ limit, see the right panel of Fig.~\ref{fig:FitAB}.

Note that the large support of the diffractive PDFs at large $\beta$ means that  the diffraction is enhanced in the $M_X^2/Q^2 \sim 1$ region, resulting in a smaller relative contribution of  the triple Pomeron contribution to diffraction, 
see Sec.~\ref{subsubsec:diffractive_masses}.

One should mention that both fits A and B correspond to very similar values of
$\alpha_{\Pomeron}(0)$ and $n_{\Reggeon}$:
\begin{eqnarray}
{\rm Fit \ A:}\ \alpha_{\Pomeron}(0)&=&1.118 \pm 0.008 \,, \quad  n_{\Reggeon}=(1.7\pm 0.4) \times 10^{-3}  \,, \nonumber\\
{\rm Fit \ B:}\ \alpha_{\Pomeron}(0)&=&1.111 \pm 0.005 \,, \quad  n_{\Reggeon}=(1.4\pm 0.4) \times 10^{-3}
\,.
\label{eq:data6}
\end{eqnarray}
It is important to note that these values of the Pomeron intercept 
$\alpha_{\Pomeron}(0)$ are very close to the one observed for soft
hadron-hadron interactions, $\alpha_{\Pomeron}(0)=1.0808$~\cite{Donnachie:1992ny}.
As we explained in Sec.~\ref{subsec:derivation}, this justifies the use
of the color fluctuation
 approximation for the interaction with three and
more nucleons of the nuclear target.

As seen from Fig.~\ref{fig:FitAB}, the gluon diffractive PDF is much larger than 
the quark one. We shall later show that this will lead to the prediction that the
leading twist nuclear shadowing for the gluon nuclear PDF is larger than that
for the quark nuclear PDFs.

In the analyses~\cite{Aktas:2006hy,Aktas:2006hx},
the PDFs of the subleading Reggeon exchange, $f_{j/\Reggeon}$, are taken to be those of the pion~\cite{Owens:1984zj}.
The $\beta$ and $Q^2$ dependence of $f_{j/\Reggeon}(\beta,Q^2)$ are given by the fit 
to the $\pi N \to J/\Psi X$ and $\pi N \to \mu^+ \mu^-X$ data.

\begin{figure}[h]
\begin{center}
\epsfig{file=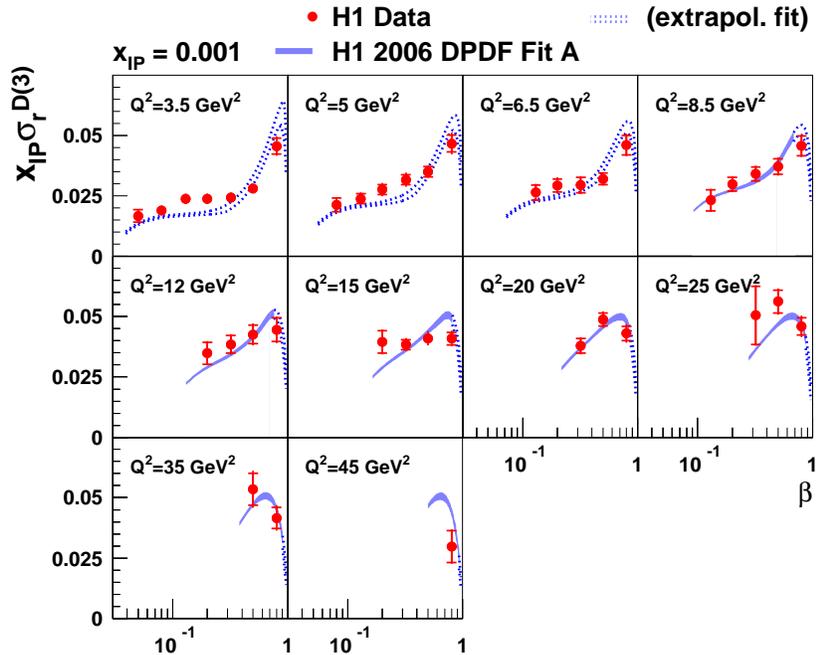,scale=0.65}
\caption{The perturbative QCD description of the H1 LRG diffractive data on
$x_{\Pomeron} \sigma_r^{D(3)} \approx x_{\Pomeron} F_2^{D(3)}$.
The figure is from Ref.~\protect\cite{Aktas:2006hy}.
Reproduced with the kind permission of the H1 Collaboration and Springer.}
\label{fig:FitAdata}
\end{center}
\end{figure}

Both fits A and B provide a good description of the H1 data on hard inclusive diffraction in DIS over the entire kinematic range~\cite{Aktas:2006hy}. 
The subleading Reggeon contribution
is required only at the large-$x_{\Pomeron}$ end of the covered range: $x_{\Pomeron}> 0.01$.
An example of the good agreement between the H1 data~\cite{Aktas:2006hy} and
its perturbative QCD description is presented in
Fig.~\ref{fig:FitAdata} (taken from Ref.~\cite{Aktas:2006hy}).
The figure shows the reduced cross section
multiplied by $x_{\Pomeron}$,
$x_{\Pomeron} \sigma_r^{D(3)} \approx x_{\Pomeron} F_2^{D(3)}$,
as a function of $\beta$ at fixed $x_{\Pomeron}=0.001$ for a wide range of $Q^2$.
The solid curves correspond to fit~A in the kinematic region used in the fit,
$Q^2 > 8.5$ GeV$^2$ and $\beta < 0.8$.
The dotted curves  correspond to the extrapolation of fit~A beyond the kinematic region used in the fit.
As one can see from Fig.~\ref{fig:FitAdata}, the QCD fits provide not only a 
good description of the data used in the fit ($Q^2 \geq 8.5$ GeV$^2$), but also of
the H1 data with 
$Q^2 < 8.5$ GeV$^2$ that were not used in the QCD fit.

One should note that the pattern of the scaling violations at large $\beta$ presented in
Fig.~\ref{fig:FitAdata} is opposite to that of the inclusive structure function 
$F_2(x,Q^2)$: it is a consequence of the large gluon diffractive PDF.

Besides the diffractive PDFs obtained by the H1 collaboration that we have just
discussed, the ZEUS collaboration performed the next-to-leading order QCD analysis~\cite{Chekanov:2009qja} 
of their own data on inclusive diffraction in DIS~\cite{Chekanov:2008fh}.
As a result, several sets of diffractive PDFs were obtained using essentially
the same method as we discussed above. 
The resulting diffractive PDFs describe well the ZEUS data
sample for $Q^2 > 5$ GeV$^2$ (only this part of the data was
used in the QCD analysis). For $Q^2 < 5$ GeV$^2$, the predictions are
extrapolated and underestimate the data.

In addition,
the ZEUS collaboration performed a QCD fit using both the inclusive diffractive
and diffractive dijet data~\cite{Chekanov:2007yw}. 
An explicit comparison of the predictions of this fit
to the H1 fit B shows that while both fits are consistent with each other and the ZEUS data
on $x_{\Pomeron} \sigma_r^{D(3)}$, the normalization of the predictions of the H1 fit B
is somewhat smaller than that of the ZEUS fit~\cite{Chekanov:2009qja}.

One should also mention that the value of $\alpha_{\Pomeron}(0)$ at low
virtualities $Q^2$ obtained by
the H1 and ZEUS analyses are very close: the H1 value of $\alpha_{\Pomeron}(0)$ in Eq.~(\ref{eq:data6}) should be compared to 
$\alpha_{\Pomeron}(0)=1.11-1.12 \pm 0.02$ obtained by ZEUS~\cite{Chekanov:2009qja}.

\subsubsection{Diffractive structure function $F_2^{D(4)}$} 
\label{subsec:f2d4}

The measurement of the $t$ dependence of hard inclusive diffraction and the structure 
function $F_2^{D(4)}$ can be performed by detecting the final state proton.
This was done using the forward proton spectrometer  
(FPS) by the H1 collaboration~\cite{Aktas:2006hx} and the leading proton
spectrometer (LPS) by the ZEUS collaboration~\cite{Chekanov:2008fh}.
In the following, we focus on the H1 results since we used the H1 Fit B as 
an input for our calculations of nuclear shadowing.

In the kinematic range $2 < Q^2 < 50$ GeV$^2$ and $x_{\Pomeron} < 0.02$, the 
$t$ dependence of $F_2^{D(4)}$ was  parameterized in a simple exponential form with
a constant slope,
\begin{equation}
F_2^{D(4)}(x,Q^2,x_{\Pomeron},t)=e^{B_{{\rm diff}} (t-t_{{\rm min}})} F_2^{D(4)}(x,Q^2,x_{\Pomeron},t_{\rm min}) \,,
\label{eq:data7}
\end{equation}
where $B_{{\rm diff}} \approx 6$ GeV$^{-2}~\cite{Aktas:2006hx}$. Note that this value is somewhat lower (but still 
consistent) than the ZEUS LPS result,
$B_{{\rm diff}}=7.0 \pm 0.3$ GeV$^{-2}$~\cite{Chekanov:2008fh}.

After the integration over $t$, the FPS data on $\sigma_r^{D(3)}$~\cite{Aktas:2006hx} can be compared
to the LRG data~\cite{Aktas:2006hy}. A point-by-point comparison shows that 
\begin{equation}
\frac{\sigma_r^{D(3)}({\rm LRG})}{\sigma_r^{D(3)}({\rm FPS})}=1.23 \pm 0.03\,({\rm stat.})
\pm 0.16\,({\rm syst.}) \,.
\label{eq:data8}
\end{equation}
Equation~(\ref{eq:data8}) is interpreted as that the excess of events in the 
LRG method compared to the FPS method must come from the proton dissociation
into the states with the invariant mass $M_Y < 1.6$ GeV.

The FPS method also allows one to find the relation between the sub-leading cross sections
obtained in the two methods:
\begin{equation}
\frac{n_{\Reggeon}({\rm LRG})}{n_{\Reggeon}({\rm FPS})}=1.39 \pm 0.48\,({\rm exp.})
\pm 0.29\,({\rm model}) \,.
\label{eq:data9}
\end{equation}

\begin{figure}[h]
\begin{center}
\epsfig{file=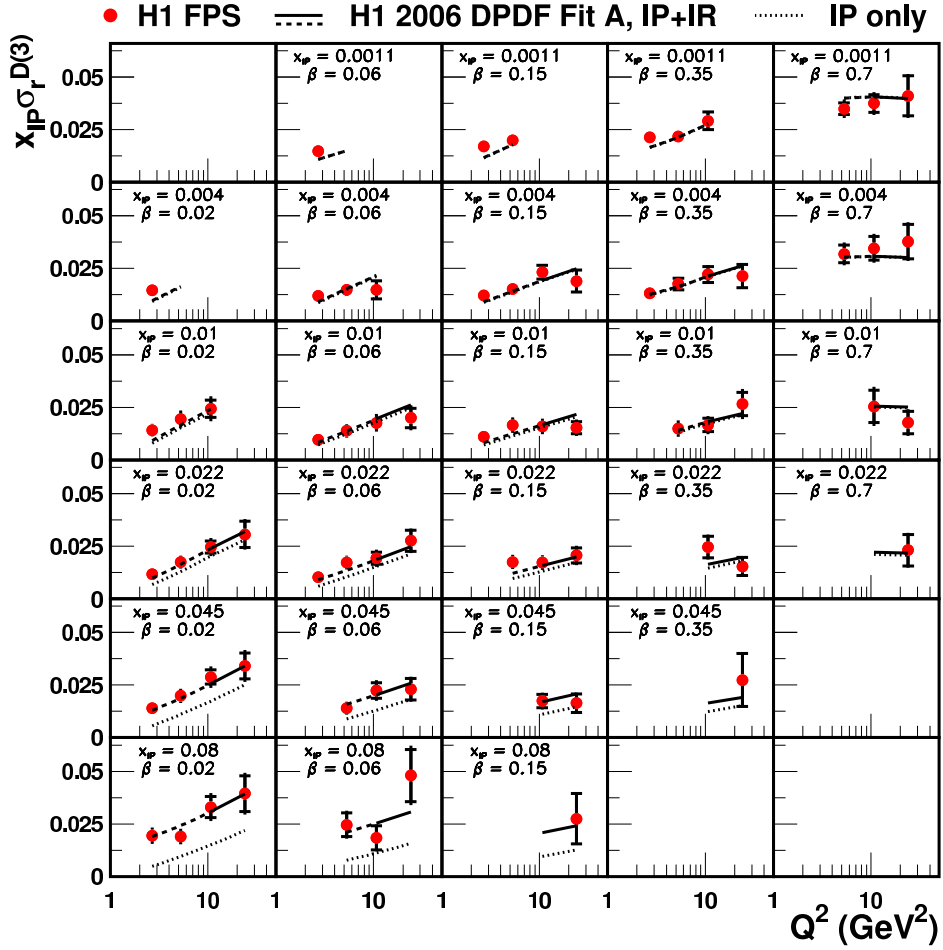,scale=0.65}
\caption{The perturbative QCD description of the H1 FPS diffractive data on
$x_{\Pomeron} \sigma_r^{D(3)} \approx x_{\Pomeron} F_2^{D(3)}$.
The figure is from Ref.~\protect\cite{Aktas:2006hx}.
Reproduced with the kind permission of the H1 Collaboration and Springer.}
\label{fig:FitAdata_scaled}
\end{center}
\end{figure}

Equations~(\ref{eq:data8}) and (\ref{eq:data9}) mean that the QCD prediction for
the diffractive structure function $F_2^{D(3)}$, which would be consistent
with the H1 FPS data~\cite{Aktas:2006hx},
 is obtained by scaling down fits A and B for the Pomeron PDFs by the factor $1.23$ and
 the constant $n_{\Reggeon}$ by the factor $1.39$.
This is illustrated in Fig.~\ref{fig:FitAdata_scaled} (taken from Ref.~\cite{Aktas:2006hx}),
where the scaled QCD predictions
are compared to the H1 FPS data. The solid curves correspond to fit~A in the kinematic 
region used in the fit (see comments for Fig.~\ref{fig:FitAdata});
the dashed curves correspond to fit A extrapolated beyond the kinematic region used
in the fit; the dotted curves correspond to the Pomeron contribution only.
Since the FPS data extend to larger values of $x_{\Pomeron}$, 
Figure~\ref{fig:FitAdata_scaled} clearly indicates the need for the sub-leading Reggeon
contribution for $x_{\Pomeron} > 0.01$.

\subsubsection{Tests of the QCD factorization using other diffractive DIS processes}

The diffractive parton
distributions (DPDFs) $f_j^{D(4)}$ are process-independent 
universal quantities that
enter the pQCD description of such  diffractive processes as
 inclusive DIS diffraction~\cite{Adloff:1997sc,Aktas:2006hy,Aktas:2006hx,Breitweg:1997aa,Breitweg:1998gc,Chekanov:2004hy,Chekanov:2008cw,Chekanov:2008fh,Chekanov:2009qja},
diffractive electroproduction of jets~\cite{Aktas:2007bv,Aktas:2007hn,Chekanov:2007yw},
diffractive photoproduction of
jets~\cite{Aktas:2007hn,Aaron:2010su,Chekanov:2007rh}, diffractive electroproductions of heavy quarks \cite{Aktas:2006up,Chekanov:2002qm},
 and diffractive photoproduction of heavy quarks \cite{Chekanov:2007pm}.
The $Q^2$ dependence of $f_j^{D(4)}$ is given by the DGLAP equations with the same
splitting functions as in the case of inclusive DIS.
Hence, 
a wide range of processes (some of them are mentioned above) can be described
from the first principles in the framework of perturbative QCD 
using universal non-perturbative DPDFs as input.

\begin{figure}[h]
\begin{center}
\epsfig{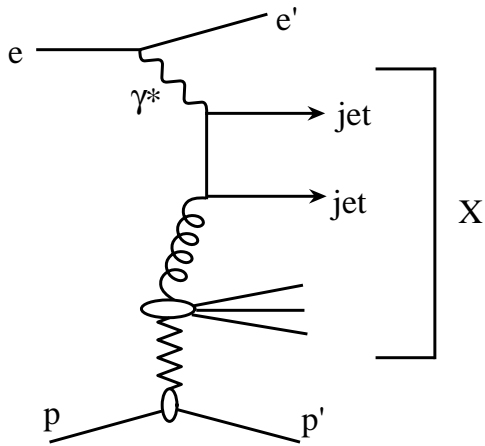}
\caption{Diffractive productions of dijets in DIS.}
\label{fig:DijetsFeynman}
\end{center}
\end{figure}

Measurements of  diffractive DIS processes serve as stringent tests of 
the QCD factorization for hard diffraction and further constrain diffractive PDFs.
One example of such a diffractive process, which predominantly probes the 
gluon diffractive PDF, is diffractive production of dijets, see 
Fig.~\ref{fig:DijetsFeynman}. The figure depicts diffractive production of dijets in
DIS. Replacing the virtual photon by the real (quasi-real) one, it is possible to
study diffractive photoproductions of dijets. In the latter process, the hard scale is given by the transverse momenta of the jets. 

Both H1 and ZEUS collaborations measured diffractive dijet production.
In detail, the H1 collaboration measured 
diffractive dijet production in DIS ($4 < Q^2 < 80$ GeV$^2$) and 
photoproduction ($Q^2 < 0.01$ GeV$^2$)
in the reaction $ep \to e\,{\rm jet_1}\,{\rm jet_2}\,XY$~\cite{Aktas:2007hn,Aaron:2010su}.
 It was found that, in DIS, the data 
are 
 described well by diffractive PDFs extracted from the fits to the H1 data
on inclusive diffraction in DIS~\cite{Aktas:2006hy,Aktas:2006hx}. The dijet data 
clearly favors fit~B, which corresponds to a smaller (compared to fit~A) gluon diffractive PDF $f_{g/\Pomeron}(\beta,Q^2)$ in the large $\beta$ limit, 
see Fig.~\ref{fig:FitAB}.

In photoproduction of dijets, theoretical predictions based on fit~B overestimate the data
by approximately a factor of two
(both for the direct and resolved contributions).
This indicates the breakdown of the QCD factorization
theorem for the photoproduction, similarly to the case of factorization breaking in
hadron-induced diffractive dijet production, see e.g., \cite{Affolder:2000vb}.
One should note that while the factorization breaking is expected for the 
resolved component of the real photon (since the resolved component consists
of hadronic fluctuations interacting with the target with typical, large
hadronic cross sections), it is surprising that the factorization is similarly 
violated for the direct component of the real photon up 
to the large transverse momenta $\sim$ 7 GeV/c~\cite{Aktas:2007hn}.

The ZEUS collaboration performed a combined QCD fit to the data on 
inclusive diffraction and diffractive dijet production in DIS~\cite{Chekanov:2007yw}.
The resulting fit provides a good description of the dijet data throughout the
whole kinematic region~\cite{Chekanov:2009qja}.
The application of the fit to the ZEUS diffractive dijet photoproduction 
data~\cite{Chekanov:2007rh} shows an adequate description of the data over the whole
$x_{\gamma}^{\rm obs}$ and $E_T^{\rm obs}$ ranges~\cite{Chekanov:2009qja}.
Hence, the ZEUS collaboration does not observe the suppression of the 
resolved component (or both the resolved and direct components), 
which appears to be in conflict with the H1 findings~\cite{Aktas:2007hn}
(see our discussion above).
Further theoretical analyses are necessary, including 
a more accurate definition of the direct and resolved processes at the 
next-to-leading (NLO) accuracy. Note that the recent analysis of Klasen and Kramer
has shown that the large majority of the H1 and ZEUS points lay below their NLO
pQCD predictions~\cite{Klasen:2010vk}.

Diffractive open charm production is
another example of  diffractive processes, where the factorization
theorem is expected to be valid. The underlying mechanism of open charm production
is given by Fig.~\ref{fig:DijetsFeynman} after the replacement of the two jets by 
$c$ and $\bar{c}$ quarks. 
The measurement of diffractive open charm ($D^{\ast}$ meson) production at HERA by the H1 
collaboration~\cite{Aktas:2006up} found a good agreement between the data and 
perturbative QCD predictions based on the H1 fits A and B, both in 
DIS and in photoproduction.
Diffractive photoproduction of $D^{\ast}$ mesons was also measured by
the ZEUS collaboration at HERA~\cite{Chekanov:2007pm}. A good agreement
between pQCD calculations and the data was found.
Also, the pQCD predictions based on the ZEUS diffractive PDFs~\cite{Chekanov:2009qja} 
provide a fair
description of the charm contribution to the diffractive structure 
function~\cite{Chekanov:2003gt}.

To summarize, the considered examples of diffractive dijet 
production and diffractive  open charm production illustrate the 
validity of the factorization theorem for  diffraction and of the concept of universal diffractive PDFs~\cite{Collins:1997sr}. 

{\it Comment.}
While the validity of the Regge  factorization supports the
dominance of the soft AJM-type configurations in diffraction, it is hardly consistent with the dominance of the pQCD Pomeron at the HERA energies. 
The latter
 hypothesis leads to $\alpha_{\Pomeron} \sim 1.25 $, which is much 
larger than that found experimentally, see Eq.~(\ref{eq:data6}).

\section{Nuclear shadowing in DIS on deuterium}
\label{sec:deuteron}

In this section, we present the application of the leading twist theory of 
nuclear shadowing to inclusive and tagged DIS on deuterium.
Numerical results presented in this section update our predictions made in~\cite{Frankfurt:2003jf,Frankfurt:2006am}.

\subsection{Nuclear shadowing for 
unpolarized deuteron structure functions}

\begin{figure}[t]
\begin{center}
\epsfig{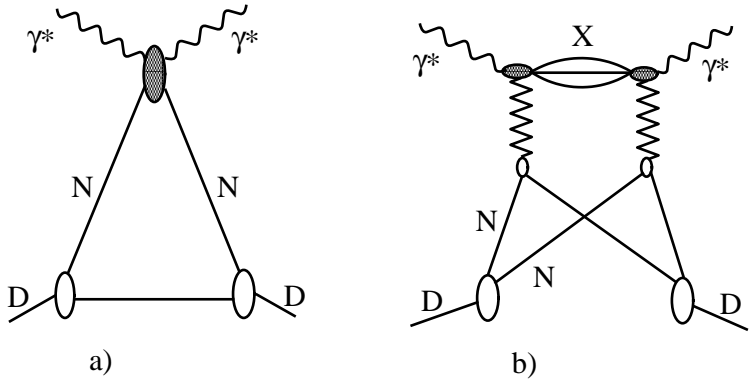}
\caption{
Graphs depicting the contributions to the zero angle $\gamma^{\ast}D$ scattering 
amplitude: 
 (a) the impulse approximation,
(b) the nuclear shadowing correction.
}
\label{fig:Deuteron_F2}
\end{center}
\end{figure}

The deuteron inclusive structure function $F_{2D}(x,Q^2)$
is proportional to the imaginary part of the  forward $\gamma^{\ast}D$ scattering amplitude, which, in the graphical form, 
receives contributions from graphs $a$ and $b$ in Fig.~\ref{fig:Deuteron_F2}:
graph $a$ is the 
impulse approximation 
contribution; graph $b$ is the nuclear shadowing correction. 
These graphs should be compared to the corresponding ones in Fig.~\ref{fig:Master1}.
Note also that graph $c$ in Fig.~\ref{fig:Master1}, which corresponds to the 
interaction with three and more nucleons, is naturally absent in the deuteron
case.

\begin{figure}[h]
\begin{center}
\epsfig{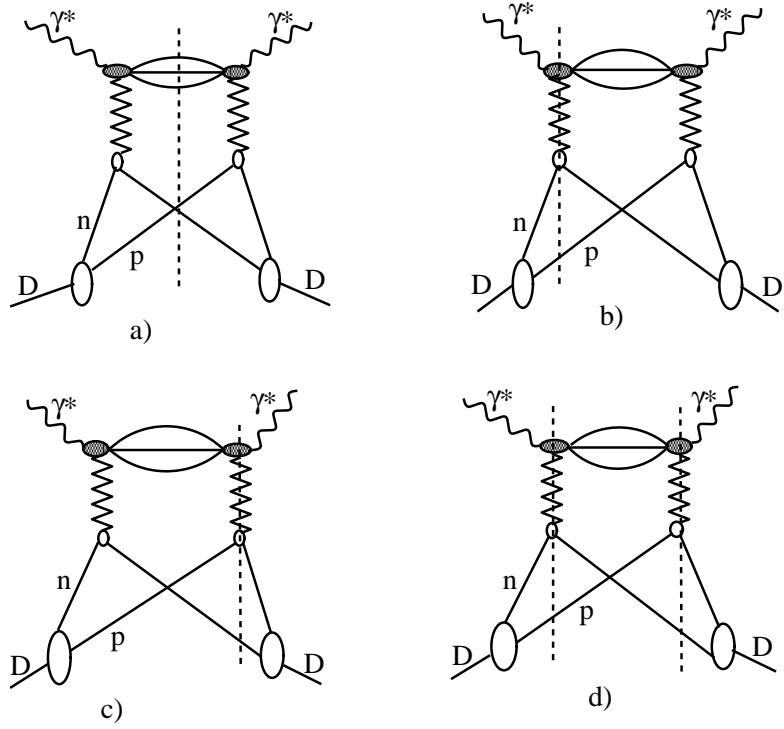}
\caption{Unitarity cuts of the imaginary part of the double scattering contribution to the forward $\gamma^{\ast}D$ scattering amplitude:
(a) diffractive cut, (b) and (c) single multiplicity cuts, 
(d) double multiplicity cut.}
\label{fig:Deuteron_AGK}
\end{center}
\end{figure}

It is important to understand and appreciate 
the fact that the shadowing correction for the total cross section as  given by graph $b$ in Fig.~\ref{fig:Deuteron_F2} corresponds to several 
distinguishable  final states 
which could be described as a 
result of the application of the Abramovsky-Gribov-Kancheli (AGK) 
cutting rules, see Sec.~\ref{subsect:agk}. 
Indeed,
the imaginary part of the double scattering contribution to the 
forward $\gamma^{\ast}D$ scattering amplitude is
given by the sum of all possible
cuts, see Fig.~\ref{fig:Deuteron_AGK}. 
In this figure, graph $a$ 
contributes
to
the diffractive final state in the reaction $\gamma^{\ast}N \to X N$
(the diffractive cut); graphs $b$ and $c$ correspond to 
the processes with single multiplicity 
in the reaction $\gamma^{\ast}N \to Y$ ($Y$ denotes all possible
final states); graph $d$ corresponds to the double
multiplicity in the reaction $\gamma^{\ast}N \to Y$.
Denoting the shadowing correction to the deuteron structure function
$F_{2D}(x,Q^2)$ as $\delta F_{2D}(x,Q^2)=F_{2D}(x,Q^2)-F_{2p}(x,Q^2)-
F_{2n}(x,Q^2)$, we find the corresponding contributions of the graphs in
Fig.~\ref{fig:Deuteron_AGK} (see also Sec.~\ref{subsect:agk}):
\begin{eqnarray}
\delta F_{2D}^a(x,Q^2)& \propto & 2 \left(\Im m D_1 \Im m D_2+\Re e D_1 \Re e D_2\right)
=2 |D_1 D_2^{\ast}|
 \,, \nonumber \\
\delta F_{2D}^{b+c}(x,Q^2) & \propto  & -8 \,\Im m D_1 \Im m D_2 \,, \nonumber \\
\delta F_{2D}^d(x,Q^2) & \propto & 4 \,\Im m D_1 \Im m D_2 \,,
\label{eq:deuteron_agk}
\end{eqnarray}
where $D_{1,2}$ denote the 
amplitude of the 
$\gamma^{\ast}+N \to X+N$
diffractive process.
Here the ratio of the contributions of the diagrams is $a:b:c:d= 1: -2:-2: 2$, which is due to the lack of the identity of  two exchanges with vacuum quantum numbers for diagrams $b$, $c$, and $d$. 
Therefore,
\begin{eqnarray}
\delta F_{2D}(x,Q^2) &=&\delta F_{2D}^a(x,Q^2)+\delta F_{2D}^{b+c}(x,Q^2)+\delta F_{2D}^d(x,Q^2) \nonumber\\
&=&-2\frac{1-\eta^2}{1+\eta^2}|D_1 D_2^{\ast}|=-
\frac{1-\eta^2}{1+\eta^2}\delta F_{2D}^a(x,Q^2) \,,
\label{eq:deuteron_agk2}
\end{eqnarray}
where $\eta=\Re e D_{1,2}/\Im m D_{1,2}$.

As we discussed in Sec.~\ref{sec:hab_pdfs}, the shadowing correction to the
nuclear structure function $F_{2A}(x,Q^2)$ is given by Eq.~(\ref{eq:m11})
in the approximation when the nucleons in the nuclear wave function are not correlated.
However, in the deuteron case, the nucleons are strongly correlated.
To calculate the contribution of graph  $b$ of Fig.~\ref{fig:Deuteron_F2},
it is convenient to recall our previous discussion that the soft (nuclear) 
part of graph  $b$ coincides with that in the shadowing correction to hadron-deuteron scattering  considered in Sec.~\ref{sec:gribov}.
The complete expression for the deuteron structure function $F_{2D}(x,Q^2)$ in the shadowing region reads~\cite{Frankfurt:2003jf,Frankfurt:2006am}:
\begin{eqnarray}
&&F_{2D}(x,Q^2)=  F_{2p}(x,Q^2)+F_{2n}(x,Q^2) \nonumber\\
&&-2 \frac{1-\eta^2}{1+\eta^2}\int_{x}^{0.1} dx_{\Pomeron}\, d k_t^2
\, F^{D(4)}_2\left(\beta, Q^2,x_{\Pomeron},t\right) \rho_D\left(4 k_t^2+4 (x_{\Pomeron} m_N)^2\right) \nonumber\\
&&=  F_{2p}(x,Q^2)+F_{2n}(x,Q^2) \nonumber\\
&&-2 \frac{1-\eta^2}{1+\eta^2} B_{\rm diff} \int_{x}^{0.1} dx_{\Pomeron}\, d k_t^2
\, F^{D(3)}_2\left(\beta, Q^2,x_{\Pomeron}\right)e^{-B_{\rm diff}\,k_t^2} \rho_D\left(4 k_t^2+4(x_{\Pomeron} m_N)^2\right) \,,
\label{eq:sh1}
\end{eqnarray}
 where the $F_{2p}(x,Q^2)+F_{2n}(x,Q^2)$ term is the impulse approximation 
(graph $a$ of Fig.~\ref{fig:Master1}); 
the term proportional to $F^{D(4)}_2$ and $F^{D(3)}_2$
 is the nuclear shadowing correction (graph $b$ of Fig.~\ref{fig:Master1}).
In Eq.~(\ref{eq:sh1}), $k_t$ is the transverse component of the momentum transfer; $|t|=k_t^2+(x_{\Pomeron} m_N)^2$; $\rho_D$ is the deuteron charge form factor of the double argument, which
can be written as an overlap between the initial and final state deuteron wave functions:
\begin{eqnarray}
\rho_D\left(4 k_t^2+4 (x_{\Pomeron} m_N)^2\right)=\int d^3 \vec{p} &&\Bigg[u(\vec{p})u(\vec{p}+\vec{k}) \nonumber \\
&&+ w(\vec{p})w(\vec{p}+\vec{k}) \left(\frac{3}{2}\frac{(\vec{p}\cdot (\vec{p}+\vec{k}))^2}{p^2 (p+k)^2}-\frac{1}{2}\right) \Bigg]  \,,
\label{eq:rho_D}
\end{eqnarray}
where $u$ and $w$ are the $S$-wave and $D$-wave components of the deuteron wave function,
respectively; $\vec{k}=\vec{k}_{t}+(x_{\Pomeron} m_N)e_z$.
 In our analysis, we use the deuteron wave function that corresponds to the
Paris nucleon-nucleon potential~\cite{Lacombe:1980dr}.
Note that the double argument of the  deuteron form factor is the consequence of the correct treatment of 
the deuteron center of mass.
One should also note that
the $t$ dependence of $F_2^{D(4)}$ cannot be neglected compared to that of
the deuteron form factor:
since the $t$ dependence of $\rho_D$ is rather moderate 
(compared to heavier nuclei), the integral in Eq.~(\ref{eq:sh1}) is
sensitive to $F^{D(4)}_2(t)$ up to $-t \leq 0.05$ GeV$^2$.

\begin{figure}[t]
\begin{center}
\epsfig{file=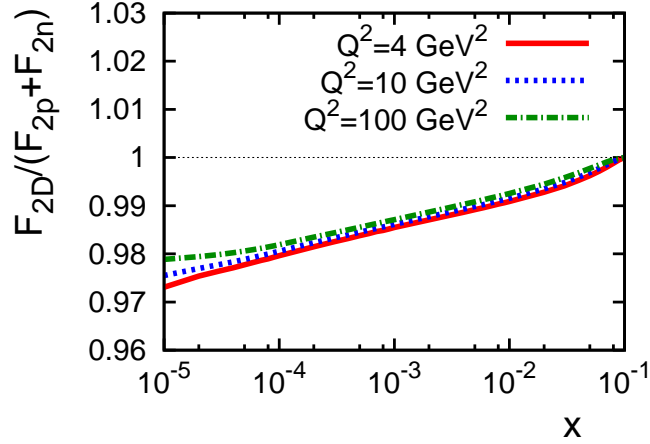,scale=1.4}
\caption{Leading twist nuclear shadowing for the deuterium structure function $F_{2D}(x,Q^2)$, see Eq.~(\ref{eq:sh1}). The ratio of the deuteron $F_{2D}(x,Q^2)$ to the sum of the free proton and neutron structure functions, $F_{2D}/(F_{2p}+F_{2n})$, as a function of Bjorken $x$ for different values of $Q^2$.}
\label{fig:Deuteron_shadowing_ratios}
\end{center}
\end{figure}

The results of the calculation of leading twist nuclear shadowing for the deuterium structure function $F_{2D}(x,Q^2)$ are presented
in Fig.~\ref{fig:Deuteron_shadowing_ratios}, where we plot the ratio of 
$F_{2D}(x,Q^2)$ given by Eq.~(\ref{eq:sh1}) to the sum of the free proton and neutron structure functions, $F_{2D}/(F_{2p}+F_{2n})$, as a function of Bjorken $x$ for different values of $Q^2$. 
The suppression of $F_{2D}/(F_{2p}+F_{2n})$ compared to unity is given by the 
nuclear shadowing correction in  Eq.~(\ref{eq:sh1}).
For the calculation of the free proton and neutron structure functions, we used as input the next-to-leading (NLO) CTEQ5M parameterization~\cite{Lai:1999wy}. 
The nucleon PDFs are evolved to the required values of $Q^2$ using the QCDNUM package~\cite{QCDNUM}. 
Note that our predictions for the magnitude of the nuclear shadowing
correction are in a very good agreement with the earlier calculation of~\cite{Edelmann:1997ik},
which considered the range of intermediate values of $Q^2$ where 
the contribution of low diffractive masses $M_X$ to nuclear shadowing is important. 

As we explained in Sec.~\ref{sec:hab_pdfs}, the use of the QCD factorization theorems
for inclusive and hard diffractive DIS allows one to generalize the shadowing
correction for the nuclear structure function $F_{2A}(x,Q^2)$ to individual
nuclear parton distributions $f_{j/A}(x,Q^2)$. In the case of the deuteron,
we obtain from Eq.~(\ref{eq:sh1}):
\begin{eqnarray}
&&f_{j/D}(x,Q^2)  =  f_{j/p}(x,Q^2)+f_{j/n}(x,Q^2) \nonumber\\
&&-2 \frac{1-\eta^2}{1+\eta^2}B_{\rm diff}\int_{x}^{0.1} dx_{\Pomeron}\, d k_t^2
\, f^{D(3)}_2\left(\beta, Q^2,x_{\Pomeron}\right) e^{-B_{\rm diff}\,k_t^2} \rho_D\left(4 k_t^2+4 (x_{\Pomeron} m_N)^2\right) \,.
\label{eq:sh2}
\end{eqnarray}

\begin{figure}[t]
\begin{center}
\epsfig{file=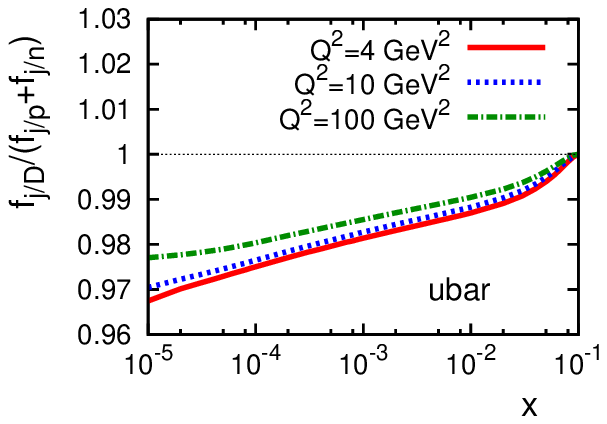,scale=1.25}
\epsfig{file=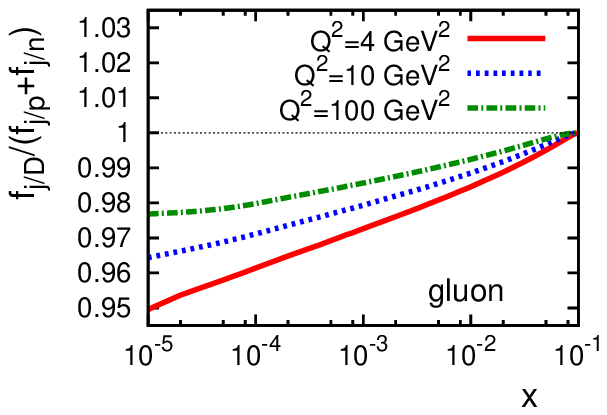,scale=1.25}
\caption{Leading twist nuclear shadowing for deuterium PDFs $f_{j/D}(x,Q^2)$, 
see Eq.~(\ref{eq:sh2}).
The ratio of $f_{j/D}(x,Q^2)$ to the sum of the free proton and neutron PDFs, 
$f_{j/D}/(f_{j/p}+f_{j/n})$, as a function of Bjorken $x$ for different values of $Q^2$.
The left panel corresponds to ${\bar u}$ quarks; the right panel corresponds to gluons. 
}
\label{fig:Deuteron_shadowing_ratios_u_gl}
\end{center}
\end{figure}

Our predictions for the leading twist nuclear shadowing for the deuterium PDFs 
$f_{j/D}(x,Q^2)$ are presented in Fig.~\ref{fig:Deuteron_shadowing_ratios_u_gl}.
In this figure, we plot the ratio
$f_{j/D}/(f_{j/p}+f_{j/n})$, where $f_{j/p}$ and $f_{j/n}$ are flavor $j$ PDFs of the free proton and neutron, respectively, as a function of Bjorken $x$ for different
values of $Q^2$. The left panel corresponds to ${\bar u}$ quarks; the right panel corresponds to gluons. 
Since the ratio of the gluon diffractive PDF to the usual gluon  PDF
is much larger than that for quarks,
see Fig.~\ref{fig:FitAB}, we predict the larger nuclear shadowing effect for
the gluon nuclear PDFs compared to that for the
quark nuclear PDFs
which reflects the stronger interaction in the gluon channel.

The case of DIS on deuterium presents a very important testing ground for the leading twist theory of nuclear shadowing since the shadowing correction can be calculated in a model-independent way, without the necessity to model multiple 
scatterings. Conversely, precise measurements of the deuterium structure function
$F_{2D}(x,Q^2)$ at low $x$ and $Q^2$ of the order of a few GeV$^2$ will 
constrain the nucleon diffractive structure function $F_2^{D(4)}$. 

DIS on deuterium is the main source of information on the neutron
structure function $F_{2n}(x,Q^2)$. The measurement of $F_{2n}(x,Q^2)$
using the deuteron beams is one of the important components of the 
planned physics program of the future Electron-Ion Collider~\cite{Deshpande:2005wd,eic}. 
The main goal of such a measurement is to study the flavor dependence
of parton distributions in a wide kinematic region, including small $x$.
The correct extraction of $F_{2n}(x,Q^2)$ from the deuteron data requires
an account of the nuclear shadowing correction.
The effect of nuclear shadowing on the extraction of $F_{2n}(x,Q^2)$ from 
$F_{2D}(x,Q^2)$ can be estimated as follows.
Since the diffractive structure function $F_2^{D(4)}$ is known with the accuracy of 
approximately 20\%, the accuracy of our calculations of nuclear shadowing
is $20 \times 0.025=0.5$\% at the smallest values of $x$. Correspondingly
the theoretical uncertainty of the $F_{2n}/F_{2p}$ ratio extracted from the
deuteron data  will be $2 \times 0.5=1$\%,
 which is likely to be smaller
than possible experimental systematic errors (the additional factor of two
comes from the fact that the calculated shadowing correction is per nucleon).

In this review, we concentrate on unpolarized scattering, singlet nuclear
parton distributions (unpolarized sea quarks and gluons) and 
nuclear shadowing driven by the 
vacuum 
exchange. 
The leading twist theory of nuclear shadowing can be generalized to non-singlet structure functions and parton distributions, 
where the shadowing correction is given by the interference between 
the 
vacuum and non-vacuum
exchanges, see Sec.~\ref{subsec:valence}.
However, the corresponding diffractive PDFs are not known and one has to model them.
Important examples of the processes, where nuclear shadowing is given by the 
interference between the 
vacuum and non-vacuum
exchanges,
include
shadowing in the valence quark channels and polarized DIS.
If one uses an eikonal-type approximation for these channels, one finds 
the much stronger nuclear shadowing for the non-vacuum channels since the scattering off the nucleus at small impact parameters does not contribute to the difference of the 
two cross sections, for instance, the $W^{+} A $ and $W^{-}A$ 
cross sections~\cite{Frankfurt:1988nt}.
However, large shadowing for the valence quarks 
combined with the baryon charge sum rule may lead to a substantial enhancement of 
the 
valence quark distribution 
around $x\sim 0.1$,
 which is probably not supported by the data.

For light nuclei, the eikonal-type approximation
just mentioned leads to 
the shadowing correction
in the non-singlet case 
which 
is approximately the factor of two  as large as that 
in the singlet case. 
Examples of 
such calculations
 of nuclear shadowing for various non-singlet observables include the difference 
between the $^3$He and $^3$H structure functions, $F_{2}^{^3{\rm He}}-F_2^{^3{\rm H}}$, 
with the application to the Gottfried sum rule~\cite{Guzey:2001tc},
polarized structure functions of deuterium~\cite{Frankfurt:2006am,Edelmann:1997ik},
the difference between the $^3$He and $^3$H
polarized structure functions and the Bjorken sum rule~\cite{Frankfurt:1996nf},
the polarized structure functions of  $^3$He~\cite{Frankfurt:1996nf,Guzey:1999rq,Bissey:2001cw},
$^7$Li~\cite{Guzey:1999rq} and $^6$LiD~\cite{Guzey:2000wh}.

The discussion in Sec.~\ref{subsec:valence} suggests that the eikonal approximation 
is likely to overestimate the nuclear shadowing effects in the scaling limit in non-singlet channels, although these effects could be large at moderate $Q^2$ corresponding to enhanced higher twist effects. Clearly further experimental
studies of nuclear shadowing in non-vacuum channels are highly desirable.

\subsection{Nuclear shadowing in 
tagged DIS processes on deuterium}  

In the previous subsection, we discussed the extraction of the neutron
structure function $F_{2n}(x,Q^2)$ from the deuteron inclusive structure
function $F_{2D}(x,Q^2)$. A strategy complimentary to the inclusive 
measurement of $F_{2D}(x,Q^2)$ is the use of the neutron tagging
by detecting a slow spectator proton.
Conversely, the proton structure function $F_{2p}(x,Q^2)$ can be studied 
by the proton tagging. The usefulness of the tagged DIS on deuterium for the extraction of 
$F_{2n}(x,Q^2)$ at large $x$ was discussed in~\cite{Melnitchouk:1996vp}.
In our analysis, we concentrate on the small $x$ region of nuclear shadowing.

By tagging the final proton in DIS on deuterium, one measures the
$\gamma^{\ast} D \to p X$ cross section, $d \sigma^{\gamma^{\ast} D \to p X}/d^3 p$, which can be expressed in terms of the tagged deuteron structure function
$F_{2D}(x,Q^2,\vec{p})$, where $\vec{p}$ is the momentum of the final
proton. The kinematic proportionality factor between 
$d \sigma^{\gamma^{\ast} D \to p X}/d^3 p$ and $F_{2D}(x,Q^2,\vec{p})$ is the same as the one between $\sigma^{\gamma^{\ast} D \to X}$ and $F_{2D}(x,Q^2)$. 

\begin{figure}[t]
\begin{center}
\epsfig{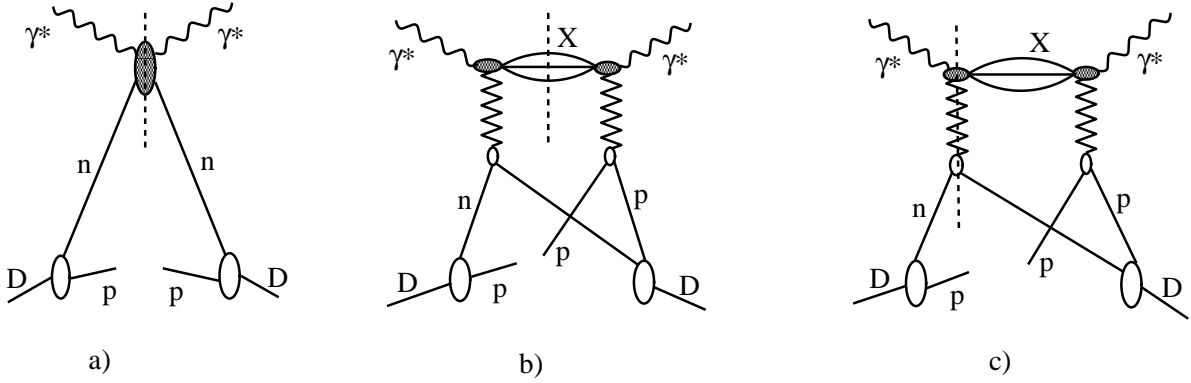}
\caption{
Graphs depicting the contributions to the 
deuteron tagged structure function $F_{2D}(x,Q^2,\vec{p})$: (a) the impulse contribution,
(b)+(c) the nuclear shadowing correction.
}
\label{fig:Deuteron_tagged_F2_diagram}
\end{center}
\end{figure}
In the graphical form, the tagged deuteron structure function is presented in 
Fig.~\ref{fig:Deuteron_tagged_F2_diagram}: graph $a$ gives the impulse approximation; 
graphs $b$ and $c$ is the nuclear shadowing correction.
The expression for $F_{2D}(x,Q^2,\vec{p})$ corresponding to the sum of graphs $a$,
$b$, and $c$ in Fig.~\ref{fig:Deuteron_tagged_F2_diagram}
 reads~\cite{Frankfurt:2003jf,Frankfurt:2006am} [compare to Eq.~(\ref{eq:sh1})]:
\begin{eqnarray}
F_{2D}(x,Q^2,\vec{p})&=&\left(1+\frac{q_z}{q_0}\frac{p_z}{m_N}\right)^{\alpha_{\Pomeron}(0)-1}
[u^2(p)+w^2(p)] F_{2n}(\tilde{x},Q^2) \nonumber\\
&-&\frac{3-\eta^2}{1+\eta^2}\int_{x}^{0.1} dx_{\Pomeron}\, \frac{d^2 \vec{k}_t}{\pi}
\, F^{D(4)}_2\left(\beta, Q^2,x_{\Pomeron},t\right)  \Bigg[u(\vec{p})u(\vec{p}+\vec{k}) \nonumber\\
 &+& w(\vec{p})w(\vec{p}+\vec{k}) \left(\frac{3}{2}\frac{(\vec{p}\cdot (\vec{p}+\vec{k}))^2}{p^2 (p+k)^2}-\frac{1}{2}\right) \Bigg]
 \,,
\label{eq:tagged_deuteron_F2}
\end{eqnarray}
where $\tilde{x}=(1-p_z/m_N)x$ which takes into account the effect of Fermi motion.
Since the second term in Eq.~(\ref{eq:tagged_deuteron_F2}) is 
a correction, the effect of Fermi motion can be safely neglected there.
In Eq.~(\ref{eq:tagged_deuteron_F2}), the first line is the impulse approximation
to $F_{2D}(x,Q^2,\vec{p})$; the second and third lines is the shadowing correction.
The $(1+q_z/q_0 \, p_z/m_N)^{\alpha_{\Pomeron}(0)-1}$ factor
reflects 
the
different invariant energies of the virtual photon-deuteron and the virtual photon-neutron
interactions---$(1+q_z/q_0 \, p_z/m_N)^{\alpha_{\Pomeron}(0)-1}$
is the flux factor of the interacting neutron
($q_0$ and $q_z$ are the energy and momentum [along the $z$-axis] of the virtual photon, respectively;
$p_z$ is the $z$-component of the neutron momentum).
In our derivation, we neglected corrections of the order of ${\cal O}(p^2/m_N^2)$
and higher, which is standard for the non-relativistic treatment of the
deuteron wave function.

The coefficient $(3-\eta^2)/(1+\eta^2)$, which determines the weight
of the shadowing correction, is a direct consequence of the application of
the AGK rules. Indeed, in the case of the proton tagging, only the diffractive 
cut (graph $b$ in Fig.~\ref{fig:Deuteron_tagged_F2_diagram})
and single unitarity cut (graph $c$ in Fig.~\ref{fig:Deuteron_tagged_F2_diagram}) are allowed.
(If both ladders are cut, the hadron final state contains no slow nucleons, and there are no spectators from the deuteron wave function.)
The respective contributions of 
these graphs
to the shadowing correction
$\delta F_2^D(x,Q^2,\vec{p})$ read [compare to the case of inclusive structure
function, see Eq.~(\ref{eq:deuteron_agk})]:
\begin{eqnarray}
\delta F_{2D}^{(b)}(x,Q^2,\vec{p})& \propto &  \left(\Im m D_1 \Im m D_2+\Re e D_1 \Re e D_2\right)
= |D_1 D_2^{\ast}|
 \,, \nonumber \\
\delta F_{2D}^{(c)}(x,Q^2,\vec{p}) & \propto  & -4 \,\Im m D_1 \Im m D_2 \,.
\label{eq:deuteron_tagged_agk}
\end{eqnarray}
Therefore,
\begin{eqnarray}
\delta F_{2D}(x,Q^2,\vec{p}) &=&\delta F_{2D}^{(b)}(x,Q^2,\vec{p})+\delta F_{2D}^{(c)}(x,Q^2,\vec{p}) \nonumber\\
&=&-\frac{3-\eta^2}{1+\eta^2}|D_1 D_2^{\ast}|=-\frac{3-\eta^2}{1+\eta^2}
\delta F_{2D}^{(b)}(x,Q^2,\vec{p}) \,.
\label{eq:deuteron_tagged_agk2}
\end{eqnarray}
One has to note that, in general, the tagged deuteron structure function
receives additional contributions from the triple and quadruple interactions
with the target nucleons (not shown in Fig.~\ref{fig:Deuteron_tagged_F2_diagram}).
However, at small values of the spectator nucleon momentum,
$p \le \sqrt{\epsilon m_N}=64$ MeV/c, where 
$\epsilon=2.2$
 MeV is the deuteron binding energy,
it is legitimate to keep only the single and double scattering terms shown in
Fig.~\ref{fig:Deuteron_tagged_F2_diagram}.
 At larger spectator momenta, 
the contributions of the triple and quadruple interactions with the target 
are no longer suppressed by the small parameter
 $p / \sqrt{\epsilon m_N}$  and, hence, should be included.
 This will introduce a certain model dependence since those terms
are not simply related to the elementary diffractive structure
function $F_2^{D(4)}$.
One should emphasize that this is only the case for the tagged structure function:
the triple and quadruple interaction terms cancel in the inclusive structure
function, which is unambiguously expressed in terms of the
nucleon diffractive structure function, see Eq.~(\ref{eq:sh1}).

As one can see from Eq.~(\ref{eq:tagged_deuteron_F2}), nuclear shadowing suppresses the spectrum of the produced protons. We quantify this effect by considering the ratio
$R(x,Q^2,\vec{p})$:
\begin{equation}
R(x,Q^2,\vec{p})=\frac{F_{2D}(x,Q^2,\vec{p})}{F_{2D}^{\rm IA}(x,Q^2,\vec{p})} \,,
\label{eq:R_deuteron}
\end{equation}
where $F_{2D}(x,Q^2,\vec{p})$ is given by Eq.~(\ref{eq:tagged_deuteron_F2}), and
$F_{2D}^{\rm IA}(x,Q^2,\vec{p})$ is the impulse approximation to $F_{2D}(x,Q^2,\vec{p})$
[the first line of Eq.~(\ref{eq:tagged_deuteron_F2})],
\begin{equation}
F_{2D}^{\rm IA}(x,Q^2,\vec{p})=\left(1+\frac{q_z}{q_0}\frac{p_z}{m_N}\right)^{\alpha_{\Pomeron}(0)-1}
[u^2(p)+w^2(p)] F_{2n}(\tilde{x},Q^2) \,.
\label{eq:tagged_deuteron_F2_IA}
\end{equation}
For the neutron structure function $F_{2n}(x,Q^2)$, we use the next-to-leading
order CTEQ5M parameterization~\cite{Lai:1999wy}; $\alpha_{\Pomeron}(0)=1.111$.

\begin{figure}[t]
\begin{center}
\epsfig{file=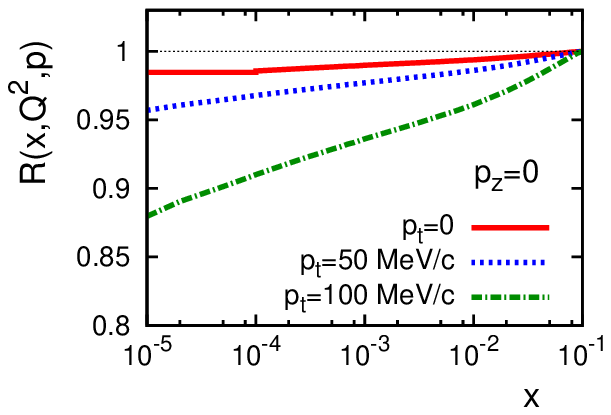,scale=1.25}
\epsfig{file=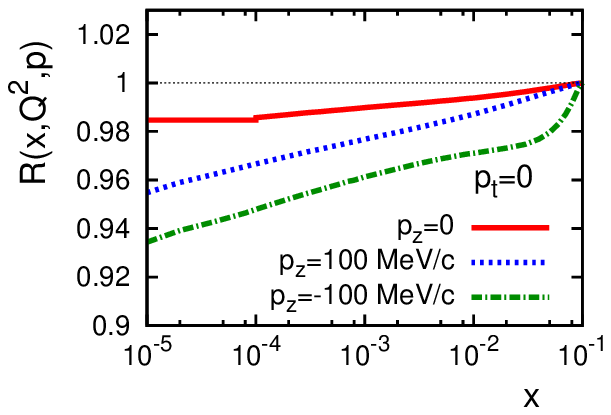,scale=1.25}
\caption{Nuclear shadowing in the tagged deuteron structure function
$F_{2D}(x,Q^2,\vec{p})$. The ratio $R(x,Q^2,\vec{p})$ of
Eq.~(\ref{eq:R_deuteron}) as a function of Bjorken $x$ at fixed $Q^2=4$ GeV$^2$.
The left panel corresponds to $p_z=0$ and $|\vec{p}_{t}|=0$, 50, and 100 MeV/c; the right panel corresponds to  $|\vec{p}_{t}|=0$
and $p_z=\pm 100$ MeV/c.
}
\label{fig:Deuteron_tagged_F2}
\end{center}
\end{figure}

Figure~\ref{fig:Deuteron_tagged_F2} presents the ratio $R(x,Q^2,\vec{p})$ of
Eq.~(\ref{eq:R_deuteron}) as a function of Bjorken $x$ at fixed $Q^2=4$ GeV$^2$.
The left panel corresponds to the case of the zero longitudinal momentum of the final proton,
$p_z=0$, and different values of the transverse momentum, $|\vec{p}_{t}|=0$, 50, and 100 MeV/c; the right panel corresponds to  $|\vec{p}_{t}|=0$
and $p_z=\pm 100$ MeV/c.
As one can see from Fig.~\ref{fig:Deuteron_tagged_F2}, nuclear shadowing decreases
the ratio  $R(x,Q^2,\vec{p})$ as $p_{\perp}$ is increased. Larger values of 
$p_{t}$ correspond to smaller transverse distances between the proton and
neutron, which leads to larger nuclear shadowing. 

A comparison of  Figs.~\ref{fig:Deuteron_tagged_F2} and 
\ref{fig:Deuteron_shadowing_ratios}
shows that the shadowing correction to the tagged deuteron structure function is much larger than that for the inclusive structure function. 
This happens
 because of the following two features of the tagged case.
First, the weight of the shadowing correction is larger because of
the AGK cutting rules (see the discussion above).
Second, the impulse approximation decreases with increasing $|\vec{p}|$ faster
than the shadowing term, which enhances the relative magnitude of
nuclear shadowing.

In addition to nuclear shadowing, the spectrum of the produced protons is also
modified by the proton-neutron final state interactions (FSI).
FSI partially cancel the effect of nuclear shadowing and decrease the deviation of
the ratio $R(x,Q^2,\vec{p})$ from unity~\cite{Frankfurt:2006am}.

From the experimental point of view, several strategies of the extraction of 
the neutron structure function $F_{2n}(x,Q^2)$ from the deuteron data 
using the proton tagging are possible. 
As our analysis above shows, by selecting only very low $p_t$ protons,
the distortion of the proton spectrum by the nuclear shadowing 
and FSI effects will be minimal. However, the drawback of this approach
is the gross loss of statistics.
The other, more promising approach is to measure the $p_t$ dependence of the spectrum 
up to $p_t \sim 200$ MeV/c, which 
will allow one to use most of the spectator protons. 
Provided the good momentum resolution of the  proton spectrometer, 
one would be able to make (longitudinal) momentum 
cuts to suppress/increase  the shadowing effect and, thus, one
would have an opportunity to independently study the enhanced nuclear shadowing.
In general, the tagged method will allow one to extract $F_{2n}(x,Q^2)$ 
from deuteron data with an accuracy  at  the level of the fraction of a percent.

Note also that, in general, the proton spectrum receives an additional contribution
that 
predominantly originates from the diffractive scattering off the proton with large $p_{t}$,
$|\vec{p}_{t}| \geq 300$ MeV/c. As a result of the AGK cancellation, this contribution
is given by the corresponding impulse approximation. The contribution has a broad
$p_{t}$ distribution, $\propto \exp(-B p_t^2)$ with $B \approx 7$ GeV$^2$, and can 
be subtracted using the measurement at, for instance,
$|\vec{p}_{t}| \geq 300$ MeV/c.

One can also use the simultaneous tagging of protons and neutrons, when 
both the neutron and proton  are detected in the reactions
 $\gamma^{\ast} D \to n X$ and $\gamma^{\ast} D \to p X$.
In this case, the effects of
nuclear shadowing and FSI cancel in the ratio $\sigma^{\gamma^{\ast} D \to n X}/\sigma^{\gamma^{\ast} D \to p X}$
and the main errors in the measurement of $F_{2n}(x,Q^2)$ will come from
 the determination of relative efficiencies of the proton and  neutron taggers.

Another possibility to extract the $F_{2n}(x,Q^2) /F_{2p}(x,Q^2)$ ratio 
is from comparing the rate of the tagged proton scattering events with the neutron spectator 
to inclusive $eD$ scattering.  Such a strategy has certain merits
as it avoids the issue of luminosity and does not require a leading proton spectrometer.
The disadvantage of this 
method 
is the sensitivity to the nuclear shadowing and FSI effects and errors in the acceptance of
the neutron detector.  One possible way to deal with the latter problem will be to perform 
measurements at very small $x$ and large
energies,  where the $ep$ and $en$ cross sections are equal 
to better than a fraction of 1\% and, hence, one would be able to cross-check
the  acceptances of the proton and neutron detectors.
Note also that taking proton data from an independent run will 
potentially lead to another
set of issues such as relative luminosity, the use of different
 beam energies, etc., 
which will likely introduce additional systematic error at the level of 1\%.

In our analysis, we neglected possible non-nucleonic components of the
deuteron wave function such as kneaded (six-quark) components, etc.
Current estimates put an upper limit on the probability of
such components at the
level of less than one percent, and they are expected to modify predominantly
the high-momentum
component of the deuteron wave function. Hence, the non-nucleonic 
components should give a
very small (less than one percent) correction for the spectator momenta  $\le 200$ MeV/c.
 Moreover, the experiments at
the  future Electron-Ion Collider
looking for production of  baryons such as $\Delta$ and 
$N^{\ast}$ in the spectator 
kinematics (the spectator Feynman $x_L(\Delta,N^{\ast})\ge 0.5$) would allow one 
to put a more stringent
upper  limit on (or to discover) non-nucleonic components of the deuteron wave
function.
 
Finally, we would like to mention that similarly to the tagged deuteron structure function
$F_{2D}(x,Q^2,\vec{p})$, nuclear shadowing and final-state interactions modify
the cross section of diffractive deuteron disintegration
$\gamma^{\ast} D \to X\, p\, n$.

\section{Phenomenology of nuclear shadowing for nuclear structure functions and parton distributions 
for medium and heavy nuclei}
\label{sec:phen}

In this section, we present predictions of the leading twist theory of  
nuclear shadowing for the nuclear PDFs and 
nuclear structure functions 
for 
a wide range of nuclei. We discuss theoretical uncertainties of our predictions as well 
as the uncertainties due the experimental knowledge of diffraction  in $ep$ DIS. 
(The largest 
experimental uncertainty is associated with the uncertainty in the $t$ dependence 
of the proton diffractive structure function $F_2^{D(4)}$.)
We also present predictions for impact parameter dependent
nPDFs, which are required in the description of hard processes in 
hadron-nucleus and nucleus-nucleus scattering.

We compare our predictions for nPDFs to those resulting from the global fits to 
fixed-target data on DIS on nuclei. 
We also make a comparison between our results on
$F_{2A}(x,Q^2)$ and the nuclear longitudinal structure function $F_{L}^A(x,Q^2)$ 
with the predictions obtained within the eikonal approximation.

Also, we compare our results with the available fixed-target data on DIS off various
nuclear targets. We demonstrate that the data contain a significant higher twist 
component, which might invalidate the extraction of leading-twist nuclear 
PDFs from the fixed-target data.

\subsection{Predictions for nuclear parton distributions}
\label{subsec:nPDFs}

\subsubsection{General remarks}

Nuclear parton distributions (nPDFs) at a fixed low scale $Q^2_0$ 
($Q^2_0=4$ GeV$^2$ in our case)
and for small values of  Bjorken $x$, $x \leq 0.1$,
$f_{j/A}(x,Q_0^2)$, 
are given  by Eq.~(\ref{eq:m13master}). 
This equation defines the input for the subsequent
next-to-leading order (NLO) DGLAP evolution of nPDFs to arbitrary
 high $Q^2> Q^2_0$. 
In the numerical analysis of Eq.~(\ref{eq:m13master}), we
used the following input.

For diffractive PDFs $f_j^{D(3)}$, we used
the H1 diffractive fit~B~\cite{Aktas:2006hy,Aktas:2006hx}. 
We assumed that the leading Pomeron
and the subleading Reggeon contribute {\it independently} to nuclear shadowing.
Note that the latter contribution is small and contributes sizably only for
 $x_{\Pomeron} > 0.05$.
As explained in Sect.~\ref{subsec:diffdata}, the FPS H1 data requires that
the Pomeron contribution is reduced by the factor 1.23 and 
the Reggeon contribution is reduced by the factor 1.39.
For the slope of the $t$ dependence of the quark and gluon diffractive PDFs, we
used $B_{{\rm diff}}=6$ GeV$^{-2}$, see Eq.~(\ref{eq:m13_b}) and the 
corresponding discussion.

Our numerical analysis shows that the subleading Reggeon contribution to the shadowing
correction is negligible. Not only the Reggeon contribution is numerically small
by itself [see Eq.~(\ref{eq:data6})], but it is also additionally 
 suppressed by the 
$(1-i\eta)^2/(1+\eta^2)$ factor in Eq.~(\ref{eq:m13master}) because
$|\eta| \approx 1$ for the Reggeon trajectory~\cite{Frankfurt:2003zd}.
Therefore, only the leading Pomeron term
contributes to the resulting nuclear PDFs.

We would like to point out that our present conclusion about the smallness of the nuclear
shadowing correction coming from the subleading Reggeon trajectory differs from
our original findings~\cite{Frankfurt:2003zd}. 
In Ref.~\cite{Frankfurt:2003zd}, we used the results of the older H1 analysis of 
inclusive diffraction at HERA, which did not report the normalization constant 
of the Reggeon contribution,
$n_{\Reggeon}$,  see Eq.~(\ref{eq:data6}), and, as a result,
we significantly overestimated
 $n_{\Reggeon}$ in~\cite{Frankfurt:2003zd}.

In the present calculation, for the ratio of the real to 
imaginary parts of the diffractive amplitude, $\eta$,
 we use its relation to the intercept of the effective Pomeron trajectory, 
$\alpha_{\Pomeron}(0)$, using the Gribov-Migdal 
relationship~\cite{Gribov:1968uy} between the imaginary and real parts of the 
scattering amplitude
at high energies,
\begin {equation}
\eta \approx \frac{\pi}{2}\big(\alpha_{\Pomeron}(0)-1\big)=0.174 \,,
\end{equation}
where the H1 Fit~B value for $\alpha_{\Pomeron}(0)=1.111$ was employed, see Eq.~(\ref{eq:data6}).

For the nuclear density $\rho_A(r)$, we used 
the two-parameter Fermi
parameterization~\cite{DeJager:1987qc}.
For the usual nucleon PDFs, $f_{j/N}$, we used the NLO CTEQ5M parameterization~\cite{Lai:1999wy}.

Two important remarks are in order.
First, while the leading twist theory of nuclear shadowing is
applicable to partons of all flavors,
at present
we are unable to  
make quantitative
 predictions for nuclear shadowing of valence quarks 
since the subleading Reggeon contribution, which gives rise 
to nuclear shadowing of nuclear valence PDFs, is
essentially unknown, but it is 
strongly suppressed (see the discussion above) since
inclusive diffraction is dominated 
by the $C$-even Pomeron exchange.
In principle,  a significant Reggeon contribution could arise from the 
Reggeon-Reggeon-Pomeron (RRP) interference term.
However, the magnitude of such a contribution is not known,
although it was included in the fits 
to the HERA diffractive data.
In any case, the absence of diffractive final states in $\gamma^{\ast}$-``valence quark''
interactions leads to a strong suppression of nuclear shadowing in this channel,
see the discussion in Sec.~\ref{subsec:valence}.
In practical terms, all this means that   
Eq.~(\ref{eq:m13master}) should be applied to evaluate nuclear shadowing for the 
sea quarks and gluons only.

Second, since diffractive PDFs of the nucleon, $f_j^{D(4)}$, represent next-to-leading
order (NLO) 
distributions, our predictions for nuclear PDFs $f_{j/A}$ are also NLO predictions, 
which correspond to QCD observables calculated to the NLO accuracy.

\subsubsection{Color fluctuation approximation for multiple interactions in 
leading twist theory of nuclear shadowing}
\label{subsubsec:color_fluct}

In the derivation of our master equation~(\ref{eq:m13master}), we used the
color fluctuation approximation, which enabled us to express the interaction with
$N \geq 3$ nucleons of the nuclear target in terms of a single cross section 
$\sigma_{\rm soft}^j(x,Q^2)$. In our numerical analysis, we use two different models for
$\sigma_{\rm soft}^j(x,Q^2)$ that we present below.
They can be interpreted as giving the higher and lower values of nuclear shadowing
in our approach.
This reflects the objective reality that currently 
the value of $\sigma_{\rm soft}^j(x,Q^2)$ has large
uncertainties. However, as we pointed out in 
Sec.~\ref{sec:hab_pdfs}, one possible strategy could be that 
once nuclear shadowing is measured on one nucleus,
$\sigma_{\rm soft}^j(x,Q^2)$ can be extracted from that measurement and predictions for 
other nuclei can be made with very little theoretical uncertainty.
Note also that $\sigma_{\rm soft}^j(x,Q^2)$ can be fixed from the studies of coherent diffraction
with nuclei, see Sec.~\ref{sec:final_states}.

\underline{Model 1 (FGS10\_H): Larger nuclear shadowing}

In the first model for $\sigma_{\rm soft}^j(x,Q^2)$, we use as a reference 
the effective cross section $\sigma_2^j(x,Q^2)$, see Eq.~(\ref{eq:m17}).
The resulting model for the gluon and quark channels reads:
\begin{eqnarray}
&&\sigma_{\rm soft}^{g ({\rm H})}(x,Q_0^2)=\left\{\begin{array}{ll}
\sigma_2^g(x,Q_0^2) \,, & x \leq x_4= 10^{-4} \,, \\
\sigma_2^g(x_4,Q_0^2) (\frac{x_4}{x})^{0.06} \,, & 10^{-4} \leq x \leq x_2=10^{-2} \,, \\
\sigma_2^g(x_4,Q_0^2) (\frac{x_4}{x_2})^{0.06} \left(\frac{0.1-x}{0.1-x_2}\right) \,, & 10^{-2} \leq x \leq 0.1 \,,
\end{array} \right. \nonumber\\
&&\sigma_{\rm soft}^{q({\rm H})}(x,Q_0^2)=\left\{ \begin{array}{ll}
\kappa(x) \sigma_2^q(x,Q_0^2) \,, & x \leq 10^{-4} \,, \\
\kappa(x_0) \sigma_2^q(x_0,Q_0^2) (\frac{x_0}{x})^{0.06} \,, & 10^{-4} \leq x \leq 10^{-2} \,, \\
\kappa(x_0) \sigma_2^q(x_0,Q_0^2) (\frac{x_0}{x_2})^{0.06}\left(\frac{0.1-x}{0.1-x_2}\right) \,, & 10^{-2} \leq x \leq 0.1 \,,
\end{array} \right.
\label{eq:sigma3_model1}
\end{eqnarray}
where $x_4=10^{-4}$ and $x_2=10^{-2}$.
The superscript ''(H)'' indicates that the resulting nuclear shadowing corresponds to
the higher value of shadowing in our approach.
In the following, predictions for nuclear shadowing made with 
the effective cross section $\sigma_{\rm soft}^{j ({\rm H})}$ will be referred to and labeled as
''FGS10\_H''.

Below we explain the motivation and building blocks used in Eq.~(\ref{eq:sigma3_model1}).
In the gluon channel, the interaction at $Q^2=Q_0^2=4$ GeV$^2$ and $x \leq 10^{-4}$ is 
rather 
close to the maximally allowed by unitarity (the black disk regime, 
see Sec.~\ref{sec:bdr}). Therefore, the cross section (color) fluctuations are small
and it is a good approximation to use $\sigma_{\rm soft}^{g(\rm H)}(x,Q_0^2) \approx \sigma_2^g(x,Q_0^2)$.
For the larger values of $x$, $10^{-2} > x > 10^{-4}$, we impose the soft energy 
dependence for $\sigma_{\rm soft}^{g(\rm H)}(x,Q_0^2)$. 
The exponent $0.06$ is taken to be equal to the exponent of the energy dependence
of the
$\sigma_{\rm soft}^{j ({\rm L})}$ cross section
 [see the details below when we
discuss the second model for $\sigma_{\rm soft}^j(x,Q^2)$].
For even larger values of $x$, $x > 0.01$, the cross section fluctuation formalism
 is not 
applicable since different fluctuations are not coherent. However, for $x > 0.01$,
the interaction with $N \geq 3$ nucleons gives a negligible correction to nuclear shadowing
and, hence, the  issue of 
modeling of
$\sigma_{\rm soft}^j(x,Q^2)$ is
unimportant.
Therefore, we simply assume that $\sigma_{\rm soft}^{j(\rm H)}(x,Q^2)$ 
linearly decreases on the interval
$10^{-2} \leq x \leq 0.1$ and vanishes at $x=0.1$.

In the quark channel, the interaction has not reached the black disk regime, the color
fluctuations are sizable, and $\sigma_{\rm soft}^{q(\rm H)}(x,Q_0^2) > \sigma_2^q(x,Q_0^2)$ because of the
dispersion of the distribution over cross section $P_j(\sigma)$.
This is modeled by the dimensionless coefficient $\kappa$ ($\kappa>1$) 
that we evaluate within the dipole model,
\begin{equation}
\kappa(x)=\frac{\langle \sigma_{q\bar q N}^3(x,d_{\perp}^2,m_i)\rangle}{\langle \sigma_{q\bar q N}^2(x,d_{\perp}^2,m_i)\rangle}\Bigg/ \frac{\langle \sigma_{q\bar q N}^2(x,d_{\perp}^2,m_i)\rangle}{\langle \sigma_{q\bar q N}(x,d_{\perp}^2,m_i)\rangle}
 \,,
\label{eq:kappa}
\end{equation}
where the brackets denote the integration 
with the weight given by
the square of the virtual photon wave function:
\begin{equation}
\langle \sigma_{q\bar q N}^n(x,d_{\perp}^2,m_i)\rangle=\int d\alpha ~d^2 d_{\perp} \sum_{i}~|\Psi(\alpha,Q^2,d_{\perp}^2,m_{i})|^2 \sigma_{q\bar q N}^n(x,d_{\perp}^2,m_i) \,,
\end{equation}
where $|\Psi|^2$ is the square of the virtual photon wave function (the probability for 
the virtual photon to fluctuate into a $q {\bar q}$ state); $\sigma_{q\bar q N}$ is the
$q {\bar q}$-nucleon cross section. For the details of the dipole formalism relevant
for the present results, see Sec.~\ref{subsec:eikonal}.
In the considered kinematics, $1.37 \leq \kappa \leq 1.55$. 

\underline{Model 2 (FGS10\_L): Lower nuclear shadowing}

\begin{figure}[t]
\begin{center}
\epsfig{file=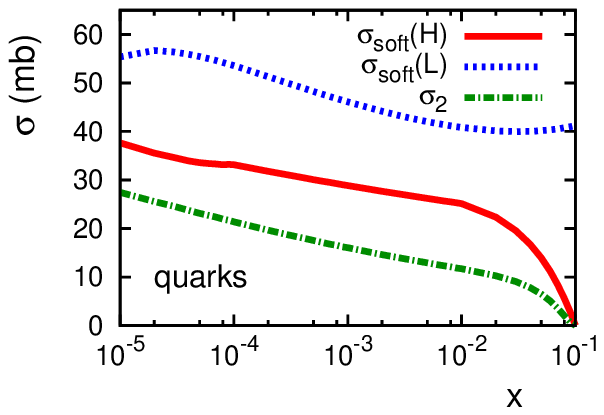,scale=1.25}
\epsfig{file=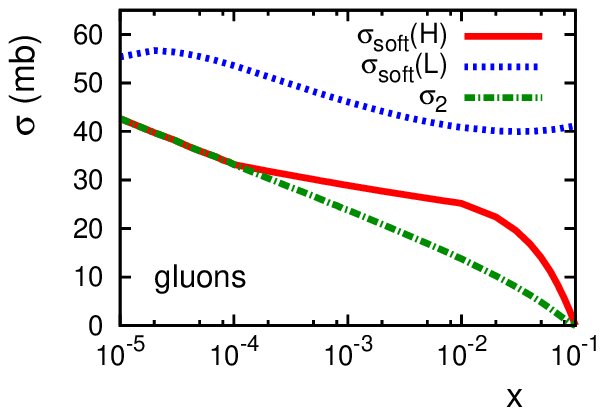,scale=1.25}
\caption{The cross sections 
$\sigma_{\rm soft}^{j(\rm H)}$, $\sigma_{\rm soft}^{j(\rm L)}$, and
$\sigma_2^j(x,Q_0^2)$ 
as functions of Bjorken $x$ at fixed $Q_0^2=4$ GeV$^2$.
The left panel corresponds to the $\bar{u}$-quark; the right panel corresponds to gluons.
}
\label{fig:sigma3_2009}
\end{center}
\end{figure}

In the second model for $\sigma_{\rm soft}^j(x,Q^2)$, we assume that the relevant 
distribution 
$P_j(\sigma)$
is given by
the distribution over cross sections for the pion, $P_{\pi}(\sigma)$, 
i.e., $P_j(\sigma) \approx P_{\pi}(\sigma)$. The form of $P_{\pi}(\sigma)$ is rather
well-known~\cite{Guzey:2009jr,Blaettel:1993rd}:
\begin{equation}
P_j(\sigma)=P_{\pi}(\sigma)=N e^{-\frac{(\sigma -\sigma_0)^2}{(\Omega \sigma_0)^2}} \,.
\label{eq:Ppion}
\end{equation}
The parameters $N$, $\sigma_0$ and $\Omega$ are constrained by following requirements:
\begin{eqnarray}
&&\int_0^{\infty} d \sigma P_{\pi}(\sigma) = 1 \,,
\nonumber\\
&&\int_0^{\infty} d \sigma P_{\pi}(\sigma) \sigma \equiv \langle \sigma \rangle
= \sigma_{\rm tot}^{\pi N}(W^2) \,,
\nonumber\\
&&\int_0^{\infty} d \sigma P_{\pi}(\sigma) \sigma^2 \equiv \langle \sigma^2 \rangle= \left(\sigma_{\rm tot}^{\pi N}(W^2)\right)^2 \left(1+\omega_{\sigma}^{\pi N}(W^2)\right) \,,
\label{eq:Psigma_moments}
\end{eqnarray}
where $\sigma_{\rm tot}^{\pi N}$ is the total pion-nucleon cross section.
The parameter $\omega_{\sigma}^{\pi N}$ characterizes the dispersion of the distribution
$P_{\pi}(\sigma)$:
\begin{equation}
\omega_{\sigma}^{\pi N}=\frac{\langle \sigma^2 \rangle-\langle \sigma \rangle^2}{\langle \sigma \rangle^2} \,,
\label{eq:omega_sigma_def}
\end{equation}
 Both $\sigma_{\rm tot}^{\pi N}$ and $\omega_{\sigma}^{\pi N}$ depend on the
pion-nucleon invariant energy squared $W^2$,
$W^2=Q^2/x-Q^2+m_N^2$.
In our numerical analysis, we 
used the Donnachie-Landshoff parameterization for 
$\sigma_{\rm tot}^{\pi N}$~\cite{Donnachie:1992ny}:
\begin{equation}
\sigma_{\rm tot}^{\pi N}(W^2)=\frac{1}{2}\left(\sigma_{\rm tot}^{\pi^+ N}
+\sigma_{\rm tot}^{\pi^- N}
\right)=13.63 \,(W^2)^{0.0808}+31.79\,(W^2)^{-0.4525} \ {\rm mb} \,.
\label{eq:sigma_pion}
\end{equation}
Note that in our calculations, we effectively use only the first term in Eq.~(\ref{eq:sigma_pion}),
see also Fig.~\ref{fig:sigma3_2009}.

The parameter $\omega_{\sigma}^{\pi N}$ decreases with an increase of energy~\cite{Guzey:2005tk}.
At the pion beam energy of $E_{\pi} \approx 300$ GeV (which corresponds to
$W^2 \approx 600$ GeV$^2$), one has 
$\omega_{\sigma} \approx 0.4$~\cite{Blaettel:1993ah,Blaettel:1993rd}. 
At the CDF energy of $W^2=(546\,{\rm GeV})^2 \approx 3 \times 10^{5}$ GeV$^2$,
we take $\omega_{\sigma}^{\pi N} \approx 0.16 \times (3/2)=0.24$.
In this estimate, we used the relation between $\omega_{\sigma}^{\pi N}$,
the parameter $\omega_{\sigma}^{pp}$ for the proton projectile, and the total pion-nucleon
and proton-proton cross sections,
\begin{equation}
\omega_{\sigma}^{\pi N}=\frac{\sigma_{\rm tot}^{p p}}{\sigma_{\rm tot}^{\pi N}}\,\omega_{\sigma}^{pp} \,,
\end{equation}
as well as
the result that $\omega_{\sigma}^{pp}=0.16$ at the CDF energy~\cite{Blaettel:1993ah},
and the constituent quark model counting rule 
$\sigma_{\rm tot}^{p p}/\sigma_{\rm tot}^{\pi N}=3/2$~\cite{Blaettel:1993rd}.
Assuming a simple linear interpolation between the two energies, we arrive at the
following model for $\omega_{\sigma}^{\pi N}$ (for $W_2^2 \geq W^2 \geq W_1^2$):
\begin{equation}
\omega_{\sigma}^{\pi N}(W^2)=0.4 -0.16\, \frac{W^2-W_1^2}{W_2^2-W_1^2} \,,
\label{eq:omega_sigma}
\end{equation}
where $W_1^2=600$ GeV$^2$ and $W_2^2=3 \times 10^{5}$ GeV$^2$.
Equations~(\ref{eq:Psigma_moments}), (\ref{eq:sigma_pion}) and (\ref{eq:omega_sigma})
fully determine $P_{\pi}(\sigma)$ and its energy (Bjorken $x$) dependence.
In summary, the second model for the effective rescattering cross section reads 
[see also Eq.~(\ref{eq:s_soft})]:
\begin{equation}
\sigma_{\rm soft}^{j(\rm L)}(W^2)=\frac{\int_{0}^{\infty} d \sigma P_{\pi}(\sigma)\sigma^3}{\int_{0}^{\infty} d \sigma P_{\pi}(\sigma)\sigma^2} \,.
\label{eq:sigma_L}
\end{equation}
The superscript "(L)" indicates that the resulting nuclear shadowing corresponds to the lower
limit on nuclear shadowing predicted in our leading twist approach.
Predictions for nuclear shadowing made with 
the effective cross section $\sigma_{\rm soft}^{j ({\rm L})}$ will be referred to as ''FGS10\_L''.

Figure~\ref{fig:sigma3_2009} presents $\sigma_{\rm soft}^{j(\rm H)}(x,Q_0^2)$ (model 1)
and $\sigma_{\rm soft}^{j(\rm L)}(W^2)$ (model 2)
as functions of Bjorken $x$ at fixed $Q_0^2=4$ GeV$^2$. 
(Note that for the latter cross section, $W^2=Q_0^2/x-Q_0^2+m_N^2$.)
 For comparison and completeness,
we also give $\sigma_2^j(x,Q_0^2)$ 
which is relevant for the calculation of nuclear shadowing in the quasi-eikonal
approximation (see Fig.~\ref{fig:LT2009_ca40_qe}). 
The left panel 
of Fig.~\ref{fig:sigma3_2009}
corresponds to the $\bar{u}$-quark; the right panel corresponds to gluons.
Note that $\sigma_{\rm soft}^{j(\rm L)}(W^2)$ is flavor-independent.

The difference between the approximation 
when one uses $\sigma_2^j(x,Q_0^2)$ as the effective
rescattering cross section and the color fluctuation approximation (models 1 and 2) 
is the amount
of point-like (very weakly interacting) configurations (PLC) in the virtual photon wave function.   
The both approximations can be considered as generalizations of the QCD-improved aligned 
jet model (AJM), where one has two components---the strongly interacting AJM component and 
a PLC. Note also that in 
general
the fraction of PLC decreases with increasing energy.

\subsubsection{Large $\beta$ diffraction dominates nuclear shadowing down to $x \sim 10^{-4}$}
\label{subsubsec:diffractive_masses}
 
The effective cross section $\sigma_{2}^j(x,Q^2)$ determines the
magnitude of nuclear shadowing when only the interaction with two nucleons of
the target is important. This is the case for the deuteron and 
heavy nuclei
in the low-nuclear density limit.
In the following, 
we examine what values of the diffractive 
masses $M_X$, or 
what values of $\beta=Q^2/(Q^2+M_X^2)$, dominate the integrand of the expression 
for $\sigma_{2}^j(x,Q^2)$ in Eq.~(\ref{eq:m17}).
This question is important in relation to the issue of the
 applicability of our leading twist approach based on 
the DGLAP evolution.

\begin{figure}[h]
\begin{center}
\epsfig{file=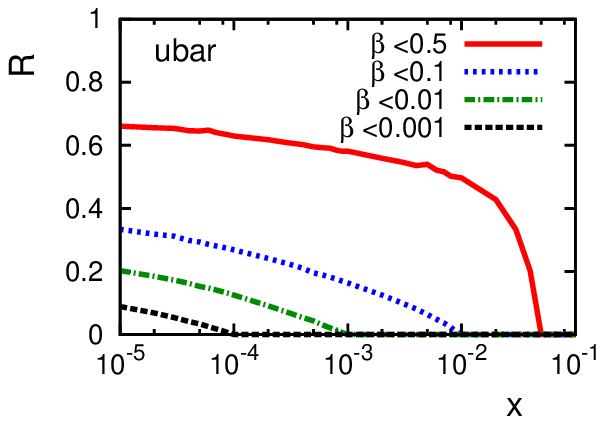,scale=1.25}
\epsfig{file=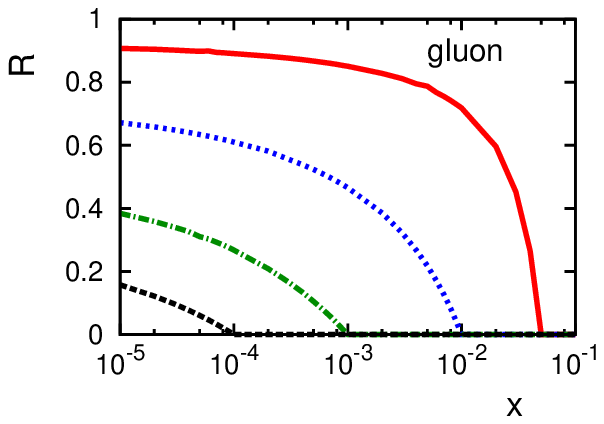,scale=1.25}
\caption{The ratio $R$ of Eq.~(\ref{eq:r_masses}) at $Q_0^2=4$ GeV$^2$. 
The solid curves correspond to $\beta_{{\rm max}}=0.5$;
the dotted curves correspond to $\beta_{{\rm max}}=0.1$; 
the dot-dashed curves correspond to $\beta_{{\rm max}}=0.01$; 
the short-dashed curves correspond to $\beta_{{\rm max}}=0.001$.}
\label{fig:r}
\end{center}
\end{figure}

To quantify the contributions of different regions of integration 
over $\beta$ to $\sigma_{{2}}^j(x,Q^2)$,
we introduce the ratio $R$ defined as follows:
\begin{equation}
R(\beta_{{\rm max}},x) \equiv \frac{\int_x^{0.1}
d x_{\Pomeron}  \beta f_{j/ N}^{D(3)}(\beta,Q_0^2,x_{\Pomeron}) \Theta(\beta_{{\rm max}}-\beta)}
{\int_x^{0.1}
d x_{\Pomeron}  \beta f_{j/ N}^{D(3)}(\beta,Q_0^2,x_{\Pomeron})} \,.
\label{eq:r_masses}
\end{equation}
The ratio $R$ for the $\bar{u}$-quark and gluon channels at $Q_0^2=4$ GeV$^2$ is presented in 
Fig.~\ref{fig:r}. In the figure, the solid curves correspond to $\beta_{{\rm max}}=0.5$;
the dotted curves correspond to $\beta_{{\rm max}}=0.1$; 
the dot-dashed curves correspond to $\beta_{{\rm max}}=0.01$; 
the short-dashed curves correspond to $\beta_{{\rm max}}=0.001$.

One can infer from Fig.~\ref{fig:r}
the relative contributions of
different 
$\beta $-regions to $\sigma_{{2}}^j(x,Q^2)$ and, hence, to nuclear shadowing.
For instance, for 
$x \leq 10^{-5}$,
 the $\beta \leq 0.001$-region contributes 
 to nuclear shadowing at most 9\% in the quark channel and 16\% in the gluon channel.
 This estimate suggests that even for such small values of
Bjorken $x$, various small-$x$ effects, which are not 
included in the DGLAP picture, should not lead to significant corrections in the evaluation of 
nuclear PDFs.

Another conclusion is that the diffractively produced masses 
$M_X^2 \approx Q^2 (1-\beta)/\beta$ can be large. 
At very high energies (small $x$), one enters the regime analogous to the
triple Pomeron limit of hadronic physics, which allows for $\beta \ll 1$.
This contribution (neglecting the large-$\beta$ contribution) to the nuclear structure
functions at extremely small $x$ was evaluated in the Color Glass Condensate framework, see, e.g., Ref.~\cite{McLerran:2003yx}.

\subsubsection{Nuclear antishadowing and DGLAP evolution}
\label{subsubsect:antishadowing}

By construction, 
Eq.~(\ref{eq:m13master}) does not describe nuclear modifications of PDFs for
 $x > 0.1$, where such effects as nuclear antishadowing and the
 EMC effect take place. 
However, we need to know nuclear PDFs at our chosen input scale
 $Q_0^2=4$ GeV$^2$ for a wide range of the values of Bjorken
$x^{\prime}$, $ x \leq x^{\prime} \leq 1$, since we use those nPDFs as an input 
for the Dokshitzer-Gribov-Lipatov-Altarelli-Parisi
(DGLAP) evolution  to higher $Q^2 > Q_0^2$.

The DGLAP evolution equations for PDFs $f_j$ of any target
(we use the nucleus)  read~\cite{dglap}:
\begin{eqnarray}
&&\frac{d\,f_{j/A}^{ns}(x,Q^2)}{d \log Q^2}=\frac{\alpha_s(Q^2)}{2 \pi} \int^1_x \frac{d x^{\prime}}{x^{\prime}}
P_{qq}\left(\frac{x}{x^{\prime}}\right) f_{j/A}^{ns}(x^{\prime},Q^2) \,, \nonumber\\
&&\frac{d}{d \log Q^2} 
\left(\begin{array}{c}
f^s_A(x,Q^2) \\
f_{g/A}(x,Q^2)
\end{array} \right)=\frac{\alpha_s(Q^2)}{2 \pi} \int^1_x \frac{d x^{\prime}}{x^{\prime}}
\left(
\begin{array}{cc}
P_{qq}\left(\frac{x}{x^{\prime}}\right) & P_{qg}\left(\frac{x}{x^{\prime}}\right) \\
P_{qg}\left(\frac{x}{x^{\prime}}\right) & P_{gg}\left(\frac{x}{x^{\prime}}\right)
\end{array}
\right)
\left(\begin{array}{c}
f_A^s(x^{\prime},Q^2) \\
f_{g/A}(x^{\prime},Q^2)
\end{array} \right) \,,
\label{eq:dglap}
\end{eqnarray}
where $\alpha_s(Q^2)$ is the QCD running coupling constant;
$P_{ij}$ are the splitting functions, which give the probability for parton 
$j$ with the longitudinal momentum fraction $x^{\prime}$ to 
radiate parton $i$ with the momentum fraction $x$.
The so-called non-singlet and singlet combinations of PDFs are denoted as
$f^{ns}_{j/A}$ and $f_A^s$, respectively. 

Equations~(\ref{eq:dglap}) are integro-differential equations, which can be 
numerically solved by,
e.g., the so-called brute force method. In this method, starting from the input
$f_{j/A}(x,Q_0^2)$ at the initial evolution scale $Q_0^2$, one obtains 
$f_{j/A}(x,Q^2)$ at $Q^2=Q_0^2+\Delta Q^2$ ($\Delta Q^2$ is small)
by integrating the right-hand side of 
Eqs.~(\ref{eq:dglap}). 
The procedure is repeated until the desired value of
$Q^2$ is reached. 
In our numerical analysis, we use the 
QCDNUM Fortran program which uses the brute force method~\cite{QCDNUM}.

It is clear from the above discussion and from Eqs.~(\ref{eq:dglap})
that in order to determine $f_{j/A}(x,Q^2)$ at any $Q^2$,
one needs to know $f_{j/A}(x^{\prime},Q_0^2)$ for all $x \leq x^{\prime} <1$.
In other words, we need to extend our calculations for nPDFs beyond the 
$x <0.1$ region and also to adopt an external model for the nuclear valence PDFs.

In our analysis,  
we adopt the general qualitative picture of nuclear modification of PDFs developed 
in~\cite{Frankfurt:1988nt,Frankfurt:1988zg,Frankfurt:1990xz}, which was later confirmed
 by the results of global fits to fixed-target DIS on nuclei 
data~\cite{Eskola:1998iy,Eskola:1998df,Eskola:2002us,Paukkunen:2010qi,Hirai:2001np,Hirai:2004wq,Hirai:2007sx,deFlorian:2003qf,Li:2001xa,Eskola:2003cc,Eskola:2007my,Eskola:2008ca,Eskola:2009uj,Schienbein:2009kk}.
In our implementation, the picture of nuclear PDFs at the initial
scale $Q_0^2=4$ GeV$^2$
looks as follows.
Antiquarks (sea quarks) in nuclei are shadowed for $x \leq 0.1$ 
according to Eq.~(\ref{eq:m13master}). For $x > 0.1$, we take
$f_{\bar{q}/A}/(Af_{\bar{q}/N}) =1$.
Gluons in nuclei are also shadowed for $x \leq 0.1$ 
according to Eq.~(\ref{eq:m13master}). In addition to nuclear shadowing,
the nuclear gluon PDF is enhanced (antishadowed).
The antishadowing of gluons is constrained using the momentum sum rule,
\begin{equation}
\sum_{j=q,{\bar q},g} \int_{0}^1 dx x f_{j/A}(x,Q^2)=1 \,. 
\label{eq:msr}
\end{equation}
In our numerical analysis, we model the gluon antishadowing at the initial
scale $Q_0^2=4$ GeV$^2$ in the interval
 $0.03 \leq x \leq 0.2$ using the following simple form,
\begin{eqnarray}
\frac{f_{g/A}(x,Q_0^2)}{Af_{g/N}(x,Q_0^2)}&=&
\frac{f_{g/A}(x,Q_0^2)}{Af_{g/N}(x,Q_0^2)}[{\rm given}\ {\rm by}\ {\rm Eq.}(\ref{eq:m13master})]
\nonumber\\
&+&\Theta(0.03 \leq x \leq 0.2) N_{{\rm anti}}\,(0.2-x)(x-0.03) \,,
\label{eq:anti}
\end{eqnarray}
where the free parameter $N_{{\rm anti}}$ is found by requiring the conservation of the momentum sum rule, see Eq.~(\ref{eq:msr}). Table~\ref{table:Nanti} summarizes
the used numerical values of $N_{{\rm anti}}$ for different nuclei.
Note that Eq.~(\ref{eq:msr}) constrains $N_{{\rm anti}}$ rather weakly: large
variations of $N_{{\rm anti}}$ lead to insignificant changes in the momentum sum rule.
Hence, the values of $N_{{\rm anti}}$ in Table~\ref{table:Nanti} should not be taken 
too literally---the momentum sum rule is a poor way to constrain antishadowing.
\begin{table}[h]
\begin{tabular}{|c|c|}
\hline
Nucleus & $N_{{\rm anti}}$ \\
\hline
$^{12}$C & 5 \\
$^{40}$Ca & 20 \\
$^{110}$Pd & 25 \\
$^{197}$Au & 30 \\
$^{208}$Pb & 30 \\
\hline
\end{tabular}
\caption{The parameter $N_{{\rm anti}}$ that controls the magnitude of antishadowing,
see Eq.~(\ref{eq:anti}), as a function of the nuclear atomic number $A$.}
\label{table:Nanti}
\end{table}

In Eq.~(\ref{eq:anti}), the first term is evaluated using Eq.~(\ref{eq:m13master}); on the $0.1 \leq x \leq 0.2$ interval, it is set to unity.
In addition, 
we assume that the gluon PDF is not modified for $x> 0.2$, 
$f_{g/A}/(Af_{g/N}) = 1$.

As was mentioned above, we do not attempt to give numerical predictions for
valence nuclear PDFs. Instead, we use the results of the global QCD fit
of Eskola and collaborators~\cite{Eskola:1998iy,Eskola:1998df}.

\subsubsection{Input nuclear charm quark PDF}

The QCD analysis of the H1 diffractive 
data~\cite{Breitweg:1998gc,Adloff:1997sc,Aktas:2006hy,Aktas:2006hx}
assumes that the charm quark diffractive PDF is zero at the initial scale $Q_0$.
As a result, the naive application of our master Eq.~(\ref{eq:m13master})
for the calculation of nuclear shadowing for charm quarks gives no
nuclear shadowing in this channel, $f_{c/A}(x,Q_0^2)/[A f_{c/N}(x,Q_0^2)]=1$.
However, there is no reason for such an approximation.

At small values of Bjorken $x$, charm quarks are mostly produced via
the QCD evolution because of the $g \to c \bar{c}$ splitting.
Thus, nuclear shadowing 
for
the charm quarks at some $x$ and $Q^2=Q^2_{0}$
originates from nuclear shadowing of gluons at larger $x$ and a certain $Q^2_{\rm eff} > Q^2_0$.
In our analysis, we use the following model~\cite{Frankfurt:2002kd}:
\begin{equation}
\frac{f_{c/A}(x,Q_{0}^2)}{A f_{c/N}(x,Q_{0}^2)}=\frac{f_{g/A}(2\,x,Q_{{\rm eff}}^2)}{A f_{g/N}(2\,x,Q_{{\rm eff}}^2)} \ ,
\label{shc}
\end{equation}
where $Q^2_{{\rm eff}}= 4m_c^2+Q_{0}^2=11$ GeV$^2$ ($m_c=1.3$ GeV).

In practice, the following procedure was performed. First, using Eq.~(\ref{eq:m13master}) and the QCD evolution, we find the gluon PDF, assuming
no nuclear shadowing for charm quarks.
Second, we use Eq.~(\ref{shc}) to determine the charm nuclear PDF at the 
input scale $Q_0^2$. Third, we repeat the QCD evolution, this time with the shadowed 
charm quarks, which gives us the final result for nuclear PDFs of all
flavors and for all scales.

\subsubsection{Predictions for nuclear PDFs}
\label{subsubsect:predictions}

We present the results of our calculations of nuclear PDFs in terms 
of the following ratios:
\begin{eqnarray}
R_j=\frac{f_{j/A}(x,Q^2)}{A f_{j/N}(x,Q^2)} \,, \nonumber\\
R_{F_2}=\frac{F_{2A}(x,Q^2)}{A F_{2N}(x,Q^2)} \,, 
\label{eq:ratios}
\end{eqnarray}
where $A f_{j/N} \equiv Z f_{j/p}+N f_{j/n}$;  $A F_{2N} \equiv ZF_{2p}+NF_{2n}$; 
the subscripts $p$ and $n$ refer to the proton and neutron, respectively.
For the PDFs and structure functions of the neutron, we used the charge symmetry, e.g.,
$f_{u/n}(x,Q^2)=f_{d/p}(x,Q^2)$.

\begin{figure}[t]
\begin{center}
\epsfig{file=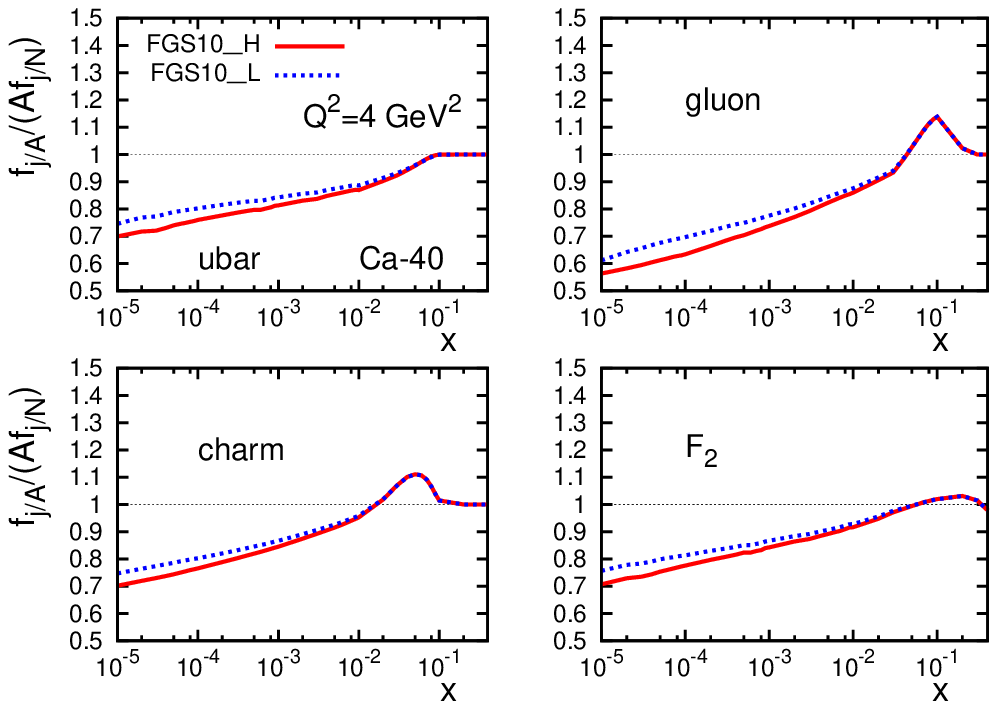,scale=1.29}
\epsfig{file=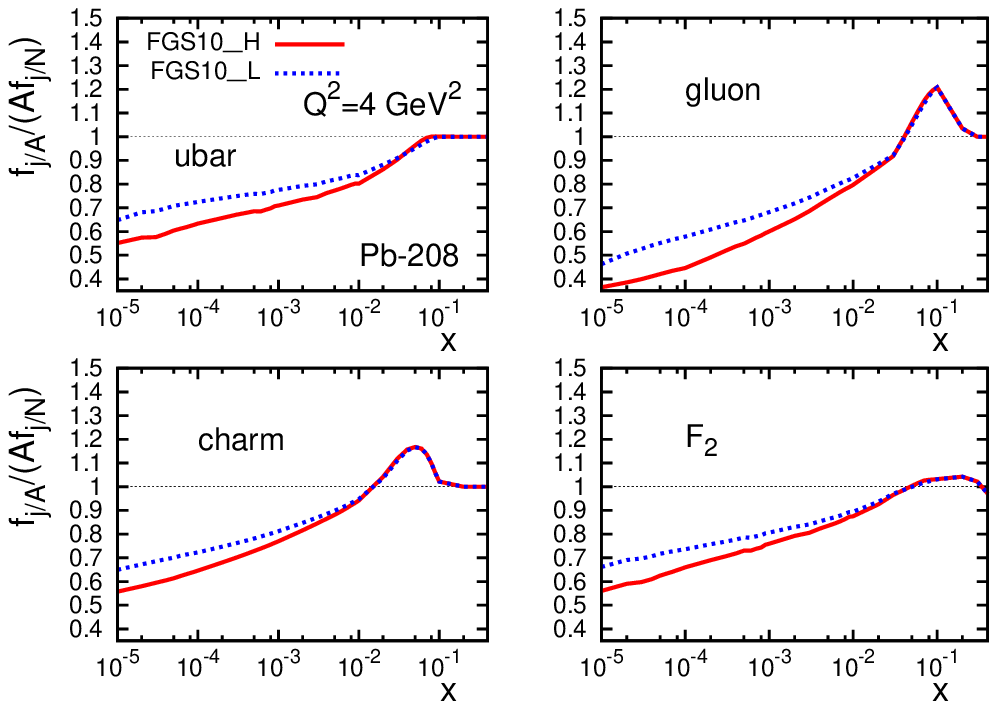,scale=1.29}
\caption{
Predictions for nuclear shadowing at the input scale $Q_0^2=4$ GeV$^2$.
The ratios $R_j$ ($\bar{u}$ and $c$ quarks and  gluons) and $R_{F_2}$
as functions of Bjorken $x$ at $Q^2=4$.
The four upper panels are for $^{40}$Ca; the four lower panels are for $^{208}$Pb.
Two sets of curves correspond to models FGS10\_H and FGS10\_L (see the text).
}
\label{fig:LT2009_ca40_models12}
\end{center}
\end{figure}

\begin{figure}[t]
\begin{center}
\epsfig{file=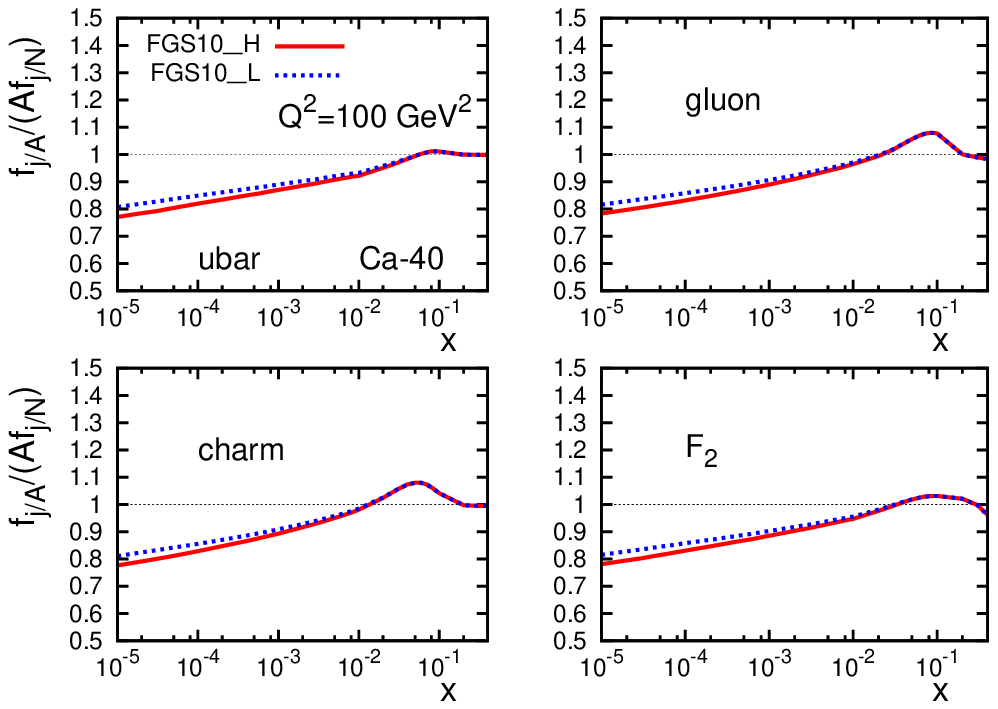,scale=1.29}
\epsfig{file=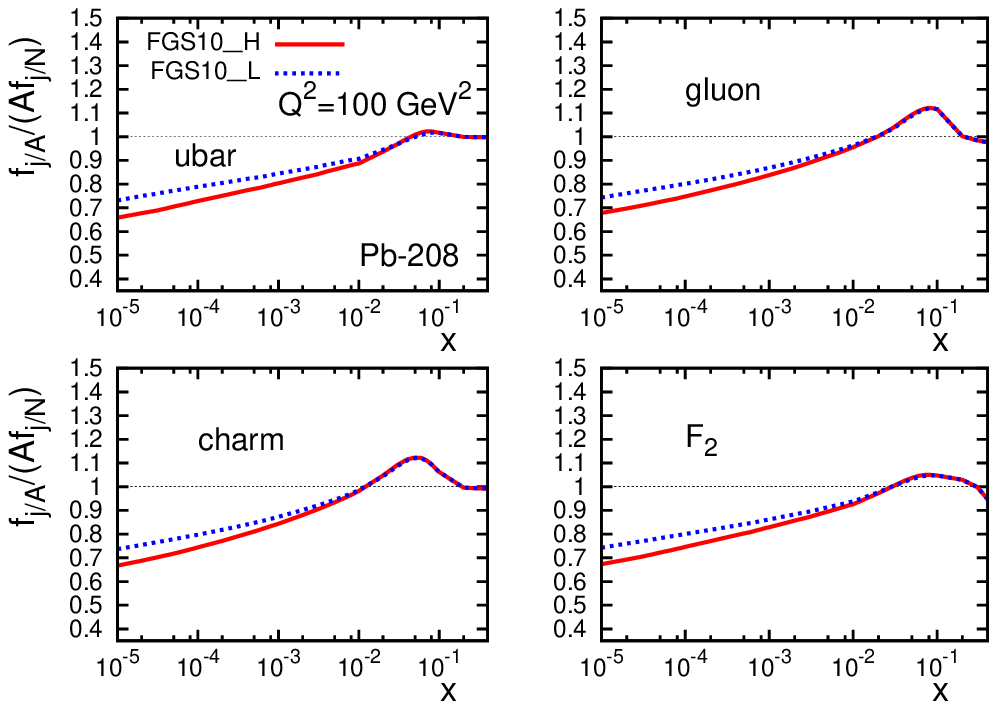,scale=1.29}
\caption{The same as in Fig.~\ref{fig:LT2009_ca40_models12}, but the ratios are evaluated at
$Q^2=100$ GeV$^2$.}
\label{fig:LT2009_ca40_models12_Q2_100}
\end{center}
\end{figure}

\begin{figure}[t]
\begin{center}
\epsfig{file=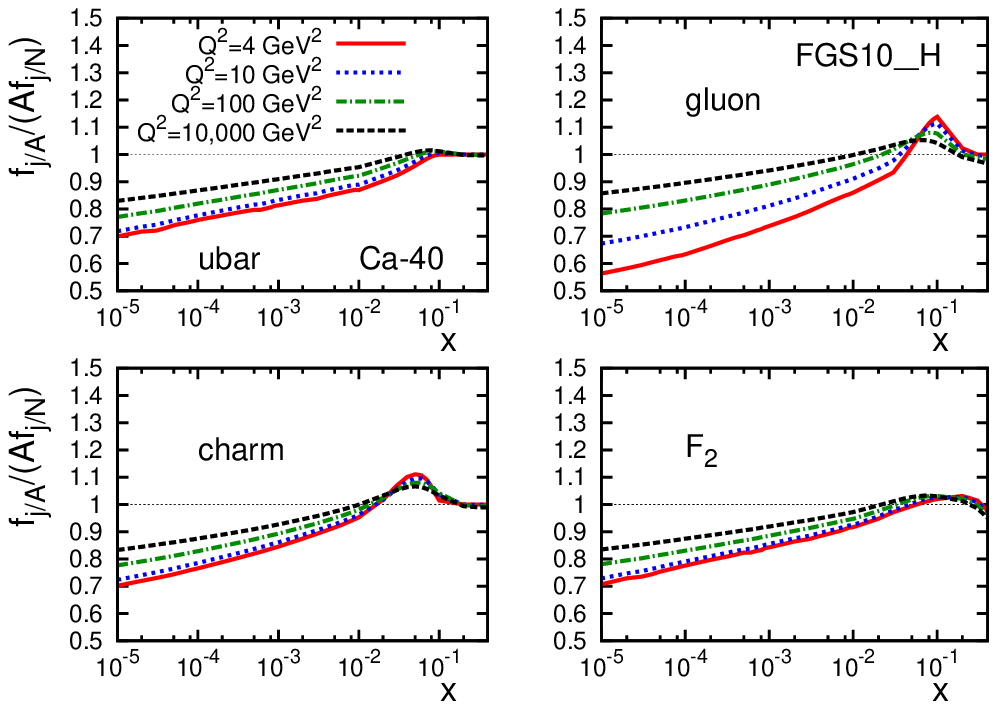,scale=1.29}
\epsfig{file=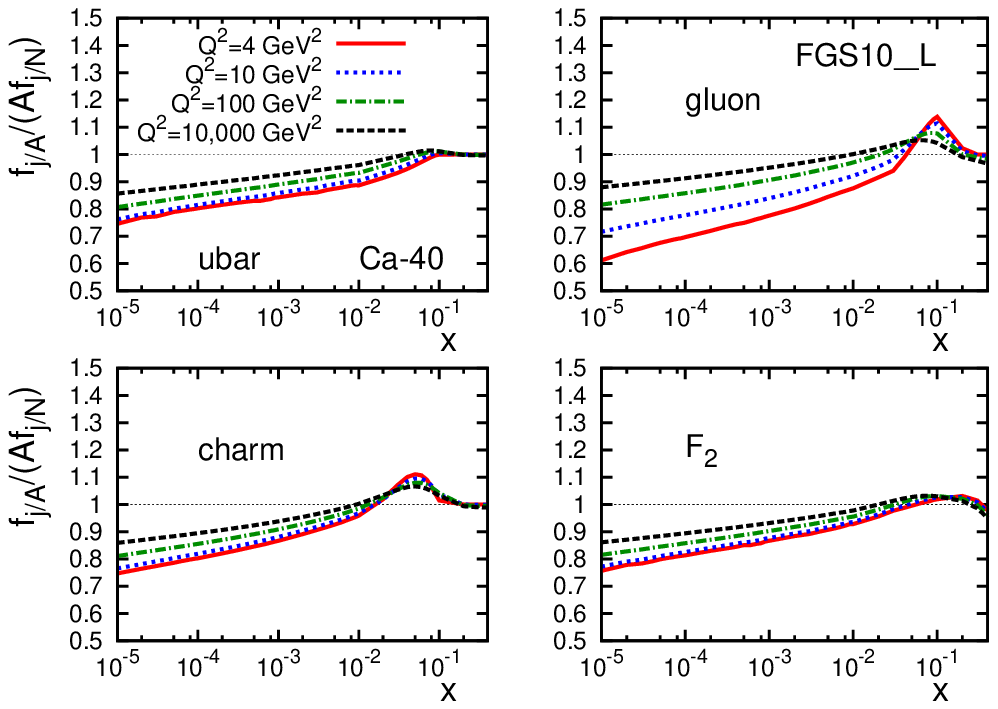,scale=1.29}
\caption{Prediction for nuclear PDFs and structure functions for $^{40}$Ca.
The ratios $R_j$ ($\bar{u}$ and $c$ quarks and  gluons) and $R_{F_2}$
as functions of Bjorken $x$ at $Q^2=4$, 10, 100 and 10,000 GeV$^2$.
The four upper panels correspond to FGS10\_H; the four lower panels correspond
to FGS10\_L.
}
\label{fig:LT2009_ca40}
\end{center}
\end{figure}

\begin{figure}[h]
\begin{center}
\epsfig{file=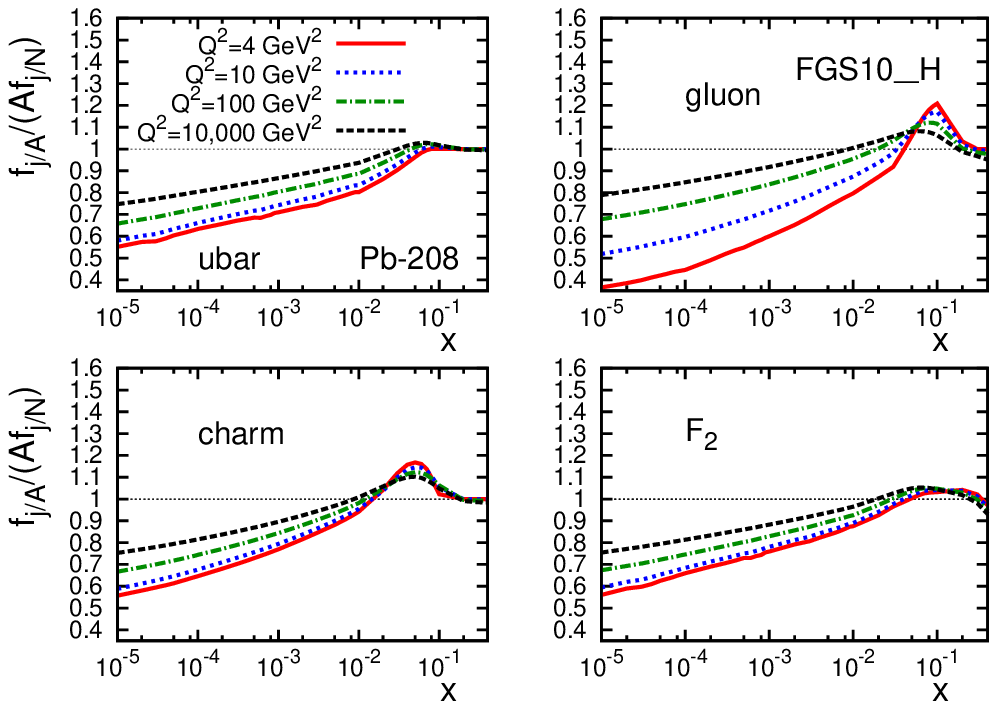,scale=1.25}
\epsfig{file=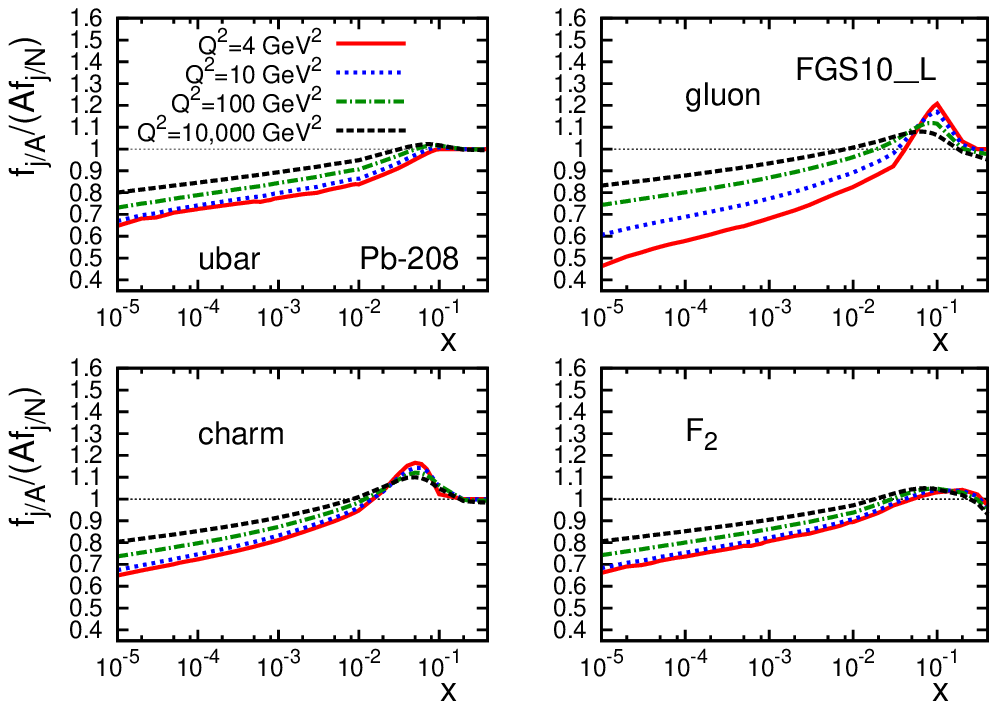,scale=1.25}
\caption{Prediction for nuclear PDFs and structure functions for $^{208}$Pb.
The ratios $R_j$ ($\bar{u}$ and $c$ quarks and  gluons) and $R_{F_2}$
as functions of Bjorken $x$ at $Q^2=4$, 10, 100 and 10,000 GeV$^2$.
The four upper panels correspond to FGS10\_H; the four lower panels correspond
to FGS10\_L.
}
\label{fig:LT2009_pb208}
\end{center}
\end{figure}

Figures~\ref{fig:LT2009_ca40_models12}, \ref{fig:LT2009_ca40_models12_Q2_100}, \ref{fig:LT2009_ca40}, 
and \ref{fig:LT2009_pb208}
present our predictions 
for $R_j$ and $R_{F_2}$ for the nuclei of $^{40}$Ca and $^{208}$Pb.

In Figs.~\ref{fig:LT2009_ca40_models12} and \ref{fig:LT2009_ca40_models12_Q2_100},
we compare predictions 
made using
the two models for the effective cross section $\sigma_{\rm soft}^j$
that we discussed in Sec.~\ref{subsubsec:color_fluct}.
The curves labeled ''FGS10\_H'' correspond to the calculation with 
$\sigma_{\rm soft}^{j(\rm H)}(x,Q_0^2)$
given by Eq.~(\ref{eq:sigma3_model1}); 
the curves labeled ''FGS10\_L'' correspond to the calculation with 
$\sigma_{\rm soft}^{j(\rm L)}(x,Q_0^2)$ of Eq.~(\ref{eq:sigma_L}).
(The effective cross sections $\sigma_{\rm soft}^{j(\rm H)}(x,Q_0^2)$ and
$\sigma_{\rm soft}^{j(\rm L)}(x,Q_0^2)$ are compared in Fig.~\ref{fig:sigma3_2009}.)
 The curves in Fig.~\ref{fig:LT2009_ca40_models12}
correspond to the input scale $Q_0^2=4$ GeV$^2$; 
the curves in Fig.~\ref{fig:LT2009_ca40_models12_Q2_100} correspond to $Q^2=100$ GeV$^2$.
The four upper panels are for $^{40}$Ca; the four lower panels are for $^{208}$Pb.
One can see from Fig.~\ref{fig:LT2009_ca40_models12} that the difference in the 
predictions of nuclear shadowing in the two models is not large
(it is smaller that the uncertainty associated with the diffractive slope $B_{\rm diff}$).
 Moreover, as
one can see from Fig.~\ref{fig:LT2009_ca40_models12_Q2_100}, the difference between
the two models decreases with increasing $Q^2$.

In Figs.~\ref{fig:LT2009_ca40} and \ref{fig:LT2009_pb208}, $R_j$ and $R_{F_2}$ are 
given as functions of Bjorken $x$
for $Q^2=4$ GeV$^2$ (input) and $Q^2=10$, 100 and 10,000 GeV$^2$
(after the QCD evolution). 
For the $R_j$ ratio, the predictions are
given for the $\bar{u}$ and $c$ quarks and 
gluons. 
For $R_{F_2}$,
next-to-leading (NLO) nuclear and nucleon structure functions
are used.

Several features of our predictions need to be
pointed out. First, at the input scale, $Q_0^2=4$ GeV$^2$, and also
after the evolution to not very large $Q^2$, 
nuclear shadowing in the gluon channel is larger than 
in the quark channel. This is a natural consequence of
the fact that the gluon diffractive PDF is much larger than the quark ones,
see Fig.~\ref{fig:FitAB}. As one increases $Q^2$, nuclear 
shadowing in the gluon channel decreases faster than in the quark channel and 
rapidly becomes compatible to that in the quark channel. 
This is the effect of antishadowing for the gluon nPDF, which feeds into the 
QCD evolution equations and reduces nuclear shadowing in the gluon channel 
for
$Q^2 > Q_0^2$.

Second, the $Q^2$ evolution of $R_j$ and $R_{F_2}$ is slow.
This is a manifestation of the leading twist nature of nuclear shadowing
in our approach.

Third, nuclear shadowing is larger for heavier nuclei.

Figures~\ref{fig:LT2009_ca40} and  \ref{fig:LT2009_pb208}
give just 
several
examples of our predictions for nuclear PDFs and nuclear structure 
functions. The complete set of predictions, which involve the nuclei of
$^{12}$C, $^{40}$Ca, $^{110}$Pd, $^{197}$Au and $^{208}$Pb and cover the wide
kinematics range of $10^{-5} \leq x \leq 0.95$ and $4 \leq Q^2 \leq 10,000$ GeV$^2$,
can be found at {\tt http://www.jlab.org/\~{}vguzey}.
In addition, these predictions and the predictions of all groups performing global fits to
nuclear PDFs can be conveniently and easily obtained using the online generator of nuclear 
PDFs at
{\tt http://lapth.in2p3.fr/npdfgenerator}.

\newpage

\newpage

\newpage

\clearpage

\newpage

\newpage

\subsection{Nuclear shadowing in longitudinal structure function $F_L^A(x,Q^2)$}

\begin{figure}[h]
\begin{center}
\epsfig{file=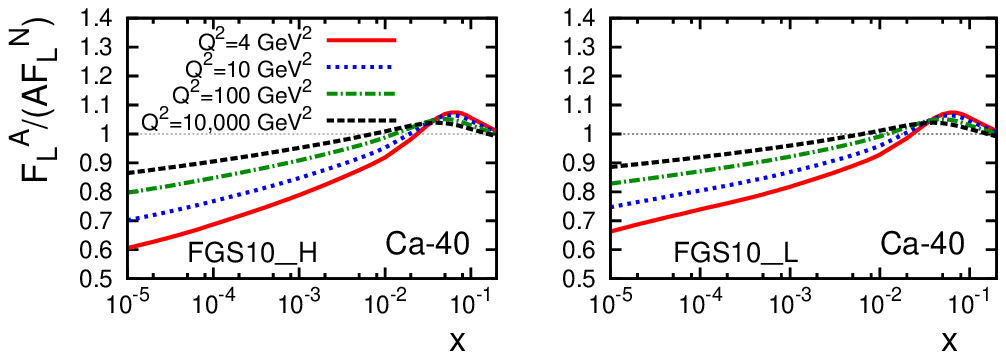,scale=1.4}
\epsfig{file=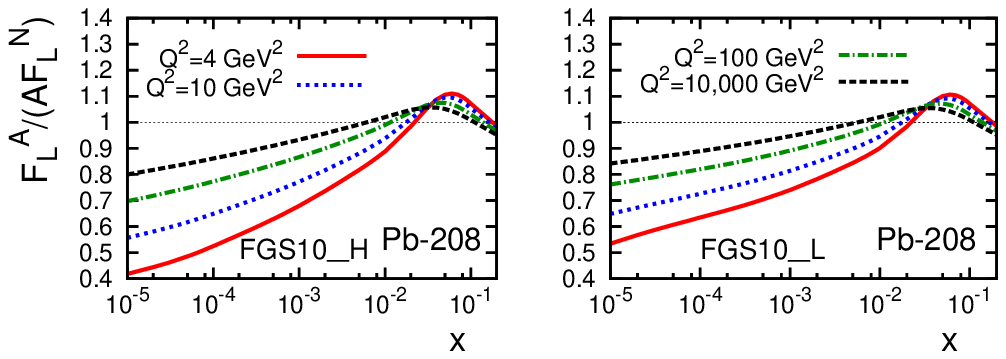,scale=1.4}
\caption{Nuclear shadowing for the longitudinal structure function 
$F_L(x,Q^2)$. The ratio of the nuclear to 
nucleon longitudinal
structure functions, $F_L^A(x,Q^2)/[A F_L^N(x,Q^2)]$, as a function of Bjorken $x$ 
for different values of $Q^2$.
The upper row of panels corresponds to $^{40}$Ca; the lower row is 
for $^{208}$Pb.
Two sets of curves correspond to models FGS10\_H and FGS10\_L.
}
\label{fig:F_L}
\end{center}
\end{figure}
The longitudinal structure function $F_L(x,Q^2)$ is sensitive to the gluon distribution
at small $x$. To the leading order in the strong coupling constant $\alpha_s$, it reads~\cite{Brock:1993sz}:
\begin{eqnarray}
F_L(x,Q^2)=\frac{2\alpha_s(Q^2)}{\pi}\int_x^1 \frac{dy}{y}\left(\frac{x}{y}\right)^2 \sum_{q}^{n_f} &&e_q^2 \Bigg[\left(1-\frac{x}{y}\right) yg(y,Q^2) \nonumber\\
&&+\frac{2}{3}\left(yq(y,Q^2)+y\bar{q}(y,Q^2)\right)\Bigg] \,,
\label{eq:F_L}
\end{eqnarray}
where the sum runs over quark flavors; $n_f$ is the number of active flavors at given 
$Q^2$.

Figure~\ref{fig:F_L} presents our predictions for the ratio of the nuclear to 
nucleon longitudinal
structure functions, $F_L^A(x,Q^2)/[A F_L^N(x,Q^2)]$, as a function of Bjorken $x$ at
different values of $Q^2$. 
The upper row of panels corresponds to $^{40}$Ca; the lower row is 
for $^{208}$Pb.
The two sets of curves correspond to models FGS10\_H and FGS10\_L.
As one can see from the figure, the amount of nuclear
shadowing for $F_L^A(x,Q^2)$ is compatible with that of the nuclear gluon PDF, see
Figs.~\ref{fig:LT2009_ca40} and \ref{fig:LT2009_pb208}. 

As we explained 
in the Introduction,
the measurement of the longitudinal structure function in inclusive DIS 
with nuclei presents a new and promising opportunity to determine the 
nuclear gluon parton distribution.

Large nuclear shadowing effects in the nuclear longitudinal structure function, 
which are similar in magnitude to the large nuclear shadowing in the nuclear gluon 
distribution, were also predicted in the approach based on nPDFs extracted from the 
global QCD fits to the available data
for $x\ge 10^{-2}$ and various guesses about the behavior of nPDFs 
at smaller $x$~\cite{Armesto:2010tg}. 

Another important quantity related to the longitudinal structure function
is the ratio of the virtual photon-target cross sections for the longitudinal
and transverse polarizations of the virtual photon, 
\begin{equation}
R \equiv \frac{\sigma_L}{\sigma_T}=\frac{F_L(x,Q^2)}{F_{2}(x,Q^2)-F_L(x,Q^2)} \,.
\label{eq:RA}
\end{equation}
Below we present our predictions for the super-ratio $R_A/R_N$, which is the ratio
of the nuclear to the nucleon ratios $R$:
\begin{eqnarray}
\frac{R_A}{R_N} &\equiv &  \frac{F_L^A(x,Q^2)}{F_{2A}(x,Q^2)-F_L^A(x,Q^2)}
 \frac{F_{2N}(x,Q^2)-F_L^N(x,Q^2)}{F_L^N(x,Q^2)}  \nonumber\\
&=& \frac{F_L^A(x,Q^2)}{A F_L^N(x,Q^2)} \frac{A F_{2N}(x,Q^2)}{F_{2A}(x,Q^2)}
\frac{1-F_L^N(x,Q^2)/F_{2N}(x,Q^2)}{1-F_L^A(x,Q^2)/F_{2A}(x,Q^2)} \,.
\label{eq:RA_super}
\end{eqnarray}
The advantage of considering the super-ratio $R_A/R_N$ is that this quantity
is essentially insensitive to the value of the elementary ratio $R_N$. 

\begin{figure}[t]
\begin{center}
\epsfig{file=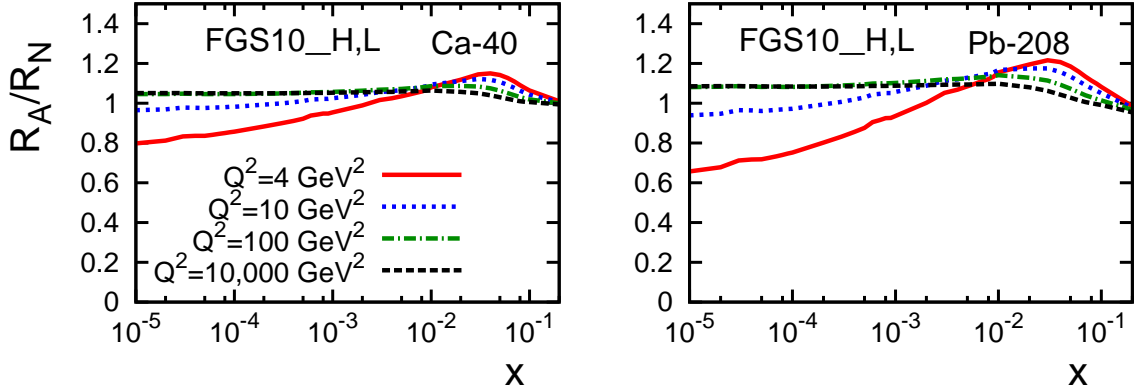,scale=1.5}
\caption{The super-ratio $R_A/R_N$
of Eq.~(\ref{eq:RA_super}) as a function of Bjorken $x$ for
different values of $Q^2$.
Models FGS10\_H and FGS10\_L give numerically indistinguishable predictions.
}
\label{fig:RA_2010}
\end{center}
\end{figure}

Figure~\ref{fig:RA_2010} presents our predictions for $R_A/R_N$
of Eq.~(\ref{eq:RA_super})
 for $^{40}$Ca and $^{208}$Pb
for four different values of $Q^2$ as a function of Bjorken $x$.
Both models FGS10\_H and FGS10\_L give numerically indistinguishable predictions for 
$R_A/R_N$. Also, as one can see from Fig.~\ref{fig:RA_2010}, 
the predicted $A$ dependence of $R_A/R_N$ is rather weak, but still non-negligible.
(This also naturally applies to the ratio $R_A$.)

The trend of the $x$ behavior of $R_A/R_N$ can be understood as follows.
For small $x$, $x \leq 10^{-3}$, and not too large $Q^2$, $Q^2 \leq 10$ GeV$^2$,
the suppression of $F_L^A/(A F_L^N)$ due to nuclear shadowing is larger than that of
$F_{2A}/(A F_{2N})$ (the nuclear gluon PDF is shadowed more that the quark nuclear PDFs), 
which makes $R_A/R_N < 1$. 
As one increases $x$, antishadowing begins to play a role, which makes $F_L^A/(A F_L^N)>1$, 
see Fig.~\ref{fig:F_L}. As a result, $R_A/R_N > 1$ for approximately
$5 \times 10^{-3} \leq  x \leq 0.2$.

One has to note that as an input for our calculation of the $R$ factor,
we use the nucleon longitudinal structure function $F_L^N(x,Q^2)$ that we calculate using
the CTEQ5M parton distributions. 
A comparison of our predictions for $F_L^N(x,Q^2)$
to the ZEUS~\cite{Chekanov:2009na} and H1~\cite{H1Collaboration:2010ry} data on 
$F_L^N(x,Q^2)$ shows that our predictions somewhat overestimate the data.
At the same time, NLO and NNLO predictions made with contemporary parton distributions
describe the data reasonably well~\cite{H1Collaboration:2010ry}.
An inspection shows that the CTEQ5M gluon distribution at small $x$ is significantly 
larger than, e.g., the CT10 gluon distribution~\cite{Lai:2010vv}
which explains our overestimate of the HERA data on $F_L^N(x,Q^2)$.

\subsection{Energy and $Q^2$ dependence of nuclear shadowing}

\begin{figure}[t]
\begin{center}
\epsfig{file=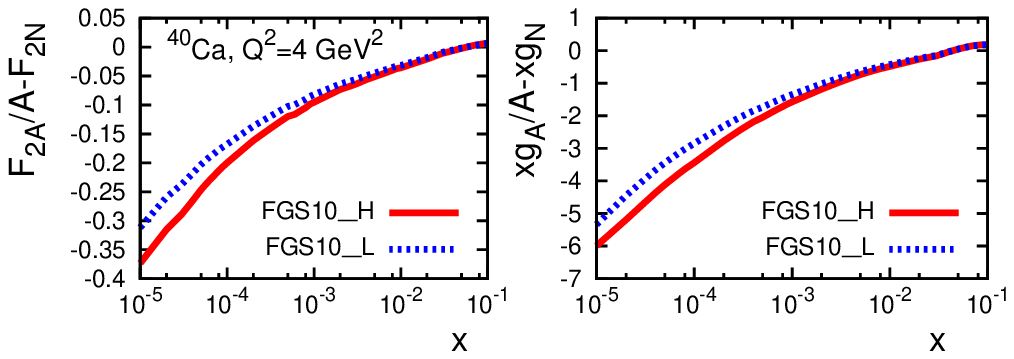,scale=1.25}
\epsfig{file=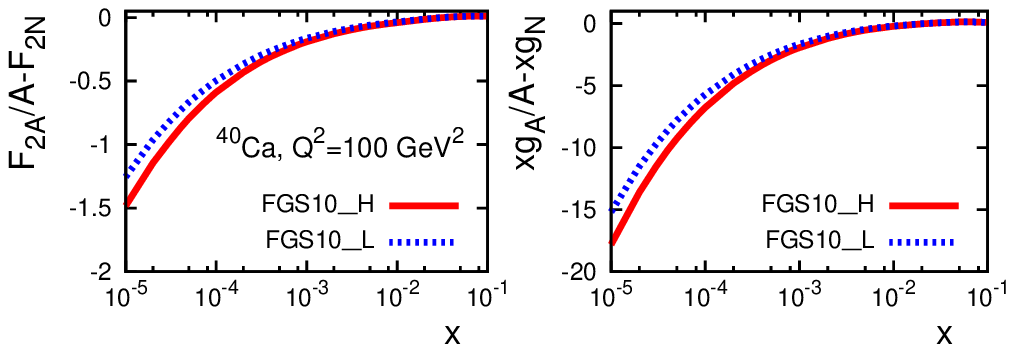,scale=1.25}
\epsfig{file=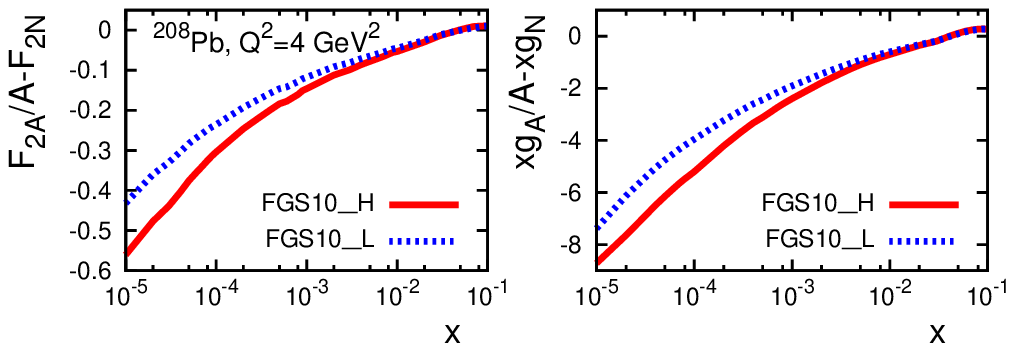,scale=1.25}
\epsfig{file=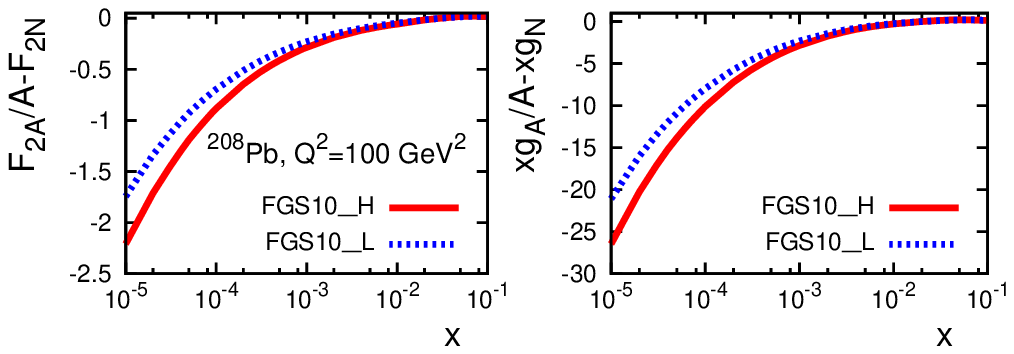,scale=1.25}
\caption{The Bjorken $x$ dependence of nuclear shadowing.
The shadowing corrections per nucleon, $\delta F_{2A}(x,Q^2)/A$ and $\delta g_{A}(x,Q^2)/A$,
see Eq.~(\ref{eq:deltaF2}), as functions of Bjorken $x$ at $Q^2=4$ GeV$^2$ and $Q^2=100$ GeV$^2$
for $^{40}$Ca (four upper panels) and $^{208}$Pb (four lower panels).
The two sets of curves correspond to models FGS10\_H and FGS10\_L.
}
\label{fig:LT2009_difference_fit}
\end{center}
\end{figure}

It is also important to study the energy dependence (the dependence on Bjorken $x$)
and $Q^2$ dependence  of nuclear shadowing. 
In the following, we consider
the shadowing corrections to the structure function $F_{2A}(x,Q^2)$ and to the 
gluon distribution $g_A(x,Q^2)$ defined respectively as
\begin{eqnarray}
\delta F_{2A}(x,Q^2) &\equiv& F_{2A}(x,Q^2)-ZF_{2p}(x,Q^2)-(A-Z) F_{2p}(x,Q^2) \,,
 \nonumber\\
\delta xg_{A}(x,Q^2) &\equiv& xg_{A}(x,Q^2)-Axg_N(x,Q^2) \,.
\label{eq:deltaF2}
\end{eqnarray}
These quantities are less sensitive to uncertainties in the nucleon PDFs at $x\sim 10^{-4}$ and $Q^2\sim 4$ GeV$^2$.
Figure~\ref{fig:LT2009_difference_fit} presents $\delta F_{2A}(x,Q^2)/A$ and 
$\delta xg_{A}(x,Q^2)/A$ as 
a function of Bjorken $x$ at two values of $Q^2$, 
$Q^2=4$ GeV$^2$ and $Q^2=100$ GeV$^2$.
 The two top rows of panels 
correspond to $^{40}$Ca; the two bottom rows correspond to $^{208}$Pb. 
The solid curves correspond to FGS10\_H; the dotted curves correspond to FGS10\_L.

For small values of $x$, $x <10^{-3}$, the curves in Fig.~\ref{fig:LT2009_difference_fit} 
can be economically parameterized by a simple analytical expression:
\begin{eqnarray}
\delta F_{2A}(x,Q^2)/A & \approx& \delta N_{F_2} \left(\frac{0.00001}{x}\right)^{0.25} 
\,, \nonumber\\
\delta xg_{A}(x,Q^2)/A & \approx& \delta N_g \left(\frac{0.00001}{x}\right)^{0.25} \,,
\label{eq:deltaF2_approx}
\end{eqnarray}
with the coefficients $\delta N_{F_2}$ and $\delta N_g$ summarized in Table~\ref{table:Energy_dependence_shadowing}.
 
\begin{table}[h]
\begin{tabular}{|c|c|c||c|c|}
\hline
& $^{40}$Ca & $^{40}$Ca & $^{208}$Pb & $^{208}$Pb \\
\hline
$Q^2=4$ GeV$^2$ & FGS10\_H & FGS10\_L & FGS10\_H & FGS10\_L \\
\hline
$\delta N_{F_2}$ & $-0.37$ & $-0.31$ & $-0.56$ & $-0.43$ \\
$\delta N_g$ & $-6.0$ & $-5.3$ & $-8.7$ & $-7.4$ \\
\hline \hline
$Q^2=100$ GeV$^2$ & FGS10\_H & FGS10\_L & FGS10\_H & FGS10\_L \\
\hline
$\delta N_{F_2}$ & $-1.48$ & $-1.25$ & $-2.21$ & $-1.74$ \\
$\delta N_g$ & $-17.8$ & $-15.2$ & $-26.5$ & $-21.1$ \\
\hline 
\end{tabular}
\caption{The parameters $\delta N_{F_2}$ and $\delta N_g$ of Eq.~(\ref{eq:deltaF2_approx}).}
\label{table:Energy_dependence_shadowing}
\end{table}

The numerical value of the exponent $\lambda=0.25$ in 
Eq.~(\ref{eq:deltaF2_approx}) can be understood as follows.
The $x$ dependence of nuclear shadowing at small $x$ is primarily driven by 
the $x_{\Pomeron}$ dependence of the Pomeron flux 
$f_{\Pomeron/p}(x_{\Pomeron}) \propto 1/x_{\Pomeron}^{(2 \alpha_{\Pomeron}-1)}
\propto 1/x_{\Pomeron}^{1.22}$. Therefore, in the very small $x$ limit, one expects from
Eq.~(\ref{eq:m13master}) that, approximately,
\begin{eqnarray}
\delta F_{2A}(x,Q^2)/A & \propto &  \left(\frac{1}{x} \right)^{0.22} \,, \nonumber\\
\delta xg_{A}(x,Q^2)/A & \propto &  \left(\frac{1}{x} \right)^{0.22} \,,
\label{eq:deltaF2_approx2}
\end{eqnarray}
which is consistent with our 
numerical
result in Eq.~(\ref{eq:deltaF2_approx}).

When we present our predictions for nuclear shadowing in the form of the ratios 
of the nuclear to nucleon PDFs, it is somewhat difficult to see the 
leading twist nature of the predicted nuclear shadowing because of the rapid
$Q^2$ dependence of the free nucleon structure functions and PDFs.
In order to see the leading twist nuclear shadowing
more explicitly, one should examine the absolute values of the 
shadowing corrections.

\begin{figure}[t]
\begin{center}
\epsfig{file=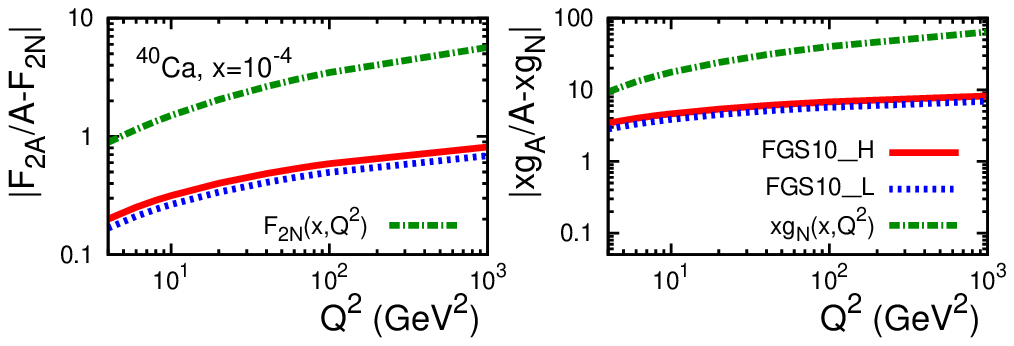,scale=1.25}
\epsfig{file=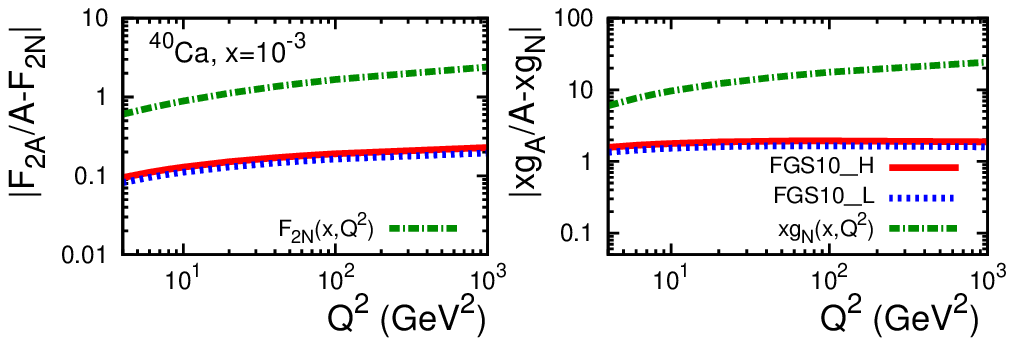,scale=1.25}
\epsfig{file=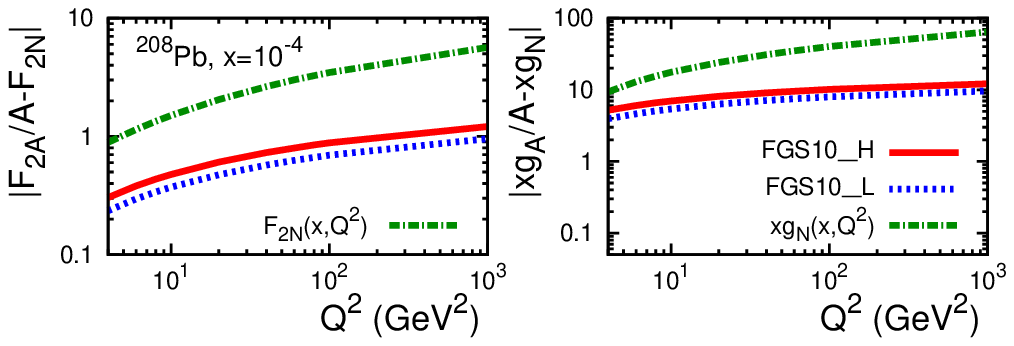,scale=1.25}
\epsfig{file=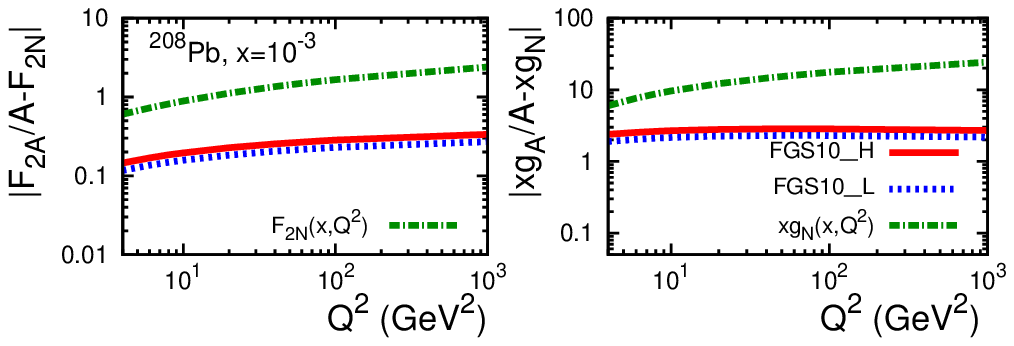,scale=1.25}
\caption{The $Q^2$ dependence of nuclear shadowing.
The shadowing corrections per nucleon, $\delta F_{2A}(x,Q^2)/A$ and 
$\delta g_{A}(x,Q^2)/A$,
see Eq.~(\ref{eq:deltaF2}), as functions of $Q^2$ at fixed $x=10^{-4}$ and
$x=10^{-3}$ for $^{40}$Ca (four upper panels) and $^{208}$Pb (four lower panels).
The solid curves is the result of 
FGS10\_H;
the dotted curves correspond to 
FGS10\_L.
For comparison, the dot-dashed curves present the impulse (unshadowed) contributions to 
$F_{2A}(x,Q^2)/A$ and $xg_{A}(x,Q^2)/A$, which are equal to
$F_{2N}(x,Q^2)$ and $xg_N(x,Q^2)$,
respectively.
}
\label{fig:LT2009_difference_fit_Q2dep}
\end{center}
\end{figure}

Figure~\ref{fig:LT2009_difference_fit_Q2dep}
presents $|\delta F_{2A}(x,Q^2)/A|$ and $|\delta xg_{A}(x,Q^2)/A|$
as functions of $Q^2$ at fixed $x=10^{-4}$ (first and third rows) and $x=10^{-3}$
(second and fourth rows)
for $^{40}$Ca (four upper panels) and $^{208}$Pb (four lower panels).
The solid curves correspond to 
FGS10\_H;
 the dotted curves correspond to 
FGS10\_L.
Also, for comparison, 
presented by
 the dot-dashed curves, we give the impulse (unshadowed) contributions to 
$F_{2A}(x,Q^2)/A$ and $xg_{A}(x,Q^2)/A$, which are equal to 
$F_{2N}(x,Q^2)$ and $xg_{N}(x,Q^2)$, respectively.

One can draw several conclusions by examining the curves in Fig.~\ref{fig:LT2009_difference_fit_Q2dep}. First, at $x=10^{-4}$ and $x=10^{-3}$, 
the difference in the predicted nuclear shadowing in models FGS10\_H and 
FGS10\_L
 (the difference between
the solid and dotted curves) is essentially zero. 
Second, $|\delta F_{2A}(x,Q^2)|$ and $|\delta xg_{A}(x,Q^2)|$ do not disappear as 
$Q^2$ is increased, which means that the considered nuclear shadowing is not a higher 
twist ($1/Q^2$ power-suppressed) phenomenon.
Third, the deviation of the dot-dashed curves
from the solid and dotted curves increases as $Q^2$ is increased. This, again, points to
the leading twist nature of nuclear shadowing, which may not be clearly seen in the ratios
of the nuclear to nucleon structure functions and PDFs.

\clearpage

\clearpage

\subsection{The $A$ dependence of nuclear shadowing}

\begin{figure}[h]
\begin{center}
\epsfig{file=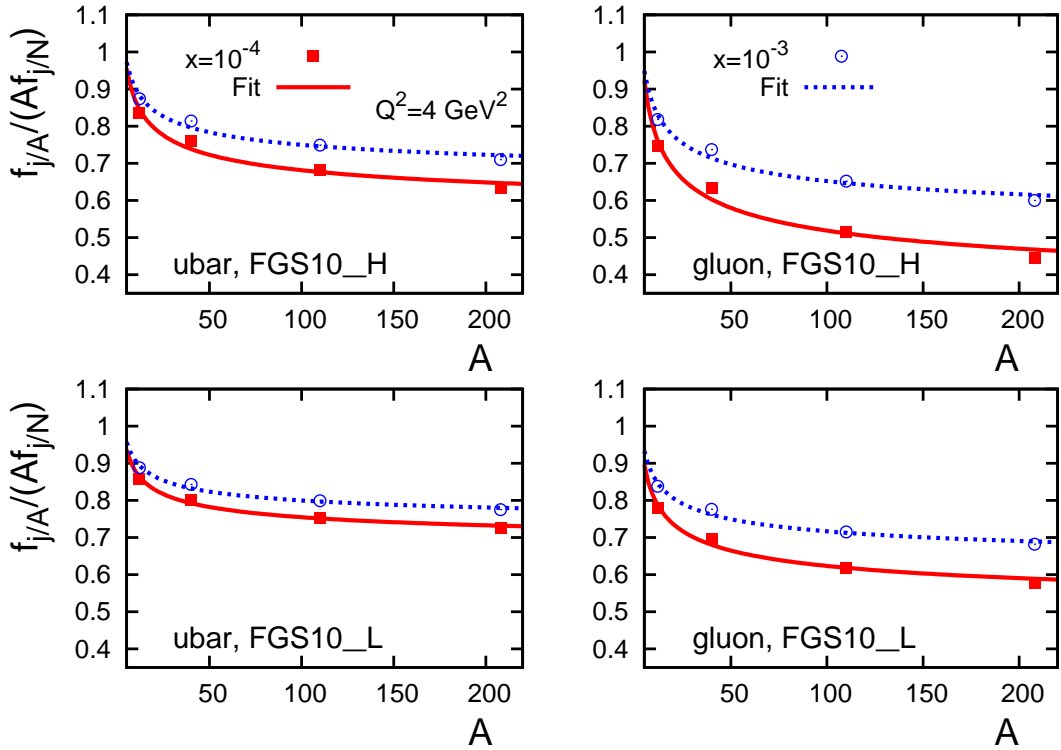,scale=1.4}
\caption{The $A$ dependence of nuclear shadowing.
The points (squares for $x=10^{-4}$ and open circles for $x=10^{-3}$) 
are the results of our calculations
for $f_{j/A}(x,Q^2)/[Af_{j/N}(x,Q^2)]$ for $^{12}$C, $^{40}$Ca, $^{110}$Pd, and 
$^{208}$Pb;
the smooth curves is a two-parameter fit of Eq.~(\ref{eq:Shadowing_Adep}).
}
\label{fig:Shadowing_Adep}
\end{center}
\end{figure}

Using our predictions for nuclear PDFs, 
one can also examine the resulting $A$ dependence of nuclear shadowing.
An example of this is presented in Fig.~\ref{fig:Shadowing_Adep}, where we plot
the $f_{j/A}(x,Q^2)/[Af_{j/N}(x,Q^2)]$ ratio as a function of the atomic mass number $A$ for
two values of $x$: $x=10^{-4}$ and $x=10^{-3}$. All curves correspond to our input scale
$Q^2=Q_0^2=4$ GeV$^2$. 
The points (squares  for $x=10^{-4}$ and open circles for $x=10^{-3}$)
are the results of our calculations
for $f_{j/A}(x,Q^2)/[Af_{j/N}(x,Q^2)]$ for $^{12}$C, $^{40}$Ca, $^{110}$Pd, and 
$^{208}$Pb;
the smooth curves is a two-parameter fit (for $A \geq 12$) in the form~\cite{Frankfurt:1988nt}:
\begin{equation}
\frac{f_{j/A}(x,Q^2)}{Af_{j/N}(x,Q^2)}=\lambda+a(1-\lambda)/A^{1/3} \,,
\label{eq:Shadowing_Adep}
\end{equation} 
where $\lambda$ and $a$ are free parameters of the fit; they are summarized in 
Table~\ref{table:Shadowing_Adep}.
Note that the fit in Eq.~(\ref{eq:Shadowing_Adep}) is designed for $A \geq 12$.
The form of the fit in Eq.~(\ref{eq:Shadowing_Adep}) corresponds to the 
simple physical picture that assumes that the virtual photon wave function contains only two
essential components: a point-like configuration (PLC) that has a small cross section and, hence,
cannot be shadowed, and an effective large-size configuration, which is a subject to full-fledged 
nuclear shadowing.

\begin{table}[h]
\begin{tabular}{|c|c|c||c|c|}
\hline
& $\lambda_{\rm gluon}$ & $a_{\rm gluon}$ & $\lambda_{\rm ubar}$ & $a_{\rm ubar}$ \\
\hline
$x=10^{-4}$ &      &      &      &     \\
FGS10\_H    & 0.28 & 1.53 & 0.52 & 1.55 \\
FGS10\_L    & 0.46 & 1.38 & 0.65 & 1.40 \\
\hline
$x=10^{-3}$ &      &      &      &     \\
FGS10\_H    & 0.48 & 1.54 & 0.62 & 1.58 \\
FGS10\_L    & 0.59 & 1.43 & 0.70 & 1.46 \\
\hline
\end{tabular}
\caption{The parameters $\lambda$ and $a$ of Eq.~(\ref{eq:Shadowing_Adep}) 
that fits the $A$ dependence of the $f_{j/A}(x,Q^2)/[Af_{j/N}(x,Q^2)]$ ratio.
}
\label{table:Shadowing_Adep}
\end{table}

The magnitude of the parameter $\lambda$ is related to the size of the unshadowed
PLC in the virtual photon wave function: the larger $\lambda$
corresponds to the higher probability of the PLC.
It is larger for the quark channel than for the gluon one. Also, as one increases energy
(decreases $x$), the probability of the PLC expectedly decreases.
One can also see that model FGS10\_H corresponds to the smaller $\lambda$ compared to model
FGS10\_L, which is understandable since FGS10\_H provides larger nuclear shadowing
than FGS10\_L (the probability of the PLC in FGS10\_H is suppressed compared to FGS10\_L).

The second term in Eq.~(\ref{eq:Shadowing_Adep}) is related to the probability of large-size
configurations in the virtual photon wave function. These configurations are strongly shadowed
and one expects that the resulting shadowing correction should behave as $A^{2/3}$, i.e., the
$A$ dependence of the second term in the $f_{j/A}(x,Q^2)/[Af_{j/N}(x,Q^2)]$ ratio should
behave as $1/A^{1/3}$.

\subsection{Impact parameter dependent nuclear PDFs}
\label{subsec:impact}

Predictions of the leading twist theory of nuclear shadowing for nPDFs can be readily 
generalized to predict the dependence of nuclear PDFs on the impact parameter $b$. 
The impact parameter dependent nPDFs, 
 $f_{j/A}(x,Q^2,b)$, can be introduced by the following relation~\cite{Frankfurt:2002kd}:
\begin{equation}
\int d^2 \vec{b} f_{j/A}(x,Q^2,b)=f_{j/A}(x,Q^2) \,.
\label{eq:impact}
\end{equation}
Removing the integration over the impact parameter $\vec{b}$ in our master
equation~(\ref{eq:m13master}),
one immediately obtains the nuclear PDFs as functions of $x$ and $b$:
\begin{samepage}
\begin{eqnarray}
&&x f_{j/A}(x,Q_0^2,b)= A\,T_A(b) x f_{j/N}(x,Q_0^2) \nonumber\\
&-&8 \pi A(A-1) B_{{\rm diff}}\,\Re e \frac{(1-i\eta)^2}{1+\eta^2} \int^{0.1}_x d x_{\Pomeron}
\beta f_j^{D(3)}(\beta,Q_0^2,x_{\Pomeron}) \nonumber\\
&\times&  
\int^{\infty}_{-\infty}d z_1 \int^{\infty}_{z_1}d z_2 \,\rho_A(\vec{b},z_1) \rho_A(\vec{b},z_2) \,
e^{i (z_1-z_2) x_{\Pomeron} m_N} 
e^{-\frac{A}{2} (1-i\eta) \sigma_{\rm soft}^j(x,Q_0^2) \int_{z_1}^{z_2} dz^{\prime} \rho_A(\vec{b},z^{\prime})} \,,
\label{eq:impact2}
\end{eqnarray}
\end{samepage}
where $T_A(b)=\int_{-\infty}^{\infty} dz \rho_A(\vec{b},z)$.
Note that the presence of the factor $T_A(b)$ in Eq.~(\ref{eq:impact2}) 
is required by the condition of Eq.~(\ref{eq:impact}).
The impact parameter dependent nPDFs, $f_{j/A}(x,Q^2,b)$, have the  meaning 
of the probability to find parton $j$ at the impact parameter $b$ at the 
resolution scale $Q^2$. 
In deriving Eq.~(\ref{eq:impact2})
the finite size of the nucleon was neglected as compared to the nucleus size. 

As we will discuss in Sec.~\ref{subsec:exclusive},
 our impact parameter dependent nuclear PDFs
are nothing else but the diagonal nuclear generalized parton distributions,
\begin{equation}
f_{j/A}(x,Q^2,b)=H_A^j(x,\xi=0,b,Q^2) \,.
\label{eq:diagonal_gpds}
\end{equation}

\begin{figure}[t]
\begin{center}
\epsfig{file=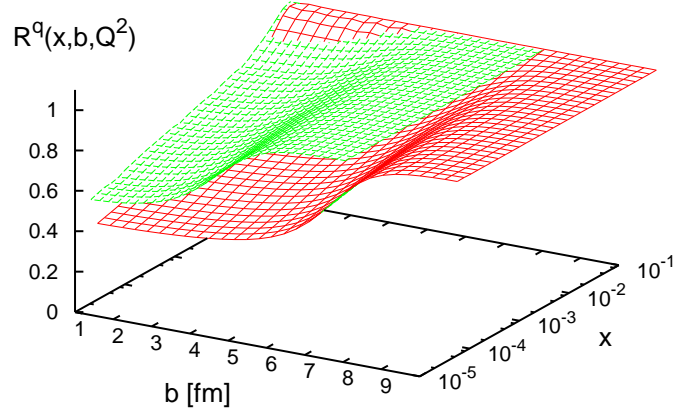,scale=1.}
\epsfig{file=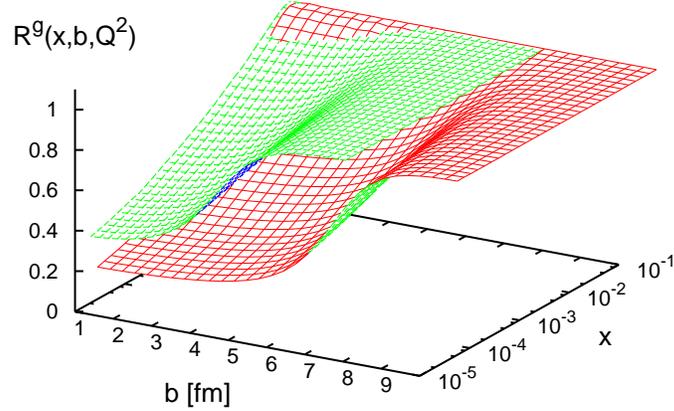,scale=1.}
\caption{Impact parameter dependence of nuclear shadowing for $^{40}$Ca 
(upper green surfaces) and $^{208}$Pb (lower red surfaces).
The graphs show the ratio $R^j(x,b,Q^2)$ of Eq.~(\ref{eq:ngpd4}) as a function of
$x$ and the impact parameter  $|\vec{b}|$ at $Q^2=4$ GeV$^2$.
The top panel corresponds to $\bar{u}$-quarks; the bottom panel corresponds to gluons.
For the evaluation of nuclear shadowing, model 
FGS10\_H
was used (see the text).
}
\label{fig:impact_dependence}
\end{center}
\end{figure}

Let us now discuss  the spatial image of nuclear shadowing.
This can be done by considering the ratio $R^j(x,b,Q^2)$:
\begin{equation}
R^j(x,b,Q^2)=\frac{f_{j/A}(x,Q^2,b)}{A\,T_A(b) f_{j/N}(x,Q^2)}=
\frac{H_A^j(x,\xi=0,b,Q^2)}{A\,T_A(b) f_{j/N}(x,Q^2)} \,.
\label{eq:ngpd4}
\end{equation}
The ratio $R^j(x,b,Q^2)$ of Eq.~(\ref{eq:ngpd4}) for $^{40}$Ca (upper green surfaces) and
$^{208}$Pb (lower red surfaces) as a function of $x$ and $|\vec{b}|$
is presented in Fig.~\ref{fig:impact_dependence}.
The top panel corresponds to $\bar{u}$ quarks; the bottom panel corresponds to gluons.
All surfaces correspond to $Q^2=4$ GeV$^2$ and to model 
FGS10\_H
 of nuclear shadowing (see the previous 
discussion).
Note that in the absence of nuclear shadowing, $R^j(x,b,Q^2)=1$.

Several features of Fig.~\ref{fig:impact_dependence} deserve a discussion.
First, as one can see from  Fig.~\ref{fig:impact_dependence},
the amount of nuclear shadowing---the suppression of $R^j(x,b,Q^2)$
compared to unity---increases as one decreases $x$ and $b$. 
 Second, nuclear shadowing for gluons is larger than for quarks.
Third, nuclear shadowing induces non-trivial correlations between $x$ and $b$ in the nuclear GPD $H_{A}^{j}(x,0,\vec{b},Q^2)$, even if such correlations were absent in the free nucleon GPD.
[In Eq.~(\ref{eq:impact2}) we neglected the $x$-$b$ correlations in the nucleon 
GPDs by neglecting the $t$ dependence of $H_N^j(x,0,t,Q^2)$
and using $H_N^j(x,0,t,Q^2) \approx f_{j/N}(x,Q^2)$.
]

\begin{figure}[h]
\begin{center}
\epsfig{file=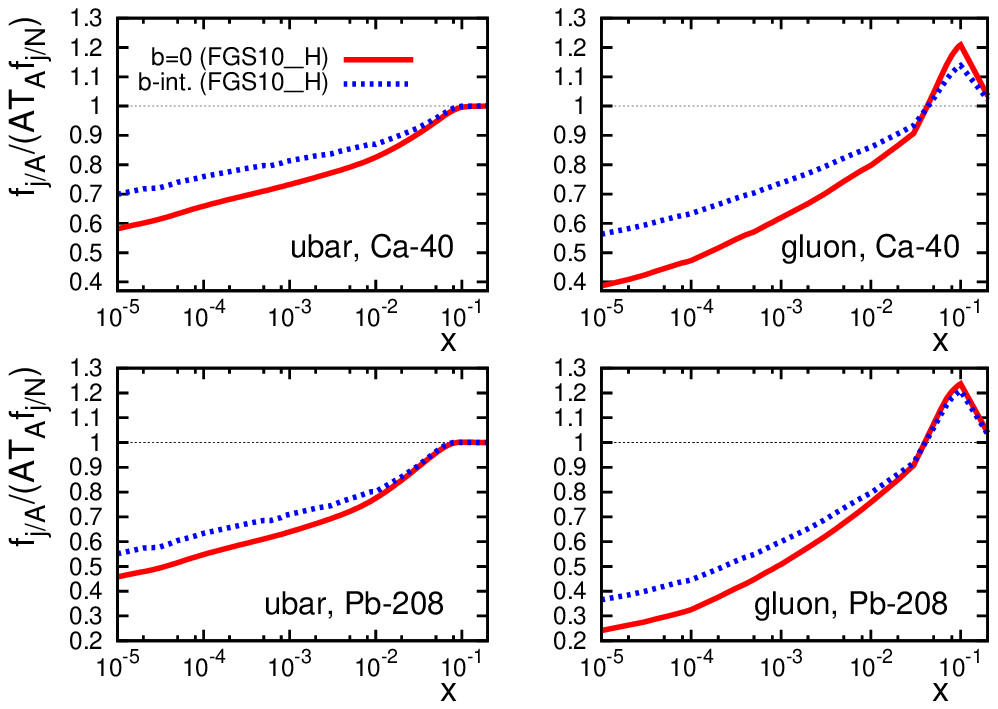,scale=1.4}
\caption{The ratio $f_{j/A}/(A T_A(b) f_{j/N})$ as a function of $x$.
The solid curves correspond to the central impact parameter ($b=0$); 
the dotted curves are for the nPDFs integrated over all $b$ (the same as in 
Figs.~\ref{fig:LT2009_ca40} and \ref{fig:LT2009_pb208}).
All curves correspond to $Q_0^2=4$ GeV$^2$ and to model FGS10\_H.
}
\label{fig:LT2009_ca40_impact}
\end{center}
\end{figure}

To make the discussed features of the spatial image of nuclear GPDs presented in Fig.~\ref{fig:impact_dependence}
more transparent, 
it is instructive to consider various
two-dimensional slices of Fig.~\ref{fig:impact_dependence}. 
In Fig.~\ref{fig:LT2009_ca40_impact}, we present $f_{j/A}/(A T_A(b) f_{j/N})$
as a function of $x$ at the central impact parameter ($b=0$) by the solid 
curves. 
For comparison, the dotted curves present the corresponding
results for the $b$-integrated nPDFs (i.e., usual nPDFs), see Figs.~\ref{fig:LT2009_ca40}
and \ref{fig:LT2009_pb208}.
 All curves correspond to our input
scale $Q_0^2=4$ GeV$^2$ and to model FGS10\_H. 
Note that since nuclear shadowing depends on the impact parameter, so should antishadowing.
We constrain the amount of antishadowing by requiring the conservation of the momentum 
sum rule locally in the impact parameter $b$ [compare to Eq.~(\ref{eq:msr})]:
\begin{equation}
\sum_{j=q,{\bar q}} \int_{0}^1 dx x f_{j/A}(x,Q^2,b)+ \int_{0}^1 dx x g_A(x,Q^2,b)=1 \,. 
\label{eq:msr_b}
\end{equation}
As a result, the parameter $N_{\rm anti}$ that controls the amount of antishadowing 
[see Eq.~(\ref{eq:anti})]
depends on $b$: $N_{\rm anti}$ decreases from its maximal value
at $b=0$ (the corresponding values of $N_{\rm anti}$
are numerically close to those given in Table~\ref{table:Nanti}) to $N_{\rm anti} = 0$ for
large $b$.

As can be seen from Fig.~\ref{fig:LT2009_ca40_impact}, nuclear shadowing
is larger at small impact parameters than that in the case when one integrates over
all $b$. This is a natural
consequence of the fact that the density of nucleons is larger in
the center of the nucleus.

\begin{figure}[h]
\begin{center}
\epsfig{file=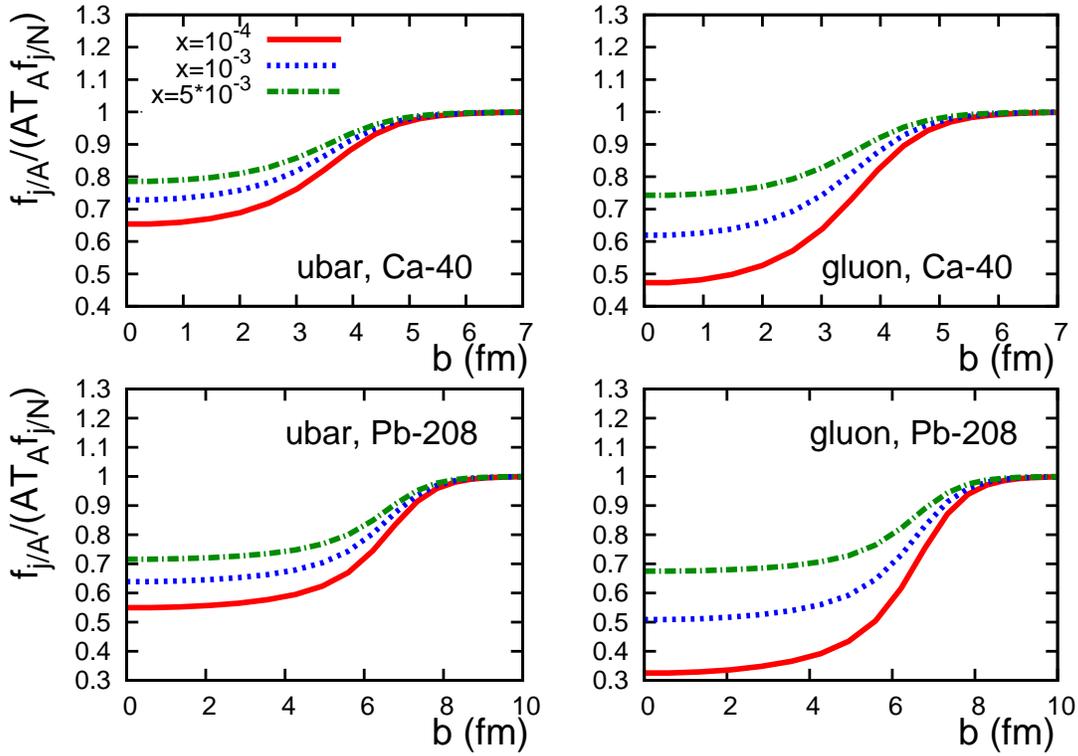,scale=1.4}
\caption{The ratio $f_{j/A}/(A T_A(b) f_{j/N})$ as a function of the impact
parameter $b$ for fixed values of $x=10^{-4}$, $x=10^{-3}$, and $x=0.005$.
All curves correspond to 
FGS10\_H
and $Q_0^2=4$ GeV$^2$.}
\label{fig:LT2009_ca40_impact_Bdep}
\end{center}
\end{figure}

In Fig.~\ref{fig:LT2009_ca40_impact_Bdep}, we plot $f_{j/A}/(A T_A(b) f_{j/N})$
as a function of the impact parameter $b$ for three different values of 
$x$, $x=10^{-4}$, $x=10^{-3}$, and $x=0.005$. All curves correspond to model 
FGS10\_H
and 
$Q_0^2=4$ GeV$^2$. As one see from the figure, nuclear shadowing 
for gluons is larger than for quarks in essentially an entire region of $b$.

DIS off nuclear targets involves usual nPDFs that are integrated over all impact 
parameters $b$. However, using the fact the nuclear shadowing is local in the impact
parameter [nuclear shadowing depends only on the nuclear density at a given $b$ and 
will be same for two different nuclei, $A_1$ and $A_2$, for the range of 
impact parameters satisfying the condition 
$A_1T_{A_1}(b_1)=A_2T_{A_2}(b_2)$],
one can 
enhance the contribution of small $b$ by considering special linear combinations
of the structure functions (parton distributions) of different nuclei. In particular, 
one can effectively eliminate the contribution of single and double scattering, and, thus,
essentially subtract the contribution of the nuclear edge (leave in mostly the contribution
of the nuclear center) by considering, e.g.,  the following combination:
\begin{equation}
F_{2A}(x,Q^2)-\lambda_{A/A_0} F_{2A_0}(x,Q^2)-(A-\lambda_{A/A_0} A_0)F_{2N}(x,Q^2) \,,
\label{eq:strikman_trick1}
\end{equation}
where $A$ refers to a heavy nucleus; $A_0$ refers to a light nucleus 
(such as $^4$He and $^{12}$C); the parameter $\lambda_{A/A_0}$ is defined as
\begin{equation}
\lambda_{A/A_0} \equiv \frac{\int d^2b\, A^2 T_A^2(b)}{\int d^2b\, A_0^2 T_{A_0}^2(b)} \,.
\label{eq:strikman_trick2}
\end{equation}
Since the expansion of the expression in
Eq.~(\ref{eq:strikman_trick1}) in the number of interactions with the 
target nucleons starts from the term proportional to $T_A^3(b)$, 
the combination in Eq.~(\ref{eq:strikman_trick1}) has the support for 
the values of $b$ that are more central (smaller) 
than those for the unsubtracted $F_{2A}(x,Q^2)$.

The dependence of nPDFs on the impact parameter and, thus, 
our predictions for nuclear shadowing as a function of the impact parameter $b$
can be probed in proton-nucleus ($pA$) and nucleus-nucleus ($AA$) collisions,
where the centrality (the impact parameter $b$) is defined
by the multiplicity of binary collisions. Examples of the application of 
the impact parameter dependent nPDFs involve inclusive production of pions~\cite{Vogt:2004hf}
and $J/\psi$~\cite{Vogt:2004dh,Vogt:2005ia} in $dA$ and $AA$ collisions at RHIC and in $pA$ and $AA$ collisions at the LHC~\cite{Abreu:2007kv}, 
where collisions with different centrality are selected using, e.g., 
the number of wounded nucleons.

Another opportunity to study the impact parameter dependence of nuclear shadowing
is provided by hard exclusive processes with nuclei such as deeply virtual Compton
scattering (DVCS) and vector meson electroproduction
(see the detailed discussion in Sec.~\ref{subsec:exclusive}).
 The amplitudes of these reactions are
expressed in terms of the convolution of the corresponding hard scattering coefficient
functions and nuclear generalized parton distributions (GPDs).
While in general those GPDs are complicated unknown distributions, 
at high energies (small Bjorken $x$), 
it is a good
approximation~\cite{Frankfurt:1997ha} to use the (nuclear) GPDs 
in the $\xi=0$ limit 
as the initial condition for the $Q^2$ evolution.
This is because the initial condition is dominated 
by the large parton light-cone fractions $x_{i} \gg \xi$ and the dependence on
$\xi$ is not evolved; the off-diagonal effects (the effects of $\xi \neq 0$)
are accounted for by the off-diagonal modifications of the QCD evolution kernels.

The leading twist nuclear shadowing in nuclear GPDs was analyzed in Ref.~\cite{Goeke:2009tu}.
It was found that nuclear shadowing leads to clear experimental signatures: 
the shift toward smaller $|t|$ of the DVCS differential cross section and dramatic oscillations of the beam-spin DVCS asymmetry. These effects can be interpreted as the fact that
the transverse size of parton distributions---as probed by hard probes---increases in nuclei.
We shall
address this issue in some detail in the following section.

Figures~\ref{fig:impact_dependence}, 
\ref{fig:LT2009_ca40_impact}, and  \ref{fig:LT2009_ca40_impact_Bdep}
present only several examples of our predictions for impact-parameter dependent nuclear PDFs. 
The complete set of predictions
for a wide range of nuclei
($^{12}$C, $^{40}$Ca, $^{110}$Pd, $^{197}$Au and $^{208}$Pb) and
 the
kinematics range $10^{-5} \leq x \leq 0.95$ and $4 \leq Q^2 \leq 10,000$ GeV$^2$
can be found at 
{\tt http://www.jlab.org/\~{}vguzey}.

\subsection{The transverse size of parton distributions in nuclei}
\label{subsec:transverse_size}

As we discussed in Sec.~\ref{subsec:impact}, the leading twist nuclear shadowing, which
increases with decreasing the impact parameter $b$, leads to distinct experimental
signatures that can be interpreted as the increase of the transverse size of the
parton distributions in nuclei compared to the free nucleon.
To quantify the effect, one can introduce the average transverse size of the 
parton distribution of flavor $j$ in terms of the corresponding impact 
parameter dependent nuclear PDFs:
\begin{equation}
\langle b_j^2 \rangle  \equiv \frac{\int d^2b \,b^2 f_{j/A}(x,Q^2,b)}{\int d^2b\, 
f_{jA}(x,Q^2,b)} \,.
\label{eq:b2}
\end{equation}
For comparison, the transverse size of the nuclear PDFs in the absence of nuclear
shadowing (in the impulse approximation) is 
\begin{equation}
\langle b_j^2 \rangle_{\rm imp} =\frac{\int d^2b \,b^2 A T_A(b) f_{j/N}(x,Q^2)}{\int d^2b\, A T_A(b) f_{j/N}(x,Q^2)}=\int d^2b \, b^2 T_A(b) = \frac{R_A^2}{6}  \,.
\label{eq:b2_imp}
\end{equation}

\begin{figure}[h]
\begin{center}
\epsfig{file=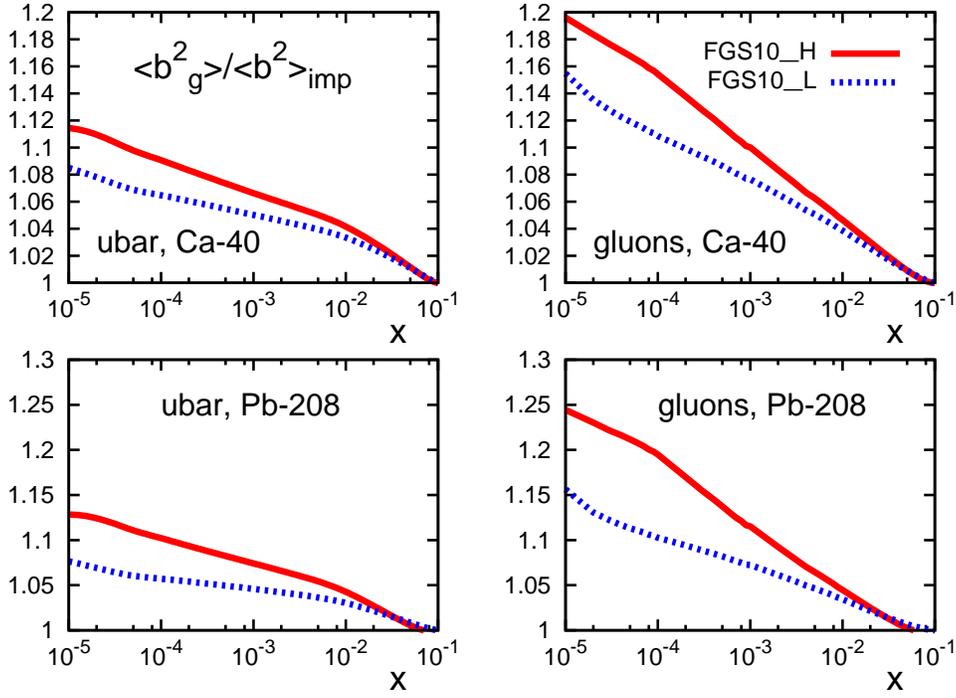,scale=1.3}
\caption{The ratio $\langle b_j^2 \rangle/\langle b_j^2 \rangle_{\rm imp}$ 
for the ${\bar u}$ quarks and gluons in $^{40}$Ca and $^{208}$Pb as a function of
Bjorken $x$ at $Q^2=4$ GeV$^2$. 
The two sets of curves correspond to models FGS10\_H and FGS10\_L.
}
\label{fig:gluon_size_pb208}
\end{center}
\end{figure}

The last equality follows from
the relation between $T_A(b)$ and the nuclear radius
$R_A$ that parameterizes the nuclear form factor at small $t$, 
$F_A(t)=\exp(-R_A^2 t/6)$. It is given as a reference point---in our analysis 
we used $T_A(b)$ obtained from the nuclear density $\rho_A(r)$~\cite{DeJager:1987qc}.
Note that in deriving Eq.~(\ref{eq:b2_imp}) we neglected the weak $t$ dependence 
of the free nucleon GPDs compared to 
the strong $t$ dependence of the nuclear form factor. In this approximation, $\langle b_j^2 \rangle_{\rm imp}$ is flavor-independent.

The ratio $\langle b_j^2 \rangle/\langle b_j^2 \rangle_{\rm imp}$ for 
the ${\bar u}$-quarks and gluons in $^{40}$Ca and $^{208}$Pb as a function of
Bjorken $x$ at $Q^2=4$ GeV$^2$ is presented in Fig.~\ref{fig:gluon_size_pb208}. 
The solid curves correspond to model FGS10\_H;
the dashed curves correspond to model 
FGS10\_L.
As one can see from the figure, the leading twist nuclear shadowing leads to an increase 
of the transverse size of shadowed parton distributions in nuclei.

\subsection{The role of the finite coherence length}

\begin{figure}[h]
\begin{center}
\epsfig{file=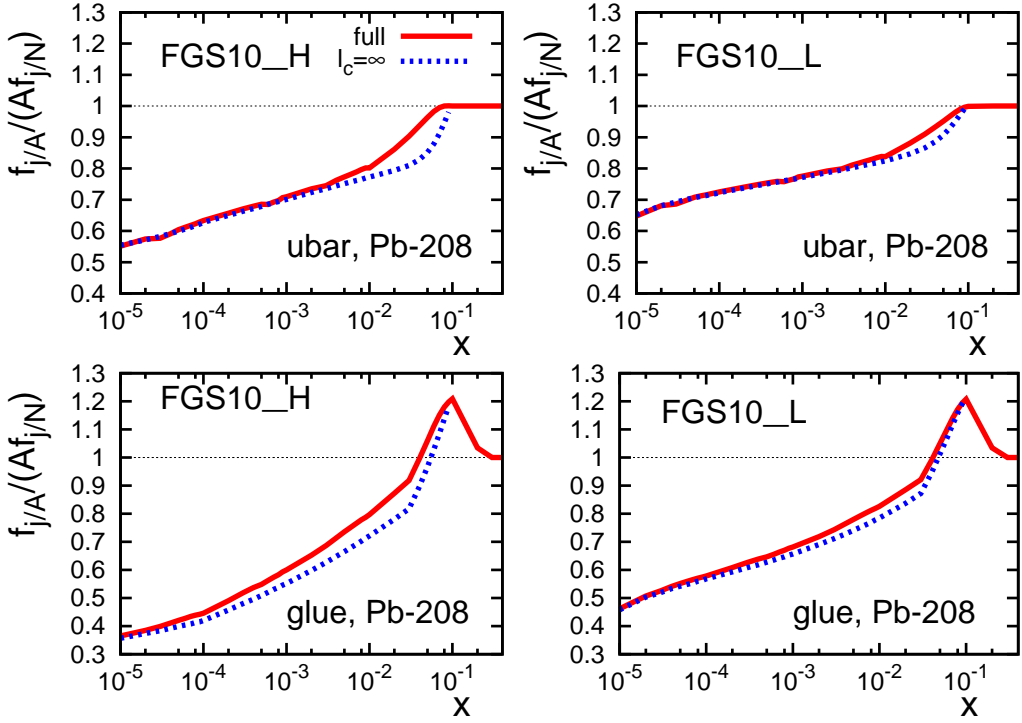,scale=1.35}
\caption{The ratios of $\bar{u}$-quark and gluon PDFs in $^{208}$Pb at $Q^2=4$ GeV$^2$.
The solid curves correspond to the full calculation using 
Eq.~(\ref{eq:m13master}); the dashed curves correspond to the calculation
with $e^{i(z_1-z_2) x_{\Pomeron} m_N}=1$ in Eq.~(\ref{eq:m13master}) or,
equivalently, to the calculation using Eq.~(\ref{eq:m13master_approx}).
}
\label{fig:LT2009_pb208_nolc}
\end{center}
\end{figure}

The effect of the non-zero longitudinal momentum transfer 
$\Delta_{\gamma^{\ast} X}$, see Eq.~(\ref{eq:m4}), or, in other words, the effect
of the finite coherence length $l_c \propto 1/\Delta_{\gamma^{\ast} X} \neq \infty$, 
on nuclear shadowing is given by the factor
$e^{i(z_1-z_2) x_{\Pomeron} m_N}$ in Eq.~(\ref{eq:m13master}).
At small values of Bjorken $x$, this factor can be neglected and after the integration by 
parts two times, Eq.~(\ref{eq:m13master}) can be cast in a much simpler form, see
Eq.~(\ref{eq:m13master_approx}).

Figure~\ref{fig:LT2009_pb208_nolc} compares the results of the full calculation
using Eq.~(\ref{eq:m13master}) (solid curves, same as in Figs.~\ref{fig:LT2009_pb208}) with the calculation using the approximate expression of  Eq.~(\ref{eq:m13master_approx})
(dashed curves).
The presented results correspond to
the ratios of $\bar{u}$-quark and gluon PDFs in $^{208}$Pb at $Q^2=4$ GeV$^2$.
The effect of antishadowing is added as explained in Sect.~\ref{subsubsect:antishadowing}.
As one can see from Fig.~\ref{fig:LT2009_pb208_nolc}, the effect of the finite
coherence length $l_c$ ($l_c \neq \infty$) can be safely neglected all the 
way up to at least
$x=10^{-2}$ for quarks. For gluons, the effect of the finite $l_c$ is somewhat larger, 
though it is still a small correction for $x\le 5 \times 10^{-3}$.

\subsection{Comparison of the color fluctuation and quasi-eikonal approximations}
\label{subsec:cf_qe}

\begin{figure}[t]
\begin{center}
\epsfig{file=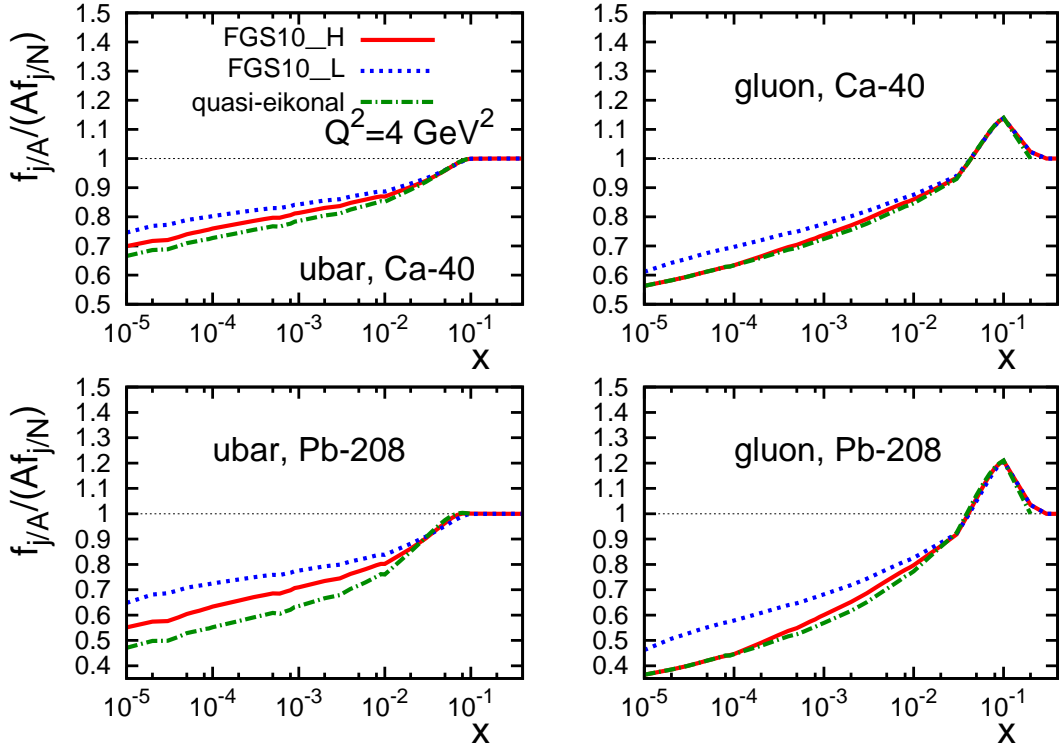,scale=1.4}
\caption{Comparison of the color fluctuation and quasi-eikonal approximations.
The solid and dotted curves correspond to the color fluctuation approximation (same as in 
Figs.~\ref{fig:LT2009_ca40} and \ref{fig:LT2009_pb208});
the dot-dashed curves correspond to the quasi-eikonal approximation.
}
\label{fig:LT2009_ca40_qe}
\end{center}
\end{figure}

Our predictions for nuclear PDFs, which are based on the color fluctuation approximation for
multiple interactions (models FGS10\_H and FGS10\_L),
see Eq.~(\ref{eq:m13master}), 
can be compared to the calculation
of nuclear PDFs in the quasi-eikonal approximation [the approximation when one
uses $\sigma_2^j$ instead of $\sigma_{\rm soft}^j$ in Eq.~(\ref{eq:m13master})].
This is presented in Fig.~\ref{fig:LT2009_ca40_qe}, where the results 
of models 
FGS10\_H and FGS10\_L
(the same curves as in Figs.~\ref{fig:LT2009_ca40} and \ref{fig:LT2009_pb208}) 
are compared to the dot-dashed curves corresponding
to the quasi-eikonal approximation. All curves correspond to $Q^2=4$ GeV$^2$.

To understand the results presented in Fig.~\ref{fig:LT2009_ca40_qe}, 
it is useful to recall the relative magnitude of the effective cross sections
$\sigma^{j(\rm H)}_{\rm soft}$, $\sigma^{j(\rm L)}_{\rm soft}$ and
$\sigma^{j}_{2}$ presented in Fig.~\ref{fig:sigma3_2009}.
In the quark channel, $\sigma^{j(\rm L)}_{\rm soft}>\sigma^{j(\rm H)}_{\rm soft}>\sigma^{j}_{2}$,
and, hence, $f_{j/A}^{\rm FGS10\_L}>f_{j/A}^{\rm FGS10\_H}>f_{j/A}^{\rm qe}$.
In the gluon channel, the trend is quite similar with the only exception that for small $x$,
$\sigma^{j(\rm H)}_{\rm soft}$ and $\sigma^{j}_{2}$
are very close ($\sigma^{j(\rm H)}_{\rm soft}=\sigma^{j}_{2}$ 
for $x \leq 10^{-4}$ by construction),
which means that $f_{j/A}^{\rm FGS10\_H} \approx f_{j/A}^{\rm qe}$
($f_{j/A}^{\rm FGS10\_H} = f_{j/A}^{\rm qe}$ for $x \leq 10^{-4}$ by construction).

\subsection{Uncertainties of predictions of the leading twist theory of nuclear shadowing}
\label{subsect:uncertainties}

\begin{figure}[h]
\begin{center}
\epsfig{file=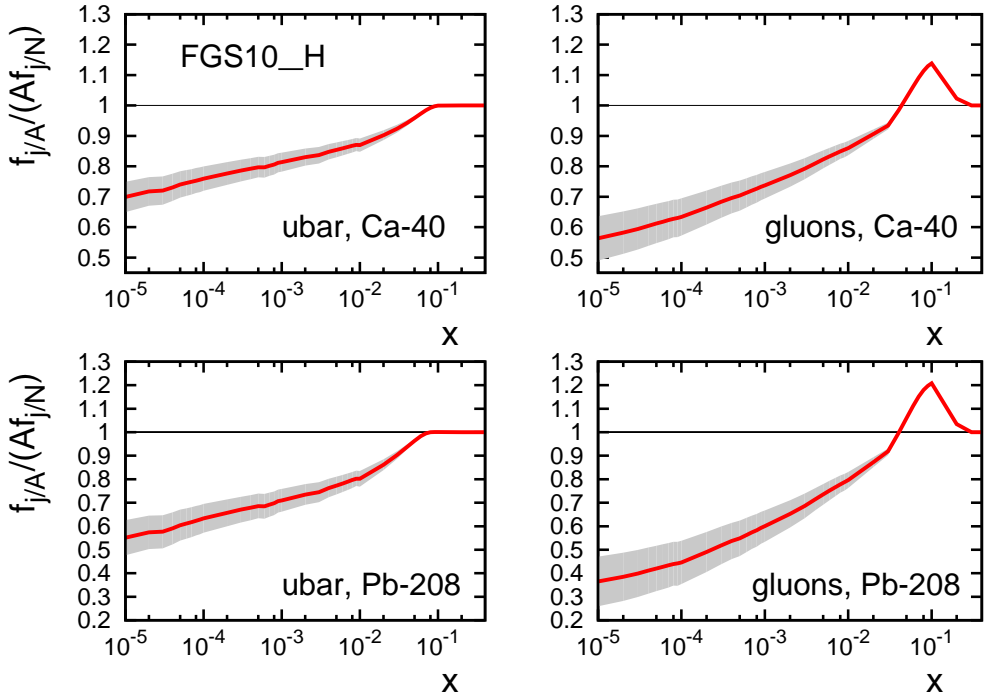,scale=1.3}
\caption
{Nuclear PDFs calculated with our standard choice $B_{{\rm diff}}=6$ GeV$^{-2}$
(solid curves) and with $B_{{\rm diff}}=5$ GeV$^{-2}$ and $B_{{\rm diff}}=7$ 
GeV$^{-2}$ that correspond to the upper and lower boundaries of the shaded areas,
respectively.
All curves correspond to model FGS10\_H and $Q_0^2=4$ GeV$^2$.}
\label{fig:LT2007_ca40_uncert2}
\end{center}
\end{figure}

Our predictions for nuclear PDFs obtained in the framework of the 
leading twist theory of nuclear shadowing contain certain uncertainties. 
They include:
\begin{itemize}
\item[(i)]
The experimental uncertainty in the slope of the $t$ dependence of the 
diffractive structure function $F_2^{D(4)}$ reported by the H1
collaboration, $B_{{\rm diff}}=6 \pm 1.6$ GeV$^{-2}$~\cite{Aktas:2006hx}
(Note that the ZEUS LPS value of $B_{{\rm diff}}$ is somewhat larger,
$B_{{\rm diff}}=7.0 \pm 0.3$ GeV$^{-2}$~\cite{Chekanov:2008fh}.);
\item[(ii)] The theoretical uncertainty related to the choice of the input scale $Q_0^2$;
\item[(iii)] The uncertainty related to the color-fluctuation approximation 
for the interaction with  $N\ge 3$ nucleons. This uncertainty manifests itself in the 
necessity to use two different models for the rescattering cross section $\sigma_{\rm soft}^j$, 
which we called $\sigma_{\rm soft}^{j(\rm H)}$ and $\sigma_{\rm soft}^{j(\rm L)}$.
This leads to the spread in the predictions for nuclear shadowing for small values 
of $x$ (scenarios FGS10\_H and FGS10\_L), 
see the discussion and results in Sec.~\ref{subsubsec:color_fluct}, 
\ref{subsubsect:predictions}, and \ref{subsec:cf_qe}.
\item[(iv)] The uncertainty related to the choice of the nucleon PDFs, primarily the gluon PDF, 
at $x\sim 10^{-4}$ and $Q^2= 4$ GeV$^2$.
\end{itemize}
The largest uncertainty among the 
first three 
that we just
mentioned is the statistical
error in the value of $B_{{\rm diff}}$, $B_{{\rm diff}}=6 \pm 1.6$ GeV$^{-2}$, extracted from the 
H1 data taken with the forward proton spectrometer~\cite{Aktas:2006hx}.
(Note that this value of $B_{{\rm diff}}$ is somewhat lower than the ZEUS LPS result,
$B_{{\rm diff}}=7.0 \pm 0.3$ GeV$^{-2}$~\cite{Chekanov:2008fh}. The two values are
still consistent with each other within errors.)
To assess the uncertainty of our predictions related to the
experimental uncertainty in $B_{{\rm diff}}$, we vary the used value of 
$B_{{\rm diff}}$ and
repeat our calculations of
nuclear PDFs using $B_{{\rm diff}}=6-1=5$ GeV$^{-2}$ and
$B_{{\rm diff}}=6+1=7$ GeV$^{-2}$.
In Fig.~\ref{fig:LT2007_ca40_uncert2}, 
we present the resulting nuclear PDFs in FGS10\_H model: the central solid curves correspond
to our standard choice $B_{{\rm diff}}=6$ GeV$^{-2}$ (same as in
Figs.~\ref{fig:LT2009_ca40} and \ref{fig:LT2009_pb208});
the shaded areas represent the theoretical uncertainty related to
the experimental uncertainty in  $B_{{\rm diff}}$ and
fill in the area between the predictions with $B_{{\rm diff}}=5$ GeV$^{-2}$
(upper boundary) and $B_{{\rm diff}}=7$ GeV$^{-2}$ (lower boundary).
The effect of the variation of $B_{{\rm diff}}$ in the calculation with
model FGS10\_L is similar to the one presented in Fig.~\ref{fig:LT2007_ca40_uncert2}. 

Next we turn to the uncertainty related to our choice of the input evolution
scale $Q_0^2$. This uncertainty can be assessed as follows.
Instead of taking the initial scale 
$Q_0^2=4$ GeV$^2$,
the leading twist nuclear PDFs can be evaluated at a different low 
input scale, for
instance, at $Q_0^2=2.5$ GeV$^2$. Then one can perform the QCD evolution 
from $Q_0^2=2.5$ GeV$^2$
to $Q^2=4$ GeV$^2$ and compare the result with the direct calculation at 
$Q_0^2=4$ GeV$^2$.
Potential differences 
between the two results characterize the theoretical uncertainty related to the
choice of $Q_0^2$.
Since our results for the double scattering term ($N=2$) do not depend on
the choice of $Q_0^2$, this  analysis effectively checks how strongly the higher order terms 
($N\ge 3$) are modified
by the DGLAP evolution.
This provides another way to 
access
the role of the color fluctuations 
due to
the interplay between the small-size and large-size configurations in the virtual
photon wave function.

\begin{figure}[h]
\begin{center}
\epsfig{file=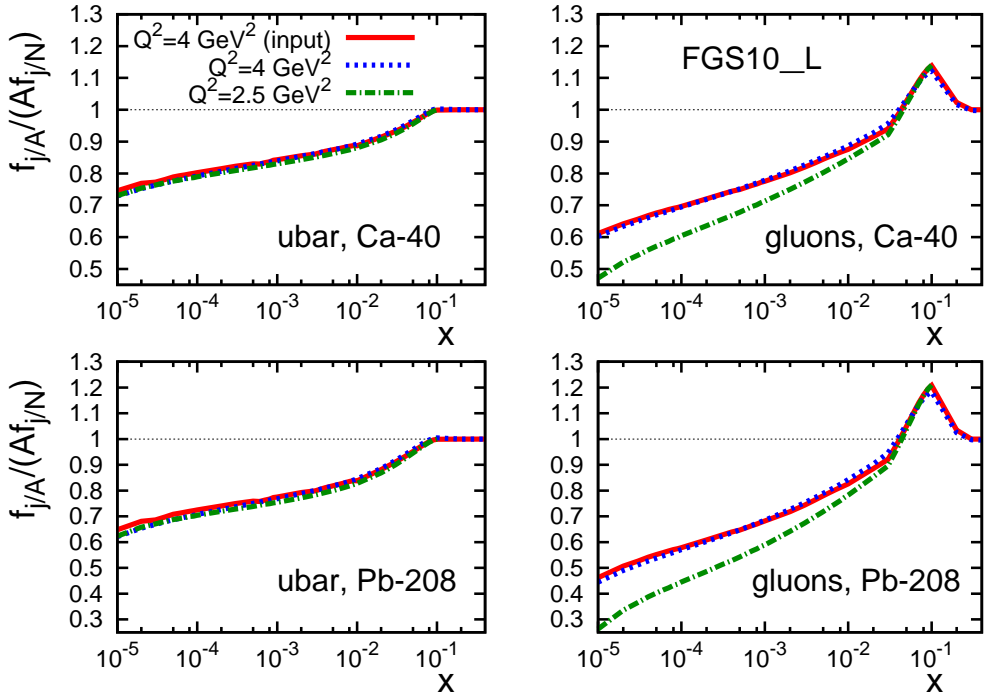,scale=1.3}
\caption{Theoretical uncertainty associated with the choice of the
input evolution scale $Q_0^2$.
The solid and dot-dashed curves are the results
of the direct calculation of nuclear PDFs using Eq.~(\ref{eq:m13master}) 
at $Q_0^2=4$ GeV$^2$ and $Q_0^2=2.5$ GeV$^2$, respectively.
Performing the QCD evolution 
from $Q_0^2=2.5$ GeV$^2$ to $Q_0^2=4$ GeV$^2$, one obtains the dotted curves,
which are indistinguishable from the solid curves. All curves correspond to model 
FGS10\_L.
}
\label{fig:LT2007_ca40_uncert1}
\end{center}
\end{figure}
The results of this procedure are presented in Fig.~\ref{fig:LT2007_ca40_uncert1}.
In this figure, the solid and dot-dashed curves present the results
of the direct calculation of nuclear PDFs at $Q_0^2=4$ GeV$^2$
and $Q_0^2=2.5$ GeV$^2$, respectively. 
One can see that the 
$Q_0^2$ dependence
of quark PDFs is very weak 
for the $Q_0^2=2.5 \div 4 $ GeV$^2$ range, while the gluon shadowing changes quite significantly in this $Q^2$ interval. 
Performing the QCD evolution 
from $Q_0^2=2.5$ GeV$^2$ to $Q^2=4$ GeV$^2$, one obtains the dotted curves,
which are indistinguishable from the solid ones. 
Therefore, the 
uncertainty related to our choice of $Q_0^2$ is negligibly small.
All curves in Fig.~\ref{fig:LT2007_ca40_uncert1} correspond to
model FGS10\_L.

The fact that nuclear shadowing in the gluon channel rather rapidly 
decreases after a very short 
$Q^2$ evolution is a consequence of
the presence of antishadowing and large-$x$ non-shadowed region (which
corresponds to point-like configurations 
in the virtual photon).
Indeed, because of the character of QCD evolution, 
nPDFs at small $x$ and $Q^2 > Q_0^2$ 
originate
from the larger values of Bjorken $x$, $x_0 > x$, 
at the input scale $Q_0^2$. Therefore, the effects of antishadowing and
the presence of non-shadowed point-like configurations 
(for $x > 0.2$ in our approach) feed into the QCD evolution and 
decrease nuclear shadowing after a few steps of the QCD evolution.
For an addition discussion of QCD evolution of nPDFs, see 
Sect.~\ref{subsec:qcd_curve}.

We also stress that in the limit of the low nuclear density when the interaction with
only two nucleons of the target is important (the deuteron target is the best
example), the modeling of multiple rescatterings using the color fluctuation approximation
is not needed, and, as a result, one can evaluate nuclear PDFs using Eq.~(\ref{eq:m13master}) at any scale $Q^2$ without the need of  QCD evolution.

\begin{figure}[h]
\begin{center}
\epsfig{file=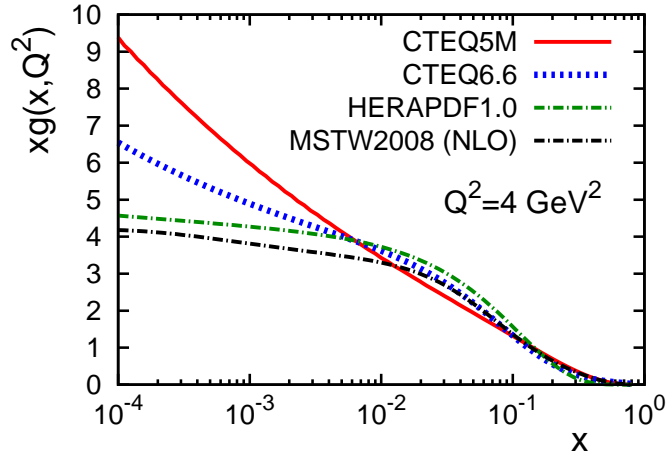,scale=1.3}
\caption{Comparison of several current parameterizations of the gluon PDF in the nucleon at $Q^2=4$
GeV$^2$ as a function of $x$.
The solid curve is CTEQ5M~\cite{Lai:1999wy} (used in this review); 
the dotted curve is CTEQ6.6~\cite{Nadolsky:2008zw}; the green dot-dashed curve is
HERAPDF1.0~\cite{:2009wt}; the black dot-dashed curve is NLO MSTW2008~\cite{Martin:2010db}.
}
\label{fig:gluon_pdf_2011}
\end{center}
\end{figure}

\begin{figure}[h]
\begin{center}
\epsfig{file=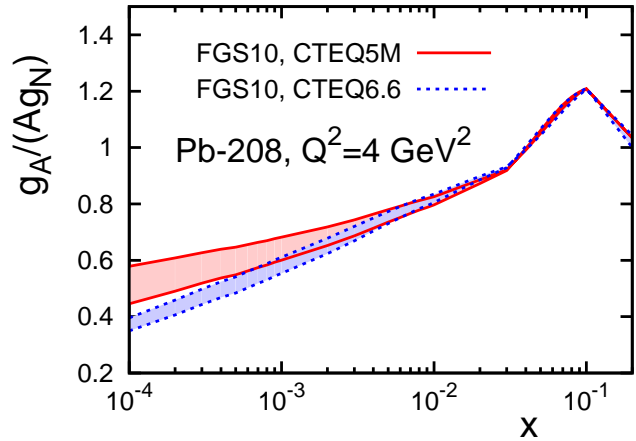,scale=1.4}
\caption{Predictions for $g_A(x,Q_0^2)/[Ag_N(x,Q_0^2)]$ for $^{208}$Pb at $Q_0^2=4$ GeV$^2$
as a function of $x$  calculated 
using CTEQ5M (upper band, our standard choice in this review) and CTEQ6.6 (lower band) 
parameterizations of the proton gluon PDF. 
}
\label{fig:shadow_pb208_cteq66}
\end{center}
\end{figure}

The HERA experiments do not cover a sufficiently large range of $Q^2$ for
$x\sim 10^{-4}$ to extract the nucleon gluon PDF for  $Q^2\sim 4$ GeV$^2$. 
(In the sea quark channel, the kinematic coverage of small $x$ region is much better
and, hence, the quark PDFs are known with much higher precision.)
This can be seen from a comparison of the current fits to the data 
(CTEQ5M~\cite{Lai:1999wy}, CTEQ6.6~\cite{Nadolsky:2008zw}, HERAPDF1.0~\cite{:2009wt}, and
NLO MSTW2008~\cite{Martin:2010db}), see Fig.~\ref{fig:gluon_pdf_2011}, where they are presented as functions of $x$ at $Q^2=4$
GeV$^2$. 
While the choice of the nucleon gluon PDF does not impact very strongly the difference of nuclear and nucleon PDFs~ (see Eq.~(\ref{eq:deltaF2})) since it only affects the value of $\sigma_{\rm soft}^{j(\rm H)}$, 
the effect is more significant for the ratios of the gluon PDFs in nuclei and the nucleon.
This happens because the leading twist theory of nuclear shadowing leads to different shadowing for the same diffractive PDFs and different nucleon PDFs and, hence, the use of our  nuclear shadowing ratios requires specifying 
the parameterization of the nucleon PDFs.
An example of this is presented in Fig.~\ref{fig:shadow_pb208_cteq66}, where
we compare our predictions for $g_A(x,Q_0^2)/[Ag_N(x,Q_0^2)]$ for $^{208}$Pb at $Q_0^2=4$ GeV$^2$ calculated 
using CTEQ5M (upper band, our standard choice in this review) and CTEQ6.6 (lower band) 
parameterizations of the proton gluon PDF. The upper boundary of each band corresponds to
model FGS10\_L; the lower boundary corresponds to FGS10\_H. 
As one can see from Fig.~\ref{fig:shadow_pb208_cteq66}, the uncertainty
associated with the choice of the gluon PDF is not large (it is smaller than
the uncertainty in the slope $B_{\rm diff}$---compare Fig.~\ref{fig:shadow_pb208_cteq66}
to Fig.~\ref{fig:LT2007_ca40_uncert2}) and essentially disappears 
for $x\ge 10^{-3}$.
 For the quark case, 
the discussed uncertainty is practically absent for all $x$.

In summary, theoretical uncertainties of our predictions for
the leading twist nuclear shadowing in nPDFs, which are 
related to the structure of color fluctuations, appear to be smaller than 
the uncertainty of the experimental input due to
the experimental uncertainty in the value of the diffractive slope
$B_{\rm diff}$
and, in the case of gluons, 
the uncertainties in the nucleon gluon PDF
at $x\sim 10^{-4}$.

\subsection{Predictions for leading twist nuclear shadowing at the leading-order accuracy}

\begin{figure}[h]
\begin{center}
\epsfig{file=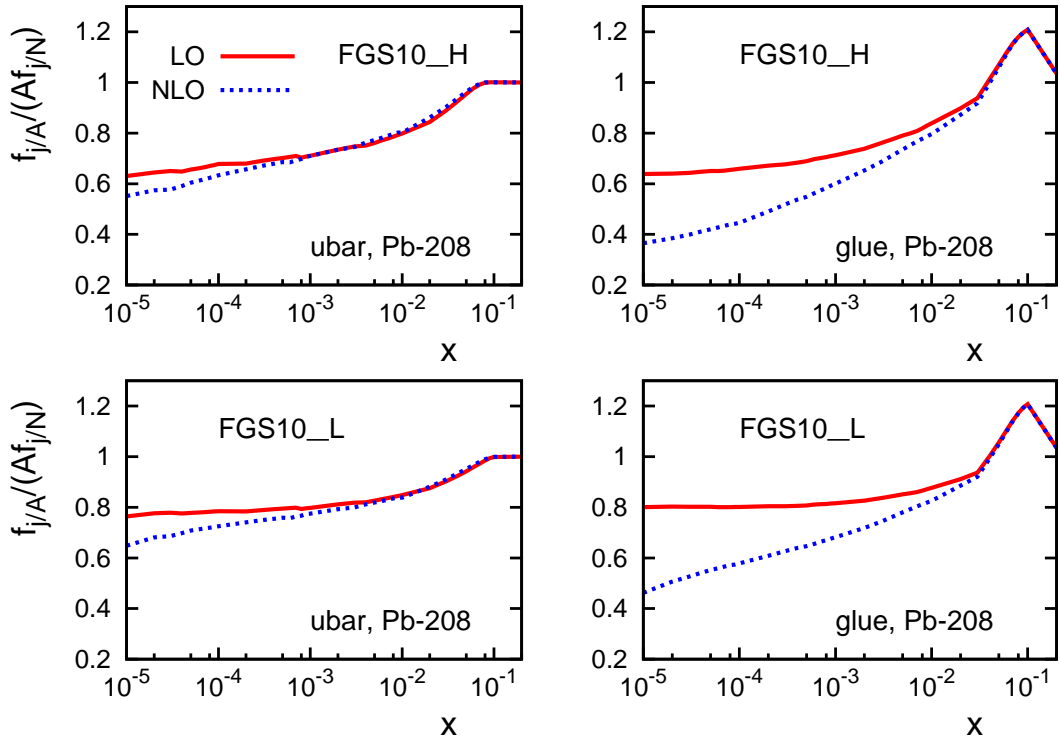,scale=1.4}
\caption{Predictions for $f_{j/A}(x,Q_0^2)/[A f_{j/N}(x,Q_0^2)]$ 
for $^{208}$Pb at $Q_0^2=4$ GeV$^2$ at LO (solid curves) and NLO (dotted curves).
}
\label{fig:FGS10_LO_pb208_5l}
\end{center}
\end{figure}

In this review, our predictions for nuclear PDFs and structure functions are given at 
the next-to-leading order (NLO) accuracy in the strong coupling constant $\alpha_s$. However, 
since the QCD factorization theorems that we used are valid at any order, we can also
make predictions at the leading-order (LO) accuracy. 
Indeed,
using LO diffractive PDFs of the proton~\cite{Newman_private} 
in conjunction with LO free proton PDFs in our master
Eq.~(\ref{eq:m13master}), we can readily make predictions for LO nuclear PDFs.
An example of this
is shown in Fig.~\ref{fig:FGS10_LO_pb208_5l} where the solid curves present the  
$f_{j/A}(x,Q_0^2)/[A f_{j/N}(x,Q_0^2)]$ ratios of LO nuclear and free proton PDFs 
for $^{208}$Pb at $Q_0^2=4$ GeV$^2$.
The left panels correspond to $\bar{u}$ quarks; the right panels correspond to the 
gluon channel; the upper row of panels is for model FGS10\_H, while the lower row is 
for FGS10\_L. For the free proton PDFs, we used the 
CTEQ5L parameterization~\cite{Lai:1999wy}.
For comparison, our predictions for $f_{j/A}(x,Q_0^2)/[A f_{j/N}(x,Q_0^2)]$ 
at the NLO accuracy are given by the dotted curves 
(same curves as in Fig.~\ref{fig:LT2009_ca40_models12}).

As one can see from  Fig.~\ref{fig:FGS10_LO_pb208_5l}, the predicted shadowing is
somewhat smaller at the LO accuracy. 
An inspection reveals that this is mostly due to the fact that the LO CTEQ5L
$\bar{u}$ quark and gluon PDFs  are larger than the corresponding NLO CTEQ5M ones.

\begin{figure}[h]
\begin{center}
\epsfig{file=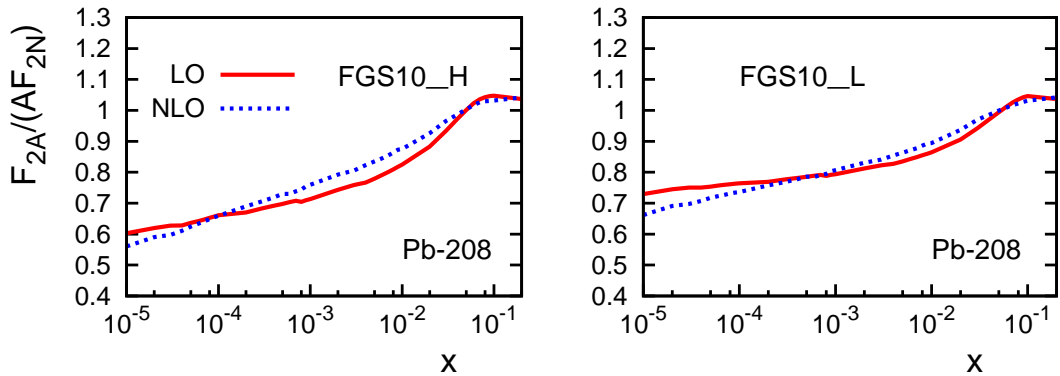,scale=1.4}
\caption{Predictions for $F_{2A}(x,Q_0^2)/[A F_{2N}(x,Q_0^2)]$ 
for $^{208}$Pb at $Q_0^2=4$ GeV$^2$ at LO (solid curves) and NLO (dotted curves) accuracy.
}
\label{fig:FGS10_LO_pb208_5l_f2}
\end{center}
\end{figure}

One should note that while the difference between the LO and NLO predictions for the  
$f_{j/A}(x,Q_0^2)/[A f_{j/N}(x,Q_0^2)]$ ratio for different parton flavors presented
in Fig.~\ref{fig:FGS10_LO_pb208_5l} is sizable, it is much smaller for the ratio of the
structure functions $F_{2A}(x,Q_0^2)/[A F_{2N}(x,Q_0^2)]$, 
see Fig.~\ref{fig:FGS10_LO_pb208_5l_f2}. 
This is due to the facts that i) the values of the 
proton structure function $F_{2N}(x,Q_0^2)$ at LO and NLO accuracy are very close
because both fits reproduce the same inclusive data, ii)
the values of the proton diffractive structure function $F_{2N}^{D(3)}(x,Q_0^2,x_{\Pomeron})$
at LO and NLO are numerically close since, like in the inclusive case, the LO and NLO
fits are constrained to reproduce the same diffractive DIS data.

While for many observables the NLO (and higher) accuracy is the state of the art,
there are situations when one cannot avoid using LO nuclear PDFs. 
One notable example is the application of the dipole formalism to processes with
nuclear targets which requires LO nuclear PDFs as input.

\subsection{The double scattering contribution to nuclear shadowing vs.~the full result}
\begin{figure}[h]
\begin{center}
\epsfig{file=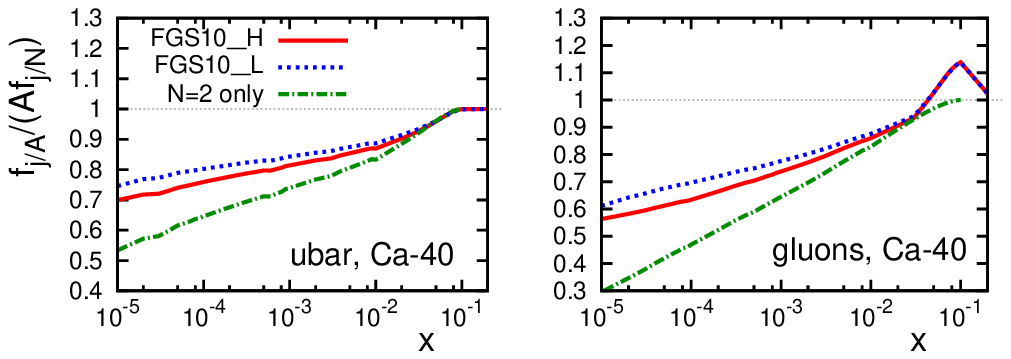,scale=1.4}
\epsfig{file=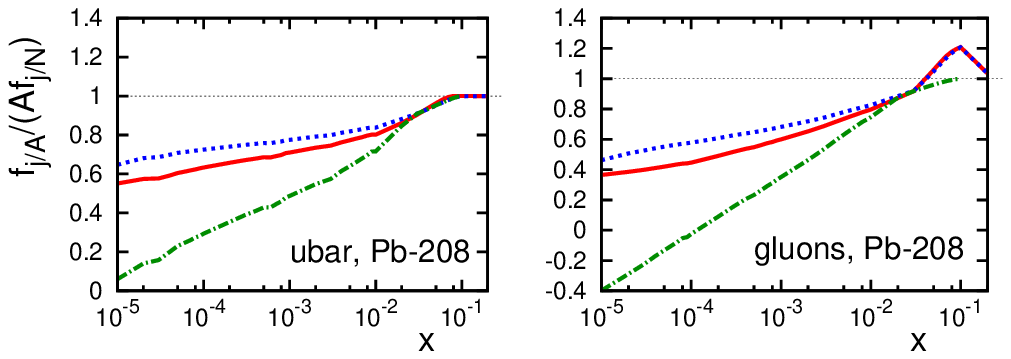,scale=1.4}
\caption{Comparison of the full calculation of $f_{j/A}(x,Q_0^2)/[A f_{j/N}(x,Q_0^2)]$
using the color-fluctuation approximation (solid and dotted curves, same as in Figs.~\ref{fig:LT2009_ca40} and \ref{fig:LT2009_pb208}) to the calculation when only the 
interaction with two nucleons is retained (dot-dashed curves).
All curves correspond to $Q_0^2=4$ GeV$^2$.}
\label{fig:shadow_ca40_2009_2only}
\end{center}
\end{figure}
To better understand the theoretical uncertainty associated with modeling
the multiple interactions using the color-fluctuation approximation,
it is important to compare the full calculation of nuclear shadowing to the calculation,
where only the double rescattering contribution to nuclear shadowing is retained. 
The latter corresponds to setting $\sigma_{\rm soft}^j(x,Q^2)=0$ in Eq.~(\ref{eq:m13master}).
This comparison is presented in Fig.~\ref{fig:shadow_ca40_2009_2only}. 
All curves correspond
to $Q^2=4$ GeV$^2$.
The solid (FGS10\_H) and dotted (FGS10\_L)
curves are the results of the full calculation using the color-fluctuation approximation for the interaction with $N \geq 3$ nucleons, the same as in Figs.~\ref{fig:LT2009_ca40}
and \ref{fig:LT2009_pb208}. The dot-dashed curves are obtained by keeping only the interaction
with two nucleons in the calculation of the shadowing correction. Naturally, the results
presented by the dashed curves are model-independent and do not rely on any approximations
to model the interaction with $N \geq 3$ nucleons.

As one can see from Fig.~\ref{fig:shadow_ca40_2009_2only}, since 
$\sigma_{\rm soft}^j(x,Q^2)$ decreases with increasing $x$ very slowly and remains large,
the effect of the interaction with $N \geq 3$ nucleons remains important up to rather large values of $x$. One can approximate the full result by the interaction with only two
nucleons of the nuclear target only for $x > 0.01$.

\subsection{Comparison with the nPDFs obtained from the DGLAP fits to the data} 
\label{subsec:comparison_to_global}

\begin{figure}[h]
\begin{center}
\epsfig{file=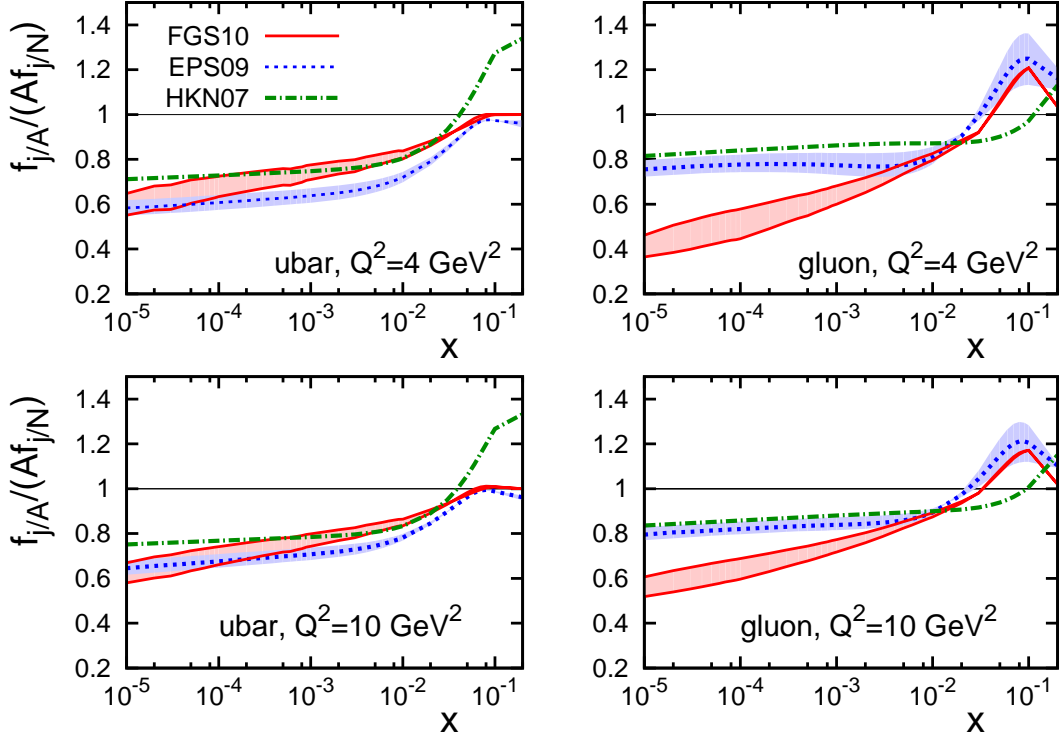,scale=1.4}
\caption{Comparison of predictions of the leading twist theory of nuclear shadowing 
[the area bound by the two solid curves corresponding to
models FGS10\_H (lower boundary) and FGS10\_L (upper boundary)], 
the EPS09 fit (dotted curves
and the corresponding shaded error bands)~\cite{Eskola:2009uj}, 
and the HKN07 fit (dot-dashed curves)~\cite{Hirai:2007sx}.
The 
NLO
$f_{j/A}(x,Q^2)/[A f_{j/N}(x,Q^2)]$ ratios for
the ${\bar u}$-quark and gluon distributions
in $^{208}$Pb are plotted as functions of $x$ at $Q^2=4$ GeV$^2$ (upper panels) and $Q^2=10$ GeV$^2$ (lower panels).}
\label{fig:LT2009_comparison_2010}
\end{center}
\end{figure}

Our predictions for the leading twist next-to-leading (NLO) order nPDFs can be 
compared to those obtained from the DGLAP fits to the available data~\cite{Eskola:1998iy,Eskola:1998df,Eskola:2002us,Paukkunen:2010qi,Hirai:2001np,Hirai:2004wq,Hirai:2007sx,deFlorian:2003qf,Li:2001xa,Eskola:2003cc,Eskola:2007my,Eskola:2008ca,Eskola:2009uj,Schienbein:2009kk}, see also the discussion in Sec.~\ref{sec:hab_intro}.
An example of such a comparison is presented in Fig.~\ref{fig:LT2009_comparison_2010}.
In this figure, we compare our predictions for the ${\bar u}$-quark and gluon distributions
in $^{208}$Pb in the leading twist theory of nuclear shadowing 
[the shaded area bound by the two solid curves corresponding to
models FGS10\_H (lower boundary) and FGS10\_L (upper boundary)], 
the EPS09 fit  
(dotted curves and the 
corresponding shaded error bands)~\cite{Eskola:2009uj}, 
and the HKN07 fit (dot-dashed curves)~\cite{Hirai:2007sx};
all curves correspond to the NLO accuracy.
The ratios of the nuclear to nucleon PDFs are plotted as a function of $x$ at two fixed
values of $Q^2$: $Q^2=4$ GeV$^2$ (upper panels) and $Q^2=10$ GeV$^2$ (lower panels).

As one can see from Fig.~\ref{fig:LT2009_comparison_2010}, the three compared approaches
give rather close values for nuclear shadowing in the sea-quark channel for a wide range of $x$, $10^{-5} \le x \le 0.02-0.03$. For larger $x$, the HKN07 fit deviates from
the other two due to the assumed antishadowing for the sea quarks.

In the gluon channel, our approach suggests much larger shadowing at $Q^2=4$ GeV$^2$
than that suggested by the extrapolation of 
the EPS09 and HKN07 results. Here, however, one has to make a 
distinction. While the shadowing in the gluon channel is insignificant in the HKN07
fit for all $Q^2$ scales, at the
input scale $Q_0^2=1.69$ GeV$^2$, the EPS09 fit suggests very large gluon shadowing 
with the very large theoretical uncertainty~\cite{Eskola:2009uj}.
This is a consequence of the fact that the available data cannot constrain the 
nuclear gluon PDF at small $x$.
(Note also that the large  gluon shadowing in the EPS09 fit is mostly 
driven by the RHIC data
which are not in the kinematics where the leading twist pQCD is applicable, see
the discussion in Sec.~\ref{sec:bdr}.)
Indeed, since the relevant nuclear data for $Q^2 \geq 4$ GeV$^2$ exist only for 
$x \geq10^{-2}$, one is forced to assume the dominance of the LT approximation down to 
$Q^2 \approx 1$ GeV$^2$ and use ad hoc assumptions about nuclear PDFs for smaller $x$ 
where they
are not constrained by the data. When these data are not included in the fit, the 
resulting error band is huge.

\begin{figure}[t]
\begin{center}
\epsfig{file=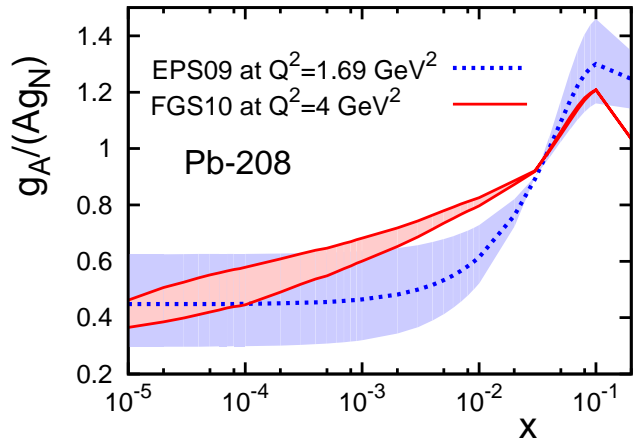,scale=1.4}
\caption{The ratio of the gluon distributions
in $^{208}$Pb and the nucleon, $g_{A}(x,Q^2)/[A g_{N}(x,Q^2)]$, as a function of $x$
for the EPS09 fit at $Q^2=1.69$ GeV$^2$  (the dotted curve with the shaded error band) 
and in the leading twist theory of nuclear shadowing at $Q^2=4$ GeV$^2$
(the shaded area spanned by the two solid curves, the same as in Fig.~\ref{fig:LT2009_comparison_2010}).}
\label{fig:LT2009_comparison_2010_2}
\end{center}
\end{figure}

To illustrate this point, in Fig.~\ref{fig:LT2009_comparison_2010_2} we present 
the ratio of the gluon distributions
in $^{208}$Pb and in the nucleon, $g_{A}(x,Q^2)/[A g_{N}(x,Q^2)]$, as a function of $x$
for the EPS09 fit at $Q^2=1.69$ GeV$^2$  (the dotted curve with the shaded error band) 
and for our leading twist theory of nuclear shadowing at $Q^2=4$ GeV$^2$
(the shaded area spanned by the two solid curves, the same as in Fig.~\ref{fig:LT2009_comparison_2010}). As one can see from Fig.~\ref{fig:LT2009_comparison_2010_2}, the predicted amounts of nuclear shadowing
in the gluon channel for $x < 10^{-3}$ are similar in the two approaches.
However, after short evolution in $Q^2$ from $Q_0^2=1.69$ GeV$^2$ to
$Q^2=4$ GeV$^2$, the shadowing in the gluon channel in the EPS09 fit
significantly reduces and 
becomes noticeably smaller than in our LT approach (compare the solid and dotted curves
in the right column of panels in Fig.~\ref{fig:LT2009_comparison_2010}). 

We point out, again, that nuclear shadowing in the gluon channel 
is essentially unconstrained by the fixed-target data. 
The future  Electron-Ion Collider, with its deep reach in the nuclear shadowing
region and a large lever arm in $Q^2$ should significantly improve our knowledge
of the gluon parton distribution in nuclei.

Recently nuclear PDFs have also been extracted using neutrino 
DIS data and combining the neutrino and lepton DIS data~\cite{Schienbein:2009kk,Schienbein:2007fs,Kovarik:2010uv,Kovarik:2010vw,Hirai:2009mq,Paukkunen:2010hb,Paukkunen:2010qj}. 
At the moment,
the results of such extractions are controversial: while the analyses of
Refs.~\cite{Schienbein:2009kk,Schienbein:2007fs,Kovarik:2010uv,Kovarik:2010vw} seem to indicate that the nuclear
corrections are different between the charged and neutral lepton DIS, 
the analyses of Refs.~\cite{Hirai:2009mq,Paukkunen:2010hb,Paukkunen:2010qj} find no such difference.

\subsection{Comparison to  the soft QCD model of \cite{Capella:1997yv,Armesto:2003fi,Armesto:2010kr}}
\label{subsec:comp_to_Capella}

An approach to nuclear shadowing that is based on the Gribov-Glauber theory of nuclear
shadowing and that has certain similarity to our leading twist approach has been
proposed
and developed in Refs.~\cite{Capella:1997yv,Armesto:2003fi,Armesto:2010kr}. 
It also starts with the Gribov relation between diffraction and shadowing for 
$F_{2A}(x,Q^2)$ for the interaction with
two nucleons and employs the 
 phenomenological Regge-motivated model for the diffractive structure
function $F_2^{D(3)}$ which contains both the leading twist and higher twist
contributions.
Since the approach effectively
includes both leading and higher twist contributions to nuclear shadowing (via the
use of the all-twist parameterization of diffraction), it provides a good
description of the fixed-target data on nuclear shadowing which is predominantly in
the kinematics where only interactions with two nucleons contribute, see the
discussion in Sec.~\ref{subsec:comparison}.

To sum up the multiple interactions with $N \geq 3$ nucleons of the nuclear target, the
fan diagram approximation is used (the Schwimmer model~\cite{Schwimmer:1975bv}). 
Such a model assumes
the dominance of large-mass diffraction, $M^2\gg Q^2$, while we find that $M^2\sim 
Q^2$ dominate in a wide range of $x$.  Also, the use of this model for large $Q^2$
results in the expressions which do not satisfy DGLAP equations even for large $Q^2$
and do not allow one to determine nuclear PDFs for individual parton flavors.

In the recent paper~\cite{Armesto:2010kr}, the authors adopted to 
some extent our QCD factorization
approach and used diffractive PDFs to calculate nuclear PDFs. However, to evaluate
nuclear shadowing as a function of $Q^2$, the authors of~\cite{Capella:1997yv,Armesto:2003fi,Armesto:2010kr} apply an equation
similar in the spirit to our master Eq.~(\ref{eq:m13master}) for all 
$Q^2$. As we explain in Sec.~\ref{sec:hab_pdfs},
the application of Eq.~(\ref{eq:m13master}) at large $Q^2$ 
violates the QCD evolution because one then
ignores the increase of the color fluctuations induced by the QCD evolution. 
[We use Eq.~(\ref{eq:m13master}) only at an input scale $Q_0^2 = 4$ GeV$^2$; 
the subsequent $Q^2$ dependence of
nuclear PDFs is given by the usual DGLAP equations.) In addition, neglecting proper
QCD evolution, one neglects the contribution of larger $x$ effects---antishadowing and
EMC effects---to the small-$x$ region.

\subsection{The leading twist theory of nuclear shadowing vs.~dipole model eikonal approximation}
\label{subsec:eikonal}

\subsubsection{The dipole model eikonal approximation}

Besides the leading twist theory of nuclear shadowing,
there is a broad class of models of nuclear shadowing, which are based on 
the so-called  eikonal  
approximation~\cite{Nikolaev:1990yw,Frankfurt:1995jw,Kopeliovich:1995yr,Kopeliovich:2000ra,Gotsman:1999vt,Gotsman:2000fy}.
The eikonal approximation in nuclear DIS  is based on the assumption that 
the virtual photon-nucleus cross section can be written as the convolution of
the probability of the transition of 
the virtual photon into a quark-antiquark pair ($q \bar{q}$ dipole) with 
the (exponential) factor describing the $q \bar{q}$ dipole-nucleus 
scattering.
The exponential factor is a result of the eikonalization of the multiple $q \bar{q}$-nucleon 
scattering series, which is done in the spirit of the Glauber model.

\begin{figure}[t]
\begin{center}
\epsfig{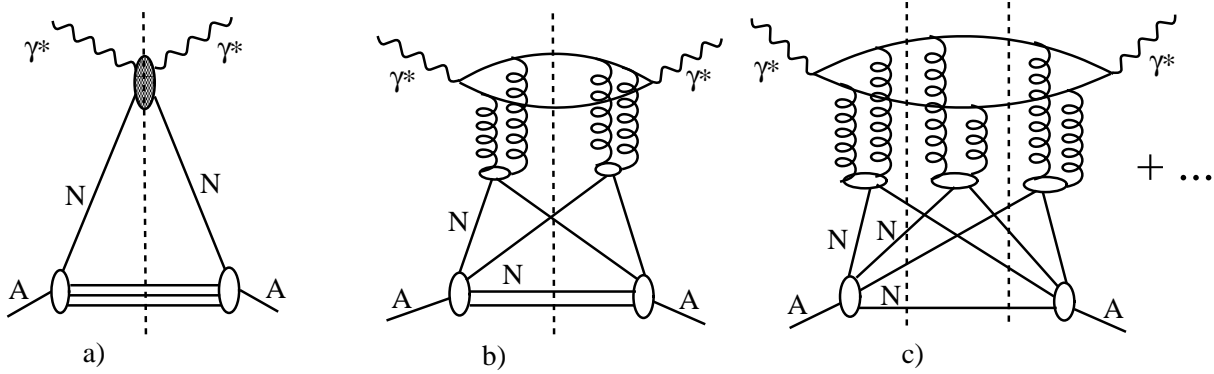}
\caption{
Graphs corresponding to 
the virtual photon-nucleus cross section in the eikonal approximation.
Graph $a$ gives the impulse approximation;
graphs $b$ and $c$ give the shadowing correction arising from the interaction with
two and three nucleons of the target, respectively.}
\label{fig:eikonal}
\end{center}
\end{figure}

The graphical representation of the virtual photon-nucleus cross section 
in the eikonal approximation is given in Fig.~\ref{fig:eikonal}
(the vertical dashed lines denote the unitary cuts).
The graphs in Fig.~\ref{fig:eikonal} should be compared to the corresponding graphs 
of the leading twist
theory of nuclear shadowing 
in Fig.~\ref{fig:Master1}. In Fig.~\ref{fig:eikonal}, graph $a$ is the
impulse approximation, which is the same as in the leading twist
approach.
 Graphs $b$ and $c$ give the shadowing correction 
arising from the interaction with
two and three nucleons of the target, respectively.
The two-gluon exchange is the symbolic notation for the $q {\bar q}$ dipole-nucleon interaction.
Graphs corresponding to the interaction with four and more nucleons 
are not shown, but are assumed.
Note that the dipole model approximation violates the energy-momentum conservation
in the case of the interaction with more than two nucleons, see the discussion in 
Sec.~\ref{subsec:cem}.

Within the eikonal approximation, the expression for the nuclear inclusive 
structure function $F_{2A}(x,Q^2)$ reads, see, e.g., \cite{Frankfurt:2002kd},
\begin{samepage}
\begin{eqnarray}
&&F_{2A}(x,Q^2) =A F_{2N}(x,Q^2)- \frac{Q^2}{4 \pi^2 \alpha_{\rm em}} \Re e \Bigg[\int_0^1 d\alpha ~d^2 d_{\perp} \sum_{i}~|\Psi(\alpha,Q^2,d_{\perp}^2,m_{i})|^2 \nonumber\\
& \times &
\frac{A(A-1)}{2} (1-i\eta)^2
\int d^2\vec{b} \int_{-\infty}^{\infty} dz_1 \int_{z_{1}}^{\infty} dz_2 
\Big[\sigma_{q\bar q N}(x,Q^2,d_{\perp}^2,m_i)\Big]^2 \rho_A(\vec{b},z_1) \rho_A(\vec{b},z_2) \nonumber\\
&\times &
e^{i(z_1-z_2)2x m_N} 
e^{-\frac{A}{2}(1-i\eta) \sigma_{q\bar q N}(x,Q^2,d_{\perp}^2,m_i) \int^{z_{2}}_{z_{1}} dz \rho_{A}(z)} \Bigg] \ .
\label{eq:eik1}
\end{eqnarray}
\end{samepage}
In Eq.~(\ref{eq:eik1}), $\alpha_{\rm em}$ is the fine-structure constant;
$\alpha$ is the fraction of the photon
longitudinal momentum carried by $q$ or $\bar{q}$; $d_{\perp}$ is the transverse diameter of 
the $q \bar{q}$-system; $m_{i}$ is the mass of the constituent quark
of flavor $i$; $\eta$ is the ratio of the real to imaginary parts of $q \bar{q}$-nucleon 
scattering amplitude; $|\Psi|^2$ is the probability of the virtual photon-$q\bar{q}$
transition (the square of the  
effective
light-cone wave functions of the virtual photon);
$\sigma_{q\bar q N}$ is the $q\bar{q}$-nucleon cross section, which
is schematically denoted by the two-gluon exchange in Fig.~\ref{fig:eikonal}.

The square of the 
effective
unpolarized light-cone wave function of the
virtual photon, $|\Psi|^2$, can be written as the sum of the squares of the wave function for the 
transversely-polarized photon, $|\Psi_T|^2$, and 
the effective wave function for the longitudinally-polarized photon,
$|\Psi_L|^2$, i.e.,
$|\Psi|^2=|\Psi_T|^2+|\Psi_L|^2$, where
\begin{eqnarray}
|\Psi_T(\alpha,Q^2,d_{\perp}^2,m_{i})|^2&=&\frac{6~\alpha_{\rm em}}{4 \pi^2}  e_{i}^2
\left[\left(\alpha^2+(1-\alpha)^2\right)\epsilon_{i}^2 ~K_{1}^2(\epsilon_{i}\, d_{\perp})+
m_{i}^2 K_{0}^2(\epsilon_{i}\, d_{\perp}) \right] \,, \nonumber\\
|\Psi_L(\alpha,Q^2,d_{\perp}^2,m_{i})|^2&=&\frac{6~\alpha_{\rm em}}{\pi^2} 
 e_{i}^2 Q^2 \alpha^2(1-\alpha)^2K_{0}^2(\epsilon_{i}\, d_{\perp})
 \,.
\label{wf}
\end{eqnarray}
In Eq.~(\ref{wf}),
$K_{0}$ and $K_{1}$ are the modified Hankel functions; $\epsilon_{i}^2=Q^2\alpha(1-\alpha)+m_{i}^2$.
Following the analysis in Ref.~\cite{Frankfurt:2002kd}, 
we include four 
quark flavors
and take
 $m_{u}=m_{d}=m_{s}=300$ MeV and $m_{c}=1.5$ GeV.
Note that the effective $\Psi_L$ differs from the light-cone wave function of the longitudinal photon. The additional factor of $Q$ results from the exact cancellation of
the  components of the longitudinal photon polarization vector that increase with
energy,
which follows from the conservation of the electromagnetic current for the whole amplitude.
(For the discussion of conceptual differences between the leading twist and dipole model eikonal approximations to nuclear shadowing, see Sec.~\ref{subsubsec:com}.)

Taking the longitudinally-polarized virtual photon, 
one readily obtains the expression for the longitudinal nuclear structure function
$F_L^A(x,Q^2)$:
\begin{samepage}
\begin{eqnarray}
&&F_{L}^A(x,Q^2)=A F_{L}^N(x,Q^2)- \frac{Q^2}{4 \pi^2 \alpha_{\rm em}} \Re e \Bigg[\int_0^1 d\alpha ~d^2 d_{\perp} \sum_{i}~|\Psi_L(\alpha,Q^2,d_{\perp}^2,m_{i})|^2 \nonumber\\
& \times &
\frac{A(A-1)}{2} (1-i\eta)^2
\int d^2\vec{b} \int_{-\infty}^{\infty} dz_1 \int_{z_{1}}^{\infty} dz_2 
\Big[\sigma_{q\bar q N}(x,Q^2,d_{\perp}^2,m_i)\Big]^2 \rho_A(\vec{b},z_1) \rho_A(\vec{b},z_2) \nonumber\\
&\times &
e^{i(z_1-z_2)2x m_N} 
e^{-\frac{A}{2}(1-i\eta) \sigma_{q\bar q N}(x,Q^2,d_{\perp}^2,m_i) \int^{z_{2}}_{z_{1}} dz \rho_{A}(z)} \Bigg] \ .
\label{eq:eik1_L}
\end{eqnarray}
\end{samepage}
\noindent
For comparison and completeness, we also give the expressions for the free nucleon structure functions in the dipole 
approximation~\cite{Nikolaev:1990yw}:
\begin{eqnarray}
&&F_{2N}(x,Q^2)=\frac{Q^2}{4 \pi^2 \alpha_{\rm em}}\int_0^1 d\alpha ~d^2 d_{\perp} \sum_{i}~|\Psi(\alpha,Q^2,d_{\perp}^2,m_{i})|^2 \sigma_{q\bar q N}(x,Q^2,d_{\perp}^2,m_i) \,,
\nonumber\\
&&F_{L}^N(x,Q^2)=\frac{Q^2}{4 \pi^2 \alpha_{\rm em}}\int_0^1 d\alpha ~d^2 d_{\perp} \sum_{i}~|\Psi_L(\alpha,Q^2,d_{\perp}^2,m_{i})|^2 \sigma_{q\bar q N}(x,Q^2,d_{\perp}^2,m_i) \,.
\label{eq:dipole_free}
\end{eqnarray}

In Eqs.~(\ref{eq:eik1}) and (\ref{eq:eik1_L}), we
take $\eta=0.25$, see, e.g., \cite{Bauer:1977iq}.
In addition, the model includes the simplifying assumption
 that 
the invariant mass of all $q \bar{q}$ dipoles 
is the same and approximately 
equals
  $Q$.
Therefore,  the diffractive light-cone fraction $\beta = Q^2/(Q^2 + M_X^2) \approx 0.5$
and, hence, $x_{\Pomeron} \approx 2 x$. This explains the argument of the
first exponential factor in Eqs.~(\ref{eq:eik1}) and (\ref{eq:eik1_L}).
 
In the very low-$x$ limit, the expressions for $F_{2A}(x,Q^2)$ and $F_{L}^A(x,Q^2)$
in the eikonal approximation can be simplified. Neglecting the $e^{i(z_1-z_2)2x m_N}$
factor in
 Eqs.~(\ref{eq:eik1}) and (\ref{eq:eik1_L}), integrating these equations
by parts two times, and using the dipole formalism expressions for $F_{2N}(x,Q^2)$
and $F_{L}^N(x,Q^2)$, one obtains [compare to Eqs.~(\ref{eq:fluct1})
and (\ref{eq:m13master_approx})]:
\begin{samepage}
\begin{eqnarray}
F_{2A}(x,Q^2) &=&\frac{Q^2}{4 \pi^2 \alpha_{\rm em}} \Re e \Bigg[\int_0^1 d\alpha ~d^2 d_{\perp} \sum_{i}~|\Psi(\alpha,Q^2,d_{\perp}^2,m_{i})|^2 \nonumber\\
&\times& 2\int d^2 b \left(1-e^{-\frac{A}{2}(1-i \eta) \sigma_{q\bar q N}(x,Q^2, d_{\perp}^2,m_i)T_A(b)}\right)\Bigg] \,, \nonumber\\
F_{L}^A(x,Q^2) &=&\frac{Q^2}{4 \pi^2 \alpha_{\rm em}} \Re e \Bigg[\int_0^1 d\alpha ~d^2 d_{\perp} \sum_{i}~|\Psi_L(\alpha,Q^2,d_{\perp}^2,m_{i})|^2 \nonumber\\
&\times& 2\int d^2 b \left(1-e^{-\frac{A}{2}(1-i \eta) \sigma_{q\bar q N}(x,Q^2,d_{\perp}^2,m_i) T_A(b)}\right)\Bigg] \,.
\end{eqnarray}
\end{samepage}

\subsubsection{The dipole cross section}

The dipole
cross section $\sigma_{q\bar q N}$
plays a central role in the dipole formalism and in the quasi-eikonal approximation.
We use the dipole model of McDermott, Frankfurt, Guzey 
and Strikman (MFGS)~\cite{McDermott:1999fa}.
In this approach, the $q{\bar q}$-nucleon dipole cross section is constructed in a piecewise form, 
where each piece corresponds to the particular range of the dipole transverse size $d_{\perp}$
with the corresponding physics motivation. For small dipole sizes, the dipole cross section is dictated by perturbative QCD. As one increases the dipole size, the dynamics becomes non-perturbative and it is no longer reasonable to think of $d_{\perp}$ as a size of a simple $q{\bar q}$ pair. Instead, it is better to think of it as corresponding to the typical transverse size of a
complicated non-perturbative system, which in general contains many constituents.
When the dipole size becomes as large as the transverse separation of quarks in a pion,
$d_{\perp}=d_{\pi} \approx  0.65$ fm, one expects that the dipole cross section should be comparable to the
soft pion-nucleon cross section, 
$\sigma^{\pi N}_{\rm tot}$.
The intermediate region between 
the small dipole sizes and $d_{\perp}= 0.65$ fm
is least known. For the lack of the physical motivation, the dipole cross section in this region 
is modeled by the  smooth interpolation between  the small-$d_{\perp}$ perturbative and the 
large-$d_{\perp}$ non-perturbative regions.

The explicit expression for the dipole cross section, $\sigma_{q{\bar q}N}$,
has the following form~\cite{McDermott:1999fa}:
\begin{equation}
\sigma_{q{\bar q}N}(x,Q^2,d_{\perp},m_i)=\left\{ \begin{array}{ll}
\frac{\pi^2}{3} d_{\perp}^2 \alpha_s(\frac{\lambda}{d_{\perp}^2}) x^{\prime} g(x^{\prime},\frac{\lambda}{d_{\perp}^2}) \,, & d_{\perp} \leq \min\{d_{\rm crit},d_{Q_0}\} \\
\left(\sigma_{\pi N}(x)-\sigma_{q{\bar q}}(x^{\prime},Q^2,d_{\rm crit})\right) H(d_{\perp})
\\
\quad \quad  +\sigma_{q{\bar q}}(x^{\prime},Q^2,d_{\rm crit}) \,, &  \min\{d_{\rm crit},d_{Q_0}\} \leq d_{\perp} \leq d_{\pi} \\
\sigma_{\pi N}(x) \frac{1.5 \, d_{\perp}^2}{d_{\perp}^2+d_{\pi}^2/2} \,,&
d_{\perp} \geq d_{\pi} \,.
\end{array} \right.
\label{eq:sigma_dipole}
\end{equation}
In Eq.~(\ref{eq:sigma_dipole}), the perturbative dipole cross section 
in the region
$d_{\perp} \leq \min\{d_{\rm crit},d_{Q_0}\}$
is proportional to the dipole size squared
$d_{\perp}^2$, the gluon density of the target (nucleon or nucleus)
 $g(x^{\prime},\lambda/d_{\perp}^2)$, and the strong coupling constant
$\alpha_s(\lambda/d_{\perp}^2)$.
(The result that $\sigma_{q\bar q N} \approx {\rm const}\cdot d_{\perp}^2$ 
was first obtained by F.~Low in the two-gluon exchange model in 
1975~\cite{Low:1975sv}.)
The gluon density and the strong coupling constant are probed at the effective
scale $Q_{\rm eff}^2=\lambda/d_{\perp}^2$, where $\lambda \approx 10$ in the original MFGS 
analysis~\cite{McDermott:1999fa}. It was later discovered that the inclusive cross sections
are rather insensitive to the precise value of $\lambda$ in the range
$\lambda=4-15$ and that the lower value of $\lambda=4$  appears to be favored by the 
$J/\psi$ photoproduction data~\cite{Frankfurt:2000ez}.
In  this work, we use $\lambda=4$.
Note also that since the perturbative expression for $\sigma_{q{\bar q}N}$
is valid to the leading order (LO) accuracy,
both $g(x^{\prime},\lambda/d_{\perp}^2)$  and $\alpha_s(\lambda/d_{\perp}^2)$ are
taken at the LO accuracy.

The light-cone fraction $x^{\prime}$, at which the target gluon density is probed, is larger 
than Bjorken $x$. This accounts 
for the fact that the virtual photon directly couples only to the quarks, which 
originate through the DGLAP evolution from the gluons with the larger light-cone 
momenta. In the MFGS model, one uses~\cite{McDermott:1999fa}:
\begin{equation}
x^{\prime}=x \left(1+\frac{4 m_i^2}{Q^2} \right)\left(1+0.75 \frac{\lambda}{Q^2\, d_{\perp}^2} \right) \,.
\label{eq:xprime}
\end{equation}
It is the dependence of $x^{\prime}$ on $Q^2$ that introduces the explicit dependence
of the dipole cross section on the virtuality $Q^2$, which is rather insignificant.
Note also that $x^{\prime}$ depends on the constituent quark mass $m_i$.

At fixed small $d_{\perp}$, as one decreases $x$
the perturbative dipole cross section grows
very rapidly as a result of the steeply rising gluon density.
If unchecked, this will eventually lead to a contradiction since the
perturbative contribution to the
dipole cross section is an {\it inelastic} cross section which should not exceed
the typical soft inelastic meson-nucleon cross section at the corresponding energy,
which we take to be 
the inelastic pion-nucleon cross section. 
Therefore, 
at given Bjorken $x$, we apply the perturbative 
expression for the dipole cross section up to some critical dipole size,
$d_{\rm crit}$, which is defined by the following equation:
\begin{equation}
\frac{\pi^2}{3} d_{\rm crit}^2 \alpha_s(\frac{\lambda}{d_{\rm crit}^2}) x^{\prime} g(x^{\prime},\frac{\lambda}{d_{\rm crit}^2})=\frac{1}{2} \sigma^{\pi N}_{\rm tot}(W^2) \,,
\label{eq:dcrit}
\end{equation}
where $W^2=Q^2/x-Q^2+m_N^2$.
In the right-hand side of Eq.~(\ref{eq:dcrit}), we used the fact that the 
maximal
value of the inelastic pion-nucleon cross section is reached in the black disk
limit and is equal to one half of the total cross section.
For the latter, we use the Donnachie-Landshoff parameterization~\cite{Donnachie:1992ny}, see Eq.~(\ref{eq:sigma_pion}).

The effective scale $Q_{\rm eff}^2=\lambda/d_{\perp}^2$ in 
Eqs.~(\ref{eq:sigma_dipole}) and (\ref{eq:dcrit})
cannot be too small 
to guarantee 
 that
the perturbative description is applicable. 
Therefore, $Q_{\rm eff}^2 \geq Q_0^2$, where $Q_0^2$ is the initial evolution scale.
Using the CTEQ5L parameterization of the gluon density 
with $Q_0^2=1$ GeV$^2$~\cite{Lai:1999wy} that 
was used in the original analysis of Ref.~\cite{McDermott:1999fa} and more recent 
parameterization, e.g.,
MSTW 2008 with $Q_0^2=1$ GeV$^2$~\cite{Martin:2009iq} and GJR 2008 with $Q_0^2=0.3$ GeV$^2$~\cite{Gluck:2007ck},
we obtain the following range of small dipole sizes for
which the perturbative
description of the dipole cross section can be applied: 
\begin{equation}
d_{\perp} \leq d_{Q_0} = \sqrt{\frac{\lambda}{Q_0^2}}= \left\{\begin{array}{cc}
0.39 \ {\rm fm} \,, & {\rm CTEQ5L},\, {\rm MSTW 2008} \,, \\
0.72 \ {\rm fm} \,, & {\rm GJR2008} \,.
\end{array}
\right.
\label{eq:d_Q0}
\end{equation}
It turns out that the considered parameterizations lead to the similar  
$\sigma_{q {\bar q}N}$ cross section. Therefore, for definiteness, in our numerical analysis
below, we shall use the GJR 2008 parameterization with $Q_0^2=0.3$ 
GeV$^2$~\cite{Gluck:2007ck}.
This choice of $Q_0^2$ allows one to generate a smooth extrapolation
to the non-perturbative region.
Note that since the initial evolution scale is very low such that
$d_{Q_0} > d_{\pi}$, we shall apply the perturbative
description of the dipole cross section only
for $d_{\perp} \leq d_{\rm crit}$.

In the intermediate region, $d_{\rm crit} \leq d_{\perp} \leq d_{\pi}$, 
we model the dipole cross section in a simple form that smoothly interpolates 
between the small-$d_{\perp}$ and 
large-$d_{\perp}$ pieces of the dipole cross section. 
The interpolating function is arbitrary, see the discussion in~\cite{McDermott:1999fa}.
In this work, we use the following smooth parameterization:
\begin{equation}
H(d_{\perp})=\frac{e}{e-1}\left[1-\exp(-H_1(d_{\perp})) \right] \,,
\label{eq:H}
\end{equation}
where
\begin{equation}
H_1(d_{\perp})=\frac{d_{\perp}-d_{\rm crit}}{d_{\pi}-d_{\rm crit}} \,.
\label{eq:H1}
\end{equation}

Finally, for very large dipole sizes, $d_{\perp} > d_{\pi}$, we impose a residual slow growth
with  increasing $d_{\perp}$, see the last line of Eq.~(\ref{eq:sigma_dipole}).

\begin{figure}[h]
\begin{center}
\epsfig{file=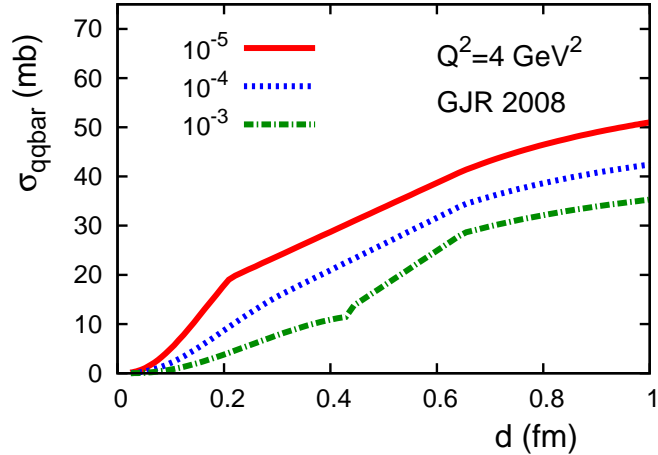,scale=1.2}
\caption{The dipole cross section, $\sigma_{q {\bar q}}(x,Q^2,d_{\perp},m_i)$, in the MFGS dipole model with the GJR 2008 parameterization of the gluon density~\cite{Gluck:2007ck} 
as a function of the dipole size $d_{\perp}$.
All curves correspond to $m_i=300$ MeV and $Q^2=4$ GeV$^2$.
}
\label{fig:sigma_dipole}
\end{center}
\end{figure}

Figure~\ref{fig:sigma_dipole} presents the $q{\bar q}$ dipole cross section in the MFGS dipole 
model with the GJR 2008 parameterization of the gluon density~\cite{Gluck:2007ck} as a function of the dipole size $d_{\perp}$. All curves correspond to the light quark cross sections ($m_i=300$ MeV) and $Q^2=4$ GeV$^2$ (note that the dependence on $Q^2$ is weak).

\subsubsection{Predictions for nuclear structure functions 
$F_{2A}(x,Q^2)$ and $F_{L}^A(x,Q^2)$ in the dipole model eikonal approximation}

\begin{figure}[t]
\begin{center}
\epsfig{file=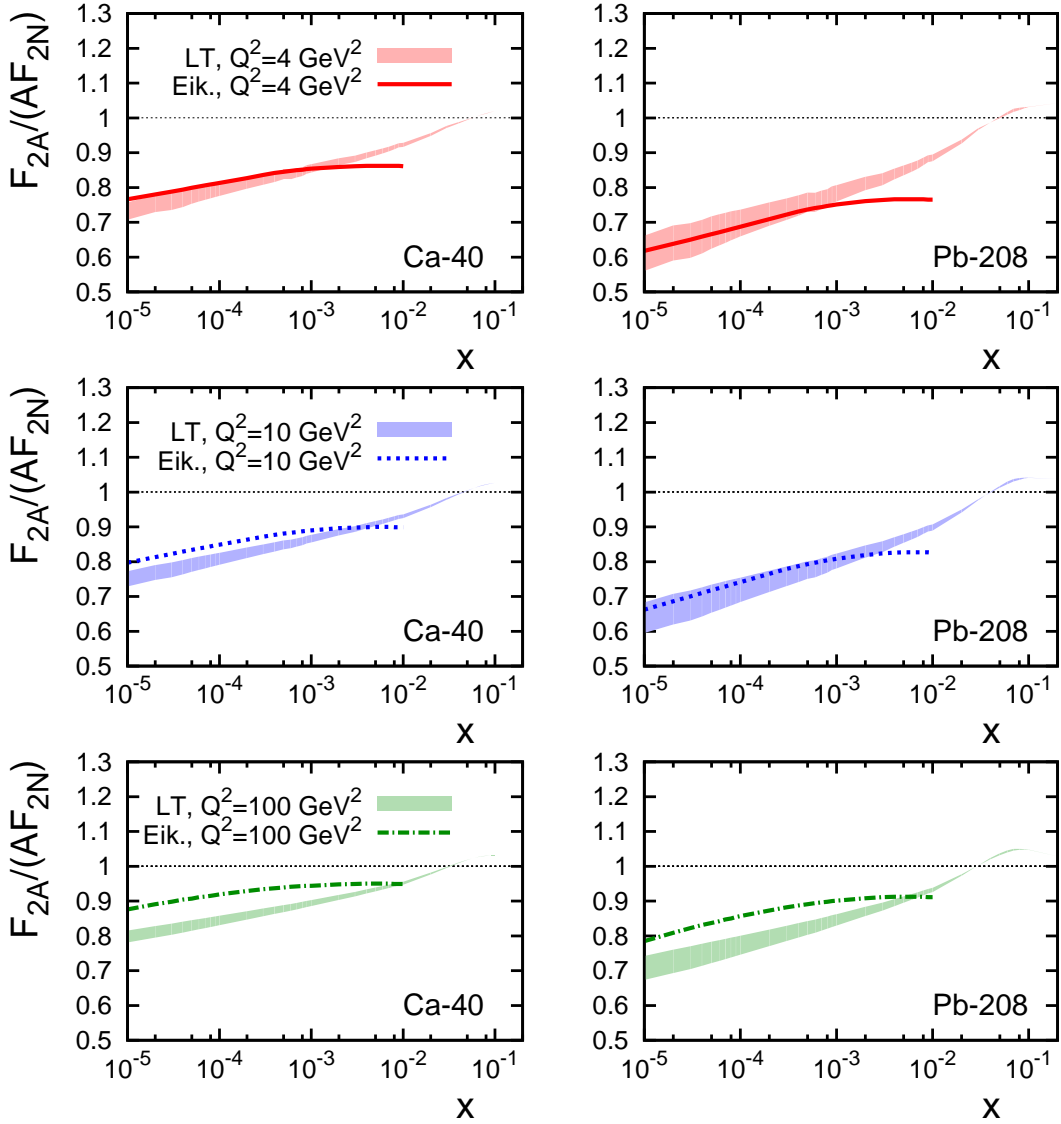,scale=1.4}
\caption{The ratio of the nuclear to nucleon structure functions 
$F_{2A}(x,Q^2)/[A F_{2N}(x,Q^2)]$ as a function of Bjorken $x$. 
The results of the calculation using the dipole model eikonal approximation, see Eqs.~(\ref{eq:eik1})
and (\ref{eq:dipole_free}), are given by the curves. The 
shaded bands
represent the corresponding predictions of the leading twist theory of nuclear
shadowing taken from Figs.~\ref{fig:LT2009_ca40} and \ref{fig:LT2009_pb208}.
}
\label{fig:eikonal_F2}
\end{center}
\end{figure}

\begin{figure}[h]
\begin{center}
\epsfig{file=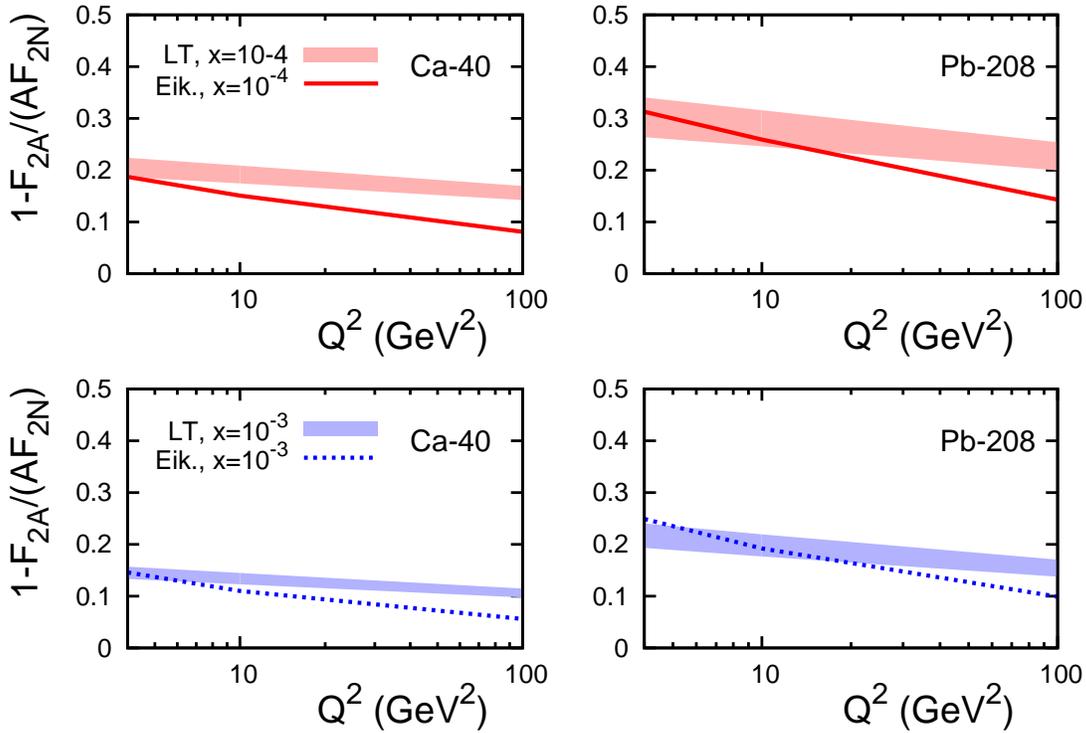,scale=1.4}
\caption{The shadowing correction to the nuclear structure function $F_{2A}(x,Q^2)$, 
$1-F_{2A}(x,Q^2)/[A F_{2N}(x,Q^2)]$, as a function of $Q^2$ for $x=10^{-4}$ and 
$x=10^{-3}$.  
The curves correspond to the dipole model eikonal approximation;
the shaded bands are the results of the leading twist theory of nuclear shadowing.
}
\label{fig:eikonal_F2_Q2dep}
\end{center}
\end{figure}

\begin{figure}[t]
\begin{center}
\epsfig{file=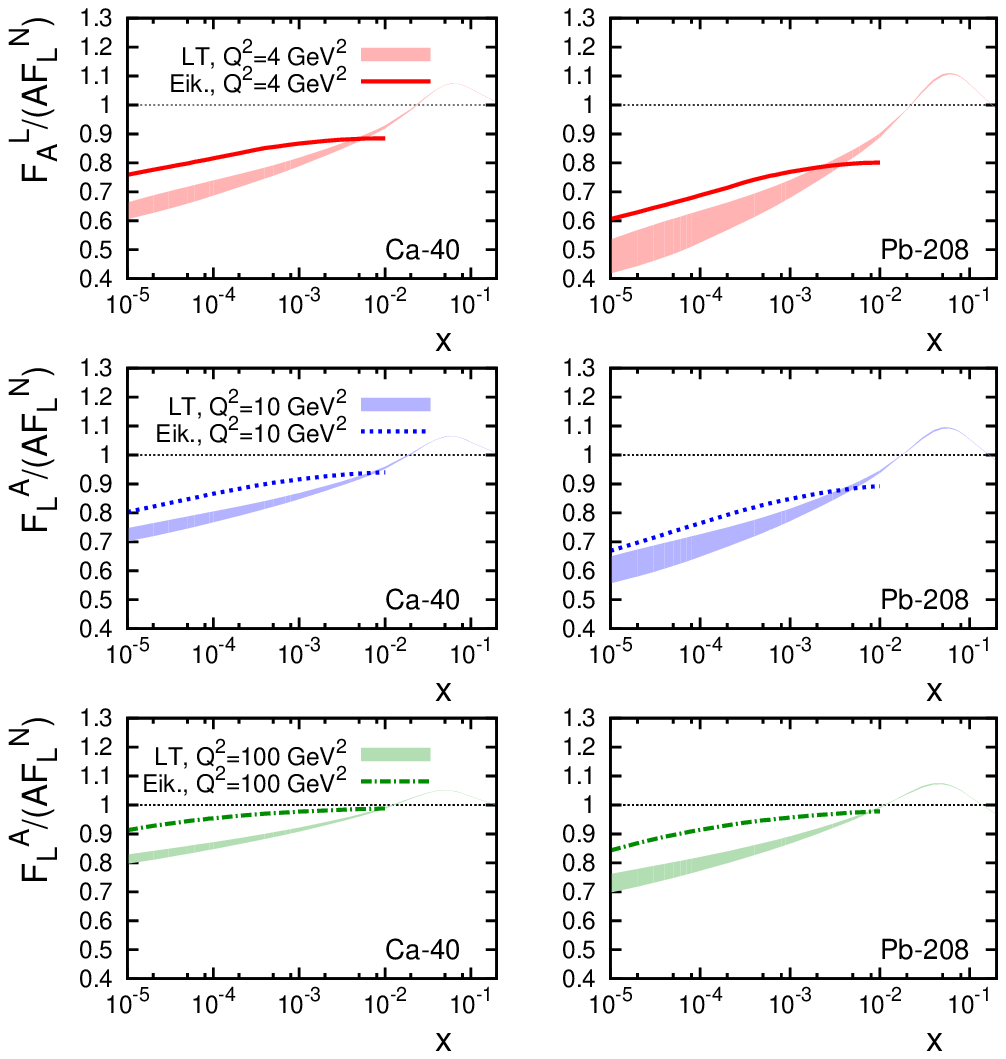,scale=1.4}
\caption{The ratio of the nuclear to the nucleon longitudinal structure functions, 
$F_{L}^A(x,Q^2)/[A F_{L}^N(x,Q^2)]$, as a function of Bjorken $x$.
The curves represent the results of the dipole model eikonal approximation; the shaded bands
correspond to the leading twist theory of nuclear shadowing.
}
\label{fig:eikonal_FL}
\end{center}
\end{figure}

\begin{figure}[h]
\begin{center}
\epsfig{file=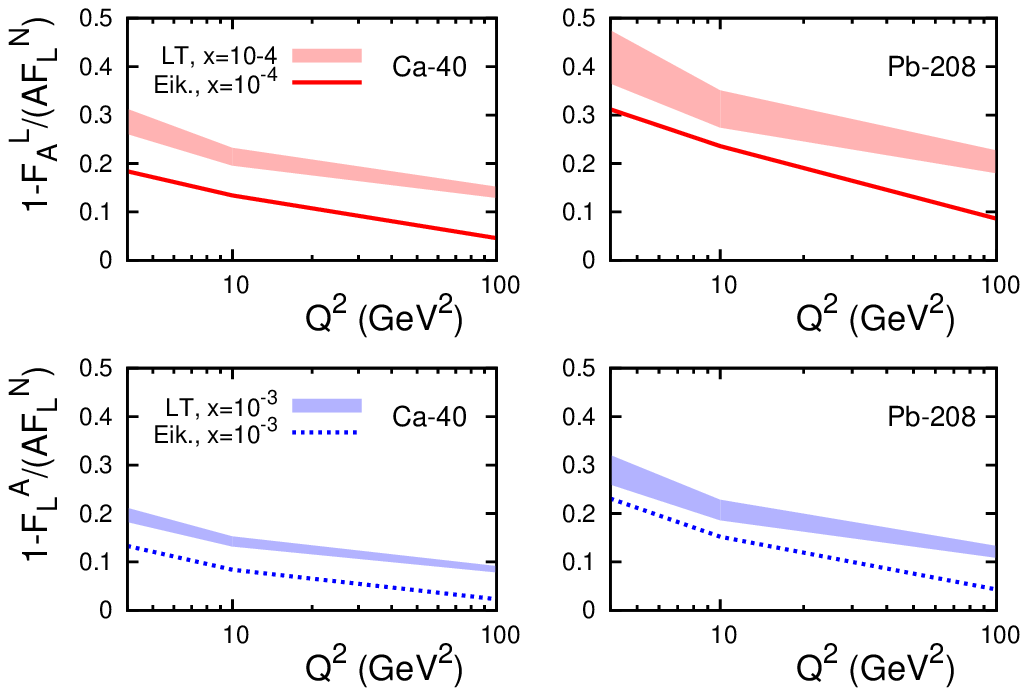,scale=1.4}
\caption{The shadowing correction to the nuclear longitudinal structure function $F_{L}^A(x,Q^2)$, 
$1-F_{L}^A(x,Q^2)/[A F_{L}^N(x,Q^2)]$, as a function of $Q^2$ for $x=10^{-4}$ and 
$x=10^{-3}$.  
The curves correspond to the dipole model eikonal approximation;
the shaded bands are the results of the leading twist theory of nuclear shadowing.
}
\label{fig:eikonal_FL_Q2dep}
\end{center}
\end{figure}

Predictions for nuclear shadowing in the nuclear structure functions 
$F_{2A}(x,Q^2)$ and $F_{L}^A(x,Q^2)$ are presented
in Figs.~\ref{fig:eikonal_F2} and \ref{fig:eikonal_F2_Q2dep}
 and Figs.~\ref{fig:eikonal_FL} and \ref{fig:eikonal_FL_Q2dep}, respectively. 
In Fig.~\ref{fig:eikonal_F2},
we give the ratio of the nuclear to nucleon structure functions 
$F_{2A}(x,Q^2)/[A F_{2N}(x,Q^2)]$ as a function of Bjorken $x$.
The results of the calculation using the dipole model eikonal approximation, see Eqs.~(\ref{eq:eik1})
and (\ref{eq:dipole_free}), are presented by the curves. 
Different curves correspond to three different values of $Q^2$,
$Q^2=4$, 10, and 100 GeV$^2$. 
For comparison, 
we also give the corresponding predictions of the leading twist theory of nuclear
shadowing taken from Figs.~\ref{fig:LT2009_ca40} and \ref{fig:LT2009_pb208}, which
are given by the 
shaded bands (the upper boundaries of the bands correspond to model FGS10\_L;
the lower boundaries correspond to model FGS10\_H).

Note that we apply the dipole model eikonal approximation only for $x \leq 0.01$. For $x > 0.01$, the dipole model
eikonal approximation is not applicable since the $q{\bar q}$ 
fluctuations of the virtual photon 
are no longer coherent at the distance scale $\sim R_A$.
In addition
to the effect of the finite $l_c$, the naive application of Eqs.~(\ref{eq:eik1}) and (\ref{eq:eik1_L}) for $x > 0.01$ is bound to significantly
overestimate the amount of nuclear shadowing since the effects of antishadowing and
QCD evolution are not included in the eikonal approximation.

As one can see from Fig.~\ref{fig:eikonal_F2}, the amount of nuclear shadowing is rather
similar in the dipole model eikonal approximation and the 
FGS10\_L version of the
leading twist theory of nuclear shadowing
at $Q^2=4$ and 10 GeV$^2$ and for a wide range of $x$;
model FGS10\_H
 predicts noticeably larger nuclear shadowing for $x < 5 \times 10^{-3}$. 
As one increases $Q^2$, e.g., $Q^2 \geq 10$ GeV$^2$,
nuclear shadowing decreases much faster in the dipole model 
eikonal approximation than in our 
leading twist approach (both in 
FGS10\_L and FGS10\_H).
 This trend
agrees with our earlier analysis~\cite{Frankfurt:2002kd} and can be explained as
follows.
While the $Q^2$ behavior of 
$F_{2A}(x,Q^2)/[A F_{2N}(x,Q^2)]$ within the leading twist approach is governed by the 
QCD DGLAP evolution equation, and is therefore logarithmic, $F_{2A}(x,Q^2)/[A F_{2N}(x,Q^2)]$
  in the eikonal approximation decreases with increasing $Q^2$ (much) faster, 
which is dictated largely by the $Q^2$ dependence of the virtual photon light-cone wavefunction. 
(We discuss the difference between the two approaches in more detail
in Sec.~\ref{subsubsec:com}.)

Note that in the dipole eikonal approximation, the main contribution to nuclear
shadowing for $F_{2A}(x,Q^2)$ originates from large $q {\bar q}$ dipoles, i.e., from
the contribution for which the virtual photon wave function is 
non-perturbative and the description in terms of the $q {\bar q}$ dipoles is at best
effective. Hence, it is a recast of the aligned jet contribution.

To illustrate the different $Q^2$ behavior of nuclear shadowing in the leading twist 
theory and eikonal approximation, in Fig.~\ref{fig:eikonal_F2_Q2dep} we present
the shadowing correction, $1-F_{2A}(x,Q^2)/[A F_{2N}(x,Q^2)]$, as a function of $Q^2$ for two
fixed values of $x=10^{-4}$ and $x=10^{-3}$. The curves correspond to the eikonal approximation;
the shaded bands are the results of the leading twist theory of nuclear shadowing.
As one can see from Fig.~\ref{fig:eikonal_F2_Q2dep}, with an increase of $Q^2$,
nuclear shadowing decreases noticeably faster in the eikonal approximation than in our
leading twist approach. 

In Fig.~\ref{fig:eikonal_FL}, we present our results for nuclear shadowing for the
ratio of the longitudinal structure functions, $F_{L}^A(x,Q^2)/[A F_{L}^N(x,Q^2)]$.
The labeling of the curves is the same as in Fig.~\ref{fig:eikonal_F2}.
One can see from  Fig.~\ref{fig:eikonal_FL} that the eikonal approximation predicts significantly 
smaller nuclear shadowing than the leading twist theory.
This can be qualitatively explained by the observation that the small-size
configurations of the virtual photon wave function---which are only weakly shadowed---give
a more important contribution to $F_{L}^A(x,Q^2)$ than to $F_{2A}(x,Q^2)$ in the 
dipole formalism. This leads to the nuclear shadowing in $F_{L}^A(x,Q^2)$ that is both small and
decreases rapidly with increasing $Q^2$, i.e., in the eikonal approximation
nuclear shadowing in $F_{L}^A(x,Q^2)$ is essentially a higher-twist effect.
At the same time, in our leading twist approach, both small-size and large-size 
fluctuations of the virtual photon contribute to nuclear shadowing in $F_{L}^A(x,Q^2)$
and make it a sizable effect at all $Q^2$.

To illustrate our discussion of the different $Q^2$ behavior of nuclear shadowing 
in the longitudinal structure function $F_{L}^A(x,Q^2)$
in the leading twist theory and eikonal approximation, 
in Fig.~\ref{fig:eikonal_FL_Q2dep} we present
the shadowing correction, $1-F_{L}^A(x,Q^2)/[A F_{L}^A(x,Q^2)]$, as a function of $Q^2$ for two
fixed values of $x=10^{-4}$ and $x=10^{-3}$. The curves correspond to the eikonal approximation;
the shaded bands are the results of the leading twist theory of nuclear shadowing.
As one can see from Fig.~\ref{fig:eikonal_FL_Q2dep}, with an increase of $Q^2$,
nuclear shadowing decreases noticeably faster in the eikonal approximation than in our
leading twist approach; the absolute value of the shadowing correction is also significantly
smaller in the eikonal approximation than in our leading twist approach.  

\subsubsection{Conceptual differences between the leading twist and 
dipole model eikonal approaches to 
nuclear shadowing}
\label{subsubsec:com}

In this subsection, we discuss conceptual differences between 
the leading twist and dipole model eikonal approaches to nuclear shadowing 
in DIS on nuclei. 
The two key differences that make the two approaches so distinct are related. They are: 
\begin{itemize}
\item[(i)]{Different space-time evolution of the scattering process.
The eikonal model approach strongly violates energy-momentum conservation; 
the color fluctuation approximation has no such a defect.
}
\item[(ii)]{The neglect of the $|q \bar{q} g \rangle$ component (and higher Fock states) of the virtual photon wave function in the dipole model eikonal approximation
and the neglect of the proper QCD evolution (see the discussion in Sec.~\ref{subsec:comp_to_Capella}).
In addition, the Sudakov form factor which arises in QCD calculations because of the gluon radiation (which 
leads to approximate Bjorken scaling) is neglected within the eikonal approximation. 

}
\end{itemize}
Strictly speaking, the eikonal approximation has been derived only in the framework of
non-relativistic quantum mechanics, where the number of interacting particles is conserved during collisions and the collisions are elastic. 
In this case, the approximation that the Fock states of the incoming high-energy virtual photon can be considered frozen would be justified, and the procedure of eikonalization can be 
unambiguously justified.

In contrast, in a quantum field theory such as QCD, the number of bare particles is not conserved.
In other words, the number of effective degrees of freedom, or relevant Fock states, 
in the photon wave function depends on $x$ and $Q^2$.
For example, the interacting $|q \bar{q} \rangle$ Fock state radiates gluons, 
thus creating and mixing with $|q \bar{q} g \rangle$, $|q \bar{q} g \dots g \rangle$ states.
This mixing is properly taken into account by the QCD evolution in the leading twist approximation.

One immediate consequence of the contrasting pictures of the 
space-time evolution in the
leading twist and dipole model eikonal approaches is the $Q^2$ dependence of nuclear shadowing. 
As $Q^2$ increases, the Fock components of the virtual photon with an 
increasing number
of gluons, $|q \bar{q} g \dots g \rangle$, become important
for nuclear shadowing. This follows straightforwardly from the connection
between nuclear shadowing and gluon-dominated diffraction, as found in ZEUS
and H1 experiments at HERA. As a result, using the factorization theorem,
the $Q^2$ dependence of nuclear shadowing is governed by the DGLAP evolution
equation within the leading twist approach.

The dipole model eikonal approximation 
underestimates nuclear shadowing at large $Q^2$---this can be seen in Figs.~\ref{fig:eikonal_F2} and \ref{fig:eikonal_F2_Q2dep}.
(As we mentioned above, this manifests even stronger 
in the case of the longitudinal
structure function $F_{L}^{A}(x,Q^2)$, 
see Figs.~\ref{fig:eikonal_FL} and \ref{fig:eikonal_FL_Q2dep}.)
This happens because 
the dipole model eikonal approximation includes only 
the $q \bar q$-component of the virtual photon wave function
and
neglects diffractively
produced inelastic states, such as $q \bar q g$, $q \bar q gg$, etc.
 To reproduce the correct $Q^2$
behavior of 
nuclear shadowing, which is governed by the DGLAP evolution equation, 
one should include the complete set of Fock states, 
i.e., a $Q^2$-dependent number of 
constituents, as well as the QCD evolution trajectories starting at 
large $x \geq 0.1$, where the nuclear PDFs are not screened in the leading twist
approach. (For the discussion of QCD trajectories, see Sec.~\ref{subsec:qcd_curve}.)

It is also worth mentioning that the lack of separation over twists in the dipole
eikonal model precludes a simple connection between the nuclear effects in
DIS and other hard processes, such as
the production of jets in $\gamma^{\ast} T \to jet_1 +jet_2 +X$ and in
$pA\to jet_1+jet_2 +X$, etc.

Also, 
there are several technical problems with the implementation 
of the eikonal approximation. Firstly, 
in the kinematics
where the elastic and inelastic $q \bar{q}$-nucleon cross sections are 
compatible, the use of 
the inelastic
$\sigma_{q \bar{q}N}$ 
cross section
alone would significantly 
underestimate nuclear shadowing.

Secondly, to reproduce nuclear shadowing at the higher end
of the shadowing region, $0.01 \leq x \leq 0.1$, one needs to take into account 
the non-zero longitudinal momentum transfer to the nucleus through the factor 
$\exp (i 2 x m_{N} (z_{1}-z_{2}))$. In order to arrive at this factor in the eikonal 
approximation, one needs to make a bold assumption that all essential Fock 
states of the virtual photon have the same invariant mass of the order 
of $Q$.

Thirdly, in the target infinite momentum frame, the main source of the disappearance of nuclear
shadowing with an increase of $Q^2$ at fixed $x$ is the mixing between the small-$x$ 
and large-$x$ contributions, which occurs due to the DGLAP evolution. This effects is 
absent in the dipole eikonal approximation.

\subsection{QCD evolution trajectories}
\label{subsec:qcd_curve}

The $Q^2$ evolution of nuclear PDFs is governed by the DGLAP evolution equations,
see Eq.~(\ref{eq:dglap}). The general trend of the DGLAP $Q^2$ evolution is well
known. As $Q^2$ increases, the parton densities shift toward lower values of $x$
because of the emission of softer partons. Therefore, the evolution proceeds along
a trajectory
 in the $x-Q^2$ plane, which extends from low $Q^2$ and high $x$ toward
large $Q^2$ and small $x$. The detailed knowledge of this 
trajectory is very important.
It enables, for example, to estimate the influence of the input PDFs at the
initial evolution scale $Q_0^2$ on the result of the QCD evolution to higher 
scales $Q^2$ and to judge as to what region of $x$ at $Q_0^2$ contributes to
the PDFs after the evolution. Also, an understanding of the QCD evolution 
trajectory
is relevant for the studies of the applicability of the leading twist QCD
evolution.

To numerically study the trajectory
in the $x-Q^2$ plane along which the DGLAP
evolution proceeds, we adopt the following algorithm~\cite{Frankfurt:2000ty}.
At the input scale, $Q_0^2=4$ GeV$^2$, we pick an arbitrary value of $x_0$, which
will serve as the starting point of the evolution 
trajectory,
$(x_0,Q_0^2)$. For any $Q^{\prime 2} > Q_0^2$, we find $x^{\prime}$, $x^{\prime} < x_0$, 
by requiring that half of $f_{j/A}(x^{\prime},Q^{\prime 2})$ comes from 
the DGLAP evolution equations where the lower limit of integration is
$x_0$ instead of $x^{\prime}$ [compare to Eq.~(\ref{eq:dglap})]:
\begin{equation}
\frac{1}{2}f_{j/A}(x^{\prime},Q^{\prime 2})=f_{j/A}(x^{\prime},Q^{\prime 2})_{|x_0} \,,
\label{eq:curve_condition}
\end{equation}
where 
\begin{eqnarray}
&&\frac{d\,f_{j/A}^{ns}(x^{\prime},Q^{\prime 2})_{|x_0}}{d \log Q^{\prime 2}}=\frac{\alpha_s(Q^{\prime 2})}{2 \pi} \int^1_{x_0} \frac{d y}{y}
P_{qq}\left(\frac{x^{\prime}}{y}\right) f_{j/A}^{ns}(y,Q^{\prime 2})_{|x_0} \,, \nonumber\\
&&\frac{d}{d \log Q^{\prime 2}} 
\left(\begin{array}{c}
f^s_A(x^{\prime},Q^{\prime 2}) \\
f_{g/A}(x^{\prime},Q^{\prime 2})
\end{array} \right)_{|x_{0}}=\frac{\alpha_s(Q^{\prime 2})}{2 \pi} \int^1_{x_0} \frac{d y}{y}
\left(
\begin{array}{cc}
P_{qq}\left(\frac{x^{\prime}}{y}\right) & P_{qg}\left(\frac{x^{\prime}}{y}\right) \\
P_{qg}\left(\frac{x^{\prime}}{y}\right) & P_{gg}\left(\frac{x^{\prime}}{y}\right)
\end{array}
\right)
\left(\begin{array}{c}
f_A^s(y,Q^{\prime 2}) \\
f_{g/A}(y,Q^{\prime 2})
\end{array} \right)_{|x_{0}} \,.
\label{eq:dglap_curve}
\end{eqnarray}
This procedure allows us to determine point $(x^{\prime},Q^{\prime 2})$ on the curve
starting at $(x_0,Q_0^2)$. By scanning the desired range of values of $Q^{\prime 2}$,
we obtain the entire QCD evolution trajectory.
Our prescription for the determination of the evolution trajectory
provides the quantitative measure of the essential 
integration region in 
$x$ which contribute to the parton density for $Q^2 > Q_0^2$ 
within the DGLAP approximation.

\begin{figure}[h]
\begin{center}
\epsfig{file=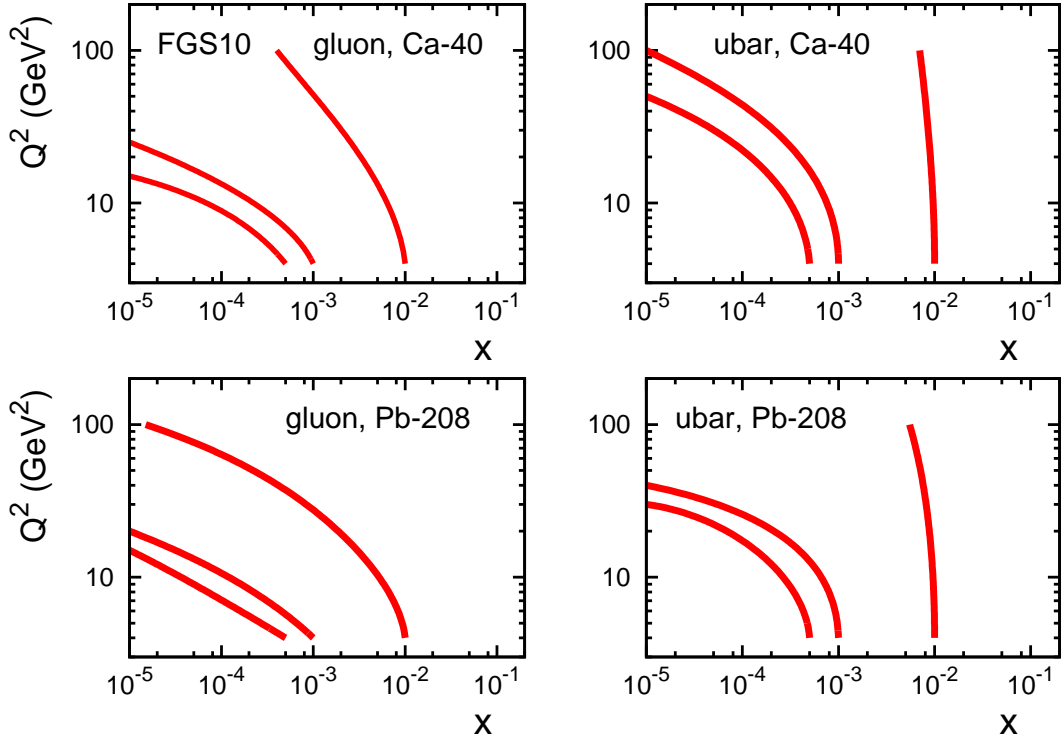,scale=1.4}
\caption{The QCD evolution trajectories determined by Eqs.~(\ref{eq:curve_condition})
and (\ref{eq:dglap_curve}) and discussed in the text.
The curves for models FGS10\_H and FGS10\_L are indistinguishable.}
\label{fig:curve}
\end{center}
\end{figure} 
\begin{figure}[h]
\begin{center}
\epsfig{file=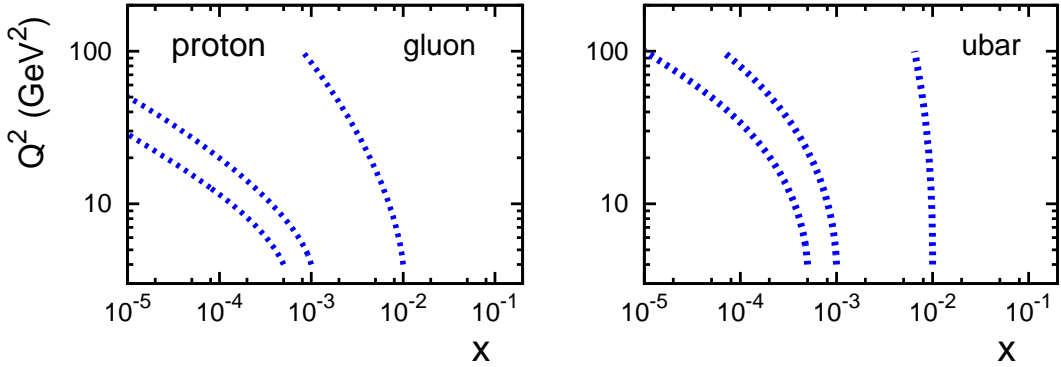,scale=1.4}
\caption{The QCD evolution trajectories for the free proton.}
\label{fig:curve_proton}
\end{center}
\end{figure} 

The resulting QCD evolution trajectories for three
different values of the initial $x_0$, 
$x_0=0.01$, $10^{-3}$, and $5 \times 10^{-4}$,
are presented in Fig.~\ref{fig:curve} ($^{40}$Ca and $^{208}$Pb) and 
Fig.~\ref{fig:curve_proton} (free proton).
(In Fig.~\ref{fig:curve}, the trajectories for models FGS10\_H and FGS10\_L are indistinguishable.)
As one can see from the figures, the trajectories 
for the gluon and quark channels are different. 
Also, the trajectories starting at $x_0 < 0.01$ for the nuclei and free proton
are different, too. 
These features are related and we address them below.

As we progressively decrease the starting point $x_0$, we increase the right
hand side of Eq.~(\ref{eq:dglap_curve}) and, hence, 
$f_{j/A}(x^{\prime},Q^{\prime 2})_{|x_0}$.
Therefore, the solution of Eq.~(\ref{eq:curve_condition}) at fixed given 
$Q^{\prime 2}$ can be found at progressively smaller values of $x^{\prime}$.
Since the gluon parton densities are larger than the quark ones, the
found values of $x^{\prime}$ in the gluon channel are smaller than those
in the quark channel, i.e., the trajectories in the gluon channel bend
toward smaller $x$ stronger than those in the quark channel.
This is true both for nuclei and free proton (see Figs.~\ref{fig:curve}
and \ref{fig:curve_proton}).

Turning to the comparison between the nuclear and free nucleon cases, 
we notice that at the starting point $x_0 < 0.01$, where the nuclear 
modifications of parton distributions are small, the trajectories 
for $^{40}$Ca, $^{208}$Pb, and proton are very similar.
(The only exception is the gluon channel for $^{208}$Pb.)
As we decrease $x_0$ and $x^{\prime}$, we move deeper in the 
shadowing region. This means that at the same small $x_0$, 
the solution of Eq.~(\ref{eq:curve_condition}) in the nuclear case can be 
found at smaller $x^{\prime}$ than in the free proton case in order
to compensate for the suppression of $f_{j/A}(x^{\prime},Q^{\prime 2})$
due to nuclear shadowing.
As a result, the trajectories in the nuclear case bend toward smaller $x$ 
stronger than those in the free proton case.

It is also instructive to compare $\ln (Q^2/Q_0^2)$ and $\ln (x/x_0)$ for
different trajectories. One can see from Figs.~\ref{fig:curve}
and \ref{fig:curve_proton}
that for a large range of $x_0 > 10^{-3}$, these
logarithms are comparable. This indicates that for this range of $x_0$, the double-log
approximation should work well. At the same time, $\ln(x/x_0)$ becomes
more important than $\ln(Q^2/Q_0^2)$ for gluons in nuclei for $x_0 < 10^{-3}$.
However, even in this case, $\alpha_s \ln (x /x_0) \ll 1$ so that corrections to 
the NLO DGLAP should remain modest.

We will discuss in Sec.~\ref{sec:bdr} that for sufficiently small $x$ and moderate $Q_0^2$,
the DGLAP approximation may break down due to proximity to the black disk regime.  
The inspection of Fig.~\ref{fig:curve}
 shows that for 
very small $x$, the increase of $Q^2$  leads to 
the
dominance of the trajectories 
that avoid the black disk region, so 
that
with an increase of $Q^2$ (for fixed $x$) one would reach the kinematics where 
predictions based on  the DGLAP approximation 
are valid.

\subsection{Comparison of predictions of the leading twist theory of nuclear shadowing 
 with fixed-target data}
\label{subsec:comparison}

\begin{figure}[h]
\begin{center}
\epsfig{file=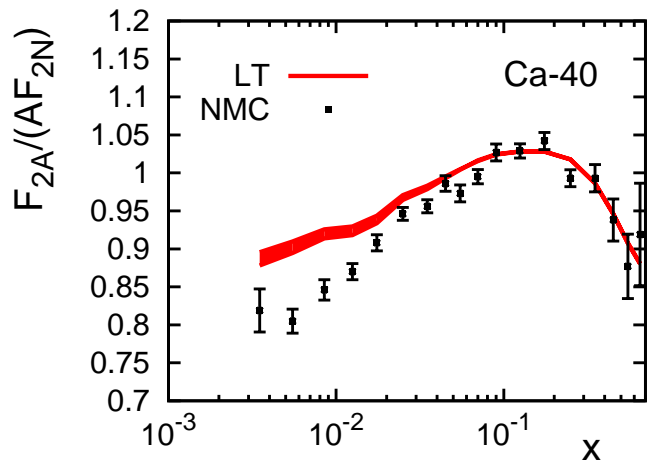,scale=1.45}
\caption{The ratio of the nuclear ($^{40}$Ca) to nucleon structure functions
$F_{2A}(x,Q^2)/[A F_{2N}(x,Q^2)]$ 
as a function of $x$. The data points are the NMC data~\cite{Amaudruz:1995tq}.
The solid band is the 
prediction of the leading twist theory of nuclear shadowing.
}
\label{fig:F2_ca40_nmc}
\end{center}
\end{figure}

\begin{figure}[t]
\begin{center}
\epsfig{file=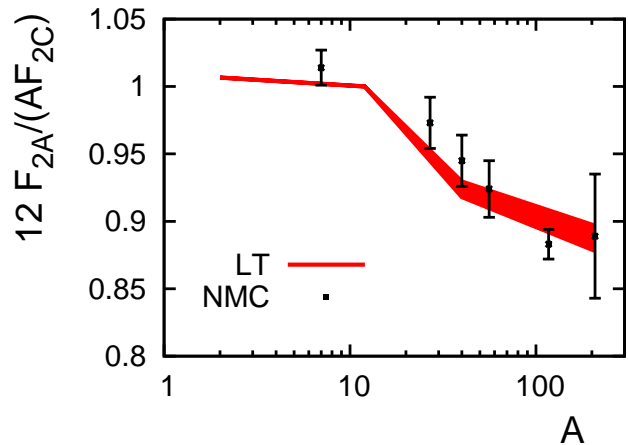,scale=1.4}
\caption{The $12 F_{2A}(x,Q^2)/[A F_{2C}(x,Q^2)]$ ratio as a function 
of the atomic number $A$. The data points are the NMC data~\cite{Arneodo:1996rv}.
The solid band is the 
prediction of the leading twist theory of nuclear shadowing.
All data points correspond to $x=0.0125$ and $Q^2=3.4$ GeV$^2$.
}
\label{fig:nmc_Adep}
\end{center}
\end{figure}

Predictions of the leading twist theory of nuclear shadowing can be compared
to the available measurements of $F_{2A}(x,Q^2)$ in nuclear DIS with fixed 
targets~\cite{Aubert:1983xm,Bodek:1983qn,Bodek:1983ec,Arnold:1983mw,Dasu:1988ru,Bari:1985ga,Benvenuti:1987az,Ashman:1988bf,Arneodo:1989sy,Amaudruz:1991cc,Amaudruz:1991dj,Amaudruz:1992wn,Adams:1992vm,Adams:1992nf,Adams:1995is,Gomez:1993ri,Amaudruz:1995tq,Arneodo:1995cs,Arneodo:1995cq,Arneodo:1996rv,Arneodo:1996ru};
an example of such a comparison is presented in Figs.~\ref{fig:F2_ca40_nmc}
and \ref{fig:nmc_Adep}.

In Fig.~\ref{fig:F2_ca40_nmc},
the NMC data for $F_{2A}(x,Q^2)/[A F_{2N}(x,Q^2)]$ for
 $^{40}$Ca~\cite{Amaudruz:1995tq} is compared to the predictions
of the leading twist theory of nuclear shadowing given by the solid band 
spanning the predictions of models FGS10\_H and FGS10\_L (see Sec.~\ref{subsec:nPDFs}).
As we explained in the Introduction, the low-$x$
fixed-target nuclear DIS data points correspond to the low values of  
$Q^2$ that are (significantly) lower than our input scale $Q_0^2=4$ GeV$^2$
and the alternative possible input scale $Q_0^2=2.5$ GeV$^2$
(see Fig.~\ref{fig:LT2007_ca40_uncert1} in Sec.~\ref{subsect:uncertainties}).
Therefore, we cannot directly
compare our predictions to the data in this kinematics. Hence,
in Fig.~\ref{fig:F2_ca40_nmc}, for the first four data points with $Q^2 < 2.5$ GeV$^2$
[the average values of $Q^2$ for these points are
$\langle Q^2 \rangle=(0.60, 0.94, 1.4, 1.9)$ GeV$^2$, respectively],
our predictions are evaluated at fixed $Q^2=2.5$ GeV$^2$.

Figure~\ref{fig:nmc_Adep} presents the $12 F_{2A}(x,Q^2)/[A F_{2C}(x,Q^2)]$
ratio
as a function of the atomic number $A$. Our leading twist predictions are given by the
solid band spanning the predictions of models FGS10\_H (lower boundary) and
FGS10\_L (upper boundary).
Our predictions are evaluated at $x=0.0125$ and $Q^2=3.4$ GeV$^2$ of the NMC 
data points~\cite{Arneodo:1996rv} used for comparison in the figure.
As one can see from the figure, the leading twist theory
of nuclear shadowing reproduces the $A$ dependence of nuclear shadowing rather well. 

It is important to emphasize that one should clearly separate two issues:
(i) the validity of the Gribov-Glauber theory of nuclear shadowing for the NMC data, and 
(ii) the applicability of the leading twist approximation to the NMC fixed-target data. 
While the Gribov-Glauber theory has been shown to work with high precision in the NMC kinematics, 
the straightforward application of the leading twist theory of nuclear shadowing
to the NMC fixed-target data fails, see Fig.~\ref{fig:F2_ca40_nmc}.
Below we elaborate on these issues.

A
fairly good description of the low-$x$ and low-$Q^2$ NMC data on the
nuclear structure function $F_{2A}(x,Q^2)$
was achieved within
the Gribov-Glauber theory of nuclear shadowing~\cite{Capella:1997yv,Armesto:2003fi}.
(A recent analysis~\cite{Armesto:2010kr} in the framework of the Gribov-Glauber 
theory provided a good description of the NMC data with $Q^2 > 2$ GeV$^2$.)
 As an input for these calculations, phenomenological parameterizations 
that fit well the inclusive and diffractive structure 
functions of the nucleon
were used.
In contrast to our
 strictly leading twist analysis, the phenomenological parameterizations used 
in~\cite{Capella:1997yv,Armesto:2003fi,Armesto:2010kr}
effectively included higher twist contributions, 
see the discussion in Sec.~\ref{subsec:comp_to_Capella}.

The Gribov-Glauber theory of nuclear shadowing combined 
with phenomenological fits to inclusive diffraction in photon-nucleon scattering 
(the accurate data  on the real photon diffraction for
 the relevant energies is available, see Ref.~\cite{Chapin:1985mf}) 
provides a good description of nuclear shadowing in real photon-nucleus
scattering~\cite{Piller:1997ny,Piller:1999wx,Adeluyi:2006xy}.

The fact that the $A$ dependence of nuclear shadowing is reproduced well
by the leading twist theory of nuclear shadowing, 
see Fig.~\ref{fig:nmc_Adep}, also indicates that the Gribov-Glauber theory 
works for the NMC data.

All these arguments put together indicate that 
it is very natural to have rather significant higher twist
effects at small $Q^2$ since, for this kinematics, the contribution of 
small diffractive masses $M_X$ becomes important. Production of small 
diffractive masses $M_X$ is dominated by the production of vector mesons,
 which is definitely a higher twist phenomenon.

To model the role
 of the higher twist contribution to nuclear shadowing, we explicitly 
calculate the contribution of the $\rho$, $\phi$ and $\omega$ vector mesons 
to nuclear shadowing
using the vector meson dominance (VMD) model.
The resulting $F_{2A}^{\rm VMD}(x,Q^2)/[A F_{2N}(x,Q^2)]$ reads~\cite{Frankfurt:2003zd,Piller:1995kh}:
\begin{samepage}
\begin{eqnarray}
&&\frac{F_{2A}^{\rm VMD}(x,Q^2)}{A F_{2N}(x,Q^2)}=1-\frac{A-1}{2}\frac{Q^2 (1-x)}{\pi F_{2N}(x,Q^2)} \sum_{V=\rho,\phi,\omega} \frac{\sigma_V^2}{f_V^2} \left(\frac{m_V^2}{Q^2+m_V^2}\right)^2
H(Q^2) \nonumber\\
&& \times \int d^2 b \int^{\infty}_{-\infty} dz_1 \int^{\infty}_{z_1} dz_2  
 \rho_A(b,z_1) \rho_A(b,z_2) \cos\left(\Delta_V(z_2-z_1)\right)e^{-(A/2)\sigma_{V} \int_{z_1}^{z_2} dz \rho_A(b,z)}  \,,
\label{eq:vmd}
\end {eqnarray}
\end{samepage}
where $m_V$ is the vector meson mass;
$g_V$ is the  $V \to e^{+}e^{-}$ 
coupling constant;
$\sigma_V$ is the vector meson-nucleon total scattering cross section;
$\Delta_V=x \, m_V (1+m_V^2/Q^2)$;
$H(Q^2)=1/(1+Q^2/Q_0^2)$ is an additional damping factor suppressing 
the overlap between the virtual photon and vector meson wave functions at large $Q^2$. 
The VMD parameters $g_V$ and 
$\sigma_V$ have their usual values, see, e.g., \cite{Piller:1995kh}:
$\sigma_{\rho}=\sigma_{\omega}=25$ mb, $\sigma_{\phi}=10$ mb,
$f_{\rho}^2/(4 \pi)=2$, $f_{\omega}^2/(4 \pi)=23$, and $f_{\phi}^2/(4 \pi)=13$.
For the nucleon structure 
function $F_{2N}(x,Q^2)$ for low $Q^2$ and $x$, we used the 
NMC parameterization~\cite{Arneodo:1995cq} 
for $x > 0.006$ and the ALLM97 fit~\cite{Abramowicz:1997ms} for $x < 0.006$
(the original ALLM fit~\cite{Abramowicz:1991xz}) gives similar results).

\begin{figure}[h]
\begin{center}
\epsfig{file=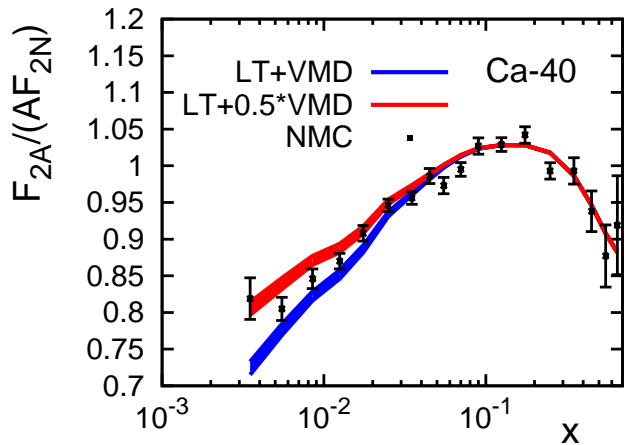,scale=1.4}
\caption{Comparison of the $F_{2A}(x,Q^2)/[A F_{2N}(x,Q^2)]$ ratio for
 $^{40}$Ca~\cite{Amaudruz:1995tq} to our predictions corresponding to 
the sum of the leading twist and VMD contributions.
The lower band corresponds to LT+VMD; the upper band corresponds to 
LT+0.5VMD (see the text).}
\label{fig:F2_ca40_nmc2}
\end{center}
\end{figure}

The inclusion of the VMD contribution significantly increases the 
nuclear shadowing correction. Adding together our leading twist and 
the VMD contributions, we obtain the predictions given by the lower solid
band in Fig.~\ref{fig:F2_ca40_nmc2}.
(Like in the case of Fig.~\ref{fig:F2_ca40_nmc}, the band corresponds to
the theoretical uncertainty of our leading twist theory predictions.)
 As one can see from
the figure, while the lower band somewhat overestimates the amount of nuclear shadowing,
the description of the data is still fairly good (except for the lowest $x$ point).

Since one does not have an 
unambiguous way to add the LT and VMD contributions, as an illustration, we consider
the scenario when the VMD contribution is added with the coefficient $1/2$.
This coefficient accounts for the 
duality between the continuum and VMD 
contributions to diffraction, 
see
also the discussion in Ref.~\cite{Frankfurt:1995jw}. The corresponding prediction
is given by the upper band in Fig.~\ref{fig:F2_ca40_nmc2}.
As one can see from the figure, the ''LT+0.5 VMD'' prescription provides a good
description of the NMC data.

Figures \ref{fig:F2_ca40_nmc} and \ref{fig:F2_ca40_nmc2} illustrate the
important qualitative phenomenon that the higher twist effects play an 
important role in nuclear shadowing in the considered kinematics.
This conclusion
is in a broad  agreement with the phenomenological approaches to nuclear shadowing
which include
 both the scaling (leading twist) and lowest mass ($\rho$, $\omega$ and $\phi$) 
vector meson (higher twist) contributions~\cite{Kwiecinski:1988ys,Badelek:1991qa,Badelek:1994qg,Piller:1995kh,Melnitchouk:1992eu,Melnitchouk:1993vc,Melnitchouk:1995am,Melnitchouk:2002ud,Bilchak:1988zn,Shaw:1993gx}.

One should also mention a very different approach to nuclear shadowing, where
nuclear shadowing is a purely higher twist effect~\cite{Qiu:2003vd}.
The analysis of~\cite{Qiu:2003vd} confirms our observation that  
the higher twist effects in the fixed-target kinematics are large.
So far the connection of the approach of~\cite{Qiu:2003vd} to the Gribov
theory is not clear. In particular, the diagrams that correspond to
 the vector meson production (which dominates the higher twist small-$x$
contribution in the Gribov theory) seem
to be neglected in~\cite{Qiu:2003vd} as a very high twist effect.
It would be interesting  to compare predictions for 
the double scattering contribution to $F_{2A}(x,Q^2)$ made using the
approach of Ref.~\cite{Qiu:2003vd} and the Gribov relation between 
shadowing and diffraction [see Eq.~(\ref{eq:m11})], which, in this limit, 
is a consequence of  
unitarity, see the discussion in  Sec.~\ref{sec:hab_pdfs}.

\subsection{The EMC effect for heavy nuclei and the Lorentz dilation of the nuclear
Coulomb field}
\label{subsec:EMC_effect}

This subsection is based on Ref.~\cite{Frankfurt:2010cb}.
In QCD one usually treats the parton wave function of a nucleus $A$ as 
built  of quarks and gluons. As a result, it satisfies the following 
momentum sum rule: 
\begin{equation}
 \int_0^1 \left[x_{A}V_{A}(x_{A}, Q^2) +x_{A}S_{A}(x_{A}, Q^2) +x_{A}G_{A}(x_{A},Q^2)\right]dx_{A}=1 \,,
 \label{eq:fracin}
 \end{equation}
where the summation over the quark flavors is assumed;
$(V_A, S_A, G_A)$ refer to the (valence quark, sea quark, gluon) distributions in the
target; $x_{A}= Q^2/(2q_0M_A)$ where $q_0$ is the virtual photon energy
and $M_A$ is the nucleus mass.
In this approximation, one neglects electromagnetic effects both in the hadron 
wave function at the initial scale of the evolution, $Q^2_0$, and in the  DGLAP QCD evolution. 

In the case of a fast particle, its Coulomb field is transformed 
into the field of equivalent photons. As a result, the photons become dynamical
degrees of freedom.
To take them into account requires the modification of the QCD 
evolution equations by including the momentum distribution of the photons, 
$P_A$, in addition to the standard contributions of quarks and gluons.
Thus,  
the presence of the photon component 
in the nuclear light-cone wave  function leads to the following modification of
the  momentum sum rule:
 \begin{equation}
 \int_0^1 \left[x_{A}V_{A}(x_{A}, Q^2) +x_{A}S_{A}(x_{A},Q^2) +x_{A}G_{A}(x_{A},Q^2)+ x_{A}P_{A}(x_{A}, Q^2)\right]dx_{A}=1 \,.
 \label{eq:fracin2}
 \end{equation}
To remove the  kinematic effects, 
it is convenient to rescale the  variables by introducing the light-cone fraction
$x$ defined as
\begin{equation}
x=Ax_{A} \,,
\label{xdef}
\end{equation}
where $A$ is the atomic mass number of the nucleus $A$. 
It satisfies the inequality $0 < x <A$ and differs from the Bjorken $x$ for the scattering off a proton, $x_p=Q^2/(2q_0m_p)$, due to the nuclear binding energy.
In terms of $x$, 
Eq.~(\ref{eq:fracin2}) reads:
 \begin{equation}
 \int_0^A \left[(1/A)(xV_{A}(x, Q^2) +xS_{A}(x, Q^2) +xG_{A}(x,Q^2))
+xP_{A}(x, Q^2)\right]dx=1 \,.
 \label{sumrule}
 \end{equation}
In this review we are interested in the $A$ dependence of nuclear PDFs,
and, hence, the electromagnetic effects at the level of the proton and 
neutron is of no relevance for our analysis since they are canceled out in the ratio 
of the nucleus and nucleon PDFs.
As a result, 
the main effect is the presence of the Coulomb coherent field of the nucleus
and
not the Coulomb fields of individual nucleons.

It is possible to calculate the contribution of the coherent field using 
Fermi-Weizsacker-Williams approximation for  the wave function of a rapid projectile with 
nonzero electric charge \cite{WW}; this is the only contribution which is proportional to $Z^2$. 
Subtracting the contribution of the individual nucleons which is proportional to 
$Z$, one finds for the photon field of a rapid
nucleus~\cite{Frankfurt:2010cb}:
\begin{equation}
\Delta xP_{A}(x,Q^2)={\alpha_{\rm em}\over \pi } {Z(Z-1)\over A}
\int d k_t^2\, k_t^2  \frac{F_A^2(k_t^2+x^2 m_N^2)}{(k_t^2+x^2 m_N^2)^2} \,,
\label{ww1}
\end{equation}
where $F_A$ is the nuclear electric form factor; $k_t$ is the transverse momentum
of the active nucleon in the nucleus.

The explicit evaluation of Eq.~(\ref{ww1}) shows that
a rather significant fraction of the nucleus light-cone momentum is
 carried by the additional  photons in the medium and heavy  nuclei.
Defining $\lambda_{\gamma}$ as
\begin{equation}
\lambda_{\gamma}=(1/A) \int_0^A \Delta xP_{A}(x, Q^2)dx \,,
\label{eq:lambda_gamma}
\end{equation}
one finds
\begin{eqnarray}
&&\lambda_{\gamma}(^4{\rm He}) = 0.03\% \,, \quad \quad \quad \lambda_{\gamma}(^{12}{\rm C}) = 0.11\% \,,\nonumber\\ 
&&\lambda_{\gamma}(^{27}{\rm Al}) = 0.21\% \,, \quad \quad \quad 
 \lambda_{\gamma}(^{56}{\rm Fe}) =0.35\% \,, \nonumber\\
&&\lambda_{\gamma}(^{197}{\rm Au}) =0.65\% \,.
\end{eqnarray}
In the discussed approximation, the $eA$ cross section could be written as a convolution
of the $ep$ cross section and the nucleon density in the nucleus. 
Performing the Taylor series expansion in powers of $k^2/m_N^2$  ($k$ is the nucleon 
momentum in the nucleus), one finds that  
the bulk effect of the additional photon component is that individual nucleons carry 
only the $(1-\lambda_{\gamma}(A))$
fraction of the nucleus light-cone momentum 
(we neglect here the effect of possible modification of the bound nucleon wave function).  
This effect can be accounted for by the following rescaling:
\begin{eqnarray}
&& xq_{A}(x, Q^2) = A x^{\prime} q_{N}(x^{\prime}, Q^2) \,, \nonumber\\
&& xg_{A}(x, Q^2) = A x^{\prime} g_{N}(x^{\prime}, Q^2) \,,
\label{EMC}
\end{eqnarray}
where 
\begin{equation}
 x^{\prime} = x/ (1-\lambda_{\gamma}(A)) \,.
\end{equation}

\begin{figure}[h]
\begin{center}
\epsfig{file=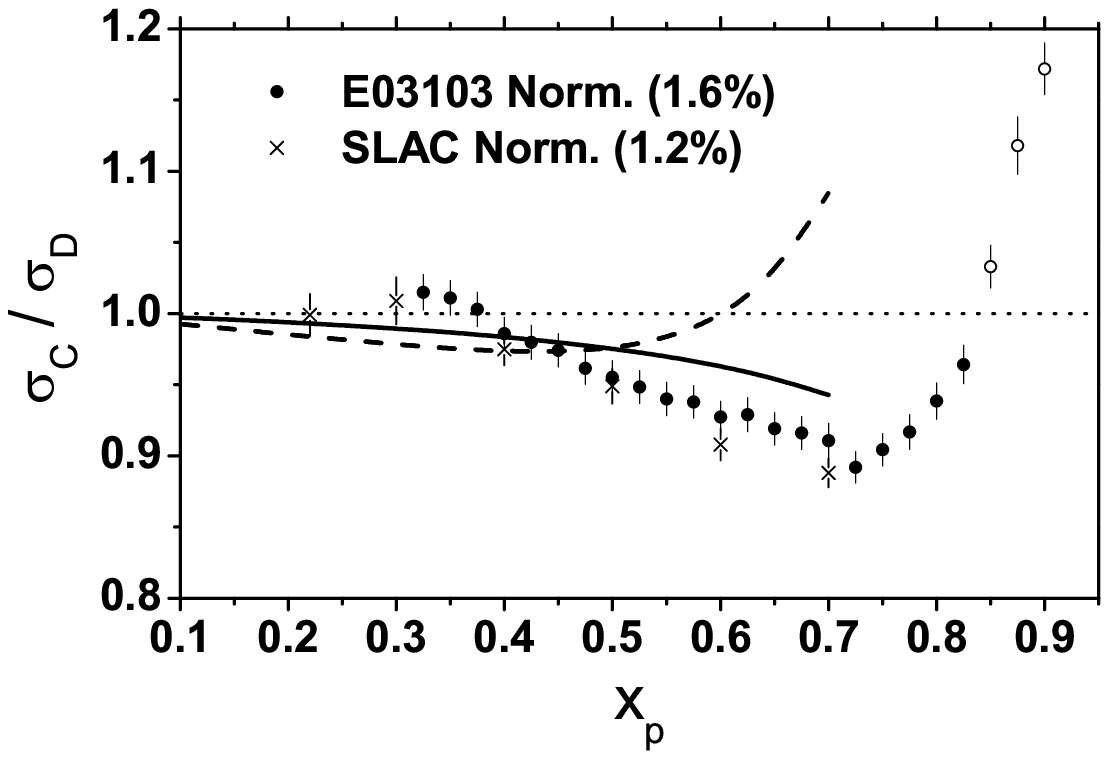,scale=0.7}
\epsfig{file=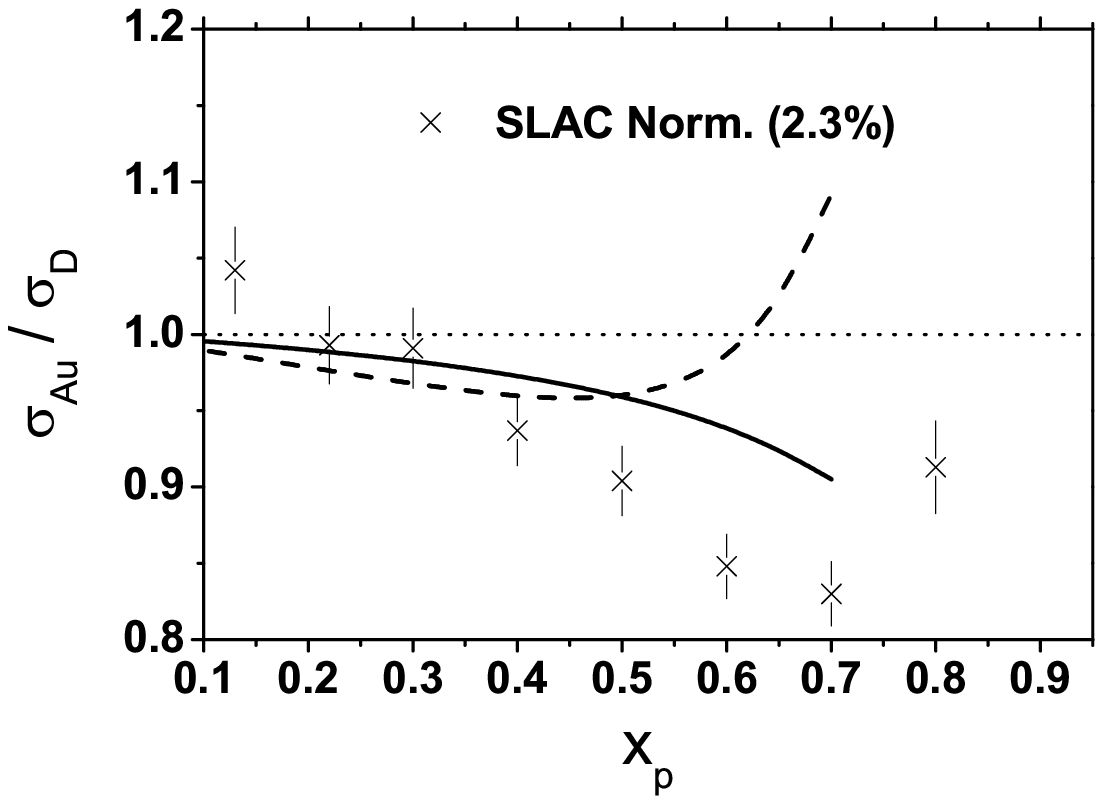,scale=0.7}
\caption{The EMC ratio as a function of $x_p$. 
The solid curve is the result that takes into account the effect of the nuclear 
Coulomb field resulting in the proper definition of $x$ calculated using Eq.~(\ref{EMC});
it is applicable for $x\le 0.7$ only.
The dashed curve includes also the Fermi motion effect. 
  The data are from~\cite{Gomez:1993ri,Seely:2009gt};   
the open circles correspond to $W<2$ GeV.}
\label{fig:EMC_photon}
\end{center}
\end{figure}
Equation~(\ref{EMC}) leads to the modification of the nuclear structure functions. 
In particular, for  the ratio $R_A^{\rm EMC}(x_p,Q^2) =(2/A) F_{2A}(x_p,Q^2)/F_{2D}(x_p,Q^2)$ 
also known as  the EMC ratio,  
it leads to the effect that explains 
a half of the effect observed in 
the data  for $x\le 0.55$
where the Fermi motion effects are small, see Fig.~\ref{fig:EMC_photon}.
Note here that for light nuclei, the main effect is due to the proper definition of 
the light-cone fraction $x$ [Eq.~(\ref{xdef})], which enters in the convolution formula,
and a small correction from the Coulomb effect. 
For heavy nuclei,  a  comparable contribution comes also from the Coulomb effect. 
Overall, our observation that the hadronic EMC effect exceeds few percent only for $x\ge 0.55$, where the structure functions are very small
and an important role is played by short-range nucleon-nucleon correlations,
confirms our starting approximation
to treat nuclei as many-nucleon systems with relativistic corrections treated 
within the light-cone many-nucleon approximation.

One can see from Eq.~(\ref{EMC}) that the discussed correction is on the scale of a fraction 
of percent for the region of $x \le 0.1$, which is the focus of the present review, 
since the parton distributions $xq(x,Q^2)$ and $xg(x,Q^2)$ are changing 
in this $x$-range rather slowly as $x^{-\lambda(Q^2)}$, with $\lambda(Q^2)$ 
varying between 0.2 and 0.4. 
The only other potentially significant effect is the change of the fraction of the momentum of the nucleus carried by the gluons. 
It is usually determined based on the application of Eq.~(\ref{sumrule}) and 
is close to $\sim $ 0.5. An account of the photons results in its reduction 
by $-2\lambda_{\gamma}(A)$, i.e., by about 1.4\% for heavy nuclei.  Only half of this reduction is accounted for by the rescaling in Eq.~(\ref{EMC}). 
The rest may somewhat reduce the enhancement of the gluon ratio at $x\sim 0.1$
(antishadowing)  which follows from the application of the momentum sum rule. 
However, as we pointed out in Sec.~\ref{subsubsect:antishadowing}, 
our estimates of the modification of the gluon PDF  for these $x$ 
have rather large uncertainties.

\section{Final states in DIS with nuclei at small $x$}
\label{sec:final_states}

In Sec.~\ref{sec:phen} we demonstrated that 
the existence of leading twist diffraction in DIS leads to the significant 
suppression of the nuclear PDFs at small $x$. In this section, we explore consequences of the leading twist shadowing  phenomenon for the final states produced
in the small $x$ processes induced by hard probes. In particular, we consider the
following three characteristics of the final states: diffraction, spectra of leading particles, and fluctuations of multiplicity at central rapidities.

\subsection{Nuclear diffractive structure functions and 
diffractive
parton distribution functions}

\subsubsection{Coherent diffraction}
\label{subsubsec:coherent_diffraction}

\begin{figure}[h]
\begin{center}
\epsfig{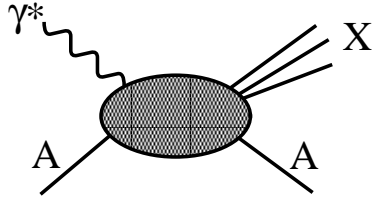}
\caption{Coherent diffractive DIS with nuclei.}
\label{fig:coh_diff_graph}
\end{center}
\end{figure}

Let us consider diffractive DIS with nuclei, $\gamma^{\ast}A \to X A^{\prime}$,
which is characterized by the presence of a rapidity gap between the products of the photon
dissociation, $X$, and the final nuclear state $A^{\prime}$. 
We first address the case of the coherent scattering when the nucleus remains intact, 
$A^{\prime}=A$, see Fig.~\ref{fig:coh_diff_graph}. 
Coherent scattering is readily amenable to the theoretical methods 
which were successful in
the case of
inclusive $eA$ scattering and
is also easier to detect experimentally in collider experiments (see the discussion in the end of this subsection).
\vspace*{0.25cm}

Squaring the $eA \to e^{\prime} X A$ amplitude, one obtains the $eA \to e^{\prime} X A$
cross section expressed in terms of the nuclear diffractive structure function
$F_{2A}^{D(4)}$ [compare to Eq.~(\ref{eq:ft1})]:
\begin{equation}
\frac{d^4 \sigma_{eA}^D}{dx_{\Pomeron}\,dt\, dx\, dQ^2}=\frac{2 \pi \alpha^2}{x Q^4} 
\left(1+(1-y)^2\right) F_{2A}^{D(4)}(x,Q^2,x_{\Pomeron},t) \,.
\label{eq:ft1_A}
\end{equation}
In Eq.~(\ref{eq:ft1_A}), we ignored the contribution of the longitudinal diffractive structure function
$F_{L,A}^{D(4)}$---its contribution is small in the considered kinematics.

From the practical point of view, it is not feasible to detect the recoil nucleus 
at the small values of $t$ characteristic for 
coherent scattering. Therefore, in the
following discussion we concentrate on the diffractive cross section integrated
over the momentum transfer $t$
which 
is expressed in terms of the nuclear diffractive structure function $F_{2A}^{D(3)}$ 
[compare to Eq.~(\ref{eq:data1})]:
\begin{equation}
F_{2A}^{D(3)}(x,Q^2,x_{\Pomeron})=\int_{-1\, {\rm GeV}^2}^{t_{{\rm min}}} dt F_{2A}^{D(4)}(x,Q^2,x_{\Pomeron},t) \,.
\label{eq:data1_A}
\end{equation}
It follows from the QCD factorization theorem for diffraction in DIS that one can
introduce the nuclear diffractive parton distributions $f_{j/A}^{D(3)}(\beta,Q^2,x_{\Pomeron})$
and express $F_{2A}^{D(3)}$ in terms of $f_{j/A}^{D(3)}(\beta,Q^2,x_{\Pomeron})$
[compare to Eq.~(\ref{eq:ft3_a})]:
\begin{equation}
F_{2A}^{D(3)}(x,Q^2,x_{\Pomeron})=\beta\sum_{j=q,\bar{q},g}  \int_{\beta}^{1} \frac{d y}{y}C_j (\frac{\beta}{y},Q^2) f_{j/A}^{D(3)}(y,Q^2,x_{\Pomeron})  \,,
\label{eq:data2_A}
\end{equation}
where $\beta=x/x_{\Pomeron}$.

\begin{figure}[h]
\begin{center}
\epsfig{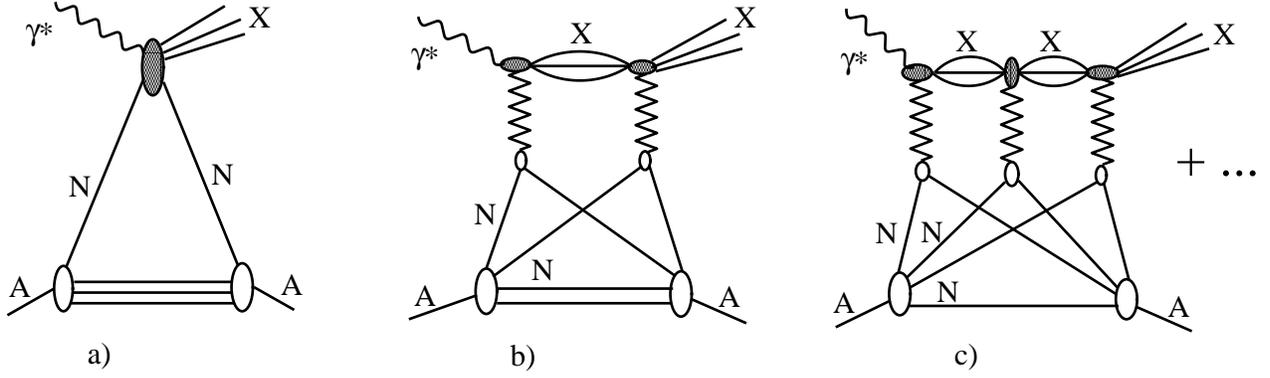}
\caption{The multiple scattering series for the $\gamma^{\ast}A \to X A$ scattering amplitude.
Graphs $a$, $b$, $c$ correspond to the interaction with one, two, and three nucleons of
the nuclear target, respectively.
 Graph $a$ is the impulse approximation; graphs $b$ and $c$ 
contribute to the shadowing correction.}
\label{fig:coh_diff_multiple}
\end{center}
\end{figure}

The derivation of the modification of the nuclear diffractive structure function and
nuclear diffractive PDFs by nuclear shadowing proceeds very similarly to
the derivation of nuclear shadowing in the usual nuclear PDFs presented 
in Sec.~\ref{sec:hab_pdfs}. 
At small $x$, the virtual photon interacts with many nucleons of the target, and the
$\gamma^{\ast}A \to X A$ scattering amplitude is given by the sum of the multiple scattering
contributions presented in Fig.~\ref{fig:coh_diff_multiple} (compare to Fig.~\ref{fig:Master1}).
Graphs $a$, $b$, $c$ correspond to the interaction with one, two, and three nucleons of
the nuclear target, respectively.
 Graph $a$ is the impulse approximation; graphs $b$ and $c$ 
contribute to the shadowing correction. Note that the interactions with four and more 
nucleons (at the amplitude level) are not shown, but 
they are
 implied.
\vspace*{0.5cm}

With help of the graphical representation for the $\gamma^{\ast}A \to X A$ amplitude in 
Fig.~\ref{fig:coh_diff_multiple} and working along the lines of the derivation of nuclear 
shadowing presented in Sec.~\ref{sec:hab_pdfs}, the expression for $f_{j/A}^{D(3)}(\beta,Q^2,x_{\Pomeron})$ can be obtained in a 
rather straightforward way. 
Using the color fluctuation formalism which takes into account inelastic
intermediate states and the non-zero longitudinal momentum transfer,
we obtain for the $\gamma^{\ast}A \to X A$ 
scattering amplitude in the impact parameter space:
\begin{equation}
\Gamma_{\gamma^{\ast} A \to X A}(b)=\langle A| \sum_i\Gamma_{\gamma^{\ast}X}(\vec{b}-\vec{r}_{i \perp}) e^{i z_i \Delta_{\gamma^{\ast}X}}
\prod_{j\neq i} \left(1-\Theta(z_j-z_i) \Gamma_X(\vec{b}-\vec{r}_{j \perp})\right)|A\rangle
\,,
\label{eq:diff_ampl}
\end{equation} 
where the brackets $\langle A| \dots |A\rangle$ denote the matrix element between the 
nuclear ground state;
$\Gamma_{\gamma^{\ast}X}$ is the $\gamma^{\ast}N \to XN$ 
scattering amplitude in the impact parameter space~(\ref{eq:m5});
$\Gamma_X$ is the diffractive state $X$-nucleon scattering amplitude;
as in Sec.~\ref{sec:hab_pdfs},
$\Delta_{\gamma^{\ast} X}  = (M_X^2+Q^2) /(2q_0)$ is the longitudinal momentum transfer 
to the nucleons.
These scattering amplitudes are proportional to 
the $\sigma_{\gamma^{\ast}N \to X N}$ and $\sigma_{\rm soft}^j$ cross sections,
respectively (see the discussion in Sec.~\ref{sec:hab_pdfs}).
Assuming that the nuclear wave function squared is given by the product of independent, one-nucleon
densities, $\rho_A$ [see Eq.~(\ref{eq:m6})], and neglecting the $t$ dependence of the 
$\gamma^{\ast}N \to X N$ diffractive amplitude compared to the nuclear form factor
[see Eq.~(\ref{eq:m5})], the integration over the positions of the nucleons in 
Eq.~(\ref{eq:diff_ampl}) can be carried out analytically with the following result:
\begin{eqnarray}
&&\Gamma_{\gamma^{\ast} A \to X A}(b)\nonumber\\
&&=A \frac{1-i \eta}{2} \sqrt{\frac{16 \pi \frac{d \sigma_{\gamma^{\ast}N \to X N}}{dt}(t_{\rm min})}{1+\eta^2}} \int dz \rho_A(b,z)
e^{i z \Delta_{\gamma^{\ast}X}} \nonumber\\
&&\times \left(1-\frac{1-i\eta}{2} \sigma_{\rm soft}^j(x,Q^2)\int_{z}^{\infty} dz^{\prime} \rho_A(b,z^{\prime})\right)^{A-1} \nonumber\\
&&=A \frac{1-i \eta}{2} 
\sqrt{\frac{16 \pi \frac{d \sigma_{\gamma^{\ast}N \to X N}(t_{\rm min})}{dt}}{1+\eta^2}}
 \int dz \rho_A(b,z)
e^{i z \Delta_{\gamma^{\ast}X}}
e^{-\frac{A}{2}(1-i\eta)\sigma_{\rm soft}^j(x,Q^2)\int_{z}^{\infty} dz^{\prime} \rho_A(b,z^{\prime})}
\,.
\label{eq:diff_ampl2}
\end{eqnarray} 
The $\gamma^{\ast}A \to X A$ cross section is expressed in terms
of the corresponding amplitude as
\begin{equation}
\sigma_{\gamma^{\ast}A \to X A}=
\int dt \, \frac{d\sigma_{\gamma^{\ast}A \to X A}}{dt}=
\int d^2 b \, |\Gamma_{\gamma^{\ast} A \to X A}(b)|^2 \,.
\label{eq:sigma_diff_A}
\end{equation} 
Substituting Eq.~(\ref{eq:diff_ampl2}) in Eq.~(\ref{eq:sigma_diff_A}), we obtain
\begin{eqnarray}
\sigma_{\gamma^{\ast}A \to X A}&=&4 \pi A^2 
\frac{d \sigma_{\gamma^{\ast}N \to X N}}{dt}(t_{\rm min})
 \int d^2 b \nonumber\\
&&\times \left| \int^{\infty}_{-\infty} dz e^{i z \Delta_{\gamma^{\ast}X}} e^{-\frac{A}{2} (1-i\eta)\sigma_{\rm soft}^j(x,Q^2) \int_{z}^{\infty} dz^{\prime}\rho_A(b,z^{\prime})} \rho_A(b,z)\right|^2 \,.
\label{eq:sigma_diff_A_2}
\end{eqnarray}
The nuclear and nucleon diffractive cross sections in Eq.~(\ref{eq:sigma_diff_A_2})
can be expressed in terms of 
the corresponding diffractive structure functions.
Therefore, Eq.~(\ref{eq:sigma_diff_A_2}) leads to the following expression for the 
nuclear diffractive structure function $F_{2A}^{D(3)}$:
\begin{samepage}
\begin{eqnarray}
F_{2A}^{D(3)}(\beta,Q^2,x_{\Pomeron})&=&4 \pi A^2  F_{2N}^{D(4)}(\beta,Q^2,x_{\Pomeron},t_{\rm min}) \int d^2 b \nonumber\\
&&\times \left| \int^{\infty}_{-\infty} dz e^{i x_{\Pomeron} m_N z} e^{-\frac{A}{2} (1-i\eta)\sigma_{\rm soft}^j(x,Q^2) \int_{z}^{\infty} dz^{\prime}\rho_A(b,z^{\prime})} \rho_A(b,z)\right|^2 \,.
\label{eq:masterD_F2}
\end{eqnarray}
\end{samepage}
Note that we expressed the longitudinal momentum transfer $\Delta_{\gamma^{\ast} X}$
in terms of $x_{\Pomeron}$, 
$\Delta_{\gamma^{\ast} X}=x_{\Pomeron} m_N$.
Using the QCD factorization theorem for diffraction~(\ref{eq:data2_A}) in
the right-hand and left-hand sides of  Eq.~(\ref{eq:masterD_F2}),
we obtain the expression for the nuclear diffractive PDFs $f_{j/A}^{D(3)}$:
\begin{eqnarray}
\beta f_{j/A}^{D(3)}(\beta,Q^2,x_{\Pomeron})&=&4 \pi A^2 \beta f_{j/N}^{D(4)}(\beta,Q^2,x_{\Pomeron},t_{\rm min}) \int d^2 b \nonumber\\
&&\times \left| \int^{\infty}_{-\infty} dz e^{i x_{\Pomeron} m_N z} e^{-\frac{A}{2} (1-i\eta)\sigma_{\rm soft}^j(x,Q^2) \int_{z}^{\infty} dz^{\prime}\rho_A(b,z^{\prime})} \rho_A(b,z)\right|^2 \,.
\label{eq:masterD_a}
\end{eqnarray}
Finally, assuming the exponential $t$ dependence of $f_{j/N}^{D(4)}$, i.e., using
Eq.~(\ref{eq:m13_c}), we obtain our final expression for the nuclear diffraction parton
distribution $\beta f_{j/A}^{D(3)}$~\cite{Frankfurt:2003gx,Guzey:2005ys}:
\begin{eqnarray}
\beta f_{j/A}^{D(3)}(\beta,Q^2,x_{\Pomeron})&=&4 \pi A^2 B_{\rm diff} \beta f_{j/N}^{D(3)}(\beta,Q^2,x_{\Pomeron}) \int d^2 b \nonumber\\
&&\times \left| \int^{\infty}_{-\infty} dz e^{i x_{\Pomeron} m_N z} e^{-\frac{A}{2} (1-i\eta)\sigma_{\rm soft}^j(x,Q^2) \int_{z}^{\infty}dz^{\prime} \rho_A(b,z^{\prime})} \rho_A(b,z)\right|^2 \,.
\label{eq:masterD}
\end{eqnarray}
The structure of the answer resembles the case of the 
diffractive productions of vector mesons 
(after the generic diffractive state $X$ is replaced by a single vector meson),
see e.g., Ref.~\cite{Bauer:1977iq}.

Equation~(\ref{eq:masterD}) should be compared to Eq.~(\ref{eq:m13master}): the both equations
are derived in the color fluctuation approximation characterized by the cross section
$\sigma_{\rm soft}^j(x,Q^2)$ that determines the strength of the multiple rescatterings.
Note also that the nuclear shadowing correction to $\beta f_{j/A}^{D(3)}$ given by Eq.~(\ref{eq:masterD}) corresponds to the diffractive unitary cut in the 
language of the AGK cutting rules, see Eq.~(\ref{eq:agk1}) and graph $a$
in Fig.~\ref{fig:AGK}.

The physics interpretation of Eq.~(\ref{eq:masterD}) is rather 
straightforward: the diffractive 
scattering takes place 
on any of $A$ nucleons of the target at
point $(\vec{b},z)$; the produced diffractive state 
gets absorbed
 on the way out with the
probability amplitude $e^{-\frac{A}{2} (1-i\eta)\sigma_{\rm soft}^j(x,Q^2) \int_{z}^{\infty}dz^{\prime} \rho_A(b,z^{\prime})}$.

In the limit of very small $x_{\Pomeron}$, the effect of the 
finite
coherent length, i.e., the  
$e^{i x_{\Pomeron} m_N z}$ factor, can be neglected and Eq.~(\ref{eq:masterD}) can be presented in
the following simplified form:
\begin{equation}
\beta f_{j/A}^{D(3)}(\beta,Q^2,x_{\Pomeron})\approx 16 \pi B_{\rm diff} \beta f_{j/N}^{D(3)}(\beta,Q^2,x_{\Pomeron}) \int d^2 \vec{b} 
\left|\frac{1-e^{-\frac{A}{2}(1-i\eta)\sigma_{\rm soft}^j(x,Q^2)T_A(b)}}{(1-i\eta) \sigma_{\rm soft}^j(x,Q^2)}
\right|^2 \,.
\label{eq:masterD_approx}
\end{equation}

In Eq.~(\ref{eq:masterD}), we neglected the possible dependence of 
$\sigma_{\rm soft}^j(x,Q^2)$ on $\beta$ (the dependence on the diffractive mass $M_X$).
Since the total probability of diffraction changes rather weakly 
as one varies the rescattering cross section, see e.g., Ref.~\cite{Frankfurt:2003wv},
this seems to be a reasonable first approximation.
At  the same time, in the region of small $\beta$ and small $x$ 
that corresponds to the 
triple Pomeron kinematics for the soft inelastic diffraction, 
one expects a suppression of diffraction
as compared to the 
color fluctuation approximation used in 
Eq.~(\ref{eq:masterD}). 
Indeed, Eq.~(\ref{eq:masterD}) evaluated at 
$Q^2=Q_0^2=4$ GeV$^2$ 
essentially corresponds to
treating diffraction as a superposition of  elastic scattering of different 
components of the virtual photon wave function. 
This is a reasonable approximation 
for the configurations
 with the masses comparable to $Q^2$. 
In 
the $\beta \ll 1$ limit (which corresponds to $M_X^2 \gg Q^2$),
 one approaches the limit analogous
to the soft triple Pomeron limit, in which case  diffraction
 off nuclei is strongly
 suppressed 
compared to 
elastic scattering,
see, e.g., Refs.~\cite{Frankfurt:1991nx,Kaidalov:2003vg}.
Hence, we somewhat overestimate diffraction for small $\beta$ and 
relatively small $Q_0^2$ scales.
At larger $Q^2$, diffraction
at small $\beta$ is dominated by the QCD evolution from 
$\beta \geq 0.1$ at $Q_0^2$ and, hence,
the  accuracy of our approximation improves.
Thus, in our numerical studies, we neglect the effect of the potential 
small-$\beta$ suppression
that we just discussed.

One can immediately see from Eq.~(\ref{eq:masterD}) that  
the Regge factorization, i.e., the factorization of $f_{j/A}^{D(3)}(\beta,Q^2,x_{\Pomeron})$
into the product of the Pomeron flux factor $f_{\Pomeron}(x_{\Pomeron})$ and
the PDFs of the Pomeron $f_{j}(\beta,Q^2)$, see Eq.~(\ref{eq:data3}), 
is not valid for the
nuclear diffractive parton distributions,
even if it approximately holds for the nucleon case.
 At fixed $x_{\Pomeron}$, the right-hand side of
Eq.~(\ref{eq:masterD}) depends not only on $\beta$, but also on  
Bjorken $x$ since the screening factor 
is given by the exponential factor containing
$\sigma_{\rm soft}^j(x,Q^2)$ which 
is a function of $x$. In addition, the right-hand side of
Eq.~(\ref{eq:masterD}) depends on the atomic mass 
number $A$ since the effect of  nuclear
shadowing increases with increasing $A$. 
The  breakdown of the Regge factorization approximation 
is a result of the  increase of nuclear shadowing both
 with the increase of the incident energy (decrease of $x$) and the atomic number $A$. 
This precludes the possibility of the scenario 
offered in Ref.~\cite{Arneodo:1996qa},
where coherent diffraction in DIS on the nucleon and nuclear targets
is provided by the same universal diffractive PDFs---a ''universal Pomeron''.

We demonstrated in Sec.~\ref{sec:phen} that the interaction with $N \geq 3$  nucleons
of the nuclear target provides a very small correction to 
nuclear shadowing 
for $x\ge 10^{-2}$.
Hence, a comparison of the diffractive cross section and the shadowing correction to the inclusive cross section
in this kinematics would provide a very stringent test of the theory 
by testing the relation: 
\begin{equation}
-\delta F_{2A}(x,Q^2)=-F_{2A}^{(b)}(x,Q^2)
= {1-\eta^2\over 1+\eta^2}  \int dx_{\Pomeron} F_{2A}^{D(3)}(x,Q^2,x_{\Pomeron}) \,.
\end{equation}
where $F_{2A}^{(b)}(x,Q^2)$ is the contribution of graph $b$ in 
Fig.~\ref{fig:Master1} to the shadowing correction
to $F_{2A}(x,Q^2)$.
Note here that the recoil effects, which are important for such $x$, do not affect  this relation.

In the collider mode,
it is rather straightforward to measure coherent diffraction by 
selecting
events with the rapidity gap and 
requiring
 that no neutrons are produced in the zero angle calorimeter (ZDC). 
Practically all events satisfying these requirements would
correspond to coherent diffraction. However,  measurements of the 
$t$ dependence would require the use of 
Roman pots at unrealistically small distances from the beam. The only exception is exclusive channels where one can measure the total transverse momentum of the produced system. In this case, coherent scattering 
can be selected using the distinctly sharp
$t$ dependence in the forward direction (forward diffractive peak), which originates
from the 
square of the nuclear form factor $F_A(t)$.

\subsubsection{Incoherent diffraction}

The coherent diffractive scattering dominates the $\gamma A \to X A^{\prime}$ diffraction
at small momentum transfers $t$ close to $t_{\rm min} \approx-x^2 m_N^2 (1+M_X^2/Q^2)$. 
For $|t| > |t_{\rm min}|$, the probability for the nucleus to stay intact rapidly decreases and
the diffraction is dominated by the incoherent final states $A^{\prime} \neq A$.
The case important for practical applications is when one sums over all products of the 
nuclear disintegration and uses the completeness of the final states 
$|A^{\prime} \rangle$.
The corresponding cross section reads, see,
e.g., \cite{Bauer:1977iq}:
\begin{eqnarray}
\sigma_{\gamma^{\ast} A \to X A^{\prime}}&=&\int d^2 \vec{b} 
\sum_{A^{\prime} \neq A} \langle A
|\Gamma_{\gamma^{\ast} A \to X A}^{\dagger}(b,r_i)|A^{\prime}\rangle 
\langle A^{\prime}|\Gamma_{\gamma^{\ast} A \to X A}(b,r_i)|A\rangle \nonumber\\
&=&\int d^2 \vec{b} \left[\langle A
||\Gamma_{\gamma^{\ast} A \to X A}(b,r_i)|^2|A\rangle- 
|\langle A|\Gamma_{\gamma^{\ast} A \to X A}(b,r_i)|A\rangle|^2\right]
\,,
\label{eq:diff_pdf_0}
\end{eqnarray}
where $\Gamma_{\gamma^{\ast} A \to X A}(b,r_i)$ is the $\gamma^{\ast} A \to X A$
scattering amplitude in the impact parameter space which also depends on the positions
of the involved nucleons,
\begin{equation}
\Gamma_{\gamma^{\ast} A \to X A}(b,r_i)=\sum_i\Gamma_{\gamma^{\ast}X}(\vec{b}-\vec{r}_{i \perp}) e^{i z_i \Delta_{\gamma^{\ast}X}}
\prod_{j\neq i} \left(1-\Theta(z_j-z_i) \Gamma_X(\vec{b}-\vec{r}_{j \perp})\right)
\,.
\label{eq:diff_pdf2}
\end{equation}
The integration over the positions of the nucleons with 
the weight given by 
the wave function squared 
of the nuclear ground state is denoted by $\langle A|\dots|A \rangle$ in 
Eq.~(\ref{eq:diff_pdf_0}).

Using Eq.~(\ref{eq:diff_pdf2}), we obtain for 
$|\Gamma_{\gamma^{\ast} A \to X A}(b,r_i)|^2$:
\begin{eqnarray}
&&|\Gamma_{\gamma^{\ast} A \to X A}(b,r_i)|^2=\sum_i |\Gamma_{\gamma^{\ast}X}(\vec{b}-\vec{r}_{i \perp})|^2
\prod_{j\neq i} \left|\left(1-\Theta(z_j-z_i) \Gamma_X(\vec{b}-\vec{r}_{j \perp})\right)\right|^2
\nonumber\\
&+&\sum_{i\neq i^{\prime}}
\Gamma^{\ast}_{\gamma^{\ast}X}(\vec{b}-\vec{r}_{i^{\prime} \perp})
\Gamma_{\gamma^{\ast}X}(\vec{b}-\vec{r}_{i \perp})
e^{i \Delta_{\gamma^{\ast}X}(z_i-z_{i^{\prime}})} \nonumber\\
&\times&
\prod_{j^{\prime}\neq i^{\prime}} \left(1-\Theta(z_{j^{\prime}}-z_{i^{\prime}}) \Gamma_X^{\ast}(\vec{b}-\vec{r}_{j^{\prime} \perp})\right) 
\prod_{j\neq i} \left(1-\Theta(z_j-z_i) \Gamma_X(\vec{b}-\vec{r}_{j \perp})\right)
\,.
\label{eq:diff_pdf3}
\end{eqnarray}
The first term in Eq.~(\ref{eq:diff_pdf3}), which we shall denote
$|\Gamma_{\gamma^{\ast} A \to X A}^{(1)}(b,r_i)|^2$ for brevity,
 corresponds to the incoherent contribution, 
which scales as $A$ and arises from the interaction 
of the probe
with the same nucleon of the target.
The second term in Eq.~(\ref{eq:diff_pdf3}), which we shall refer to as
$|\Gamma_{\gamma^{\ast} A \to X A}^{(2)}(b,r_i)|^2$,
 is similar to the coherent contribution since
it scales as the number of the nucleon pairs, $A(A-1)$, and arises from the 
interference diagrams when the external electromagnetic probe couples to different 
nucleons.

Let us evaluate the contribution of the first term in Eq.~(\ref{eq:diff_pdf3}). 
In this calculation, from the outset, one cannot neglect the slope of the $\gamma^{\ast}N \to X N$ 
amplitude, $B_{\rm diff}$. 
Therefore, instead of approximate
Eq.~(\ref{eq:m5}), we shall use the exact expression:
\begin{equation}
\Gamma_{\gamma^{\ast}X}(\vec{b}-\vec{r}_{i \perp})=
\frac{(1-i\eta)}{4 \pi B_{\rm diff}}\sqrt{\frac{16 \pi \frac{d \sigma_{\gamma^{\ast}N \to X N}}{dt}(t_{\rm min})}{1+\eta^2}} e^{-(\vec{b}-\vec{r}_{i \perp})^2/(2 B_{\rm diff})} \,.
\label{eq:diff_pdf4}
\end{equation}
Equation~(\ref{eq:diff_pdf4}) is a standard expression for the scattering
amplitude (profile function) in the Glauber formalism.
Integrating with the nuclear density, we obtain:
\begin{eqnarray}
&&\int d^2 \vec{r}_{i \perp} \rho_A(\vec{r}_{i\perp},z_i)|\Gamma_{\gamma^{\ast}X}(\vec{b}-\vec{r}_{i \perp})|^2 
\nonumber\\
&&=\frac{1}{\pi B_{\rm diff}^2}\frac{d \sigma_{\gamma^{\ast}N \to X N}}{dt}(t_{\rm min})
 \int d^2 \vec{r}_{i \perp} \rho_A(\vec{r}_{i \perp},z_i) 
e^{-(\vec{b}-\vec{r}_{i \perp})^2/B_{\rm diff}} \nonumber\\
&\approx& 
\frac{1}{B_{\rm diff}}\frac{d \sigma_{\gamma^{\ast}N \to X N}}{dt}(t_{\rm min})
 \rho_A(b,z_i)=\sigma_{\gamma^{\ast}N \to X N}\rho_A(b,z_i)
 \,.
\label{eq:diff_pdf5}
\end{eqnarray}  

For the profile function (scattering amplitude) corresponding to the rescattering of state $X$ on the remaining 
$A-1$ nucleons of the target, we use the form similar to Eq.~(\ref{eq:diff_pdf4}):
\begin{equation}
\Gamma_{X}(\vec{b}-\vec{r}_{j \perp})=\frac{\sigma_{XN \to XN}(1-i\eta)}{4 \pi B_{X}} e^{-(\vec{b}-\vec{r}_{i \perp})^2/(2 B_{X})} \,,
\label{eq:diff_pdf7}
\end{equation}
where $\sigma_{XN \to XN}$ is the elastic $XN$ cross section and $B_X$ is its slope.
Their numerical values will be defined later on.
Working along the lines of the derivation in Eq.~(\ref{eq:diff_pdf5}), we obtain
\begin{equation}
\int d^2 \vec{r}_{j \perp} dz_j \rho_A(\vec{r}_{j \perp},z_j) 
\left|\left(1-\Theta(z_j-z_i) \Gamma_X(\vec{b}-\vec{r}_{j \perp})\right)\right|^2
=1-\int_{z_i}^{\infty} dz_j \rho_A(b,z_j) \sigma_X^{\rm inel} \,,
\label{eq:diff_pdf8}
\end{equation}
where $\sigma_X^{\rm inel}$ is the $XN$ inelastic cross section,
\begin{equation}
\sigma_X^{\rm inel}=\sigma_X-\frac{\sigma_X^2(1+\eta^2)}{16 \pi B_X} \,.
\label{eq:diff_pdf9}
\end{equation}
Using Eqs.~(\ref{eq:diff_pdf5})-(\ref{eq:diff_pdf9}) and integrating over $z_i$ 
and the impact parameter $b$,
we obtain the following compact expression for the 
contribution 
of 
$|\Gamma_{\gamma^{\ast} A \to X A}^{(1)}(b,r_i)|^2$
to
$\sigma_{\gamma^{\ast} A \to X A^{\prime}}$:
\begin{eqnarray}
&&\int d^2 \vec{b} \langle A
|\sum_i |\Gamma_{\gamma^{\ast}X}(\vec{b}-\vec{r}_{i \perp})|^2
\prod_{j\neq i} \left|\left(1-\Theta(z_j-z_i) \Gamma_X(\vec{b}-\vec{r}_{j \perp})\right)\right|^2|A\rangle\nonumber\\
&=&A \sigma_{\gamma^{\ast}N \to X N}\int d^2 b\, dz\, \rho_A(b,z) e^{-A \sigma_X^{\rm inel}
\int_z^{\infty} dz^{\prime} \rho_A(b,z^{\prime})} \,.
\label{eq:diff_pdf10}
\end{eqnarray}

Now we turn to the evaluation of the second term in Eq.~(\ref{eq:diff_pdf3}),
$|\Gamma_{\gamma^{\ast} A \to X A}^{(2)}(b,r_i)|^2$.
Assuming that the nucleons in the nuclear target are independent, it can be written in the following explicit form:
\begin{eqnarray}
&&|\Gamma_{\gamma^{\ast} A \to X A}^{(2)}(b,r_i)|^2 \equiv \sum_{i\neq i^{\prime}}
\Gamma^{\ast}_{\gamma^{\ast}X}(\vec{b}-\vec{r}_{i^{\prime} \perp})
\Gamma_{\gamma^{\ast}X}(\vec{b}-\vec{r}_{i \perp})
e^{i \Delta_{\gamma^{\ast}X}(z_i-z_{i^{\prime}})} \nonumber\\
&\times&
\prod_{j^{\prime}\neq i^{\prime}} \left(1-\Theta(z_{j^{\prime}}-z_{i^{\prime}}) \Gamma_X^{\ast}(\vec{b}-\vec{r}_{j^{\prime} \perp})\right) 
\prod_{j\neq i} \left(1-\Theta(z_j-z_i) \Gamma_X(\vec{b}-\vec{r}_{j \perp})\right)
\nonumber\\
&=&A(A-1)\Gamma^{\ast}_{\gamma^{\ast}X}(\vec{b}-\vec{r}_{2 \perp})
\Gamma_{\gamma^{\ast}X}(\vec{b}-\vec{r}_{1 \perp})e^{i \Delta_{\gamma^{\ast}X}(z_1-z_2)}
\nonumber\\
&\times& \left(1-\Theta(z_2-z_1) \Gamma_X(b-\vec{r}_{2 \perp})-\Theta(z_1-z_2) \Gamma_X^{\ast}(b-\vec{r}_{1 \perp})\right) \nonumber\\
&\times& \prod_{j \neq 1,2}\left(1-\Theta(z_j-z_1) \Gamma_X(b-\vec{r}_{j \perp})-\Theta(z_j-z_2) \Gamma_X^{\ast}(b-\vec{r}_{j \perp})\right)
\,.
\label{eq:diff_pdf11}
\end{eqnarray}
The integration over the transverse positions of the nucleons weighted with the nuclear density can be carried out using the explicit expressions for $\Gamma_{\gamma^{\ast}X}$ and $\Gamma_X$:
\begin{samepage}
\begin{eqnarray}
&&\prod_{i=1}^A\int d^2 \vec{r}_{i \perp} \rho_A(\vec{r}_{i \perp} ,z_i)|\Gamma_{\gamma^{\ast} A \to X A}^{(2)}(b,r_i)|^2 
=A(A-1) 4 \pi  \frac{d\sigma_{\gamma^{\ast}N \to XN}}{dt}(t_{\rm min})
\rho_A(b,z_1)\rho_A(b,z_2) \nonumber\\
&\times& e^{i \Delta_{\gamma^{\ast}X}(z_1-z_2)}
\left(1-\Theta(z_2-z_1)\frac{\sigma_X (1-i \eta)}{8 \pi B_X}
-\Theta(z_1-z_2) \frac{\sigma_X (1+i \eta)}{8 \pi B_X}
\right) \nonumber\\
&\times& \prod_{j \neq 1,2}\Big(\int d^2 \vec{r}_{j \perp} \rho_A(\vec{r}_{j \perp},z_j)
-\Theta(z_j-z_1)\frac{\sigma_X (1-i \eta)}{2} \rho_A(b,z_j)
\nonumber\\
&-&\Theta(z_j-z_2) \frac{\sigma_X (1+i \eta)}{2}\rho_A(b,z_j)
\Big)
\,.
\label{eq:diff_pdf12}
\end{eqnarray}
\end{samepage}
The expression in Eq.~(\ref{eq:diff_pdf12}) can be significantly simplified by 
neglecting the contribution of the real part of the $XN \to XN$ 
amplitude
in the second line 
of Eq.~(\ref{eq:diff_pdf12}):
\begin{equation}
\Theta(z_2-z_1)\frac{\sigma_X (1-i \eta)}{8 \pi B_X}+
\Theta(z_1-z_2) \frac{\sigma_X (1+i \eta)}{8 \pi B_X} \approx \frac{\sigma_X}{8 \pi B_X}
\,.
\label{eq:diff_approx}
\end{equation}
Integrating over the longitudinal coordinates of all interacting nucleons
and the impact parameter $b$, we obtain
\begin{eqnarray}
&&\int d^2 \vec{b} \,\langle A||\Gamma_{\gamma^{\ast} A \to X A}^{(2)}(b,r_i)|^2 |A\rangle
=4 \pi A (A-1) \frac{d\sigma_{\gamma^{\ast}N \to XN}}{dt}(t_{\rm min})
 \left(1-\frac{\sigma_X}{8 \pi B_X}\right) \nonumber\\
&\times& \int d^2 \vec{b} \left| \int dz \rho_A(b,z) e^{i \Delta_{\gamma^{\ast}X}z}
e^{-\frac{A}{2}(1-i\eta)\sigma_X \int_{z}^{\infty} dz^{\prime} \rho_A(b,z^{\prime})} \right|^2 \,.
\label{eq:diff_pdf13}
\end{eqnarray}
This completes the calculation of the first term, $\langle A||\Gamma_{\gamma^{\ast} A \to X A}(b,r_i)|^2|A\rangle$, in Eq.~(\ref{eq:diff_pdf_0}).
The second term in Eq.~(\ref{eq:diff_pdf_0}), $|\langle A|\Gamma_{\gamma^{\ast} A \to X A}(b,r_i)|A\rangle|^2$, corresponds to the purely coherent diffraction and was already calculated in Sec.~\ref{subsubsec:coherent_diffraction}:
\begin{eqnarray}
&&\int d^2 \vec{b} \,|\langle A|\Gamma_{\gamma^{\ast} A \to X A}(b,r_i)|A\rangle|^2
=4 \pi A^2 \frac{d\sigma_{\gamma^{\ast}N \to XN}}{dt}(t_{\rm min})
 \nonumber\\
&\times& \int d^2 \vec{b} \left| \int dz \rho_A(b,z) e^{i \Delta_{\gamma^{\ast}X} z}
e^{-\frac{A}{2}(1-i\eta)\sigma_X \int_{z}^{\infty} dz^{\prime} \rho_A(b,z^{\prime})} \right|^2 \,.
\label{eq:diff_pdf14}
\end{eqnarray}
Therefore, the $\int d^2 \vec{b} \langle A||\Gamma_{\gamma^{\ast} A \to X A}^{(2)}(b,r_i)|^2 |A\rangle$ and $\int d^2 \vec{b} |\langle A||\Gamma_{\gamma^{\ast} A \to X A}(b,r_i) |A\rangle|^2$ terms 
partially cancel each other.
Adding together Eqs.~(\ref{eq:diff_pdf10}), (\ref{eq:diff_pdf13}), and
(\ref{eq:diff_pdf14}), 
we obtain
our final expression for the 
cross section of incoherent diffraction 
$\sigma_{\gamma^{\ast}A \to X A^{\prime}}$:
\begin{eqnarray} 
\sigma_{\gamma^{\ast}A \to X A^{\prime}}&=&A \sigma_{\gamma^{\ast}N \to X N}
\int d^2 b \,dz \rho_A(b,z) e^{-A \sigma_X^{\rm inel}
\int_z^{\infty} dz^{\prime} \rho_A(b,z^{\prime})}
\nonumber\\
&-&4 \pi A\left[A\frac{\sigma_X}{8 \pi B_X}+\left(1-\frac{\sigma_X}{8 \pi B_X}\right) \right] 
\frac{d\sigma_{\gamma^{\ast}N \to X N}}{dt}(t_{\rm min})
 \nonumber\\
&\times& \int d^2 \vec{b} \left| \int dz \rho_A(b,z) e^{i \Delta_{\gamma^{\ast}X} z}
e^{-\frac{A}{2}(1-i\eta)\sigma_X \int_{z}^{\infty} dz^{\prime} \rho_A(b,z^{\prime})} \right|^2 \,.
\label{eq:diff_pdfs_final_0}
\end{eqnarray}
Expressing the diffractive cross sections in Eq.~(\ref{eq:diff_pdfs_final_0}) in terms
of the corresponding diffractive structure functions, see Eqs.~(\ref{eq:ft1_A}) and
(\ref{eq:data1_A}), we obtain the expression for the {\it incoherent}
nuclear diffractive structure function $F_{2A, {\rm incoh}}^{D(3)}$:
\begin{eqnarray} 
F_{2A, {\rm incoh}}^{D(3)}(\beta,Q^2,x_{\Pomeron})&=&AF_{2N}^{D(3)}(\beta,Q^2,x_{\Pomeron}) \int d^2 b \,dz \rho_A(b,z) e^{-A \sigma_X^{\rm inel}
\int_z^{\infty} dz^{\prime} \rho_A(b,z^{\prime})}
\nonumber\\
&-&4 \pi A\left[A\frac{\sigma_X}{8 \pi B_X}+\left(1-\frac{\sigma_X}{8 \pi B_X}\right) \right] B_{\rm diff} F_{2N}^{D(3)}(\beta, Q^2, x_{\Pomeron}) \nonumber\\
&\times& \int d^2 \vec{b} \left| \int dz \rho_A(b,z) e^{ix_{\Pomeron} m_N z}
e^{-\frac{A}{2}(1-i\eta)\sigma_X \int_{z}^{\infty} dz^{\prime} \rho_A(b,z^{\prime})} \right|^2 \,.
\label{eq:diff_pdfs_final}
\end{eqnarray}

At this point,
it is appropriate to discuss the parameters $\sigma_X$ and $B_X$.
As follows from our derivation, the rescattering cross section 
$\sigma_X$ has the same meaning as the cross section $\sigma_{\rm soft}^j$ that we introduced and discussed in Sec.~\ref{subsubsec:color_fluct}. Therefore, in our numerical analysis,
we use $\sigma_X=\sigma_{\rm soft}^q$, where $\sigma_{\rm soft}^q$ corresponds to the ${\bar u}$-quarks.
For the slope parameter $B_X$, we used $B_X=7$ GeV$^{-2}$, which is consistent with the
value of the diffractive slope $B_{\rm diff}=6-7$ GeV$^2$~\cite{Aktas:2006hx,Chekanov:2008fh}; like $B_{\rm diff}$, $B_X$
is known with a significant uncertainty (of the order of 15\%).
As a result, we observe that 
$\sigma_X/(8 \pi B_X)$ 
is not a small
parameter, $\sigma_X/(8 \pi B_X) \sim {\cal O}(1)$ ,
and also that
$\sigma_X^{\rm inel} \sim {\cal O}(\sigma_X)$.

One should note that in our calculation of incoherent diffraction, we considered the absorption
due to the inelastic piece of the diffractive state $X$-nucleon interaction. 
However, other inelastic transitions---such as, e.g., $X p \to X^{\prime} p$, 
where $X^{\prime} \neq X$---can also in principle lead to rapidity gap events. 
While these transitions are very small at $t=0$ since they are mostly spin-flip,
they will still decrease the effective absorption cross section $\sigma_X$ and will
lead to an increase of $F_{2A, {\rm incoh}}^{D(3)}$.
We expect that the overall effect will be numerically small; it will be smaller than the theoretical
uncertainty of our predictions associated with the uncertainty of the value 
of $B_X$.

A direct application of the AGK cutting rules connects the shadowing (screening) 
correction to the total cross section with
the corresponding contributions to the diffractive final states, 
which involve both 
the final states where the nucleus remains intact (coherent diffraction) and where 
it breaks up and several nucleons are produced (incoherent diffraction). 
Since the incoherent contribution constitutes 
$15 \div 25$\%
for $^{40}$Ca and $3 \div 5$\% for $^{208}$Pb
 of the coherent 
contribution for a wide range of $\sigma_X$ 
(these values refer to the studied kinematics of $Q^2=4$ GeV$^2$ and
$x_{\Pomeron}=10^{-3}$ and $x_{\Pomeron}=10^{-2}$,
see Sec.~\ref{sec:incohdiff_num_res} and Table~\ref{table:F2diff}),
one can correct for these effects.
One can  also test the importance of 
the incoherent contribution by studying the multiplicity of neutron production in 
the nuclear break-up, see e.g., Ref.~\cite{Strikman:2005ze}.

It is worth noting that we calculated the incoherent cross section for the final state
where the nucleus decays into a collection of nuclear fragments without hadron
production. On the other hand, HERA experimental data indicate that in DIS, the
ratio of the double diffraction dissociation $e+ p \to e +X +Y $ to the 
single diffraction $e+ p \to e+X +p$, $r_{\rm double}$,  is rather large:
$r_{\rm double}$ is
 of the order $\sim 0.4 $. To arrive at this estimate, one
uses the factorization approximation for $t=0$, which corresponds to
$r_{\rm double} (t=0)\sim 0.2$, and the fact that the ratio of the $t$-slopes 
of the double and single diffraction dissociation is of
 the order of two. 
Hence, one expects that $r_{\rm double}$ for nuclei will be of the same magnitude. 
It would be a challenge to separate the incoherent contribution with and without hadron
production since the hadrons will be produced predominantly at the rapidities close to
the nucleus rapidity.

\subsubsection{Numerical predictions for coherent diffraction}
\label{sec:diff_num_res}

One way to quantify the effect of nuclear shadowing on the nuclear diffractive parton
distributions is to introduce the probability of diffraction for a given parton
flavor $j$, $P_{\rm diff}^j$~\cite{Frankfurt:2003gx,Guzey:2005ys}:
\begin{equation}
P_{\rm diff}^j=\frac{\int_{x}^{0.1} d x_{\Pomeron} \,\beta f_{j}^{D(3)}(\beta,Q^2,x_{\Pomeron})}{xf_j(x,Q^2)} \,.
\label{eq:d2}
\end{equation}
For 
hard process with a 
specific trigger, the probability of  diffraction 
may be close to $R_{\rm diff}^q$ (for the measurement of the diffractive structure function $F_{2N}^{D(3)}$) or to $R_{\rm diff}^g$ ($b$-quark production). 
Also, the probability of
diffraction can have an intermediate value between $R_{\rm diff}^q$ and $R_{\rm diff}^g$, for instance, for the $s$-quark production.

\begin{figure}[h]
\begin{center}
\epsfig{file=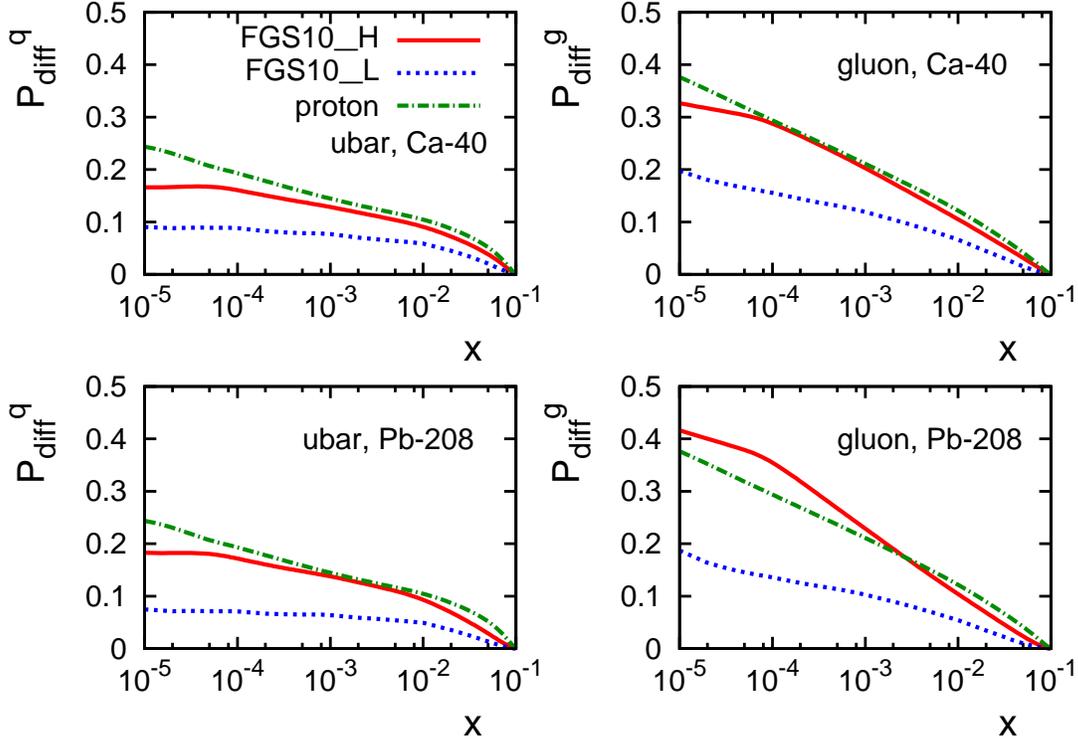,scale=1.4}
\vskip 0cm
\caption{The probability of diffraction, $P_{\rm diff}^j$ of Eq.~(\ref{eq:d2}), as a function of Bjorken $x$ at $Q^2=4$ GeV$^2$. 
The solid and dotted curves correspond to models 
FGS10\_H and FGS10\_L,
respectively.
For comparison, $P_{\rm diff}^j$ for the proton 
is given by the dot-dashed curves.
The left panels correspond to the ${\bar u}$-quark
channel; the right panels correspond to the gluon channel. The upper row of panels is
for $^{40}$Ca; the lower row is for $^{208}$Pb.}
\label{fig:Pdiff}
\end{center}
\end{figure}

Figure~\ref{fig:Pdiff} presents $P_{\rm diff}^j$ of Eq.~(\ref{eq:d2}) as 
a function of Bjorken $x$ at $Q^2=4$ GeV$^2$, where the nuclear diffractive PDF
$f_{j/A}^{D(3)}(\beta,Q^2,x_{\Pomeron})$ is calculated using Eq.~(\ref{eq:masterD}).
The two sets of curves correspond to models 
FGS10\_H and FGS10\_L.
Also, for comparison,
we present $P_{\rm diff}^j$ for the proton by the dot-dashed curves.
The two left panels correspond to the ${\bar u}$-quark
channel; the two right panels correspond to the gluon channel. The upper row of panels is
for $^{40}$Ca; the lower row is for $^{208}$Pb.

The results presented in Fig.~\ref{fig:Pdiff} merit a discussion. First,
the probability of diffraction in the gluon channel is larger (by approximately a
factor of two) than that in the quark channel. This is a direct consequence
of the very large gluon diffractive PDF of the nucleon. Second,
the probability of diffraction for nuclei is smaller than that for the free proton
for FGS10\_L and compatible with the free proton for FGS10\_H.
This is a consequence of the different values of the cross section
$\sigma_{\rm soft}^j(x,Q^2)$ in models FGS10\_L and FGS10\_H, which determines the magnitude of 
the multiple interactions in 
our treatment of coherent diffraction in DIS with nuclei in the color fluctuation
approximation.
Since $\sigma_{\rm soft}^j(x,Q^2)$ is smaller in model FGS10\_H than in model FGS10\_L,
the survival probability of the diffractive state $X$ 
[the numerator of $P_{\rm diff}^j$ in Eq.~(\ref{eq:d2})] is larger 
in the FGS10\_H case. 
In addition,
the smaller  $\sigma_{\rm soft}^j(x,Q^2)$ in model  FGS10\_H
leads to the smaller nuclear PDFs (larger nuclear shadowing), which further increases
$P_{\rm diff}^j$ in model  FGS10\_H compared to the FGS10\_L case.
As a result, $P_{\rm diff}^j$ is 
significantly 
larger in model FGS10\_H than in model FGS10\_L.

\begin{figure}[h]
\begin{center}
\epsfig{file=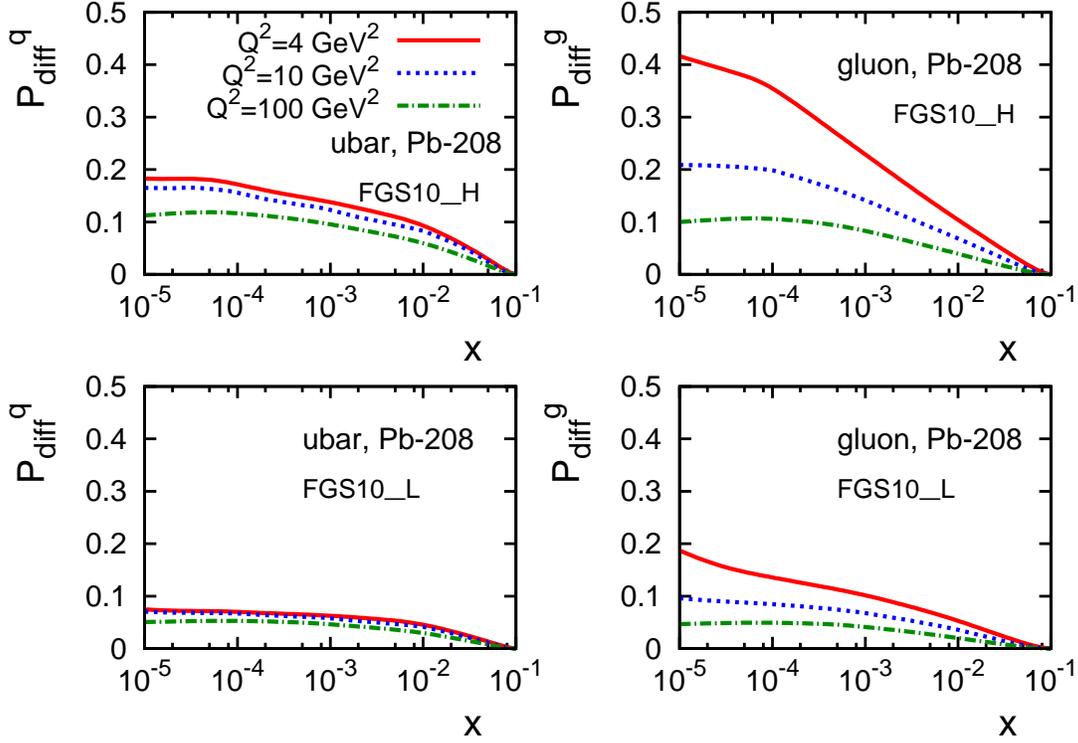,scale=1.4}
\vskip 0cm
\caption{The probability of diffraction, $P_{\rm diff}^j$ of Eq.~(\ref{eq:d2}), as a function of Bjorken $x$ for different values of  $Q^2$.
All curves correspond to $^{208}$Pb.}
\label{fig:Pdiff_Q2dep}
\end{center}
\end{figure}
Figure~\ref{fig:Pdiff_Q2dep} presents our predictions for the probability of diffraction
$P_{\rm diff}^j$ as a function of $x$ for different values of $Q^2$: 
$Q^2=4$, 10, and 100 GeV$^2$. One can see from the figure that the $Q^2$ dependence
of $P_{\rm diff}^j$ is faster in the gluon channel than in the quark one.
This trend can be understood by recalling that usual nuclear PDFs enter the denominator
of Eq.~(\ref{eq:d2}) and
noticing that the $Q^2$ evolution increases usual nuclear PDFs
faster in the gluon channel than in the quark one.

While the color fluctuation approximation is our main approach to the 
treatment of the multiple interactions, one can also 
evaluate $P_{\rm diff}^j$ in the quasi-eikonal approximation. 
The latter is done by replacing
$\sigma_{\rm soft}^j(x,Q^2)$ in Eq.~(\ref{eq:masterD}) by $\sigma_2^j(x,Q^2)$,
see Eq.~(\ref{eq:m17}) and also Fig.~\ref{fig:sigma3_2009}.
Since $\sigma_2^j(x,Q^2) < \sigma_{\rm soft}^j(x,Q^2)$, 
the absorption of the diffractive state $X$ is not as large as in the 
color fluctuation approximation
and, as a result, the probability of diffraction should be significantly larger.

\begin{figure}[h]
\begin{center}
\epsfig{file=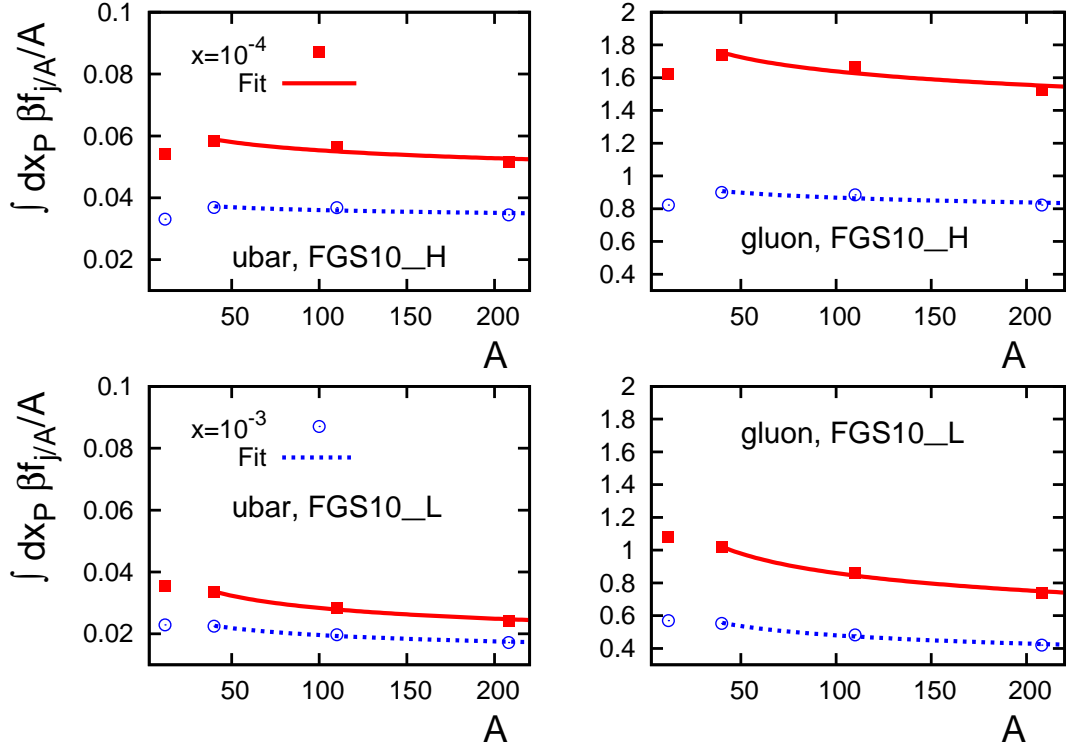,scale=1.4}
\vskip 0cm
\caption{The $A$ dependence of nuclear diffractive PDFs.
The points (squares and open circles) are the results of our calculations
of  $\int_{x}^{0.1} d x_{\Pomeron} \beta f_{j}^{D(3)}(\beta,Q^2,x_{\Pomeron})/A$ 
for $^{12}$C, $^{40}$Ca, $^{110}$Pd, and 
$^{208}$Pb;
the smooth curves is a two-parameter fit of Eq.~(\ref{eq:Diffraction_Adep_fit}).
}
\label{fig:Diffraction_Adep}
\end{center}
\end{figure}

The trend of the $A$ dependence of the probability of diffraction $P_{\rm diff}^j$ is rather non-trivial since
it comes from different $A$ dependences of the numerator and denominator in Eq.~(\ref{eq:d2}).
To disentangle the two and also to better understand the role of nuclear shadowing
in nuclear diffractive PDFs, it is useful to study the $A$ dependence of 
the numerator of Eq.~(\ref{eq:d2}). An example of this is presented in Fig.~\ref{fig:Diffraction_Adep},
where we plot $\int_{x}^{0.1} d x_{\Pomeron} \beta f_{j}^{D(3)}(\beta,Q^2,x_{\Pomeron})/A$
as a function of $A$ at $Q^2=Q_0^2=4$ GeV$^2$.
In the figure, the points (squares for $x=10^{-4}$ and open circles
for $x=10^{-3}$) are the results of our explicit calculations for
$^{12}$C, $^{40}$Ca, $^{110}$Pd, and $^{208}$Pb;
the smooth curves is a two-parameter fit
to the $A \geq 40$ points in the following form:
\begin{equation}
\frac{1}{A}\int_x^{0.1} dx_{\Pomeron} \beta f_{j}^{D(3)}(\beta,Q^2,x_{\Pomeron})=a_1 A^{1/3(1-a_2)} \,,
\label{eq:Diffraction_Adep_fit}
\end{equation}
where $a_1$ and $a_2$ are the free parameters of the fit. The resulting values of 
$a_1$ and $a_2$ are summarized in Table~\ref{table:Diffraction_Adep}.

\begin{table}[h]
\begin{tabular}{|c|c|c||c|c|}
\hline
& $a_{1,\rm ubar}$ & $a_{2,\rm ubar}$ & $a_{1,\rm gluon}$ & $a_{2,\rm gluon}$ \\
\hline
$10^{-4}$   &                   &                   &  &\\
FGS10\_H    & 0.0758 & 1.205 & 2.307 & 1.223 \\ 
FGS10\_L    & 0.0675 & 1.565 & 2.048 & 1.566 \\
\hline 
$10^{-3}$   &                   &                   & &\\
FGS10\_H    & 0.0426 & 1.109 & 1.090 & 1.149 \\ 
FGS10\_L    & 0.0404 & 1.471 & 1.008 & 1.484 \\

\hline 
\end{tabular}
\caption{The parameters $a_1$ and $a_2$ of the fit of Eq.~(\ref{eq:Diffraction_Adep_fit}).}
\label{table:Diffraction_Adep}
\end{table}

The parameterization in Eq.~(\ref{eq:Diffraction_Adep_fit}) interpolates 
between the two limiting regimes:\\
 $\int_x^{0.1} dx_{\Pomeron} \beta f_{j}^{D(3)}/A \propto A^{1/3}$ corresponding to the (unshadowed) impulse approximation and $\int_x^{0.1} dx_{\Pomeron} \beta f_{j}^{D(3)}/A \propto A^{-1/3}$ corresponding to full-fledged nuclear
shadowing.
As one can see from Table~\ref{table:Diffraction_Adep} and also from Fig.~\ref{fig:Diffraction_Adep},
$\int_x^{0.1} dx_{\Pomeron} \beta f_{j}^{D(3)}(\beta,Q^2,x_{\Pomeron})/A$ 
decreases as $A$ is increased. 
This is consistent with the expectation of the onset of the full-fledged nuclear shadowing that
reduces the $A$ dependence of the $t$-integrated diffractive parton distributions 
(structure functions, cross sections) from $A^{4/3}$ (impulse approximation) to $A^{2/3}$.

Note that apparently $^{12}$C is too light for this trend to present; we did not include the 
$A=12$ point in the fit of Eq.~(\ref{eq:Diffraction_Adep_fit}).

Having addressed the $A$ dependence of the nuclear diffractive PDFs, 
we can now better understand the $A$ dependence 
and the absolute value of the probability of diffraction $P_{\rm diff}^j$.
Since both $\int_x^{0.1} dx_{\Pomeron} \beta f_{j}^{D(3)}(\beta,Q^2,x_{\Pomeron})/A$
and $f_{j/A}(x,Q^2)/A$ are rather flat functions of $A$, see Eqs.~(\ref{eq:Shadowing_Adep})
and (\ref{eq:Diffraction_Adep_fit}), $P_{\rm diff}^j$ very weakly depends on $A$ for
$A \geq 40$ (see Fig.~\ref{fig:Pdiff}).
As to the absolute value of $P_{\rm diff}^j$, nuclei do not seem to enhance the probability
of diffraction compared to the free proton case. This is a result of the strong leading 
twist nuclear shadowing that significantly suppresses nuclear diffractive PDFs
and slows down the onset of the black disk limit, where $P_{\rm diff}^j$
is supposed to approach 1/2.

\begin{figure}[t]
\begin{center}
\epsfig{file=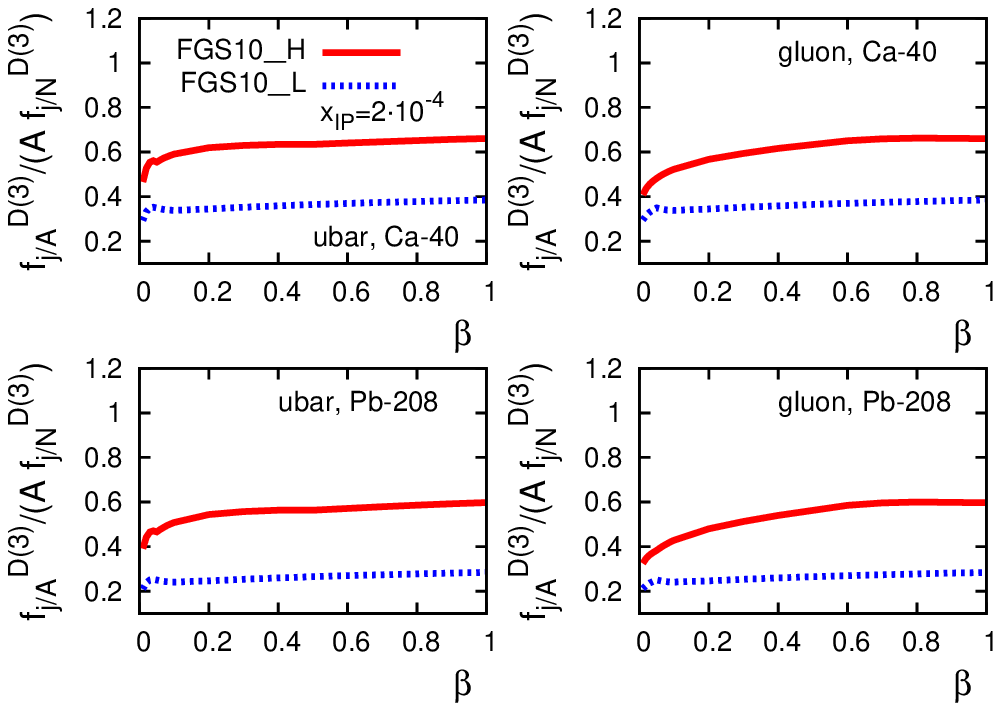,scale=1.25}
\epsfig{file=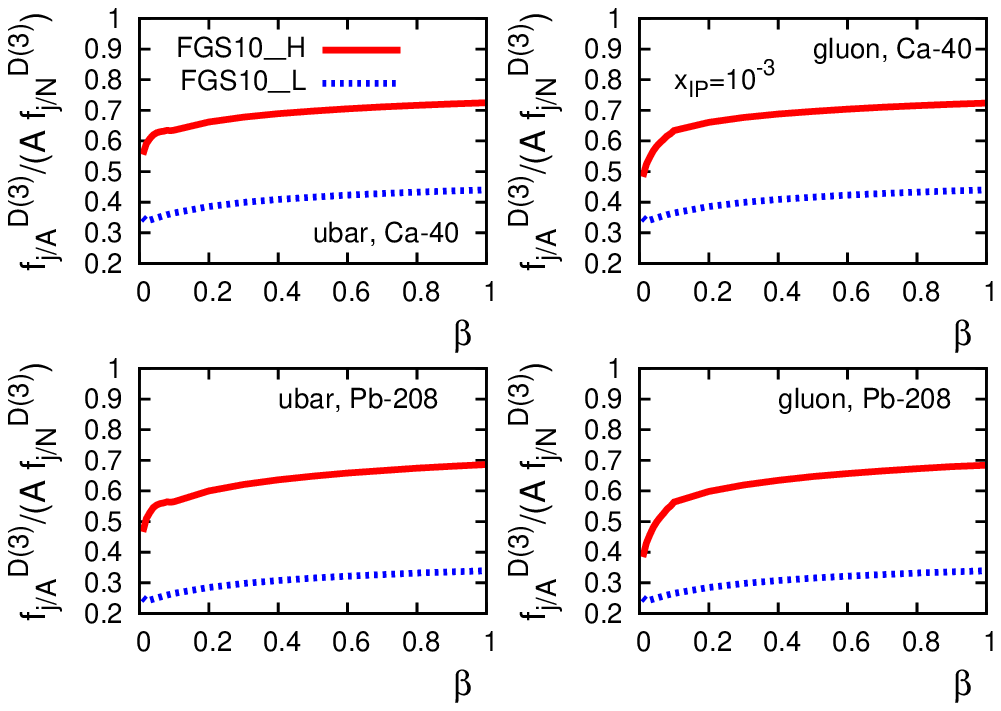,scale=1.25}
\vskip 0cm
\caption{The $f_{j/A}^{D(3)}(\beta,Q^2,x_{\Pomeron})/[Af_{j/N}^{D(3)}(\beta,Q^2,x_{\Pomeron})]$ ratio as a function of $\beta$ at $x_{\Pomeron}=2 \times 10^{-4}$ (four upper panels) and $x_{\Pomeron}=10^{-3}$ 
(four lower  panels) at $Q^2=4$ GeV$^2$ and for  $^{40}$Ca and $^{208}$Pb.
The left column of panels corresponds to the ${\bar u}$-quark density;  the right column
corresponds to the gluon density. 
The solid and dashed curves correspond to models FGS10\_H and FGS10\_L, respectively.
}
\label{fig:diffraction_beta}
\end{center}
\end{figure}

\begin{figure}[t]
\begin{center}
\epsfig{file=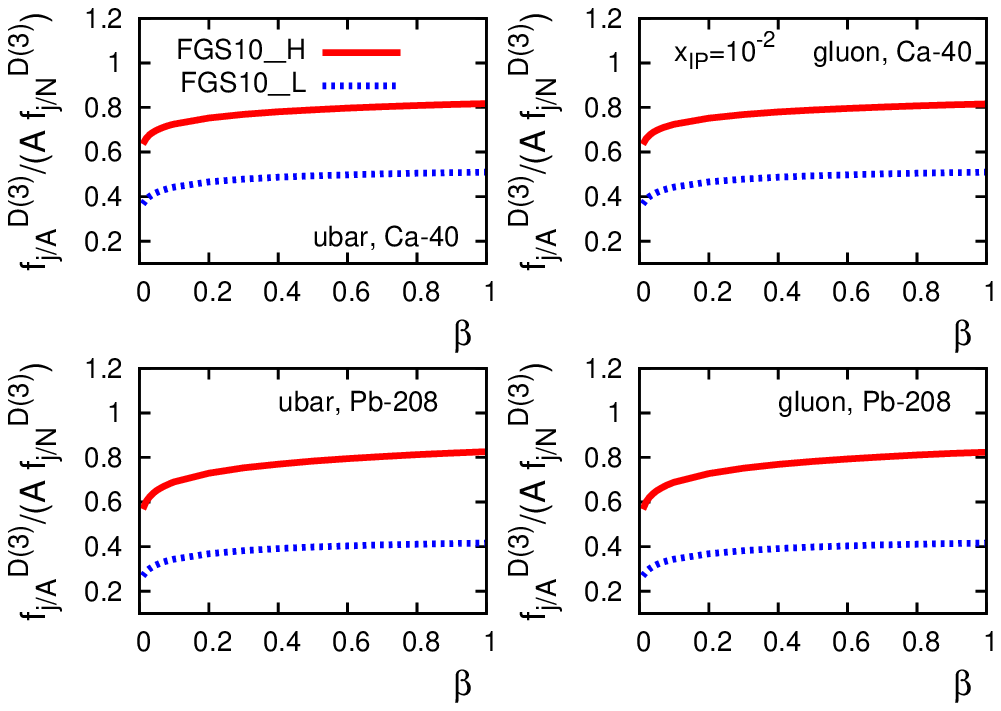,scale=1.25}
\epsfig{file=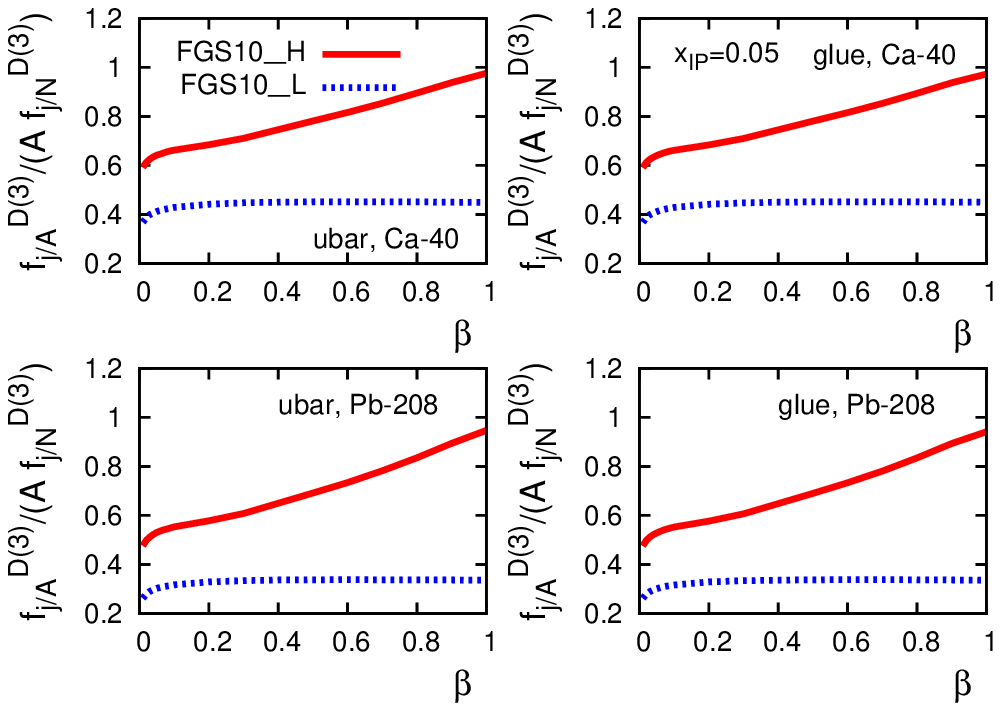,scale=1.25}
\vskip 0cm
\caption{The $f_{j/A}^{D(3)}(\beta,Q^2,x_{\Pomeron})/[Af_{j/N}^{D(3)}(\beta,Q^2,x_{\Pomeron})]$ ratio as a function of $\beta$ at $x_{\Pomeron}=10^{-2}$ (four upper panels) and $x_{\Pomeron}=0.05$ 
(four lower  panels) at $Q^2=4$ GeV$^2$ and for  $^{40}$Ca and $^{208}$Pb.
For the rest of the legend, see Fig.~\ref{fig:diffraction_beta}.
}
\label{fig:diffraction_beta_2}
\end{center}
\end{figure}

Next we present our predictions for the nuclear diffractive PDFs $f_{j/A}^{D(3)}$.
To this end, it is convenient to present our results in terms of the ratio of the nuclear to 
free proton diffractive PDFs $f_{j/A}^{D(3)}/(Af_{j/N}^{D(3)})$.
Figures~\ref{fig:diffraction_beta} 
and \ref{fig:diffraction_beta_2} 
 show the $f_{j/A}^{D(3)}/(Af_{j/N}^{D(3)})$ ratio 
for $^{40}$Ca and $^{208}$Pb at $Q^2=4$ GeV$^2$
as a function 
of the light-cone fraction $\beta$ at fixed $x_{\Pomeron}=2 \times 10^{-4}$ 
(four upper panels of Fig.~\ref{fig:diffraction_beta}), $x_{\Pomeron}=10^{-3}$
(four lower panels of Fig.~\ref{fig:diffraction_beta}),
$x_{\Pomeron}=10^{-2}$
(four upper panels of Fig.~\ref{fig:diffraction_beta_2}),
and $x_{\Pomeron}=0.05$ (four lower panels of Fig.~\ref{fig:diffraction_beta_2}). 
The solid and dotted curves correspond to models 
FGS10\_H and FGS10\_L,
respectively
(see the text). The left columns of the panels correspond to the ${\bar u}$-quark; the right columns
of the panels correspond to the gluons.

\begin{figure}[t]
\begin{center}
\epsfig{file=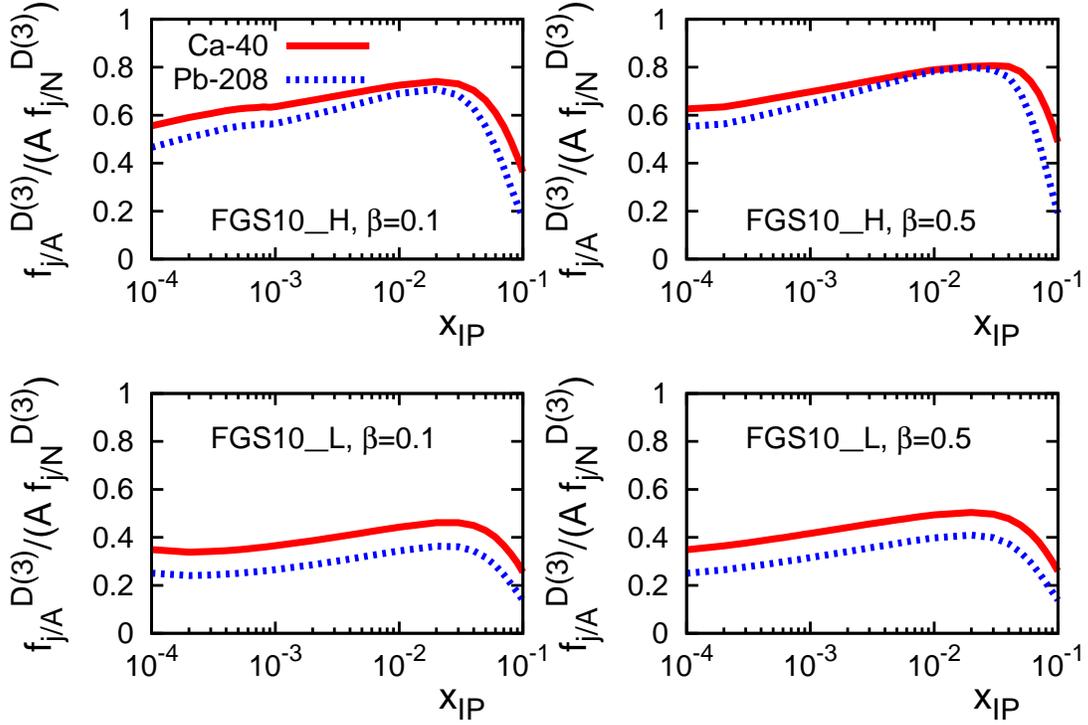,scale=1.4}
\vskip 0cm
\caption{The $f_{j/A}^{D(3)}(\beta,Q^2,x_{\Pomeron})/[Af_{j/N}^{D(3)}(\beta,Q^2,x_{\Pomeron})]$
ratio
 as a function of $x_{\Pomeron}$ at fixed $\beta=0.1$ and $\beta=0.5$ and at $Q^2=4$ GeV$^2$. 
All curves correspond
to the ${\bar u}$-quark parton distribution.}
\label{fig:diffraction_xpom}
\end{center}
\end{figure}

\begin{figure}[t]
\begin{center}
\epsfig{file=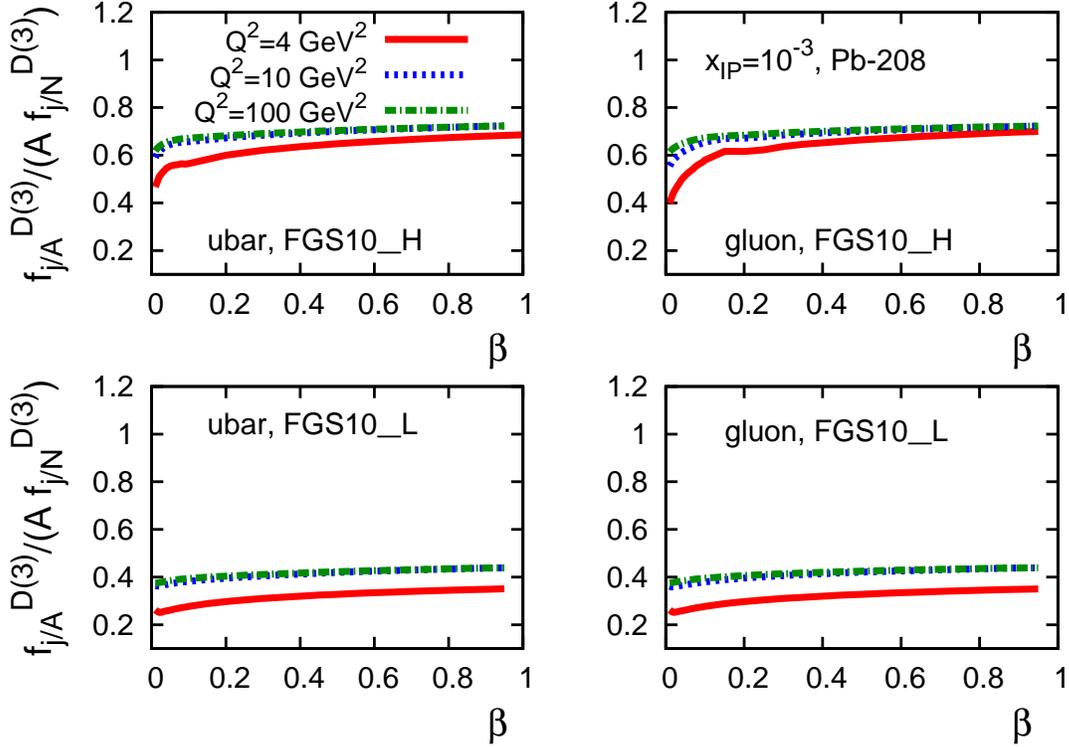,scale=1.4}
\vskip 0cm
\caption{The $Q^2$ dependence of the  $f_{j/A}^{D(3)}(\beta,Q^2,x_{\Pomeron})/[Af_{j/N}^{D(3)}(\beta,Q^2,x_{\Pomeron})]$ ratio
as a function of $\beta$ at fixed $x_{\Pomeron}=10^{-3}$ for $^{208}$Pb.}
\label{fig:diffraction_beta_Q2dep}
\end{center}
\end{figure}

Several features of Figs.~\ref{fig:diffraction_beta} and \ref{fig:diffraction_beta_2} 
deserve a discussion.
\begin{itemize}
\item[(i)] At small fixed $x_{\Pomeron}$,
the dependence of $f_{j/A}^{D(3)}/(Af_{j/N}^{D(3)})$ on $\beta$ is rather weak since it enters
only through the rescattering cross section $\sigma_{\rm soft}^j(x=x_{\Pomeron}\beta,Q_0^2)$, see Eq.~(\ref{eq:masterD}). 
\item[(ii)] By the same token, the dependence of $f_{j/A}^{D(3)}/(Af_{j/N}^{D(3)})$ on
$x_{\Pomeron}$ at fixed $\beta$ is also very weak (compare the upper panels with 
the respective lower panels). See also Fig.~\ref{fig:diffraction_xpom}.
\item[(iii)]
Since $\sigma_{\rm soft}^{j(\rm H)}$ is smaller than $\sigma_{\rm soft}^{j(\rm L)}$,
the nuclear diffractive PDFs are larger in model FGS10\_H (the solid curves lie
above the dotted ones). Note also that $f_{j/A}^{D(3)}/(Af_{j/N}^{D(3)})$ is flavor-independent
in model FGS10\_L (dotted curves) since the corresponding $\sigma_{\rm soft}^{j(\rm L)}$ is 
taken to be the same for all parton flavors.
\item[(iv)] The nuclear dependence of $f_{j/A}^{D(3)}/(Af_{j/N}^{D(3)})$
is rather weak, too. 
See also Fig.~\ref{fig:Diffraction_Adep} and our discussion above.
\item[(v)] Our analysis shows that the $Q^2$ dependence of $f_{j/A}^{D(3)}/(Af_{j/N}^{D(3)})$, 
which results from the separate DGLAP evolution of $f_{j/A}^{D(3)}$ and $f_{j/N}^{D(3)}$, is also
very insignificant. See Fig.~\ref{fig:diffraction_beta_Q2dep}.
\end{itemize}

\clearpage

\clearpage

As an illustration of the $x_{\Pomeron}$ dependence, in 
Fig.~\ref{fig:diffraction_xpom}, we present $f_{j/A}^{D(3)}/(Af_{j/N}^{D(3)})$ as a function of $x_{\Pomeron}$ at fixed $\beta=0.1$ and $\beta=0.5$ and $Q^2=4$ GeV$^2$. All curves correspond
to the ${\bar u}$-quark parton distribution. [The flavor dependence of $f_{j/A}^{D(3)}/(Af_{j/N}^{D(3)})$ is weak.]
At small $x_{\Pomeron}$, 
$x_{\Pomeron} \leq 0.05$,
$f_{j/A}^{D(3)}/(Af_{j/N}^{D(3)})$ very weakly depends on 
$x_{\Pomeron}$; 
the values of $f_{j/A}^{D(3)}/(Af_{j/N}^{D(3)})$ are naturally the same
as in Figs.~\ref{fig:diffraction_beta} and \ref{fig:diffraction_beta_2}.
 As one increases $x_{\Pomeron}$,
$f_{j/A}^{D(3)}/(Af_{j/N}^{D(3)})$ initially increases 
because $\sigma_{\rm soft}^j(x,Q^2)$ decreases, which
leads to a smaller suppression of the diffractively produced state $X$ by the multiple interactions
with the target nucleons. However, as $x_{\Pomeron}$ becomes larger than approximately
$0.02$, the coherence of
the nucleus is destroyed by the $e^{ix_{\Pomeron} m_N z}$ exponent 
in Eq.~(\ref{eq:masterD}) and
nuclear coherent diffraction rapidly vanishes. 

Next we 
present our predictions
for the $Q^2$ dependence of $f_{j/A}^{D(3)}/(Af_{j/N}^{D(3)})$.
Figure~\ref{fig:diffraction_beta_Q2dep} shows 
$f_{j/A}^{D(3)}/(Af_{j/N}^{D(3)})$ as a function of $\beta$ at small fixed value of
$x_{\Pomeron}=10^{-3}$ for $^{208}$Pb for three different values of $Q^2$: $Q^2=4$, 10 and 100 GeV$^2$. 
As one can see from the figure, the $Q^2$ dependence almost completely cancels in the ratio of the nuclear and nucleon diffractive PDFs.

A few words of explanation about our procedure is in order here.
At fixed $x_{\Pomeron}$, we calculated $f_{j/A}^{D(3)}(\beta,Q^2,x_{\Pomeron})$
as a function of $\beta$ at fixed $Q_0^2=4$ GeV$^2$ using Eq.~(\ref{eq:masterD}).
The result was used as an input for the DGLAP evolution equations~(\ref{eq:dglap}) to higher 
$Q^2$ scales, $Q^2=10$ and 100 GeV$^2$. The $Q^2$ evolution of the free proton diffractive
PDFs, $f_j^{D(3)}(\beta,Q^2,x_{\Pomeron})$, was performed separately. Having the nuclear and nucleon
diffractive PDFs at desired values of $Q^2$, we formed the $f_{j/A}^{D(3)}/(Af_{j/N}^{D(3)})$ ratio
presented in Fig.~\ref{fig:diffraction_beta_Q2dep}.

\begin{figure}[t]
\begin{center}
\epsfig{file=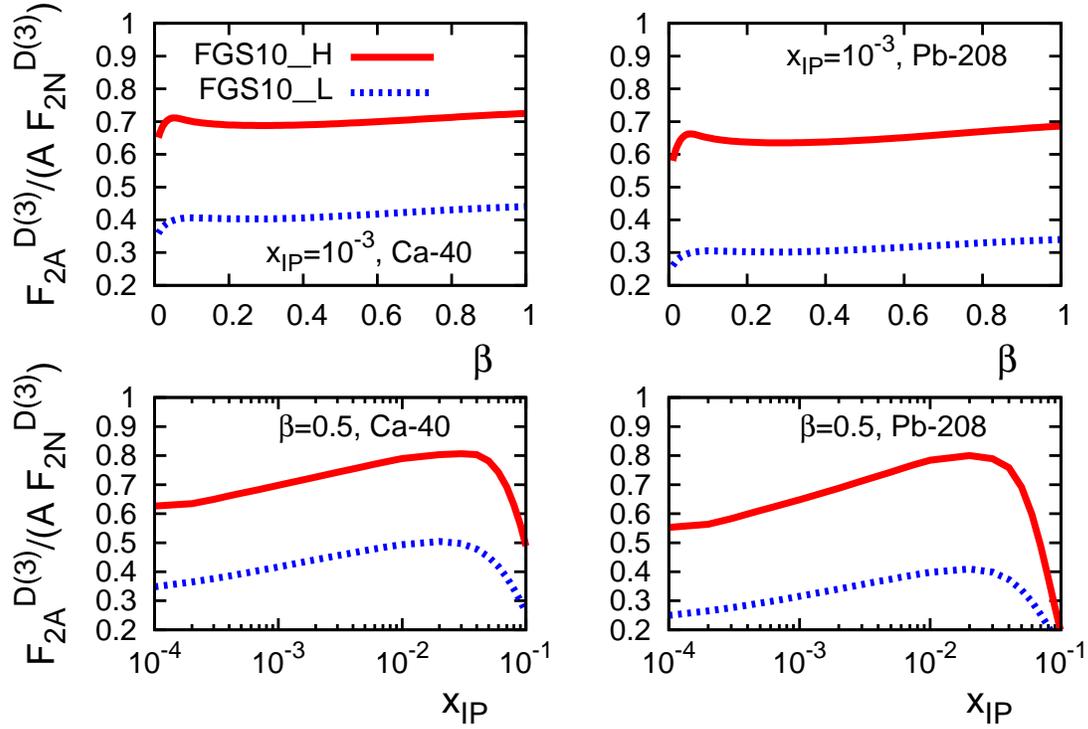,scale=1.4}
\vskip 0cm
\caption{
Predictions
for the ratio of the nuclear to nucleon diffractive structure functions,  $F_{2A}^{D(3)}/(AF_{2N}^{D(3)})$, 
at $Q_0^2=4$ GeV$^2$.}
\label{fig:diffraction_F2_2010}
\end{center}
\end{figure}

Our predictions for $f_{j/A}^{D(3)}/(Af_{j/N}^{D(3)})$ in 
Figs.~\ref{fig:diffraction_beta}, \ref{fig:diffraction_beta_2}, and \ref{fig:diffraction_xpom}
are very weakly flavor-dependent. Nevertheless, 
$f_{j/A}^{D(3)}/(Af_{j/N}^{D(3)})$ is not exactly equal to the ratio of the 
nuclear to nucleon diffractive structure functions, $F_{2A}^{D(3)}/(AF_{2N}^{D(3)})$. 
Therefore, we separately show our predictions for the NLO nuclear and nucleon 
diffractive structure 
functions and present an example of the resulting $F_{2A}^{D(3)}/(AF_{2N}^{D(3)})$
ratio in Fig.~\ref{fig:diffraction_F2_2010}.
As one can see in the figure, the predicted $F_{2A}^{D(3)}/(AF_{2N}^{D(3)})$ 
are very similar 
to $f_{j/A}^{D(3)}/(Af_{j/N}^{D(3)})$ from
Figs.~\ref{fig:diffraction_beta} and \ref{fig:diffraction_xpom}.

Our predictions for $F_{2A}^{D(3)}/(AF_{2N}^{D(3)})$
can be directly compared to the predictions for $F_{2A}^{D(3)}/(AF_{2N}^{D(3)})$ 
made in the framework of the color dipole model~\cite{Kowalski:2008sa}. 
We begin with the $\beta$ dependence and compare
the two upper panels of Fig.~\ref{fig:diffraction_F2_2010} to Fig.~7 of Ref.~\cite{Kowalski:2008sa}. 
We observe that the predicted shape of $F_{2A}^{D(3)}/(AF_{2N}^{D(3)})$ 
as a function of $\beta$ is similar in the two approaches, 
especially for $^{40}$Ca.
Moreover, for model 
FGS10\_H
 and $^{40}$Ca, our predictions are also close in  
the absolute value to those of Ref.~\cite{Kowalski:2008sa}.
At the same time, for $^{208}$Pb, 
our prediction for $F_{2A}^{D(3)}/(AF_{2N}^{D(3)})$ is that 
it mildly increases with increasing $\beta$ and takes on the values in the interval 
$0.6 \leq F_{2A}^{D(3)}/(AF_{2N}^{D(3)}) \leq 0.7$ 
(FGS10\_H)
and
$0.25 \leq F_{2A}^{D(3)}/(AF_{2N}^{D(3)}) \leq 0.35$ 
 (FGS10\_L),
while the prediction of~\cite{Kowalski:2008sa} is that $F_{2A}^{D(3)}/(AF_{2N}^{D(3)})$ rather rapidly
changes from 0.6 to 1.2 between $\beta=0$ and $\beta=0.3$ and further 
grows and becomes approximately
 1.4 as
$\beta$ approaches unity.

Turning to the $x_{\Pomeron}$ dependence of $F_{2A}^{D(3)}/(AF_{2N}^{D(3)})$, 
the two lower panels of Fig.~\ref{fig:diffraction_F2_2010} are to be
compared to Fig.~11 of Ref.~\cite{Kowalski:2008sa}. 
For $x_{\Pomeron} \leq 0.01$, the
two approaches predict 
the similar shape of $F_{2A}^{D(3)}/(AF_{2N}^{D(3)})$ 
as a function of $x_{\Pomeron}$. 
For $^{40}$Ca and model FGS\_L, the predictions of the two approaches are also close
in the absolute values.
At the same time, the trend of the 
$A$ 
dependence is opposite:
our leading twist approach predicts that $F_{2A}^{D(3)}/(AF_{2N}^{D(3)})$ for $^{40}$Ca is slightly larger than that for $^{208}$Pb,
while $F_{2A}^{D(3)}/(AF_{2N}^{D(3)})$ for $^{208}$Pb is noticeably larger than that
for $^{40}$Ca in Ref.~\cite{Kowalski:2008sa}.
Also,
we predict smaller values of $F_{2A}^{D(3)}/(AF_{2N}^{D(3)})$ for 
$^{208}$Pb
compared to the curves in Fig.~11 of Ref.~\cite{Kowalski:2008sa}.
While the  
$x_{\Pomeron} > 0.01$ region is not shown in Ref.~\cite{Kowalski:2008sa}, we predict a dramatic decrease
of $F_{2A}^{D(3)}/(AF_{2N}^{D(3)})$ for $x_{\Pomeron} > 0.01$
as a consequence of the decrease of 
the coherence length
in this region (see the discussion above).

One should also mention that one expects the different $Q^2$ dependences of $F_{2A}^{D(3)}/(AF_{2N}^{D(3)})$ in the leading twist approach and color dipole model:
while the $Q^2$ dependences of $F_{2A}^{D(3)}/(AF_{2N}^{D(3)})$ is very slow 
(logarithmic) in the 
leading twist theory (see, e.g., Fig.~\ref{fig:diffraction_beta_Q2dep}), 
it is faster in the color dipole model due to the eventual 
dominance of small-size configurations in the virtual photon wave function at large $Q^2$. 

Also, 
unlike the leading twist theory of nuclear shadowing that
allows one to predict nuclear diffractive PDFs of separate flavors, the dipole model does not
have a consistent way to predict, e.g., the gluon nuclear diffractive PDF.
(The same is also true for the usual nuclear PDFs, see the discussion in Sec.~\ref{sec:phen}.)

\subsubsection{Numerical predictions for incoherent diffraction}
\label{sec:incohdiff_num_res}

In this subsection, we present our numerical predictions for hard incoherent diffraction in $eA$ DIS.
The corresponding incoherent structure function, $F_{2A,{\rm incoh}}^{D(3)}$,
 is given by Eq.~(\ref{eq:diff_pdfs_final}). The essential input for this equation
is the rescattering cross section $\sigma_X$ and the corresponding slope $B_X$. As we discussed above,
we take $\sigma_X=\sigma_{\rm soft}^{q(\rm H)}(x,Q_0^2)$ [$\sigma_{\rm soft}^{q(\rm H)}(x,Q_0^2)$
corresponds to the ${\bar u}$ quark and model FGS10\_H, see
Eq.~(\ref{eq:sigma3_model1}) and Fig.~\ref{fig:sigma3_2009}] and
$B_X=7$ GeV$^{-2}$.

\begin{figure}[t]
\begin{center}
\epsfig{file=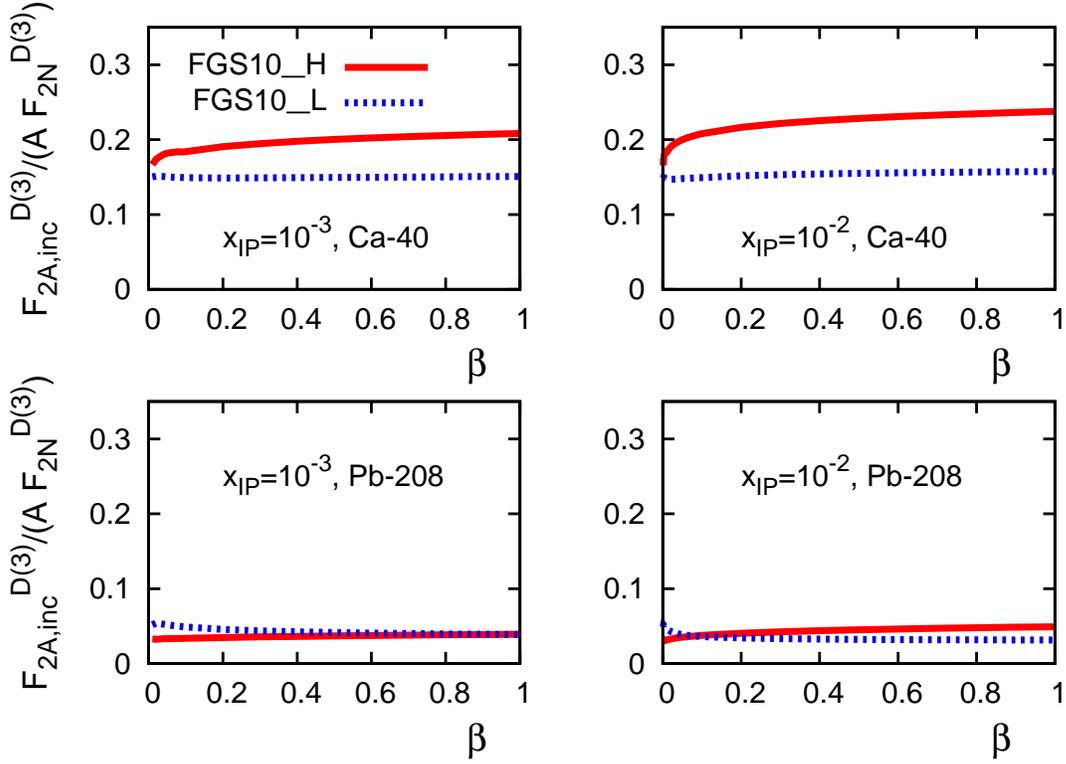,scale=1.4}
\vskip 0cm
\caption{The ratio of the incoherent nuclear and free nucleon diffractive structure
functions, $F_{2A,{\rm incoh}}^{D(3)}/(AF_{2N}^{D(3)})$, 
as a function of $\beta$ at fixed $x_{\Pomeron}=10^{-3}$ and $10^{-2}$ and $Q^2=4$ GeV$^2$.}
\label{fig:diffraction_INCOH_2010}
\end{center}
\end{figure}

As an illustration of the resulting incoherent diffractive structure function, 
in Fig.~\ref{fig:diffraction_INCOH_2010} we show the 
$F_{2A,{\rm incoh}}^{D(3)}/(AF_{2N}^{D(3)})$ ratio as a function of $\beta$ at fixed
$x_{\Pomeron}=10^{-3}$ and $10^{-2}$ and $Q^2=4$ GeV$^2$. The two upper panels correspond to $^{40}$Ca; the two lower panels are for $^{208}$Pb.

Our predictions for $F_{2A,{\rm incoh}}^{D(3)}/(AF_{2N}^{D(3)})$ can be compared to
the coherent case presented in Fig.~\ref{fig:diffraction_F2_2010} (two upper panels).
While the shapes of the $\beta$ dependence of $F_{2A,{\rm incoh}}^{D(3)}/(AF_{2N}^{D(3)})$
and $F_{2A}^{D(3)}/(AF_{2N}^{D(3)})$ are very similar, the incoherent diffractive 
structure functions
are smaller than the coherent ones: incoherent diffraction
constitutes approximately 15-25\% for $^{40}$Ca and 3-5\% for $^{208}$Pb
of the coherent diffraction (depending on the choice of model FGS10\_H or FGS10\_L). 
The $F_{2A,{\rm incoh}}^{D(3)}/F_{2A}^{D(3)}$ ratio at 
$\beta=0.5$, $Q_0^2=4$ GeV$^2$, and $x_{\Pomeron}=10^{-3}$ and $10^{-2}$
(the $\beta$-dependence is weak)
is summarized in Table~\ref{table:F2diff}.
\begin{table}[h]
\begin{tabular}{|c|c|c|}
\hline
$A$/model & $F_{2A,{\rm incoh}}^{D(3)}/F_{2A}^{D(3)}$, $x_{\Pomeron}=10^{-3}$ 
& $F_{2A,{\rm incoh}}^{D(3)}/F_{2A}^{D(3)}$, $x_{\Pomeron}=10^{-2}$
\\
\hline
$^{40}$Ca, FGS10\_H & 0.20 & 0.23\\
$^{40}$Ca,  FGS10\_L & 0.15 & 0.16\\
$^{208}$Pb, FGS10\_H & 0.037 & 0.045\\
$^{208}$Pb, FGS10\_L & 0.041 & 0.032 \\
\hline
\end{tabular}
\caption{The $F_{2A,{\rm incoh}}^{D(3)}/F_{2A}^{D(3)}$ ratio
at $\beta=0.5$, $Q_0^2=4$ GeV$^2$, and $x_{\Pomeron}=10^{-3}$ and $10^{-2}$.
The ratio is weakly $\beta$-dependent.}
\label{table:F2diff}
\end{table}

When comparing our results for incoherent diffraction to those of 
Ref.~\cite{Kowalski:2008sa}, one has to keep in mind that the break-up channel 
in Ref.~\cite{Kowalski:2008sa} corresponds to the sum of the coherent and incoherent channels
in this work (one needs to add the corresponding curves in Figs.~\ref{fig:diffraction_INCOH_2010}
and \ref{fig:diffraction_F2_2010}). 
The conclusion that can be drawn is similar to the one we already presented and discussed
above:
our leading twist approach and the color dipole formalism of Ref.~\cite{Kowalski:2008sa}
 predict a very similar shape of the $\beta$ dependence of the
$(F_{2A}^{D(3)}+F_{2A,{\rm incoh}}^{D(3)})/(AF_{2N}^{D(3)})$ ratio.
Moreover,
for $^{40}$Ca and model FGS10\_H, the predictions of the both approaches are also close
in the absolute values. For model FGS10\_L and $^{40}$Ca and for the both models
(FGS10\_H and FGS10\_L) and  $^{208}$Pb,
our predictions
for $(F_{2A}^{D(3)}+F_{2A,{\rm incoh}}^{D(3)})/(AF_{2N}^{D(3)})$ are smaller than 
the corresponding
$F_{2A}^{D(3)}/(AF_{2N}^{D(3)})$ in the break-up channel in Ref.~\cite{Kowalski:2008sa}.

\subsection{Exclusive diffraction at small $x$}
\label{subsec:exclusive}

In this subsection, we consider hard coherent exclusive production of 
real photons (deeply virtual  Compton scattering, DVCS) and vector  mesons ($J/\psi$, $\rho$, $\dots)$ off nuclei:
\begin{equation} 
\gamma^{\ast} +A \to \gamma (J/\psi, \rho, \dots) +A \,.
\label{eq:hard_exclusive_processes}
\end{equation}
The QCD factorization theorems 
proved
for  the exclusive
meson production by longitudinally polarized virtual
photons~\cite{Brodsky:1994kf,Collins:1996fb}
and 
for
the photon production  initiated by transversely polarized virtual photon 
(DVCS)~\cite{Collins:1998be}  allow one to express the 
amplitudes of the processes in Eq.~(\ref{eq:hard_exclusive_processes})
as the convolution of the hard interaction block, the meson $q\bar q$ wave function (in the case of meson production), and the generalized parton distributions (GPDs) of the target. 
The proofs 
of the  QCD factorization for these processes 
were based on the dominance in the considered processes of the
contribution of the point-like component of the wave function of 
highly virtual 
longitudinally polarized photon. Derived formulas
have demonstrated 
that the GPDs enter the  description of a wide range of 
hard exclusive
processes in an universal way.

In the case of the hard exclusive processes initiated by transversely polarized photons, the QCD factorization theorem is also applicable, although for larger $Q^2$ than in the  case of longitudinally polarized photons. This is because the contribution of non-perturbative QCD (the aligned jet model) is suppressed by the Sudakov form factor
only. 
This form factor is not compensated by processes with gluon radiation since the amplitudes for such processes
are suppressed by the overlapping integral with the wave function of the vector meson which is mostly non-perturbative. The HERA data indicate that the ratio of the transverse and longitudinal 
cross sections of $\rho$ meson production does not change with $W$ or with $t$ indicating that the squeezing of the $q\bar q$ dipole is similar in the two cases. This suggests that the Sudakov radiation in these processes starts 
as early as 
at $k_t \sim \Lambda_{QCD}$.

Evolution equations for bilocal operators,  whose matrix elements are now referred to as GPDs, were first studied 
in~\cite{Bukhvostov:1985rn}.  
GPDs
were used in~\cite{Mueller:1998fv} to parameterize the matrix elements of bilocal operators between hadronic states with non-equal momenta (non-forward matrix
elements), which appear in the QCD description of hard exclusive processes (DVCS, productions of mesons by longitudinally polarized photons, etc.)  Over the last ten years, the subject of GPDs has been one of the most active   fields in hadronic physics, see~\cite{Ji:1998pc,Goeke:2001tz,Diehl:2003ny,Belitsky:2005qn} 
for the reviews.

\begin{figure}[h]
\begin{center}
\epsfig{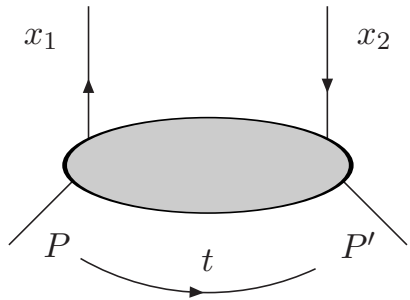}
\caption{The kinematic dependence of generalized parton distributions.}
\label{fig:diagram_gpds}
\end{center}
\end{figure}

GPDs generalize usual PDFs and, in general, 
depend on two-light cone fractions $x_1$ 
and $x_2$ of the partons emitted/absorbed by the target,  the invariant momentum transfer $t=(P^{\prime}-P)^2$
with $P^{\prime}$ and $P$ the final and initial momenta of the target, respectively, and the resolution scale
$Q^2$, see Fig.~\ref{fig:diagram_gpds}. 
\vspace*{0.5cm}
 The direct calculation of the amplitudes of the hard exclusive processes in question shows that 
\begin{equation}
x_1={Q^2+M^2\over \nu} \,, \quad \quad \, x_2={M^2 - M^2_V\over \nu} \,,
\end{equation}
where $M^2$ is the invariant mass squared in the quark loop in the transition 
$\gamma^{\ast} \to V$ (we consider here the case of small $x$ where the amplitude is predominantly imaginary);
 $\nu= W^2 + Q^2 - m_T^2$. 
 
 In the following we will use the variables $x_{\pm} = (x_1\pm x_2)/2$.
 In the Bjorken limit of fixed $x_B=Q^2/\nu$ and  $Q^2\to \infty$, 
 the light-cone fraction $x_{-}$ is fixed by the 
external kinematics:
\begin{equation}
x_- =\frac{Q^2+M^2_V}{\nu} \to x_B/2 \,. 
\end{equation}
Moreover, for the leading contribution to the imaginary part, the  
light-cone fraction $x_{+}$ is also constrained:
\begin{equation}
x_+\to \, x_B/2 \,.
\label{xplus}
\end{equation}

In the literature, the symmetric variables of Ji~\cite{Ji:1998pc} are often used. They are related to $x_{\pm}$ as
 \begin{equation}
x={x_+ \over 1 -x_-}\,,  \quad \quad \xi= {x_- \over 1- x_-}=\frac{x_B}{2-x_B} \,.
\end{equation}

In  this review, we will refer to the Bjorken variable  $x_B$ simply as $x$ where it 
does not cause confusion.

Note that in the limit $x\to 0$ and $Q^2 = const \gg \Lambda_{QCD}^2$,
the relation between $x_{+} $ and $x$ is qualitatively different 
from Eq.~(\ref{xplus}). Indeed,
an increase of the transverse momenta of the partons in the overlapping integral for the process $\gamma^{\ast} + T \to \gamma + T$  
with an increase of energy~\cite{Blok:2009cg} leads to $M^2 \gg Q^2$ and, hence,  
results in the following
relations:
\begin{equation}
x_1\to x_2 \,, \quad \quad x_+\, \to \, {M^2\over \nu} \,, \quad \quad  x\ll x_+ \,.
\end{equation}

\subsubsection{Nuclear GPDs at small $x$ and impact parameter dependent nuclear PDFs}

The number of GPDs depend on the spin of the target: for the spinless target,
one has one twist-two chirally-even GPD $H^j$ (we follow here the notations of~\cite{Diehl:2003ny}).

In Ref.~\cite{Goeke:2009tu}, the leading twist theory of nuclear shadowing was generalized to the non-forward kinematics and the expression for the nuclear GPD  $H$ at small $x$ was derived. In the 
$x_-=0$ 
limit,  the nuclear 
diagonal GPD   
 reads:
\begin{eqnarray}
H_{A}^{j}(x,&x_-=0&,t,Q_0^2)=A F_A(t) H_N^j(x,x_-=0,t,Q_0^2) \nonumber\\
&-&
\frac{A(A-1)}{2} 
\, 16 \pi B_{\rm diff} \, \Re e \Bigg\{\frac{(1-i \eta)^2}{1+\eta^2}
 \int d^2 \vec{b}\, e^{i \vec{\Delta}_{\perp} \cdot \vec{b}}
\int^{\infty}_{\infty} dz_1 \int^{\infty}_{z_1} dz_2
\int_{x}^{0.1} dx_{\Pomeron} 
\nonumber\\
& \times & \rho_A(b,z_1) \rho_A(b,z_2)  \,e^{i m_N x_{\Pomeron}(z_1-z_2)}
e^{-\frac{A}{2} (1-i \eta) \sigma_{\rm soft}^j(x,Q_0^2) \int^{z_2}_{z_1}dz^{\prime} \rho_A(\vec{b},z^{\prime})}
\nonumber\\
& \times &
\frac{1}{x_{\Pomeron}} f_{j}^{D(3)}(\beta,Q_0^2,x_{\Pomeron})
 \Bigg\}
 \,,
\label{eq:xiAlimit}
\end{eqnarray}
\noindent
where $H_N^j$ is the GPD of the free nucleon. As in the case of the calculation of the nuclear PDFs, 
$H_{A}^{j}$ in Eq.~(\ref{eq:xiAlimit}) is evaluated at $Q^2=Q_0^2$.
The $t$ dependence of the rescattering contribution (shadowing correction) 
originates mostly  from the 
overlap of the nuclear wave functions, with
an additional small correction due to the $t$ dependence of the elementary amplitudes
 which enter with $t_{\rm eff} \sim t/N^2$
for the interaction with $N$ nucleons.  This is a rather small effect that can effectively be 
taken into account with a good accuracy by using in Eq.~(\ref{eq:xiAlimit}) the nuclear matter 
density rather than the distribution of the point-like nucleons. 
This is because the $t$ dependence of the elementary 
GPDs in  the case of sea quark distribution is close to that of the 
proton e.m. form factor.
In the gluon channel, the $t$ dependence of the elementary 
GPD is given by the two-gluon form factor.

A more accurate treatment would require taking into account 
 the $x-b$ correlations in the nucleon 
GPDs. One should also note that the effects of the $t$ dependence of the elementary GPDs are more important for the lightest nuclei like $^2$H and $^4$He. 
A generalization for this case is straightforward since the $t$ dependence of the two dominant terms
(impulse approximation and double scattering) can be determined in a model-independent way.
 
\begin{figure}[h]
\begin{center}
\epsfig{file=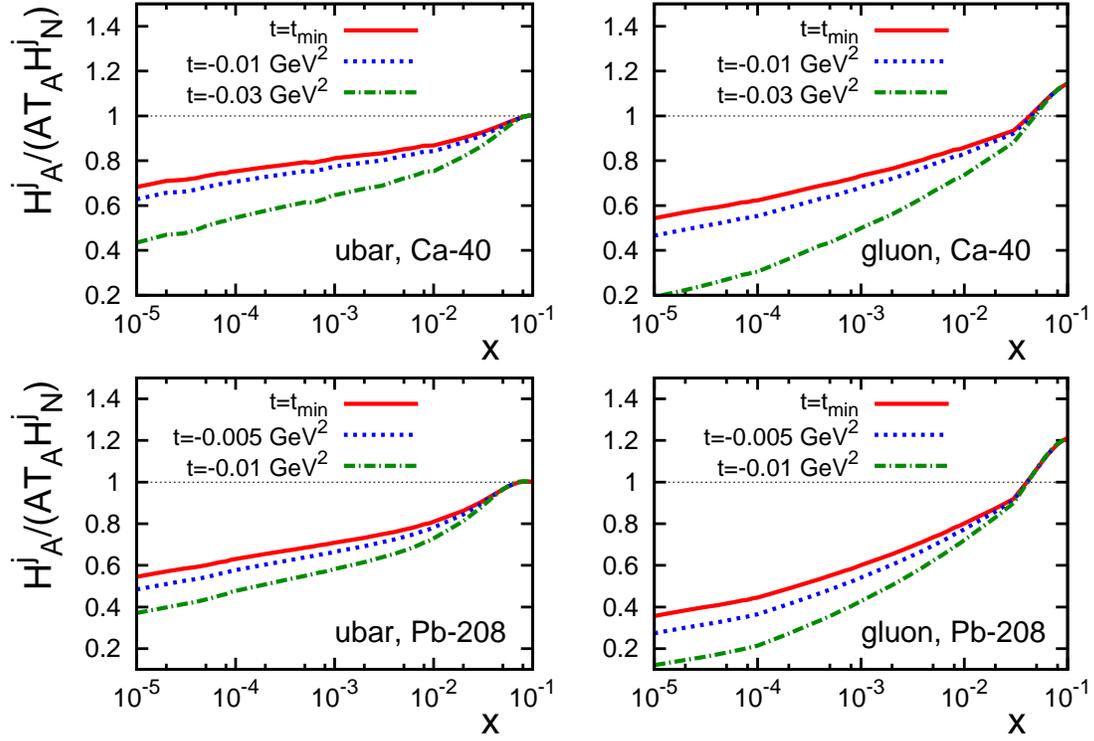,scale=1.4}
\caption{The ratio $H_{A}^{j}(x,x_{-}=0,t)/[A F_A(t) H_N^j(x,x_{-},t)]$ 
as a function of $x$ for different values
of $t$. All curves correspond to $Q^2=Q_0^2=4$ GeV$^2$ and model FGS10\_H.
}
\label{fig:IMP_ca40_2011}
\end{center}
\end{figure}

Figure~\ref{fig:IMP_ca40_2011} presents our predictions for the ratio $H_{A}^{j}(x,x_-=0,t)/[A F_A(t) H_N^j(x,x_-=0,t)]$ as a function of $x$ for different values  of $t$. The two left panels correspond to the 
${\bar u}$-quark distributions;
the two right panels correspond to the gluon distributions.
All curves correspond to $Q^2=Q_0^2=4$ GeV$^2$ and model FGS10\_H.
Since the $t$ dependence of the shadowing correction to $H_{A}^{j}(x,x_{-}=0,t)$
[the second term in Eq.~(\ref{eq:xiAlimit})] is significantly
slower than that of the impulse  approximation [the first term in Eq.~(\ref{eq:xiAlimit})], the effect of nuclear shadowing
expectedly increases as $|t|$ is increased.
Note also that for the $t$ dependence of the forward limit of the nucleon GPDs $H_N^j(x,x_{-}=0,t)$,
we use the exponential parameterization with the slope $B_{\rm DVCS}=5.45 \pm 0.19 \pm 0.34$ GeV$^{-2}$~\cite{Aaron:2007cz}:
\begin{equation}
H_N^j(x,x_-=0,t,Q^2)=e^{-\frac{1}{2}B_{\rm DVCS} |t|} H_N^j(x,x_-=0,t=0,Q^2)=e^{-\frac{1}{2}B_{\rm DVCS} |t|} f_{j/N}(x,Q^2) \,.
\label{eq:tdep_DVCS_elem}
\end{equation}
We also neglect a small difference in the $t$ dependence of the gluon and quark GPDs. 

In quantum mechanics and quantum field theory, it is convenient to consider 
high energy scattering amplitudes in the impact parameter representation by making 
the Fourier transform from the momentum representation to the 
coordinate representation. 
Since the amplitudes of hard exclusive process depend on $t$ only  through the GPDs, 
it is natural to consider  the GPDs in the impact parameter space.
In the $x_-=0$ limit,
the momentum transfer $t$ is purely transverse,
$t=-\Delta^2_{\perp}$, and one obtains the nuclear GPD in the mixed
momentum-impact parameter representation
(see also \cite{Burkardt:2002hr}):
\begin{equation}
H_A^j(x,x_{-}=0,b,Q^2)=\int \frac{d^2 \vec{\Delta}_{\perp}}{(2 \pi)^2}\,
e^{-i \vec{\Delta}_{\perp} \cdot \vec{b}}\, H_A^j(x,x_{-}=0,t=-\Delta^2_{\perp},Q^2)\,.
\label{eq:ngpd2}
\end{equation}
Substituting Eq.~(\ref{eq:ngpd2}) in Eq.~(\ref{eq:xiAlimit}), one obtains
the expression for the $x_-=0$ limit of the nuclear GPD $H_A^j$ at small $x$ in the
impact parameter space:
\begin{samepage}
\begin{eqnarray}
H_{A}^{j}(x,&x_-=0&,b,Q_0^2)=A T_A(b) f_{j/N}(x,Q_0^2) \nonumber\\
&-&
\frac{A(A-1)}{2} 
\, 16 \pi B_{\rm diff} \, \Re e \Bigg\{\frac{(1-i \eta)^2}{1+\eta^2}
\int^{\infty}_{\infty} dz_1 \int^{\infty}_{z_1} dz_2
\int_{x}^{0.1} dx_{\Pomeron} 
\nonumber\\
& \times & \rho_A(b,z_1) \rho_A(b,z_2)  \,e^{i m_N x_{\Pomeron}(z_1-z_2)}
e^{-\frac{A}{2} (1-i \eta) \sigma_{\rm soft}^j(x,Q_0^2) \int^{z_2}_{z_1}dz^{\prime} \rho_A(\vec{b},z^{\prime})}
\nonumber\\
& \times &
\frac{1}{x_{\Pomeron}} f_{j}^{D(3)}(\beta,Q_0^2,x_{\Pomeron})
 \Bigg\}
 \,.
\label{eq:ngpd3}
\end{eqnarray}
\end{samepage}
Note that in the first line of Eq.~(\ref{eq:ngpd3}), 
we neglected the weak $t$ dependence of the free nucleon 
GPD compared to the steep
$t$ dependence of the nuclear form factor (see the discussion above) 
and used the fact that 
$H_N^j(x,0,0,Q^2)=f_{j/N}(x,Q^2)$.
A comparison of Eq.~(\ref{eq:ngpd3}) to Eq.~(\ref{eq:impact2}) 
shows that, indeed,  $f_{j/A}(x,Q^2,b)=H^j_A(x,\xi=0,b,Q^2)$
which demonstrates consistency of our treatment of nuclear shadowing for $H^j_A$ and for nuclear PDFs.

In general, GPDs do not lend themselves to the probabilistic
interpretation. However, in the $x_-=0$ limit, they do: $H_A^j(x,x_-=0,b,Q^2)$ is
the probability to find the parton of flavor $j$ with the light-cone fraction
$x$ and the transverse distance $b$ from the transverse center of 
momentum of the nucleus~\cite{Feynman,Burkardt:2002hr}.
In fact, one can formally prove that the diagonal GPDs in the 
$b$-space are positive-definite~\cite{Pobylitsa:2002iu}.
The equivalence of the impact parameter dependent nuclear PDFs and the $x_-=0$ limit
of nuclear GPDs in the impact parameter space 
implies that the spacial image of the 
nuclear GPDs in this limit is the same as for nuclear PDFs and is given by 
 Fig.~\ref{fig:impact_dependence}.

\subsubsection{Modeling nuclear GPDs and the role of skewness}

Equation~(\ref{eq:xiAlimit}) 
defines our expression for the nuclear
GPD $H_{A}^j(x_+,x_-,t,Q^2)$ in the $x_-=0$ limit at the initial evolution scale $Q_0^2$.
However, in general, experimental observables measured in hard exclusive processes~(\ref{eq:hard_exclusive_processes}) probe a much more complicated function, 
namely, the integration of perturbatively calculable coefficient functions 
(scattering kernels) with the GPD $H_{A}^j(x_+,x_-,t,Q^2)$ over the entire region of
the light-cone variable $x$, $0 \leq x \leq 1$.
At the same time, at high energies (small $x_B$), the situation simplifies 
because the high-energy scattering amplitudes are predominantly imaginary; 
the imaginary part of the $\gamma^{\ast} +A \to \gamma (J/\psi, \rho, \dots) +A$ scattering amplitudes
can be approximated
in terms of the diagonal GPDs.

Several theoretical ideas  were suggested which help to evaluate non-diagonal GPDs at 
small $x_B$. 
It was observed in~\cite{Frankfurt:1997ha} that GPDs at small $x$ and sufficiently large $Q^2$ are  calculable through diagonal GPDs. The reason is that $x_1-x_2 \approx 2\xi$ is conserved in the kernel of 
the QCD evolution equations for the GPDs. 
At the same time, $x_+$ evolves similarly to the case of the QCD evolution equations for 
usual PDFs. An analysis of the trajectories for the QCD evolution shows that for sufficiently large $Q^2$, dominant trajectories correspond to $x_+$ at $Q_0^2$ that are much larger than $x_-$ 
(see the discussion in Sec.~\ref{subsec:qcd_curve}).
Hence, GPD at  small $x$ and large $Q^2$ can be calculated through the diagonal GPDs at the initial $Q_0^2$.
Note, however, that there is a substantial $Q^2$ interval where there is significant sensitivity 
to the boundary condition.

There are also additional considerations applicable
for moderate $Q^2$.  Let us first consider the case of DVCS $\gamma^{\ast} +N \to \gamma +N$. 
We argued before that the dominant configurations in the photon wave function, which dominate 
the interaction with the nucleon for  $Q^2 \sim$ few GeV$^2$, are aligned jet configurations. 
If we neglect contributions of other configurations, we can write explicitly the integral over the momenta in the quark loop and notice that this part of the  interaction 
is not sensitive to the skewness. As a result, 
the main difference between the diagonal $\gamma^{\ast}+N \to \gamma^{\ast}+N$ amplitude at $t=0$ 
(which is expressed through the total cross section of the $\gamma^{\ast} N$ scattering) and the corresponding  DVCS amplitude is due to different energy denominators, 
which are equal to $1/(Q^2+ M^2_{q\bar q})$ for the $\gamma^{\ast} \to q\bar q $ transition
and $1/M^2_{q\bar q}$ for the 
 $\gamma \to q\bar q $ transition
($M^2_{q\bar q}$ is the invariant mass squared of the $q{\bar q}$ system).
Therefore, one finds that
\begin{equation} 
R=\frac{\Im m T(\gamma^{\ast}p \to \gamma p)_{|t=0}}{\Im m T(\gamma^{\ast}p \to \gamma^{\ast} p)_{|t=0} }  \approx 2 \,.
\label{eq:R_DVCS}
\end{equation}
On the other hand, at the leading order accuracy,
$R$ is equal to the ratio of the GPDs at the cross-over point $x_+=x_-$ and the usual PDFs:
\begin{equation}
R=\frac{\sum_q e_q^2 \left[H^q(x_-,x_-,t=0,Q^2)+H^{\bar q}(x_-,x_-,t=0,Q^2)\right]}{\sum_q e_q^2 \left[q(2x_-,Q^2)+{\bar q}(2x_-,Q^2)\right]} \,,
\label{eq:R_DVCS2}
\end{equation}
where $\sum_q e_q^2$ is the sum over all active quark flavors (quarks and antiquarks)
with the weight given by the quark charge squared $e_q^2$;
the denominator is evaluated at Bjorken $2x_{-} \approx 2 \xi \approx x$. 
Therefore, $R$ directly constrains the GPDs at the cross-over line $x_+=x_-$ ($x_2=0$)
and is very sensitive to the $x_- \to 0$ behavior of the GPDs, see
the discussion in~\cite{Kumericki:2007sa}.
The QCD evolution leads to a slow increase of $R$ with $Q^2$. 
$R$ has been measured in the free proton case at HERA~\cite{Aaron:2007cz,Schoeffel:2007dt}; it
agrees well with the predictions of~\cite{Frankfurt:1997at,Freund:2002qf}. 

A useful, though rough, approximation for the gluon GPD, which is relevant for the exclusive  vector meson production, is to use the symmetry of the $\gamma^{\ast} + T \to V + T $ amplitude 
with respect to the transposition $x_1\nu \to x_2 \nu$~\cite{Brodsky:1994kf}. 
Odd powers of $x_-$ 
do not contribute and one obtains:
\begin{eqnarray}
H^g(x_1,x_2,...) & =&  H^g((x_+ + x_-)/2, (x_+ + x_-)/2, ...) \nonumber\\
&+&
{x_-^2\over 2} {\partial^2\over \partial^2 x_-} 
H^g(x_++x_-, x_+-x_-)_{\left | x_-=0 \right.} \,.
\end{eqnarray}
Taking $H^g(x_1, x_2) \propto 1/(x_1\, x_2)^{n/2}$, where 
$H^g(x, x) \propto  1/x^n$ and $n\sim 0.2 \div 0.3$, we can estimate 
\begin{equation}
H^g(x_1, x_2) \approx H^g(x_+, x_-=0)\left(1+ \frac{x_-^2}{x_+^2}\right)^{n/2} \,.
\end{equation}
Hence, we see that for a wide range of skewness, 
one can estimate the non-diagonal gluon GPD as the diagonal gluon GPD at the average point $x=x_+$. 
For example, for $x_2=0$ 
which enters the description of the exclusive vector meson production at large $Q^2$,
\begin{equation}
H^g(x_1, x_2=0) \approx H^g(x_1/2, x_1/2)(1+ n/2) \,.
\end{equation}
 
In summary,  at high energies (small $\xi$)
and in 
the leading logarithmic approximation (LLA), generalized parton distributions at an 
input scale $Q_0^2$ can be approximated well by the usual parton distributions~\cite{Frankfurt:1997ha}:
\begin{eqnarray}
&& H_{A}^q(x_+,x_-, t=0,Q_0^2)=q_A(x,Q_0^2) \,, \nonumber\\
&& H_{A}^g(x_+, x_-,t=0,Q_0^2)=g_A(x,Q_0^2) \,,
\label{eq:LLA_approx}
\end{eqnarray}
where $x=x_+$. 
Note that we do not introduce the additional factor of $x$ in the second line of Eq.~(\ref{eq:LLA_approx}) as was done, e.g., in Refs.~\cite{Diehl:2003ny}
and \cite{Frankfurt:1997ha};  this factor 
is essentially a matter of convention. The form of Eq.~(\ref{eq:LLA_approx})
allows us to use the diagonal  approximation reproducing the boundary condition for 
$R \approx 2$
(see Eq.~(\ref{eq:R_DVCS})) (though this condition  is expected to be violated for very small 
$x$, see the discussion above).

In Eq.~(\ref{eq:LLA_approx}) we did not consider the $t$ dependence.
However, it can be straightforwardly restored since it does not conflict with the
LLA. Therefore, we have the following final relation between the small-$\xi$ GPDs
and the impact parameter dependent PDFs:
\begin{eqnarray}
H_{A}^j(x_+,x_-,t,Q_0^2)&=&H_{A}^j(x,x_-=0,t,Q_0^2)=
\int d^2 \vec{b} \, e^{i \vec{\Delta}_{\perp} \cdot \vec{b}} H_A^j(x,x_-=0,b,Q_0^2)
\nonumber\\
&=&\int d^2 \vec{b} \, e^{i \vec{\Delta}_{\perp} \cdot \vec{b}} f_{j/A}(x,Q_0^2,b)
 \,.
\label{eq:LLA_approx2}
\end{eqnarray}
Note that the right-hand side of Eq.~(\ref{eq:LLA_approx2}) is known: it is predicted by
the leading twist theory of nuclear shadowing, see Eq.~(\ref{eq:xiAlimit}) 
and (\ref{eq:ngpd3}).
Note also that as the value of $Q^2$ is increased, the accuracy of 
Eq.~(\ref{eq:LLA_approx2}) worsens since the DGLAP evolution for GPDs introduces additional
skewness (dependence on $\xi$).
Equation~(\ref{eq:LLA_approx2}) defines the nuclear GPDs that we shall use below
to make predictions for various observables in hard exclusive reactions with nuclei.

The skewness ratio $R$ can also be introduced for nuclear targets and separate parton
flavors:
\begin{equation}
R_{j/A} \equiv \frac{H^j_A(x_-,x_-,t=0,Q_0^2)}{f_{j/A}(2x_-,Q_0^2)}=
\frac{H^j_A(x_-,x_-=0,t=0,Q_0^2)}{f_{j/A}(2x_-,Q_0^2)}=\frac{f_{j/A}(x_-,Q_0^2)}{f_{j/A}(2x_-,Q_0^2)} \,.
\label{eq:R_DVCS3}
\end{equation}
Here we used Eq.~(\ref{eq:LLA_approx2}) and the fact that in the forward limit
($x_-\to 0$ and $t \to 0$), the GPD $H$ reduces to the usual PDFs.
Hence, the skewness ratio $R_{j/A}$ is given in terms of the usual nuclear PDFs
evaluated at different light-cone fractions.

\begin{figure}[h]
\begin{center}
\epsfig{file=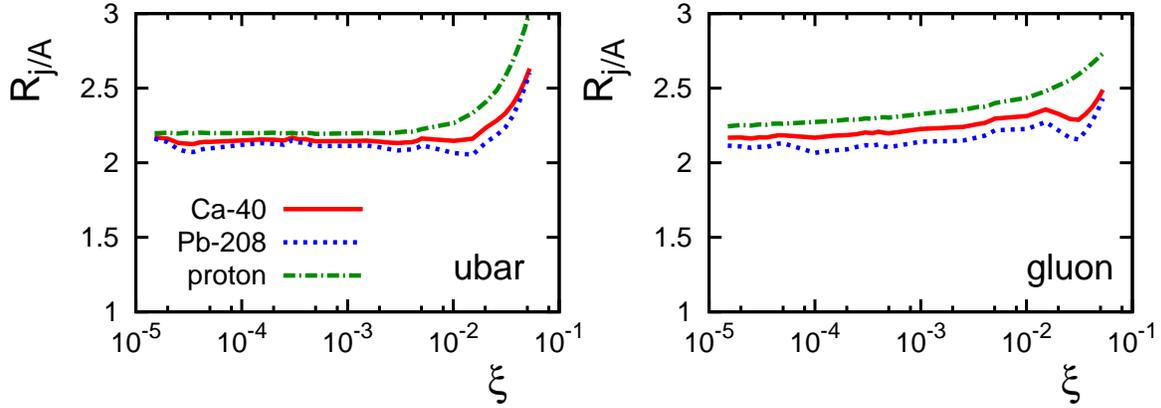,scale=1.5}
\caption{The skewness ratio $R_{j/A}$ of Eq.~(\ref{eq:R_DVCS3}) as a function of 
$\xi$ at $Q_0^2=4$ GeV$^2$ for $^{40}$Ca (solid curves), 
$^{208}$Pb (dotted curves) and free proton (dot-dashed curves).
The left panel is for the ${\bar u}$-quark distribution; the right panel is for the
gluon distribution.
}
\label{fig:GPDs_R}
\end{center}
\end{figure}

Figure~\ref{fig:GPDs_R} presents our predictions for $R_{j/A}$ of Eq.~(\ref{eq:R_DVCS3})
as a function of $\xi$ at $Q_0^2=4$ GeV$^2$.
The solid curves correspond to $^{40}$Ca; the dotted curves correspond to
$^{208}$Pb; models FGS10\_H and FGS10\_L give numerically indistinguishable
predictions for $R_{j/A}$. For comparison, we also give the skewness ratio
$R_{j/A}$ for the free proton target as dot-dashed curves.
In Fig.~\ref{fig:GPDs_R}, the left panel corresponds to ${\bar u}$-quarks; the 
right panel corresponds to the gluon channel.
As one can see from the figure, $R_{j/A}$ depends weakly on the atomic mass number
and the parton flavor. Also, $R_{j/A}$ is a weak function of $x_-$ for
$10^{-5} \leq x_- \leq 10^{-2}$.

The numerical value of $R_{j/A}$ is very important for the phenomenology of 
GPDs. The fact that $R_{j/A}$ for the proton is of the order of $2-2.5$ is in
agreement with the analysis of the HERA data on the skewness ratio 
$R$ of Eq.~(\ref{eq:R_DVCS})~\cite{Aaron:2007cz,Schoeffel:2007dt},
the aligned jet type model for the proton GPDs at the input scale~\cite{Frankfurt:1997at,Freund:2002qf},
and the phenomenological parameterization of proton GPDs as the conformal partial wave decomposition~\cite{Kumericki:2007sa} (see the discussion 
in Ref.~\cite{Kumericki:2007sa}).
The value of $R_{j/A}$ is somewhat smaller for nuclei than for the free nucleon
since nuclear shadowing tames the increase of nuclear PDFs with decreasing 
Bjorken $x$. Note that the discussed model may overestimate 
$R(Q_0^2)$. Indeed, as we discussed above, $R\sim 2$ corresponds to the 
dominance of the AJM contribution with small transverse momenta. Nuclear shadowing  reduces this contribution to nuclear PDFs as compared to the contribution of components with large transverse momenta, for which $R$ is closer to unity. 
Hence, the effect of the reduction of $R_{j/A}$ with an increase of $A$ and 
 a decrease of $x$ may be somewhat larger than presented in Fig.~\ref{fig:GPDs_R}.

\subsubsection{Leading twist nuclear shadowing and coherent nuclear DVCS}

\begin{figure}[h]
\begin{center}
\epsfig{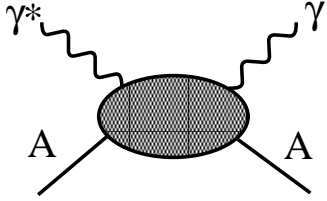}
\caption{Coherent deeply virtual Compton scattering (DVCS) with nuclei.}
\label{fig:coh_dvcs_graph}
\end{center}
\end{figure}

Having defined and discussed the expression for the nuclear GPDs at small $\xi$ 
[see Eq.~(\ref{eq:LLA_approx2})], we can now form predictions for various observables
measured in high-energy hard exclusive processes with nuclei~(\ref{eq:hard_exclusive_processes}).
(In this review, we consider only {\it diffractive} hard exclusive processes.) 
The cleanest ways to access GPDs is via DVCS; below we focus on
unpolarized coherent nuclear DVCS, $\gamma^{\ast}+A \to \gamma +A$, 
see Fig.~\ref{fig:coh_dvcs_graph}.
\vspace*{0.35cm}

At the photon level, the $\gamma^{\ast}+A \to \gamma +A$
cross section reads, see, e.g.,~\cite{Belitsky:2001ns}:
\begin{equation}
\frac{d \sigma_{\rm DVCS}}{dt}=\frac{\pi \alpha_{\rm em}^2 x^2(1-\xi^2)}{Q^4 \sqrt{1+\epsilon^2}}|{\cal A}_{\rm DVCS}(\xi,t,Q^2)|^2 \,,
\label{eq:sigma_dvcs}
\end{equation}
where $\alpha_{\rm em}$ is the fine-structure constant; 
$\epsilon^2=4 x_B^2 m_N^2/Q^2$;
${\cal A}_{\rm DVCS}$ is the DVCS amplitude
(more precisely, it is the so-called Compton form factor).
At high energies (small Bjorken $x$ and $x_-$), 
${\cal A}_{\rm DVCS}$ is predominantly imaginary.
At the leading order in the strong coupling constant $\alpha_s$
(the handbag approximation), the imaginary part of ${\cal A}_{\rm DVCS}$ is given in terms of the 
quark nuclear GPDs at the $x_-=x_+=\xi$ cross-over line,
\begin{equation}
\Im m {\cal A}_{\rm DVCS}(\xi,t,Q^2)=-\pi \sum_q e_q^2 \left[H_A^q(\xi, \xi,t,Q^2)+H_A^{\bar q}(\xi,\xi,t,Q^2)\right] \,,
\label{eq:sigma_dvcs2}
\end{equation}
where $e_q$ are the quark charges; $H_A^q(\xi,\xi,t,Q^2)$ are given by 
Eq.~(\ref{eq:LLA_approx2}). Note that Eq.~(\ref{eq:sigma_dvcs2}) involves
the $q+{\bar q}$ singlet combination of the quark GPDs.

\begin{figure}[h]
\begin{center}
\epsfig{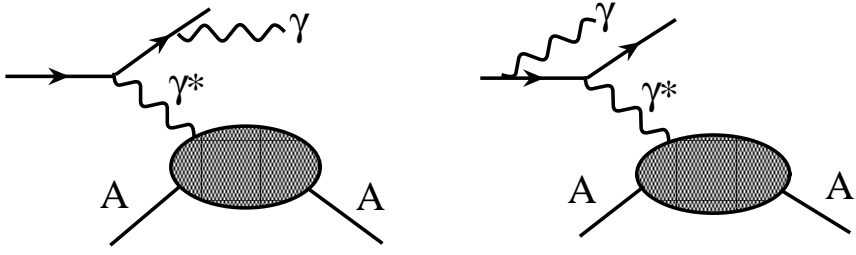}
\caption{Coherent Bethe-Heitler (BH) process with nuclei.}
\label{fig:coh_bh_graph}
\end{center}
\end{figure}

The DVCS process interferes and competes with the purely electromagnetic
Bethe-Heitler (BH) process, see Fig.~\ref{fig:coh_bh_graph}.
\vspace*{0.35cm}

The BH cross section at the photon level can be written in the following
form~\cite{Belitsky:2001ns}:
\begin{equation}
\frac{d \sigma_{\rm BH}}{dt}=\frac{\pi \alpha_{\rm em}^2}{4 Q^2 t (1+\epsilon)^{5/2} (1-y-y^2/2)}
\int^{2 \pi}_{0} \frac{d \phi}{2 \pi} \frac{1}{{\cal P}_1(\phi) {\cal P}_2(\phi)}
|{\cal A}_{\rm BH}(\xi,t,Q^2,\phi)|^2 \,,
\label{eq:sigma_bh}
\end{equation}
where $y=(q \cdot P_A)/(k \cdot P_A)=Q^2/(x s)$ is the fractional energy loss of the 
incoming lepton with momentum $k$ ($q$ is the momentum of the virtual photon,
$P_A$ is the momentum of the incoming nucleus, $s$ is the total invariant energy
squared); 
$\phi$ is the angle between the lepton and hadron scattering planes; 
 ${\cal P}_1(\phi)$ and  ${\cal P}_2(\phi)$ are proportional to the lepton
propagators; $|{\cal A}_{\rm BH}(x,t,Q^2)|^2$ is the BH amplitude squared.
The expressions for ${\cal P}_{1,2}(\phi)$ and $|{\cal A}_{\rm BH}(x,t,Q^2)|^2$ 
can be found in Refs.~\cite{Belitsky:2001ns,Belitsky:2000vk}.
Note that $|{\cal A}_{\rm BH}(x,t,Q^2)|^2$ is proportional to the nuclear
electric form factor squared, $|F_A(t)|^2$, and the nucleus charge
squared, $Z^2$.

Note that at high energies (small Bjorken $x$), 
the $\gamma^{\ast}+A \to \gamma +A$
amplitude is predominantly imaginary. The contribution of the interference
between the DVCS and BH amplitudes is proportional to the real part of the
DVCS amplitude, sizable and concentrated at small $t$. However, after the integration
over $\phi$, the interference term essentially disappears and, thus, 
can be safely neglected.

Integrating the differential cross sections in Eqs.~(\ref{eq:sigma_dvcs}) and 
(\ref{eq:sigma_bh}) over $t$, one obtains the corresponding $t$-integrated cross
sections:
\begin{eqnarray}
\sigma_{\rm DVCS}&=&\int^{t_{\rm min}}_{t_{\rm max}} dt \frac{d \sigma_{\rm DVCS}}{dt} \,,
\nonumber\\
\sigma_{\rm BH}&=&\int^{t_{\rm min}}_{t_{\rm max}} dt \frac{d \sigma_{\rm BH}}{dt} \,,
\label{eq:sigma_tint}
\end{eqnarray}
where $t_{\rm min} \approx -x^2 m_N^2$ (the exact expression can be found in~\cite{Belitsky:2001ns}); ${t_{\rm max}}=-1$ GeV$^2$
(in practice, one can take $|{t_{\rm max}}|$ much smaller for heavy nuclei, e.g.,
$\approx -0.1$ GeV$^2$ for $^{208}$Pb.)

\begin{figure}[h]
\begin{center}
\epsfig{file=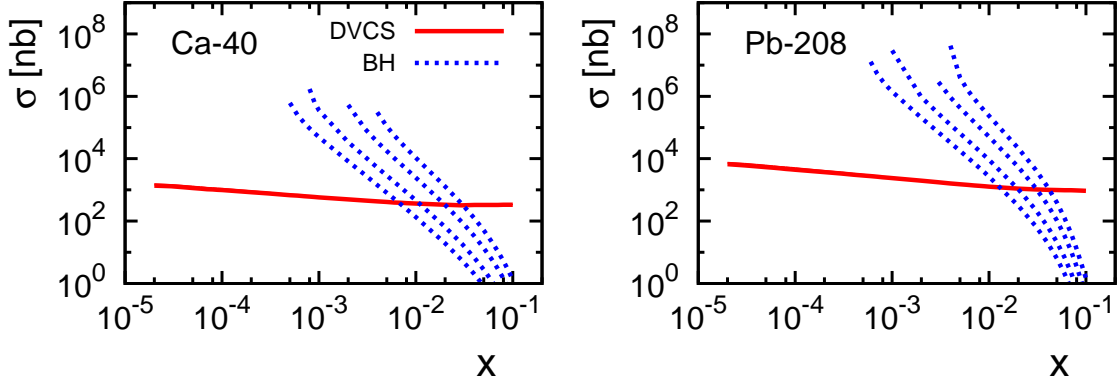,scale=1.5}
\caption{The DVCS (solid curves) and Bethe-Heitler (dotted curves) 
cross sections~(\ref{eq:sigma_tint}) as functions of Bjorken $x$ at $Q_0^2=4$ GeV$^2$.
The left panel corresponds to
$^{40}$Ca;
the right panel corresponds to $^{208}$Pb. 
Four curves for the BH cross section correspond to the four values of $\sqrt{s}$ 
in Table~\ref{table:y} (the rightmost dotted curve corresponds to the lowest
$\sqrt{s}$; the leftmost dotted curve corresponds to the largest $\sqrt{s}$).
}
\label{fig:DVCS_2011}
\end{center}
\end{figure}

Figure~\ref{fig:DVCS_2011} presents our calculations for the $t$-integrated
DVCS (solid curves) and BH (dotted curves) 
cross sections as a function of $x$ at fixed $Q^2=Q_0^2=4$ GeV$^2$.
The left panel corresponds to
$^{40}$Ca;
the right panel corresponds to $^{208}$Pb. 
Note that at the given large scale along the $y$-axis, 
models FGS10\_H and FGS10\_L give indistinguishable predictions.

The BH cross section explicitly depends on the energy of the process
through the variable $y$. In Fig.~\ref{fig:DVCS_2011}, the four
values of the BH cross section correspond to the values of
$y$, $y=Q^2/(xs)$, that correspond to the energy settings presented in
Table~\ref{table:y}. These energies of the lepton and hadron beam are the
discussed energy settings of a future Electron-Ion Collider.
The rightmost dotted curve corresponds to the lowest
$\sqrt{s}$; the leftmost dotted curve corresponds to the largest $\sqrt{s}$.
The dotted curves extend from large $x$ down to the smallest possible 
$x$ defined by the condition that $y \geq 0.95$.
 
\begin{table}[h]
\begin{tabular}{|c|c|c|c|c|c|c|}
\hline
$E_l$ (GeV)& $E_N$ (GeV) & $\sqrt{s}$ (GeV) & $E_N$ (GeV) & $\sqrt{s}$ (GeV)
& $E_N$ (GeV) & $\sqrt{s}$ (GeV)
\\
& in $^{208}$Pb & & in $^{40}$Ca & & free nucleon &
\\
\hline
11 & 24  & 32 & 30 & 36  & 60 &  51\\
11 & 100 & 66 & 125 & 74 & 250 & 105 \\
5  & 100 & 44 & 125 & 50 & 250 & 71  \\
20 & 100 & 90 & 125 & 100 & 250 & 141 \\
\hline
\end{tabular}
\caption{The lepton and hadron beam energies and the corresponding $\sqrt{s}$
that correspond to the BH cross section presented in Figs.~\ref{fig:DVCS_2011}
and \ref{fig:DVCS_proton_2011}.
}
\label{table:y}
\end{table}

\begin{figure}[h]
\begin{center}
\epsfig{file=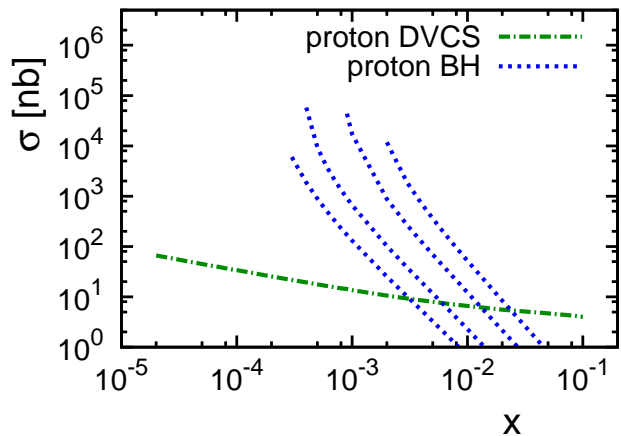,scale=1.4}
\caption{The free proton DVCS (dot-dashed curve) and Bethe-Heitler (dotted curves) 
cross sections as functions of Bjorken $x$ at $Q_0^2=4$ GeV$^2$.
Four curves for the BH cross section correspond to the four values of $\sqrt{s}$ 
in Table~\ref{table:y} (the rightmost dotted curve corresponds to the lowest
$\sqrt{s}$; the leftmost dotted curve corresponds to the largest $\sqrt{s}$).
}
\label{fig:DVCS_proton_2011}
\end{center}
\end{figure}

For comparison with the nuclear case, we also present the free proton DVCS
and BH cross sections as functions of $x$ at $Q_0^2=4$ GeV$^2$ in Fig.~\ref{fig:DVCS_proton_2011}.
The dot-dashed curve corresponds to the DVCS cross section;
the four dotted curves correspond to the BH cross section 
evaluated at the four values of  $\sqrt{s}$ presented
in Table~\ref{table:y} (the rightmost dotted curve corresponds to the lowest
$\sqrt{s}$; the leftmost dotted curve corresponds to the largest $\sqrt{s}$).

In our calculations, we used the 
standard expressions for the DVCS and BH cross
sections~\cite{Belitsky:2001ns}. For the free proton GPDs, 
we used Eq.~(\ref{eq:tdep_DVCS_elem})
along with the leading order CTEQ5L parameterization for
$f_{j/N}(x,Q_0^2)$~\cite{Lai:1999wy}.

As one can see from Fig.~\ref{fig:DVCS_2011}, in the considered kinematics (note 
that we considered the kinematics of a future EIC) and for the considered
medium-heavy ($^{40}$Ca) and heavy ($^{208}$Pb) nuclei,
the BH cross section is much larger than the DVCS cross section for small $x$.
One of the main reasons for this is the dramatic enhancement of the BH cross section
at small $t\approx t_{\min}$ by the factor $1/t$, see Eq.~(\ref{eq:sigma_bh}).
Therefore, in order to extract a small DVCS signal on the background of the dominant
BH contribution, one needs to consider the observable differential in $t$ 
and generally stay away from $t \approx t_{\min}$.

\begin{figure}[t]
\begin{center}
\epsfig{file=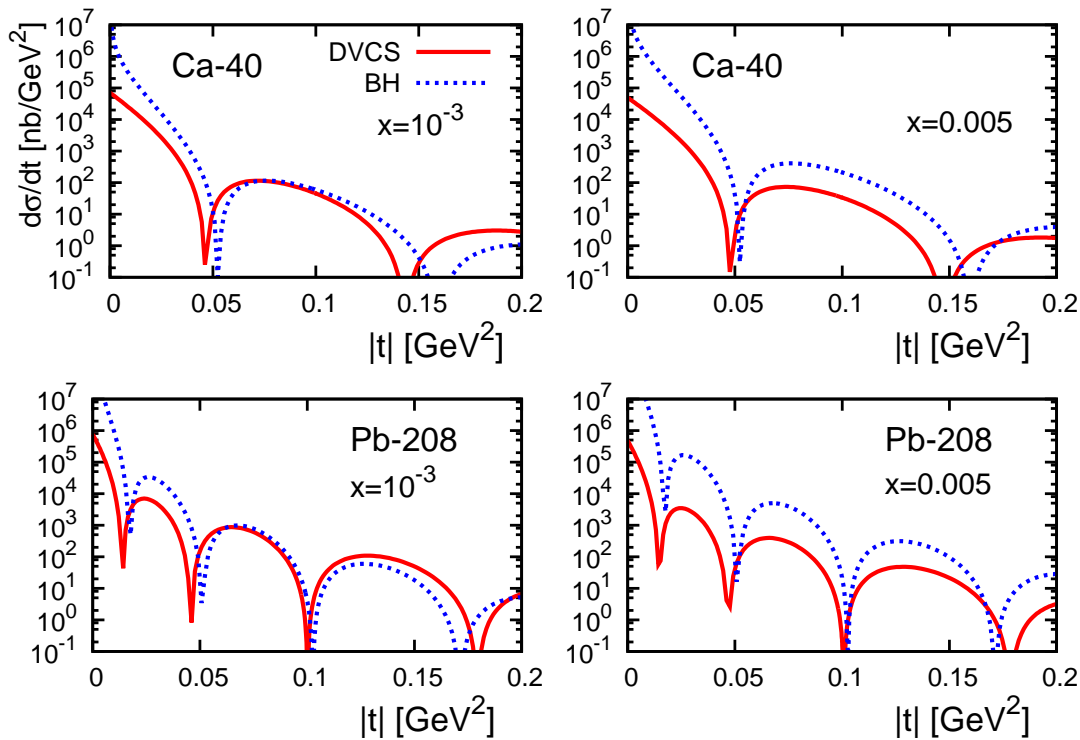,scale=1.4}
\caption{The DVCS (solid curves) and Bethe-Heitler (dotted curves) cross sections, see Eqs.~(\ref{eq:sigma_dvcs}) and (\ref{eq:sigma_bh}),  as functions of $|t|$ at $Q_0^2=4$ GeV$^2$ and $x=10^{-3}$ and $x=5 \times 10^{-3}$.
The upper panels corresponds to $^{40}$Ca; the lower panels corresponds to $^{208}$Pb. 
}
\label{fig:DVCS_2011_tdep}
\end{center}
\end{figure}

Figure~\ref{fig:DVCS_2011_tdep} presents our predictions for the differential
DVCS cross section (solid curves) and BH cross section (dotted curves), see Eqs.~(\ref{eq:sigma_dvcs}) and (\ref{eq:sigma_bh}), as functions of the momentum transfer $|t|$ at fixed $Q_0^2=4$ GeV$^2$ and $x=10^{-3}$ and $x=5 \times 10^{-3}$.
The two upper panels correspond to $^{40}$Ca; the two bottom panels 
corresponds to $^{208}$Pb.
For $x=10^{-3}$ (two left panels), 
the BH contribution is evaluated assuming a high-energy EIC setting with
$E_l=20$ GeV and $E_N=125$ GeV (for $^{40}$Ca) and $E_N=100$ GeV (for $^{208}$Pb).
For $x=5 \times 10^{-3}$ (two right panels),
the BH contribution is calculated using a low-energy EIC setting with
$E_l=11$ GeV and $E_N=30$ GeV (for $^{40}$Ca) and $E_N=24$ GeV (for $^{208}$Pb).

As one can see from Fig.~\ref{fig:DVCS_2011_tdep}, the $t$ dependence of the
DVCS and BH differential cross sections has the characteristic shape 
of the nuclear form factor squared, with distinct minima and maxima.
However, the minima of the DVCS cross section are shifted toward smaller $t$:
this is the effect of the leading twist nuclear shadowing in quark nuclear GPDs.
The shift of the minima toward smaller $t$ can be interpreted as an increase of 
the transverse size of the distributions of quarks in nuclei---this is exactly the
phenomenon that we discussed and quantified in Sec.~\ref{subsec:transverse_size}.

One has to note that the shift of the position of the minima in 
Fig.~\ref{fig:DVCS_2011_tdep} is very small. One way to enhance the effect is to
use lighter nuclei, such as, e.g., $^{4}$He and $^{12}$C (see Fig.~\ref{fig:DVCS_ALU_c12} and discussion below).
Also, the shift of the minima presented in Fig.~\ref{fig:DVCS_2011_tdep}
is smaller than that predicted
in Ref.~\cite{Goeke:2009tu} where a different model 
for nuclear shadowing in quark nuclear GPDs was used
(the larger shadowing in Ref.~\cite{Goeke:2009tu} leads to the larger shift of the minima).

While an access to the effect of nuclear shadowing in nuclear GPDs 
through the measurement of the DVCS cross section
seems to be problematic
(the shifts of the minima are probably too challenging to measure experimentally),
an interesting possibility is offered by the measurement of DVCS cross section asymmetries~\cite{Goeke:2009tu}.
These asymmetries are proportional to the interference between the DVCS
and BH amplitudes and, thus, use the large and well known BH amplitude to amplify the
generally smaller DVCS amplitude. 

One example is the beam-spin asymmetry, $A_{\rm LU}$, measured with the longitudinally 
polarized lepton beam and  unpolarized nuclear target.
In the leading twist approximation, $A_{\rm LU}$ for a spinless nuclear target reads~\cite{Goeke:2009tu}:
\begin{equation}
A_{\rm LU}(\phi)=-
\frac{8K(2-y) Z F_A(t)\Im m {\cal A}_{\rm DVCS}(\xi,t,Q^2)}
{\frac{1}{x}|{\cal A}_{\rm BH}|^2+\frac{x t {\cal P}_1(\phi) {\cal P}_2(\phi)}{Q^2}4 (1-y+y^2/2)|\Im m {\cal A}_{\rm DVCS}|^2} \sin \phi \,,
\label{eq:ALU}
\end{equation}
where $K \propto \sqrt{t_{\rm min}-t}$ is the kinematic factor~\cite{Belitsky:2001ns}; 
$Z$ is the nuclear charge; $F_A(t)$ is the nuclear electric form factor;
$\Im m {\cal A}_{\rm DVCS}$ is given by Eq.~(\ref{eq:sigma_dvcs2}); $\phi$ is the angle between the 
lepton and hadron scattering planes. The overall minus sign corresponds to the negatively
charged lepton beam.
To consistently work to the leading twist accuracy, one should use only the leading twist contributions
to ${\cal P}_1(\phi)$, ${\cal P}_2(\phi)$ and $|{\cal A}_{\rm BH}|^2$. However, in the considered 
kinematics, $Q^2=4$ GeV$^2$, $x=10^{-3}$ and small $t$, the higher twist effects are either absent
(at $\phi=90^{0}$) or numerically insignificant. Therefore, in the evaluation of 
$A_{\rm LU}$, we use the standard expressions for ${\cal P}_1(\phi)$, ${\cal P}_2(\phi)$ and $|{\cal A}_{\rm BH}|^2$~\cite{Belitsky:2001ns,Belitsky:2000vk}.

\begin{figure}[h]
\begin{center}
\epsfig{file=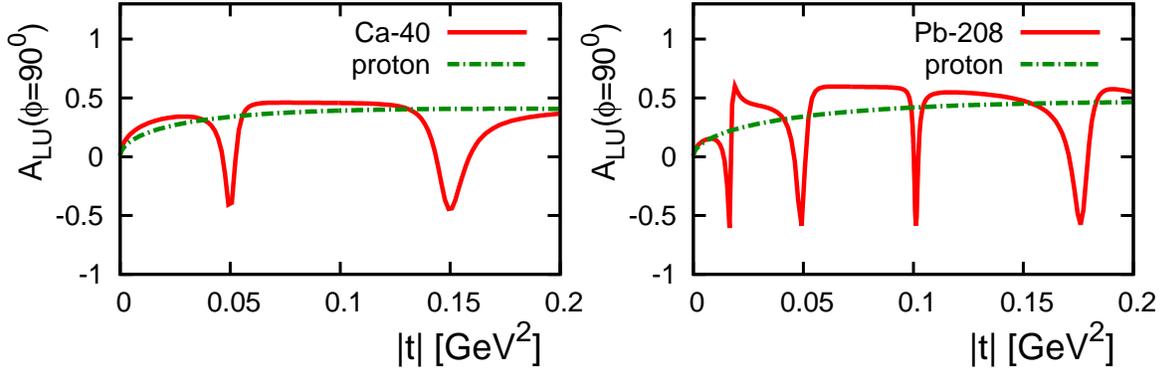,scale=1.5}
\caption{The DVCS beam-spin asymmetry at $\phi=90^{0}$, $A_{\rm LU}(\phi=90^{0})$,
as a function of $|t|$ at fixed $Q_0^2=4$ GeV$^2$ and $x=10^{-3}$.
The solid curves in the left and right panels correspond to $^{40}$Ca and $^{208}$Pb, respectively.
For comparison, $A_{\rm LU}(\phi=90^{0})$ for the free
proton is given by the dot-dashed curve.
}
\label{fig:DVCS_ALU}
\end{center}
\end{figure}

\begin{figure}[h]
\begin{center}
\epsfig{file=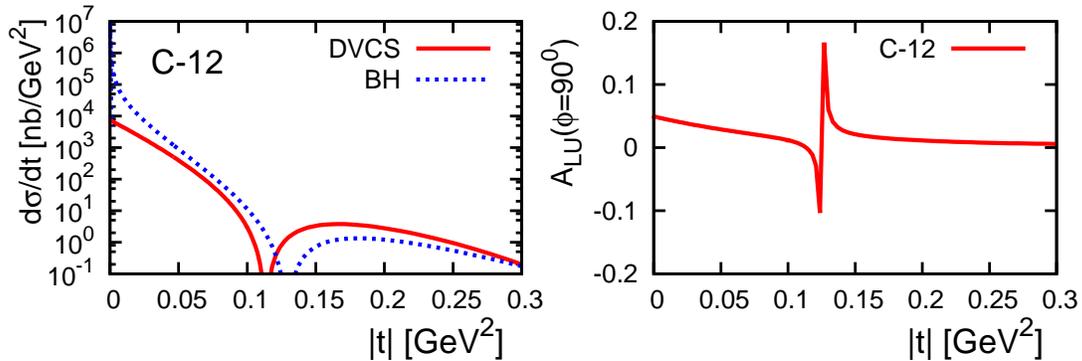,scale=1.4}
\caption{Left panel: The DVCS and BH cross sections for $^{12}$C
as functions of $|t|$ at $Q^2=4$ GeV$^2$ and $x=10^{-3}$.
Right panel: The DVCS beam-spin asymmetry for $^{12}$C, $A_{\rm LU}(\phi=90^{0})$,
as a function of $|t|$ at $Q^2=4$ GeV$^2$ and $x=10^{-3}$.
}
\label{fig:DVCS_ALU_c12}
\end{center}
\end{figure}

Figure~\ref{fig:DVCS_ALU} presents our predictions for $A_{\rm LU}(\phi=90^{0})$ as a function
of $|t|$ at fixed $Q_0^2=4$ GeV$^2$ and $x=10^{-3}$. The solid curves correspond to 
$^{40}$Ca (left panel) and $^{208}$Pb. For comparison, $A_{\rm LU}(\phi=90^{0})$ for the free
proton is given by the dot-dashed curve.
As in Fig.~\ref{fig:DVCS_2011_tdep}, the BH contribution 
to $A_{\rm LU}(\phi=90^{0})$
is evaluated assuming a high energy EIC setting with
$E_l=20$ GeV and $E_N=125$ GeV (for $^{40}$Ca) and $E_N=100$ GeV (for $^{208}$Pb).

Our predictions for $A_{\rm LU}(\phi=90^{0})$ for nuclei are rather remarkable. The sole reason
for the dramatic oscillations of $A_{\rm LU}(\phi=90^{0})$ as a function of $|t|$ is nuclear shadowing
that shifts the $t$ behavior of the DVCS amplitude relative to the BH one. Loosely speaking,
$A_{\rm LU}(\phi=90^{0})$ is given by the ratio of the imaginary part of the DVCS amplitude and
the BH amplitude, i.e., by the ratio of the solid and dotted curves from Fig.~\ref{fig:DVCS_2011_tdep}.

In more detail, the trend of the $t$ behavior of  $A_{\rm LU}$ in Fig.~\ref{fig:DVCS_2011_tdep}
can be explained as follows. At small $t$, $t \approx t_{\rm min}$, 
$A_{\rm LU}$ is vanishingly small due
to the kinematic factor $K$. As $|t|$ is increased, the kinematic factors rapidly increase 
$A_{\rm LU}$. At the same time, nuclear shadowing works in the opposite direction
and decreases $\Im m {\cal A}_{\rm DVCS}$, see Fig.~\ref{fig:IMP_ca40_2011}.
Near the position of the first minimum of the nuclear form factor,
$|t| \approx 0.05$ GeV$^2$ for $^{40}$Ca and $|t| \approx 0.01$ GeV$^2$ for $^{208}$Pb,
$\Im m {\cal A}_{\rm DVCS}$ changes sign and $A_{\rm LU}$ goes through zero and eventually
 reverses its sign. (Note that at these values of $|t|$, the nuclear electric form
factor $F_A(t)$ is still positive.)
As $|t|$ is further increased, $|\Im m {\cal A}_{\rm DVCS}|$ increases, which 
increases $|A_{\rm LU}|$. As $|t|$ is increased even further, the nuclear form factor
$F_A(t)$ changes sign and makes $A_{\rm LU}$ again positive.
The asymmetry stays positive until $\Im m {\cal A}_{\rm DVCS}$ changes the sign
near the second minimum of the nuclear form factor.
Then, the mechanism of the oscillations just described repeats itself.
In summary, the dramatic oscillations of $A_{\rm LU}$ as a function of $|t|$
is a result of the leading twist nuclear shadowing that causes the $t$ dependence
of the shadowing correction to $\Im m {\cal A}_{\rm DVCS}$ to be different from
that of the impulse approximation to $\Im m {\cal A}_{\rm DVCS}$.
(If the effect of shadowing is neglected, the $t$ dependence of the DVCS and BH
amplitudes is the same and $A_{\rm LU}$ behaves as in the free proton case, see
the dot-dashed curve in Fig.~\ref{fig:IMP_ca40_2011}.)

As we remarked earlier, in order to enhance the effect of the relative shift 
of the position of the minima of the differential DVCS and BH cross sections, 
one can consider light nuclei, such as, e.g., $^{4}$He and $^{12}$C.
Figure~\ref{fig:DVCS_ALU_c12} presents an example of such a study for 
$^{12}$C. In the left panel of the figure, the DVCS (solid curve) and BH
(dotted curve) cross sections are plotted as a function $|t|$ at fixed
$Q^2=4$ GeV$^2$ and $x=10^{-3}$. As one can see from the panel, the relative
shift of the position of the first minimum of the differential DVCS and BH 
cross sections for $^{12}$C is significantly larger than that for $^{40}$Ca
and $^{208}$Pb.
In the right panel of Fig.~\ref{fig:DVCS_ALU_c12}, the beam-spin asymmetry at
the angle $\phi=90^{0}$,
$A_{\rm LU}(\phi=90^{0})$, is plotted as a function of $|t|$ at 
$Q^2=4$ GeV$^2$ and $x=10^{-3}$. The asymmetry reveals the dramatic pattern of 
oscillations, similarly to the case of the heavier nuclei presented in
Fig.~\ref{fig:DVCS_ALU} and already discussed.

The DVCS beam-spin asymmetry enhances and accesses the imaginary part of the
DVCS amplitude. Its real part can be probed by considering the DVCS asymmetry
measured with the lepton beams of opposite signs (the beam-charge asymmetry)
or by studying the $\phi$-angle modulation of the unpolarized cross section.
While the imaginary part of the DVCS amplitude accesses GPDs at the
$x=\xi$ cross-over line, see Eq.~(\ref{eq:sigma_dvcs2}), the real part involves the GPDs
in the entire region of $x$, and, hence, is sensitive to the whole host of
nuclear effects (shadowing, antishadowing, EMC-type suppression, etc.).
The pattern of nuclear modifications of the real part of the nuclear DVCS
amplitude is quite different from the one for the imaginary part;
it was presented and discussed in Refs.~\cite{Freund:2003wm,Freund:2003ix}.

{\it Comment}. The differential measurements of coherent nuclear 
DVCS for small $t$ appear 
very challenging since the final nucleus is emitted at very small angles. 
At the same time, it is possible to suppress the inelastic contribution (at least for the heavy nuclei)  in the collider kinematics since the break-up of the nucleus is accompanied by the emission of neutrons which are easily detected by a zero-degree calorimeter.

\subsubsection{Predictions for the ratio of vector meson ($J/\psi$, $\Upsilon$) production} 

In the leading twist approach to hard exclusive processes,
exclusive electroproduction of heavy vector mesons ($J/\psi$, $\Upsilon$)
with nuclear targets probes the nuclear gluon GPD.
Similarly to the case of DVCS discussed above, at high energies, 
the theoretical description can be simplified and the answer is expressed in terms
of the usual nuclear gluon PDF. In particular, in the leading logarithmic approximation,
the cross section for coherent production of longitudinally polarized vector mesons in the
$t=0$ limit reads~\cite{Frankfurt:1995jw}:
\begin{equation}
\frac{d \sigma_{\gamma^{\ast}_LA \to V A}}{dt}|_{t=0}=\frac{4 \pi^3 \Gamma_V M_V}{3 \alpha_{\rm em}Q^6} \eta_V^2 T(Q^2) |1+i \beta|^2 \alpha_s(Q^2) [x g_A(x,Q^2)]^2 \,,
\label{eq:sigma_VM}
\end{equation}
where $\Gamma_V$ is the $V \to e^{+}e^{-}$ decay width;
$M_V$ is the vector meson mass; $\beta$ is the ratio of real to imaginary parts of 
the $\gamma^{\ast}A \to V A$ scattering amplitude;
the factor $T(Q^2)$ accounts for pre-asymptotic effects, i.e., $T(Q^2 \to \infty) = 1$. 
 The parameter $\eta_V$ is defined in terms
of the $q{\bar q}$ component of the vector meson light-cone wave function,
$\Phi_V^{q {\bar q}}$, 
\begin{equation}
\eta_V=\frac{1}{2} \frac{\int \frac{dz d^2 k_t}{z(1-z)}\Phi_V^{q {\bar q}}(z,k_k)}{\int dz d^2 k_t \Phi_V^{q {\bar q}}(z,k_k)} \,.
\end{equation}
In
the ratio of the vector meson production on the nucleus and the free proton, all fine details
of Eq.~(\ref{eq:sigma_VM}) to a good accuracy cancel and one obtains:
\begin{equation}
R_{\rm VM} \equiv \frac{d \sigma_{\gamma^{\ast}_LA \to V A}/dt(t=0)}{A^2 d \sigma_{\gamma^{\ast}_L N \to V N}/dt(t=0)}=\left(\frac{g_A(x,Q^2)}{A g_N(x,Q^2)}\right)^2 \,.
\label{eq:R_VM}
\end{equation}
Therefore, the ratio $R_{\rm VM}$ is a sensitive direct measure of the effects of nuclear shadowing
and antishadowing in the nuclear gluon distribution.
We neglected here the effect of skewness which can be accounted for 
by substituting $x$ in Eq.~(\ref{eq:R_VM}) by $x_+$.
The ratio $r=x_+/x$ depends on $Q^2$ and the mass of the produced vector meson. For example, for the photoproduction 
of $J/\psi$, $r\approx 1$ and it slowly decreases with $Q^2$ to $r \approx 1/2$, while for  the $\Upsilon$ case, $r\sim 1/2$ already for $Q^2 =0$~\cite{Frankfurt:1997fj}.

\begin{figure}[h]
\begin{center}
\epsfig{file=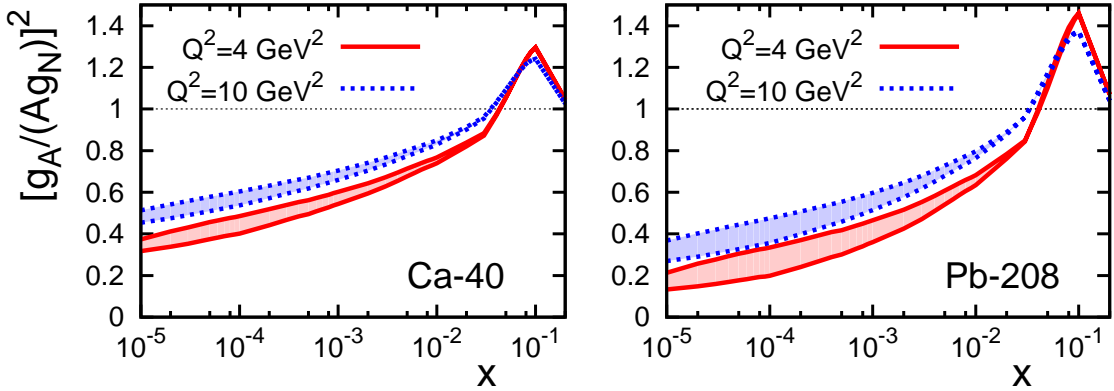,scale=1.5}
\caption{The ratio $R_{\rm VM}=(g_A(x,Q^2)/[A g_N(x,Q^2)])^2$ as a function of $x$ at
$Q^2=4$ GeV$^2$ (solid curves) and $Q=10$ GeV$^2$ (dotted curves). 
The shaded bands reflect the theoretical uncertainty of our predictions.}
\label{fig:LT2009_pb208_g2}
\end{center}
\end{figure}

Figure~\ref{fig:LT2009_pb208_g2} presents our predictions for the ratio
$R_{\rm VM}=[g_A(x,Q^2)/(A g_N(x,Q^2))]^2$ as a function of $x$ for $Q^2=4$ and 10 GeV$^2$. 
The shaded bands reflect the theoretical uncertainty of our predictions: 
the lower boundaries of the 
bands correspond to model FGS10\_H; the upper boundaries correspond to model FGS10\_L.
As one can see from Fig.~\ref{fig:LT2009_pb208_g2}, the suppression of $R_{\rm VM}$ for small $x$
due to nuclear shadowing
and the enhancement around $x \approx 0.1$ due to antishadowing are very significant.

Since the $t$ dependence of the nuclear gluon distribution is known, 
one can also study the $t$ dependence of hard exclusive heavy 
vector meson
production. Again, it is convenient to present the results in terms of the ratio
of the production on the nucleus and on the nucleon, $R_{\rm VM}(t)$:
\begin{equation}
R_{\rm VM}(t) \equiv \frac{d \sigma_{\gamma^{\ast}_LA \to V A}/dt}{A^2 d \sigma_{\gamma^{\ast}_L N \to V N}/dt} \approx
\frac{d \sigma_{\gamma^{\ast}_LA \to V A}/dt}{A^2 d \sigma_{\gamma^{\ast}_L N \to V N}/dt(t=0)}
=\left(\frac{g_A(x,Q^2,t)}{A g_N(x,Q^2)}\right)^2 
\,,
\label{eq:R_VM_t}
\end{equation}
where $g_A(x,Q^2,t)=H_A^g(x,\xi=0,t,Q^2)$ given
by Eq.~(\ref{eq:xiAlimit}). Note that in Eq.~(\ref{eq:R_VM_t}), we neglected the
weak $t$ dependence of the nucleon gluon GPD compared to that of the nucleus one and
assumed that the entire $t$ dependence comes from the gluon nuclear GPD.
The latter is a good approximation for heavy nuclei.

\begin{figure}[h]
\begin{center}
\epsfig{file=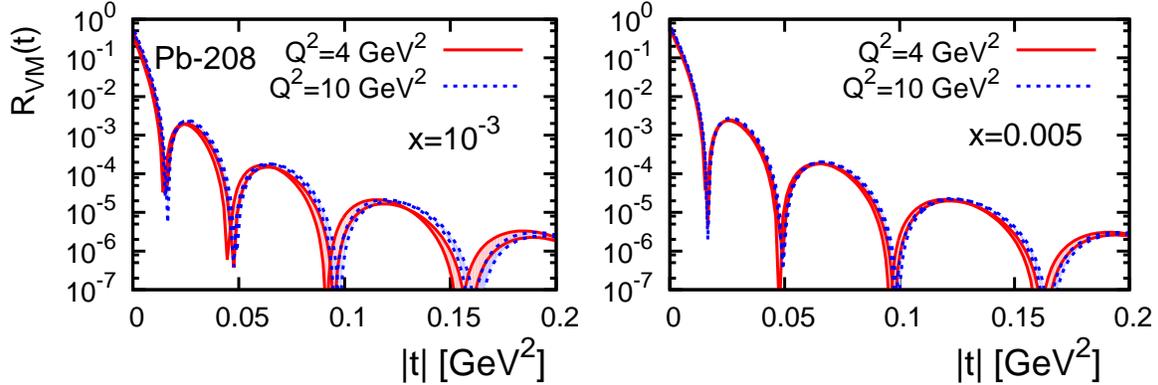,scale=1.5}
\caption{The ratio $R_{\rm VM}(t)=(g_A(x,Q^2,t)/[A g_N(x,Q^2)])^2$ 
of Eq.~(\ref{eq:R_VM_t}) as a function of $|t|$ at
$Q^2=4$ and 10 GeV$^2$ and $x=10^{-3}$ and $x=5 \times 10^{-3}$
The shaded bands reflect the theoretical uncertainty of our predictions.
}
\label{fig:JPsi_tdep}
\end{center}
\end{figure}

Figure~\ref{fig:JPsi_tdep} presents our predictions for the ratio $R_{\rm VM}(t)$
of Eq.~(\ref{eq:R_VM_t}) for $^{208}$Pb as a function of $|t|$ at  
$Q^2=4$ and 10 GeV$^2$ and $x=10^{-3}$ and $x=5 \times 10^{-3}$.
The shaded bands (barely distinguishable at the given scale along the $y$-axis) 
span the predictions for the gluon nuclear shadowing made using models FGS10\_H
and  FGS10\_L.
One can see from Fig.~\ref{fig:JPsi_tdep} that the $Q^2$ and $x$ dependence of 
$R_{\rm VM}(t)$ is rather weak. Also, the $t$ behavior of $R_{\rm VM}(t)$ is very 
similar to that of $d\sigma_{\rm DVCS}/dt$ in Fig.~\ref{fig:DVCS_2011_tdep};
the 
difference is that $R_{\rm VM}(t)$ probes the gluon nuclear GPD, while 
$d\sigma_{\rm DVCS}/dt$ probes the quark nuclear GPDs.
Since the effect of shadowing is larger in the gluon channel, the shift of the
positions of the minima toward smaller $t$ of $d \sigma_{\gamma^{\ast}_LA \to V A}/dt$
with respect to the impulse approximation is larger than the shift between the coherent
nuclear DVCS and BH cross sections presented in  Fig.~\ref{fig:DVCS_2011_tdep}.
 
Photoproduction of $J/\psi$ on nucleons and nuclei in the dipole formalism
was discussed in~\cite{Caldwell:2010zza}. It was suggested that  high precision measurement of the transverse momenta of $J/\psi$ would allow one to determine the  $t$ dependence of the $\gamma + A \to J/\psi + A $ amplitude  and, hence, determine 
the transverse distribution of gluons in nuclei using a Fourier transform of the amplitude. 
Unfortunately the processes $\gamma + A \to J/\psi+ A^{\ast}$, where $A^{\ast}$ denotes excited nuclear states which decay into the ground state and a photon (photons), 
dominate the cross section beyond the first minimum 
making it very difficult to observe the coherent channel beyond 
the first minimum~\cite{White:2010tu}.
At the same time, if the precision measurements of the $t$ dependence at $ t\sim 0$ 
are feasible, it would be possible to check the change of the $t$ slope as compared to the impulse approximation, which we find to be of the order of $10-15$\% depending on $x$, 
see the discussion of $\langle b^2 \rangle$ in Sec.~\ref{subsec:transverse_size}.

{\it Comment.}
In this section, we estimated the ratio of the vector meson production on the 
nucleus and the free proton, $R_{VM}$, using next-to-leading order (NLO) nuclear and proton gluon 
distributions, where the latter was given by the CTEQ5M fit. Our estimate is subject to several
theoretical uncertainties including (i) the implementation of the effect of skewness, and
(ii) the uncertainty in the choice of a parameterization of the gluon distribution 
in the free proton. Our initial analysis shows that the latter uncertainty dominates. 
Moreover, while the relevant formulas were derived in the leading-order (LO) approximation, 
we assumed that the form of the expression for $R_{VM}$ is the same in the NLO approximation.
Hence, we used the NLO CTEQ5M parameterization for the gluon density 
which corresponds to the energy-dependence of the cross section of electroproduction of
vector mesons on a nucleon which is close to the experimentally observed one.

\subsection{Nuclear effects in inclusive leading hadron spectra in $eA$ collisions}

Numerous data on hadron-nucleus scattering at fixed-target
energies indicate that the multiplicities of leading hadrons,
$N_A(z)$,
\begin{equation}
N_A(z)\equiv {1 \over \sigma_{\rm inel}(aA)}{d\sigma(z)^{a+A\to h +X}\over dz} \,,
\label{eq:Nz}
\end{equation}
strongly decrease with an increase of $A$. 
In definition~(\ref{eq:Nz}),
$z$ is the light-cone fraction of the 
projectile ``$a$'' momentum carried by hadron ``$h$''.
On the contrary, 
for inclusive hadron production in DIS, 
the QCD factorization theorem implies
that no such dependence 
should be present  
in the kinematics where the leading twist contribution dominates. 
This indicates that there should be an interesting 
transition from the soft physics dominating the interactions of
 real photons with nuclei  to the hard physics
in the inclusive hadron production in the DIS kinematics. 
(At sufficiently high energies, the transition  to the leading twist regime will be delayed by the presence of a significant $Q^2$ range, where the interaction is 
close to the black disk limit---see the discussion in Sec.~\ref{sec:bdr}.) 
The transition between the soft and hard regimes
should be
manifested in the disappearance
of the $A$ dependence of the leading spectra at large $z$:
\begin{equation}
N_A(z,Q^2)=N_N(z,Q^2), ~{\rm for}~ z\ge 0.2, Q^2 \ge ~{\rm few} ~{\rm GeV}^2 \,. 
\end{equation}
At small $x$, a new interesting phenomenon 
should emerge in DIS within the LT approximation
because of
 the presence of diffraction and nuclear shadowing for smaller $z$.
Indeed, 
diffraction originates from the presence 
of partons with relatively small
virtualities in the $\gamma^{\ast}$ wave function, which screen the color of the leading parton(s)
 with large  virtualities
and can rescatter elastically from  the target (several target nucleons in
the case of a nuclear target). Inelastic interactions of these soft partons
with several nucleons should lead to a
plenty of new revealing phenomena in small-$x$ $eA$ DIS,
which  resemble hadron-nucleus scattering,  but with a
shift in the rapidity from $y_{\rm max}({\rm current})$.
(It is important that far from the BDR, the pQCD part in not shadowed.)
This shift can be 
expressed through the average masses of the hadron states
produced in the diffraction:
\begin{equation}
y_{\rm soft\,partons} \sim  y_{\rm max}- \ln \left(\left<M^2_X\right>/\mu^2\right) \,,
\label{gapy}
\end{equation}
where $\mu \sim 1$ GeV is the typical soft  QCD scale. 
In the notation used for diffraction in DIS, one has
$\left<M^2_X\right>= Q^2 \left<(1/\beta - 1)\right>$, where $\beta$ 
is the light-cone fraction of the interacting parton with respect to the Pomeron
exchange, see Eq.~(\ref{eq:ft2}).
Using the information on 
the $\beta$ distributions measured at HERA 
(2006 H1 fit B~\cite{Aktas:2006hy}), 
one concludes that  for  moderate $Q^2$,
 $\left<M^2_X\right> \sim 3\, Q^2$.
Note also that the values of $\left<M^2_X\right>$ are different for the quark
and gluon channels: $\left<M^2_X\right> \sim 2\,  Q^2$ for the quarks
and $\left<M^2_X\right> \sim 6\, Q^2$ for the gluons.

The partons with  the rapidities  given by Eq.~(\ref{gapy}) 
 interact in multiple collisions and
loose their energy, which leads to a dip in the ratio 
$\eta_A(y)\equiv {N_A(y)/N_p(y)}$. At the same time, these multiple 
interactions should generate larger multiplicities at smaller rapidities.
The application of the Abramovsky-Gribov-Kancheli (AGK) rules~\cite{Abramovsky:1973fm} indicates that 
for $y \le y_{\rm soft\,partons} -\Delta$, where $\Delta = 2-3$,
the hadron multiplicity in the case of DIS off nuclei will be enhanced
by the factor $\eta_A(y)$: 
\begin{equation}
\eta_A(y)={AF_{2N}(x,Q^2) \over F_{2A}(x,Q^2) } \equiv \frac{A}{A_{\rm eff}} \,.
\end{equation}
One should note that
this estimate neglects the effects of the energy conservation and the increase of multiplicities at central rapidities with an increase of $W$. In the case of deuteron-gold collisions, 
the observed increase of multiplicity is smaller than that given by 
the AGK rules by a factor of 0.7, cf.~the discussion in
Ref.~\cite{Frankfurt:2007rn}, which is in line with the account of energy-momentum
conservation.

\begin{figure}[h] 
\begin{center} 
\epsfig{file=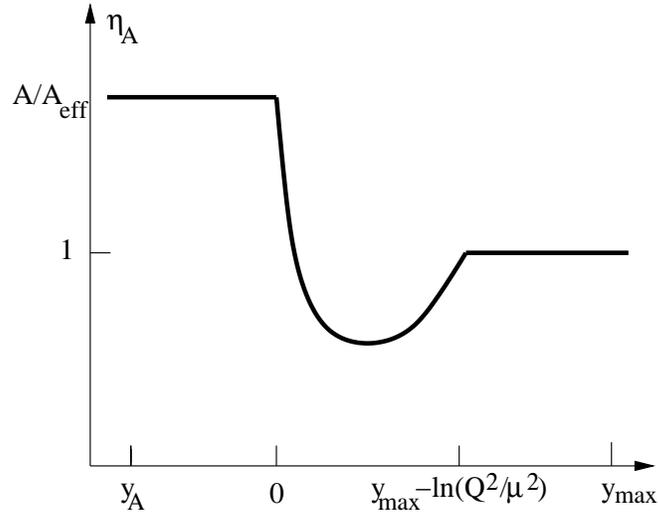,scale=0.45}
\vskip 0cm
\caption{A sketch of the ratio of the nuclear to free proton hadron multiplicities,
$\eta_A(y)=N_A(y)/N_p(y)$ as a function of the hadron rapidity.}
\label{ydist}
\end{center}
\end{figure}

At the rapidities close to the nuclear rapidities, a further increase of 
$\eta_A(y)$ is possible because of the  formation of hadrons inside the nucleus.
A sketch of the expected rapidity dependence of $\eta_A(y)$ is presented 
in Fig.~\ref{ydist}.

The discussed phenomenon should be more pronounced for collisions at central impact parameters leading to a correlation between the number of particles produced in the nucleus fragmentation region and the depletion at the rapidities given by Eq.~(\ref{gapy}).

With an increase of $Q^2$ at fixed $x$, an increasingly larger fraction of collisions occurs due to the scattering off the partons that had large enough $x$ at the initial $Q^2$ scale where nuclear shadowing is absent.
For these events, multiple interactions are absent and 
do 
not contribute significantly to the processes where several nucleons of the nucleus are wounded. Hence, though the fraction of the events with  long-range correlations described above should drop with $Q^2$ similarly to the decrease of the overall shadowing effect, the strength of
the correlations in the events, where several nucleons are wounded, would remain strong.

\subsection{Hadron production at central rapidities}

We discussed in Sec.~\ref{subsect:agk} that the application of the AGK cutting 
rules allows one to write the inelastic (non-diffractive) cross section as a sum of the positive cross sections with exactly $j$ nucleons involved in the inelastic interactions, $\sigma_j$, see Eq.~(\ref{eq:bertocchi_b}).
In the approximation when the fluctuations of the effective rescattering 
cross section $\sigma_X$  are neglected, 
the probability of the interaction with $j$ nucleons is~\cite{Bertocchi:1976bq}
[see Eq.~(\ref{eq:bertocchi_b})]:
\begin{equation}
p_j \equiv \frac{\sigma_j}{\sigma_{\rm summed}^{hA, {\rm inel}}}={\frac{A!}{(A - j)! j!} \int d^2 b \, [x(b)]^{j} [1 - x(b)]^{A - j} \over  
\int d^2 b \,[1- (1 - x(b))^A]} \,,
\label{pj}
\end{equation}
where $x(b)=\sigma_{X}^{\rm inel} T_A(b)$; $\sigma_{X}^{\rm inel}$
 is the non-diffractive component of $\sigma_{X}$, which
is given by Eq.~(\ref{eq:diff_pdf9}); $T_A(b)$ is the nuclear optical density.
To account for  the  fluctuations of $\sigma_{X}^{\rm inel}$, 
one would have to add the integral over $\sigma_{X}^{\rm inel}$
 with the measure $P(\sigma_{X}^{\rm inel})$ 
both in the numerator and denominator of Eq.~(\ref{pj}). 
This would obviously lead to a broader distribution over the number of ''wounded'' nucleons (see below).

The average number of the ''wounded'' nucleons, $\nu=\sum^{A}_{n = 1} j p_j$, satisfies  the relation which is an example of the so-called AGK cancellation~\cite{Bertocchi:1976bq},
\begin{equation} 
\nu \equiv \sum^{A}_{n = 1} j p_j={A\sigma_{X}^{\rm inel} \over 
\int d^{2} b \,[1- (1 - x(b))^A]} \,,
\label{eq:nu}
\end{equation}
where the denominator has the meaning of the inelastic 
cross section for the  interaction of the diffractive  configuration,
$\sigma_{\rm summed}^{hA, {\rm inel}}$.
Since the fraction of diffractive events (which also include the processes with the break-up of the nucleus) 
in $eA$ scattering is larger than in $eN$ scattering, the number of wounded nucleons
$\nu$ is somewhat larger  than the  $A\sigma_{\rm tot}^{\gamma^{\ast} N}/\sigma_{\rm tot}^{\gamma^{\ast} A}$ ratio.

In addition, one also expects a number of phenomena 
resulting from the existence of
the long-range correlations in rapidity. These include: \\
(a) Local fluctuations of the multiplicity
in the central rapidity region, e.g., the observation of a broader 
distribution of the number of particles per unit of rapidity due to the 
fluctuations of the number of wounded nucleons~\cite{Frankfurt:1991nx}.
These fluctuations should be larger for the hard processes induced by gluons,
for example, for the direct photon production of two high $p_t$ dijets.\\
(b) The correlation of the central multiplicity with the multiplicity of 
neutrons in the forward
neutron detector.

To illustrate the effect of the broadening of the 
distribution over multiplicities, we calculate the 
$A$ dependence of the multiplicity distribution 
for a fixed rapidity range. We use the H1 analysis of
charged particle multiplicities in DIS at HERA
in the pseudorapidity range $1 \le \eta^{\ast} \le 2$ 
for events without rapidity gaps~\cite{Aid:1996cb}.
The multiplicity distribution is characterized by the 
probabilities to produce $n$ particles, $P_n$, which can be described by the negative
binomial distribution:
\begin{equation}
P_n(k,\langle n \rangle)=\frac{k(k+1) \dots (k+n-1)}{n!} \left(\frac{\langle n \rangle}{\langle n \rangle+k} \right)^n \left(\frac{k}{\langle n \rangle+k} \right)^k \,.
\label{eq:Pn}
\end{equation}
In Eq.~(\ref{eq:Pn}), $\left<n\right>$ is the average multiplicity; 
the parameter $k$ characterizes the dispersion of the distribution over $P_n$.
Both $\left<n\right>$ and $k$ are found from fitting to the data.
In particular, for the H1 data~\cite{Aid:1996cb} and for $1 \le \eta^{\ast} \le 2$ and
$80 \leq W \leq 115$ GeV, $\left<n\right>=2.52 \pm 0.10$ and $1/k=0.285 \pm 0.080$.

The probabilities $P_n$ can also be described using the generating function $G(z)$,
\begin{equation}
G(z)=\left[1 +r(1-z)\right]^{-k}\equiv \sum_{n=0}^{\infty}
P_n(k,\langle n \rangle) \,z^n \,,
\end{equation}
where $r=\left<n\right>/k$.

To estimate the probabilities $P_n$ for nuclear targets, one 
notices that the generating function for the production of particles in the interaction with $m$ nucleons,
$G_m(z)$, is simply given by 
\begin{equation}
G_m(z)=\left[G(z)\right]^m \,.
\label{gener}
\end{equation}
Therefore, the probability to produce $n$ particles in lepton-nucleus DIS, $P_n^A$,
can be introduced by the following relation,
\begin{equation}
G_m(z)=\left[1 +r(1-z)\right]^{-km}\equiv \sum_{n=0}^{\infty}
P_n^A(k^{\prime},\langle n^{\prime} \rangle) \,z^n \,,
\label{gener_A}
\end{equation}
where $k^{\prime}= m k $ and $\langle n^{\prime} \rangle=m \langle n \rangle$.
 As follows from Eq.~(\ref{gener_A}),
the explicit expression for $P_n^A$ is given by Eq.~(\ref{eq:Pn}) after the replacement
$k \to mk$ and $\langle n \rangle \to m \langle n \rangle$.

In our numerical analysis, we neglected the fluctuations in the number of the nucleons
participating in the particle production and used 
$m=\nu$, where $\nu$ is the average number of the wounded nucleons given by
Eq.~(\ref{eq:nu}). To evaluate $\nu$, we used $\sigma_{X}^{\rm inel} =24$ mb 
[see Eq.~(\ref{eq:diff_pdf9})], which
corresponds to $\sigma_{\rm soft}^{q(\rm H)}=29$ mb (the effective rescattering cross section in the sea-quark channel, model FGS10\_H) at $Q^2=4$ GeV$^{2}$ and $x\sim 10^{-3}$. This value of Bjorken $x$
approximately corresponds to the kinematics of the H1 analysis of $P_n$ in the $ep$ case discussed above~\cite{Aid:1996cb}.

\begin{figure}[t] 
\begin{center} 
\epsfig{file=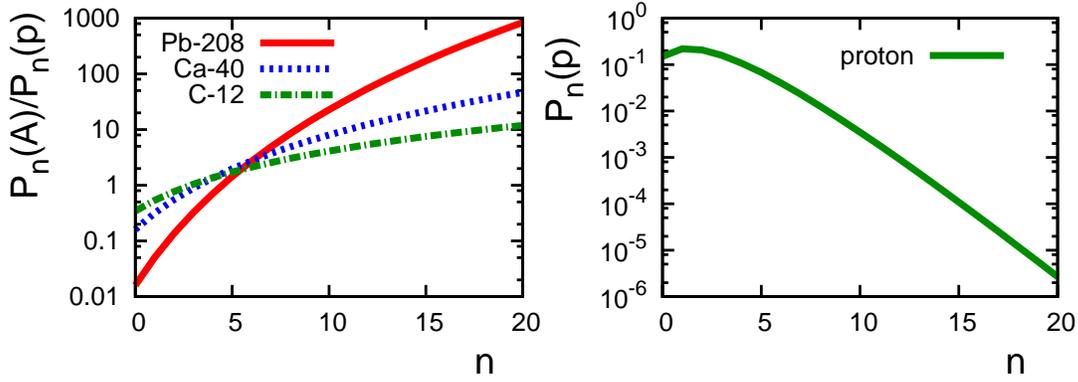,scale=1.4}
\vskip 0cm
\caption{The multiplicity distributions of Eqs.~(\ref{pj})-(\ref{gener_A})
as a function of the number of produced particles $n$.
Left panel: The ratio of the nucleus to proton probabilities
$P_n^A/P_n$ as a function of $n$.
Right panel: The absolute value of the free proton $P_n$ as a function of $n$.
The curves correspond to
the fixed pseudorapidity interval $1 \le \eta^{\ast} \le 2$, 
$Q^2=4$ GeV$^{2}$ and $x\sim 10^{-3}$.}
\label{multi}
\end{center}
\end{figure}

The results of the calculation of the multiplicity 
distributions for events without rapidity gaps
 using Eqs.~(\ref{pj})-(\ref{gener_A}) are presented in Fig.~\ref{multi}.
In the left panel of the figure, we examine the $A$ dependence
and plot the ratio of the nucleus to proton probabilities
$P_n^A/P_n$ as a function of $n$.
One can see from  Fig.~\ref{multi} that a much broader distribution 
over multiplicity is predicted
for heavy nuclei. Measurements of such distributions would
 serve as a complementary 
(to the measurement of the diffractive cross sections) probe of the dynamics 
of nuclear shadowing. 
While the nuclear enhancement of $P_n^A/P_n$ for large $n$ is very large, the
absolute value of the probabilities is tiny. It is illustrated in the right 
panel of  Fig.~\ref{multi} where we plot the free proton $P_n$ as a function of $n$. 

With an increase of $Q^2$, the pattern of 
fluctuations in the number of collisions will become more involved since the 
fluctuations of $\sigma_{X}^{\rm inel}$ should increase because of the mixing of 
the unshadowed and shadowed contributions. This will effectively correspond to 
the presence of the fluctuations with both small and large cross sections, 
$\sigma_{X}^{\rm inel}\sim 0$ and
$\sigma_{X}^{\rm inel} \sim \sigma^{X}_{\rm inel}(Q_0^2)$, 
respectively.

The results of the calculation of the fraction of events with enhanced 
central multiplicities that we just discussed should be considered as an estimate only. 
The eikonal formulas do not properly take into account the energy-momentum conservation. 
In particular,  quark-gluon 
configurations of the projectile propagating through the nucleus 
are counted $n$-times in the $n$-''Pomeron''
cuts. Since the multiplicity in the 
central region depends on energy, accounting for this effect would lead to 
some suppression of events with double, triple and higher multiplicity.

Another effect neglected in 
our analysis
 is the 
gluon radiation by a quark propagating through the nuclear medium. 
It is a relatively small effect in the leading twist approximation, for a review, see 
Ref.~\cite{Baier:2000mf}. However, with the onset of the black disk regime, 
the pattern changes quite strongly, see the discussion in Sec.~\ref{sec:bdr}.

\section{Leading twist nuclear shadowing and suppression of hard coherent 
diffraction in proton-nucleus scattering}
\label{sec:pA}

In the last two decades intensive studies of hard processes with rapidity gaps 
were performed at $p\bar p$ colliders at CERN and FNAL, for a review, see Ref.~\cite{Goulianos:2004as}. These processes are usually  referred to as hard diffractive processes. Examples of such types of processes studied at the Tevatron include
the production of two jets, $Z$, and $W$ bosons in the reaction
\begin{equation}
p+ \bar p \to 2\, {\rm jets}\, (Z,\,W,...) + X + {\rm gap} + p \,.
\label{gappp}
\end{equation}
The probability of the presence of rapidity gaps in the events with hard 
subprocess, $P_{\rm gap}$, is rather small. 
At the Tevatron, applying a cut on Feynman $x$ of the recoiled proton, $x_F(p) \ge 0.93$,
one finds that 
$P_{\rm gap} \approx 0.01$~\cite{Goulianos:2004as,Goulianos:1995wy}.
(The cross section of each hard diffractive channel constitutes
approximately 1\% of the contribution of
the corresponding channel to the inclusive $p\, \bar{p}$ cross section.)
In such processes, multiple soft and hard interactions are not suppressed and the QCD factorization theorem, which is valid for hard diffraction in DIS~\cite{Collins:1997sr}, 
is not applicable. Hence, one expects a strong suppression of 
$P_{\rm gap}$ compared to the naive application of the QCD factorization theorem with
the diffractive PDFs measured at HERA. Qualitatively, 
the breakdown of factorization is due to the dominance of small impact parameters in 
the diffractive PDFs and nearly 100\% absorption in $pp$ scattering at 
such
impact parameters. The magnitude of the suppression---a factor of 5 $\div $ 10---is in agreement with such expectations, see e.g., the discussion in Ref.~\cite{Frankfurt:2004ti}.
The factorization breaking can also be explained by the absorptive effects
associated with multi-Pomeron exchanges that
 make the gap survival very unlikely in the case of hadronic  
collisions~\cite{Kaidalov:2001iz}.

Since we developed the theory of nuclear diffractive PDFs
(see Sec.~\ref{sec:final_states}), we can make predictions for the rates of the $pA$ reaction analogous to the elementary reaction in Eq.~(\ref{gappp}):
\begin{equation}
p+ A \to 2\, {\rm jets}\, (Z,\,W,...) + X + {\rm gap} + A \,.
\label{gappA}
\end{equation}
This involves the combination of the effects of soft absorption in $pA$ 
scattering and the model of the diffractive PDFs
as well as the use of the leading twist theory of nuclear shadowing.
Using hard coherent production of two jets in proton-nucleus scattering
as an example,
we demonstrate that soft multiple 
rescatterings lead to the factorization breaking of hard diffraction in proton-nucleus  
scattering which is larger than that in hadron-nucleon scattering.
We also compare the hard diffractive and 
electromagnetic (ultraperipheral)
mechanisms of 
the reaction~(\ref{gappA})
in the
RHIC and LHC kinematics and make quantitative estimates of the feasibility to
study the high-energy real photon-proton scattering at RHIC and the LHC.
Numerical results presented in this section update the results of our earlier
work on this subject~\cite{Guzey:2005ys}.

\subsection{Soft coherent diffraction off nuclei}
\label{subsec:soft_coherent_diffraction}

Before proceeding to calculate the gap survival probability, we need to briefly review 
main elements of the theory of inelastic coherent diffraction off nuclei. 
The Gribov formalism does not make specific predictions for this process as it requires a more detailed information about the hadron--multi-Pomeron vertices. 

To address this problem one can use the concept of color fluctuations which we already discussed in the context 
of multiple rescatterings in the calculation of nuclear PDFs, see Sec.~\ref{sec:phen}.
One introduces a distribution over the strength of interactions in the projectile $h$,  $P_h(\sigma)$. 
The probability $P_h(\sigma)$
is constrained by the following relations:
\begin{eqnarray}
&&\int d \sigma P_h(\sigma)=1 \,, \nonumber\\
&&\int d \sigma \sigma P_h(\sigma)=\sigma_{\rm tot} \,, \nonumber\\
&&\omega_{\sigma} \equiv \frac{\int d \sigma\, \sigma^2\, P_h(\sigma)}{\left[\int d \sigma\,
\sigma\, P_h(\sigma) \right]^2}-1 
= \frac{d\sigma(h+p\to M_X +p)/dt}{d\sigma(h+p\to h +p)/dt}(t=0) \,.
\label{eq:Pconst}
\end{eqnarray}
In Eq.~(\ref{eq:Pconst}), the first line is the probability conservation sum rule; the
second line is the relation to the total cross section; the third line is the relation
between the parameter $\omega_{\sigma}$ that characterizes the dispersion of the distribution
$P_h(\sigma)$ and the ratio of the differential and elastic cross sections.
In addition, pQCD constrains the $\sigma\to 0$ behavior of $P(\sigma)$. This information 
and Eq.~(\ref{eq:Pconst})
in combination with the information on coherent inelastic diffraction in 
hadron-deuteron scattering allows one to reconstruct $P_h(\sigma)$ for 
hadron-proton interactions.

Using the color fluctuation formalism it is possible to derive the expression for
the cross section of 
inelastic coherent diffraction off nuclei 
(diffractive dissociation)~\cite{Frankfurt:2000ty,Frankfurt:1993qi,Guzey:2005tk,Strikman:1995jf}:
\begin{equation}
\sigma_{DD}^{hA}=\int d^2 \, b \, \left(
  \int d \sigma P_h(\sigma)\left|\Gamma_A(b,\sigma) \right|^2 -
 \left| \int d \sigma P_h(\sigma)\Gamma_A(b,\sigma)\right|^2 \right) \,.
\label{scf6}
\end{equation}
In Eq.~(\ref{scf6}),
$b$ is the impact parameter; 
$\Gamma_A(b,\sigma)$ is the amplitude of interaction of
 the projectile in the configuration with the cross section $\sigma$ with the nucleus at the impact parameter $b$, cf.~\cite{Glauber:1970jm}:
\begin{equation}
\Gamma_A(b,\sigma)=1-\exp \left(-\frac{A}{2} \, \sigma \, T_A(b) \right)
 \,,
 \label{scf5}
\end{equation}
where $T_A(b)=\int dz \,\rho_A(b,z)$ and 
$\rho_A$ is the nuclear density normalized to unity.
The energy-dependence of $\sigma_{DD}^{hA}$ is determined by the 
energy-dependence of $P_h(\sigma)$, 
which is modeled based on the available experimental information and 
extrapolations to the LHC  energies~\cite{Guzey:2005tk}.
Note that in Eq.~(\ref{scf5}),  we neglected the slope of the elementary
hadron-nucleon scattering amplitude compared to the 
slope of the nuclear form factor
and
assumed that the elementary scattering amplitude is purely imaginary, 
which is a good approximation at high energies.
Calculations based on Eq.~(\ref{scf5}) describe well the data on the coherent diffraction at 
the fixed-target energies, which are unfortunately rather limited, see Ref.~\cite{Frankfurt:2002kd} 
for a review.

The notion of $P_h(\sigma)$ 
provides 
 a compact phenomenological description
of soft coherent diffraction in hadron-nucleon and
hadron-nucleus scattering.
The function $P_h(\sigma)$ describes the probability that the incoming hadron 
interacts with target nucleons with a given cross section $\sigma$.
In other words, $P_h(\sigma)$ describes cross section fluctuations
in the energetic projectile. 
As follows from Eq.~(\ref{scf6}),
ignoring cross section fluctuation, i.e., setting 
$P_h(\sigma) \propto \delta(\sigma-\sigma_{{\rm tot}})$, would result in 
the unphysical result $\sigma_{DD}^{hA}=0$ that once again 
illustrates that diffraction is not possible without fluctuations 
of the strength of the interaction.

The formalism of cross section fluctuations is 
rather general 
and is based on the space-time picture of
the strong interaction that we discussed in Sec.~\ref{subsect:gribov_picture}.
In this picture, the incoming hadron
is represented by a coherent superposition of eigenstates of the
scattering operator. Since different eigenstates correspond to
different $\sigma$, the scattered state is in general different from
the incoming state, but it has the same quantum numbers.
This corresponds to the process of diffractive dissociation.
We also used the formalism of cross section fluctuations in our analysis
of nuclear shadowing in $eA$ DIS in Sec.~\ref{sec:hab_pdfs} and \ref{sec:phen}.

One should note that the formalism of cross section fluctuations 
implicitly uses the assumption of the completeness of the scattering
eigenstates and, hence, it is 
applicable only at $t \approx 0$. At $t \neq 0$, the diffractive
final state can be produced as a result of 
the elastic scattering of a constituent of the projectile off the target even if there are no fluctuations of the strength of the projectile-target interaction. 
A simple example is scattering of the deuteron off the proton target in the impulse approximation. For $t=0$, the cross section of the process $^2Hp \to pnp $ is equal to zero due 
to orthogonality of the $^2$H and continuum wave functions, 
while for finite $t$, the cross section is 
$\propto 1-F^2_{\rm D}(t)$,  where $F_{\rm D}(t)$ is the deuteron form factor.

The function $P_h(\sigma)$ is different for different projectiles:
protons, pions, photons. 
(The pion case was considered in Sec.~\ref{subsubsec:color_fluct}.)
In the case of the proton projectiles (especially at the collider energies),  
$P_h(\sigma)$ has a narrow
dispersion around $\langle \sigma \rangle=\sigma_{{\rm tot}}$, where $\sigma_{{\rm tot}}$
is the total proton-nucleon cross section. Hence,
a good approximation to Eq.~(\ref{scf6}) can be given in the following form: 
 \begin{equation}
 \sigma^{hA}_{DD} \approx {\omega_{\sigma}\sigma_{{\rm 
 tot}}^2 A^2 \over 4}
  \int d^2b  \,T^2_A(b)e^{-A \left<\sigma\right> T(b)} \,.
  \label{eq:approx}
  \end{equation}
  Equation~(\ref{eq:approx}) can be interpreted as follows. 
  Since the fluctuations are small, we need to take them into account only in the interaction with one of the nucleons at a given impact parameter $b$.
  The corresponding scattering amplitude squared
(in the impact parameter space) 
is proportional to $A^2 \omega_{\sigma}\,\sigma_{{\rm tot}}^2\,T^2_A(b)$. 
On the way through the nucleus, the fluctuation 
is partially absorbed (suppressed) with the $\langle \sigma \rangle=\sigma_{\rm tot}$
characteristic cross section. 
The corresponding soft suppression factor can be read off directly from 
Eq.~(\ref{eq:approx}):
\begin{equation}
T_{{\rm soft}}^{\,pA}=\exp\left(-A \sigma_{{\rm tot}}\, T_A(b) \right) \,.
\label{eq:T_soft}
\end{equation}
 One can see from Eq.~(\ref{scf6}) that only scattering off the periphery of the nucleus 
contributes to the cross section of diffraction. 
Numerical results indicate a strong drop of soft diffraction with an increase of energy
as a result of the decrease of cross section fluctuations (dispersion $\omega_{\sigma}$).
For instance, $\omega_{\sigma}$ decreases from $\omega_{\sigma}=0.15$ at the Tevatron energies
($\sqrt{s}=546$ and 1800 GeV) to  $\omega_{\sigma}=0.1$
at $pA$ scattering energies at the LHC ($\sqrt{s} \approx 9$ TeV) and 
further down to 
$\omega_{\sigma}=0.065$ at $pp$ scattering energies at the LHC 
($\sqrt{s}=14$ TeV)~\cite{Guzey:2005tk}.

\subsection{Hard coherent diffraction in $pA$ scattering}
\label{hardpA}

In the case of hard coherent $pA$ diffraction, we need to take into account 
two effects: 
the reduction of the soft diffraction due to the suppression of the fluctuations in a wide range of impact parameters and the  reduction of the nuclear diffractive PDFs due to the shadowing effects which we discussed in Sec.~\ref{sec:final_states}.
The soft suppression factor, $T_{\rm soft}^{pA}$, should be compared to the factor that 
suppresses hard coherent diffraction in $\gamma^{\ast}A$ DIS. 
Let us recall Eq.~(\ref{eq:masterD_approx}) that expresses the nuclear diffractive
PDF $f_{j/A}^{D(3)}$ in the very high-energy limit, which allows one to neglect
the effect of the finite coherence length.
Equation~(\ref{eq:masterD_approx})  can be interpreted as follows. 
The incoming virtual photon fluctuates into its hard diffractive 
component long time before the photon interacts with the target.
The hard diffractive component elastically rescatters on 
the target nucleus, which gives the suppression factor 
$[1-\exp(-\frac{A}{2}\sigma_{\rm soft}^j(x,Q^2)\, T_A(b))]^2$, and emerges as the
final hard diffractive state. 
(Remember that $\sigma_{\rm soft}^j(x,Q^2)$ describes the strength of the interaction of the configurations which are involved in the hard diffraction in DIS.)
Note that we set $\eta=0$ in Eq.~(\ref{eq:masterD_approx})
to be consistent with our discussion of soft diffraction.

It is important to note that the expression for 
$f_{j/A}^{D(3)}$ in Eq.~(\ref{eq:masterD_approx}) corresponds 
to diffractive dissociation of the virtual photon 
[the corresponding expression for the hadron-nucleus case is given by 
Eq.~(\ref{eq:approx})]
 since the elastic contribution to DIS is absent (suppressed by the 
smallness of $\alpha_{{\rm e.m.}}$). 
Therefore, the analogy between Eqs.~(\ref{eq:approx}) and (\ref{eq:masterD_approx})
enables us to introduce the attenuation factor characterizing the suppression of
hard coherent diffraction in DIS on nuclear targets
due to the effect of  nuclear shadowing (multiple interactions with the target nucleons),
\begin{equation}
T_{{\rm hard}}^{\,\gamma^{\ast} j/A}=\exp\left(-A \sigma_{\rm soft}^j(x,Q)\, T_A(b) \right) \,.
\label{eq:T_hard}
\end{equation}

As we discussed in Sec.~\ref{subsec:derivation}, in general, the calculation of $T_{{\rm hard}}^{\,\gamma^{\ast} j/A}$
is model-independent only for the interaction
with two nucleons. For the interaction with $N \ge 3 $ nucleons,
 we implicitly used the 
color fluctuation approximation in Eq.~(\ref{eq:T_hard}). 
This approximation is equivalent to the observation of the small dispersion of 
$P_h(\sigma)$
used in the derivation of Eq.~(\ref{eq:approx}), see details in 
Sec.~\ref{subsec:derivation}, and takes into account the presence of the point-like configurations in the interaction of a hard probe with the nucleon which contribute to the total cross section but not to the diffractive cross section.

\subsection{Suppression factor for hard proton-nucleus coherent diffraction}
\label{subsec:suppression_hard_pA}

As an example of the hard coherent diffractive process on heavy 
nuclear targets, we consider
the hard coherent diffractive production of two jets in the reaction
$p+A \to 2\,{\rm jets}+X+A$. In this process, $A$ denotes the nucleus; 
$X$ denotes the soft diffractive component; 
the invariant mass of the jets provides the hard scale.

The $p+A \to 2\,{\rm jets}+X+A$ cross section
can be readily obtained by 
generalizing the well-known expression for
the dijet inclusive cross section in hadron-hadron scattering~\cite{Ellis} 
and by introducing the new quantity, the screened nuclear diffractive PDFs
$\tilde{f}_{j/A}^{D(3)}$:
\begin{samepage}
\begin{eqnarray}
&&\frac{d^3 \sigma^{p+A \to 2\,{\rm jets}+X+A}}{d x_1 \,d p_T^2 \,d x_{\Pomeron}} 
\nonumber\\
&& \propto \sum_{i,j,k,l=q,\bar{q},g} f_{i/p}(x_1,Q_{{\rm eff}}^2)\tilde{f}_{j/A}^{D(3)}(x_2,Q_{{\rm eff}}^2,x_{\Pomeron}) \overline{\sum}|{\cal M}(ij \to kl)|^2 \frac{1}{1+\delta_{kl}} 
\,,
\label{eq:cs_hard}
\end{eqnarray}
\end{samepage}
where $f_{i/p}$ are the usual proton PDFs;
 $\overline{\sum}|{\cal M}(ij \to kl)|^2$ are the invariant matrix elements for
two-to-two parton scattering given in Table~7.1 of~\cite{Ellis};
$x_1$ and $x_2$ are the light-cone momentum fractions of the active quarks of 
the proton and the nucleus, respectively;
$p_T$ is the transverse momentum of each of the jets in the final state;
$Q_{{\rm eff}}^2$ is the effective hard scale of the process.
For the simplification of our analysis, 
we consider only the case of the $90^{0}$ hard
 parton scattering in the center of mass frame, which constrains $x_1$ (as a function
of $x_2=\beta \,x_{\Pomeron}$) and 
$Q_{{\rm eff}}^2$:
\begin{equation} 
x_1=\frac{4 \, p_T^2}{\beta x_{\Pomeron} \,s} \,, \quad \quad Q_{{\rm eff}}^2=4\, p_T^2 \,, 
\label{eq:x1}
\end{equation} 
where $\sqrt{s}$ is the proton-nucleon invariant mass.
The term ''screened PDF'' means that this parton distribution contains
certain soft suppression effects, i.e., the screened PDF is suppressed compared to
the analogous PDF extracted from hard processes.

\begin{figure}[t]
\begin{center}
\epsfig{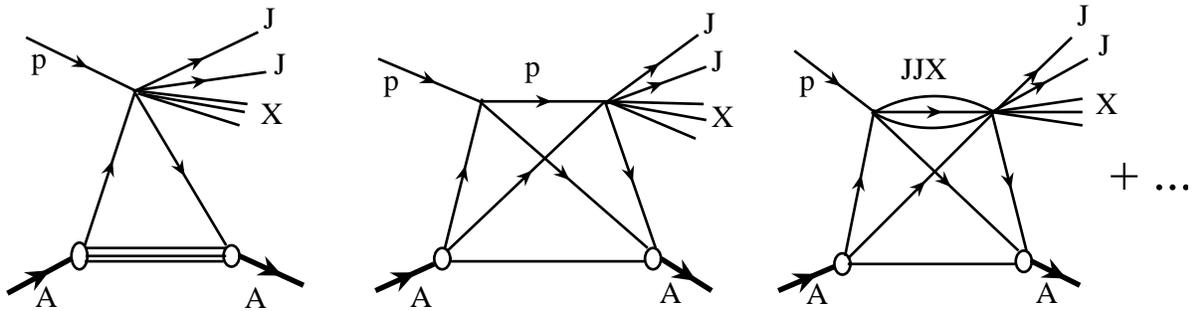}
\vskip 0cm
\caption{Graphs representing first terms of the multiple
scattering series for the $p+A \to 2\,{\rm jets}+X+A$ scattering amplitude.}
\label{fig:2jets_FG}
\end{center}
\end{figure}

The derivation of the expression for the screened nuclear diffractive PDFs, $\tilde{f}_{j/A}^{D(3)}$, is carried out
similarly to the derivation of Eq.~(\ref{eq:approx}),
 see also Fig.~\ref{fig:2jets_FG}. 
The final expression reads:
\begin{eqnarray}
\hspace{-0.5cm}
\tilde{f}_{j/A}^{D(3)}(x,Q^2,x_{\Pomeron})
 &=&  4\,\pi A^2 \tilde{f}_{j/N}^{D(4)}(x,Q^2,x_{\Pomeron},t_{\rm min}) \int d^2 b \,T_A^2(b)\,
e^{-A (\sigma_{{\rm tot}}(s)+\sigma_{\rm soft}^j(x,Q^2)) T_A(b)} \nonumber\\
&=&  4\,\pi A^2 B_{\rm diff} \tilde{f}_{j/N}^{D(3)}(x,Q^2,x_{\Pomeron}) \int d^2 b \,T_A^2(b)\,
e^{-A (\sigma_{{\rm tot}}(s)+\sigma_{\rm soft}^j(x,Q^2)) T_A(b)}
\,,
\label{eq:npdf_effective}
\end{eqnarray} 
where $\tilde{f}_{j/N}^{D(4)}$ is the screened diffractive PDF of the nucleon, 
which enters the QCD description of the $p+p \to 2\,{\rm jets}+X+p$ reaction
[see Eqs.~(\ref{eq:diffractive_slope2}) and (\ref{eq:screen}) and 
their discussion below];
$\sigma_{{\rm tot}}(s)$ is the total proton-nucleon cross section.
In Eq.~(\ref{eq:npdf_effective}), we neglected the
slope and the real part of the elementary 
$p+N \to 2\,{\rm jets}+X+N$
scattering amplitude and a small longitudinal momentum
transfer in the  $p+N \to 2\,{\rm jets}+X+N$ vertex.
In the second line of Eq.~(\ref{eq:npdf_effective}), 
we assumed that the $t$ dependence of the screened
diffractive PDFs is the same as that of the usual ones, see Eq.~(\ref{eq:data7})
 and used the following relation:
\begin{equation}
\tilde{f}_{j/N}^{D(4)}(x,Q^2,x_{\Pomeron},t_{\rm min})=B_{\rm diff}
\tilde{f}_{j/N}^{D(3)}(x,Q^2,x_{\Pomeron}) \,.
\label{eq:diffractive_slope2}
\end{equation}

For comparison with the nuclear case [see Eq.~(\ref{eq:cs_hard})], we also
present the cross section of 
the hard coherent diffractive dijet production in proton-proton scattering:
\begin{eqnarray}
&&\frac{d^3 \sigma^{p+p \to 2\,{\rm jets}+X+p}}{d x_1 \,d p_T^2 \,d x_{\Pomeron}} 
\nonumber\\
&&\propto \sum_{i,j,k,l=q,\bar{q},g} f_{i/p}(x_1,Q_{{\rm eff}}^2)
\tilde{f}_{j/N}^{D(3)}(\beta,Q_{{\rm eff}}^2,x_{\Pomeron})
 \overline{\sum}|{\cal M}(ij \to kl)|^2 \frac{1}{1+\delta_{kl}} \nonumber\\
&&\equiv  r_{{\rm h}} \sum_{i,j,k,l=q,\bar{q},g} f_{i/p}(x_1,Q_{{\rm eff}}^2)
f_{j/N}^{D(3)}(\beta,Q_{{\rm eff}}^2,x_{\Pomeron})
 \overline{\sum}|{\cal M}(ij \to kl)|^2 \frac{1}{1+\delta_{kl}}
\,.
\label{eq:cs_hard2}
\end{eqnarray}
The screened diffractive PDF of the nucleon, $\tilde{f}_{j/N}^{D(3)}$,
is introduced through 
the suppression factor $r_{{\rm h}}$ and the usual diffractive PDF of the nucleon
$f_{j/N}^{D(3)}$:
\begin{equation}
\tilde{f}_{j/N}^{D(3)} \equiv r_{{\rm h}}\,f_{j/N}^{D(3)} \,.
\label{eq:screen}
\end{equation}
The suppression factor $r_{{\rm h}}$ takes into
account the significant factorization breaking in hard hadron-hadron diffraction
(see the discussion in the beginning of this section).
In our numerical analysis, we used the following model for $r_{{\rm h}}$:
\begin{equation}
r_{{\rm h}}=\frac{0.75}{N(s)}=0.75\,\left(\int^{0.1}_{1.5/s} dx_{\Pomeron} \int^{0}_{-\infty} dt \, f_{\Pomeron / p}(x_{\Pomeron},t)\right)^{-1} \,.
\label{eq:goulianos}
\end{equation}
This expression is based on the phenomenological model of~\cite{Goulianos:1995wy},
 which describes the suppression of 
 diffraction at the Tevatron ($\sqrt{s}=546$ and 1800 GeV) by 
rescaling the Pomeron flux, $f_{\Pomeron / p}(x_{\Pomeron},t)$, 
by the factor $N(s)$. In Eq.~(\ref{eq:goulianos}), the Pomeron flux is given
by the standard expression:
\begin{equation}
f_{\Pomeron / p}(x_{\Pomeron},t)=\frac{1}{x_{\Pomeron}^{1+2\,\epsilon+2\,\alpha^{\prime}t}} 
\frac{\beta^2_{\Pomeron pp}(t)}{16 \pi} \,,
\label{eq:pomeron_flux}
\end{equation}
where $\epsilon=0.1$; $\alpha^{\prime}=0.25$ GeV$^{-2}$;
$\beta_{\Pomeron pp}(t)$ is the $\Pomeron pp$ form factor~\cite{Goulianos:1995wy}.

We also introduced the additional factor of $0.75$ in Eq.~(\ref{eq:goulianos})
in order to phenomenologically take into account the observation that the effects of
factorization breaking should be larger in the 
elementary diffractive PDFs at $t=0$ [see Eq.~(\ref{eq:npdf_effective})] 
than in the $t$ integrated diffractive PDFs [see Eq.~(\ref{eq:goulianos})].

The application of Eq.~(\ref{eq:goulianos}) at the RHIC and LHC energies gives:
\begin{eqnarray}
r_{{\rm h}} &=& \frac{1}{4.2} \,, \quad {\rm RHIC} \,, \nonumber\\
r_{{\rm h}} &=& \frac{1}{16.0} \,, \quad {\rm LHC} \,.
\label{eq:goulianos3}
\end{eqnarray}
Note that in the simple model that we use here,  the amount of the suppression does not depend on the light-cone fraction $\beta$. However, the recent NLO QCD analysis~\cite{Klasen:2009bi} found that $r_{\rm h}$ in the $p\bar p$ scattering decreases with an increase of $\beta $ for large $\beta$. 
This effect would further suppress 
the hard diffraction mechanism as compared to the electromagnetic one (see discussion below).   

Returning to the case of hadron-nucleus scattering and the discussion of Eq.~(\ref{eq:npdf_effective}),
it is important to emphasize that in the case of hard coherent proton-nucleus
diffraction, the nuclear suppression factor, $T_{{\rm hard}}^{\,pA}$, is the
product of the soft and hard suppression factors introduced previously,
\begin{equation}
T_{{\rm hard}}^{\,pA}=T_{{\rm soft}}^{\,pA}\ T_{{\rm hard}}^{\,\gamma^{\ast}A} \,.
\label{eq:T_hard_pa}
\end{equation}
This can be understood from Fig.~\ref{fig:2jets_FG}, which represents 
first terms of the multiple scattering series
for the $p+A \to 2\,{\rm jets}+X+A$ scattering amplitude. 
The rescattering cross section in the middle graph is 
$\sigma_{\rm tot}(s)$;
the rescattering cross section in the right
graph is $\sigma_{\rm soft}^j(x,Q^2)$ (in the color fluctuation approximation).
Therefore, the resulting nuclear 
attenuation, which results from the sum of the middle and right graphs, is 
driven by the $\sigma_{\rm tot}(s)+\sigma_{\rm soft}^j(x,Q^2)$ effective cross section.
After the eikonalization (summing all graphs corresponding to the interaction with
all nucleons of the target), one obtains Eq.~(\ref{eq:T_hard_pa}).

Equation~(\ref{eq:npdf_effective}) can be interpreted 
in two complimentary ways.
On the one hand, one can start from soft diffractive dissociation
of protons on heavy nuclei, see Eq.~(\ref{eq:approx}).
Since we are interested in the hard diffractive component of the 
diffractive dissociation cross section, one has to take into account
the additional suppression of the nuclear diffractive PDFs given by the
$T_{{\rm hard}}^{\,\gamma^{\ast}A}$ factor. 
As a result, one arrives at 
Eq.~(\ref{eq:T_hard_pa}).
On the other hand, one can 
start from the expression for  inclusive diffraction of protons
on nuclei, which is proportional to the nuclear diffractive 
PDFs~(\ref{eq:masterD_approx}).
Since the final diffractive state contains a soft component, which is 
partially absorbed by the nucleus, one should take into account this suppression by
introducing  the $T_{{\rm soft}}^{\,pA}$ factor, which represents the probability 
of the absence of soft inelastic interactions at a given impact parameter $b$.

The comparison of the nuclear screened diffractive PDFs $\tilde{f}_{j/A}^{D(3)}$
to the nucleon screened diffractive PDFs $\tilde{f}_{j/N}^{D(3)}$, i.e.,
the comparison of the diffractive production of two jets in the $pA$ and $pp$ cases,
can be quantified  
by introducing the factor $\lambda^j$,
\begin{equation}
\lambda^j(x,Q^2) \equiv \frac{\tilde{f}_{j/A}^{D(3)}(x,Q^2,x_{\Pomeron})}{\tilde{f}_{j/N}^{D(3)}(x,Q^2,x_{\Pomeron})}= 4 \pi A^2 B_{{\rm diff}} \int d^2 b \,T_A^2(b)\, e^{-A (\sigma_{{\rm tot}}(s)+\sigma_{\rm soft}^j(x,Q^2)) T_A(b)} \,.
\label{eq:lambda_parton}
\end{equation}
In Eq.~(\ref{eq:lambda_parton}), $\sigma_{\rm soft}^j(x,Q^2)$ is the effective 
cross section
introduced and discussed in Sec.~\ref{subsubsec:color_fluct}, see Fig.~\ref{fig:sigma3_2009}.
For the 
proton-nucleon cross section, $\sigma_{{\rm tot}}(s)$, we use the Donnachie-Landshoff 
parameterization~\cite{Donnachie:1992ny}:
\begin{equation}
\sigma_{\rm tot}(s)=21.7 \,s^{0.0808}+56.08\,s^{-0.4525} \ {\rm mb} \,.
\label{eq:sigma_pp}
\end{equation}
Certain features of Eq.~(\ref{eq:lambda_parton}) deserve a discussion.
First,
while the diffractive PDFs depend separately on $\beta$ and
  $x_{\Pomeron}$, the
factor $\lambda^j$ depends only on their
 product $x=\beta \,x_{\Pomeron}$
in our approach [compare to the $f_{j/A}^{D(3)}/(A f_{j/N}^{D(3)})$ ratio
that also depends only on Bjorken $x$ in our approach].
Second, $\lambda^j$ is weakly flavor-independent 
because $\sigma_{{\rm tot}}(s) \gg |\sigma_{\rm soft}^g-\sigma_{\rm soft}^q|$
(again, it should be compared to
the $f_{j/A}^{D(3)}/(A f_{j/N}^{D(3)})$ ratio which is also weakly flavor-independent).
In addition, since the slope $B_{{\rm diff}}$ is independent on the 
hard scale $Q^2$ in our approach, $\lambda^j$ has a very weak dependence 
on $Q^2$.

\begin{figure}[t]
\begin{center}
\epsfig{file=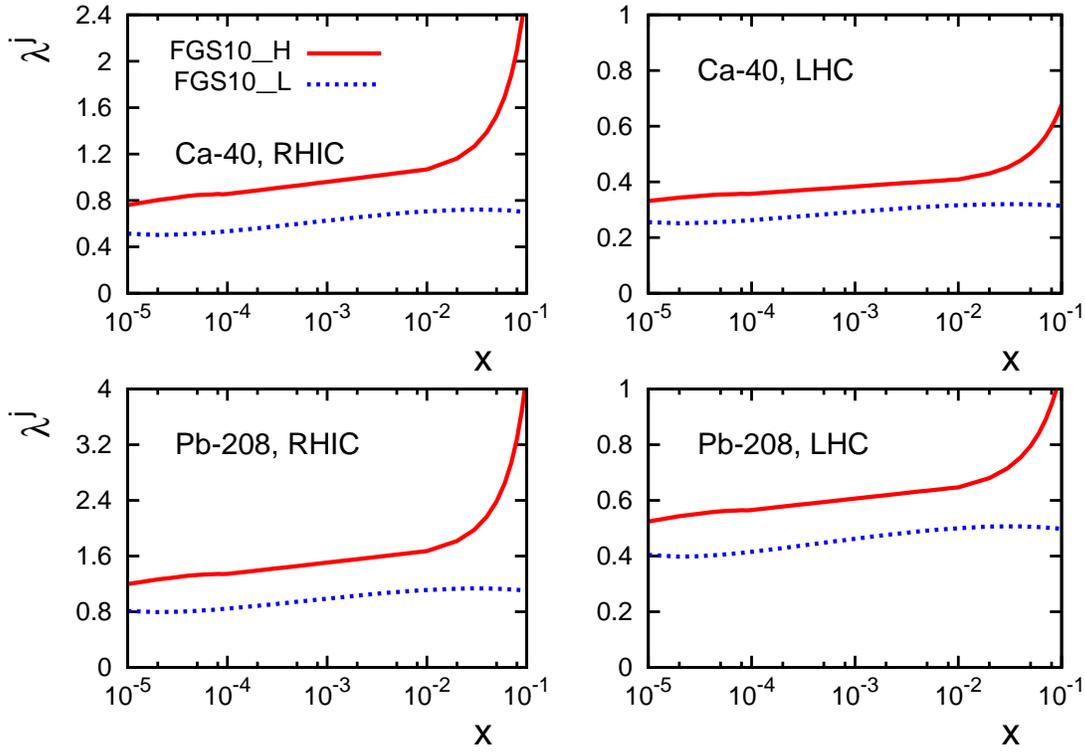,scale=1.4}
\caption{
The ratio of the nuclear to proton screened diffractive PDFs, 
 $\lambda^j=\tilde{f}_{j/A}^{D(3)}/\tilde{f}_{j/N}^{D(3)}$, see Eq.~(\ref{eq:lambda_parton}), as
a function of Bjorken $x$ at $Q^2=4$ GeV$^2$
in the RHIC and LHC kinematics.
The solid curves correspond to model FGS10\_H; the dotted curves correspond to model
FGS10\_L.
}
\label{fig:lambda_parton}
\end{center}
\end{figure}

\begin{figure}[t]
\begin{center}
\epsfig{file=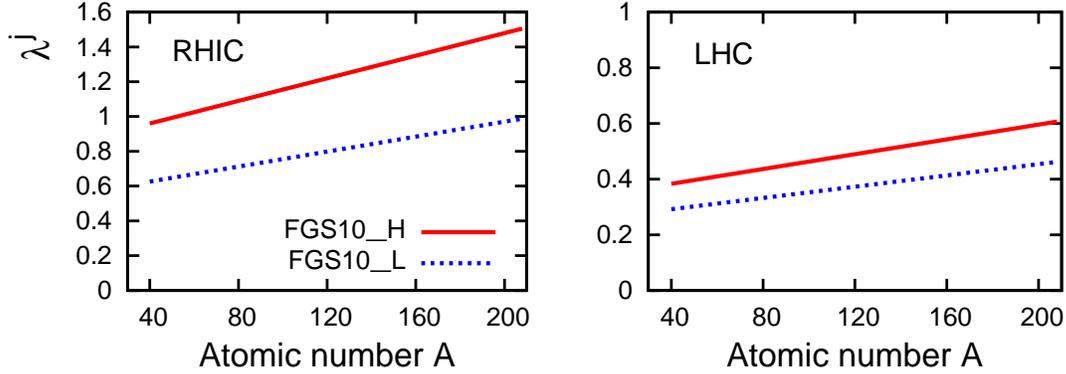,scale=1.4}
\caption{The ratio of the nuclear to proton screened diffractive PDFs, 
 $\lambda^j=\tilde{f}_{j/A}^{D(3)}/\tilde{f}_{j/N}^{D(3)}$, as
a function of the atomic number $A$ at $x=10^{-3}$ and $Q^2=4$ GeV$^2$
in the RHIC and LHC kinematics.  
The solid curves correspond to model FGS10\_H; the dotted curves correspond to model
FGS10\_L.
}
\label{fig:lambda_parton_adep}
\end{center}
\end{figure}

Figures~\ref{fig:lambda_parton} and \ref{fig:lambda_parton_adep} present
the results for the ratio of the nuclear to proton screened diffractive PDFs, 
 $\lambda^j=\tilde{f}_{j/A}^{D(3)}/\tilde{f}_{j/N}^{D(3)}$, see Eq.~(\ref{eq:lambda_parton}),
 in the RHIC and LHC kinematics.
The RHIC kinematics corresponds to $\sqrt{s}=200$ GeV;
the LHC kinematics corresponds to $\sqrt{s} \approx 9.9$ TeV per nucleon for proton-$^{40}$Ca
collisions and $\sqrt{s} \approx 8.8$ TeV per nucleon for proton-$^{208}$Pb
collisions, see Table~\ref{table:gamma} and Ref.~\cite{Morsch}.

Figure~\ref{fig:lambda_parton} shows $\lambda^j$ as a function of
 Bjorken $x$ 
at $Q^2=4$ GeV$^2$ for $^{40}$Ca and $^{208}$Pb.
 The solid curves correspond to model FGS10\_H and
the dotted curves correspond to model  FGS10\_L.
All curves are for the ${\bar u}$-quark flavor; the 
predictions for the gluon channel differ from the presented
ones insignificantly.
Despite the fact that  $\lambda^j$ is of the order of unity, the 
corresponding suppression of hard diffraction is very large because in the
absence of the suppression, nuclear diffractive PDFs 
would have been
 enhanced compared to
the nucleon diffractive PDFs by the very large factor
 $f_{j/A}^{D(3)}/f_{j/N}^{D(3)} \propto A^{4/3}$.

Figure~\ref{fig:lambda_parton_adep} presents the $A$ dependence of  
$\lambda^j$ at fixed $x=10^{-3}$ and $Q_0^2=4$ GeV$^2$, i.e., at fixed 
$\sigma_{\rm soft}^j(x,Q_0^2)$.
 As seen from
Fig.~\ref{fig:lambda_parton_adep}, the 
$A$ dependence of $\lambda^j$ is rather slow. A simple fit gives
that $\lambda \propto A^{0.28}$ at both the RHIC and LHC.

\subsection{Hard diffraction and ultraperipheral proton-nucleus collisions}
\label{sec:ultraperipheral}

\begin{figure}[h]
\begin{center}
\epsfig{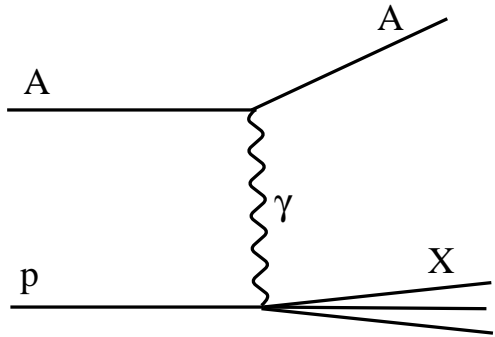}
\caption{The ultraperipheral $p+A \to X+A$ scattering.}
\label{fig:ultraperipheral}
\end{center}
\end{figure}

In proton--heavy-nucleus collisions (for example, $^{208}$Pb),
most of the diffractive events ($\sim 80\%$) will be generated by the
scattering of the proton off the coherent nuclear Coulomb field at large
impact parameters,
$p+A \to p+\gamma+A \to X+A$~\cite{Guzey:2005tk}, see
Fig.~\ref{fig:ultraperipheral}.
\vspace*{0.5cm}
These ultraperipheral proton-nucleus
collisions open a possibility for studies of hard photon-proton interactions
 at extremely high energies 
and allow one to probe the gluon density in the proton at the values of 
Bjorken $x$, which are a factor of ten smaller (for the same virtuality)
 than those probed at HERA~\cite{Baur:2001jj,Bertulani:2005ru,Strikman:2005yv}.

In Sec.~\ref{subsec:suppression_hard_pA}, 
we studied  jet production 
in hard coherent proton--heavy-nucleus diffraction, $p+A \to 2\,{\rm jets}+X+A$;
in this section, we 
compare it to  the production
of hard jets by the photon-proton interaction, where the photon is 
coherently produced by the elastically recoiled nucleus,
$p+A  \to p+\gamma+A \to 2\,{\rm jets}+X+A$.
This corresponds to the situation when the generic final state $X$ in
Fig.~\ref{fig:ultraperipheral} contains a hard two-jet component and
a soft remaining part $X$.

Qualitatively, we expect that the rate of the hard dijet
production through the hard diffractive mechanism will be 
significantly smaller than that  due to the 
 ultraperipheral mechanism
because of the 
following two suppression effects.
First, hard diffractive dijet production is suppressed by the $\lambda^j$ 
factor discussed in Sec.~\ref{subsec:suppression_hard_pA}, 
see Figs.~\ref{fig:lambda_parton}
and \ref{fig:lambda_parton_adep}.
 Second, the shapes of the parton distributions in the photon and 
screened nuclear 
diffractive PDFs are rather different. In the photon case, 
the  dominant contribution to the photon PDFs comes from the $\beta \sim 1$ region
 corresponding to the kinematics where a pair of jets is at the rapidities
 close to the gap.
In 
the screened nuclear diffractive PDFs at large
 virtualities,
which are relevant for the measurements at the LHC,
the main contribution comes from small $\beta$, see Fig.~3 
of Ref.~\cite{Frankfurt:2003gx}.

As we just mentioned, hard coherent dijet production in proton-nucleus
scattering can proceed via the e.m.~mechanism, when
the nucleus coherently emits a quasi-real photon which interacts with the proton
and diffractively produces two hard jets,
$p+A  \to p+\gamma+A \to 2\,{\rm jets}+X+A$,
see Fig.~\ref{fig:ultraperipheral}.
The 
corresponding cross section can be written as a sum of the
resolved and direct photon contributions (the separation into the 
resolved and direct components is only meaningful in the leading-order
calculation):
\begin{eqnarray}
\hspace{-0.75cm}
\frac{d^3 \sigma^{p+A \to 2\,{\rm jets}+X+A}_{{\rm em}}}{d x_1 d p_T^2 d x_{\Pomeron}}  &\propto& r_{{\rm em}}\sum_{i,j,k,l=q,\bar{q},g} f_{i/p}(x_1,Q_{{\rm eff}}^2)\frac{dN_{\gamma}(x_{\Pomeron})}{d x_{\Pomeron}}
f_{j/\gamma}(\beta,Q_{{\rm eff}}^2) \overline{\sum}|{\cal M}(ij \to kl)|^2 \frac{1}{1+\delta_{kl}} \nonumber\\
& +& \sum_{i,j,k,l=q,\bar{q},g} f_{i/p}(x_1,Q_{{\rm eff}}^2)\frac{dN_{\gamma}(x_{\Pomeron})}{dx_{\Pomeron}} \delta(\beta-1) \overline{\sum}|{\cal M}(i\gamma \to kl)|^2 \frac{1}{1+\delta_{kl}}
\,,
\label{eq:cs_em}
\end{eqnarray}
where $dN_{\gamma}(x_{\Pomeron})/dx_{\Pomeron}$ is the flux of equivalent photons \cite{Baur:2001jj} expressed 
in terms of $x_{\Pomeron}$ instead of the photon energy $\omega$ 
($\omega=x_{\Pomeron} p_{\rm lab}$, where $p_{\rm lab}$ is the momentum
of the nucleus in the laboratory frame);
$f_{j/\gamma}$ is the PDF of the real photon;  $\overline{\sum}|{\cal M}(i\gamma \to kl)|^2$
are the invariant matrix elements for the direct photon-parton scattering,
 see Table 7.9 in
\cite{Ellis}; $r_{{\rm em}}$ is a phenomenological factor describing the
factorization breaking for the resolved (hadron-like) component of
the real photon. The exact value of  $r_{{\rm em}}$ is uncertain: It
ranges from $r_{{\rm em}}=0.34$~\cite{Klasen:2004ct}
 to $r_{{\rm em}} \approx 1$
with large errors~\cite{Chekanov:2001bw}. Since our analysis is a simple leading-order 
estimate, we conservatively take $r_{{\rm em}}=0.5$.

The flux of equivalent photons approximately 
equals~\cite{Bertulani:1987tz}:
\begin{eqnarray}
\frac{dN_{\gamma}(x_{\Pomeron})}{dx_{\Pomeron}} &=& \frac{Z^2 \alpha_{{\rm em}} \omega}{\pi^2 \gamma^2} \int_{|b| \geq R_A} d^2b \left[K_1^2 \left(\frac{\omega |b|}{\gamma}\right)+\frac{1}{\gamma^2}
K_1^2 \left(\frac{\omega |b|}{\gamma}\right) \right]_{\big|\omega=x_{\Pomeron} p_{\rm lab}} \nonumber\\
&=& \frac{2 Z^2 \alpha_{{\rm em}}}{\pi x_{\Pomeron}} \left[x K_0(x)K_1(x)+\frac{x^2}{2}
\left(K_0^2(x)-K_1^2(x)\right)\right] \,,
\label{eq:flux}
\end{eqnarray}
where $Z$ is the nuclear charge; $\gamma$ is the Lorentz factor
of the fast moving nucleus; $R_A$ is the effective nuclear radius, 
$R_A=1.145\,A^{1/3}$; $p_{{\rm lab}}$ is the momentum of the nucleus in 
the laboratory frame. Table~\ref{table:gamma} summarizes the values of 
$p_{\rm lab}$ and $\gamma$ that we used in our analysis, see also \cite{Baltz:2007kq}.
In the table, $\sqrt{s_{NN}}$ is the invariant energy of the $pA$ collision per nucleon;
$\gamma=\sqrt{s_{NN}}/(2 m_N)$; the proton beam energy is 250 GeV for RHIC and 7 TeV 
for the LHC (we assume the maximal energy for the latter).
\begin{table}[h]
\begin{tabular}{|c|c|c|c||c|c|c|}
\hline
Nucleus & $p_{\rm lab}$, RHIC & $\sqrt{s_{NN}}$, RHIC & $\gamma$, RHIC & 
$p_{\rm lab}$, LHC &
$\sqrt{s_{NN}}$, LHC & $\gamma$, LHC \\
\hline
$^{40}$Ca & 125 GeV & 354 GeV & 187 & 3.5 TeV & 9.9 TeV & 5280 \\ 
$^{208}$Pb & 100 GeV & 316 GeV & 169 & 2.76 TeV & 8.8 TeV & 4690 \\
\hline
\end{tabular}
\caption{The values of the momentum of the nucleus in the laboratory frame, $p_{\rm lab}$,
the invariant energy of the $pA$ collision per nucleon, $\sqrt{s_{NN}}$,
and the Lorentz factor of the fast moving nucleus, $\gamma$, in the RHIC and LHC 
kinematics used in our analysis. Note that we use that the proton beam energy is 250 GeV for RHIC and 7 TeV for the LHC. 
}
\label{table:gamma}
\end{table}

To quantify the comparison between the hard and e.m.~mechanisms of the 
$p+A \to 2\,{\rm jets}+X+A$ process, we introduce the ratio $R$:
\begin{equation} 
R(\beta,x_{\Pomeron},p_T)=\frac{d^3 \sigma_{\rm hard}^{p+A \to 2\,{\rm jets}+X+A}}{d x_1\,d p_T^2\,d x_{\Pomeron}} \Bigg/
 \frac{d^3 \sigma^{p+A \to 2\,{\rm jets}+X+A}_{{\rm em}}}{d x_1 \,d p_T^2 \,d x_{\Pomeron}} \,,
\label{eq:r}
\end{equation}
where the numerator and denominator are given by Eqs.~(\ref{eq:cs_hard})
and (\ref{eq:cs_em}), respectively
(we introduced the subscript ''hard'' for the cross section in Eq.~(\ref{eq:cs_hard})),
with the equal coefficients of proportionality.
In the simplified kinematics that we use, at given $p_T$ and $x_{\Pomeron}$, the ratio
$R$ depends only on $\beta$.

\begin{figure}[h]
\begin{center}
\epsfig{file=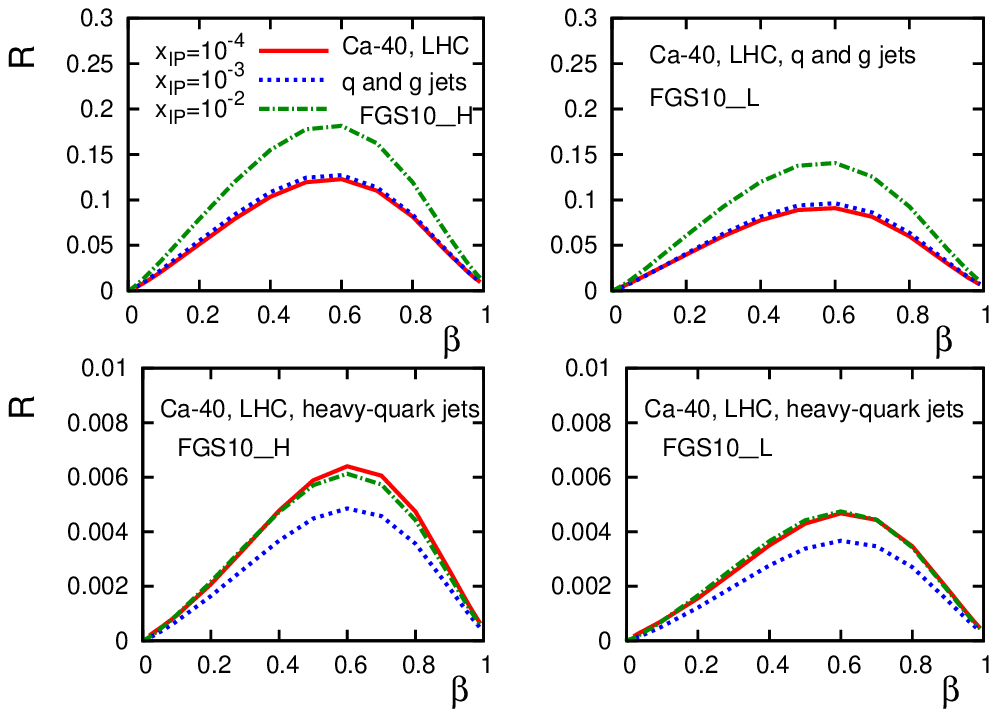,scale=1.3}
\epsfig{file=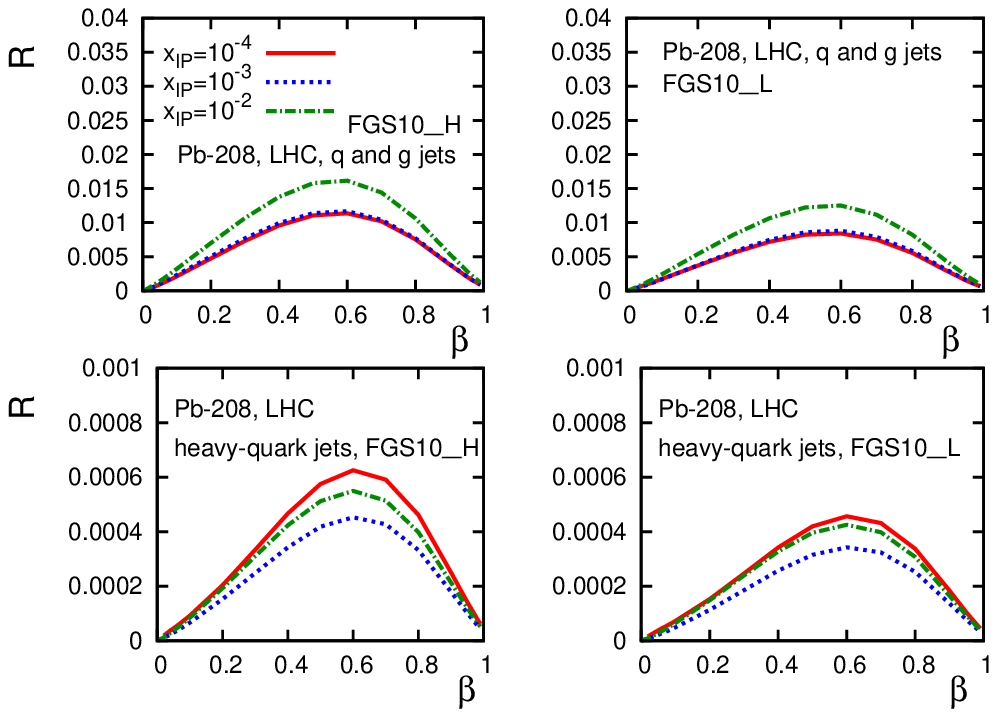,scale=1.3}
\caption{The suppression of hard diffractive dijet (quark and
gluon jets and heavy-quark jets) production compared to e.m.~coherent dijet production in
proton-nucleus scattering at the LHC. The suppression factor $R$ of Eq.~(\ref{eq:r})
at $p_T=5$ GeV and $x_{\Pomeron}=10^{-4}$, $10^{-3}$ and $10^{-2}$ 
as a function of $\beta$.
}
\label{fig:jets_all}
\end{center}
\end{figure}

In our analysis, we considered two cases: The dijet production summed over gluon and quark jets
and the production of two heavy-quark jets ($c$ and $b$ quarks). 
The resulting ratios $R$ at $p_T=5$ GeV and 
$x_{\Pomeron}=10^{-4}$, $10^{-3}$ and $10^{-2}$ as functions of $\beta$ are presented
in Fig.~\ref{fig:jets_all}. 
The left column of panels correspond to model FGS10\_H; the right column of panels
correspond to FGS10\_L. The four upper panels are for $^{40}$Ca; the four lower 
panels are for $^{208}$Pb. All curves correspond to the LHC kinematics, 
see Table~\ref{table:gamma}.

\clearpage

The results presented in Fig.~\ref{fig:jets_all} 
deserve a detailed discussion. The dependence of the ratio $R$ for the quark
and gluon jets 
on $x_{\Pomeron}$ 
can be explained as follows.
The main  contribution to the 
$x_{\Pomeron}$ dependence of $R$ at fixed $\beta$ comes from the changing 
of $x_1$.
As $x_{\Pomeron}$ is decreased, $x_1$ is increased, 
see Eq.~(\ref{eq:x1}),
which diminishes the role
played by the gluons in the projectile. As explained in the following, it
is the gluon contribution that affects $R$ most significantly. 
Hence, $R$ decreases with
decreasing $x_{\Pomeron}$. 
Note that 
the dependence of the diffractive PDFs on $x_{\Pomeron}$,
$f_{j/N}^{D(3)}(\beta,x_{\Pomeron},Q_{{\rm eff}}^2) 
\propto 1/x_{\Pomeron}^{1+2\,\epsilon}$, see Eq.~(\ref{eq:pomeron_flux}), 
is similar to the $1/ x_{\Pomeron} \ln(1/x_{\Pomeron})$-behavior
 of the e.m.~cross section. Therefore, these two factors 
weakly affect the $x_{\Pomeron}$ dependence of
$R$.

The dependence of $R$ on $\beta$ is rather fast
and reflects different
shapes of the proton diffractive PDFs and the PDFs of the real
photon. While the proton diffractive PDFs times $\beta$ are flat in the $\beta \to 0$
limit, the photon PDFs times $\beta$ grow. This explains why $R$ approaches
zero when $\beta$ is small. In the opposite limit, $\beta \to 1$,
diffractive PDFs 
are small and the e.m.~contribution wins over due to the 
non-vanishing
direct photon contribution, i.e., $R \to 0$ as $\beta \to 1$.

In Fig.~\ref{fig:jets_all}, the ratio $R$ at its peak is much larger
for the production of quark and gluon jets than for the production
of heavy-quark jets.
An examination shows that
 this effect is due to the large gluon diffractive 
PDF, which in tandem with
the large $gg \to gg$ hard parton invariant matrix element~\cite{Ellis},
 works to increase $R$ in the presence of the gluon jets.

One should also note that the ratio $R$ is much larger for $^{40}$Ca than for $^{208}$Pb.
This is because
the $^{40}$Ca flux of the equivalent photons, which
is proportional to $Z^2$ [see Eq.~(\ref{eq:flux})], is 16 times smaller than that for $^{208}$Pb.

We used the following input in 
our numerical analysis of the ratio $R$. We used the LO parameterization
of the real photon PDFs from Ref.~\cite{Gluck:1991jc}. We have also checked that the use
of a different parameterization~\cite{Abramowicz:1991yb}
 leads to rather similar 
predictions.

The suppression factor  $\lambda^j$~(\ref{eq:lambda_parton}),
which implicitly enters Eq.~(\ref{eq:r}) at the scale $Q^2=Q_{{\rm eff}}^2=4\,p_T^2=100$ GeV$^2$,
very weakly depends on $Q^2$. Therefore, for $\lambda^j$, we used the results
of our calculation at $Q^2=4$ GeV$^2$ presented in Fig.~\ref{fig:lambda_parton}.

The $\delta$-function for the direct photon contribution was numerically modeled in
 the following simple form:
\begin{equation}
\delta(\beta-1)=\frac{1}{\pi} \frac{\epsilon}{(\beta-1)^2+\epsilon^2} \,, \quad {\rm with} \quad
 \epsilon=0.01 \,.
\end{equation}

\begin{figure}[t]
\epsfig{file=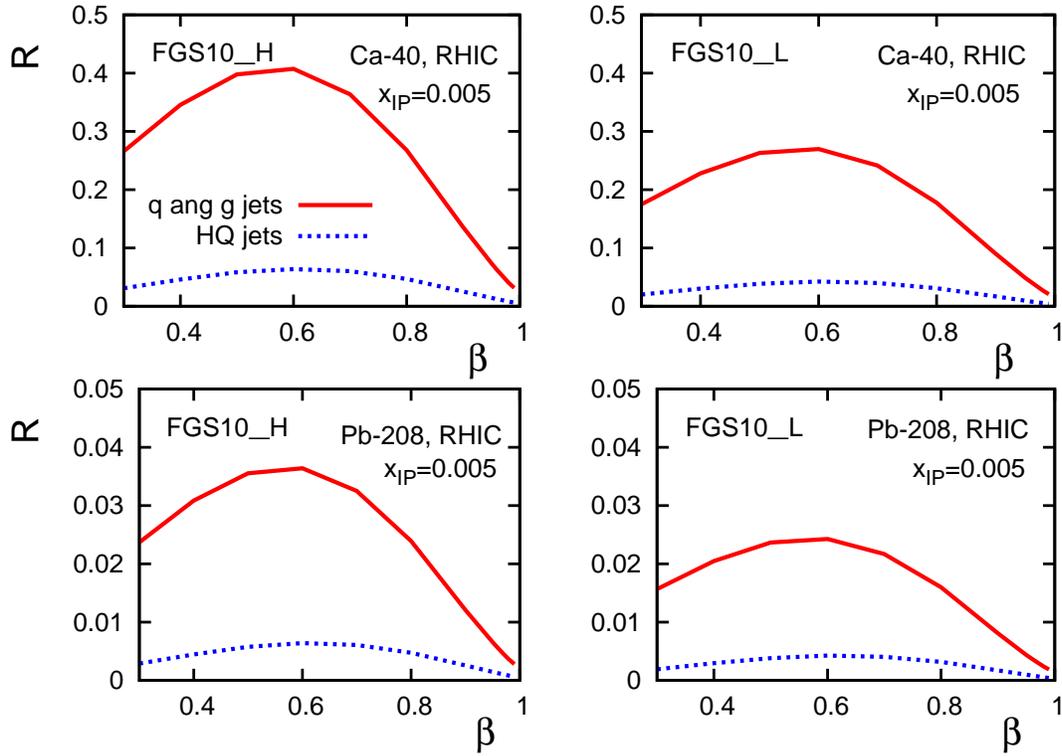,scale=1.4}
\caption{The suppression factor $R$ of Eq.~(\ref{eq:r}) in the RHIC
kinematics and at  $p_T=5$ GeV and $x_{\Pomeron}=5 \times 10^{-3}$ 
as a function of $\beta$. The solid curves correspond to 
quark and gluon jets; the dotted curves correspond to heavy-quark jets.
The left panels correspond to FGS10\_H model; the right panels correspond to
FGS10\_L model.
}
\label{fig:rhic}
\end{figure}

Besides the LHC, RHIC also has the potential to measure hard diffraction in
proton-nucleus scattering.
We consider a typical example of the corresponding 
RHIC kinematics with 250 GeV protons 
scattering on 
the beams of $^{40}$Ca and $^{208}$Pb, see Table~\ref{table:gamma}.
Producing sufficiently high 
diffractive masses, e.g., $M_X^2=500$ GeV$^2$, one accesses the 
typical kinematics of 
hard diffraction, $x_{\Pomeron}=5 \times 10^{-3}$ and $\beta > 0.3$.
Note also that the suppression of hard diffraction at RHIC is approximately 
four times smaller
than at the LHC, see Eq.~(\ref{eq:goulianos3}).

We calculated  the suppression factor $R$ of Eq.~(\ref{eq:r}) in the considered 
RHIC kinematics at $p_T=5$ GeV. The resulting values of $R$ as a function of
$\beta$ are presented in Fig.~\ref{fig:rhic}.
The solid curves correspond to 
quark and gluon jets; the dotted curves correspond to heavy-quark jets.
The left panels correspond to FGS10\_H model; the right panels correspond to
FGS10\_L model.
As seen from Fig.~\ref{fig:rhic}, the factor $R$ at RHIC is larger than
that at the LHC. This is mostly a consequence of the decrease
 of the flux of equivalent photons when going from the LHC to RHIC 
kinematics.

Our results presented in this subsection can be summarized as follows.
For proton-$^{208}$Pb scattering at the LHC, hard diffraction is suppressed compared to
the e.m.~contribution, especially at $x_{\Pomeron}=10^{-4}$ and large $\beta$, e.g., for
$\beta > 0.8$, see Fig.~\ref{fig:jets_all}.
 The suppression is very strong for the production of heavy-quark
jets.
The physical reason for the suppression is the strong coherent Coulomb field of 
$^{208}$Pb, which enhances the e.m.~mechanism of hard diffraction.

Replacing $^{208}$Pb by $^{40}$Ca, the hard diffractive mechanism 
becomes almost compatible to
the e.m.~one in the case of the production of quark and gluons jets.
However, like in the case of $^{208}$Pb, the production of heavy-quark jets is
dominated by the e.m.~mechanism.

As a result of the smaller Lorentz dilation factor $\gamma$ at RHIC,
 the factor $R$ at the
 RHIC kinematics is larger than at the LHC.

Our results suggest the following experimental strategies. First, the use of heavy nuclei
 in $pA$ scattering at the LHC will provide a clean method to study  hard
 real photon-proton scattering at the energies exceeding the HERA energies by the
 factor of ten. Second, taking lighter nuclei and choosing the appropriate kinematics,
where the e.m.~contribution 
can be controlled, one can effectively study the factorization
breaking in nuclear diffractive PDFs. Third, in the same kinematics, a comparison
of the dijet diffractive production to the heavy-quark-jet diffractive production
will measure the nuclear screened diffractive gluon PDF. It  can be compared to the nuclear
diffractive PDFs, which will be measured in nucleus-nucleus ultraperipheral
collisions at the LHC and  
which could also be measured in $eA$ coherent diffraction in DIS at an Electron-Ion Collider.

\section{The black disk regime}
\label{sec:bdr}

\subsection{Introduction}
\label{subsec:intro}

The aim of this section is to quantify the kinematical domain of small  $x$ and large $Q^2$ where  the onset 
of a new QCD regime of the strong interaction with a small coupling constant---black disk regime (BDR)---occurs and its role in the nuclear shadowing phenomenon in DIS. We outline  properties of this new QCD regime 
which are strikingly different from those of the pQCD regime and soft QCD regime,  compare theoretical expectations with experimental data obtained at HERA, RHIC,   and Tevatron and make predictions for the LHC.

The approximation of the low parton densities  is applicable in the kinematical domain  
where the nuclear shadowing effect in the interaction of a hadron projectile with a two-nucleon system represents a correction to the total cross section.  (In the case of a heavy nucleus, nuclear shadowing is larger due to the larger  number of nucleons at the same impact parameter.)  In this kinematics and  in the target rest frame,
a tiny ($\approx 4 k_{t,\rm soft}^2/Q^2$) fraction of the phase volume kinematically allowed for the
light-cone wave function of the  virtual photon is occupied by its non-perturbative component   which 
dominates  nuclear shadowing effects
($k_{t,\rm soft}$ is a typical soft momentum of the parton constituents of the virtual photon).
Its  dependence on $Q^2$ and $x$ is properly taken into account by QCD evolution equations, see the discussion in the preceding sections.  
At the same time, the interaction of the most of configurations of
 the virtual photon wave function, whose phase volume is $1- 4k_{t,\rm soft}^2/Q^2$, 
is not shadowed --- the color transparency phenomenon.
 The color transparency phenomenon has been observed in fixed-target experiments at FNAL
 and TJNAF,  
for overview and references, see, e.g., \cite{Strikman:2007nv,Strikman:2009gp}. 
However, the color  transparency phenomenon 
for the interaction with a target of the quark-gluon configurations 
with a small 
but fixed transverse size 
in the photon wave function 
along with the regime of small parton densities 
disappear at sufficiently small $x$ as a consequence of specific properties of pQCD.

Perturbative QCD predicts a rapid increase of the structure functions and parton distributions with an increase 
of energy, which was observed at FNAL and HERA, see Sec.~\ref{subsec:bdr_observation}.
At sufficiently high energies, this increase comes in conflict with the
probability conservation.  It follows from the probability conservation that the cross section for the scattering of a wave packet of quarks and gluons off a nucleon (nucleus)  at a given impact parameter $b$ cannot increase with an increasing energy forever since it is bound from above. The physical meaning of this boundary is that the absorption cannot exceed 100\%, see Sec.~\ref{subsec:geometry_bdr}.
Indeed, the probability for a wave packet of quarks and gluons to interact inelastically with the target cannot exceed unity, or, equivalently, the  probability not to interact at a given $b$ is positive. This regime is referred to as the black disk regime (BDR) because it has  the features of scattering (diffraction) of light on a completely absorbing disk in the classical wave optics.  Within the BDR, at sufficiently small $x$ the most part of the virtual photon wave function participates in the nuclear shadowing phenomenon.

In the BDR,  the structure function of a hadron target at a given impact parameter $b$ is given
by the convolution of the virtual photon light-cone wave function with the cross section of interaction of the produced quark and gluon wave packet with the target. For a transversally polarized virtual photon, 
$F_{T}(x,Q^2,b)\le c Q^2 \ln(x_0/x)$, i.e., $F_{T}(x,Q^2,b)$ is allowed to only slowly (logarithmically) increase with increasing energy.   Note that such a 
behavior is realized in pQCD 
assuming the $k_t$ factorization in the 
DGLAP approximation of pQCD as well as in the resummation models (the term is explained in Sec.~\ref{subsec:bdr_observation})~\cite{LipatovFadin,Ciafaloni:1999yw,Ball:2005mj,Altarelli:2008aj}.
In this case, transverse momenta of partons in the photon wave function increase 
with energy~\cite{Blok:2009cg};
this phenomenon   is a precursor of the change  of the pQCD regime at sufficiently 
small $x$.  The origin of this increase is the same as 
for the running coupling constant in QED and QCD ---
 the  singular behavior of the light-cone wave function of the virtual photon in coordinate space (the polarization operator of the photon at large $Q^2$ is $\Pi_{\rm em}(Q^2)\propto Q^2\ln(Q^2/Q_0^2)$). 
 This property reveals qualitative and quantitative difference of the BDR regime  in QCD from the popular hypothesis of saturation, i.e., the energy independence of parton densities at extremely small $x$, 
for a recent discussion, see e.g., \cite{Gelis:2010nm}.
The theoretical boundary on the behavior of the structure functions integrated over all impact parameters  is weaker, e.g.,
$F_{T}(x,Q^2)\le c Q^2 \ln^3(x_0/x)$~\cite{Frankfurt:2001nt}.

In addition to the probability conservation, the energy-momentum conservation plays also an important 
role in the onset of the BDR.   The account of the latter effect 
 significantly reduces the number of gluons 
(gluon showers) in the multi-Regge kinematics as compared to 
leading $\log(x)$ approximations and resummation models, see the discussion in Sec.~\ref{subsec:energy-momentum}. In particular, the  energy-momentum conservation is relevant 
for the applicability of the DGLAP approximation
in the HERA kinematics where the masses $M^2\approx Q^2$ dominate  in the cross section of diffraction in  DIS.
This should be contrasted with the onset of the
triple Pomeron limit where $M^2 \gg Q^2$.
As a result, the DGLAP approximation, which accounts for the conservation of the longitudinal energy-momentum,  is better suited  to  evaluate the  probability conservation, the radius of convergence of 
pQCD series, etc.   Note however that numerical studies show that puzzles with violation of probability conservation within pQCD calculations occur at comparatively moderate 
$x$  where the DGLAP approximation is formally justified, see Sec.~\ref{subsec:energy-momentum}.

At sufficiently high energies,  
the interaction of hadronic configurations of the virtual photon will eventually reach its
geometric limit (BDR), but this will take place at different collision energies for 
different fluctuations. For instance, according to pQCD calculations, the BDR 
should initially be achieved for the interaction of colorless two-gluon ($q{\bar q}g$)  dipoles and later, 
for significantly larger energies, also for  the interaction of $q \bar q$ colorless dipoles.

The complete absorption of the projectile wave function in the BDR 
means the violation of the leading twist approximation and an onset of a new QCD regime 
with a different  continuous symmetry, see Sec.~\ref{subsec:symmetry_bdr}.  
Within the BDR, nuclear shadowing in the total cross section of DIS achieves its maximal value
allowed by probability conservation for the collisions at central impact parameters
and does not depend on the nuclear thickness; 
diffractive processes become a shadow of inelastic ones. 
Significant pre-selection ''energy losses''  can be considered as a precursor of 
the onset of BDR, see Sec.~\ref{sec:nuc:black_nuc:is}.

The BDR is well known in hadron-hadron and hadron-nucleus interactions.  For example, it takes place 
in the scattering of protons off heavy nuclei and in elastic $pp$ collision at collider energies at the zero impact parameter,
for a  review and references,  see~\cite{Block:2006hy}.
A new distinctive feature 
of high energy QCD is that the BDR is expected for the interaction of 
small-size colorless quark-gluon wave packets 
(e.g., bound quarkonium states like $J/\psi$) which interact rather weakly at medium energies.

The distinctive property of the BDR that the interaction of small-size
dipoles remains strong leads to a number of features of the onset of 
the BDR that can be used as experimental signals.
By comparing the properties of the BDR with the available data, one finds 
evidence  that an onset of the BDR for hard processes 
has been  observed indirectly in diffraction in DIS (i.e., for the 
$q{\bar q} g$ colorless dipole scattering off nucleons) ~\cite{Frankfurt:2007rn} and 
directly  in  $dA$ collisions at RHIC, see the discussion in Sec.~\ref{subsec:post-selection2_bdr}.

\subsection{BDR {\it vs.}~weak density limit}
\label{subsec:comparison_bdr}

For the accurate calculation of nuclear shadowing in the weak 
density limit, it was sufficient to evaluate the interaction of the projectile 
with 2-3 nucleons. On the contrary, the calculations of shadowing  
in the vicinity of the BDR requires evaluation and summing of the whole series of multiple collisions. 
(We draw attention to the fact that  the concept of series over  multiple collisions becomes ill-defined  in the kinematics close to the BDR  since each term in this series becomes large and not under control.) 

To visualize the challenging theoretical phenomena,
we  consider the dipole scattering off a nuclear target  
within the optical model with  the potential $V=\sigma \rho_A$  and ignore temporarily 
its fundamental flaws. Here $\sigma$ is the dipole-nucleon cross section (rapidly increasing with energy)
and  $\rho_A$ is the nuclear density. 
At sufficiently high energies, the dipole-nucleus inelastic
cross section becomes equal $\pi R^2_A$ + terms $\propto R_NR_A \ln^2(x_0/x)$ , where $R_A(R_N)$ is the nuclear (nucleon) radius, since the optical potential $V$ increases with 
an increase of energy, i.e., the cross section becomes independent of the optical potential $V$.    
Naturally, this answer is strongly different from what one would obtain in the weak  density limit.  One 
can use this model as the guide to pQCD. In the kinematics, where the  effective parameter characterizing 
the perturbative QCD series, $\eta$,
\begin{equation}
\eta= \alpha_s(Q^2_0)(N_c/2\pi)\ln(Q^2/Q^2_0)\ln(x_0/x)\ge 1 \,,
\label{eq:eta_convergence}
\end{equation} 
the pQCD series looses 
the advantage of an asymptotic series, which was ensured at moderate $x$ by the smallness of 
the running coupling constant, and becomes a series with the finite radius of convergence. 
Thus, the theoretical challenge is how to define the series properly or to find a 
more effective theoretical framework because at very large energies (in the kinematics of the BDR), 
cross sections will become independent of the coupling constant and all traces of the 
weak coupling regime will disappear.

\subsection{Formal definition of BDR}
\label{subsec:geometry_bdr}

To formulate the condition of probability conservation,  it is instructive to consider  the amplitude $A(s,t)$  of elastic scattering of a projectile (e.g., a heavy $q{\bar q}$ pair) off a hadron (nucleus) target in the impact parameter representation:
 \begin{equation}
A(s,b)= {1\over 2s} \int {d^2q_t\over (2\pi)^2}\, e^{i\vec{q}_t \cdot  \vec b}\, A(s,t) \,,
 \label{impact}
 \end{equation}
where the momentum transfer is transverse, and   ${q}^2_t=-t$.  
Since the  angular momentum is conserved in the scattering process, 
the impact parameter $b$ is conserved as well for $s\gg m_i^2$ ($m_i$ 
denote the masses of the involved particles) and the scale of the interaction.   
The inclusion of the spins of the colliding particles leads to 
trivial modifications of the above formula
and to the appearance of the spin-flip amplitudes.

In the case of scattering of a heavy quarkonium,  the exact S-matrix 
unitarity condition reads, see e.g., Ref.~\cite{eden}:
\begin{equation}
\Im m  A(s,b)=\frac{1}{2}\left|A(s,b)\right|^2+~{\rm positive~terms} \,.
\end{equation}

The rather stringent unitarity condition emerges in the case of high energies, 
where the interactions are driven by the inelastic interactions, and the 
scattering amplitudes are predominantly imaginary. 
Indeed, defining  the profile function as
\begin{equation}
\Gamma(s,b)=-i A(s,b)\,,
\end{equation}
one can express the total, elastic and inelastic cross sections in 
terms of the profile function $\Gamma(s,b)$ as:
\begin{equation}
\left. \begin{array}{l} 
\sigma_{\rm tot}(s) \\[1ex]
\sigma_{\rm el}(s) \\[1ex]
\sigma_{\rm inel}(s)
\end{array}
\right\}
\;\; = \;\; \int d^2 b \; \times 
\left\{ \begin{array}{l} 
2 \, \Re e  \, \Gamma^{} (s,b) \\[1ex]
|\Gamma^{} (s,b)|^2 \\[1ex]
1 - |1-\Gamma^{} (s,b)|^2  \,.
\end{array}
\right.
\label{unitarity}
\end{equation}
Since the high energy dynamics is driven by the inelastic processes, 
$\sigma_{\rm el} \le \sigma_{\rm inel}$. Also, at high energies 
$ \Re e A(s,b) / \Im m A(s,b) \ll 1$. Under these conditions, 
Eq.~(\ref{unitarity}) leads to the restriction:
\begin{equation}
\Im m A(s,b)\le 1 \,.
\label{upperlimit}
\end{equation}
In the case of the inelastic cross section, the integrand $\Gamma^{\rm inel} (s,b) \equiv \left[ 1 - |1-\Gamma^{} (s,b)|^2 \right]$ corresponds to the {\it probability of the inelastic interaction} for a given impact parameter $b$. Since the strength of the inelastic interaction increases with energy, at high enough energies, the wave packet will interact  inelastically with the target with the probability close to unity and, hence, 
\begin{equation}
\Gamma(s,b)=1 \,.
\label{absorption0}
\end{equation}
This equation provides the formal definition of the BDR.
The generalization of this condition to the interaction of a wave packet of quarks and gluons gives:
\begin{equation}
\Gamma(s,b^2,d^2_{\perp})=1 \,,
\label{absorption}
\end{equation}
where $d_{\perp}$ is the transverse size of the projectile wave packet.

Exploring the conservation of probability, we will often  use a
colorless dipole as a projectile.   The concept of the dipole wave function 
is well-defined in QCD in the following two important cases. 
First, within the validity of the concept of the LT approximation, 
explicitly taking into account different components of the projectile (virtual photon, etc.) wave function or including them instead
into the interaction produces the same results. Second, it is the case of the strong interaction of a projectile with a target. 
It has been understood long ago that the sum over the contributions 
of different components of the virtual photon wave function squared and
integrated over the allowed phase volume gives the polarization operator of the photon 
with bare  and highly virtual quarks. Hence, in the case when the interaction becomes strong, the
use of the concept of a dipole is justified.  The conservation  of probability for the interaction of such a dipole has approximately 
the same form as  the unitarity of the S-matrix in the  $s$-channel and all restrictions valid for the scattering of heavy quarkonium states  should be close to the ones  for the scattering of such a wave packet.

It is worth emphasizing that although formally the expression for 
$\sigma_{\rm el}$ in Eq.~(\ref{unitarity}) is dominated by the contribution of small $b$, it should be understood in the sense of 
the Babinet's principle for the wave diffraction in the classical electrodynamics: the scattering from the complementary screens 
(an opaque absorbing disk vs.~a hole of the same geometry) are identical. It is the surrounding  wave that generates the elastic scattering in the proximity of the BDR. 
To check this interpretation of
the expression for $\sigma_{\rm el}$, one can consider the propagation of a deuteron through a heavy nucleus at $b\sim 0$.  Selecting a typical event where, e.g., a proton interacts inelastically with several nucleons, one can calculate the probability 
for the neutron to pass through the nucleus without inelastic interactions. The answer is obviously zero, while a naive application of Eq.~(\ref{unitarity}) 
would give a number of the order of unity.

The contribution of small $b$ to $\sigma_{\rm inel}$
is determined by the value of $1-\left|\Gamma(s,b)\right|^2$.
As one can see from Eq.~(\ref{unitarity}),  the BDR corresponds to the regime when $\sigma_{\rm el} = \sigma_{\rm inel}$. The interpretation of $\sigma_{\rm el} = \sigma_{\rm inel}$ in the language of the
Babinet's principle of complimentary screens is that the amount of the 
absorbed light (inelastic cross section) is equal to the amount of
light that passed around the target without interactions (elastic cross section).  
The latter is equal to the amount of light that passes through a hole  
in the shape of the target.

\subsection{Observations of an onset of the BDR}
\label{subsec:bdr_observation}

Theoretical methods for the evaluation of nuclear shadowing in DIS at moderately small $x$ are 
based on the leading twist approximation, which is justified by the difference in
scales characterizing soft and hard interactions, and resulting QCD factorization theorems.
As we have already mentioned, with an increase of the collision energy (decrease of $x$), 
the interaction at a fixed resolution scale $Q^2$ becomes stronger and the hierarchy over 
the powers of $1/Q^2$  disappears. This is a reflection of the fact that in all currently used 
approximations to pQCD---DGLAP and resummation models (see below)---the 
total inelastic cross section of the interactions of a small color dipole with a hadronic target is proportional to the gluon density of the target [Eq.~(\ref{eq:sigma_dipole})], which increases with a decrease of 
$x$ approximately as a power of $1/x$. 

A note on terminology is in order here. It has recently been 
understood that the accurate account of the double logarithmic terms 
is important for the improvement of the  calculations of small $x$ behavior of DIS,  see, e.g., Refs.~\cite{LipatovFadin,Ciafaloni:1999yw,Ball:2005mj,Altarelli:2008aj}.  Therefore, in this review, we refer to all such approaches as  {\it resummation models}.

As  a  result of the fast increase of the gluon density at small $x$,   
all the approximations just mentioned should break down at 
sufficiently high energies since  they lead to the  probability of the inelastic interactions at a fixed impact parameter $b$  exceeding  unity.  An important role in 
the evaluation  is played by the fact that the gluon distribution is more narrow in the impact parameter space
than that given by e.m. radius of a nucleon~\cite{annual}.  
This conclusion is based on the analysis of the gluon core 
of the target  probed in hard exclusive processes and the QCD factorization theorem for hard exclusive 
processes~\cite{Brodsky:1994kf,Collins:1996fb}, 
which allows one
to fit the two-gluon form factor, $F_{2g}(t)$, in the following simple form, 
$F_{2g}(t)=1/(1-t/m^2_g)^2$ where $m_{2g}\approx 1$ GeV.
The knowledge of the two-gluon form factor allows one to evaluate the probability 
of hard processes at a given impact parameter.

Analyses of special processes observed at HERA, RHIC and Tevatron (FNAL) data indicate that 
an onset of the violation of the LT approximation occurs  in the kinematical domain  
achieved at the existing  generation of accelerators. 
The signals of an onset of the BDR include:
\begin{itemize}
\item[(i)]
The gluon distribution which is large and rapidly increases with 
energy, see, e.g., Ref.~\cite{Lai:1999wy},  corresponds to the probability of the interaction of the color singlet gluon dipole at the 
zero impact parameter which is  $\approx 1/2$, see, e.g., Ref.~\cite{Rogers:2003vi}.
\item[(ii)]
The significant diffractive gluon density~\cite{Aktas:2006hy,Aktas:2006hx}  
corresponds to $\sigma_{\rm el}/\sigma_{\rm tot}\approx 1/2$ for the interaction of the color singlet gluon dipole at central impact parameters,   where $\sigma_{\rm el}$ and $\sigma_{\rm tot}$ refer 
to the respective integrands in Eq.~(\ref{unitarity}).
\item[(iii)]
In the elastic $pp$ collisions at Tevatron, the partial waves for zero impact parameters
 are close to unity.
However, at present it is unclear whether this is the BDR for hard or soft QCD interactions. 
If  hard QCD interactions dominate, one expects an increase of the hadron transverse momenta as a function of the centrality of a collision.
\item[(iv)]
Significant ''fractional energy losses'' in the fragmentation region 
have recently been discovered at RHIC in the hard processes 
$d+A\to \pi(2\pi) +X$, for the discussion and references, 
see Sec.~\ref{subsec:post-selection_bdr} and \ref{subsec:post-selection2_bdr}.
\end{itemize}
All these observations taken together indicate that the onset of a new QCD  regime---we refer to it as the black disk regime (BDR)---occurs for the scattering of colorless gluon dipole off a proton 
(at zero impact parameters) on the verge of the kinematics 
of HERA and, in the certain kinematics of small $x$, also at the LHC.  
We stress that in the HERA kinematics, as well as in the kinematics of the future  Electron-Ion Collider (EIC), the full-fledged BDR has not been and will not be realized. Instead, only an onset (approach) to the full BDR may take place. The kinematics achieved at the LHC in 
$pp$ and $pA$  collisions for  $x\le 10^{-5}$ and in the ultraperipheral processes in  
heavy ion collisions as well as dimuon pair production for $x<10^{-5}$ will allow for an onset of
the BDR.

\subsection{The onset of BDR in the dipole model: inclusive scattering}
\label{subsec:bdr_inclusive}

\subsubsection{The dipole model with the impact parameter dependence}

The dipole model for $\gamma_T^{\ast}$-nucleon (nucleus)  scattering 
in the impact parameter representation allows for a quantitative 
analysis of the  restrictions on the region of applicability of pQCD 
due to probability conservation since it gives a smooth interpolation between the calculable pQCD regime, model-dependent soft QCD regime and the BDR hard regime.   
Our interest is in the kinematics  of  the large $Q^2={\rm const}$ and  sufficiently small $x$.

The dipole approximation assumes that the $\gamma^{\ast}$ wave function is given by the superposition of $q\bar q$ dipoles of transverse diameters $d_{\perp}$ and that higher order components are included in the interaction\footnote{In the case of a dipole of the small transverse size, the dipole model of the inelastic DIS cross section is equivalent to the leading order DGLAP and BFKL~\cite{bfkl} approximations and resummation models  for this cross section in the
target rest frame. To prove this, one can evaluate the dominant Feynman diagrams by taking  the residues over the energies carried by the quark-antiquark pair in the photon fragmentation region. Note also that the infinite momentum frame structure function arises when the residues are taken over the fractions of the photon momentum 
carried by any parton in the direction of the target momentum.}.
Note that such a procedure breaks down in the vicinity of the 
BDR where the number of constituents is regulated by the target. 
However, in the BDR, another approximation is useful. 
In particular, one can
sum over the contributions of different components of the photon wave function, 
account for the  lack of nondiagonal transitions~\cite{Gribov:1968gs}, 
and use the completeness of QCD states. The resulting 
cross section is expressed in terms of the $q\bar q$ component of the photon wave function where both $q$ and $\bar q$ are strongly virtual --- a result reminiscent of the dipole model.
  
As we repeatedly discussed in this review, 
in the target rest frame, the incoming photon interacts with the target via its partonic fluctuations.  Since the interaction time is much shorter than the lifetime of  the fluctuations at small $x$, the DIS amplitude can be factorized  in three factors: one describing the formation of the fluctuation, another -- the hard interaction with the target, and the final factor describing the formation of the hadronic final state. 

Within this high-energy factorization framework, the {\it  inelastic} contribution to the transverse and longitudinal  structure functions,
$F_T$ and $F_L$, respectively,
 at a given impact parameter $b$ can be written in the following factorized  form:
\begin{equation}
F_{T,L}^{\rm inel} (x,Q^2,b)=\frac{Q^2}{4 \pi^2 \alpha_{\rm em}}2 \int dz\,d^2 d_{\perp} \sum_i |\Psi_{T,L}(z,Q^2,d_{\perp}^2,m_i^2)|^2
\Gamma^{\rm inel}(s, d_{\perp}, x,b)  \, ,
\label{eq:nuc:black}
\end{equation}
where $\Psi_{T,L}(z,Q^2,d_{\perp}^2,m_i^2)$ are the $q \bar q$ 
components of light-cone wave functions
of the transversely and longitudinally polarized virtual photon,
respectively, see Eq.~(\ref{wf});
$z$ is the photon momentum fraction carried by one of
the dipole constituents; $s$ is the invariant
energy of the dipole-target system [$s = (P + q)^2$ for
the $q {\bar q}$ dipole].
The dependence of $\Gamma^{\rm inel}$ on $z$ is rather weak and will be neglected  in our analysis below. (Note that $\Psi_{T,L}$ are usually called the light-cone wave functions, although $\Psi_{L}$ arises as the result of cancellation  between the time and longitudinal components of the electromagnetic current.)

The {\it  inelastic} contribution to the
structure functions of a target is obtained by the 
integration over the impact parameter $b$:
\begin{equation}
F_{T,L}^{\rm inel}(x,Q^2)=\int d^2 b\, F_{T,L}^{\rm inel} (x,Q^2,b) \,.
\label{eq:sf_bdep}
\end{equation}

In the LT approximation, the inelastic impact factor (profile function)
is given by the following expression,
see e.g., \cite{Frankfurt:1995jw,McDermott:1999fa,Brodsky:1994kf}:
\begin{equation}
2 \Gamma^{\rm inel}(x,b,d_{\perp})_{\rm pQCD} 
= \frac{\pi^2 F^2}{4} \, d_{\perp}^2 \, \alpha_s
(Q_{\rm eff}^2) \, x^{\prime} g_T (x^{\prime},Q_{\rm eff}^2,b) \, ,
\label{gammainel}
\end{equation}
where $F^2$ is the Casimir operator of the color group, 
$F^2(\rm triplet)=4/3$ and $F^2(\rm octet)=3$;
$g_T(x,Q_{\rm eff}^2,b)$ 
is the generalized gluon distribution of a given target $T$
(in the $\xi=0$ limit) that
was introduced and discussed  in Sec.~\ref{subsec:impact}. 
When the target is a nucleus, $g_T (x,Q_{\rm eff}^2,b)$
is the impact parameter dependent nuclear PDF (nuclear GPD), see Eq.~(\ref{eq:ngpd3}) and Fig.~\ref{fig:impact_dependence}.
In the case of the scattering from the  nucleon,
$g_T (x,Q_{\rm eff}^2,b)$ is given by the product of the structure function and the two-dimensional Fourier  transform of the two-gluon
 form factor~\cite{Rogers:2003vi,Frankfurt:2002ka} (see the discussion below). The integration of $2 \Gamma^{\rm inel}(x,b,d_{\perp})_{\rm pQCD}$ over the impact parameter
gives the perturbative (small-$d_{\perp}$) part of the dipole cross section used in Sec.~\ref{subsec:eikonal}, see Eq.~(\ref{eq:sigma_dipole}):
\begin{equation}
2 \int d^2b \,\Gamma^{\rm inel}(x,b,d_{\perp})_{\rm pQCD} 
= \frac{\pi^2}{3} \, d_{\perp}^2 \, \alpha_s
(Q_{\rm eff}^2) \, x^{\prime} g_T(x^{\prime},Q_{\rm eff}^2) \,.
\label{eq:psigma}
\end{equation}
The unitarity relation~(\ref{unitarity}) provides the relation  between 
$\Gamma^{\rm inel}(b)$ and $\Gamma(b)$ that can be employed 
[in combination with Eq.~(\ref{eq:psigma})] to calculate $\Gamma(b)$.

In the limit of small dipoles, where $\Gamma^{\rm inel}(x,b,d_{\perp})_{\rm pQCD}$ is small
and far away from the BDR, 
$\Gamma(x,b,d_{\perp})_{\rm pQCD} \approx \Gamma^{\rm inel}(x,b,d_{\perp})_{\rm pQCD}$
and $\sigma_{\rm tot}(s) \approx \sigma_{\rm inel}(s)$.
As the size of the dipole increases, the non-perturbative dynamics starts to play a role
and one has to model $\Gamma(x,b,d_{\perp})$. 
One effect is an increase of the number of constituents. 
We take this effect into account phenomenologically  by requiring  the 
effective interaction to match the understood properties of the interaction.  In particular, we use
the interpolation formula in the impact parameter representation 
that coincides with the pQCD formula in  the small-$d_{\perp}$ limit 
and smoothly matches the non-perturbative soft QCD physics for
large $d_{\perp}$ (cf.~Sec.~\ref{subsec:eikonal}):
\begin{equation}
2\Gamma(x,b,d_{\perp})=\sigma_{q{\bar q}N}(x,Q^2,d_{\perp},m_i) f(x,d,b) \,,
\label{eq:Gamma_proton}
\end{equation}
where $\sigma_{q{\bar q}N}(x,Q^2,d_{\perp},m_i)$ is the dipole cross section of
Eq.~(\ref{eq:sigma_dipole}); $f(x,d,b)$ is the Fourier transform of the form factor 
$f(x,d_{\perp},t)$ introduced in Ref.~\cite{Rogers:2003vi} ($f(x,d_{\perp},t)$ is a generalization of the two-gluon form factor discussed in Ref.~\cite{Frankfurt:2002ka})
and measured in the hard diffractive processes at HERA:
\begin{equation}
f(x,d_{\perp},b)=\int \frac{d^2 q_{\perp}}{(2 \pi)^2} e^{i q_{\perp} b} f(x,d_{\perp},t=-q_{\perp}^2) \,.
\label{eq:f_b} 
\end{equation}
The form factor $f(x,d_{\perp},t)$ satisfies the condition  $f(x,d_{\perp},t=0)=1$, 
which leads to  $\int d^2b f(x,d_{\perp},b)=1$ and
automatically ensures  that Eq.~(\ref{eq:Gamma_proton}) is consistent with Eq.~(\ref{unitarity}).
The form factor $f(x,d_{\perp},t)$ is modeled as the product of three functions describing 
the contributions to the overall $t$ dependence coming from the target, projectile and
Gribov diffusion~\cite{Rogers:2003vi}:
\begin{equation}
f(x,d_{\perp},t)=\frac{1}{[1-t/M^2(d_{\perp}^2)]^2}\frac{1}{1-t/m_2^2 \,d_{\perp}^2 /d_{\pi}^2}\,
e^{\alpha^{\prime} t\,d_{\perp}^2 /d_{\pi}^2 \log(x_0/x)} \,,
\label{eq:f_t} 
\end{equation}
where the effective mass squared $M^2(d_{\perp}^2)$ is defined as:
\begin{equation}
M^2(d_{\perp}^2)=\left\{\begin{array}{ll}
m_1^2-(m_1^2-m_0^2) \frac{d_{\perp}^2}{d_{\pi}^2} \,, & d_{\perp} \leq d_{\pi} \,,
\\
m_0^2 \,, & {\rm otherwise} \,.
\end{array} \right.
\label{eq:M2_eff}
\end{equation}
The parameters in Eqs.~(\ref{eq:f_t}) and (\ref{eq:M2_eff}) are
$m_0^2=0.7$ GeV$^2$ (from the fits to the nucleon form factor),
$m_1^2=1.1$ GeV$^2$ (from the fits to diffractive $\rho$, $\omega$ and $J/\psi$ electroproduction),
$m_2^2=0.6$ GeV$^2$, 
$d_{\pi}=0.65$ fm, $\alpha^{\prime}=0.25$ GeV$^{-2}$, and $x_0=0.01$.
The last factor in Eq.~(\ref{eq:f_t}) is set to unity for $x > x_0$.

\begin{figure}[t]
\begin{center}
\epsfig{file=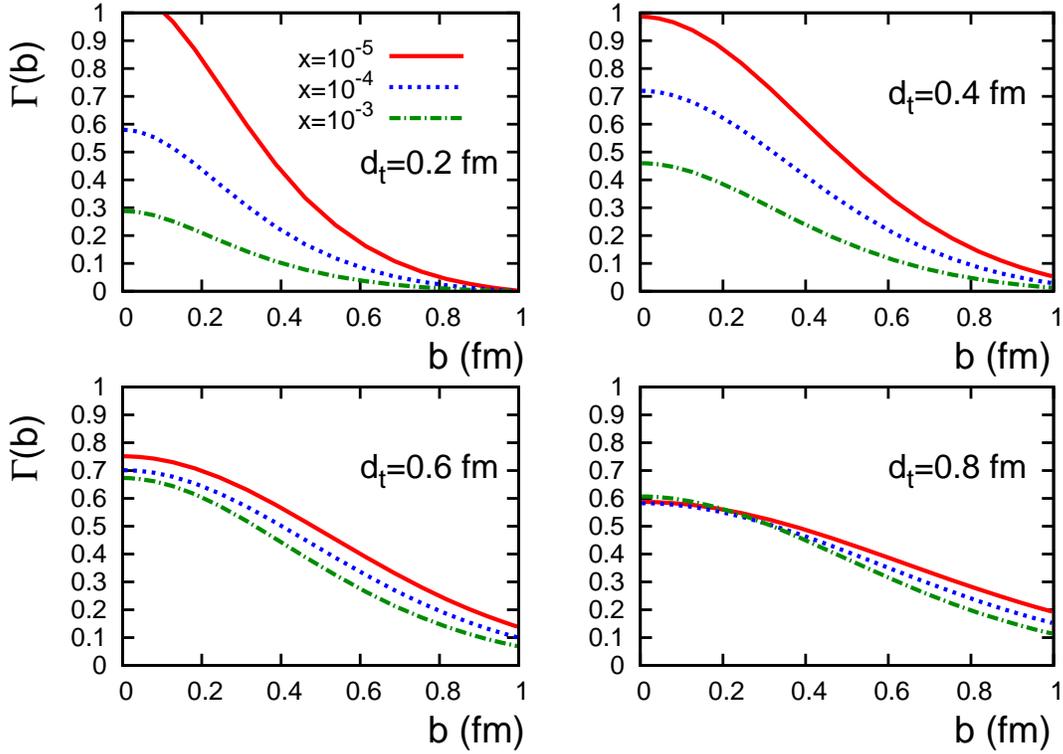,scale=1.4}
\caption{The impact factor for the nucleon of Eq.~(\ref{eq:Gamma_proton}) as a function of the impact 
parameter $b$ for several fixed dipole sizes $d_{\perp}$.
}
\label{fig:Gamma_dipole_2010}
\end{center}
\end{figure}

Figure~\ref{fig:Gamma_dipole_2010} presents the
results of the calculation of the impact factor for the nucleon 
of Eq.~(\ref{eq:Gamma_proton}), 
$\Gamma(x,b,d_{\perp})$, as a function of the impact parameter $b$ for four fixed dipole
sizes, $d_{\perp}=0.2$, 0.4, 0.6, and 0.8 fm.
All curves correspond to $Q^2=4$ GeV$^2$.

In the case of nuclei, the impact factor can be calculated adopting the procedure of   Ref.~\cite{Rogers:2003vi}. For small dipoles, 
$d_{\perp} \leq d_0=0.2$ fm, according to the QCD factorization theorem,  the dipole cross
section and the resulting impact factors are obtained by 
replacing the gluon distribution
in the proton in Eq.~(\ref{gammainel}) by that in the nucleus: 
\begin{equation}
2 \Gamma_A(x,b,d_{\perp})_{\rm pQCD} 
= \frac{\pi^2 F^2}{4} \, d_{\perp}^2 \, \alpha_s
(Q_{\rm eff}^2) \, x^{\prime} g_{A} (x^{\prime},Q_{\rm eff}^2,b) \, .
\label{gammainel_A}
\end{equation}
The impact parameter dependent gluon distribution in nuclei
entering Eq.~(\ref{gammainel_A})
is given by Eq.~(\ref{eq:impact2}) and is illustrated in Figs.~\ref{fig:impact_dependence},
\ref{fig:LT2009_ca40_impact}, and \ref{fig:LT2009_ca40_impact_Bdep}.

As we explained in detail in Sec.~\ref{subsec:eikonal}, as one increases the transverse dipole
size, the dynamics of the dipole-target interaction becomes progressively non-perturbative and
for sufficiently large dipoles, $d_\perp \approx d_{\pi}=0.65$ fm, it is completely
determined by the soft interactions. In particular, it is reliable to assume that
at $d_\perp=d_{\pi}$ fm, the nuclear impact factor can be evaluated in the Glauber multiple 
scattering formalism:
\begin{equation}
\Gamma_A(x,b,d_{\perp})=1-e^{-\frac{1}{2}A \sigma_{\pi N}(x) T_A(b)} \,,
\label{gammainel_A_pi}
\end{equation}
where $\sigma_{\pi N}(x)$ is the total pion-nucleon cross section.
In the intermediate region, $d_0=0.2\, {\rm fm} \leq d_{\perp} \leq d_{\pi}=0.65$ fm, 
the nuclear impact factor
is modeled by interpolating between the pQCD expression of Eq.~(\ref{gammainel_A}) and
the calculation at $d_\perp=0.65$ fm of Eq.~(\ref{gammainel_A_pi}):
\begin{equation}
\Gamma_A(x,b,d_{\perp})=\left[\Gamma_A(x,b,d_{\pi})-\Gamma_A(x,b,d_0)\right]
\frac{d^2_{\perp}-d_0^2}{d^2_{\pi}-d_0^2}+\Gamma_A(x,b,d_0) \,.
\label{eq:Gamma_A_inter}
\end{equation}
Finally, for the dipoles with the size $d > d_{\pi}$, the nuclear impact factor is given
by Eq.~(\ref{gammainel_A_pi}), where the total pion-nucleon cross section is allowed to
slowly grow as
\begin{equation}
\sigma_{\pi N}(x,d_{\perp})=\sigma_{\pi N}\frac{1.5 \,d_{\perp}^2}{d_{\perp}^2+d^2_{\pi}/2} \,.
\label{eq:sigma_piN_modif}
\end{equation}
The interpolation is chosen so that for $d_{\perp}^2= d^2_{\pi}$, the cross section 
is equal to $\sigma_{\pi N}$. It also takes into account the presence of the fluctuations 
of the strength of the $\pi N$ interaction leading to $\sigma > \sigma_{\pi N}$ 
as indicated by the presence of the inelastic $\pi N $ diffraction, see the discussion 
in Sec.~\ref{subsec:soft_coherent_diffraction}.

\begin{figure}[t]
\begin{center}
\epsfig{file=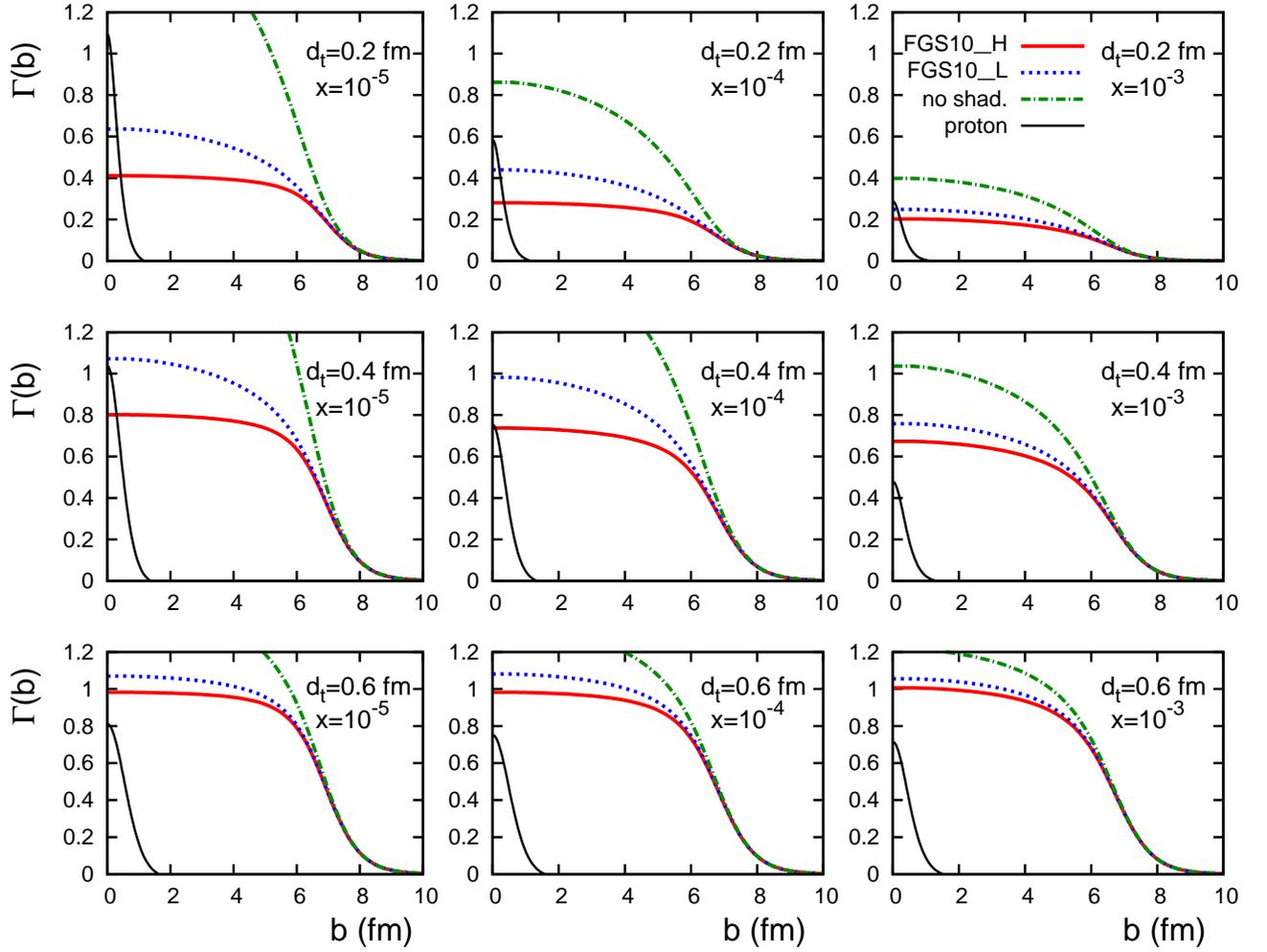,scale=1.25}
\caption{The impact factor $\Gamma_A(x,b,d_{\perp})$ for $^{208}$Pb at $Q^2=4$ GeV$^2$
as a function of the impact 
parameter $b$ for different values of $x$ and dipole sizes $d_{\perp}$.
The solid (red) curves correspond to model FGS10\_H; the dotted curves correspond to FGS10\_L. For comparison, we also give the impulse approximation predictions for 
$\Gamma_A(x,b,d_{\perp})$ by the
dot-dashed curves and the free proton $\Gamma(x,b,d_{\perp})$ by the 
thin solid (black) curves.
}
\label{fig:Gamma_dipole_pb208_2010}
\end{center}
\end{figure}

The resulting nuclear impact factor is presented in Fig.~\ref{fig:Gamma_dipole_pb208_2010}.
We plotted $\Gamma_A(x,b,d_{\perp})$ for $^{208}$Pb 
as a function of the impact parameter $b$ for different values
of Bjorken $x$ ($x=10^{-5}$, $10^{-4}$, and $10^{-3}$)
and dipole sizes $d_{\perp}$ ($d_{\perp}=0.2$, $0.4$, and $0.6$ fm).
All curves correspond to $Q^2=4$ GeV$^2$ 
(the dependence on $Q^2$ is weak and enters through the definition of $x^{\prime}$ in the dipole
cross section). 
The solid (red) curves correspond to the nuclear shadowing for the gluon distribution
in model FGS10\_H; the dotted curves correspond to FGS10\_L. For comparison, 
we also give the dot-dashed curves corresponding to the nuclear gluon distribution that is not
shadowed (impulse approximation) and the thin solid (black) curves corresponding to the free proton case (same as in
Fig.~\ref{fig:Gamma_dipole_2010}).

The results presented in Fig.~\ref{fig:Gamma_dipole_pb208_2010} deserve a discussion.\\
(i) Model FGS10\_H, which corresponds to larger shadowing than model FGS10\_L, 
naturally leads to
the smaller value of $\Gamma_A(x,b,d_{\perp})$. The difference between the two models
increases as one decreases $x$ and the amount of nuclear shadowing.\\
(ii) In both FGS10\_H and FGS10\_L models, nuclear shadowing at central impact 
parameters is very large.
Therefore, at small $b$, the solid and dotted curves lie significantly below the 
dot-dashed curves. When $b$ increases and becomes compatible to the nuclear size, $R_A \approx 7$ fm for $^{208}$Pb, nuclear shadowing begins to rapidly disappear and
all nuclear curves converge.\\
(iii) The $b$ dependence of all nuclear curves is flat for $b < 7$ fm, which is essentially
determined by the nuclear optical density $T_A(b)$ and the nuclear size 
$R_A$. At the same time, the $b$ dependence of the free nucleon impact factor is determined
by the form factor $f(x,d_{\perp},b)$ of Eq.~(\ref{eq:f_b}), whose characteristic scale is
the effective radius of the gluon distribution in the nucleon, which is 
smaller than 1 fm. This effective radius decreases as $d_{\perp}$ is decreased, see 
Eq.~(\ref{eq:f_b}), and, hence, the proton $\Gamma(x,b,d_{\perp})$ increases.
Therefore, at very small dipole sizes, the proton $\Gamma(x,b,d_{\perp})$ can exceed the 
nuclear impact factor $\Gamma_A(x,b,d_{\perp})$.

As we explained above, unitarity of the scattering matrix places the model-independent
constraint of the impact factor: $\Gamma(x,b,d_{\perp})\leq 1$. As one can see from
Figs.~\ref{fig:Gamma_dipole_2010} and \ref{fig:Gamma_dipole_pb208_2010}, this 
unitarity constraint is not always satisfied in the dipole formalism, unless special 
measures are taken; we will not discuss ways and means to correct the dipole formalism
to make it to comply with the unitarity constraint. 

It follows from the QCD  factorization theorem that the interaction 
with a target of 
a colorless two-gluon dipole or a dipole, where a gluon is substituted by a $q \bar q$ pair,  
is larger than that for a $q \bar{q}$ dipole by the factor $F^2(8)/F^2(3)=9/4$, 
where $F^2$ is the Casimir operator of color group $SU(3)_c$. 
Hence, the interaction of such dipoles with nucleons and nuclei 
reaches the BDR at 
significantly lower energies.  One example of such processes is diffraction into large masses 
($M^2\gg Q^2$) in DIS  where the dominant role is played by the 
$q {\bar q}g$   component of the photon light-cone wave function.

\subsubsection{Nuclear enhancement of the dipole cross section}

Nuclei provide a better arena to
study the dynamics of high parton densities than the free nucleon since (i) the number of
partons in the transverse slice of a nucleus is enhanced by the number of nucleons
and, (ii)  the distribution of nuclear matter over impact parameters is almost flat.
In the language of the impact factors, the nuclear 
$\Gamma_A(x,b,d_{\perp})$ is enhanced compared to the free proton 
$\Gamma(x,b,d_{\perp})$, but the enhancement is partially masked by nuclear shadowing.

To quantify this enhancement, we introduce
the ratio of the nuclear to free proton impact factors, $R_{\Gamma}$:
\begin{eqnarray}
R_{\Gamma}(x,Q^2, b,d_{\perp})& \equiv &\frac{\Gamma_A(x,b,d_{\perp})}{\Gamma(x,b,d_{\perp})}
\nonumber\\
&=&\frac{g_A(x^{\prime},Q_{\rm eff}^2,b)}{f(x,d_{\perp},b)g_N(x^{\prime},Q_{\rm eff}^2)} =
\frac{A T_A(b)r_g(x^{\prime},Q_{\rm eff}^2,b)}{f(x,d_{\perp},b)} \,.
\label{eq:Oompf}
\end{eqnarray}
The second line is valid only for small dipoles, $d_{\perp} < 0.2$ fm, 
when the impact factors can be calculated in
pQCD, see Eqs.~(\ref{gammainel}) and (\ref{gammainel_A}).
In this case, 
$r_g(x,Q^2,b) \equiv g_A(x,Q^2,b)/(A T_A(b) g_N(x,Q^2))$ is the factor characterizing
the impact parameter dependent nuclear shadowing in the gluon channel, 
see Figs.~\ref{fig:impact_dependence},
\ref{fig:LT2009_ca40_impact}, and \ref{fig:LT2009_ca40_impact_Bdep};
$f(x,d_{\perp},b)$ is the free nucleon two-gluon form factor, 
see Eqs.~(\ref{eq:f_b}) and
(\ref{eq:f_t}). The $R_{\Gamma}$ factor somewhat depends on the used model for the 
dipole cross that enters through the model-dependent definitions of $x^{\prime}$ and
$Q^2_{\rm eff}$. Still, the strongest model-dependence of the $R_{\Gamma}$ factor in Eq.~(\ref{eq:Oompf}) comes from modeling of the nucleon two-gluon form factor $f(x,d_{\perp},b)$.
We also emphasize that $R_{\Gamma}$ depends strongly on the impact parameter $b$ 
through the rapid $b$ dependence of $f(x,d_{\perp},b)$, see 
Fig.~\ref{fig:Gamma_dipole_pb208_2010}.

\begin{figure}[h]
\begin{center}
\epsfig{file=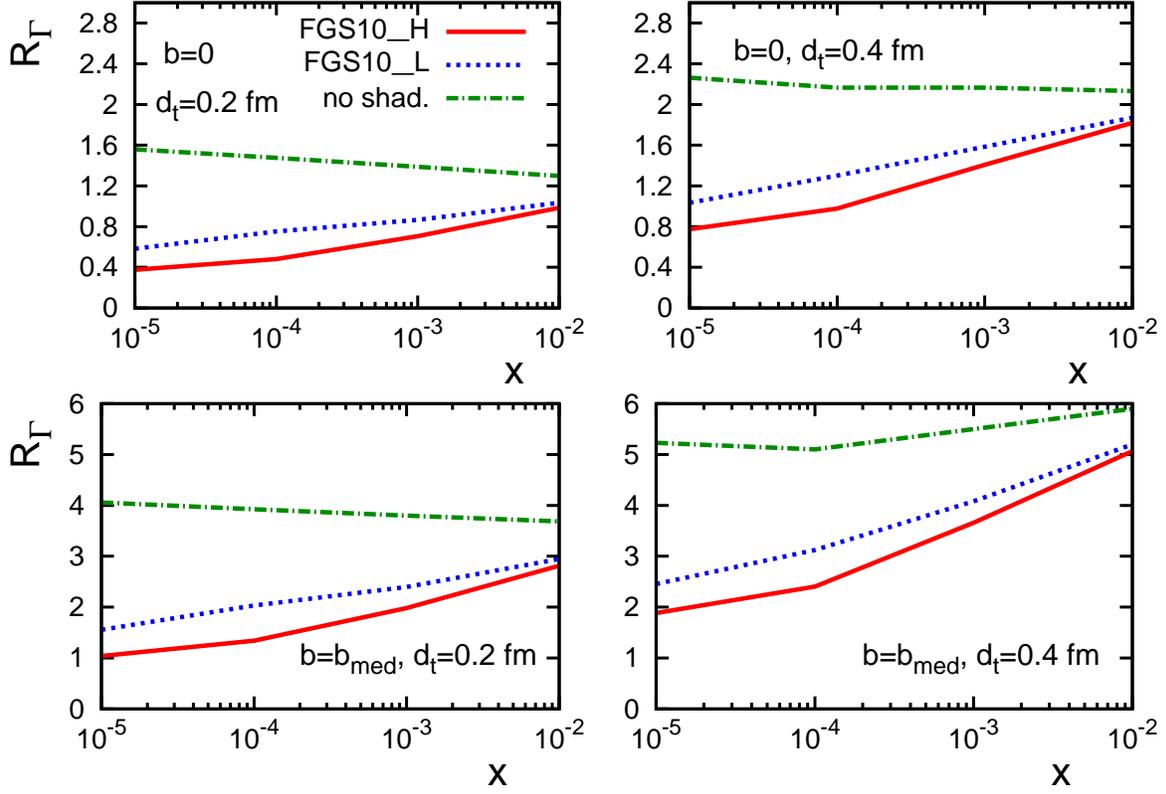,scale=1.5}
\caption{The $R_{\Gamma}$ factor of Eq.~(\ref{eq:Oompf}) for $^{208}$Pb 
at  $Q^2=4$ GeV$^2$ 
as a function of Bjorken $x$ for two values of the dipole size, $d_{\perp}=0.2$ and 0.4 fm, and two values of the
impact parameter $b$. 
The upper panels correspond to the central impact parameter $b=0$;
the lower panels correspond to the median impact parameter $b_{\rm med}$ 
[see Eq.~(\ref{eq:b_med})].
}
\label{fig:Oompf_b_all}
\end{center}
\end{figure}

Figure~\ref{fig:Oompf_b_all} presents our predictions for $R_{\Gamma}$
of Eq.~(\ref{eq:Oompf})
for $^{208}$Pb at $Q^2=4$ GeV$^2$ as a function of Bjorken $x$;
the results are given for two selected dipole sizes, $d_{\perp}=0.2$ and 0.4 fm,
and two values of the impact parameter $b$ (see explanations below).
The solid curves correspond to model FGS10\_H; the dotted curves correspond to FGS10\_L;
the dot-dashed curves correspond to the nuclear gluon PDF in the impulse approximation
(no shadowing).

Evidently (cf. Fig.~\ref{fig:Gamma_dipole_pb208_2010}), the lower limit on the value of $R_{\Gamma}$ is achieved at the impact parameter $b=0$;  this case is presented in
the two upper panels of Fig.~\ref{fig:Oompf_b_all}.
As one can see from Fig.~\ref{fig:Oompf_b_all}, without the effect of
the leading twist nuclear shadowing, $R_{\Gamma}(b=0) \approx 2$.
The strong nuclear shadowing in the gluon channel at small $x$ reduces the  enhancement factor from $R_{\Gamma}(b=0) \approx 2$ down to $R_{\Gamma}(b=0) \approx 0.4-1.2$ at $x=10^{-5}-10^{-4}$. Therefore, for studies of the effects of high parton densities, which are
probed when  the values of the impact factor are close to unity,
heavy nuclei at central impact parameters
do not give any advantage over the free nucleon at $b=0$. The advantage of nuclear targets is that the physics of large parton densities occupies a large volume in the impact parameter space.

The shapes of the $b$ dependence of the nuclear 
$\Gamma_A(x,b,d_{\perp})$ and free nucleon 
$\Gamma(x,b,d_{\perp})$ are very different:   while $\Gamma_A(x,b,d_{\perp})$ is almost a step-function, the nucleon 
$\Gamma(x,b,d_{\perp})$ is a steep function peaking at $b=0$. 
Another possibility to compare the nucleus and nucleon impact factors is to  compare them at a certain median value $b_{\rm med}$ that we define by the following relations:
\begin{samepage}
\begin{eqnarray}
2 \pi \int_0^{b_{\rm med}} db\, b f(x,d_{\perp},b)=\frac{1}{2} \,, \nonumber\\
2 \pi \int_0^{b_{\rm med}} db\,b T_A(b)= \frac{1}{2} \,.
\label{eq:b_med}
\end{eqnarray}
\end{samepage}
Equation~(\ref{eq:b_med}) corresponds to the situation  when no selection  of impact parameters is performed and, hence,  it is natural to compare nucleon and nucleus cases for such $b_{\rm med}$ that half of the gluon density is located at $b> b_{\rm med}$.

The values of $b_{\rm med}$ resulting from solving Eq.~(\ref{eq:b_med})
are different for the nucleus and free nucleon. For instance,
for the nucleon, $b_{\rm med}$ varies between $b_{\rm med}=0.41$ fm at $d_{\perp}=0.2$ fm and
$x=0.01$ and $b_{\rm med}=0.73$ fm at $d_{\perp}=0.6$ fm and
 $x=10^{-5}$ (at fixed $d_{\perp}$, the
$x$ dependence of $b_{\rm med}$ is weak).
For the nucleus of $^{208}$Pb, $b_{\rm med}=2.5$ fm.
While the nucleon $\Gamma(x,b=b_{\rm med},d_{\perp})$ is 
significantly smaller than
$\Gamma(x,b=0,d_{\perp})$ (by at least a factor of two), 
the nuclear $\Gamma_A(x,b=b_{\rm med},d_{\perp})$ is 
essentially the same as $\Gamma_A(x,b=0,d_{\perp})$. 
Therefore, $R_{\Gamma}(b_{\rm med})$
should be larger than $R_{\Gamma}(b=0)$ by at least a factor of two.
The enhancement factor $R_{\Gamma}$
at $b=b_{\rm med}$ as a function of Bjorken $x$ is presented in the two lower panels of Fig.~\ref{fig:Oompf_b_all}.

While our numerical results for $R_{\Gamma}$ in the unrealistic scenario when the 
leading twist nuclear shadowing is neglected (dot-dashed curves in Fig.~\ref{fig:Oompf_b_all})
are in a quantitative agreement with the nuclear enhancement of the saturation scale $Q_s^2$ discussed in Ref.~\cite{Kowalski:2007rw}, our principal predictions for $R_{\Gamma}$
in the framework of the leading twist nuclear shadowing (solid and dotted curves in
in Fig.~\ref{fig:Oompf_b_all}) correspond to a significantly smaller enhancement than in the model of~\cite{Kowalski:2007rw}.

\subsubsection{Probability conservation as the constraint on the region of applicability
of the LT approximation}

The application of QCD factorization theorems, LO and NLO approximations  
leads to the contradiction with the probability 
conservation due to the non-linear relations between QCD Green functions.  
Within these approximations, the violation of the probability conservation for the interaction of small dipoles follows from the mismatch between the inelastic cross section increasing with energy  
$\propto xg_T(x,Q^2)$ and the elastic/diffractive cross section 
$\propto [xg_T(x,Q^2)]^2$, where $g_T(x,Q^2)$ is the gluon density of the target.  The mismatch progressively increases with a decrease of $x$ and ultimately breaks down 
the condition that 
$\sigma_{\rm el}\le \sigma_{\rm inel}$. Note that a more accurate treatment would involve generalized parton distributions (GPDs) for the description of diffractive processes. However, 
the account of the QCD evolution demonstrated that this will not change significantly 
the $x$ and $Q^2$ dependencies, but will somewhat increase the absolute value of the diffractive cross section, see, e.g., \cite{Frankfurt:2000ty}.

The strongest constraint on the kinematical region of
applicability of the LT approximation 
follows from  the requirement of probability conservation at small impact parameters. 
In the derivation, one uses the impact parameter representation of the dipole-target scattering amplitude~(\ref{impact}) and determines how close the resulting impact factor  
$\Gamma(x,b,d_{\perp})$ is to unity.

\begin{figure}[t]
\begin{center}
\epsfig{file=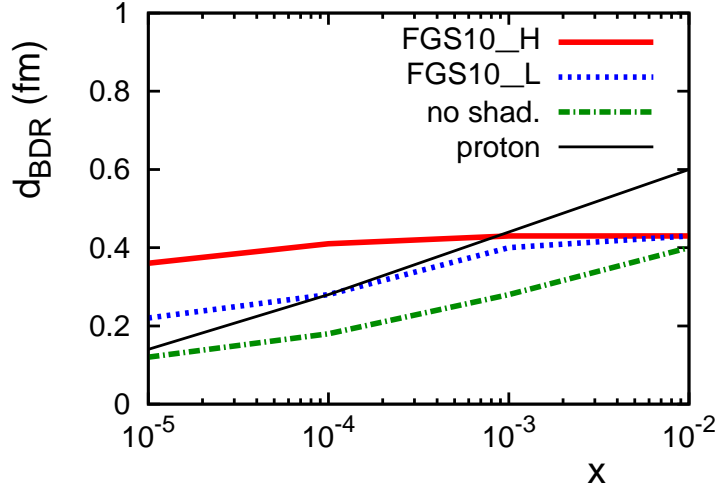,scale=1.4}
\caption{The critical transverse dipole size $d_{\rm BDR}$ of Eq.~(\ref{eq:d_BDR})
for $^{208}$Pb at $Q^2=4$ GeV$^2$
as a function of Bjorken $x$
at the central impact parameter $b=0$.
The thick solid curve corresponds to model FGS10\_H; the dotted curve corresponds to FGS10\_L;
the dot-dashed curve corresponds to the nuclear gluon PDF in the impulse approximation
(no shadowing). For comparison, the thin solid curve gives $d_{\rm BDR}$ for the free proton case.
}
\label{fig:d_BDR}
\end{center}
\end{figure}

Let us begin our quantitative consideration from the analysis of 
the interaction of a colorless $q \bar q$ dipole with a hadron (nucleus) target.  
It was explained above that the probability conservation    places
the  model-independent  constraint of the impact factor, 
$\Gamma(x,b,d_{\perp})\leq 1$. As one can see from Figs.~\ref{fig:Gamma_dipole_2010} and \ref{fig:Gamma_dipole_pb208_2010}, this constraint is not satisfied at very small $x$ in the dipole model 
or, more generally, in pQCD.  However, the $q\bar q$ dipole-nucleon interaction still
remains rather far from the unitarity limit for small dipoles with $d_{\perp} \le 0.3$ fm 
(i.e., for the dipoles of the size comparable to the size of  $J/\psi$ or smaller) 
practically in the whole range of energies available at HERA, except at very small $b$, which contribute very little to the total cross section. 
This constraint can be made more explicit by introducing the critical dipole size, $d_{\rm BDR}$,
for which
the impact factor $\Gamma$ is rather close to unity, e.g.:
\begin{equation}
\Gamma(x,b,d_{\rm BDR})=\frac{3}{4} \,.
\label{eq:d_BDR}
\end{equation}
Figure~\ref{fig:d_BDR} presents $d_{\rm BDR}$ [the solution of Eq.~(\ref{eq:d_BDR})]
for $^{208}$Pb at $Q^2=4$ GeV$^2$
as a function of Bjorken $x$
at the central impact parameter $b=0$.
The thick solid curve corresponds to model FGS10\_H; the dotted curve corresponds to FGS10\_L;
the dot-dashed curve corresponds to the nuclear gluon PDF in the impulse approximation
(no shadowing). Also, for comparison, the thin solid curve gives the solution of Eq.~(\ref{eq:d_BDR}) for the free proton case.

The results presented in Fig.~\ref{fig:d_BDR} deserve a discussion.\\
(i) The nuclear shadowing in the gluon channel is larger in model 
FGS10\_H than that in model FGS10\_L. 
Therefore, the impact factor in model FGS10\_H is smaller than that 
in model FGS10\_L and, hence, the transverse sizes solving Eq.~(\ref{eq:d_BDR}),
$d_{\rm BDR}$,  are larger for  FGS10\_H than for FGS10\_L.\\
(ii)
By turning off the leading twist nuclear shadowing,
one significantly increases the nuclear impact factor, which can now reach its limiting 
value for rather small dipole sizes (the dot-dashed curve in Fig.~\ref{fig:d_BDR}).
Remarkably, for  $x \leq 10^{-4}$, the corresponding $d_{\rm BDR} <0.2$ fm, which means that 
it is determined using the perturbative expression~(\ref{gammainel_A}), i.e., in a model-independent way.
Note that in the limit of large $x$, $x \geq 0.01$, the effect of nuclear shadowing  is small and all nuclear curves in Fig.~\ref{fig:d_BDR} converge to the regime of small parton density.

An increase of the impact parameter away from $b=0$ in Eq.~(\ref{eq:d_BDR}) and in Fig.~\ref{fig:d_BDR}
 will lead to (i) an increase 
of $d_{\rm BDR}$ (models FGS10\_L and FGS10\_H), and (ii) the disappearance of the reaching of the BDR regime in the free proton case.

For $x \leq 10^{-3}$, we approximately parameterize
 the $x$ (energy) dependence of $d_{\rm BDR}$ in the following simple form:
\begin{equation}
d_{\rm BDR}(x) = (x/x_0)^n d_{\rm BDR}(x_0) \,,
\label{eq:d_BDR_fit}
\end{equation}
where $x_0=0.001$.
By fitting to the points in Fig.~\ref{fig:d_BDR}, we find  that
$n \approx 0.04$ for model FGS10\_H, $n \approx 0.13$ for model FGS10\_H,
$n \approx 0.18$ for the no-shadowing case,
and $n \approx 0.21$ for the free proton case.
Note that these
values correspond to $Q^2=4$ GeV$^2$. As one increases $Q^2$, the exponent $n$ in 
Eq.~(\ref{eq:d_BDR_fit}) should also increase.
Therefore, after the BDR has been reached for the dipoles of sufficiently small sizes
for which the pQCD approach is applicable for moderate $x$, 
the further expansion of the BDR to smaller $d_{\perp}$ becomes rather rapid.

The situation with the unitarity constraint is different in the gluon channel 
since the interaction of a colorless dipole built of color octet constituents   
is significantly   stronger than for the color singlet 
$q {\bar q}$ dipole.  In particular, in the pQCD regime, 
the inelastic cross section is stronger by the factor of 9/4 and by the factor of 81/16 in 
the elastic channel. 
The inelastic impact factor for the colorless dipole with color octet constituents,
 $\Gamma^{gg}_{\rm inel}$, can be obtained by the rescaling of the inelastic 
$q {\bar q}$ dipole impact factor:
 \begin{equation}
 \Gamma^{gg,\rm inel}(x,b,d_{\perp}) = \frac{9}{4}\, \Gamma^{\rm inel}(x,b,d_{\perp}) \,,
 \label{qtog}
 \end{equation}
 in the kinematics where the resulting $\Gamma^{gg,\rm inel}(x,b,d_{\perp})$ is 
still  smaller than unity. 

Equation~(\ref{qtog}) allows us to quantitatively study the approach to the BDR 
in the gluon channel when the probability of inelastic interactions is close to unity.
For instance, we can define the critical size $d_{BDR}^{\prime}$
for which
$\Gamma^{gg}_{\rm inel}(x,b,d_{\rm BDR}^{\prime})=3/4$, or equivalently, the
total impact factor 
$\Gamma^{gg}(x,b,d_{\rm BDR}^{\prime}) =1/2$, 
see Eqs.~(\ref{unitarity}) and (\ref{gammainel_A}).
It follows from 
Eq.~(\ref{unitarity})  that this value of $\Gamma^{gg}(x,b,d_{\perp})$ is reached for $\Gamma(x,b,d_{\perp})= 1- \sqrt{2/3}  \approx 0.18$. 
As one can see from Fig.~\ref{fig:Gamma_dipole_2010}, this value of 
$\Gamma(x,b,d_{\perp})$ is reached  in a rather wide range of $b$ and $x$ at $Q^2= 4$ GeV$^2$, which is near  
the edge of the HERA  kinematics.

One should also note that the fact that $\Gamma^{gg}(x,b,d_{\perp}) \gg \Gamma(x,b,d_{\perp})$  
implies the significantly larger diffraction in the gluon channel compared to the quark one for moderate values of $Q^2$,  $Q^2\sim 4$ GeV$^2$, which is indeed 
indicated by the analysis of the HERA inclusive diffraction  
data---see the discussion in Sec.~\ref{subsec:ft}.
 
To estimate the proximity to the BDR in the gluon channel in the case of scattering off nuclei, one can again use the rescaling of the $q{\bar q}$-nucleus impact factor, Eq.~(\ref{qtog}), along
with the model for $\Gamma_A(x,b,d_{\perp})$, see the discussion in Sec.~\ref{subsec:bdr_inclusive}. The onset of the BDR in the gluon channel 
can be inferred from Fig.~\ref{fig:Gamma_dipole_pb208_2010}, in a direct analogy
to the case of the $q{\bar q}$ dipole discussed above.

We can also introduce the maximal  transverse momentum of the gluon, for which the interaction is close to the BDR: 
\begin{equation}
p_{t,{\rm BDR}} \equiv {\pi\over 2 d_{\rm BDR}^{\prime}} \,. 
\end{equation}
For $p_t \le p_{t,{\rm BDR}}$, the interaction is close to the BDR and, hence, the   average transverse momenta are also close to 
$p_{t,{\rm BDR}}$.  In the opposite limit, the condition for the applicability of the LT approximation is that 
\begin{equation}
Q^2_0 \gg 4  p_{t,{\rm BDR}}^2 \,.
\end{equation}
Finally, we remark that  the restrictions due to the probability conservation are less severe for  inclusive structure functions where the region around $b=0$ gives only a  small contribution to 
the integrated quantity. These restrictions play a more important role 
in the description of the structure of the final states in the new particle production, especially in multiparton interactions since the dijet production is dominated  by the scattering at small impact parameters. 

\subsubsection{BDR predictions for inclusive structure functions}
\label{subsec:bdr_sf}

We demonstrated above that when the BDR is reached, it is reached only for a small fraction of all dipoles. However, if we consider asymptotically large energies,  $d_{\rm BDR}$ will progressively decrease and one will encounter the situation when 
the dominant contribution to $F_{T,L}^{\rm inel} (x,Q^2,b)$ 
comes from the region $d_{\perp} \ge d_{\rm BDR}$ for which 
$\Gamma_{\rm inel}(x,b,d_{\perp})=1$. If this limit is reached for all essential impact parameters $b$, the dipole interacts with the
nucleon (nucleus) with the maximal total cross section allowed by the conservation of probability.

In the preceding sections we explained that within the BDR, essential impact parameters increase  with
an increase of energy. The BDR expression for the dipole-nucleon cross has the following form:
\begin{eqnarray} 
\sigma_{q{\bar q} N}(x,Q^2,d_{\perp})&=&  \Theta (d_{\perp}-d_{\rm BDR})2 \pi R_N^2 (1+c_N\ln^2(x_0/x)) 
+\Theta (d_{\rm BDR}-d_{\perp})\sigma_{q{\bar q} N}^{LT} \nonumber\\
&=&\Theta (d_{\perp}-d_{\rm BDR})
2\pi R^2_{N,{\rm eff}} +\Theta (d_{\rm BDR}-d_{\perp})\sigma_{q{\bar q} N}^{LT}  \,,
\label{eq:sigma_dipole_max_N}
\end{eqnarray}
where $c_N$ is a constant;  $R_N$ is the nucleon radius. To 
simplify formulas we introduced the quantity $R^2_{N,{\rm eff}}$.   
To account for the
interaction of the dipoles whose size is sufficiently small,  
we included the leading twist contribution of 
$\sigma_{q{\bar q} N}^{LT}=\sigma_{q{\bar q} N}^{\rm inel}$~(\ref{gammainel}).
This LT contribution  follows from the QCD factorization theorem and rapidly increases with energy; it was discussed in the preceding sections dealing with the small parton density limit.

The cross section for the dipole scattering off a nuclear target has the same form as for the nucleon target except for specifics related to the large nuclear radius $R_A$:
\begin{eqnarray} 
\sigma_{q {\bar q} A}(x,Q^2,d_{\perp}) &=&
\Theta(d_{\perp}-d_{\rm BDR})2 \pi [R_A^2 +cR_AR_N\ln^2(x_0/x)]+\Theta (d_{\rm BDR}-d_{\perp}) \sigma_{q{\bar q} A}^{LT} \nonumber\\
&=&\Theta(d_{\perp}-d_{\rm BDR})2\pi R^2_{A,{\rm eff}}+
\Theta (d_{\rm BDR}-d_{\perp}) \sigma_{q{\bar q} A}^{LT} \,,
\label{eq:sigma_dipole_max}
\end{eqnarray}
where $R_A$ is the nucleus radius.
In the above formula, we account for the fact that  the interaction of the virtual photon at the impact parameters close to the nucleus edge produces the cross section close to that for a nucleon target up to the factor accounting for the number of nucleons within the nuclear edge. 
To simplify the formula, we introduce $R_{A,{\rm eff}}$ --- the effective radius of the interaction which increases with energy; $c$ 
is a constant.  The LT contribution $\sigma_{q{\bar q} A}^{LT}$ has been discussed in the review with respect to the weak parton density limit  including nuclear shadowing phenomenon.

In the momentum representation, the onset of the BDR for pQCD interactions at extremely small $x$ corresponds to the situation when the average transverse momenta in the quark loop in the photon wave function are (much) larger than the photon virtuality $Q$, 
which clearly corresponds to a breakdown of the DGLAP and BFKL approximations.  This result follows from the application of the $k_t$ factorization theorem to small $x$ processes.
  Note that the onset of this regime can already be seen in the DGLAP 
approximation as the average transverse momenta of quarks in the quark loop attached to the photon gradually increase with a decrease of $x$~\cite{Blok:2009cg}.

Using Eq.~(\ref{eq:nuc:black}) and the discussion following it, the BDR prediction for the inclusive structure functions of heavy nuclei reads:
\begin{eqnarray}
F_{T,L}^A(x,Q^2)&=&\frac{Q^2}{4 \pi^2 \alpha_{\rm em}}\int_0^1 dz\,
\int_{d^2_{\rm BDR}}^{d_{\rm soft}^2} d^2 d_{\perp} \sum_i |\Psi_{T,L}(z,Q^2,d_{\perp}^2,m_i^2)|^2 \,2 \,\pi R_{A,{\rm eff}}^2 \nonumber\\ &+&F^{LT,A}_{T,L}(x,Q^2_{\rm eff})\,.
\label{eq:sf_A_b1}
\end{eqnarray}
The BDR is present on the interval $d_{\rm BDR} \leq d_{\perp} \leq d_{\rm soft}$,
where $d_{\rm BDR}$ is the boundary of the kinematical region where the 
interaction of the dipole is still black and is calculable in pQCD; $d_{\rm soft}$ is a scale of the order of
$1/\Lambda_{\rm QCD}$, which cuts from above the integral over the dipole sizes.
The last term in the above equation is the LT contribution from the dipoles in the wave function of the virtual photon of the transverse sizes $d_{\perp} \le d_{\rm BDR}$ whose value and properties  follow 
from the QCD factorization theorem; 
$Q^2_{\rm eff} \ge Q^2_{BDR}$ and is calculable using the above formulas for the dipole-nucleus interaction. To shorten formulae 
in the following discussion, we will often omit the LT term whose properties were thoroughly discussed in the review.
To obtain a direct correspondence with the Feynman diagrams, it is convenient to   work in the momentum representation for the 
$q{\bar q}$ component of the photon light-cone wave function,
\begin{equation}
\Psi_{T,L}(z,Q^2,k_{\perp}^2,m_i^2)=\int d^2 d_{\perp} e^{-i d_{\perp} k_{\perp}}\Psi_{T,L}(z,Q^2,d_{\perp}^2,m_i^2) \,,
\end{equation}
where we suppressed the indices labeling the photon and quark helicities.
The squared light-cone wave functions of the transversely and longitudinally-polarized photon 
in the momentum space read, see, e.g., \cite{Gieseke:1999ub}:
\begin{eqnarray}
|\Psi_{T}(z,Q^2,k_{\perp}^2,m_i^2)|^2&=&6 \alpha_{{\rm em}} e_i^2 \left(z^2+(1-z)^2 \right) \frac{k_{\perp}^2}{(k_{\perp}^2+Q^2 z (1-z) +m_i^2)^2} \,, \nonumber\\
|\Psi_{L}(z,Q^2,k_{\perp}^2,m_i^2)|^2&=&24 \alpha_{{\rm em}} e_i^2 z^2(1-z)^2 \frac{Q^2}{(k_{\perp}^2+Q^2 z (1-z) +m_i^2)^2} \,.
\end{eqnarray}
Equation~(\ref{eq:sf_A_b1}) can be identically rewritten in the momentum representation in the following
form:
\begin{equation}
F_{L,T}^A(x,Q^2)=\frac{Q^2}{4 \pi^2 \alpha_{\rm em}}\int_0^1 dz\,
\int_0^{k_{t, {\rm max}}^2} \frac{d^2 k_{\perp}}{(2 \pi)^2} \sum_i |\Psi_{L,T}(z,Q^2,k_{\perp}^2,m_i^2)|^2 \,2 \,\pi R_{A,{\rm eff}}^2  \,,
\label{eq:sf_A_b2}
\end{equation}
where 
$k_{\perp}$ has the meaning of the transverse momentum of each quark (orthogonal to the momentum
of the virtual photon). Thus, the quark constituent of the $q{\bar q}$ dipole is characterized by the 
longitudinal momentum fraction $z$ and the transverse momentum $k_{\perp}$, while the antiquark is characterized by $1-z$ and $-k_{\perp}$.  In Eq.~(\ref{eq:sf_A_b2}), 
we implicitly used the orthogonality of the wave functions of the eigenstates  of the QCD Hamiltonian with different energies.  As a consequence, 
the interaction in the BDR is diagonal in $z$ and $k_{\perp}$, c.f.~\cite{Gribov:1968gs}.
(Note that $\Psi_L$ results from the cancellation of the components of the polarization
vector of a longitudinally polarized photon that separately linearly increase with energy.)

Next, it is convenient to define the mass squared of the $q\bar q$ system, 
\begin{equation}
M^2= \frac{k_{\perp}^2 + m_q^2}{z(1-z)} \,,
\label{msq}
\end{equation}
and the angle $\theta$ between the direction of the momentum of the quark in 
the center of mass frame and the photon direction (the transverse plane is 
defined to be perpendicular to this). Neglecting the quark mass $m_i^2$ compared 
to $M^2$ and  $Q^2$, one obtains: 
\begin{equation}
\sin \theta=\frac{2 p_t}{\sqrt{M^2}}, \quad z= \frac{1}{2}(1 + \cos \theta)\, .
\label{zvar}
\end{equation}
Changing the variables $(|k_{\perp}|^2,z)$ to $(M^2,\cos \theta)$, we readily obtain from
Eq.~(\ref{eq:sf_A_b2}) the BDR predictions for the nuclear inclusive structure 
functions~\cite{Frankfurt:2001nt}:
\begin{equation}
F^{A}_T (x,Q^2)=\int_0^{M^{2}_{\rm max}} dM^2 \,{2\pi R_{A,{\rm eff}}^2\over 12 \pi^3}\,
{Q^2M^2\rho(M^2)\over (M^2+Q^2)^2} \int_{-1}^1 d \cos \theta ~{3\over 8}(1+\cos^2\theta) \, ,
\label{gr2}
\end{equation}
\begin{equation}  
F_L^{A} (x,Q^2)=\int_0^{M^{2}_{\rm max}} dM^2 \,
{2\pi R_{A,{\rm eff}}^2\over 12 \pi^3}\, {Q^4\rho(M^2)\over (M^2+Q^2)^2} \int_{-1}^1 d \cos \theta ~{3\over 4}\sin^2\theta \, ,
\label{gr3}
\end{equation}
where 
\begin{equation}
\rho(M^2)=
\sigma^{e^+e^- \to \mbox{{\small hadrons}}}/\sigma^{e^+e^-\to \mu^+\mu^-}
\approx N_c \sum_i e_i^2 \, .
\label{eq:rho_ee}
\end{equation}
In Eq.~(\ref{eq:rho_ee}), $N_c=3$ is the number of colors; the last equality holds to 
the leading order  in the strong coupling constant $\alpha_s$.
The inclusion of $\rho(M^2)$ in Eqs.~(\ref{gr2}) and (\ref{gr3})
corrects the dipole formulas for higher order corrections (in $\alpha_s$) contributing at a given 
$M^2$. Moreover,  the BDR expressions  are insensitive to the number of constituents in the photon wave function.
This is due to the theorem---which is well known in QED and QCD---that as a consequence of gauge 
invariance, the sum over the amplitudes squared for the transitions $\gamma^{\ast} \to q +{\bar q}+ g$ gives 
the polarization operator of the photon with off-shell quarks, i.e.,  $\rho(M^2)$ .  We point out again that in the BDR,  as a result of the orthogonality of the
eigenstates of QCD Hamiltonian with different energies,
the non-diagonal transitions between the states 
with different $M^2$ in Eqs.~(\ref{gr2}) and (\ref{gr3})
vanish~\cite{Gribov:1968gs}.

The integration over $\theta$ in Eqs.~(\ref{gr2}) and (\ref{gr3}) can now  be performed analytically
and one obtains the BDR predictions for the transverse 
and longitudinal structure functions~\cite{Frankfurt:2001nt}:
\begin{equation}
F^{A}_T (x,Q^2)= {2\pi R_{A,{\rm eff}}^2\over 12 \pi^3}Q^2
\rho(M_{\rm max}^2) \ln (M_{\rm max}^2/m_0^2) = {2\pi R_{A,{\rm eff}}^2\over 12 \pi^3}Q^2\rho(M_{\rm max}^2)  \ln (x_0/x) \,,
\label{eq:BDR_FT}
\end{equation}
\begin{equation}
F^{A}_L (x,Q^2)= {2\pi R_{A,{\rm eff}}^2\over 12 \pi^3}Q^2\rho(M_{\rm max}^2) \,.
\end{equation}
When one takes into account only the suppression induced by the square of the nuclear form factor in the rescattering amplitude, then $M^{2}_{\rm max}\le W^2/(m_N R_A)$, and Eqs.~(\ref{gr2}) and (\ref{gr3}) 
coincide with the original result of~\cite{Gribov:1968gs},
except for the account of the effective nuclear radius $R_{A,{\rm eff}}$ increasing with energy.  
However, in QCD, because of the color transparency, 
$M^{2}_{\rm max}$ is determined by the unitarity constraint and is substantially smaller than $W^2/(m_N R_A)$.
 Since $M^2_{\rm max} \propto Q^2 x^{-n}$ with $n \approx 0.1$,  the proportionality of $F^{A}_T (x,Q^2)$ to   $\ln(x_0/x)$ derived by   Gribov is retained, but the numerical coefficient is much smaller.

It is important to emphasize that the contribution of small transverse size configurations (which have not reached the BDR)
remains significant in a wide range of $x$ and $Q^2$. Hence,
studies of the total cross sections are a rather ineffective way to 
search for the onset of the BDR. In particular,
it may be rather difficult to distinguish the BDR from the DGLAP approximation with different initial conditions.

Note that the BDR will reveal itself  in the interaction of the 
$q{\bar q}g$ component of the photon wave function
with a hadron (nucleus) target at smaller energies compared to the
$q{\bar q}$ component. In the leading twist approximation, it will be double counting to consider the  $q{\bar q}g$ component since it is included in the interaction; in the BDR, the situation is different.

In what follows we shall demonstrate that studies of
DIS final states provide a number of
clear signatures of the onset of the BDR, which will be
qualitatively different from the leading
twist regime. For simplicity, we will assume that the BDR is
reached for a significant
part of the cross section and, hence, restrict our discussion to
DIS on a large nucleus so that
edge effects (which are important in the case of scattering off a nucleon)  can be neglected.

\subsection{Diffractive final states}
\label{sec:nuc:black_nuc:dfs}

In Eqs.~(\ref{gr2}) and (\ref{gr3}), the mass $M$ is the mass of the 
diffractively produced state.
Removing the integral over $M^2$ and noticing that in the BDR
diffraction constitutes 50\% of all events, we readily obtain 
predictions for the diffractive 
structure functions (the spectrum of the masses of diffractively 
produced states) from 
Eqs.~(\ref{gr2}) and (\ref{gr3}):
\begin{equation}
{dF_T^{D(3)}(x,Q^2,M^2)\over dM^2}=
{\pi R_{A,{\rm eff}}^2\over 12 \pi^3}
{Q^2M^2\rho(M^2)\over (M^2+Q^2)^2} \,,
\end{equation}
\begin{equation}
{dF_L^{D(3)}(x,Q^2,M^2)\over dM^2}=
{\pi R_{A,{\rm eff}}^2\over 12 \pi^3}
{Q^4\rho(M^2)\over (M^2+Q^2)^2} \, .
\label{eq:nuc:black_nuc:dfs:1}
\end{equation}
Moreover, the spectrum of hadrons in the center of mass of
the diffractively produced system should be the same as
in $e^+e^-$ annihilation.
Hence, the dominant diffractively produced final state will have two jets
with a distribution over the center of mass emission angle proportional to
$1+\cos^2 \theta$ for the transverse case and 
proportional to $\sin^2 \theta$ for the
longitudinal case~\cite{Frankfurt:2001nt}:
\begin{equation}
{dF_T^{D(3)}(x,Q^2,M^2)\over dM^2 d\cos \theta}=
\frac{3}{8} (1 + \cos^2 \theta) {\pi R_{A,{\rm eff}}^2\over 12 \pi^3}
{Q^2M^2\rho(M^2)\over (M^2+Q^2)^2} \ ,
\label{eq:nuc:black_nuc:dfs:2}
\end{equation}
\begin{equation}
{dF_L^{D(3)}(x,Q^2,M^2)\over dM^2d\cos \theta}=\frac{3}{4}
\sin^2 \theta {\pi R_{A,{\rm eff}}^2\over 12 \pi^3} {Q^4\rho(M^2)\over (M^2+Q^2)^2} \ .
\label{eq:nuc:black_nuc:dfs:3}
\end{equation}

The transverse momentum of the produced jet, $p_{t}=k_{\perp}$, and the longitudinal fraction
of the photon energy carried by the jet, $z$,  are related to
the diffractive mass $M$ and the angle $\theta$ 
by Eq.~(\ref{zvar}).
Hence, in the BDR, diffractive production of high
$p_t$ jets is strongly enhanced:
\begin{eqnarray}
&&\left<p_t^2(jet)\right>_T
=3M^2/20 \ , \nonumber\\
&& \left<p_t^2(jet)\right>_L
=M^2/5 \ .
\label{eq:nuc:black_nuc:dfs:5}
\end{eqnarray}
This should be compared to the leading twist approximation
where $\langle p_t^2(jet) \rangle \propto \ln Q^2$.

The relative rate and distribution of the jet variables for the three
jet events (originating from $q\bar q g $ configurations) will also be
the same as in $e^+e^-$ annihilation and, hence, is given by the standard
expressions for the $e^+e^- \to q\bar q g$ subprocess 
(see e.g., Ref.~\cite{Ellis2}).
The  inelastic 
cross section of the interaction of a color octet dipole is enhanced in pQCD
by the factor of 9/4 as compared to the interaction of the triplet 
$q {\bar q}$ dipole. 
As a result, the onset of the  BDR occurs earlier for processes dominated by the 
octet configurations,  such as the  emission of gluons at large angles relative to $q$ and $ \bar q$. Consequently,  
for $k_t$ in the vicinity of the BDR, one expects  the enhanced production of 
the star-like three jet events.

An important advantage of the diffractive BDR signal is that
the discussed features of the diffractive final states should hold for
$M^2 \leq Q^2_{\rm BDR}$ ($Q_{\rm BDR}$ is a characteristic scale of the BDR),
even if $Q^2 \geq Q^2_{\rm BDR}$
because the configurations with the transverse momenta
$p_t \leq Q_{\rm BDR}/2$ still interact in the black disk regime
(and correspond to the fluctuations of the transverse size for which the
interaction is already black/maximal).

Next we turn to exclusive vector meson production, which  in the BDR is in a sense a resurrection of  the original vector meson dominance model~\cite{Sakurai} without off-diagonal transitions. 
The amplitude of the vector meson-nucleus interaction is proportional
to $2\pi R_{A,{\rm eff}}^2$ (since each configuration in
the virtual photon interacts with the same BDR cross section).
This is markedly different from the requirements~\cite{Frankfurt:1997zk} for matching of 
generalized vector dominance models (VMD) (see, e.g., \cite{VDM})
with QCD in the scaling limit. In generalized VMD models, 
the matrix elements for 
 the non-diagonal transitions between vector mesons
are large and lead to strong cancellations mimicking the approximate scaling.

In the BDR, we can factorize out the dipole-nucleus cross section 
from the overlap integral between the  virtual photon and vector meson wave functions
and obtain the dominant contribution to the differential
cross section of the electroproduction of vector mesons~\cite{Frankfurt:2001nt}:
\begin{eqnarray}
{d\sigma^{\gamma^{\ast}_{T} +A\to V+A} \over dt} &=&
{M_V^2\over Q^2} {d\sigma^{\gamma^{\ast}_{L} +A\to V+A} \over dt}
\nonumber\\
&=&
{(2\pi R_{A,{\rm eff}}^2)^2\over 16\pi}{3 \Gamma_V M_V^3 \over \alpha_{\rm em}
(M_V^2 + Q^2)^2 } \frac{4~\left(J_1(\sqrt{|t|}R_{A,{\rm eff}}^2)\right)^2}{|t|R_{A,{\rm eff}}^2} \,,
\label{vm}
\end{eqnarray}
where $\Gamma_V$ is the electronic decay width $V\to e^+e^-$;
$M_V$ is the vector meson mass;
$\alpha_{\rm em}$ is the fine-structure constant; $J_1(x)$ 
is the Bessel function.
Therefore, the parameter-free BDR prediction that at large $Q^2$
the vector meson electroproduction initiated by longitudinally-polarized photons
behaves as $1/Q^2$ is 
in a stark contrast with the $1/Q^6$  asymptotic behavior 
predicted in perturbative QCD~\cite{Frankfurt:1995jw,Brodsky:1994kf,Frankfurt:1997fj}. 
This makes this reaction very attractive for scanning the transition 
from the BDR at small $Q^2$ to the pQCD regime at large $Q^2$.

\subsection{Post-selection mimicking fractional parton energy 
losses in inclusive spectra in the vicinity of BDR}
\label{sec:nuc:black_nuc:is}

In the leading twist approximation, the QCD factorization theorem 
is valid and  leads to the prediction of universal spectra of leading particles (independent of the target)  for the scattering off partons of the same flavor. Fundamentally,  this can be explained by the fact that, in the Breit frame, the  fast parton which is hit by the photon carries practically all of the photon's  light-cone momentum 
($z \to 1$).   As a result of the QCD evolution, 
this parton acquires the virtuality $\sim \sqrt {Q^2 Q_0^2}$
and a rather large    transverse momentum $k_{t}^2$ (which is 
still $\ll Q^2$). So,  in pQCD, quarks and gluons
emitted in the process of QCD evolution and 
fragmentation of highly virtual partons 
carry all the photon momentum.

In contrast, in the BDR, the projectile interacts with several nucleons at the same impact parameter. Hence, it follows from the energy-momentum conservation that different subgroups of partons within the wave function of the photon should interact with 
different nucleons. [These configurations are rather similar to the 
ones dominating diffractive scattering, see e.g., Eqs.~(\ref{eq:nuc:black_nuc:dfs:2}) and (\ref{eq:nuc:black_nuc:dfs:3}).]
The total momentum carried by all subgroups is equal to the photon momentum. 
As a result, the fraction  of the photon momentum carried by any subgroup is less than 
unity. Thus,  the target post-selects different configurations in the projectile depending on the number of nucleons at the same impact parameter and the strength of the interaction. Fragmentation of an individual  subgroup into hadrons produces leading hadrons with 
total energies equal to the energy of the subgroup.  Hence, in the BDR case, the spectrum of leading hadrons (in  the direction of the virtual photon) is expected to be noticeably more depleted  than in the pQCD regime.   Although no energy losses occur for the partons propagating through the target, the selection of subgroups of partons effectively looks as significant fractional energy losses.

Neglecting the contribution of $q\bar q g$ configurations to the photon wave function,
the leading hadron spectrum  
can be obtained from 
the $\theta$ dependence in Eqs.~(\ref{eq:nuc:black_nuc:dfs:2}) and (\ref{eq:nuc:black_nuc:dfs:3}).
For fixed $M^2$,
the jet distribution over the light-cone fraction $z$ has the following dependence~\cite{Frankfurt:2001nt}:
\begin{eqnarray}
&&{d\sigma_T \over d z}\propto 1+(2z-1)^2 \,, \nonumber\\
&&{d\sigma_L \over d z}\propto ~z(1-z) \, ,
\label{eq:nuc:black_nuc:dfs:6}
\end{eqnarray}
where the subscripts denote the transversely and longitudinally polarized photons.

If no special L-T separation procedure is undertaken, at small $x$ one
actually measures $\sigma_L+\epsilon\sigma_T$
($\epsilon$ is the photon polarization). In this case,
combining Eqs.~(\ref{eq:nuc:black_nuc:dfs:2}) and
(\ref{eq:nuc:black_nuc:dfs:3}) we find:
\begin{equation}
{d(\sigma_T + \epsilon \sigma_L)\over d z}\propto {M^2\over 8 Q^2}
(1+(2z-1)^2) + \epsilon z(1-z) \, .
\end{equation}

In the approximation of Eq.~(\ref{eq:nuc:black_nuc:dfs:6}), the inclusive spectrum 
of leading hadrons can be estimated assuming the  independent fragmentation of the 
quark and antiquark with the virtualities $\geq Q^2$ and $z$ and  $p_t$ distributions given by 
Eqs.~(\ref{eq:nuc:black_nuc:dfs:2}) and
(\ref{eq:nuc:black_nuc:dfs:3}) (cf.~the case of
diffractive production of jets discussed above).
The independence of fragmentation is justified because 
large transverse momenta of the quarks dominate in the photon wave function  [cf.~Eqs.~(\ref{eq:nuc:black_nuc:dfs:2}) and (\ref{eq:nuc:black_nuc:dfs:3})]
and because of the weakness of the final state interaction 
between $q$ and $\bar q$ since
$\alpha_s$ is small and the rapidity interval is of the order of unity.
Obviously,
the independent fragmentation
 leads to a gross depletion of the
leading hadron spectrum 
compared to the leading twist
approximation 
where leading
hadrons are produced in the fragmentation region of the parton which carries
essentially all momentum of the virtual
photon. Qualitatively, this pattern is similar
to the one expected in
the soft regime of the strong interaction since the spectrum of hadrons produced in hadron-nucleus
interactions is much softer than that in hadron-nucleon interactions.
In addition, taking into account 
the production of multi-jet states---like $q\bar q g$---will
further enhance 
the departure from the leading twist picture of fragmentation.

Neglecting the gluon emission in the photon wave function, we readily obtain
for the total differential multiplicity of leading hadrons in the BDR:
\begin{eqnarray}
&&{d N^{\gamma_{T}^{\ast}/h}\over dz}=
2\int_z^1 dy D^{q/h}(z/y,Q^2)\,{3\over 4}(1+(2y-1)^2)\,, \nonumber\\
&&{d N^{\gamma_{L}^{\ast}/h}\over dz}=
2\int_z^1 dy D^{q/h}(z/y,Q^2)\,6y(1-y) \,.
\label{eq:nuc:black_nuc:dfs:41}
\end{eqnarray}
In Eq.~(\ref{eq:nuc:black_nuc:dfs:41}), 
$D^{q/h}(z/y,Q^2)$ is the fragmentation function of a quark of
flavor $q$ into the hadron $h$ (more accurately, it is the sum of the
quark and antiquark fragmentation functions); the factor of two in front
of the integrals comes from the fact that we included two quark flavors and assumed 
that $D^{u/h}(z/y,Q^2)=D^{d/h}(z/y,Q^2)$.
The contribution of heavier flavors has been neglected.

\begin{figure}
\begin{center}
\epsfig{file=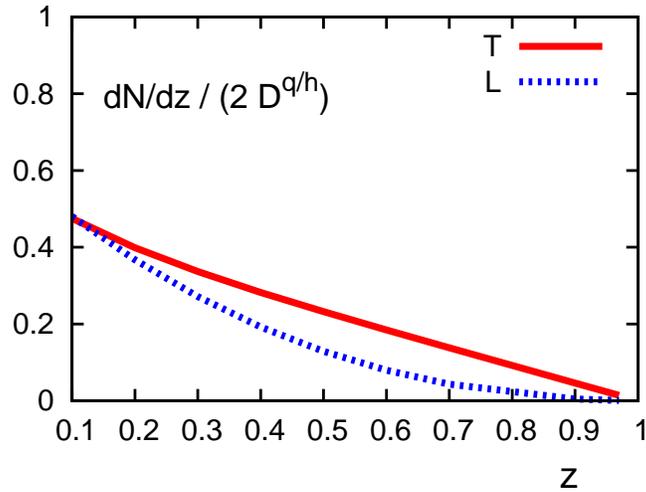,scale=1.4}
\caption{The ratio $d N^{\gamma_{T,L}^{\ast}/h} /dz/[2D^{u/h}(z,Q^2)]$
as a function of the momentum fraction $z$. 
$d N^{\gamma_{T,L}^{\ast}/h} /dz$ is the total differential multiplicity in the BDR
calculated using Eq.~(\ref{eq:nuc:black_nuc:dfs:41}); $D^{u/h}(z,Q^2)$ is the leading twist up-quark fragmentation function of Ref.~\cite{Bourhis:2000gs}. 
The solid curve corresponds to the transverse photons; the dotted curve corresponds to the 
longitudinal photons.
Both curves correspond to
$Q^2$=2 GeV$^2$.}
\label{fig:nuc:black_nuc:1}
\end{center}
\end{figure}

An illustration of the results of the calculation
of $d N^{\gamma_{T,L}^{\ast}/h} /dz$ is presented in Fig.~\ref{fig:nuc:black_nuc:1}. 
In this figure, we plot the ratio $d N^{\gamma_{T,L}^{\ast}/h} /dz/[2D^{u/h}(z,Q^2)]$
as a function of the momentum fraction $z$.
The solid curve corresponds to the transverse photons; the dotted curve corresponds to the 
longitudinal photons.
 In the absence of the BDR modifications
of the spectrum of the produced  particles, this ratio would be unity.
For $D^{u/h}(z,Q^2)$, we used the LT up-quark fragmentation function of 
Ref.~\cite{Bourhis:2000gs} at $Q^2 =2$~GeV$^2$.
One can see
from the figure that a gross violation of the
factorization theorem is expected in the BDR:
the spectrum of leading hadrons is much softer for $z > 0.1$.
Note also that the use of the leading twist fragmentation functions in
the above expression probably underestimates absorption.
Hence, the  curves in Fig.~\ref{fig:nuc:black_nuc:1} can
be considered as a conservative lower limit on the amount of the suppression due to the BDR dynamics.

With an increase of $Q^2$, we expect a further softening of the yield of leading partons  related to the change in the partonic structure of the virtual photon wave function:
progressively more configurations will contain extra hard gluons, which fragment independently in the BDR, further amplifying deviations from the standard leading twist predictions.

Another important signature of the BDR is the change of  $p_t$ distributions with  decreasing $x$ (at fixed $Q^2$). The spectrum   of leading hadrons should broaden due to the increased $p_t$ of the fragmenting partons. Hence, the most efficient strategy to look for the onset of the BDR  would be to  select leading jets 
in the current fragmentation region and examine the  $z$ and $p_t$ dependence of such jets. Qualitatively, 
the effect of broadening of the $p_t$ distributions is similar to the increase of the $p_t$ distribution in  
the Color Glass Condensate approach~\cite{McLerran:1993ni}, 
see also Refs.~\cite{Gelis:2010nm,Iancu:2003xm} for reviews.

An important advantage of inclusive scattering off a nucleus is the possibility to use simple triggers for centrality.  For example, one could use the number of nucleons emitted in the nucleus decay (soft nucleons in the nucleus rest frame) to select scattering at the central impact parameters. Such a selection allows for the effective thickness of the nucleus to be increased, 
as compared to the inclusive situation,
by a factor $\sim 1.5$ and, hence, allows for the BDR to be
reached at significantly
larger $x$.
The signal for the BDR will be a change of the spectrum with centrality of the collisions, in
contrast to the LT case where no such a correlation is expected. Note that the lack of 
the absorption 
of leading particles in DIS off nuclei at the fixed-target energies and $Q^2 \ge 2$~GeV$^2$ 
is well established experimentally, see, e.g.~\cite{Ashman:1991cx}.

To summarize, predictions for a number of simple final state
observables in the
black disk regime (BDR) limit are distinctly different from those made in the leading twist
approximation and, hence, will provide model-independent tests of
the onset of the BDR.

\subsection{Post-selection phenomenon as a precursor of the BDR} 
\label{subsec:post-selection_bdr}

It has been explained above that a rapid increase of the amplitudes of hard processes
with increasing energy leads at some $x\le x_{cr}$ 
($x_{cr}$ is a certain critical value of Bjorken $x$) 
to 
significant higher twist (HT) terms  in the cross sections of the hard processes, i.e., to the violation of the QCD predictions based on factorization theorems. An important physical manifestation of this increase is the suppression of leading jet  (high $p_t$ hadron) production as compared to the LT approximation. (The simplest example of this phenomenon is production of the leading hadrons in the BDR for the $\gamma^{\ast}A$ scattering; it was considered in 
Sec.~\ref{sec:nuc:black_nuc:is}.) 
In this subsection we explain that this phenomenon leads to deviations from the LT approximation in the kinematics, where the effects for inclusive cross section are still small, which serve as a precursor of the BDR.

The evaluation of the suppression of the yield of leading partons  
(high $p_t$ hadrons)  within the classical theory gives the suppression  factor at a fixed impact parameter $b$,
$R_{\rm supp}(b)$: 
\begin{equation}
R_{\rm supp}(b)=e^{ -AT_A(b)\,\sigma_{\rm inel}} \,,
\end{equation}
which should be
compared to the impulse approximation expectation of the yield $\propto AT_A(b)$.
The suppression is large for central impact parameters where 
$AT(b) \sim 2$ fm$^{-2}$;  even for $\sigma_{\rm inel} =10$ mb, one obtains 
$R_{\rm supp}(b) \approx 1/7$.

At the same time, the eikonal approximation in the form given by 
the classical or non-relativistic quantum mechanics is inapplicable in 
a quantum field theory and, in particular,  in QCD where all amplitudes 
are predominantly imaginary and elastic processes are the shadows of
inelastic ones. It strongly violates the energy-momentum constraint and should be modified.  Indeed,
it follows from the combination of the topology of
the  dominant HT effects  and the constraints due to energy-momentum conservation that several valence partons of the projectile should participate in the hard high energy process,
which is different from a one-parton situation in the LT approximation 
where HT effects are absent,  see the 
discussion in Sec.~\ref{subsec:cem}.  

It was argued in~\cite{Frankfurt:2001nt} that in the black disk regime  
the energy of the projectile splits before the collision---the post-selection phenomenon.  The presence of the strong gluon fields in 
the target selects different manifolds of quark and gluon configurations  in the projectile wave function much more strongly than a target with moderate gluon fields. In Sec.~\ref{sec:nuc:black_nuc:is} we considered the  simplest   example of the  inclusive production of the leading hadrons in DIS for $Q\le 2 p_{t,{\rm BDR}}$. The interactions with the target are not suppressed 
up to   $p_t \sim p_{t,{\rm BDR}}$, leading to the selection of configurations in 
$\gamma^{\ast}$ where the longitudinal fractions carried by the quark and antiquark are comparable. 
Thus, the  post-selection phenomenon is well suited for the experimental detection of the kinematical region (where $x$ is not far from  $x_{cr}$), where HT effects are significant, 
i.e., in the kinematics where the strength of the interaction approaches the BDR strength.
  
In the case of a parton  of a hadron  projectile propagating through 
the nucleus near the BDR, i.e., in the projectile fragmentation region, the
effective energy losses were estimated in Ref.~\cite{Frankfurt:2007rn}. 
For quarks, they  are expected to be  of the order of 10$\div$ 15\% 
in the regime of the onset of the BDR and larger deep inside this regime. 
Also, the effective energy losses  are somewhat larger for gluons as the $g\to gg$ 
splitting is more symmetric in the light-cone fractions than the $qg$ splitting.

To visualize and evaluate the post-selection phenomenon, we somewhat generalize the technology of the AGK cutting rules~\cite{Abramovsky:1973fm} to hard processes to account  
for the nontrivial constraint due to energy-momentum  conservation. For the sake of simplicity, 
we discuss here the simplest case of the  exchange 
by two ladders.  The   generalization to the general case of exchange by many ladders should
be rather straightforward.  

(i) As we already discussed in Sec.~\ref{sec:hab_pdfs}, the contribution of the double ladder exchange diagram to the total cross section $\sigma_2$ (the effect of the screening  corresponds to  
$\sigma_2 < 0$, so that $\sigma_{\rm tot}=\sigma_1 +\sigma_2 < 
\sigma_1 $) can be written as a sum of the contributions of different cuts of the double ladder diagram corresponding to three different final states: diffraction, production of the final states with single multiplicity (the cut of one of the ladders), and production of the states with double multiplicity (the cut of both ladders). Neglecting the real part of the ladder amplitude, we find 
\begin{eqnarray}
&&\sigma_2= \sigma_{\rm diff}^{(2)} + \sigma_{\rm single}^{(2)} + \sigma_{\rm double}^{(2)} \,, \nonumber\\
&&\sigma_{\rm diff}^{(2)} = - \sigma_2 \,, \nonumber\\
&&\sigma_{\rm single}^{(2)}= 4  \sigma_2 \,, \nonumber\\
&&\sigma_{\rm double}^{(2)} =-2  \sigma_2 \,.
\end{eqnarray}   
In this derivation, we ignored the  constraint due to the energy-momentum conservation 
which requires to substitute diffraction by the system 
of partons taking into account the post-selection phenomenon,  see the discussion below.  
     
(ii) The yield of jets (high $p_t$ hadrons) at zero rapidity is not screened since the contributions of  $\sigma_{\rm single}$ and $\sigma_{\rm double }$ cancel in the inclusive yield as the double cut corresponds to the double multiplicity and 
 $\sigma_{\rm single}=-2\sigma_{\rm double }$. 
Taking into account the energy-momentum conservation does not change 
this relation which is the application of the AGK combinatorics derived 
within the "Pomeron Calculus"~\cite{Gribov:1968fc} to hard processes.

(iii) The yield of the leading jets (high $p_t$ hadrons) at large rapidities is screened:
\begin{equation}
\sigma({\rm inclusive})= (\sigma_{1} -|4 \sigma_{2}|) N_{{\rm inclusive,\, single}} + 
2 |\sigma_{2}| \kappa N_{\rm inclusive,\, single} +|\sigma_2| N_{\rm diff} \,,
\end{equation}
where $N_{\rm inclusive,\,single}$ is the multiplicity of the  hadrons in the cut 
of the single ladder;  the factor $\kappa$ accounts for the suppression of 
the multiplicity for the leading particles in the double ladder cut, which reflects the sharing 
of the projectile-parton energy between the two ladders. In the kinematics corresponding to the leading particle production,  we expect that 
$\kappa \ll 1$.   Therefore, in the case of the central collisions, 
where diffraction does not contribute, we find the suppression factor close to $1 - 4 \sigma_{2}/\sigma_{1}$, which is much larger than in the total  cross section.  
Thus, HT effects should reveal themselves 
at significantly larger $x$ as compared to the BDR. In fact, 
in the kinematics where the correction to the total cross section due to the higher twist is just 25\%, 
the  single multiplicity cross section turns negative~\cite{Frankfurt:1995jw}.

(iv)
The smallness of the parameter $\kappa$ follows from the constraint due  to the energy-momentum conservation.  To evaluate $\kappa$, let us take into account the fundamental property of parton ladders that the leading  parton  within a ladder  carries on average the fraction $1- \epsilon$  of the initial parton momentum.  The double multiplicity hard processes may arise when at least two partons participate in the hard collision. 
In this case, the system of two leading partons carrying fractions $z_1$ and $z_2$ of the projectile momentum is initiated by the parton configuration in the projectile wave function where the partons carry the fraction $z_1(1+\epsilon)+z_2(1+\epsilon)$. 
This fraction  should be compared to $\langle z_1(1+\epsilon)+z_2 \rangle$ in the case of single multiplicity events.  An account of $\epsilon \neq 0 $ (see below) is important in the fragmentation region as it follows from the light-cone  momentum conservation $\sum z_i+(z_1+z_2)\epsilon =1$ and $z_i\ge 0$.

(v)
The nonzero value of $\epsilon\approx 0.1\div 0.05$ arises automatically within the LO and NLO DGLAP approximations. ($\epsilon =0.14$ 
corresponds to the distance in rapidity between the adjacent partons within the parton ladder equal to 
$\Delta y=2$, i.e., in the kinematics achieved at RHIC and HERA;  $\epsilon=0.05$ corresponds to  $\Delta y=3$ which could be achieved in the specific  kinematics at the LHC.)
The fact that
$\epsilon \neq 0$ is ignored within the leading $\alpha_s\log(1/x)$ approximation that assumes the dominance of the multi-Regge kinematics; to some extent, it is accounted for within the 
resummation models.
   
(vi) In the case of the interaction with $N$ nucleons, the suppression effect is increasing due to the AGK combinatorics and a small contribution to the cross section from two cut ladders.
(The higher order terms $\propto \sigma_3, \sigma_4$ enter 
the total cross section
with  alternating signs, which reduces the screening effect as 
compared to the double scattering approximation. At the same time, these terms 
enter  the expression for the inclusive cross section with a negative sign.)
Hence, for central collisions, the actual  ``post-selection energy loss'' should increase with an atomic number  $\propto A^{1/3}$.

Let us consider several examples.  In the case of a virtual photon projectile,
 $z_1\approx z_2\approx 1/2$ and the production of the leading hadron carrying the
fraction $\ge 1-\epsilon $ is dynamically suppressed. In the limit when the BDR contribution dominates, there are two contributions to the inclusive spectrum: one is due to the diffraction, which is expressed through the virtual photon wave function, and the other one that is due to the wave packet 
fragmentation. In the approximation that we considered in this section, 
they give equal contributions (both suppressed as compared to the LT expectation). 
The present  discussion indicates that the non-diffractive component with $k_t < k_{t,\rm BDR}$  
should be further  suppressed  as compared  to the estimate presented in 
Sec.~\ref{sec:nuc:black_nuc:is}, which originated solely from the structure 
of the virtual photon wave function. However, the 
configurations interacting inelastically with nearly the BDR strength will interact with a large probability via several ladders, leading to a further suppression of the forward spectrum. 

Our second example is the proton-nucleus scattering.  
A new feature here is that it is possible  to select non-diffractive interactions at different impact parameters. Comparing the yield of leading particles in this case to the impulse approximation expectation, we observe that  the contribution of $N$-ladder interactions  
leads to the significantly larger ''post-selection energy loss'' $\approx (N-1)\epsilon$.

Thus the post-selection phenomenon corresponding to significant effective fractional  energy losses is definitely present in QCD.
However, to calculate the post-selection phenomenon   more accurately, one needs to model LT and HT effects.

\subsection{Evidence for post-selection 
effect in the forward pion production in the deuteron-Gold collisions at RHIC}
\label{subsec:post-selection2_bdr}

The leading hadron production for  $p_t \sim {\rm few}$ GeV/c
in hadron-nucleus scattering at high energies 
can be used as a sensitive test of the  onset of the BDR dynamics. In this limit, 
pQCD provides a good description of the forward single inclusive pion production in $pp$  
scattering at the RHIC energies~\cite{Werner}. At the same time, a 
 comparison of the pQCD calculations with the data  shows that  such calculations 
grossly overestimate the cross section of the pion production in $dAu$ collisions in the same kinematics. The analysis of~\cite{Guzey:2004zp} has demonstrated that the dominant mechanism of 
the single pion production in the $NN$ collisions in the kinematics which was studied at RHIC 
is scattering of the leading quark of the nucleon  off the gluons of the target  with the median value of $x$ 
in the range $x_g \sim 0.01 \div 0.03$ depending on the rapidity of the pion.  
The nuclear gluon density for such $x_g$ is known to be close to the incoherent sum of the gluon fields of the individual nucleons since the coherent length in the interaction is rather modest for such distances (see, e.g., Fig.~\ref{fig:LT2009_pb208}). As a result, the leading twist  nuclear shadowing effects can explain only a very small fraction of  the observed suppression~\cite{Guzey:2004zp} and one needs a novel dynamical  mechanism to suppress the generation of pions in such collisions. 
In particular, it was pointed out in~\cite{Guzey:2004zp} that the fractional energy losses on the scale of $10\div 15\%$ give a correct magnitude of the suppression of the inclusive spectrum due to a steep fall of the cross section with $x_F$, which is consistent with the estimates within the post-selection 
dynamics.
 
An important additional information comes from the studies 
of the correlation of the leading pion with the pion produced at the central rapidities, 
$\eta \sim 0$~\cite{star,phenix2}, which corresponds to the kinematics which receives   the dominant contribution from the scattering off gluons with $x_g\sim 0.01 \div 0.02$. 
The rate of the correlations for $pp$ scattering is consistent with pQCD expectations.
An extensive analysis performed in~\cite{Frankfurt:2007rn} has demonstrated that the 
strengths of ''hard forward pion''  -- ''hard $\eta \sim 0$ pion''  correlations in 
$dAu$ and $pp$ scattering are similar; a  rather small difference in the pedestal originates from the multiple soft collisions. The smallness of the increase of the soft pedestal as compared to $pp$ collisions unambiguously  demonstrates that the dominant source of the leading pions is the $dAu$  scattering 
at large impact parameters. 
This conclusion is supported by the experimental observation~\cite{Rakness} that the associated multiplicity of soft hadrons in events with forward pions is a factor of two smaller than in the minimal bias $dAu$ events. The reduction by a 
 factor of two is consistent with the estimate of~\cite{Frankfurt:2007rn} 
based on the analysis of the soft component of $\eta =0$ production for the forward pion trigger.
 Overall these data indicate that 
 (i) the dominant source of the forward pion production 
 is $2\to 2$ mechanism, (ii) production is dominated by the projectile scattering at large impact parameters, (iii) the proportion of small $x_g$ contribution in the inclusive rate is approximately the same for $pp$ and $dAu$ collisions. 
 
 The lack of an additional suppression of the $x_g \sim 0.01$ contribution to the double inclusive spectrum as compared to the suppression of the inclusive spectrum is explained in the post-selection dynamics  since the pions with $\eta \sim 0$ are produced in the fragmentation of the gluons
 with  relatively small momenta  in the nucleus rest frame, putting these gluons far away from the BDR.
 
 It is difficult to reconcile  the enumerated  features of the forward pion production data  with the 
 $2\to 1 $ mechanism of Ref.~\cite{Kharzeev} inspired by the Color Glass Condensate model. 
In the scenario of~\cite{Kharzeev}, the  incoherent $2\to 2$ mechanism is neglected
and  a strong suppression of the recoil pion production is predicted. 
Also, it leads to the dominance of the central impact parameters and, hence, to
a larger multiplicity for the central hadron production in the events with the forward pion trigger. 
The observed experimental pattern indicates that the models that neglect the
contribution of the $2\to 2$ mechanism and consider only  $2\to 1$ processes, see, e.g., 
Ref.~\cite{Dumitru:2005gt}, strongly overestimate  the contribution of the $2\to 1$ mechanism to the inclusive cross section.

 Additional information comes from the recent studies~\cite{starqm09,phenixqm09} of the production of two forward neutral pions  in $pp$ and $dA$ scattering. 
This kinematics strongly enhances the contribution of small $x$ in the target~\cite{Guzey:2004zp}.
One leading  pion serves as a trigger and the second leading pion has somewhat smaller 
longitudinal and transverse momenta. 
The data indicate a strong suppression of the back-to-back production of pions in the central $dAu$ collisions. Also,  a large fraction   of the double inclusive cross section  is isotropic in the azimuthal angle $\Delta\varphi$ between the two pions.  
In order to understand the origin of the suppression and other features of the data,  
there has been performed a study~\cite{SV} which is summarized below.
  
\begin{figure}[t]  
   \centering
   \includegraphics[width=1.0\textwidth]{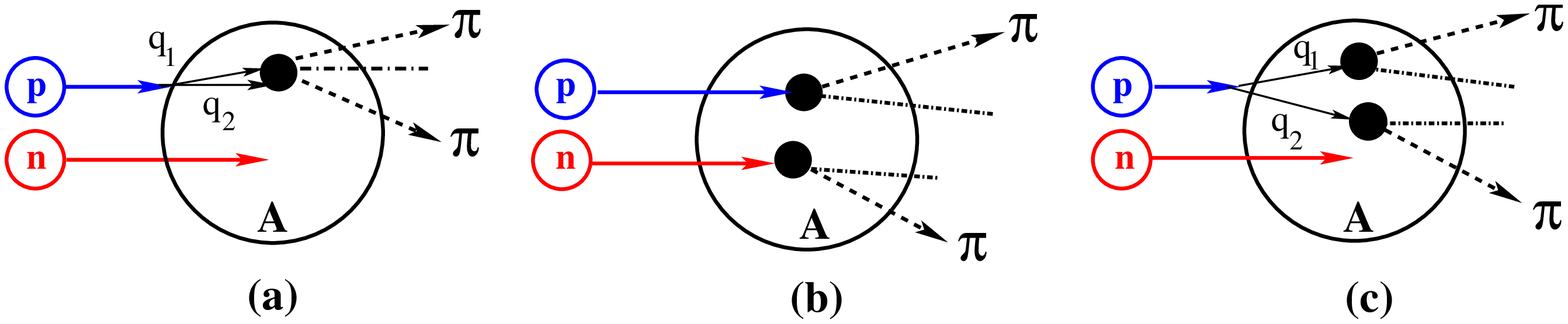} 
   \caption{Three  double parton mechanisms of dipion production in $dAu$ collisions.}
    \label{sketch}
 \end{figure}

First, it was demonstrated that for the discussed forward kinematics, the binary hard collisions are dominated by very large $x\sim 0.7 \div 0.8$. 
As a result, the double scattering mechanism, where two quarks with smaller $x$ scatter off the gluons 
of the target, gives an important contribution, see Fig. \ref{sketch} for the case of deuteron-nucleus scattering.
The double scattering mechanism leads to the contribution to the cross section which does not depend on $\Delta\varphi$.

In  the case of $pp$ scattering, the  numerical analysis of~\cite{SV} indicates  that in the RHIC kinematics, the double and single contributions are comparable. 
 This gives a natural explanation to the   experimental observation that in $pp$ scattering,
 the   number of events in the pedestal is somewhat larger than in  the peak around 
$\Delta\varphi \sim \pi$,  which is dominated by the LT contribution. 
 
In the case of $dAu$ collisions, the double parton interaction contribution for collisions at the central impact parameters is strongly enhanced by the combinatorial 
 factor [in addition to the mechanism analogous to the one for $pp$ scattering shown 
in Fig.\ref{sketch}(a)],
 it is possible to emit pions in the interactions of quarks of the proton and neutron  
with two different nucleons of the nucleus [Fig.~\ref{sketch}(b)] or in the interactions of quarks 
of one of the nucleons with two different nucleons of the nucleus [Fig.~\ref{sketch}(c)]. 
In the impulse approximation,  the enhancement factor 
 for the multiplicity of the events in the pedestal in $dAu$ collisions as compared to the 
$pp$ collisions, $r_A$, is proportional to the nucleus thickness, $T_A(b)$,  
times the combinatorial factor due to a larger number of 
quark-quark pairs in two nucleons than in one nucleon. As a result,  
one finds $r_A\sim 10$ for central $dAu$ collisions with the typical  pseudorapidities $\eta_{1,2}$ 
of two pions probed in the RHIC experiment; $r_A$ increases with an increase of the 
rapidities of the pions.
 
Since the single inclusive pion spectrum for $\eta_2 \sim 2 \div 3$ is suppressed by a factor of the order $R_A(b)= 1/3 \div 1/4$, one finds for the ratio of the pedestals in $dAu$ and $pp$:
\begin{equation}
R_{\rm pedestal}= r_A R_A(b) \sim 2.5  \div 4 \,,
\end{equation}
which should be compared with the experimental value of $R_{\rm pedestal}\sim 3$. 
Hence, the double scattering mechanism with a strong suppression of pion production at small impact parameters  naturally explains the magnitude of the enhancement of the pedestal in central 
$dAu$ collisions.

If  most of the pedestal in the kinematics studied at RHIC is due to the double parton mechanism,  the uncertainties in the estimate of the rates due to this mechanism   and uncertainties in the strength of the suppression of the single inclusive forward pion spectrum at $b\sim 0$ would make it  very difficult to subtract this contribution with a precision necessary to find out whether all pedestal is due to double parton  mechanism or there is a room for  a small contribution of the $2\to 1$ broadening mechanism  as assumed  in~\cite{Albacete:2010pg}.

The suppression of the away peak originating from the LT contribution is  due to 
the following two effects: (i) the gluon shadowing 
  for $x\sim 10^{-3}$,
$b\le  3$ fm and $Q^2 \sim {\rm few}$ GeV$^2$ reduces the cross section by a factor of about two 
(see Fig.~\ref{fig:LT2009_ca40_impact_Bdep}), (ii) a stronger effect of the
effective fractional energy losses due to larger $x$ of the quark in the LT mechanism than in the double parton mechanism leads to the suppression factor of the order of two~\cite{SV}. 
Combined together these effects result in  a suppression of the order of four as compared to the single pion trigger and overall suppression of the order of ten.  This is consistent with the experimentally observed magnitude of the suppression.

Note, however, that for the suppression of such a magnitude, the treatment of the nucleus
that 
neglects fluctuations of the positions of the nucleons in nuclei may be an oversimplification. 
In particular, it does not take into account a possibility of 
the ''punch through'' process in which a quark scatters off one nucleon but does not encounter any extra nucleons at its impact parameter. The probability of such collisions at $b\sim 0$ 
for interaction with Au nucleus is of the order 
$5 \div 10\%$~\cite{Alvioli:2009ab}. 
Thus, these collisions may be responsible for a large fraction of the away peak.

In summary, the RHIC data clearly indicate the break-down of pQCD in 
the scattering off heavy nuclei at central impact parameters in the kinematics, 
where pQCD works for the $pp$ scattering.  
The post-selection effect (effective fractional energy losses) for the propagation of the leading partons is the only proposed mechanism  that  allows one to explain the suppression of the $2\to 2$ mechanism for production of leading pions, which dominates in the $pp$ scattering including such observations
as the survival of 
forward-central correlations, the
strength of the pedestal,  and the suppression of forward-forward correlations.
 The processes of the production of two forward pions in $pp$ and $dAu$ scattering due to 
the double parton mechanism  and large suppression of the single pion yield for the scattering at small impact parameters explain the pedestal observed at RHIC  and make it difficult to search for the signal of $2\to 1$ processes. The magnitude of the away peak is consistent with the expected magnitude of the suppression of the $2\to 2$  mechanism due to LT gluon shadowing and fractional energy losses due to the post-selection mechanism.
   
 In the rest frame of the nucleus, the RHIC kinematics corresponds to the interaction of
 a quark with the energy of the order of 10 TeV and virtuality corresponding to 
$p_t \sim 1 \div 2$~GeV/c. Such quarks  can resolve the gluon fields with 
$x \sim 3 \times 10^{-4}$. This gives an independent estimate of the kinematics 
where the BDR sets in for the interaction of quarks. 
It corresponds to  somewhat higher $x$ than the ones we found in the analysis 
of the impact factors in Sec.~\ref{subsec:bdr_inclusive}.
 However, as we mentioned in Sec.~\ref{sec:nuc:black_nuc:is}, the single particle spectrum is more sensitive to non-linear effects than the total cross section since the shadowing effects 
are amplified in this case.

The observation of the post-selection effects at RHIC in the discussed above
kinematics has  important implications for $AA$ and $pA$ collisions at the LHC.
Indeed, a parton carrying the fraction $x_F$ of the projectile hadron momentum
dissociates into $N$ partons long before a collision with a target. So, 
in the vicinity of the BDR,
the
energy carried by each of these partons can be estimated  as
$\approx x^{\prime}_F/2$ assuming dominance of double inelastic collisions, i.e.,
$N=2$.  Thus, in this case, a
leading parton carries only the fraction  $\approx x^{\prime}_F/2$,     which is
significantly smaller than the  corresponding fraction $x^{\prime}_F$ expected
within the  LT  approximation. Furthermore, the  increase of the collision
 energy and the atomic number of the target  lead to an increase of the
effective number of inelastic collisions of a parton, $N(b,x)$.  
Hence,  the fraction of projectile hadron momentum carried by
the interacting parton  should  decrease with an increase of the collision energy
tending  to   $\approx 1/N(b,s)$ --- the number of nucleons at a given impact
parameter. This picture follows
from the generalization of the Gribov-Glauber approximation to inelastic
processes at high energies which includes an account  of the energy-momentum
conservation. This generalization is called the color fluctuation approximation in 
this review.

Thus, pre-selection effects should be more important  at the LHC than at RHIC.
   To visualize the expected effects, for numerical estimates we use the analysis
of the $dAu$ RHIC data described above. Indeed, energies of the partons of 10
TeV  and above and $p_t \sim 1 \div 2$ GeV/c (for which the post-selection
effect is large) correspond to a wide range of rapidities
  in the nucleus rest frame  for collisions at
the LHC energies. For example, for the
full LHC energy for proton-$^{208}$Pb collisions of $\sqrt{s_{NN}}$=8.8 TeV, 
the post-selection effect for $p_t\le 2$ GeV/c should extend  to 
$y_{\rm max}-y\le 8$, or $x_F> 3\times 10^{-4}$, for quarks  and 
to a broader range for gluons.


\subsection{Implications of energy-momentum conservation for the onset of BDR}
\label{subsec:energy-momentum}

In Sec.~\ref{subsec:cem}, we explained  that the eikonal approximation strongly violates the energy-momentum conservation. Here we analyze,  albeit briefly, implications of the energy-momentum conservation for the  dynamics of an onset of the BDR for hard processes.  

The energy-momentum conservation restricts a possible number 
of gluons radiated within the multi-Regge kinematics. This number can 
be estimated from the condition that the distance in rapidity between the adjacent gluons in a pQCD ladder should be at least $\ge 2-3$ units. This estimate shows 
that in the whole kinematical region covered by HERA and 
most of kinematics of hard processes at the
LHC (with an exception of a special kinematical region),  the number
of gluons in the multi-Regge kinematics, $N$, is small. It is related to 
the span in the rapidity for a given accelerator, $\delta y$, by the following relation:
\begin{equation}
\delta y=2+3(N+1)  +2 \,,
\label{eq:rapidity_span}
\end{equation}
where the factors of two is an estimate of length in rapidity of each fragmentation region.
Let us give  illustrative estimates of the production of  $N$ gluons in the 
multi-Regge kinematics at different accelerators:
\begin{itemize}
\item[(i)]
$\delta y({\rm HERA})=\ln(Q/xm_N)\approx 9.2$  for $Q^2=4$ GeV$^2$.
Therefore, $N \approx 1$. 
\item[(ii)]   
At the LHC for the full energy $\sqrt s=(7+7)$ TeV, the rapidity span is
$\delta y=\ln(s/m^2_N)=18.7$.  Then, for instance, if a $q\bar q$ dipole in the 
light-cone wave function of proton carries the proton momentum fraction
$z\approx 0.1$ and the partons have the relative momentum 
$p^2_t=$few GeV$^2$,  the number of allowed gluons is $N\approx  3$. 
\item[(iii)]
In the ultraperipheral collisions at the LHC, $N\approx 2$.
\end{itemize}
This restriction on $N$ explains the experimental observation of the
dominance of large $\beta=Q^2/(Q^2+M^2_X)$
in the diffraction in DIS, i.e., the contribution of multi-Regge kinematics is a small correction, see the discussion in Sec.~\ref{subsubsec:diffractive_masses}.  A practical conclusion from this is that production of gluons in the multi-Regge kinematics can be 
taken into account by calculating only a few Feynman diagrams with the emission of few gluons. 

Thus, the energy-momentum conservation ensures the dominance of the double logarithmic approximation of pQCD and the DGLAP saddle point in the inverse Mellin transform of  the proton structure functions in a wide kinematical domain, including the kinematics achieved at HERA and most of the kinematics of the LHC.  Moreover, 
as a consequence of the account of the energy-momentum conservation,
the multiplicity of gluons in multi-Regge kinematics is  significantly smaller than that assumed 
within the BFKL and Color Glass Condensate (CGC) approaches.

However, at extremely large energies, where  the  parameter 
$\alpha_s(Q^2_0)(N_c/2\pi)\ln(x_0/x)\ge 1$, the  DGLAP approximation becomes questionable and this kinematics may appear to be a window to the gluon radiation in the multi-Regge kinematics, see
also the discussion in Sec.~\ref{subsec:comparison_bdr}.

\subsection{BDR as new phase transition with the change of continuous symmetry}
\label{subsec:symmetry_bdr} 

A direct consequence of the  condition of complete absorption, 
Eq.~(\ref{absorption}),  is the  complete disappearance of approximate Bjorken scaling for the
$\gamma^{\ast}+T\to X$ DIS cross section:
this cross section is independent of $Q^2$ within the BDR. 
Such a behavior means  the breakdown of the  scale invariance characteristic for the parton 
model and the dynamical two-dimensional conformal invariance characteristic for the 
DGLAP~\cite{Gross:1973id}   and BFKL~\cite{Lipatov:1993qn}  approximations.   
Thus, the black disk  and pQCD regimes have  different  continuous symmetries, 
and the transition  from the pQCD regime to the BDR resembles a phase transition with 
a change of the continuous symmetry.

A hadronic state produced in DIS off a nucleon target within the BDR is a result of 
fragmentation of three phases of QCD. 
(These phases depend on the impact parameter of the collision $\vec b= \vec \rho_1 -\vec \rho_2$,
where $\rho_i$ are transverse coordinates  of the center of mass of the hadrons.)
  At small impact parameters $b$, it is  
the BDR phase  without  the  two-dimensional conformal symmetry;  at larger $b$, it is
the pQCD regime with the unbroken chiral  and two-dimensional conformal symmetries; and at even larger impact parameters, it is the soft QCD regime with the spontaneously broken chiral symmetry, 
where the conformal symmetry is probably also spontaneously broken.   An important role of transitions 
with the change of continuous symmetries has  important 
implications for high energy hadron-hadron interactions.

\noindent 
(i) For $b\sim 0$  in the cylinder (in transverse coordinate-rapidity space) centered around  
$\vec r = (\vec \rho_1 +\vec \rho_2)/2$, the BDR interactions dominate and suppress 
soft QCD interactions generating the Froissart limit of soft QCD at high energies.  The radius of the cylinder grows with an increase of energy, while the BDR  scale at given $\rho_i$ grows with  the incident energy as well.
Note here that the interactions in $pp$ scattering at small impact parameters become practically black for small $b$, although it is not a priori  clear whether pQCD or non-perturbative QCD is responsible for this.

\noindent 
(ii) With an increase of $b$, the BDR region occupies an almond-shaped transverse area centered around $\vec r$,
which is surrounded by the pQCD phase and far out by the non-perturbative phase.

\noindent 
(iii) For $b > 2r_N$, for a wide range of energies most of the interaction region  is in  the non-perturbative phase with the spontaneously broken chiral symmetry with a small 
pQCD cylinder around small $\vec r$  and practically no BDR region.

\subsection{BDR {\it vs.}~DGLAP approximation regime}

An attractive  feature of the BDR for a large  nucleus is that a  new phase of QCD, with a small 
coupling constant and  the continuous symmetry different from that of pQCD, should manifest 
itself in $ep$, $pA$, and $AA$ collisions for  a wide  range of impact parameters and
in a broad range of phenomena.

We have demonstrated in the previous subsections that 
predictions for many observables in the BDR  are qualitatively 
different from those in the leading twist regime. 
We summarize here the most striking of these predictions.
\begin{itemize}
\item[(i)]
The QCD factorization theorem is violated in the vicinity of the BDR. 
Therefore, the pQCD and BDR phases have
different continuous symmetries, see Sec.~\ref{subsec:symmetry_bdr}.
Also,  the independence of the form of leading jets from a target is violated,
see Sec.~\ref{sec:nuc:black_nuc:is}.
In the vicinity of the BDR, 
a major new effect is the post-selection  of projectile configurations which mimics fractional energy losses and leads to the suppression of the leading hadron production for   
$p_t \le$ the BDR scale~\cite{Frankfurt:2007rn}, see Sec.~\ref{subsec:post-selection_bdr}.
\item[(ii)]
Compared to the DGLAP approximation,  the BDR predicts a more
rapid $Q^2$ dependence and a different---but still rather rapid---increase with energy of structure functions 
of a nucleon:   $F_T^N(x,Q^2) \propto Q^2 R_N^2 \ln (x_0/x)(1+c_N \ln^2 (x_0/x))$ (the coefficient of proportionality is given in Eq.~(\ref{eq:BDR_FT})].  

\item[(iii)]
The structure function of a heavy nucleus  should increase with increasing energy more slowly than the 
nucleon one: $F_T^A(x,Q^2)= Q^2 \ln (x_0/x)(R_A^2+c R_A R_N \ln^2 (x_0/x))$ as a consequence of the large nucleus radius (see the discussion in Sec.~\ref{subsec:bdr_sf}).
Thus, the ratio of the nuclear to nucleon structure 
functions decreases with increasing $(x_0/x)$: 
\begin{equation}
F_{2A}/(A^{2/3} F_{2N}) \propto 1 / (1+c_N\ln^{2} (x_0/x)) +c A^{-1/3} \,.
\end{equation}
It would be a difficult problem to distinguish experimentally 
such a behavior from the  leading twist nuclear shadowing phenomenon.
\item[(iv)]
The cross section of diffraction should reach nearly half of the total cross section.
\item[(v)]
The cross section of exclusive vector meson electroproduction should
have a much weaker $Q^2$ dependence and be calculable in a 
model-independent way, see the discussion in 
Sec.~\ref{sec:nuc:black_nuc:dfs}.
\end{itemize}

\subsection{BDR {\it vs.}~parton saturation hypothesis}

Properties of the BDR have certain similarities, as well as a number of qualitative and quantitative differences, from the predictions based on  the  parton saturation hypothesis (the Color Glass Condensate framework---CGC).
 On the one hand, various models using the features of the CGC explore the approximation of large gluon density evaluated within the framework  of the LO BFKL approach together with the  elastic  eikonal approximation to stop the increase of structure functions with an increasing energy and, hence, some effects 
due to blackening of the interaction should be present within such models.   At the same time, 
it has been found that corrections to the LO BFKL approximation are huge and comparable with the leading term. Thus, it  should be significantly improved by a more accurate account of double logarithmic terms making 
it more close to the NLO DGLAP  approximation (resummation models) \cite{LipatovFadin,Ciafaloni:1999yw,Ball:2005mj,Altarelli:2008aj}. 
So far the resummation models have not been implemented within CGC approaches.

On the other hand, in contrast to the LO BFKL approximation, the DGLAP approximation used in the evaluation of an onset of the BDR conserves longitudinal energy and momentum. Moreover,
in contrast to the CGC models,
the color fluctuations approach (Sec.~\ref{sec:hab_pdfs}) together with the DGLAP approximation accounts for the energy-momentum conservation for multiple scattering processes  and, therefore, leads to a different physical picture and predicts different phenomena.

\begin{itemize}
\item[(i)] 
The BDR predicts that structure functions of a hadron (nucleus) at a fixed impact parameter  
should increase with energy  as, e.g., 
$F_{T}(x,b,Q^2)\propto \ln(x_0/x)$ as a consequence of the ultraviolet divergency of hadron  renormalization of the electric charge~\cite{Gribov:1968gs}.  Moreover, within the  BDR  structure functions of a hadron integrated over impact parameters  should increase with energy as $F_{T}\propto \ln^3(x_0/x)$.  
On the contrary, the  CGC models assume that ``saturated''  parton distributions and the composition of partons within a projectile photon  should be energy independent, see e.g.,~\cite{GolecBiernat:1998js}.
\item[(ii)]   
The BDR predicts that the structure functions  of a heavy nucleus  should depend on energy more slowly at achievable energies than the nucleon structure function  since the nucleon impact parameter distribution within a heavy nucleus is  flat in a wide range of impact parameters.  Note that in practice,  this prediction of the BDR
competes with the nuclear shadowing phenomenon.
On the contrary, the  CGC models assume the ``saturated'' parton distribution of a hadron (nucleus) and, therefore, the ratio of the nucleus to nucleon structure functions should be energy independent. 
\item[(iii)]  
CGC models and, in particular  the Balitsky-Kovchegov  
approximation~\cite{Balitsky:1995ub,Kovchegov:1999yj}, assume that branching of the parton showers in multi-Regge kinematics plays the  dominant role practically at any energies.
It follows from the account of the energy-momentum conservation  that 
an onset of multi-Regge showers  requires significantly larger 
energies than those required for an onset of the BDR. In particular, at the energies achievable at accelerators, the dominant diffractive processes correspond to production of masses $M^2\approx Q^2$ and slowly increasing with energy. This theoretical observation is in accordance with the HERA data on diffractive processes, 
see the discussion in Sec.~\ref{subsubsec:diffractive_masses}  and in particular 
Fig.~\ref{fig:r}.
\item[(iv)] 
CGC models  often employ the elastic eikonal approximation for the evaluation of hard processes, 
which violates the energy-momentum conservation, see Sec.~\ref{subsec:cem}.  
An analysis of the implications of the energy-momentum conservation finds that the 
application of the elastic eikonal approximation to high energy processes in the kinematics where inelastic processes dominate leads to serious inconsistencies. 
\item[(v)]
To evaluate nuclear shadowing and an onset of the BDR, in this review  we use the  semiclassical approximation 
in the form appropriate for a quantum field theory, which we called
the ``color fluctuations approach''. 
Because of the proper account of the diffractive processes, this approximation allowed us to implement the energy-momentum conservation in the multiple collisions in the regime where amplitudes are predominantly absorptive. This is achieved by summing over the contributions of degenerate trajectories in the allowed phase space  instead of taking into account just one trajectory within   the elastic
eikonal approximation familiar from  one-dimensional quantum mechanics and often
used within the CGC.   (Recall that the  semiclassical approximation to a many-body
quantum mechanical system, where variables in the Hamiltonian cannot be separated,  includes the 
sum over degenerate trajectories in the allowed phase space.) 
\item[(vi)] 
An account of the energy-momentum conservation leads to 
the effects mimicking energy losses. Such effects  do not exist within the eikonal approximation but are characteristic for an onset of the BDR.
\item[(vii)]
Recent CGC studies also began to include the dependence
of the interaction scale on the impact parameter in $ep$ and $eA$ scattering.  
(It has not been extended to the case of
 $pp$, $pA$, and $AA$  collisions yet.)  However, as we discussed
 in Sec.~\ref{subsec:symmetry_bdr},    at high energies and 
at large impact parameters, the non-perturbative QCD  phase, where the
chiral symmetry is spontaneously broken, should dominate and the pQCD phase would populate a
moderate $b$ region. At central impact parameters (which can be as large as $R_A$ for nuclear targets), the
BDR phase with the broken two-dimensional conformal symmetry should dominate at sufficiently high energies.

\item[(viii)]
Transverse momenta of partons within the projectile dipole,
i.e., in the region of current fragmentation,
are increasing with energy  as a consequence of the increase of pQCD interactions
with an increase of energy. Numerical calculations give for 
$\sigma_L$ for $Q^2=10$ GeV$^2$:  $k^2_t/k^2_0=(s/s_0)^{n}$  with $n\approx 0.04$ and slowly growing 
with $Q^2$ within LO DGLAP~\cite{Blok:2009cg}.   
An increase with energy of parton transverse momenta in DIS in the current
fragmentation region demonstrates the ambiguity of leading log approximations in the small $x$ regime since at the fixed-order in the coupling constant,  there is no such an effect.  At the same time, both the BDR and CGC predict an increase of parton momenta in the center of rapidity with an increase of energy.

\item[(ix)]
In this review, we discuss properties of the BDR inspired by hard QCD phenomena only. 
Soft QCD phenomena are suppressed by the choice of kinematics or accounted for within the 
framework of QCD factorization theorems, see the discussion of the weak parton density limit. On the contrary, models 
of CGC  explore eikonalization of  the LO BFKL approach to describe  both soft and hard QCD  phenomena.
\end{itemize}

Thus, the BDR and CGC models account for different phenomena and, therefore, make different predictions for the domain covered by HERA, RHIC, and the LHC.

\section{Conclusions}
\label{sec:conclusions}

In this review, we considered two distinctive regimes of nuclear shadowing: (i) the regime
of moderately small $x$ and not very large parton densities where the leading twist 
approximation (the QCD factorization theorem)
is applicable, and (ii) the regime of very small $x$ and large nuclei
(central impact parameters) where the leading twist (LT) approximation dramatically
breaks down and is replaced by the black disk regime (BDR) of the strong interaction.

In the leading twist regime, combining the Gribov technique 
developed earlier
for hadron-deuteron scattering,   the QCD factorization theorems for DIS, and 
QCD analyses 
of the HERA data on 
diffraction  in lepton-proton DIS, we developed the theory of 
the nuclear shadowing dynamics in the leading twist approximation.
In the
review, the model-dependent contribution of the interactions with $N \ge  3$ nucleons of the
nuclear target is treated using the 
semiclassical approximation in the form suited for a quantum field theory. We call this
approximation 
to high energy processes in QCD 
{\it the color fluctuation approximation};
it takes into account the phenomenological observation that diffraction in $ep$ DIS is
dominated by large-size fluctuations of the virtual photon wave function. (The color fluctuation 
approach allows one to overcome the principal difficulty of the treatment of 
multiple  inelastic interactions within the elastic eikonal approximation, 
i.e., in the form derived within the framework of one-dimensional non-relativistic quantum mechanics
which neglects the constraints of the energy-momentum conservation.)
Theoretical analyses of diffraction DIS $ep$ HERA data demonstrate the dominance of large
$\beta= Q^2/(Q^2 + M^2_X)$, where $M_X$ is the diffractively produced mass, which unambiguously
supports the dominance of the regime resembling 
Gribov-Glauber 
multiple rescatterings for
$x \ge 10^{-4}$ and forms the basis for the quantitative calculations of nuclear shadowing.

Using the framework of the leading twist theory of nuclear shadowing, we make predictions
for a number of quantities that can be accessed in DIS and hard photoproduction  with nuclei. 
These include, but not
limited to: (i) sea quark and gluon parton distributions and structure functions for a wide
range of nuclei (from deuterium to Lead) for $10^{-5} \le x < 1$ and  $4 \le Q^2 \le 10^4$ GeV$^2$, (ii)
nuclear impact parameter dependent PDFs and generalized parton distributions (GPDs)
in nuclei and cross sections of hard exclusive processes (deeply virtual Compton scattering,
exclusive production of $J/\psi$) with nuclei, (iii) nuclear diffractive PDFs and coherent and
incoherent nuclear structure functions. It will be possible to test a number of our predictions in 
the near future using ultraperipheral heavy ion collisions at the LHC, with further tests in the 
proton-nucleus collisions at the LHC. High precision tests will be possible  at a future
Electron-Ion Collider (EIC). The EIC is an ideal machine to study many predictions of
the leading twist theory of nuclear shadowing discussed in this review.

As $x$ becomes very small and for 
hard processes off
heavy nuclei at central impact parameters, the pQCD
amplitudes of hard processes that rapidly increase with energy reach their maximal values
dictated by the
probability conservation at a fixed impact parameter
in the $s$-channel and one enters the 
black disk regime (BDR)
 of the
strong interactions. The existence of this regime and the corresponding master formulas
with the predictions for hard processes follow from the conservation of probability for
the collisions at a fixed impact parameter. In particular, nuclear shadowing in the total
cross section of DIS, cross sections of hard diffractive (inclusive and exclusive) processes,
etc., are directly and model-independently calculable within the BDR. 
The characteristic features 
of the BDR include 
the forever increase of structure functions of nucleons and nuclei with an increase of 
energy, 
disappearance of the approximate scale invariance (approximate Bjorken scaling) 
and dynamical two-dimensional conformal 
symmetry characteristic for pQCD, the effective restoration of vector meson dominance for 
the exclusive production of vector mesons,  suppression of the leading hadrons in DIS final states, 
the post-selection  ''effective fractional energy losses", etc.

The QCD phenomena discussed in the review have serious implications for the final states
produced in hadron-nucleus and nucleus-nucleus collisions. In particular, the produced
matter is characterized by the interplay of three phases of QCD matter with different
continuous symmetries: the non-perturbative QCD phase with spontaneously broken chiral
symmetry, the pQCD phase with two-dimensional conformal invariance, and the BDR.
A variety of signals distinctive for the onset of the BDR---a new QCD phase---which are feasible for the 
experimental observation at the LHC is suggested.
Forward production of hadrons/jets in $pA$ collisions at RHIC and the LHC would provide first tests 
of these expectations. Detailed studies will be possible at the Large Hadron-Electron Collider
(LHeC)  currently under discussion at CERN. 

\section*{Acknowledgments}

We  thank our collaborators who were involved in a number of studies reflected in this review: H.~Abramowicz, A.~Freund, W.~Koepf, M. McDermott, T.~Rogers, W.~Vogelsang, C.~Weiss, and M.~Zhalov. 
Our special thanks are due to the Ultraperipheral study group  and especially to R.~Vogt for investigations of the feasibility studies of the small $x$ physics in the ultraperipheral collisions at
 the LHC. 
Over many years we enjoyed numerous illuminating discussions on many of the topics discussed in the review with J.~Bjorken, S.~Brodsky, L.~McLerran, A.~Mueller, and R.~Venugopalan.  
We hope to continue these discussions for years to come.

We thank J.~Collins for the collaboration and discussions of the QCD factorization
theorems for hard diffractive processes.

We thank the H1 collaboration and Springer for allowing us to reproduce figures from
Refs.~\cite{Aktas:2006hy} and \cite{Aktas:2006hx}.

This research was partially supported by the BSF (LF and MS) and DOE (MS).

Authored by Jefferson Science Associates, LLC under U.S. DOE Contract No. DE-AC05-06OR23177. The U.S. Government retains a non-exclusive, paid-up, irrevocable, world-wide license to publish or reproduce this manuscript for U.S. Government purposes.

\bibliographystyle{h-elsevier}
\bibliography{fgs_review_main_final}

\end{document}